\documentclass[british,aps,prd,nofootinbib,showkeys,showpacs,T1,hyphens,spaces,a4paper,reprint,floatfix]{revtex4-2}%
\usepackage[T1]{fontenc}%
\usepackage{microtype}%
\usepackage[british]{babel}%
\usepackage{textgreek}%
\usepackage{setspace}%
\usepackage{enumerate}
\usepackage{csquotes}%
\usepackage[super]{nth}
\usepackage[normalem]{ulem} %

\usepackage{float}
\usepackage{graphicx}
\usepackage[inline, final]{showlabels}
\usepackage[most]{tcolorbox}
\tcbset{%
    enhanced%
    }
    \tcbuselibrary{raster} 
\usepackage{tabularray}
\UseTblrLibrary{booktabs}
\usepackage[title,titletoc]{appendix}%
\usepackage[caption=false]{subfig}
\usepackage{flafter} %

\usepackage[overload]{empheq} %
\usepackage{amssymb}%
\usepackage{bm}

\usepackage[separate-uncertainty=true,multi-part-units=single]{siunitx}
\DeclareSIUnit\parsec{pc}
\usepackage{physics}
\AtBeginDocument{\RenewCommandCopy\qty\SI}

\usepackage[nospace]{varioref}%
\usepackage[pdfencoding=auto,psdextra,unicode=true,pdfusetitle,bookmarks=true,bookmarksnumbered=false,bookmarksopen=false, breaklinks=true,pdfborder={0 0 0},backref=false,colorlinks=false] {hyperref}%
\usepackage{cleveref}%
\crefname{paragraph}{para.}{paras.}
\Crefname{paragraph}{Paragraph}{Paragraphs}
\usepackage{ellipsis} %
\usepackage{orcidlink} %

\usepackage[nonumberlist,nopostdot,numberedsection=autolabel]{glossaries-extra}%
\setabbreviationstyle[acronym]{long-short}
\makenoidxglossaries
\newacronym{rhs}{RHS}{Right-Hand Side}
\newacronym{lhs}{LHS}{Left-Hand Side}
\newacronym{cmb}{CMB}{Cosmic Microwave Background}
\newacronym{wgc}{WGC}{Weak Gravity Conjecture (\cref{sec:gravity})}
\newacronym{adscft}{AdS/CFT}{Anti\textendash de~Sitter/Conformal Field Theory}
\newacronym[\glsshortpluralkey={CFTs},longplural={Conformal Field Theories}]{cft}{CFT}{Conformal Field Theory}
\newacronym{ads}{AdS}{Anti\textendash de~Sitter}
\newacronym{dsc}{dSC}{de~Sitter Conjecture (\cref{sec:deSitter})}
\newacronym{adsdc}{AdSDC}{Anti\textendash de~Sitter Distance Conjecture (\cref{sec:AdSDC})}
\newacronym[\glsshortpluralkey={EoS},longplural={Equations of State}]{eos}{EoS}{Equation of State}
\newacronym[\glsshortpluralkey={EoM},longplural={Equations of Motion}]{eom}{EoM}{Equation of Motion}
\newacronym{bao}{BAO}{Baryon Acoustic Oscillations}
\newacronym{bhedc}{BHEDC}{Black Hole Entropy Distance Conjecture}
\newacronym{flrw}{FLRW}{Friedmann\textendash Lemaître\textendash Robertson\textendash Walker}
\newacronym[name={$\bm{\Lambda}$CDM}]{lcdm}{$\Lambda$CDM}{$\Lambda$ Cold Dark Matter}
\newacronym{geodes}{GEODEs}{GEneric Objects of Dark Energy}
\newacronym{svt}{SVT}{Scalar\textendash Vector\textendash Tensor}
\newacronym{de}{DE}{Dark Energy}
\newacronym{ede}{EDE}{Early Dark Energy}
\newacronym{dm}{DM}{Dark Matter}
\newacronym{tcc}{TCC}{Trans-Planckian Censorship Conjecture (\cref{sec:tpcc})}
\newacronym{trgb}{TRGB}{Tip of the Red Giant Branch}
\newacronym{bbn}{BBN}{Big Bang Nucleosynthesis}
\newacronym{aic}{AIC}{Akaike Information Criterion}
\newacronym{bic}{BIC}{Bayesian Information Criterion}
\newacronym{kids}{KiDS}{Kilo Degree Survey}
\newacronym{boss}{BOSS}{Baryon Oscillation Spectroscopic Survey}
\newacronym{2dflens}{2dFLenS}{2-degree Field Lensing Survey}
\newacronym{des}{DES}{Dark Energy Survey}
\newacronym{viking}{VIKING}{Visible and Infrared Survey Telescope for Astronomy (VISTA) Kilo-Degree Infrared Galaxy Survey}
\newacronym{bps}{BPS}{Bogomol'nyi\textendash Prasad\textendash Sommerfield}
\newacronym{aligo}{aLIGO}{advanced Laser Interferometer Gravitational-Wave Observatory}
\newacronym{lisa}{LISA}{Laser Interferometer Space Antenna}
\newacronym{shoes}{SH0ES}{Supernovae and $H_0$ for the Dark Energy Equation of State}
\newacronym{eboss}{eBOSS}{extended Baryon Oscillation Spectroscopic Survey}
\newacronym{holicow}{H0LiCOW}{$H_0$ Lenses in COSmological MOnitoring of GRAvItational Lenses (COSMOGRAIL)'s Wellspring}
\newacronym[\glsshortpluralkey={SNe},longplural={Supernovae}]{sn}{SN}{Supernova}
\newacronym{calpriorsnia}{CalPriorSNIa}{effective Calibration Prior on the absolute magnitude of Type Ia Supernovae}
\newacronym[\glsshortpluralkey={QFTs},longplural={Quantum Field Theories}]{qft}{QFT}{Quantum Field Theory}
\newacronym{gr}{GR}{General Relativity}
\newacronym{qg}{QG}{Quantum Gravity}
\newacronym[\glsshortpluralkey={EFTs},longplural={Effective Field Theories}]{eft}{EFT}{Effective Field Theory}
\newacronym{nec}{NEC}{Null Energy Condition}
\newacronym{ceb}{CEB}{Covariant Entropy Bound}
\newacronym{kklt}{KKLT}{Kachru\textendash Kallosh\textendash Linde\textendash Trivedi \cite{kachru_sitter_2003}}
\newacronym{pbh}{PBH}{Primordial Black Hole}
\newacronym{sm}{SM}{Standard Model}
\newacronym{kk}{KK}{Kaluza\textendash Klein}
\newacronym{lhc}{LHC}{Large Hadron Collider}
\newacronym{cy}{CY}{Calabi\textendash Yau}
\newacronym{slwgc}{sLWGC}{sub-Lattice Weak Gravity Conjecture (\cref{p:sLWGC})}
\newacronym{rfc}{RFC}{Repulsive Force Conjecture (\cref{s:rfc})}
\newacronym{twgc}{TWGC}{Tower Weak Gravity Conjecture (\cref{p:TWGC})}
\newacronym{lwgc}{LWGC}{Lattice Weak Gravity Conjecture (\cref{p:LWGC})}
\newacronym{lep}{LEP}{Large Electron\textendash Positron Collider}
\newacronym{ds}{dS}{de~Sitter}
\newacronym{ccc}{CCC}{Cosmic Censorship Conjecture}
\newacronym{qed}{QED}{Quantum Electrodynamics}
\newacronym{adm}{ADM}{Arnowitt\textendash Deser\textendash Misner \cite{arnowitt_coordinate_1961,arnowitt_dynamical_1959}}
\newacronym{bh}{BH}{Black Hole}
\newacronym{rnds}{RNdS}{Reissner\textendash Nordström\textendash de~Sitter}
\newacronym{rn}{RN}{Reissner\textendash Nordström}
\newacronym{ir}{IR}{Infrared}
\newacronym{uv}{UV}{Ultraviolet}
\newacronym{knp}{KNP}{Kim\textendash Nilles\textendash Peloso \cite{kim_completing_2005}}
\newacronym{etw}{ETW}{End of The World}
\newacronym{cp}{CP}{Charge conjugation Parity}
\newacronym{qcd}{QCD}{Quantum Chromo Dynamics}
\newacronym{gsc}{GSC}{Gravitino Swampland Conjecture (\cref{sec:gravitino})}
\newacronym{cc}{CC}{Completeness Conjecture (\cref{sec:complet})}
\newacronym{ssc}{SSC}{Species Scale Conjecture (\cref{s:ssc})}
\newacronym{ep}{EP}{Emergence Proposal (\cref{sec:emerge})}
\newacronym{flb}{FLB}{Festina Lente Bound (\cref{sec:Festina})}
\newacronym{dc}{DC}{Distance Conjecture (\cref{sec:distance})}
\newacronym{sec}{SEC}{Strong Energy Condition}
\newacronym{dbi}{DBI}{Dirac\textendash Born\textendash Infeld}
\newacronym{ew}{EW}{Electro Weak}
\newacronym{desi}{DESI}{Dark Energy Spectroscopic Instrument \cite{desi_collaboration_desi_2024,desi_collaboration_papers_2024}}
\newacronym{nanograv}{NANOGrav}{North American Nanohertz Observatory for Gravitational Waves}
\newacronym{cpl}{CPL}{Chevallier\textendash Polarski\textendash Linder \cite{chevallier_accelerating_2001,linder_exploring_2003}}
\newacronym[longplural={Grand Unified Theories}]{gut}{GUT}{Grand Unified Theory}
\newacronym{bd}{BD}{Bunch\textendash Davies}
\newacronym{gb}{GB}{Gauss\textendash Bonnet}
\newacronym{pde}{PDE}{Partial Differential Equation}
\newacronym{tpc}{TPC}{Tadpole Conjecture (\cref{s:tadpole})}
\newacronym{dgkt}{DGKT}{DeWolfe\textendash Giryavets\textendash Kachru\textendash Taylor \cite{dewolfe_type_2005}}
\newacronym{lss}{LSS}{Large-Scale Structure}
\newacronym{gw}{GW}{Gravitational Wave}
\newacronym{lvs}{LVS}{Large Volume Scenario}

\newcommand*{\defeq}{\mathrel{\vcenter{\baselineskip0.5ex \lineskiplimit0pt
                     \hbox{\scriptsize.}\hbox{\scriptsize.}}}%
                     =} %

\makeatletter
\g@addto@macro\bfseries{\boldmath}  %

\def\l@subsubsection#1#2{}
\makeatother

\newcommand*\DAlambert{\mathop{{}\Box}\nolimits} %

\DeclareMathOperator{\sign}{sgn}

\begin{document}
\title{Hitchhiker's Guide to the Swampland: The Cosmologist's Handbook to the string-theoretical Swampland Programme}%
\author{\orcidlink{0000-0002-2908-8638}Kay Lehnert}
\email{k.lehnert@protonmail.com}
\affiliation{Department of Physics, National University of Ireland, Maynooth}
\date{\today}

\begin{abstract}
	String theory has strong implications for cosmology: it tells us that we cannot have a cosmological constant, that single-field slow-roll inflation is ruled out, and that black holes decay. We elucidate the origin of these statements within the string-theoretical swampland programme.
The swampland programme is generating a growing body of insights that have yet to be incorporated into cosmological models.
Taking a cosmologist’s perspective, we highlight the relevance of swampland conjectures to black holes, dark matter, dark energy, and inflation, including their implications for scalar fields such as quintessence and axions.
Our goal is to inspire cosmological model builders to examine the compatibility of effective field theories with quantum gravitational UV completions and to address outstanding cosmological tensions such as the Hubble tension.
This comprehensive literature review presents clear definitions, cosmological implications, and the current status\,\textemdash\,including evidence and counterexamples\,\textemdash\,of the following swampland conjectures:
the \textit{anti\textendash de~Sitter distance conjecture (AdSDC)},
the \textit{completeness conjecture (CC)},
the \textit{cobordism conjecture},
the \textit{de~Sitter conjecture (dSC)},
the \textit{swampland distance conjecture (SDC)},
the \textit{emergence proposal (EP)},
the \textit{Festina Lente Bound (FLB)},
the \textit{finite number of massless fields conjecture} (or \textit{finite flux vacua conjecture (FFV)}),
the \textit{no global symmetries conjecture},
the \textit{no non-supersymmetric theories conjecture},
the \textit{non-negative null energy condition conjecture},
the \textit{positive Gauss\textendash Bonnet term conjecture},
the \textit{species scale conjecture},
the \textit{gravitino swampland conjecture (GSC)},
the \textit{tadpole conjecture},
the \textit{tameness conjecture},
the \textit{trans\textendash Planckian censorship conjecture (tPCC/TCC)},
the \textit{unique geodesic conjecture},
and the \textit{weak gravity conjecture (WGC)}, including the \textit{repulsive force conjecture (RFC)}.

\end{abstract}

\maketitle

\tableofcontents
\section{Introduction and Summary}\label{sec:intro}%
Modern-day physics is mainly described by two theoretical frameworks: \gls{gr} and \gls{qft}. A framework that is hoped to host a unified theory of \gls{qg} is string theory. String theory offers a plethora of \glspl{eft}.\footnote{
    String theory offers the notion of a theory space. Each \gls{eft} represents a position in this space, parametrised by moduli\,\textemdash\,massless scalar fields without a potential, yet with an \gls{eom} \cite{susskind_anthropic_2003}.
    Since we do not observe massless scalar fields in our Universe, it might either be the case that there are mechanisms that for instance break supersymmetry, which leaves us with massive scalar fields, or that the moduli only describe the curled-up extra-dimensions and have no counterpart in our 4D spacetime \cite{dabholkar_string_1999}.
    The resulting vacua differ from each other with regard to the shape, size \cite{susskind_anthropic_2003}, amount of supersymmetry \cite{adams_string_2014}, and content (e.g. branes and fluxes) of the internal spaces, and correspond to universes with different phenomenological properties \cite{ruiz_morse-bott_2025}.
    \enquote{The beauty of the supermoduli–space point of view is that there is only one theory but many solutions which are characterised by the values of the scalar field moduli} \cite{susskind_anthropic_2003}.
}
Not all \glspl{eft} consistently merge \gls{qft} and gravity. A valid theory of \gls{qg} is self-consistent at the low-energy \gls{ir} scale as well as at the high-energy \gls{uv} scale. It is expected that the search for such a theory can be guided by certain rules\,\textemdash\,rules that are obeyed by every valid theory of \gls{qg}. 
The set of these rules divides the theory space within string theory into a theory \textit{landscape} of valid theories and a \textit{swampland} of theories that might seem valid at low energy-scales, but break down at high energy-scales \cite{palti_swampland_2019,brennan_string_2018,vafa_cosmic_2019,vafa_string_2005,agmon_lectures_2023,schneider_trans-planckian_2021}.
The dividing rules obtained from string theory are coined \textit{swampland conjectures} \cite{brennan_string_2018}.\footnote{
    Starting to refine or base a new theory on conjectures is not unknown in physics, e.g. the formulation of the Heisenberg uncertainty principle, Bohr's correspondence principle \cite{cribiori_gravitino_2021} or the Copenhagen interpretation of quantum mechanics are foremost based on conjectures.
    }
Early work on no-go theorems in string theory was tough, so tough that \citet{kachru_bounds_2006} even reasoned that there might be a no-go theorem for no-go theorems.
Yet, the hurdles have been overcome, giving rise to the swampland programme.\footnote{
    No-go theorems about \gls{qg} were around earlier, but the \textit{Swampland Programme} was probably initiated when \citet{vafa_string_2005} coined the name. Now, it is an actively growing field.
    \citet{kinney_swampland_2021} derives, jokingly and assuming 100 kilobits of entropy per conjecture, an upper bound of \num{e117} conjectures, using the Bekenstein\textendash Hawking bound. We are not entirely sure if our review article will decrease the entropy by bringing more order to the swampland programme, or increase the entropy by adding to the confusion.
}

One approach to derive swampland conjectures is to use vacua of string theory as experimental data and derive criteria that are satisfied by all well-understood vacua. 
Criteria based on this approach are theory-driven, rather than proven by microscopic physics \cite{palti_swampland_2019}. The fewer vacua fulfil a conjecture, the weaker is the evidence for this conjecture.
Another approach is to use arguments from low-energy \gls{qg} and generalise them for higher energies \cite{palti_swampland_2019}. This includes findings and considerations regarding holography, \glspl{bh},\footnote{
    See \citet{afshordi_black_2024} for a review of historical and still open questions about \glspl{bh}.
    }
or \gls{qft} on curved space-times \cite{de_biasio_geometric_2023}.\footnote{
    \citet{conlon_putting_2019} suggest a strong relation between \gls{ads} \glspl{eft} and \glspl{cft}: They conjecture that swampland constraints on consistent \gls{ads} theories are equivalent to bootstrap constraints on the dual \gls{cft}. We will frequently refer to \glspl{cft} and use findings as supportive evidence in the following. \citet{heckman_fine_2019} conjecture that nearly all \glspl{cft} belong to the swampland. That indicates that the swampland of \glspl{eft} must be a lot bigger than the landscape.
}
Furthermore, it is also possible to derive criteria directly from microscopic physics \cite{palti_swampland_2019}. Ideally, all three approaches lead to consistent criteria. 

The conjectures themselves may expose internal inconsistencies within string theory, and many are expected to hold beyond string theory.\footnote{
    \citet{eichhorn_absolute_2024} propose the concept of the \textit{absolute} and \textit{relative} swampland: while a relative swampland produces theories that are incompatible with one set of tools, e.g. string theory or asymptotic safety \cite{bonanno_asymptotically_2017,eichhorn_asymptotically_2019,reichert_lecture_2020,pawlowski_quantum_2021,eichhorn_status_2022},
    the absolute swampland is the intersection of the relative swamplands, i.e. it contains only those \glspl{eft} that are incompatible with every Ansatz for \gls{qg}. However, if the cobordism conjecture (\cref{sec:cobordism}) is true, then there is only one theory of \gls{qg}, and no distinction between a relative and an absolute swampland has to be made. Yet, before we succeed in constructing this \gls{qg}, such a distinction can be a guiding principle: a swampland conjecture becomes more likely if it is supported by many \gls{qg} Ansätze.
    Furthermore, according to \textit{string universality} respectively the \textit{string lamppost principle} \textit{everything that can happen (in agreement with \gls{qg}) does happen in string theory}, i.e. string theory provides everything that \gls{qg} permits, and the string landscape corresponds to the \gls{qg} landscape \cite{agmon_lectures_2023,van_beest_lectures_2022} (cf. \cite{dijkgraaf_topological_2005}).
}
Therefore, the swampland conjectures might be used to debunk string theory \cite{blumenhagen_large_2018,barrau_string_2021} and as grounding rules for any kind of \gls{qg}, besides string theory \cite{cicoli_string_2023}.\footnote{\label{f:low-d_gravity}
    The swampland conjectures are not applicable if there are only 2 spacetime dimensions, as gravity is not dynamical, and their applicability to 3-dimensional theories is questionable, as gravity has no propagating degrees of freedom there either \cite{van_beest_lectures_2022}.
}

Furthermore, since \gls{qg} is essential to understand, among other things, the ((very) early) Universe and \glspl{bh}, the swampland programme opens the cosmological testing ground for \gls{qg} \cite{heisenberg_model_2021,barrau_testing_2017}.
While summaries and introductions to the string-theoretical tools used to derive and understand the swampland conjectures are, among others, presented by \citet{prieto_moduli_2024,quirant_aspects_2022,polchinski_string_1998-V1,polchinski_string_1998-V2,blumenhagen_basic_2013,agmon_lectures_2023,hebecker_lectures_2023},
and the swampland programme is for instance reviewed by \citet{vafa_string_2005,brennan_string_2018,palti_swampland_2019,agmon_lectures_2023,grana_swampland_2021,van_beest_lectures_2022,gomez_gravity_2019,vafa_swamplandish_2024,cribiori_supergravity_2023,andriot_web_2020,van_riet_beginners_2023,silk_swampland_2022,yamazaki_swampland_2019},
this work offers a comprehensive overview of the current\footnote{
    All reviewed literature was published before the year 2025.
}
state of the swampland programme with a focus on its \textit{cosmological} implications.\footnote{
    For an overview of string cosmology, we'd like to refer to the work by \citet{cicoli_string_2023}, which includes a brief chapter on swampland conjectures. However, their focus lies on the opportunities the string theoretical landscape provides for cosmology, rather than on the excluding power of the swampland conjectures, i.e. they look into \enquote{what constructions enables string theory}, rather than into \enquote{what cannot be done in \gls{qg}}.
}
Each chapter presenting a different swampland conjecture starts with a general definition of the conjecture, followed by implications for cosmology, before we conclude with more general, often string-theoretical, remarks and evidence for the conjecture.
We must caution the reader: The swampland conjectures are typically statements about what happens at the infinite boundary of moduli space, yet cosmology happens in the bulk. The applicability of the conjectures has therefore to be judged with scepticism.\footnote{
    \citet{klaewer_quantum_2021} express the notion that the conjectures should still be applicable in the bulk, under the condition that the necessary quantum corrections are taken into account.
    }
Nevertheless, the swampland conjectures can act as inspiration for model building, since the conjectures \textit{might} still apply in the bulk. 
For model building, we consider it easier to start with a strong statement to find tensions that hint at overly strong assumptions, rather than building models based on conjectures that are, from a model builder's perspective, relatively loose. We leave it up to the reader to make their own judgment (guided by the remarks we make in the text), to which degree they would like to implement the statements into their cosmological modelling. 
However, the opposite direction doesn't work: current cosmological observations cannot be used to rule out the swampland conjectures or even string theory in general, since cosmology, being far in the bulk, cannot probe the boundary of moduli space, where most of the conjectures are formulated and best tested.\footnote{
    However, \citet{halverson_tasi_2018} notes that the observation of a feature that is firmly determined to be in the swampland would indeed rule out string theory.
}

\begin{figure*}
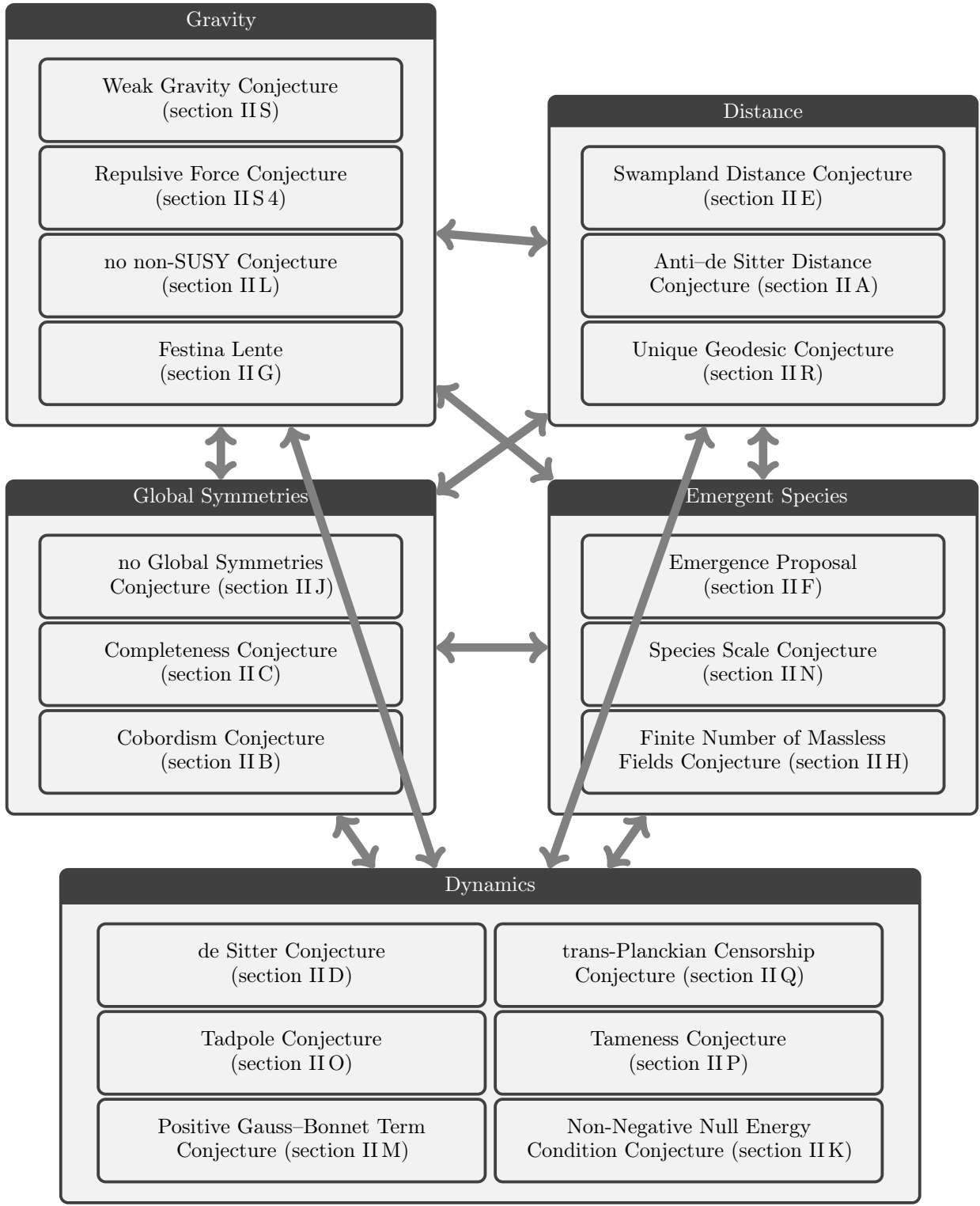

    \centering
    \begin{tcolorbox}[title=Gravity, width=.4\linewidth, nobeforeafter, remember as=NodeGravity]
        \begin{tcolorbox}[nobeforeafter]
            Weak Gravity Conjecture (\cref{sec:gravity})
        \end{tcolorbox}
        \hfill
        \begin{tcolorbox}[nobeforeafter]
            Repulsive Force Conjecture (\cref{s:rfc})
        \end{tcolorbox}
        \hfill
        \begin{tcolorbox}[nobeforeafter]
            no non-SUSY Conjecture (\cref{sec:nononSUSY})
        \end{tcolorbox}
        \hfill
        \begin{tcolorbox}[nobeforeafter]
            Festina Lente\\ (\cref{sec:Festina})
        \end{tcolorbox} 
    \end{tcolorbox}
    \hfill\vspace{12pt}
    \begin{tcolorbox}[title=Distance, width=.4\linewidth, nobeforeafter, remember as=NodeDistance]
        \begin{tcolorbox}[nobeforeafter]
            Swampland Distance Conjecture (\cref{sec:distance})
        \end{tcolorbox}
        \hfill
        \begin{tcolorbox}[nobeforeafter]
            Anti\textendash de~Sitter Distance Conjecture (\cref{sec:AdSDC})
        \end{tcolorbox} 
        \hfill
        \begin{tcolorbox}[nobeforeafter]
            Unique Geodesic Conjecture (\cref{s:UGC})
        \end{tcolorbox} 
    \end{tcolorbox}
    \hfill\vspace{12pt}
    \begin{tcolorbox}[title=Global Symmetries, width=.4\linewidth, nobeforeafter, remember as=NodeGlobal]
        \begin{tcolorbox}[nobeforeafter]
            no Global Symmetries Conjecture (\cref{sec:nGSym})
        \end{tcolorbox}
        \hfill
        \begin{tcolorbox}[nobeforeafter]
            Completeness Conjecture (\cref{sec:complet})
        \end{tcolorbox}
        \hfill
        \begin{tcolorbox}[nobeforeafter]
            Cobordism Conjecture (\cref{sec:cobordism})
        \end{tcolorbox} 
    \end{tcolorbox}
    \hfill\vspace{12pt}
    \begin{tcolorbox}[title=\ Emergent Species, width=.4\linewidth, nobeforeafter, remember as=NodeSpecies]
        \begin{tcolorbox}[nobeforeafter]
            Emergence Proposal (\cref{sec:emerge})
        \end{tcolorbox}
        \hfill
        \begin{tcolorbox}[nobeforeafter]
            Species Scale Conjecture (\cref{s:ssc})
        \end{tcolorbox} 
        \hfill
        \begin{tcolorbox}[nobeforeafter]
            Finite Number of Massless Fields Conjecture (\cref{s:fnomf})
        \end{tcolorbox} 
    \end{tcolorbox}
    \hfill\vspace{12pt}
    \begin{tcolorbox}[title=Dynamics, width=0.8\linewidth, nobeforeafter, remember as=NodeDynamics, 
    raster columns=2
    ]
        \begin{tcolorbox}[nobeforeafter, width=0.49\linewidth]
            de~Sitter Conjecture\\ (\cref{sec:deSitter})
        \end{tcolorbox}
        \begin{tcolorbox}[nobeforeafter, width=0.49\linewidth]
            trans-Planckian Censorship Conjecture (\cref{sec:tpcc})
        \end{tcolorbox} 
        \begin{tcolorbox}[nobeforeafter, width=0.49\linewidth]
            Tadpole Conjecture\\ (\cref{s:tadpole})
        \end{tcolorbox}
        \begin{tcolorbox}[nobeforeafter, width=0.49\linewidth]
            Tameness Conjecture\\ (\cref{sec:tame})
        \end{tcolorbox}
        \begin{tcolorbox}[nobeforeafter, width=0.49\linewidth]
            Positive Gauss\textendash Bonnet Term Conjecture (\cref{s:pGBTC})
        \end{tcolorbox}
        \begin{tcolorbox}[nobeforeafter, width=0.49\linewidth]
            Non-Negative Null Energy Condition Conjecture (\cref{s:nnNECC})
        \end{tcolorbox}
    \end{tcolorbox}\hspace*{0.1\linewidth}

    \tikz[overlay, remember picture] \draw[<->, line width=4pt, gray] (NodeGravity)--(NodeDistance);
    \tikz[overlay, remember picture] \draw[<->, line width=4pt, gray] (NodeGravity)--(NodeDynamics);
    \tikz[overlay, remember picture] \draw[<->, line width=4pt, gray] (NodeDistance)--(NodeDynamics);
    \tikz[overlay, remember picture] \draw[<->, line width=4pt, gray] (NodeGravity)--(NodeGlobal);
    \tikz[overlay, remember picture] \draw[<->, line width=4pt, gray] (NodeGlobal)--(NodeDynamics);
    \tikz[overlay, remember picture] \draw[<->, line width=4pt, gray] (NodeDistance)--(NodeGlobal);
    \tikz[overlay, remember picture] \draw[<->, line width=4pt, gray] (NodeGravity)--(NodeSpecies);
    \tikz[overlay, remember picture] \draw[<->, line width=4pt, gray] (NodeDynamics)--(NodeSpecies);
    \tikz[overlay, remember picture] \draw[<->, line width=4pt, gray] (NodeGlobal)--(NodeSpecies);
    \tikz[overlay, remember picture] \draw[<->, line width=4pt, gray] (NodeDistance)--(NodeSpecies); 
    \caption{The various conjectures are connected with each other, as we highlight in particular in \cref{sec:Conj-Relation}.}
    \label{fig:conjecture_relations}
\end{figure*}

In the following, we introduce our notation in \cref{s:notation}, before be present a very brief and pointed executive summary in \cref{s:exec_summary}.
We conclude this section with directions for future research in \cref{s:outlook}.
All conjectures are presented in more detail in \cref{sec:swamp}.
The conjectures are tightly connected with each other, spanning a swampland web, which we highlight in \cref{fig:conjecture_relations,sec:Conj-Relation}.

\subsection{Notation}\label{s:notation}
All abbreviations used in the text are accessible in the glossary at the end. All definitions and variables are gathered in \cref{a:sym}.

We mostly work in natural Planck units with $c=1=\hbar$. Even though $M_\textnormal{P;d=4}=1$ holds as well, we generally explicitly state it nonetheless. On the one hand, the Planck mass has a different value if we are not working in 4 dimensions. On the other hand, Planck-suppressed operators are no longer suppressed in the \gls{uv}, so the order of suppression becomes relevant.

Many of the swampland conjectures are plagued with not yet defined constants, often of $\order{1}$. Different authors sometimes use the same symbol for different conjectures, and some symbols like $\alpha$ are generally oversubscribed in the physics literature. To better distinguish between the different constants, and to have a consistent notation, we chose different letters and a Fraktur typeface for all the swampland constants:
\begin{description}
    \item[$\mathfrak{a}$] \gls{dc} constant \\
    \item[$\mathfrak{b}$] \gls{bhedc} constant (\cref{p:AdSDC_BH}) \\
    \item[$\mathfrak{d}$] \gls{adsdc} constant \\
    \item[$\mathfrak{g}$] Gravitino \gls{dc} constant (\cref{p:DC_Gravitinos}) \\
    \item[$\mathfrak{s}_1$] \gls{dsc} constant \\
    \item[$\mathfrak{s}_2$] refined \gls{dsc} constant \\
    \item[$\mathfrak{t}$] Flux-tadpole constant (\cref{s:tadpole})
\end{description}
Furthermore, there are three more symbols we'd like to specifically mention:
\begin{description}
    \item[$\mathfrak{c}$] a constant \\
    \item[$\mathfrak{p}$] a positive constant \\
    \item[$\mathfrak{l}$] an $\order{1}$ constant
\end{description}
These will be used frequently in the text. However, their value from one part of the text cannot be compared with their value from other sections, paragraphs, or even sentences: $\mathfrak{c}$ just means \textit{a} constant\,\textemdash\,any constant\,\textemdash\,and it will not be the same constant meant by $\mathfrak{c}$ somewhere else in the text. The same holds for $\mathfrak{p}\geq0$ and $\mathfrak{l}\sim\order{1}$.

\subsection{Executive Summary}\label{s:exec_summary}

\begin{description}
    \item[AdSDC] The \gls{adsdc} $m\lesssim\abs{V}^\mathfrak{d} M_\textnormal{P}^{1-2\mathfrak{d}}$ states that \gls{ads} and \gls{ds} spaces are infinitely far apart, i.e. one cannot smoothly transition from one into the other by passing through Minkowski space. This puts $\Lambda_\textnormal{s}$CDM models with a sign switch in the cosmological constant in the swampland. Furthermore, the lightest neutrino is a Dirac neutrino, and not a Majorana neutrino, if we currently are in a \gls{ds} phase.
    \item[Cobordism Conjecture] The bordism group of \gls{qg} is trivial. This implies that there are boundary-ending spacetimes, i.e. a beginning of time (or an end of it) is perfectly compatible with \gls{qg}. Furthermore, \glspl{bh} have to completely evaporate and cannot leave behind a stable remnant.
    \item[CC] The \gls{cc} states that the charge spectrum of a theory is complete, i.e. every representation of the gauge group must exist. This implies that magnetic monopoles exist.
    \item[dSC] The \gls{dsc} has strong implications for cosmology, as it sets bounds on the derivatives of scalar fields of $\abs{\nabla V}\geq V\mathfrak{s}_1/M_\textnormal{P}$ and $\text{min}\left(\nabla^i\nabla_j V\right)\leq-\mathfrak{s}_2V/M_\textnormal{P}^2$, which rules out
    a cosmological constant, as well as simple models of
    single-field inflation,
    slow-roll inflation, and
    eternal inflation.
    \item[DC]  The \gls{dc} limits scalar field ranges to $\Delta\phi\lesssim\order{10}M_\textnormal{P}$, respectively, predicts a tower of light states with exponentially suppressed masses otherwise. This rules out large-field inflation.
    \item[EP] The \gls{ep} states that there are no kinetic terms in the \gls{uv}. All dynamics in the \gls{ir} emerge due to loop corrections. It also states that the lightest towers that emerge, e.g. as predicted by the \gls{dc} or the \gls{adsdc}, are either \gls{kk} towers and therefore related to a decompactification of a dimension, or string towers, related to a fundamental string becoming tensionless. The lightest neutrino is likely a Dirac neutrino.
    \item[FLB] The \gls{flb} sets a lower bound on the mass of charged particles in \gls{ds} space of $m^4\gtrsim6\left(gqM_\textnormal{P} H\right)^2$. This has far-reaching consequences, such as potentially limiting the energy scale of inflation to \SI{e5}{\giga\electronvolt}, the cosmological constant to $\Lambda_\textnormal{cc}\lesssim m^4/4\pi\alpha\sim\num{3e-89}$, or giving millicharged particles a hard time. It is inspired by the observations that in \gls{ds} space, the area of the cosmological horizon is finite, which limits the area of (charged) \glspl{bh}.
    \item[Finiteness] the Finite Number of Massless Fields Conjecture (\cref{s:fnomf}) states that the number of massless fields in an \gls{eft} coupled to gravity is finite. This can be used to rule out the anthropic principle if there is \gls{de}\textendash\gls{dm} interaction. To invoke the anthropic principle would require so much fine-tuning that there are not enough string theory compactifications to do so.
    \item[GSC] The \gls{gsc} states that gravitinos have a positive-definite sound speed of $c_\textnormal{s}^2>0$.
    \item[no Global Symmetries Conjecture] The no Global Symmetries Conjecture (\cref{sec:nGSym}) states that there are no continuous global symmetries in \gls{qg}, yet, \glspl{eft} are allowed to have such symmetries. This requires \glspl{bh} to completely evaporate, as stable remnants would represent global charges.
    \item[nnNECC] The non-negative \gls{nec} Conjecture (\cref{s:nnNECC}) $T_{\mu\nu}l^\mu l^\nu\geq0$ states that every 4d \gls{eft} has to satisfy the \gls{nec}, i.e. parallel light rays never cross. The \gls{nec} does not necessarily hold in higher-dimensional theories, as a hierarchy of higher-order terms can lead to cancellation, such that no pathologies occur. This imposes a constraint on the string coupling, which, in turn, corresponds to the \gls{nec} in 4 flat or closed large dimensions. This rules out various wormhole solutions, time-machines, warp-drives, and also models with a cosmological bounce, \gls{de} with $\omega<-1$ (e.g. phantom fields or some models of $k$-essence), as well as phases with growing Hubble parameter, e.g. superinflation.
    \item[nnSUSYC] The no non-Supersymmetric Theories Conjecture (\cref{sec:nononSUSY}) states that there are no non-supersymmetric stable \gls{ads} vacua in 4 dimensions. The conjecture limits the mass of the lightest neutrino to $m_{\nu_1}\lesssim\Lambda_\textnormal{cc}^{1/4}$.
    \item[pGBTC] The positive \gls{gb} Term Conjecture (\cref{s:pGBTC}) $R_{\mu\nu\rho\sigma}R^{\mu\nu\rho\sigma}-4R_{\mu\nu}R^{\mu\nu}+R^2>0$ states that the \gls{gb} term is positive. In string theory, the \gls{gb} term corresponds to the inverse string tension\,\textemdash\,which cannot be negative. A negative contribution might reveal naked singularities during Hawking evaporation of \glspl{bh}, violating the strong \gls{ccc}, the cobordism conjecture, the no global symmetries conjecture, and the \gls{wgc}. The effect of the \gls{gb} term on the Hubble parameter also affects \gls{de}, as well as scalar and tensor mode dispersion, which in turn links it to inflation and to the \gls{tcc} and the \gls{dsc}, as well as to the \gls{dc}.
    \item[SSC] The \gls{ssc} $\Lambda_\textnormal{S}=M_{\textnormal{P};d}/N_\textnormal{S}^{\frac{1}{d-2}}$ says that there is a cutoff of validity for any \gls{eft} way below the Planck scale, as new dynamics appear, e.g. in the form of higher-derivative operators. This has implications for inflation (and early universe cosmology, as the number of relativistic degrees of freedom changes: $g_*$ would be higher than in the concordance model).
    \item[TPC] The \gls{tpc} $q_{D3}^\text{stabilisation}>\mathfrak{t}n_\text{mod}$ acts as supporting evidence for various other swampland conjectures.
    \item[Tameness Conjecture] The Tameness Conjecture (\cref{sec:tame}) states that an \gls{eft} has to be tame, which means that functions cannot be too wild, which rules out that functions like $\sin(1/x)$ are part of potentials, couplings, partition functions, metrics, and anything that would describe the geometry of an \gls{eft}. It also means that there can only be a finite number of fields or infinite towers of states. Ergo, it supports finiteness in \gls{qg}. 
    \item[TCC] The \gls{tcc} $a_\textnormal{f}/a_\textnormal{i}<M_\textnormal{P}/H_\textnormal{f}$ says that trans-Planckian modes (modes with a wavelength shorter than the Planck length) cannot be stretched enough to become sub-Hubble, as this would allow classical observations of trans-Planckian physics. This limits the strength of inflation and \gls{de}, either in the form of an upper limit on the number of $e$-foldings accelerated expansion can last, or e.g. in the form of an energy scale for the inflaton. The \gls{tcc} generally predicts a suppression of tensor modes, which means that an observation of tensor modes would either mean the \gls{tcc} is wrong, or that those modes were not produced during inflation, but e.g. due to non-Gausssianities or non-\gls{bd} initial states. It also sets an upper bound on the lifetime of our Universe.
    \item[UGC] The Unique Geodesic Conjecture (\cref{s:UGC}) states that in a moduli space with a given basis for physical observables, there is always a unique geodesic of finite distance between any two points in the moduli space that are not on the boundary of that space. We make the following two proposals for cosmology.
    First, eternal inflation cannot happen.
    Second, the anthropic principle cannot be invoked to explain the smallness of the cosmological constant.
    \item[WGC] The \gls{wgc} $m\leq\sqrt{\left(d-2\right)/\left(d-3\right)}gq\left(M_{\textnormal{P};d}\right)^{\frac{d-2}{2}}$ states that gravity is the weakest force. This indicates that \glspl{bh} must be able to decay. Furthermore, simple models of large-field inflation are in tension with the \gls{wgc}, and photons are likely massless.
\end{description}

\subsection{Outlook}\label{s:outlook}
The swampland programme is an actively growing field. Conjectures are refined, new conjectures are proposed, and some conjectures get proven in various special cases. However, there are no general proofs yet. The string community is actively providing and discussing various angles to find proofs as well as counterexamples to the different conjectures. However, as we highlight in \cref{fig:conjecture_relations,sec:Conj-Relation}, many of the swampland conjectures are neatly interwoven and support each other. We expect that more and tighter relations will be discovered, maybe even leading to unifications of some conjectures.
A big question for the collaboration between the string and the cosmology community remains: which and to what degree are the swampland conjectures applicable in the bulk, where cosmology takes place? Until this question is answered, the swampland conjectures can nonetheless serve as inspiration for model building.
What is there to do for the cosmology community is to keep an open ear to the string community, assessing their findings, and trying to incorporate the findings into cosmological models.
In particular, findings for inflation and \gls{de} are rich and can often directly be implemented, such as constraints on scalar field ranges coming from the \gls{dc} or on derivatives coming from the \gls{dsc} and the \gls{tcc}. Future \gls{cmb} observations will help to strengthen the bounds on inflationary models, in particular in combination with the swampland bounds on the tensor-to-scalar ratio $r_\textnormal{ts}$. Furthermore, the recently published \gls{desi} results support dynamic \gls{de}, which is also demanded by swampland conjectures, in particular the \gls{dsc}.
Particle physicists will use the \gls{lhc}, the \gls{lep}, and other experiments to put more stringent constraints on neutrinos and \gls{dm}, which can help, together with the theoretical constraints from the swampland programme, to get a more profound understanding of \gls{bbn} physics, \gls{lss} formation, and dark sector physics.
It will be interesting to see what families of cosmological models survive the observational and theoretical constraints, and also, which conjectures survive the critical assessment of the communities. We agree with \citet{silk_swampland_2022} who conclude that
\enquote{[i]t is inevitable that the various swampland conjectures will be modified, some ruled out by counter-examples, others refined by epicycles, and some, just conceivably, refined by future observations.}
Therefore: don't panic.

\section{Swampland Conjectures}\label{sec:swamp}

\subsection{Anti\textendash de~Sitter Distance Conjecture}\label{sec:AdSDC}%

In the limit $V\rightarrow0$, a family of potentials with \gls{ads} minima is accompanied by an infinite tower of states with mass scale
\begin{equation}
    m\lesssim\abs{V}^\mathfrak{d} M_\textnormal{P}^{1-2\mathfrak{d}},\label{eq:AdSDC}
\end{equation}
with $\mathfrak{d}$ of $\order{1}$ \cite{lust_ads_2019,van_riet_beginners_2023}, and likely $\frac{1}{d}\leq\mathfrak{d}\leq\frac{1}{2}$ \cite{montero_dark_2023,seo_uplift_2023,gonzalo_ads_2021,brennan_string_2018}. $V$ is often taken to be the cosmological constant $\Lambda_\textnormal{cc}$. The case $\mathfrak{d}=1/2$ is called the \textit{strong} \gls{adsdc}, which applies to supersymmetric \gls{ads} vacua and omits scale separation\footnote{
    Scale separation means that the size of the internal dimensions is much smaller than the curvature scale of the \gls{ads} space \cite{montero_pure_2023}. It is conjectured that scale separation cannot be part of supersymmetric \gls{ads} spaces \cite{montero_pure_2023}. A potential counterexample to this strong form is presented by \citet{quirant_aspects_2022} for AdS$_4$ vacua. In 10- and 11-dimensional supergravities, the mass scale of the cosmological constant corresponds to the mass scale of the first \gls{kk} mode: $m_{\Lambda_\textnormal{cc}}\sim m_\text{KK}$ \cite{apruzzi_non-supersymmetric_2021}. Another potential counterexample, without definite conclusion, is investigated by \citet{marchesano_supersymmetric_2021}.
    Furthermore, \citet{van_hemelryck_weak_2024} presents a scale-separated model where the 7-dimensional manifold has codimension-4 singularities that are expected yet not proven to be resolvable.
    \citet{junghans_o-plane_2020} presents a \gls{dgkt} model with scale separation despite being supersymmetric, which violates the strong \gls{adsdc}.
    Moreover, a perturbatively stable model with scale separation, yet with a non-universal mass spectrum of light fields, is presented in the context of massive type IIA flux compactifications of AdS$_4\times$X$_6$, where X$_6$ admits a \gls{cy} metric and O6-planes wrapping three-cycles \cite{carrasco_new_2024}. The scale separation stems from an asymmetric flux rescaling, which is, in general, not a simple symmetry at the level of 4d \glspl{eom} \cite{carrasco_new_2024}.
    \citet{petrini_ads_2013} present scale separated type IIB solutions.
    \citet{blumenhagen_kklt_2020} shows that the \gls{kklt} \gls{ads} minimum satisfies $m_\textnormal{KK}^2\sim\abs{\Lambda}^{1/3}/\log^2\left(-\Lambda\right)$, i.e. $\mathfrak{d}=1/6$ and the strong \gls{adsdc} is violated, yet there is an additional log-correction.
}
between the external and internal space \cite{apers_comments_2022,font_scale_2020,buratti_discrete_2020,nam_4d_2023,montero_pure_2023,cribiori_correspondence_2023,lust_ads_2019,bernardo_purely_2021,blumenhagen_quantum_2020,bonnefoy_swampland_2021,gonzalo_ads_2021,coudarchet_hiding_2024}.

\subsubsection{Implications for Cosmology}\label{sss_AdSDC_Cosmology}
If the $\Lambda_\textnormal{cc}\rightarrow0$ limit lies at an infinite distance locus in moduli space, this implies that a \gls{ads} space cannot transition into a Minkowski space and from there into a \gls{ds} space, since those are infinitely far apart \cite{lust_ads_2019}. This agrees with the notion from \cref{sec:deSitter} that meta-stable \gls{ds} vacua do not exist \cite{lust_ads_2019}. This also means that in a theory that is holographically dual to \gls{ads}, a \gls{ds} vacuum cannot be a state in that theory \cite{lust_ads_2019}. In the following, we highlight the important consequences of this observation for the description of \glspl{bh} and \gls{de}.

\paragraph{Black Holes}\label{p:AdSDC_BH}
A smooth interpolation from asymptotically \gls{ads} to asymptotically Minkowski is forbidden for \gls{bps} \glspl{bh} \cite{cribiori_correspondence_2023}. This leads to a \gls{bh}-specific conjecture:
A polynomial mass dependence of the form $m\sim \mathcal{S}^{-\mathfrak{b}}$ with $\mathcal{S}$ the entropy and $\mathfrak{b}>0$ a strictly positive constant, is presented as the \gls{bhedc} \cite{luben_black_2021,bonnefoy_infinite_2020,cribiori_correspondence_2023}. Since both, the entropy of \glspl{bh} and a cosmological constant, can be expressed in terms of the volume of the internal space, there is a direct relation between the \gls{adsdc} and the \gls{bhedc} \cite{cribiori_correspondence_2023,bonnefoy_infinite_2020}. %
The main idea is that the state of infinite entropy is at an infinite distance locus in the space of \gls{bh} metrics, and the \gls{dc} predicts a tower of states when approaching this locus \cite{anchordoqui_landscape_2024}.

\paragraph{Dark Dimension}\label{p:AdSDC_darksector}
Recent considerations regarding the small value of the cosmological constant in combination with the \gls{adsdc} and the species scale lead to the conclusion that a fifth, mesoscopic dimension should be part of our Universe, at a length scale proportional to $\Lambda_\textnormal{cc}^{1/4}$ in the micron range \cite{gonzalo_dark_2022,vafa_swamplandish_2024,anchordoqui_dark_2022_PBH,montero_dark_2023,blumenhagen_dark_2023,anchordoqui_landscape_2024}: %
Applying the \gls{adsdc} to the cosmological constant yields a tower of states with $m\lesssim\Lambda_\textnormal{cc}^{1/4}\sim\SI{2}{\milli\electronvolt}$, where $\mathfrak{d}=1/4$ comes from the lower limit set by the Casimir energy and experimental bounds on Newton's gravitational inverse-square law\footnote{
    See \citet{baeza-ballesteros_dynamical_2022,baeza-ballesteros_towards_2023} for proposals for more stringent experimental tests.}
\cite{lee_new_2020}.
In the infinite distance limit of a vanishing cosmological constant, the \gls{ep} informs us that there is either an infinite tower of \gls{kk} states, or an infinite tower of string oscillator modes.
The string limit is expected to have already observable effects.\footnote{
    \citet{basile_dark_2024} show that if some fine-tuning takes place, a tower of higher-spin excitations is an allowed solution in the little string scenario \cite{antoniadis_little_2001} (in this scenario, strings are very weakly coupled and resolvable near current accelerator energies).
}
Therefore, it is assumed that the \gls{kk} limit is realised.
A \gls{kk} tower corresponds to a decompactification limit, which yields an additional, fifth dimension.\footnote{
    Even though proposals with 3 macroscopic and 2 mesoscopic spatial dimensions exist (e.g. to explain the hierarchy problem with a sixth dimension, compactified at the TeV-scale, in which the \gls{sm} fields propagate as well \cite{fadafan_cosmological_2023}, or to explain the smallness of $\Lambda_\textnormal{cc}$ \cite{carroll_sidestepping_2003,aghababaie_towards_2004}), having one\,\textemdash\,and not more\,\textemdash\,extra-dimensions is favoured by several arguments:
    according to the \gls{ep}, decompactifying one dimension allows for the largest scalar field transversal for a given potential \cite{van_de_heisteeg_bounds_on_field_2023};
    if there was more than one mesoscopic extra-dimension, the \gls{lhc} would produce \glspl{bh} \cite{vafa_swamplandish_2024};
    and if there were more than two extra-dimensions, neutron stars would be heated up to temperatures above the observed lower temperature thresholds, by dark gravitons decaying into photons \cite{hannestad_supernova_2004,montero_dark_2023}.
}
The upper bound on the length scale depends on the geometry of the extra-dimension \cite{anchordoqui_searching_2024}, yet based on the mass scale given by the cosmological constant, the size of this additional dark dimension is of $\order{\SI{10}{\micro\meter}}$.\footnote{    
    \citet{branchina_does_2023} question various assumptions made in this derivation, starting from the necessity that $\mathfrak{d}=1/4$ is the only viable choice to the finding that the \gls{kk} tower must be realised.
    \citet{anchordoqui_cosmological_2023} respond to this criticism.
    \citet{branchina_dark_2025} rebut.
}
Such a dark dimension would have some welcomed properties:
it unifies \gls{dm} and \gls{de}, 
it unifies \gls{dm} with gravity\footnote{
    If \gls{dm} is in the form of graviton excitations, gravity \textit{is} \gls{dm}.
    },
and it solves the \textit{why now?}\footnote{
    Why do we live right now, when the Universe is entering a \gls{de}-dominated phase?
    }
puzzle
and explains the \textit{cosmological coincidence}\footnote{
    Why is the \gls{de} density so similar to the energy density of matter and radiation at matter\textendash radiation equality? See \citet{velten_aspects_2014} for a discussion and \citet{blau_lecture_2023} for some numerical baublery.
    }
\cite{vafa_swamplandish_2024}.
A success of the model is the predicted and observed sharp cutoff of the flux of ultra-high-energy cosmic rays \cite{anchordoqui_dark_2022_GKZ,the_pierre_auger_collaboration_observation_2008,hires_collaboration_first_2008,noble_probing_2023}.\footnote{
    See also the work by \citet{anchordoqui_aspects_2023} for testable predictions.
}
Furthermore, having a mesoscopic dimensions brings various interesting and promising consequences, for instance for \glspl{bh} \cite{anchordoqui_dark_2023,anchordoqui_dark_2024,anchordoqui_through_2024,anchordoqui_bulk_2024,anchordoqui_more_2024},
inflation \cite{anchordoqui_large_2023,antoniadis_cosmological_2023},
\gls{de} \cite{danielsson_experimental_2023,burgess_perils_2023}, and
\gls{dm} \cite{law-smith_astrophysical_2024,anchordoqui_dark_2022_PBH,gonzalo_dark_2022,anchordoqui_dark_2023,anchordoqui_fuzzy_2024}.
We present some in the following.

In the mesoscopic dark dimension proposal, \gls{dm} occurs in the form of massive spin-2 \gls{kk} modes residing in the dark dimension, called \textit{dark gravitons} \cite{gonzalo_dark_2022,anchordoqui_dark_2023}, which have a mass of less than 1\textendash\SI{100}{\kilo\electronvolt} \cite{law-smith_astrophysical_2024} and only interact gravitationally \cite{gonzalo_dark_2022}. The excited states decay, releasing photons, which heat up their surrounding, while the \gls{dm} mass decreases over time. The decay also gives kinetic kicks to the decay products, which suppresses structure formation \cite{obied_dark_2023}.\footnote{
    Observations of Milky Way satellite galaxies and simulations of isolated \gls{dm} halos limit the kick velocity to \SI{20}{\kilo\meter\per\second} \cite{mau_milky_2022} respectively to \SI{100}{\kilo\meter\per\second} \cite{anchordoqui_cosmological_2024,peter_dark-matter_2010}.
    \citet{anchordoqui_landscape_2024} mention that there could be several \gls{dm} species that are dominant during different periods in the cosmic evolution, with masses ranging from $\sim\SI{1}{\mega\electronvolt}$ at \gls{cmb} to $\SI{50}{\kilo\electronvolt}$ in the local Universe.
}
The higher the graviton mass, the higher is the production rate of photons. The observed extra-galactic background radiation sets an upper bound of about \SI{1}{\mega\electronvolt} for the graviton mass scale \cite{vafa_swamplandish_2024} and the observed cosmic ray spectrum \cite{anchordoqui_dark_2022_GKZ} fixes $m\sim\num{e3}\Lambda_\textnormal{cc}^{1/4}$.

\citet{anchordoqui_dark_2023} claim that the dark gravitons are practically indistinguishable from 5-dimensional \glspl{bh}\,\textemdash\,a statement whose validity relies on a close analogy between \glspl{bh} and bound states of gravitons \cite{dvali_black_2011}. \glspl{bh}, especially \glspl{pbh}, are often studied as \gls{dm} candidates.\footnote{
    Studying \glspl{bh} as \gls{dm} candidates comes with a benefit: \enquote{Black holes are the only dark matter candidate that is known to actually exit.} \cite{riotto_future_2024}.}
Having a fifth, mesoscopic dimension is of relevance here: if \glspl{bh} sense the fifth dimension, their horizon is larger, which reduces their temperature, and therefore the rate at which Hawking radiation \cite{hawking_particle_1975} takes place, which in turn increases the lifespan of the \glspl{bh} \cite{anchordoqui_more_2024}\,\textemdash\,this makes \glspl{pbh} in the mass range of \SI{e14}{\gram} to \SI{e22}{\gram} feasible candidates to explain \gls{dm} \cite{anchordoqui_dark_2022_PBH}.\footnote{
    This mass range assumes that the \gls{bh} is localised on the brane during the entire evaporation process\,\textemdash\,relaxing this assumption and allowing \glspl{bh} to escape into the dark dimension relaxes the allowed mass bound for \glspl{pbh} that account for the entirety of \gls{dm} to \SI{e11}{\gram} to \SI{e21}{\gram} \cite{anchordoqui_bulk_2024,anchordoqui_more_2024}.
    For 4d \glspl{pbh}, there are various mass constraints presented in the literature, e.g. various observational constraints rule out \glspl{pbh} as sole \gls{dm} particles for masses $<\SI{e17}{\gram}$ \cite{carr_primordial_2020,green_primordial_2021,villanueva-domingo_brief_2021,bagui_primordial_2023,carr_new_2010,clark_planck_2017,derocco_constraining_2019,laha_primordial_2019,dasgupta_neutrino_2020,keith_511_2021,mittal_constraining_2022,laha_integral_2020,berteaud_strong_2022,korwar_updated_2023};
    masses $>\SI{e21}{\gram}$ are ruled out by the non-observation of microlensing events (5d \glspl{pbh} have a Schwarzschild radius smaller than the wavelength of visible light, which makes geometric optics and therefore this specific mass-bound non-applicable, as wave effects dominate that suppress magnification \cite{croon_subaru_2020}\,\textemdash\,future x-ray observatories might be able to observe microlensing events that could act as supporting evidence for 5-dimensional \glspl{pbh} \cite{anchordoqui_through_2024}) \cite{anchordoqui_landscape_2024};
    and the mass range from \SI{1}{\gram} to \SI{e15}{\gram} is excluded on theoretical grounds \cite{hamaide_primordial_2024}.
    }

Instead of gravitons, the \gls{kk} tower could also give rise to axions with an axion decay constant $f_\textnormal{a}\sim\SI{e10}{\giga\electronvolt}$, which would constitute a minor portion of \gls{dm} ($\sim0.1\%\sim1\%$) and be compatible with observational constraints \cite{yang_search_2024,gendler_axions_2024}.

\citet{heckman_dark_2024} reason that the dark dimension scenario predicts a \gls{gut} scale of $\Lambda_\textnormal{GUT}\lesssim\order{\SI{e16}{\giga\electronvolt}}$ and an extra boson with mass $m\sim \qtyrange{e15}{e16}{\giga\electronvolt}$, which is above the 5-dimensional Planck scale. Therefore, this object cannot be localised (otherwise it would be a \gls{bh}), i.e. this gauge boson is a string in 5d \cite{heckman_dark_2024}.

Inflation might also take place in the fifth dimension: starting from a size of $\order{M_\textnormal{P;5}^{-1}}$, related to the 5-dimensional Planck mass $M_\textnormal{P;5}$, the extra dimension requires 42 $e$-folds to reach the current mesoscopic size \cite{anchordoqui_landscape_2024}.

A concrete string-theoretical realisation of the dark dimension proposal is presented by \citet{blumenhagen_dark_2023}, in the setting of a warped throat. However, they also point to a shortcoming, namely that in their setting, other \gls{kk} modes do not separate enough from the dark dimension, which is incompatible with astrophysical observations and the predicted size of this mesoscopic extra-dimension. \citet{schwarz_comments_2024} goes even one step further: When $E_8\cross E_8$ is described in 11-dimensional M-theory, the \nth{11} dimension is the dark dimension, the bulk in which the two \gls{etw} 9-branes live. A similar model, with a large extra-dimension, including phenomenological predictions, is described by \citet{danielsson_stringy_2023} as \textit{dark bubble cosmology}.

Despite the popularity of the dark dimension model, we see several weaknesses in this proposal: the arguments regarding neutron star heating and the absence of \glspl{bh} in the \gls{lhc} might \textit{allow} for one mesoscopic dimension, but are fulfilled for zero extra-dimensions as well. This leaves us with a somewhat weaker footing for this idea. Furthermore, a key assumption is that the \gls{sm} particles only interact with the extra-dimension gravitationally.
Some authors claim that this can emerge from string theory \cite{anchordoqui_dark_2022_PBH,montero_dark_2023}. For example, if the \gls{sm} fields are localised on a brane-defect on the 4-dimensional subspace of the 5-dimensional spacetime \cite{vafa_swamplandish_2024}. 
Other authors mention that kinetic mixing, and higher-order loops or derivatives will always induce interactions between geometrically separated sectors \cite{acharya_hidden_2019,anisimov_brane_2002}.
If the \gls{sm} fields propagated the entire spacetime, each \gls{sm} particle would be accompanied by a \gls{kk} tower\,\textemdash\,something we do not observe \cite{vafa_swamplandish_2024}. This was also criticised by \citet{mckeen_signatures_2024}: The predicted extra dimension would yield a \gls{kk} tower of sterile neutrinos, lifting the effective number of neutrinos above the observational bounds. This critique was met by \citet{anchordoqui_cosmological_2024} by introducing intra-tower neutrino decays with purely \enquote{dark-to-dark} interactions that would allow for a mesoscopic fifth dimension and a tower of light \gls{kk} (final) states that is compatible with observational constraints.\footnote{
    This is in agreement with the proposed framework of dynamical \gls{dm} by \citet{dienes_dynamical_2012}.
}
\citet{casas_small_2024} find that having only one extra large(r) dimension does not allow for small enough Dirac neutrino masses; two extra dimensions are required to be compatible with observational data, while four and six large(r) dimensions are experimentally ruled out, as these scenarios then lead to a too low \gls{qg} cutoff scale.

\citet{petretti_investigating_2024} find that the dark dimension scenario yields a worse fit to \gls{cmb} data than the baseline model: the dark dimension scenario is in tension with low-$l$ temperature data and brings an extra degree of freedom (which is punished in a Bayesian analysis). Overall, the \gls{cmb} spectrum predicted by a fifth-dimension scenario is boosted compared to the concordance model, yet more strongly for small $l$ (see figure 1 by \citet{petretti_investigating_2024}).

\paragraph{Dark Energy}
$\Lambda_\textnormal{s}$CDM models that introduce a sign switch in the cosmological constant $\Lambda_\textnormal{cc}$ (e.g. \cite{akarsu_graduated_2020,akarsu_relaxing_2021,akarsu_lambda_rm_2023,akarsu_relaxing_2023})\,\textemdash\,which indicates a transition from \gls{ds} to \gls{ads}\,\textemdash\,are ruled out by the \gls{adsdc}, as \gls{ads} and \gls{ds} spaces are an infinite distance apart \cite{lust_ads_2019}.\footnote{
    A proposal by \citet{anchordoqui_anti-sitter_2024,anchordoqui_infinite_2024} might present some remedy:
    They present a mechanism involving Casimir forces that allows a smooth transition from \gls{ads} to \gls{ds} if the vacua are meta-stable, which is accompanied by a reduction of the species scale (\cref{s:ssc}) in the strong-gravity regime. A reduction of the species scale is equivalent to an increase in the species entropy \cite{cribiori_species_2023}.
    This mechanism is also compatible with the dark dimension proposal \cite{montero_dark_2023,anchordoqui_anti-sitter_2024}.
}

\paragraph{Particle physics}\label{p:AdSDC_particles} is directly addressed by the \gls{adsdc}: If we have a scalar field $\phi$ that is rolling, a tower of light states emerges with $m_l\sim V^{1/4}\left(\phi\right)$ \cite{anchordoqui_darkdimension_2023}. Through the seesaw mechanism \cite{gell-mann_complex_2013,yanagida_horizontal_1980}, we'd also expect another particle to be present, with $m_h\sim\sqrt{m_lM_\textnormal{P}}$. If we take the scalar field to represent \gls{de} and identify the value of the potential with the \gls{de} density, then $m_l$ is roughly of the order of the neutrino mass and $m_h$ of the Higgs mass. This indicates that the swampland programme might give us some hints on how to solve the mass fine-tuning problem of particle physics \cite{vafa_cosmic_2019}.

When studying the \gls{adsdc} in combination with the no non-SUSY conjecture (\cref{sec:nononSUSY}), \citet{gonzalo_ads_2021} come to the conclusion that in theories with a positive cosmological constant, a surplus of light fermions is required, which means that in the current universe, which is in a \gls{ds} phase, the lightest neutrino is a Dirac neutrino with $m_\nu\lesssim\Lambda_\textnormal{cc}^{1/4}\sim\SI{10}{\milli\electronvolt}$ \cite{harada_further_2022,gonzalo_ads_2021,gonzalo_fundamental_2018,gonzalo_swampland_2022,ibanez_constraining_2017}.\footnote{
    Majorana and Dirac fermions have different wave equations \cite{hsieh_dirac_2024}.
    Dirac neutrinos avoid lepton number violations and have 4 degrees of freedom, which can compensate for bosonic degrees of freedom of the photon and the graviton\,\textemdash\,Majorana neutrinos have 2 degrees of freedom \cite{sabir_fermion_2025}.
    }
If the lightest neutrino was Majorana, we had a stable \gls{ads} vacuum solution, which is challenged by the \gls{adsdc} \cite{harada_further_2022,anchordoqui_dark_2024,grana_swampland_2021}.\footnote{
    There is a possibility to have Majorana neutrinos without \gls{ads} vacua: if there is a very light gravitino, its \gls{kk} tower can compensate the graviton tower and the \gls{ads} vacuum is avoided \cite{anchordoqui_dark_2024,anchordoqui_darkdimension_2023}.
    }

\subsubsection{General Remarks}
The \gls{adsdc} basically tells us that the mass of the lightest scalar should be small compared to the \gls{ads} scale \cite{gautason_tension_2019,apers_comments_2022}. It can be deduced that if there is no scale separation, the size of the \gls{ads} space is directly related to the mass of its lightest mode by $m\sim\frac{1}{l_\text{AdS}}$ \cite{blumenhagen_quantum_2020}.\footnote{
    Similarly, coined the \gls{ads} moduli conjecture \cite{blumenhagen_quantum_2020}, is the expectation that there is always a scalar such that $m_\phi^2l_\text{AdS}^2\sim\order{1}$, which is weaker than the strong \gls{adsdc}, as a single scalar suffices to obey the conjecture, instead of an entire tower of states \cite{van_riet_beginners_2023}.
    A study by \citet{ishiguro_stabilization_2024} performs a random sample of type IIB flux vacua, which peaked at $\sqrt{\abs{m_\phi^2l_\text{AdS}^2}}\sim9$.
    }

In the remainder of this section, we'd like to address various questions to shed some light on different aspects of the \gls{adsdc}.

\paragraph{How is the \gls{adsdc} motivated?}
\citet{castellano_iruv_2022} give a fundamental explanation based on the \gls{ceb}\footnote{
    \enquote{Let $A$ be the area of a connected $(D-2)$\textendash dimensional spatial surface $B$. Let $L$ be a hypersurface bounded by $B$ and generated by one of the four null congruences orthogonal to $B$. Let $S$ be the total entropy contained on $L$. If the expansion of the congruence is nonpositive (measured in the direction away from $B$) at every point on $L$, then $S\leq A/4$} \cite{bousso_holography_1999}.
    }
\cite{bousso_covariant_1999,bousso_holography_1999}: the holographic principle implies that a given volume of spacetime has an upper bound of information it can contain (on the area of the boundary), so there is a bound on the entropy that can be associated to a \gls{uv} cutoff scale for an \gls{eft}\,\textemdash\,which is identified with the species scale (\cref{s:ssc})\,\textemdash, whereas the entropy itself is associated with the entropy of a massless scalar field; taking into account the characteristic length scale of \gls{ads} space\,\textemdash\,the inverse of the maximum length scale serves as an \gls{ir} cutoff\,\textemdash\,leads to \cref{eq:AdSDC} \cite{calderon-infante_entropy_2023}. Here, we observe one form of \gls{uv}/\gls{ir} mixing: the area-to-volume ratio determines the \gls{uv} cutoff and is linked to the \gls{ir} cutoff through the length scale.

Focusing on scale separation, and therefore mostly on the strong \gls{adsdc}, \citet{coudarchet_hiding_2024} presents\,\textemdash\,besides a summary of swampland considerations against scale separation\,\textemdash\,arguments from holography and \glspl{cft}: neither scale-separated \gls{cft} duals to theories of gravity nor scale-separated spacetimes for the bulk have been constructed yet, raising suspicion that scale-separated constructions are generally unstable. Furthermore, \citet{andriot_exploring_2022,andriot_erratum_2022} rule out scale separation for \gls{ads} on nilmanifolds and Ricci-flat manifolds.

\paragraph{How is the \gls{adsdc} related to the \gls{dc}?}
Conceptually, the main difference between the \gls{adsdc} and the \gls{dc} is that the \gls{dc} deals with continuous backgrounds (even though discretised version exist, see \cref{eq:discretedistance}), whereas for the \gls{adsdc}, scale separation and disconnected backgrounds are an important topic, which also means that the scalar fields possess a potential: different vacua are \textit{labelled} by different cosmological constants / \gls{ads} radii / flux quantum numbers \cite{bonnefoy_swampland_2021}.
In the limit $V\rightarrow0$, we expect a tower of states with
\begin{equation}
    m\propto\abs{V}^\mathfrak{d}.
\end{equation}
The \gls{dc} predicts a tower of states with exponentially suppressed masses
\begin{equation}
    m(\phi)\propto \exp(-\mathfrak{a} d(\phi,\phi_0)).
\end{equation}
For $\mathfrak{a} d(\phi,\phi_0)=-\mathfrak{d}\log\abs{V}$, the two conjectures are equivalent.\footnote{
    \citet{palti_positive_2024} make a similar identification, and derive $\mathfrak{a}\simeq0.06$ for \gls{dgkt} vacua. This is below the lower bound on $\mathfrak{a}$. Either, this identification between $\mathfrak{a}$ and $\mathfrak{d}$ is too naïve, the \gls{dc} or the \gls{adsdc} are wrong, the lower bound on $\mathfrak{a}$ does not apply, or \gls{dgkt} is in the swampland (\gls{dgkt} are approximate solutions that smear O6-planes\,\textemdash\,O6-planes are localised objects and if smearing them renders them ill-suited to solve the 10d \glspl{eom} is still debated, e.g. \citet{junghans_o-plane_2020} shows that \gls{dgkt} can approximate viable type IIA supergravity solutions with arbitrary precision).
}
The link between the \gls{adsdc} and the \gls{dc} is given through the measurement of the distance:
$d(\phi,\phi_0)$ is the distance in moduli space. Let us consider
a generalisation of the geometric distance given by
\begin{equation}
    \mathfrak{d}\int_{\tau_i}^{\tau_f}\!\left(\frac{1}{\textnormal{Vol}\left(\mathcal{M}\right)}\int_\mathcal{M}\sqrt{g}g^{MN}g^{OP}\frac{\partial g_{MO}}{\partial\tau}\frac{\partial g_{NP}}{\partial\tau}\right)^{1/2}\,\mathrm{d}\tau,
\end{equation}
with $\tau$ the proper time and $g$ the Riemann metric on the manifold $\mathcal{M}$ 
\cite{gil-medrano_riemannian_1991,lust_ads_2019}, which leads to the geodesic equation
\begin{align}
    \ddot{g}&=\dot{g}g^{-1}\dot{g}&+&\frac{1}{4}\Tr{\left(g^{-1}\dot{g}\right)^2}g&-&\frac{1}{2}\Tr{g^{-1}\dot{g}}\dot{g}\nonumber\\
    &&-&\frac{1}{4}\langle\Tr{\left(g^{-1}\dot{g}\right)^2}\rangle g&+&\frac{1}{2}\langle \Tr{g^{-1}\dot{g}}\rangle \dot{g},
\end{align}
where 
\begin{equation}
    \langle X \rangle=\frac{\int_\mathcal{M}\sqrt{g}X}{V_\mathcal{M}}
\end{equation}
and dot is the derivative with respect to $\tau$ \cite{bonnefoy_infinite_2020}. 
For
\begin{equation}
    \mathrm{d}s^2=f(\tau)g_{\mu\nu}(x)\mathrm{d}x^\mu \mathrm{d}x^\nu
\end{equation}
with $f$ an arbitrary function, the geodesic equation is solved by
\begin{equation}
    f(\tau)=f_0e^{\tau/\sqrt{d_\mathcal{M}}},
\end{equation}
with $d_\mathcal{M}$ the dimension of the moduli space $\mathcal{M}$, which gives a geometric distance of
\begin{equation}
    \mathfrak{d}\sqrt{d_\mathcal{M}}\log\left(\frac{f_f}{f_i}\right).
\end{equation}
Using an \gls{ads} metric
\begin{align}
    \mathrm{d}s^2&=\frac{\left(1-d\right)\left(d-2\right)}{2\Lambda_\textnormal{cc}}\nonumber\\
    &\quad\cross\left(-\cosh^2\left(\rho\right)\mathrm{d}t^2+\mathrm{d}\rho^2+\sinh^2\left(\rho\right)\mathrm{d}\Omega^2_{d-2}\right)
\end{align}
and identifying the potential $V$ with the cosmological constant $\Lambda_\textnormal{cc}$ yields then the \gls{adsdc} (\cref{eq:AdSDC}). The choice of a metric in \gls{ads} and the corresponding moduli space is not trivial, as is highlighted by \citet{li_towards_2023}: The moduli space metric is obtained by varying the higher-dimensional Einstein\textendash Hilbert action. There are three types of variations: 
\begin{itemize}
    \item infinitesimal transverse and traceless deformations of the metric of the external \gls{ads} space,
    \item changes in the overall conformal volume factor of the internal space,
    \item variations of fluxes.
\end{itemize}
In Minkowski and \gls{ds} space, the volume variations can be factored out, such that variations along paths through moduli space with constant spacetime volume can be studied, and fluxes are only a subtle contribution that are not required to define a distance measure \cite{li_towards_2023}. For \gls{ads} space, however, these variations cannot safely be neglected, as otherwise the obtained metric might become negative, i.e. proper distance is not well-defined \cite{lust_ads_2019}. \citet{li_towards_2023} study the effect of the different contributions and find that varying only the conformal volume factor of \gls{ads} leads to a negative metric on the family of solutions, which can be remedied by including variations of the internal space (see also \citet{lust_ads_2019}). If \gls{ads} spaces with more than six dimensions are examined, the inclusion of fluxes is required as well, to obtain a positive metric that yields well-defined distance measures \cite{li_towards_2023}.\footnote{
    \citet{palti_positive_2024} raise this to another conjecture: The metric over \gls{qg} solutions calculated from a quadratic off-shell variation of the action restricted to the on-shell solutions is always positive, i.e. the landscape of theories has real proper distance measures.
}
Combining both variations is generally possible if there is no scale separation between the internal and the \gls{ads} space, i.e. the \gls{ads} radius scales like the internal volume with the parameter of the solutions. Scale separation would lead to the negative \gls{ads} contributions dominating the positive internal contributions.

The exponential mass suppression known from the \gls{dc} is also derived in the \gls{ads} context, if the domain wall tension between different field values (a quasi-distance) is used, instead of a geodesic distance measure \cite{mohseni_measuring_2024}.
This Ansatz is motivated by the cobordism conjecture (\cref{sec:cobordism}), which predicts that theories of \gls{qg} with the same dimensions are separated by domain walls of finite tension, and further supported by finiteness considerations (see \cref{s:fnomf}) \cite{mohseni_measuring_2024}. 

\citet{grana_swampland_2021} state that the \gls{adsdc} and the \gls{dc} are unified in the gravitino distance conjecture (\cref{eq:DCgravitino}) for vanishing gravitino mass. For non-vanishing gravitino mass, the \gls{ads} and the non-\gls{ads} case cannot be continuously connected, as an infinite tower of states appears, which spoils the connection. The \gls{flb} informs us that the gravitino mass has a lower, non-negative bound, i.e. this connection cannot be realised.

\citet{calderon-infante_tensionless_2024} find that 4d holographic supersymmetric \glspl{cft} with a conformal manifold and an overall free limit respectively the \gls{ads}$_5$ bulk cannot be scale separated if the \gls{dc} parameter in the bulk is $\mathfrak{a}=1/\sqrt{d-2}$ (see around \cref{eq:DC_Bound}).

\paragraph{What are the bounds on $\mathfrak{d}$?}
An argument for $\mathfrak{d}\geq1/d$ comes from considering the dual description of the theory, where particles of mass $m$ from the weakly interacting light tower contribute $m^d$ to the vacuum energy, ergo, $\Lambda_\textnormal{cc}\geq m^d\Leftrightarrow m\leq\Lambda_\textnormal{cc}^{1/d}$ \cite{vafa_swamplandish_2024}.

The specific value of $\mathfrak{d}=1/2$ is conjectured to hold in all supersymmetric \gls{ads} vacua. This strong \gls{adsdc} forbids scale separation and is found to hold in all known 10- or 11-dimensional solutions to string theory and M-theory \cite{lust_ads_2019,apers_comments_2022}. The 
\gls{kklt} scenario was considered a counterexample to the strong \gls{adsdc}, but if non-perturbative effects that lead to logarithmic corrections are considered as well, no scale separation has been found \cite{bernardo_purely_2021, blumenhagen_swampland_2019,blumenhagen_quantum_2020, gautason_tension_2019}. Also some AdS$_4$ type II vacua seem to violate the strong \gls{adsdc}, but only if backreactions in the full 10-dimensional theory are ignored \cite{font_scale_2020}.
No scale-separated vacua are found in type IIA string theory with metric fluxes \cite{prieto_moduli_2024}.
Yet, scale-separated vacua are not necessarily in the swampland, as the work by \citet{shiu_ads_2023} shows: in their setup, massive type IIA flux vacua are in agreement with the \gls{dc}. Also \citet{emelin_effective_2020} find that brane-instantons wrapped around 4-cycles allow for scale separation in \gls{ads}. A refined version of the strong \gls{adsdc}, in which scale separated \gls{ads} vacua are allowed if there is a large discrete higher form symmetry with some additional properties, is presented by \citet{buratti_discrete_2020}. \citet{apers_comments_2022} provide counterexamples to this refined version in the form of AdS$_3$ vacua that violate the refined conjecture. However, for supersymmetric AdS$_4$ vacua with discrete Z$_k$ symmetries, it seems to hold that the ratio between the \gls{kk} mass scale $m_\text{KK}$ and the cosmological constant $\Lambda_\textnormal{cc}$ is given by $m_\text{KK}\sim\left(k\Lambda_\textnormal{cc}\right)^{1/2}$ \cite{buratti_discrete_2020}.\footnote{
    Scale separation and broken supersymmetry go hand in hand: $\mathfrak{d}=1/2$ implies that $m_\text{KK}^2/\abs{\Lambda}\sim\order{1}$, such that for $m_\text{KK}\sim1/r_\textnormal{KK}$ and $\Lambda\sim 1/l_\textnormal{AdS}^2$, $l_\textnormal{AdS}\sim r_\textnormal{KK}$ holds \cite{junghans_o-plane_2020}.
}
\citet{cribiori_quantum_2023} invoke the species scale (\cref{s:ssc}) and holography to rule out scale separation for 5-dimensional supergravities.

\paragraph{What are the implications for our understanding of spacetime?}
If the \gls{adsdc} is true, the \gls{ads} space has to be strongly curved or be part of a higher-dimensional space with a limit such as $\text{AdS}_d\cross Y_p\rightarrow\text{Mink}_d$, for a $p$-dimensional internal space $Y_p$, as near flat limits are in the swampland \cite{lust_ads_2019}.

Topological massive gravity, more precisely a spin-2 truncation of a higher-spin theory with Chern\textendash Simons terms, fermions, and scalar fields, was found to be compatible with the \gls{adsdc} \cite{alvarez-garcia_swampland_2022}.

\subsubsection{Evidence}
The \gls{adsdc} was first proposed by \citet{lust_ads_2019}
and is studied in the context of
high-energy \gls{ads} spaces with an embedded low-energy \gls{ds} space \gls{eft} \cite{nam_4d_2023},
supersymmetry-breaking in type II string theory \cite{parameswaran_ds_2024},
flux compactifications of massive IIA supergravity\footnote{Supergravity is a string limit where the string coupling is weak and the length scales in the theory are much larger than the string scale \cite{van_riet_beginners_2023}.} with O6 planes \cite{apers_comments_2022},
4D type IIA orientifold flux compactifications \cite{quirant_aspects_2022},\footnote{
    Orientifold planes are objects of negative string tension and fixed position \cite{van_riet_beginners_2023}.
    }
toroidal / orbifold type IIA vacua with metric fluxes \cite{font_scale_2020},
\gls{ads} vacua as solutions to purely non-perturbative contributions to the superpotential \cite{bernardo_purely_2021},
the \gls{kklt} scenario \cite{blumenhagen_quantum_2020,lust_holography_2022},
type IIB string theory \cite{emelin_effective_2020,ishiguro_sharpening_2021,bizet_testing_2020,ishiguro_stabilization_2024,bardzell_type_2022},
11d supergravity on AdS$_4\cross S^7$ and AdS$_7\cross S^4$ \cite{li_towards_2023},
5d supergravities \cite{cribiori_quantum_2023},
arbitrary dimensions \cite{bonnefoy_swampland_2021}, 
discrete gauge symmetries \cite{buratti_discrete_2020},
non-associativity and non-commutativity \cite{bubuianu_nonassociative_2023},
the \gls{wgc} \cite{cribiori_weak_2022} (see also \cref{rel:AdSDC_WGC}),
the \gls{flb} (\cref{rel:AdSDC_FLB}),
the \gls{tpc} (\cref{rel:TPC_AdSDC}),
the no Global Symmetries Conjecture (\cref{rel:nGSC_AdSDC,sec:nGSym}),
and the no non-supersymmetric Theories Conjecture (\cref{rel:AdSDC_nnSUSYC,sec:nononSUSY}).

\subsection{Cobordism Conjecture}\label{sec:cobordism}%

The bordism group of \gls{qg} is trivial \cite{debray_chronicles_2023}:
\begin{equation}\label{eq:cobordism}
    \Omega_k^\text{QG}=0.
\end{equation}
The elements of $\Omega_k^\text{QG}$ are the equivalence classes of $\left(k=D-d\right)$-dimensional compactifications of \gls{qg} that can be connected via domain walls \cite{debray_chronicles_2023,mcnamara_cobordism_2019}. 

\subsubsection{Implications for Cosmology}
The cobordism conjecture implies certain stringy defects \cite{mcnamara_cobordism_2019} and that there has to be a boundary-ending spacetime \cite{cicoli_string_2023,buratti_dynamical_2021}.

\paragraph{Big Bang}
An open question in cosmology is if there is a beginning of time, i.e. if there is a spacelike singularity \cite{quevedo_lectures_2002,craps_big_2006}. The cobordism conjecture gives a positive answer: a consistent theory of \gls{qg} should admit spacetime-ending configurations (\gls{etw} branes), i.e. bubbles of anything \cite{friedrich_cobordism_2024} or bubbles of nothing respectively walls of nothing \cite{etxebarria_nothing_2020,ooguri_new_2017,dibitetto_nothing_2020,bomans_bubble_2022,bedroya_sitter_2020,blanco-pillado_bubbles_2016,garriga_smooth_1998} as boundaries or cobordism defects \cite{angius_dynamical_2022,angius_at_2022,blumenhagen_dynamical_2023,buratti_dynamical_2021}.\footnote{
    Intersecting \gls{etw} branes can be regarded as cosmological bubble collisions in the context of eternal inflation \cite{angius_intersecting_2023,kleban_cosmic_2011}.
}
An explicit example is worked out by \citet{friedrich_boundary_2024}: tensionless \gls{etw} branes can yield a compact, flat torus / \gls{ds} universe from nothing.
However, it is important to note that the cobordism conjecture makes no statement about the dynamics of \gls{qg} and the possibility to reach \textit{nothing} \cite{andriot_looking_2022}. The conjecture just states that the empty set (nothing) is part of the class of valid \glspl{qg} \cite{andriot_looking_2022}. Other swampland conjectures might dynamically prohibit that nothing can be reached \cite{andriot_looking_2022}.

\paragraph{Black Holes}\label{p:Cobordism_BH}
The cobordism conjecture forbids \gls{bh} remnants and therefore supports the no global symmetries conjecture (\cref{sec:nGSym}) and the \gls{wgc} \cite{andriot_looking_2022}: The interior of a \gls{bh} is described by a compact space. While the \gls{bh} evaporates, this space shrinks. This process is represented by a bordism. According to the cobordism conjecture, the initial compact space and the empty set (the nothing) are connected through this bordism, which means that remnants are not a final state, ergo global charges and therefore \gls{bh} remnants\footnote{
    See \cref{p:WGC_Remnants} on further arguments against \gls{bh} remnants.
    }
are not allowed.

\paragraph{CP Symmetry}
\gls{cp} symmetry is well-known to be broken \cite{christenson_evidence_1964}. This means that it is either not a global symmetry at all, or a gauged or broken global symmetry \cite{mcnamara_reflections_2022}.
\begin{itemize}
    \item If \gls{cp} is not a global symmetry at all, one faces the question of why it is almost always respected, except for weak interactions.
    \item \gls{cp} could also be a gauge symmetry. In this case, a domain wall would occur, where the domain wall is either inflated away \cite{dine_cp_1992} or destroyed by a dynamical process \cite{choi_is_1993}. \citet{mcnamara_reflections_2022} show that the latter is not possible, as the \gls{cp} domain wall is stable.
    \item If \gls{cp} is a global symmetry, it has to be broken (according to the no global symmetries conjecture (\cref{sec:nGSym})). In favour of \gls{cp} symmetry being a spontaneously broken symmetry \cite{hiller_solving_2001,dine_challenges_2015,albaid_strong_2015,hall_implications_2018,craig_p_2021,de_vries_minimal_2021,nir_solving_1996,aloni_spontaneous_2021,nakai_supersymmetric_2021,chakdar_collider_2013,dagnolo_finding_2016}, speaks the observation that it is only broken in the weak sector. If \gls{cp} is a spontaneously broken global symmetry, there is a stable domain wall \cite{mcnamara_reflections_2022,craig_p_2021}.
\end{itemize}
From a cosmological standpoint, a stable domain wall is problematic: it redshifts slower than matter or radiation, eventually dominating the energy content of the universe \cite{craig_p_2021}, unless the domain wall is inflated away, which requires $H_\text{infl}\lesssim m_\text{CP}$ \cite{mcnamara_reflections_2022}. Furthermore, a domain wall would carry a global charge, which is in tension with the no global symmetries conjecture (\cref{sec:nGSym}), unless it can decay in a topology-changing process \cite{mcnamara_reflections_2022}. The cobordism conjecture delivers us the perfect candidate: \gls{etw} branes. \citet{mcnamara_reflections_2022} show that a parity domain wall can dynamically turn into a pair of \gls{etw} branes in such a way, that the pair of branes carries gauge charge but no global charge. The \gls{etw} branes are \gls{uv} \gls{qg} objects that do not spoil our \gls{ir} observation that \gls{cp} is broken.

\paragraph{Inflation}
\citet{hertog_holographic_2024} study eternal inflation, applying holography and the \gls{ds}/\gls{cft} correspondence. In their model, the inflaton is described by a wave function similar to the Hartle\textendash Hawking no boundary proposal.
They show that in order for the large volume outcome to be classical, rather than dominated by quantum fluctuations, the inflaton field is restricted to sub-Planckian field ranges\footnote{
    This is a result from their holographic approach in line with the \gls{dc}.
}
around the minimum of its potential.
Holography excludes the regime of large field-ranges, which also violates the cobordism conjecture: there are no no-boundary saddle points that would allow for a cobordism to nothing. 
That the regions in tension with swampland conjectures are excluded by holography, acts as supportive evidence in favour of the swampland conjectures.

\subsubsection{General Remarks}
\paragraph{What is a (co)bordism?}
Two compact, $d$-dimensional manifolds $\mathcal{A}$ and $\mathcal{B}$ are called \textit{cobordant}, if a $\left(d+1\right)$-dimensional compact manifold $\mathcal{W}$ exists (called \textit{cobordism}), such that $\partial\mathcal{W}=\mathcal{A}\sqcup\mathcal{B}$ \cite{blumenhagen_dimensional_2023,prieto_moduli_2024}.

A cobordism is a weaker form of a diffeomorphism. While a diffeomorphism is a mapping between two manifolds on a point-level, a cobordism is a topology-changing sequence of quantum-gravitationally allowed dynamical operations to get from one manifold to another \cite{mcnamara_cobordism_2019}.

The terms \textit{bordism} and \textit{cobordism} are often used interchangeably, yet some authors make a distinction between the two concepts; we will use them interchangeably here, as \citet{tadros_low_2022} point out that the approaches are dual to each other, which should not change our findings regarding cosmology.

\paragraph{When does the Cobordism Conjecture lose validity?}
The current understanding limits the cobordism conjecture to compact spaces, i.e. in the limit Vol$_k\rightarrow\infty$, the cobordism conjecture loses validity \cite{andriot_looking_2022}.\footnote{
    In the asymptotic boundary, long-range gauge symmetries induce global symmetries, which are forbidden in \gls{qg} (see \cref{sec:nGSym}) \cite{agmon_lectures_2023}.
}

\paragraph{What role do charges play in the cobordism conjecture?}
The conjecture informs us that the cobordism charge is zero in a valid configuration \cite{buratti_dynamical_2021}. This implies that
\begin{itemize}
    \item the cobordism class for every theory of the landscape is trivial, i.e. that each quantum gravitational background is cobordant to another one, ergo there are no global symmetries \cite{velazquez_cobordism_2022,mcnamara_cobordism_2019,raman_swampland_2024,tadros_low_2022}.
    \item in a consistent theory of \gls{qg}, all compactifications are connected through manifolds, as otherwise there would be a global topological charge which corresponds to a global symmetry \cite{prieto_moduli_2024}.\footnote{
        In the gravitational path integral, we consider the different field configurations, which tend towards asymptotic boundary conditions and can therefore, according to the conjecture, always be connected asymptotically \cite{debray_chronicles_2023}.
        From an \gls{eft} perspective, the interpolating manifolds correspond to domain walls \cite{angius_intersecting_2023}.
    }
\end{itemize}
Regarding the latter point, imagine the following scenario \cite{ruiz_morse-bott_2025}:
You start with a spacetime $X_d$. From this spacetime, you replace a submanifold $\mathcal{A}_k\subset X_d$ with a submanifold $\mathcal{B}_k$ that belongs to a different cobordism class. Your new spacetime $X^\prime_d$ is no longer cobordant to $X_d$.
The cobordism class is like a label, and therefore represents a conserved, topological global charge.\footnote{
    \citet{tadros_low_2022} highlight how the global charge emerges: If the cobordism class is not the trivial one, each element of the class is a higher-dimensional manifold with a unique label. Adding fluxes or interactions will not change the label of the manifold, i.e. this cannot be gauged away, ergo it is a global symmetry.
}
This change cannot be inferred locally, far away from the change, ergo, it is not a local defect or can be gauged away\,\textemdash\,the spacetime itself has been altered, globally.
Introducing a global charge violates the no global symmetries conjecture (\cref{sec:nGSym}); ergo, only a trivial charge can be introduced, respectively the only allowed cobordism class is the trivial one (cf. \cite{rudelius_symmetry-centric_2024}).

\paragraph{What are the implications for our search for a theory of \gls{qg}?}
Assuming the \gls{wgc} and the \gls{dc} to be true would tell us that there is only a finite number of valid theories of \gls{qg}, assuming the cobordism conjecture would imply that the number of valid theories of \gls{qg} is actually \textit{one} \cite{hamada_finiteness_2022}.\footnote{
    That the number of string theories is finite is e.g. supported by the finding that the volume of moduli space is finite \cite{douglas_finiteness_2005}.
    See \cref{s:fnomf} for further comments on finiteness.
}

The cobordism conjecture is the expression of the view that \gls{qg} is unique in the sense that all \gls{qg} compactifications are equivalent to each other \cite{andriot_looking_2022}.\footnote{
    Similar in spirit is the Spinor\textendash Vector Duality, which is based on mirror symmetry and suggests that there is a single, connected moduli space for heterotic $\mathcal{N}=4$ string theory, within which all such theories are connected by orbifolds or continuous interpolation \cite{faraggi_spinorvector_2024}.
}
This goes further than the dualities well-known in string theory \cite{andriot_looking_2022}: the cobordism conjecture states that there is only a single equivalence-class of \gls{qg}, which is the trivial one, and that all consistent \gls{qg} backgrounds are representatives of this class. The cobordism conjecture explicitly allows different $k$-dimensional manifolds of different topologies. In particular, \cref{eq:cobordism} can have more than one solution. Furthermore, the bordism can be along the time direction, which makes the equivalence a dynamical process\,\textemdash\,this is typically not the case for string dualities (as these are spatial bordisms).

\subsubsection{Evidence}
The cobordism conjecture is proposed by \citet{mcnamara_cobordism_2019} and is studied in the context of
\gls{adscft} \cite{ooguri_cobordism_2020},
type IIB string theory \cite{debray_chronicles_2023,huertas_aspects_2023,dierigl_swampland_2021,dierigl_iib_2023},
$d>6$ supergravity theories \cite{montero_cobordism_2021},
Ricci flows\footnote{
    See the work by \citet{hamilton_three-manifolds_1982,perelman_entropy_2002,bykov_deformed_2021} for an introduction and current research on Ricci flows, as well as \cref{sec:distance} about the \gls{dc}, as this conjecture can be recast in terms of Ricci flows \cite{kehagias_swampland_2020,de_biasio_geometric_2023}.} \cite{velazquez_cobordism_2022},
tachyon condensation \cite{angius_dynamical_2022},
tachyon-free non-supersymmetric string theories \cite{basile_global_2024},
non-supersymmetric domain walls \cite{blumenhagen_dynamical_2022},
vacua with tadpoles \cite{buratti_dynamical_2021,blumenhagen_dynamical_2023},
Morse\textendash Bott theory \cite{ruiz_morse-bott_2025}, and
K-theory \cite{blumenhagen_dimensional_2023,blumenhagen_open-closed_2022}.
Relations to other swampland conjectures are highlighted in \cref{rel:CC_Cobord,rel:Cobord_nGSym,rel:DC_Cobordism,rel:Cobordism_nnSUSYC}.

\subsection{Completeness Conjecture}\label{sec:complet}%

The charge spectrum of a theory coupled to gravity must be complete, which means that charged matter in every representation of the gauge group must exist \cite{harlow_weak_2023,martucci_quantum_2023,palti_swampland_2019}, i.e. if there is a compact gauge group G, there are \enquote{physical states that transform in all finite-dimensional irreducible representations of G} \cite{harlow_symmetries_2019}.\footnote{
    A consequence is that while discrete gauge groups can be non-compact, continuous gauge groups must be compact to avoid global symmetries \cite{van_beest_lectures_2022} or having an infinite number of physical states \cite{tadros_low_2022}.
}

\subsubsection{Implications for Cosmology}
Magnetic monopoles must exist \cite{draper_snowmass_2022,polchinski_monopoles_2004}, but their mass could be close to the Planck mass.

\subsubsection{General Remarks}
Let us start with an example: If G = U(1), with allowed charges $Q = nq$ with $n\in\mathbb{Z}$, then there must be states with all such charges \cite{harlow_symmetries_2019}.
Take the global limit of a generalised symmetry $A_\mu\rightarrow A_\mu+\sigma_\mu$ with $\partial_{[\nu}\sigma_{\mu]}=0$, which is broken by charged matter, as the gauge covariant derivative is not invariant under this symmetry, only under the exact part, the local gauge symmetry \cite{rojo_swampland_2019}.
Another example would be $\mathbb{R}$ being the gauge group, such that irrational charge exists \cite{van_beest_lectures_2022}: Particles of rational charge cannot decay into particles with irrational charge (because of charge conservation). This implies that there is a global symmetry (see \cref{sec:nGSym} why this is problematic).
To avoid global symmetries, we expect states of all charges to be present \cite{rojo_swampland_2019}.
The mass of the states is not specified\,\textemdash\,all could have mass around the Planck mass \cite{palti_swampland_2019}. This is also true for the magnetic monopoles. Furthermore, the charged states could be multi-particle states or meta-stable bound states \cite{agmon_lectures_2023}.

\paragraph{What speaks in favour of the \gls{cc}?}
The \gls{cc} is supported by the observation that once a gauge field exists, the appropriate generalisation of the \gls{rn} solution of the gauge group exists as well \cite{harlow_symmetries_2019}. However, this observation might be deemed insufficient \cite{harlow_symmetries_2019}:
The solution can be a non-extremal two-sided wormhole with net charge 0.
If we demand that it is a charged \gls{bh}, we need charged matter to collapse into such a \gls{bh}\,\textemdash\,and if we have the charged matter, we don't need the \gls{bh} to satisfy the conjecture. If we want the conjecture to be satisfied without any a priori charged matter, it is unclear how to do this.

\paragraph{Where do the implications for the \gls{eft} come from?}
The \gls{cc} demands that for every $p$-form gauge field in \gls{qg}, there is a $p$-dimensional object charged under this symmetry. This poses consistency constraints on the worldvolume of these objects, which in turn has implications for the bulk \gls{eft}, as has been shown for supersymmetric theories in 10 \cite{kim_branes_2019}, 8 \cite{hamada_8d_2021,bedroya_compactness_2022}, 6 \cite{kim_branes_2019,lee_swampland_2019,tarazi_finiteness_2021,angelantonj_string_2020,cheng_anomalies_2022}, 5 \cite{katz_swampland_2020}, and 4 \cite{martucci_quantum_2023} dimensions as well as theories with 16 supercharges \cite{kim_four_2020}.

\subsubsection{Evidence}
The \gls{cc}, or completeness hypothesis, was first proposed by \citet{polchinski_monopoles_2004}, and is studied in the context of
F-theory \cite{lee_tensionless_2018},
type IIB $\mathcal{N}=1$ orientifolds with O3/O7 planes \cite{enriquez-rojo_swampland_2020},
4d $\mathcal{N}=2$ \glspl{eft} \cite{cecotti_special_2020},
\gls{adscft} \cite{harlow_symmetries_2019}, and
topology \cite{rudelius_topological_2020}.
Further motivation for the \gls{cc} comes from \glspl{bh} \cite{banks_symmetries_2011,harlow_universal_2022} and holography \cite{harlow_weak_2023,harlow_wormholes_2016,harlow_constraints_2019,harlow_symmetries_2019}.

The \gls{cc} is highly intertwined with the Cobordism Conjecture (\cref{sec:cobordism}) and the no Global Symmetries Conjecture (\cref{sec:nGSym}), and is therefore also related to the \gls{wgc}, which we highlight in \cref{rel:CC_Cobord,rel:CC_nGSymC,rel:WGC_CC}.

\subsection{de~Sitter Conjecture}\label{sec:deSitter}%

The absence of successful implementations of fully stable \gls{ds} vacua in string theory leads to the conjecture that there are no (meta-)stable \gls{ds} vacua.\footnote{
    See the review by \citet{anninos_sitter_2012} for a general introduction to \gls{ds} spacetimes.
}
Such vacua can be excluded by the following condition \cite{palti_swampland_2019,ooguri_distance_2019,denef_ds_2018,obied_sitter_2018,ooguri_distance_2019}:
\begin{equation}\label{eq:dSc}
\frac{\abs{\nabla V}}{V}\geq \frac{\mathfrak{s}_1}{M_\textnormal{P}}.
\end{equation}
The \gls{dsc} parameter in Planck units is $\mathfrak{s}_1\sim\order{1}$.\footnote{
    See our discussion in \cref{p:dSC_c_value} about the value of $\mathfrak{s}_1$.
    }
The gradient of the potential quite generally means $\sqrt{g^{ij}\partial_{\phi_i}V\partial_{\phi_j}V}$ \cite{andriot_sitter_2018}.
For $V\leq0$ or $M_\textnormal{P}\rightarrow\infty$, \cref{eq:dSc} is trivially satisfied \cite{ooguri_distance_2019,obied_sitter_2018,bernardo_contracting_2021}.

\Cref{eq:dSc} is a strong conjecture, as it not only rules out metastable \gls{ds} vacua but also local maxima and saddle points \cite{denef_ds_2018,roupec_flux_2021,roupec_sitter_2019}.
Various counterexamples to the conjecture in this strong form exist \cite{garg_bounds-dS_2019,quigley_gaugino_2015,polchinski_braneantibrane_2015,agrawal_cosmological_2018,hertzberg_inflationary_2007,andriot_sitter_2018,haque_minimal_2009,danielsson_towards_2009,danielsson_universal_2010,danielsson_sitter_2011,shiu_stability_2011,van_riet_classical_2012,murayama_we_2018,conlon_sitter_2018,denef_sitter_2018,roupec_flux_2021}.
For instance, the only known scalar field in our Universe\,\textemdash\,the Higgs field\,\textemdash\,violates \cref{eq:dSc}, as it has a maximum \cite{denef_ds_2018,murayama_we_2018,cicoli_sitter_2019,denef_sitter_2018,choi_ds_2018,hamaguchi_swampland_2018,cicoli_string_2023}.
A refined version %
states that either \cref{eq:dSc} is satisfied \textit{or} that
\begin{equation}\label{eq:dScrefined}
    \text{min}\left(\nabla^i\nabla_j V\right)\leq-\frac{\mathfrak{s}_2}{M_\textnormal{P}^2}V
\end{equation}
holds \cite{palti_swampland_2019},
with $\mathfrak{s}_2\sim\order{1}$ a positive constant,
and the left-hand side of \cref{eq:dScrefined} the minimum eigenvalue of the Hessian of $V$ in an orthonormal frame \cite{ooguri_distance_2019,garg_bounds_2019}.\footnote{
    For a constant field, the mass matrix is given by $m^i_j=\nabla^i\nabla_j V$, and negative eigenvalues indicate the presence of a tachyon \cite{andriot_tachyonic_2021}, i.e. an unstable extremum \cite{roupec_flux_2021}.
    To rule out tachyonic instabilities with lifetimes shorter than the age of the Universe in a highly curved spacetime, one can also focus on the mass term directly, which in this case does not automatically correspond to the lowest Eigenvalue of the Hessian, as the field curvature contributes to the mass term and $\left(\min m^2\right)/H^2\lesssim -\mathfrak{s}_2$ is the appropriate condition in this situation \cite{mizuno_hyperbolic_2019}.
}
Taken together, the two constraints ensure that only \gls{ds} minima are ruled out \cite{rasulian_swampland_2021} but not critical points in general \cite{van_beest_lectures_2022}.

We'd like to stress that even though \cref{eq:dScrefined} is the generally discussed refinement, it's not necessarily the only possibility to rule out \gls{ds} minima while allowing saddle points and maxima.\footnote{
    \citet{garg_bounds_2019} note that, in principle, the second order derivatives could be allowed to vanish at a critical point, as long as there are higher derivatives that ensure a downward flow of the trajectory. However, no examples of such critical points have been found.
    Since the $\order{1}$ parameters are supposed to ensure naturalness of the solution and avoid fine-tuning, the \gls{dsc} could also be formulated as 
    the condition $V^{(n)}/V\sim\order{1}$ with $V^{(n)}$ the highest-order \textit{non-zero} derivative \cite{garg_bounds_2019}.
}
In fact, the refined \gls{dsc} has three shortcomings:
\begin{enumerate}
    \item its limits do not lead to trivial constraints
    \begin{itemize}
        \item $M_\textnormal{P}\rightarrow\infty$ does not trivialise the constraint,
        \item $V^{\prime\prime}\rightarrow0$ leads to an inconsistency for $V>0$,
        \item $V\rightarrow0$ leads to negative mass, i.e. a tachyonic instability if there is an extremum for $V>0$ \cite{bernardo_contracting_2021}.
    \end{itemize}
    \item the two conditions are connected by a logical \textit{or}.
    \item the scenario where both $\mathfrak{s}_1$ and $\mathfrak{s}_2$ are intermediate values remains unaddressed by the conjecture; a scenario that can happen in string theory at the tree level \cite{garg_bounds_2019,blaback_accelerated_2014}.
\end{enumerate}
To address these shortcomings, \citet{andriot_further_2019,roupec_flux_2021} combine \cref{eq:dSc,eq:dScrefined}, and present the \textit{combined \gls{dsc}}\footnote{
    \Cref{eq:dSCcombi} is sometimes called the \textit{further refining de~Sitter swampland conjecture} \cite{sadeghi_sitter_2023}. We will refer to \cref{eq:dSCcombi} as the \textit{combined \gls{dsc}}.
}
\begin{equation}\label{eq:dSCcombi}
    \left(M_\textnormal{P}\frac{\abs{\nabla V}}{V}\right)^\mathfrak{c} - \mathfrak{p}_2M_\textnormal{P}^2\frac{\text{min}\nabla_i\nabla_j V}{V}\geq \mathfrak{p}_1,
\end{equation}
with $\mathfrak{c}>2$, $\mathfrak{p}_1,\,\mathfrak{p}_2>0$, and $\mathfrak{p}_1+\mathfrak{p}_2=1$, which should hold at any point in field space where the potential is positive.\footnote{
    Numerical results from flux compactifications show that $\mathfrak{p}_2\geq0.286$ holds, for which $\mathfrak{c}\leq3.03$ is required, yet higher values of $\mathfrak{p}_2$ allow for higher values of $\mathfrak{c}$ \cite{roupec_flux_2021}. 
}
In terms of the parameters $\mathfrak{s}_1$ and $\mathfrak{s}_2$, the combined \gls{dsc} can be expressed as \cite{ahmed_supersymmetric_2024}
\begin{equation}
    \mathfrak{s}_1^\mathfrak{c}+\mathfrak{p}_2\left(1+\mathfrak{s}_2\right)>1.
\end{equation}
This new relation is not exactly equal to the bounds presented by \cref{eq:dSc,eq:dScrefined} but overlapping for a large part of the parameter space \cite{roupec_flux_2021}.\footnote{
    See \cref{f:dSCcombi} in \cref{p:dSC_combined}.
    }

\subsubsection{Implications for Cosmology}
We'd like to note here that the \gls{dsc} has particularly strong implications for cosmology, while the theoretical motivation for the \gls{dsc} is comparatively weak, as we discuss in \cref{p:dSC_Counterexamples,p:dSC_Motivation,f:dS_lifetime}. Most importantly, there are no proven no-go theorems against fully stable \gls{ds} solutions. 
Furthermore, the \gls{dsc} is supposed to hold in the asymptotic region of scalar field space, yet cosmology happens in the bulk. We will see that a plethora of models are in tension with the $\order{1}$ values of $\mathfrak{s}_{1,2}$, yet would be compatible with $\mathfrak{s}_{1,2}\lesssim\order{0.1}$.
Following our spirit \textit{it is easier to loosen an overly tight bound than to tighten a loose model}, we will take the \gls{dsc} at face value and apply the scalar field boundary constraints in the bulk. Therefore, we are overly cautious.

\paragraph{Axions}\label{p:dSC_Axions} with a potential of the form
\begin{equation}
    V\sim1-\cos\left(\phi/f\right)
\end{equation}
face constraints, in particular around the maximum of the potential.
Smaller values of $\mathfrak{s}_1$ and $\mathfrak{s}_2$ increase the range of compatibility for the values of $\phi/f$.
Furthermore, the field displacement is restricted to be sub-Planckian by the \gls{dc},
while the axion decay constant $f$ is restricted to be sub-Planckian by the \gls{wgc}.
To satisfy all three conjectures simultaneously might require fine-tuning \cite{ibe_quintessence_2019} or a dynamical mechanism \cite{reig_stochastic_2021}. For example, \citet{gasparotto_cosmic_2022} use the birefringence\footnote{
    Cosmic birefringence, i.e. the rotation of linearly polarised light caused by a phase velocity difference between left- and right-handed photons, is difficult to measure, as the calibration uncertainty of Planck's polarimeter ($\delta\beta\approx\SI{0.28}{\degree}$) is of the same order as the measured effect ($\SI{0.3}{\degree}\sim\SI{0.45}{\degree}$), which does not yet take into account systematics such as foreground dust \cite{gasparotto_cosmic_2022,eskilt_improved_2022,diego-palazuelos_cosmic_2022,sullivan_planck_2025,mohammadi_cross-correlation_2023,sherwin_cosmic_2023,namikawa_cmb_2021,clark_origin_2021,nagata_numerical_2021,minami_new_2020,minami_simultaneous_2019,komatsu_new_2022}.
 } angle in the Planck 2018 data set to constrain the axion decay constant to $f\lesssim\SI{e16}{\giga\electronvolt}$, if the axion plays the role of \gls{de} and has a monodromy potential.\footnote{
    This constraint rules out the axi-Higgs proposal \cite{fung_axi-higgs_2021}, which requires a higher decay constant, yet leaves a window of compatibility for an \gls{ede} proposal where the axion couples to the photon \cite{fujita_detection_2021}.
    }
While this bound might satisfy the \gls{wgc}, the \gls{dsc} is violated for their model, as they derive an upper bound of $\mathfrak{s}_1\lesssim\num{4e-8}$, which is far below the expected $\order{1}$ value.

\citet{cicoli_sitter_2019} apply the \gls{dsc} to an axion model, and we summarise their findings in the following:
Axions as \gls{de} in the form of quintessence come with the benefit that the shift symmetry naturally prevents large quantum corrections to the potential. Moreover, ultra-light axions in the form of pseudo-scalars evade fifth-force constraints. Axions become ultra-light if the corresponding saxions ($\varphi$) receive large mass contributions from perturbative effects, as $m_a\sim\exp(-\varphi)$. A rather general form of an axion potential is
\begin{equation}
    V=\Lambda_\text{cc}^4-\sum_{i=1}^{N_a}\Lambda_i^4\cos\left(\frac{a_i}{f_i}\right)+\dots,
\end{equation}
with $\Lambda_\text{cc}$ the cosmological constant,
$N_a$ the number of ultra-light axions,
$\Lambda_i$ the scale of the non-perturbative effects that gives rise to the axion,
$a_i$ the field value of the $i$-th axion,
$f_i$ the axion decay constant,
and the ellipsis indication that this is only the leading order of the non-perturbative potential.
The potential has a minimum at $\expval{V}=\Lambda_\text{cc}^4-\sum_i\Lambda_i^4$, an inflection point at $\Lambda_\text{cc}^4$, a maximum at $\Lambda_\text{cc}^4+\sum_i\Lambda_i^4$, and $2^{N_a-1}$ saddle points.
The large Hubble friction freezes every axion into its field value after inflation, until the Hubble scale drops below the mass scale and the axion starts to roll and oscillate around its minimum.
Even though all axions in the theory are ultra-light, one will be the lightest and all others can be integrated out, which yields a slow-roll condition that depends only on the lightest axion, $a_\textnormal{l}$:
\begin{align}
    \epsilon_V&\defeq\frac{(\nabla V)^2}{V^2}\\
    &=\frac{1}{2}\left[\left(\frac{\Lambda_\textnormal{l}}{\Lambda_\text{cc}}\right)^4\frac{1}{f_\textnormal{l}}\right]^2\frac{\sin^2\left(\frac{a_\textnormal{l}}{f_\textnormal{l}}\right)}{\left(1-\left(\frac{\Lambda_\textnormal{l}}{\Lambda_\text{cc}}\right)^4\cos(\frac{a_\textnormal{l}}{f_\textnormal{l}})\right)^2}\\
    &<1.
\end{align}
Choosing different values for the model parameters leads to different verdicts about the axion models:\footnote{See also \cref{sss:WGC_Cosmology} on axions and their \gls{wgc}-compatibility.}
\begin{description}
    \item[Alignment] occurs when the minimum of the potential vanishes, i.e. $\Lambda_\text{cc}=\Lambda_\textnormal{l}$, which requires a super-Planckian axion decay constant, which violates the \gls{dc}, unless it is achieved as an effective decay constant in assisted quintessence or through other multi-field mechanisms \cite{cicoli_sitter_2019,kim_completing_2005,shiu_large_2015,dimopoulos_n-flation_2005,cicoli_n-flation_2014,reig_stochastic_2021}.
    \item[Hilltop quintessence] occurs if the region of the potential around the maximum is at a positive energy level and the field starts close to the maximum. The expectation is that, given a high enough number of axion fields, one will be around the maximum after inflation, as it is easy to displace the ultra-light fields from their minimum. However, even if that happens, it needs fine-tuning to get enough $e$-foldings while keeping the axion decay constant sub-Planckian. Hilltop quintessence violates the \gls{wgc} and \cref{eq:dSc}, but not the refined \gls{dsc}.
    Similar results with an axion close to the hilltop were derived in a \gls{kklt} scenario: in order to be compatible with the \gls{dsc}, fine-tuning, especially about the initial conditions, is necessary. Interestingly, if the fine-tuning takes place and \cref{eq:dScrefined} is satisfied, \cref{eq:dSc} is satisfied away from the saddle \cite{emelin_axion_2019}.
    
    Also \citet{cicoli_quintessence-numerically_2022-1} find that axionic quintessence hilltop inflation requires very finely tuned initial conditions, as well as a very low inflationary scale of $H_\textnormal{I}\lesssim\SI{1}{\mega\electronvolt}$.
    \item[Quasi-natural quintessence] occurs if the minimum of the potential corresponds to $\Lambda_\text{cc}$, which yields a slow-roll condition of $f_\textnormal{l}\gtrsim\left(\Lambda_\textnormal{l}/\Lambda_\text{cc}\right)^4$ that is satisfied as long as $\Lambda_\text{cc}\gg\Lambda_\textnormal{l}$. Quasi-natural quintessence violates the \gls{dsc}.
    \item[Oscillating scalars] occur when $m_a\simeq H$. The field oscillates around its minimum, which violates the \gls{dsc}.
\end{description}

\paragraph{Boson stars} are found to be compatible with the \gls{dsc} \cite{choi_probing_2019} as long as they are stable \cite{herdeiro_compact_2019}. Unstable boson star solutions belong to the swampland.

\paragraph{Dark Energy} 
can take many forms, from a cosmological constant \cite{martin_everything_2012}
to scalar fields \cite{tsujikawa_quintessence_2013,dent_time_2009,riazuelo_cosmological_2002,perrotta_extended_1999,uzan_cosmological_1999,armendariz-picon_essentials_2001,armendariz-picon_dynamical_2000,caldwell_cosmological_1998,wetterich_cosmology_1988,ratra_cosmological_1988,copeland_dynamics_2006,eskilt_cosmological_2022,akrami_multi-field_2021,dallagata_warm_2020,hardy_thermal_2020,mukhopadhyay_swampland_2019,david_marsh_swampland_2019,park_reconstructing_2021,emelin_axion_2019,odintsov_finite-time_2019,heisenberg_dark_2018,Heisenberg_Dark_II_2018,agrawal_cosmological_2018,han_quintessence_2019,wang_electroweak_2019,caldwell_limits_2005,astashenok_scalar_2012,sahni_cosmological_2002,gasperini_quintessence_2001,faraoni_inflation_2000,barreiro_quintessence_2000,wang_cosmic_2000,zlatev_quintessence_1999,carroll_quintessence_1998,schoneberg_news_2023,brax_swampland_2020,freigang_cosmic_2023,barrau_string_2021,ratra_cosmological_1988,ferreira_cosmology_1998,frieman_cosmology_1995,hebecker_f_2019,raveri_swampland_2019,akrami_landscape_2019,trivedi_implications_2021,ben-dayan_draining_2019,barrau_string_2021,colgain_testing_2019,akrami_landscape_2019,raveri_swampland_2019,bhattacharya_cosmological_2024,alestas_curve_2024,park_reconstructing_2021,payeur_swampland_2024,akrami_multi-field_2021,garg_bounds-dS_2019,eskilt_cosmological_2022,eskilt_cosmological_2022,akrami_multi-field_2021,cicoli_sitter_2019,storm_swampland_2020,cicoli_new_2020,gallego_anisotropic_2024,sharma_reconstruction_2021,han_quintessence_2019,tosone_constraints_2019,nitta_dynamical_2025,brinkmann_stringy_2022,zhai_uplifting_2020,oikonomou_occurrence_2021,cicoli_out_2020,chiang_building_2018,linder_pole_2020,yang_model-independent_2020,mishra_cosmological_2023,gasparotto_cosmic_2022,bramberger_non-singular_2019,matsui_isocurvature_2019,singh_model_2024,baldes_forays_2019,olguin-trejo_runaway_2019}
and modified gravity \cite{arjona_effective_2020,arjona_designing_2019,arjona_unraveling_2019,hu_effective_2014,gubitosi_effective_2013,clifton_modified_2012,dixit_stability_2020,arjona_machine_2021,nojiri_introduction_2007,nojiri_unified_2011,nojiri_modified_2017,olmo_palatini_2011,capozziello_extended_2011,capozziello_curvature_2002,sussman_cotton_2023,artymowski_fr_2019,brahma_dark_2019}.
A successful model of \gls{de} explains the following observational constraints \cite{artymowski_emergent_2021}:
\begin{description}
    \item[\gls{de} density  at \gls{bbn}] $\rho_\textnormal{DE}/\rho_\textnormal{r}\leq0.086$.
    \item[\gls{de} density after inflation] subdominant \gls{de}, with a matter dominated epoch between the radiation dominated epoch and today's \gls{de} dominated phase.
    \item[\gls{de} density today] $\rho_0\simeq\num{1.7e-119}M_\textnormal{P}^4$.
    \item[\gls{de} \gls{eos} today]
        $w=p/\rho\in\left(-1.14,-0.94\right)$.\footnote{
        For a representation of \gls{de} as a fluid with pressure $p$ and energy density $\rho$, parametrised by $\omega$, the \gls{eos} reads $p=\omega\rho$. \citet{kopeikin_dynamic_2014} stress that the representation of a scalar field as a fluid is a rather formal one\,\textemdash\,not all thermodynamic properties of an ideal fluid are fulfilled by a scalar field, e.g. the sound speed is capped at the speed of light, irrespective of the formal prediction using the \gls{eos}.
    }
\end{description}
The \gls{lcdm} model satisfies those constraints, but it is incompatible with the \gls{dsc}, as $\nabla V/V=0\ngtr0$ for a cosmological constant \cite{agrawal_cosmological_2018,colgain_hint_2019}.
To explain the observed accelerated expansion of our Universe with $V>0$ and $\nabla V\neq0$, \gls{de} must be dynamical, i.e. the \gls{eos} is time-dependent \cite{palti_swampland_2019,agrawal_cosmological_2018}.
Current cosmological observations give us some boundaries for the \gls{eos} parameter $\omega$. The \gls{dsc} puts further constraints on $\omega$, namely that $1+\omega\gtrsim0.15\mathfrak{s}_1^2$ should hold to be compatible with the \gls{dsc} while avoiding a suppression of \gls{lss} growth \cite{agrawal_cosmological_2018}.\footnote{
    The \gls{dsc} constrains $\omega\simeq\left(\dot{\phi}^2/2-V(\phi)\right)\left(\dot{\phi}^2/2+V(\phi)\right)$ directly through $V$ and indirectly through $\partial_\phi V \sim \DAlambert\phi\sim\ddot{\phi}\sim\dot{\phi}^2/\phi$ \cite{palti_swampland_2019}.
}

The compatibility between the \gls{dsc} and observations is discussed using model-independent reconstructions:
\begin{itemize}
    \item \citet{arjona_machine_2021} find the \gls{dsc} compatible with low redshift data.
    \item A parametrisation-dependent Bayesian machine learning algorithm that reconstructs $H(z)=H_0+H_1z^2/\left(1+z\right)$ finds that \cref{eq:dSc} is respected while \cref{eq:dScrefined} is violated \cite{elizalde_interplay_2021}.\footnote{
        The \gls{dc} is satisfied. However, the model shows a sign switch in the \gls{de} \gls{eos} in the redshift range $z\in\left[0,\,5\right]$, i.e. \gls{de} is in the phantom regime for some time.
    } 
    \item The model-independent Gaussian process used by \citet{khurshudyan_swampland_2023} shows a tension between reconstructed \gls{de} and the \gls{dsc} in the low redshift range $z\in\left[0,1\right]$.\footnote{
         Furthermore, the relative \gls{dm} energy density is strongly dependent on the used kernel, ranging from $\Omega_\textnormal{DM}\simeq\num{0.262(0.011)}$ to $\Omega_\textnormal{DM}\simeq\num{0.293(0.013)}$, while the relative energy density of radiation ranges from $\Omega_\gamma\simeq\num{0.00013(0.00002)}$ to $\Omega_\gamma\simeq\num{0.00023(0.00002)}$.
    }
    \item The claim by \citet{akrami_landscape_2019} that \gls{de} models with $\mathfrak{s}_1>1$ are ruled out with $3\sigma$ by observational constraints is not supported by a model-independent analysis of \gls{desi} data \cite{arjona_swampland_2024}.\footnote{
        Concretely, \citet{akrami_landscape_2019} test models of the form $V(\phi)\sim\exp(-\mathfrak{s}_1\phi)$, for which they find $\mathfrak{s}_1\leq1.02$ is needed to be compatible with observational constraints. They don't find cosmologically viable solutions for models with double exponential potentials in M-theory compactifications, for the $O(16) \cross O(16)$ heterotic string, and for type II string theory. Observed issues with those models stem from unknown quantum corrections, decompactification of extra-dimensions in cosmologically relevant regions of the parameter space, and fifth forces that are incompatible with the \gls{wgc}.
    }
\end{itemize}
That contradicting verdicts are derived is predicted by \citet{schoneberg_news_2023}. They identify two major pitfalls of model-independent analyses: On the one hand, the smoothness of the expansion history of the Universe has to be modelled, and the appropriate number of derivatives has to be taken into account. This emulation can differ. On the other hand, it is not trivial to enforce positivity of kinetic terms and field energy density. The results depend on the used method to achieve these. The two pitfalls together can lead to different conclusions.

Concrete models are for example assessed by \citet{schoneberg_news_2023}, who find several explicit \gls{de} potentials that are in at least slight tension with the \gls{dsc}, or \citet{freigang_cosmic_2023}, who find it challenging to fulfil the \gls{dsc}, even in multi\textendash scalar field settings. However, they \cite{freigang_cosmic_2023} conclude that having a large $\eta_V$ parameter makes it easier to satisfy the \gls{dsc}.
In the following, we make some remarks about various models that are constrained by the \gls{dsc}:
\begin{itemize}
    \item In analogy to warm inflation (see \cref{sp:dSC_Warm-Inflation}), \citet{dallagata_warm_2020} present warm \gls{de}, where an axion field is slow-rolling down a steep potential while experiencing friction induced by a coupling to a U(1) gauge field. Due to the additional friction term in the \gls{eom}, the potential can satisfy the \gls{dsc}, while the observations can be matched by a suitable choice for the axion decay constant $f$ \cite{dallagata_warm_2020}:
    \begin{align}
        S&=\int\!\sqrt{-g}\left(-\frac{\left(\partial\phi\right)^2}{2}-V-\frac{F_{\mu\nu}F^{\mu\nu}}{4}\right.\nonumber\\
        &\phantom{\ =\int\!\sqrt{-g}\left(\right.}\left.-\frac{\phi}{4f}F_{\mu\nu}\tilde{F}^{\mu\nu}\right)\,\mathrm{d}^4x\\
        0&=\partial_\tau^2\phi+\frac{2}{a}\frac{\partial a}{\partial\tau}\frac{\partial\phi}{\partial\tau}+a^2\frac{\partial V}{\partial\phi}-\frac{a^2}{f}\vec{E}\cdot\vec{B}\\
        \vec{E}\cdot\vec{B}&=\frac{1}{4\pi^2a^4}\int\!k^3\partial_\tau\abs{A}^2\,\mathrm{d}k.
    \end{align}
    The quantities have their usual meaning: $F_{\mu\nu}$ is the gauge field strength tensor, $\tilde{F}_{\mu\nu}$ its dual, $a$ is the scale factor, $\tau$ the conformal time, and $k$ is the comoving momentum of the gauge field $A$.
    The model is also studied by \citet{papageorgiou_gravitational_2020} and found to satisfy the \gls{dsc} and observational constraints, including \gls{gw} production with a frequency of $\SI{e-16}{\hertz}$.
    \item Warm \gls{de} coupled to a non-Abelian SU(2) field is studied by \citet{papageorgiou_gravitational_2020} and found to satisfy the \gls{dsc} in a range compatible with observations. While in warm U(1) \gls{de} the additional friction is caused by tachyon production and backreactions of the gauge field, in warm SU(2) \gls{de}, the friction is classical. This model shares similarities with chromonatural inflation (see \cref{p:TCC_Inflation,p:chromonatural_inflation}).
    \item Chameleon models studied by \citet{casas_cosmological_2024} enable short phases of up to one $e$-fold of accelerated expansion while keeping the field excursion sub-Planckian if heavy states are present that stabilise the scalar potential during that time. The density of the heavy states is diluted by the expansion, which limits the lifetime of this phase. They highlight that there are significant obstacles to implement such models within string theory, as these requires a flat region in the otherwise steep potential.
    
    In chameleon models studied by \citet{brax_swampland_2020}, where \gls{de} is coupled to matter such that
    \begin{align}
        m&=A(\phi)m_0\\
        \frac{\dot{\phi}^2}{2}&=V(\phi)+\omega\phi\rho_\textnormal{eff}\\
        V(\phi)&=\left(1-\omega_\phi\right)\frac{\phi_\textnormal{eff}}{2}-\left(A-1\right)\frac{\rho}{2}\\
        V_\textnormal{eff}&=V(\phi)+\left(A(\phi)-1\right)\rho_\textnormal{m},
    \end{align}
    the coupling to matter
    \begin{align}
        \beta&\defeq M_\textnormal{P}\frac{\partial\log A}{\partial\phi}
    \end{align}
    is bounded by the \gls{dsc} to
    \begin{align}
        \beta&\geq\frac{\mathfrak{s}_1\left(1-\omega_\phi\right)}{2A_\textnormal{max}}\frac{\rho_\textnormal{eff}}{\rho}-\mathfrak{s}_1\frac{\Delta A}{2A_\textnormal{max}},
    \end{align}
    where $A_\textnormal{max}$ is the maximal value $A$ can take, either because the function has a maximum or because the field displacement is capped by the \gls{dc}.
    For a \gls{de}-dominated universe with $\rho_\textnormal{eff}/\rho=\Omega_\Lambda/\Omega_\textnormal{m}$,
    very small field displacement, 
    and $\omega_\phi\approx-1$,
    they find that 
    $\beta\gtrsim\order{1}$, i.e. another $\order{1}$ bound related to the \gls{dsc} and the \gls{dc}. This puts such chameleon models with below-unity \gls{de}\textendash matter coupling and sub-Hubbleian field mass in the swampland \cite{brax_swampland_2020}.
    The observed absence of fifth-forces puts an upper bound on the coupling constant of $\beta\lesssim\num{e4}$ \cite{brax_swampland_2020}.
    Some concrete models with screened scalar-interactions in scalar\textendash tensor theories are the following \cite{brax_swampland_2020}:
    \begin{itemize}
        \item Chameleon models track the minimum of an effective potential for most of cosmic history. To be compatible with fifth-force and swampland constraints, chameleon fields must vanish in the future, as either large couplings $\beta$ are required, which would be in contradictions with the observed smallness/absence of fifth forces, or super-Planckian field excursions, which would be in contradiction with the \gls{dc}.
        \item Symmetron models have  Higgs-like potentials of the form $V(\phi)=V_0-\mu^2\phi^2/2+\mathfrak{c}\phi^4/4$ and a coupling to matter of the form $A(\phi)=1+\phi^2/m^2$ respectively $\beta\defeq M_\textnormal{P}\frac{\partial\log A}{\partial\phi}=\phi M_\textnormal{P}/m$ with the vacuum value constraint by the \gls{dsc} to be $\beta_0\geq \mathfrak{s}_1\Omega_{\Lambda,0}/\Omega_{\textnormal{m},0}$. To be compatible with fifth-force and swampland constraints, scalar fields must vanish in the future to satisfy $\phi(\rho)\rho/m^2\geq \mathfrak{s}_1V(\phi)/M_\textnormal{P}$.
        \item Dilaton models, where a dilaton tracks the minimum of the effective potential, have potentials of the form $V_0\exp(-\phi)+\order{\exp(-2\phi)}$, and  couplings $\beta\gtrsim \mathfrak{s}_1\rho_\Lambda/\left(\rho+4\rho_\Lambda\right)$ respectively coupling functions $A=l_\textnormal{s}\exp\left(\Psi(\phi)\right)/\sqrt{8\pi G_\textnormal{N}}$ with $l_\textnormal{s}$ the string scale and $\Psi(\phi)$ the matter field. Such models have a patch in parameter space that is compatible with observational constraints, the \gls{dsc}, and the \gls{dc}.\footnote{
            \gls{de} described by strongly coupled dilatons can be eternal \cite{brax_dilaton_2010}.
        }  
    \end{itemize}
    \item A $f(R)=R-\mathfrak{p}R^k$ \gls{de} model with $k\in\left(0,1\right)$ is found to yield $V^\prime/V=0$ around the minimum, which renders it incompatible with the \gls{dsc} \cite{artymowski_fr_2019}.
    \item A holographic model of emergent \gls{de}, where \gls{de} is a holographic fluid on the 4d \gls{flrw} hypersurface in a 5d Minkowski bulk, is found to be compatible with \gls{sn} Ia and $H(z)$ data as well as the \gls{dsc} up to the present epoch, but will violate the \gls{dsc} in the future \cite{cai_emergent_2019}.
\end{itemize}

\subparagraph{\gls{ede}}
In \cref{p:DC_EDE} we discuss an \gls{ede} model by \citet{mcdonough_early_2022} that reduces the Hubble tension. The model is in tension with the \gls{dc}, yet also with the \gls{dsc}, as $\nabla V/V\lesssim0.1<\mathfrak{s}_1\sim\order{1}$.

\subparagraph{Quintessence\footnote{
    The term \textit{quintessence} for a fifth energy component in our Universe, besides cold \gls{dm}, baryonic matter, photons, and neutrinos, was first used by \citet{caldwell_cosmological_1998}.
}}\label{sp:dSC_Quintessence}
The action, Friedmann equation, and \gls{eom} of quintessence
are given by
\begin{AmSalign}
    S&=\int\!\sqrt{-g}\left(\frac{R}{2}+\frac{1}{2}G_{ab}\partial_\mu\phi^a\partial^\mu\phi^b-V(\phi)\right)\,\mathrm{d}^4x\label{eq:Q-action}\\
    3H^2&=\frac{1}{2}G_{ab}\dot{\phi}^a\dot{\phi}^b+V+\rho_\textnormal{m}+\rho_\textnormal{r}\label{eq:Q-Friedmann}\\
    0&=\nabla_t\dot{\phi}^a+3H\dot{\phi}^a+G^{ab}V_{,\phi^b},\label{eq:Q-EoM}
\end{AmSalign}
with $g_{\mu\nu}$ the spacetime metric,
$G_{ab}$ the field space metric,
$R$ the Ricci scalar,
$H$ the Hubble parameter,
$\phi$ the quintessence field,
$V(\phi)$ its potential,
prime denoting a derivative with respect to the scalar field and
an overdot a derivative with respect to time,
and the Christoffel symbol in the covariant derivative is with respect to the field space metric $G_{ab}$.

To assess the viability of quintessence models in the light of the \gls{dsc}, several aspects have to be considered:
\begin{description}
    \item[Field value] The model could satisfy \cref{eq:dSc} for some part of the field trajectory and \cref{eq:dScrefined} for another part, but the parts could be disconnected, restricting the allowed field ranges and values.
    \item[\gls{dsc} parameters] Even for field values that are theoretically allowed, observational data might put bounds on the parameters $\mathfrak{s}_1$ and $\mathfrak{s}_2$.
    \item[Other swampland conjectures] The field range is restricted to remain sub-Planckian by the \gls{dc}. The \gls{dsc} parameter is restricted by the \gls{tcc}.\footnote{See \cref{p:dSC_c_value} for the \gls{tcc} bound $\mathfrak{s}_1\geq2/\sqrt{\left(d-1\right)\left(d-2\right)}$ and other bounds.}
\end{description}
The second Friedmann equation for a \gls{flrw} universe
\begin{align}
    \frac{\ddot{a}}{a}&=-\frac{1}{6}\left(2\rho_\textnormal{r}+\rho_\textnormal{m}+\rho_\phi+3p_\phi\right)\\
    p_\phi&=\frac{1}{2}\dot{\phi}^2-V
\end{align}
indicates that the potential energy of quintessence has to dominate over the other energy components of the Universe to yield accelerated expansion \cite{barrau_string_2021}.
Models that assure a late-time dominance relatively independent of initial conditions are
freezing models,\footnote{
    An example of a freezing model is the Ratra\textendash Peebles potential $V=\Lambda^{4+\mathfrak{p}}\phi^{-\mathfrak{p}}$, with $\Lambda$ some energy scale and $\mathfrak{p}>0$ \cite{barrau_string_2021,ratra_cosmological_1988}.
    However, such models are disfavoured by the \gls{dsc} \cite{tosone_constraints_2019} and violate the \gls{dsc} in the infinite future when $V^\prime/V\rightarrow0$ \cite{barrau_string_2021}.
    Another example of a freezing model is an exponential potential of the form $V\sim\exp\left(-\mathfrak{c}\phi\right)$, with $\mathfrak{c}$ a constant \cite{barrau_string_2021,ferreira_cosmology_1998}.
}
where the motion of the field slows down at late times when the potential becomes flat, and
thawing models,\footnote{
    Thawing models, where a field is initially damped by Hubble friction to $\dot{\phi}_\textnormal{i}=0$ and $\omega$ is increasing over time with $\omega_\textnormal{i}\approx-1$ \cite{storm_swampland_2020}, typically have potentials $V=V_0\cos\left(\phi/\mathfrak{c}\right)$ or $V=V_0\left(1+\cos\left(\phi/\mathfrak{c}\right)\right)$ \cite{barrau_string_2021,frieman_cosmology_1995}, and satisfy either \cref{eq:dSc} or \cref{eq:dScrefined} \cite{barrau_string_2021} for the different values that $\phi$ can take, i.e. thawing models are compatible with the \gls{dsc} in general.
}
where the field is initially damped by Hubble friction \cite{barrau_string_2021}.
In a model-agnostic reconstruction of quintessence from observational data, no preference for thawing or freezing models was derived; the data seems to indicate an oscillation of the \gls{eos} parameter around $w=-1$ \cite{park_reconstructing_2021}.
\citet{tada_quintessential_2024} derive a thawing quintessence model respectively an axion-like quintessence by a reconstruction of \gls{desi}-data for $\omega_0\omega_a$CDM. Both potentials satisfy the \gls{dsc} (as well as the \gls{dc} and the \gls{wgc}). 

Simple quintessence models are often disfavoured by theoretical considerations \cite{hebecker_f_2019} or by observational constraints, especially if the \gls{dc} is taken into account \cite{raveri_swampland_2019,akrami_landscape_2019,trivedi_implications_2021,ben-dayan_draining_2019,barrau_string_2021,brinkmann_stringy_2022} 
or $\mathfrak{s}_1\sim\sqrt{2}$ from the asymptotic limit of string theories is demanded \cite{colgain_testing_2019,akrami_landscape_2019,raveri_swampland_2019,bhattacharya_cosmological_2024,alestas_curve_2024,park_reconstructing_2021,schoneberg_news_2023}.
Many simple, exponential single-field quintessence models favour a small value of $\mathfrak{s}_1\sim\order{\num{e-1}}\text{\ \textendash\ }\order{\num{e-2}}$ to satisfy the observational constraints, which is in tension with the \gls{dsc} assumption of $\mathfrak{s}_1\sim\order{1}$.
For example, \citet{tosone_constraints_2019} find $\nabla V/V\lesssim0.31$ for thawing quintessence with exponential potentials and $\nabla V/V\lesssim0.54$ for models with double exponential potentials, yet these bounds can be eased by using different observational data sets.
It should be noted that \cref{eq:dScrefined} is often not considered in the literature cited above.
The refined \gls{dsc} \cref{eq:dScrefined} is, for example, satisfied by runaway quintessence, where a runaway scalar potential is frozen by Hubble friction near its hilltop maximum, constituting rolling quintessence \cite{olguin-trejo_runaway_2019}. Fifth forces are avoided if the modulus is localised in the extra dimensions, away from the \gls{sm} \cite{olguin-trejo_runaway_2019}. The model can be tuned to have an \gls{eos} of $\omega=-1$ in the past, from which it slowly deviates, i.e. it is currently increasing (meaning becoming less negative).

\citet{emelin_axion_2019} find axion-like quintessence with a hilltop potential to satisfy the \gls{dsc}: either the potential satisfies \cref{eq:dSc} at all times, yet violates the \gls{dc}, or the potential has a saddle point, respectively, a local minimum in the axionic direction. In the latter case, \cref{eq:dSc} is satisfied except for the saddle point, but there \cref{eq:dScrefined} is satisfied. The \gls{dc} is satisfied as well.

Focusing on \cref{eq:dSc} could be somewhat excused by the finding that the \gls{dsc} in the form of \cref{eq:dSc} favours higher values of $H_0$, whereas the refined conjecture in the form of \cref{eq:dScrefined} favours models that lead to a lower value of the Hubble parameter (compared to the \gls{lcdm} model), at least for the generic quintessence models \citet{banerjee_hubble_2021} studied. To solve the Hubble crisis, the former would be required.
However, it might be the case that single-field quintessence is not able to resolve the Hubble crisis, unless there is a non-minimal coupling \cite{banerjee_hubble_2021}. This seems to be a more general finding: A coupling is actually expected, unless it is broken by a symmetry \cite{carroll_quintessence_1998}.
Although \citet{yang_model-independent_2020} warn that the available observational data might be insufficient to make a statement about the refined \gls{dsc} \cref{eq:dScrefined}, coupled quintessence is a viable Ansatz to satisfy both \gls{dsc} criteria (\cref{eq:dSc,eq:dScrefined}) \cite{mishra_cosmological_2023}:
\citet{bruck_dark_2019,baldes_forays_2019} find that introducing a coupling helps to alleviate tensions regarding the \gls{dsc} and the \gls{dc} for \gls{de} models, and
a model-independent reconstruction by \citet{yang_model-independent_2020} finds coupled quintessence with \gls{de} coupled to \gls{dm} compatible with \cref{eq:dSc}, and presents an upper bound of $\mathfrak{s}_1\lesssim4.44$ at $z=0$ with 95\% significance (the uncoupled case of quintessence yields $\mathfrak{s}_1\lesssim1.23$ in the same study).

Besides introducing a coupling, also more complicated fields or multi-field settings can satisfy the \gls{dsc} constraints \cite{payeur_swampland_2024,akrami_multi-field_2021,garg_bounds-dS_2019,eskilt_cosmological_2022,cicoli_new_2020,gallego_anisotropic_2024}.\footnote{
    However, as \citet{brinkmann_stringy_2022} discuss, multi-field quintessence cannot start in a matter-dominated epoch and match the current \gls{de} contribution. If $\Omega_\textnormal{DE}\simeq0.7$ and $\omega_\textnormal{de}\simeq-1$ are to be matched, the quintessence fields need to start in an epoch of kinetic domination.
}
Multi-field models can circumvent the constraints for instance by keeping one field near an extremum while the other field rolls down a steep potential (such that the gradient is of $\order{1}$ in field units) \cite{ben-dayan_draining_2019}.

In multi-field quintessence, similar considerations regarding the turning rate of the fields apply as presented and referred to around \cref{eq:turningrate}: a non-geodesic motion through a curved field space allows the individual fields to not be slow-rolling, i.e. to have a steep potential that satisfies the \gls{dsc}, while at the same time showing a flat effective potential, which satisfies the observational preference for slow-rolling. \citet{cicoli_out_2020} study multi-field quintessence regarding this aspect. \citet{payeur_swampland_2024} study rapid-turn \gls{de}, a multi-field quintessence model inspired by hyperinflation, and claim that there is a small patch in the parameter space that satisfies observational constraints as well as the \gls{dsc}, the \gls{dc}, and the \gls{tcc}.
Turning allows a two-field model studied by \citet{eskilt_cosmological_2022,akrami_multi-field_2021} with a scalar field space metric of the form $G_{ab}=\text{diag}\left(1,f(r)\right)$, with $f(r)$ a function depending on the radial field $r$,\footnote{
    The model assumes a field space metric of the form $\mathrm{d}s^2=\mathrm{d}r^2+f(r,\theta)\mathrm{d}\theta^2$ and the scalars $\phi^a=(r,\theta)$. The polar coordinates $r$ and $\theta$ are coordinates in the field space, not in the physical space.
    }
to realise cosmic acceleration that effectively resembles a cosmological constant while satisfying the \gls{dsc}:
The action, Friedmann equation, and \glspl{eom} are given by \cref{eq:Q-EoM,eq:Q-action,eq:Q-Friedmann}.
In this model, field space has strongly curved trajectories and the fields spin rapidly in field space. Even though the \gls{de} background resembles \gls{lcdm}, there are differences that are potentially observable: spinning solutions enhance clustering on sub-Hubble scales.\footnote{
    A heavy mode present during inflation is usually suppressed \cite{akrami_multi-field_2021,achucarro_heavy_2012}. Yet in the late Universe, the heavy mode could enhance clustering and affect the \gls{dm} distribution due to a tachyonic instability \cite{akrami_multi-field_2021}.
    In this case though, the main mechanism for clustering comes from the light mode and is as follows \cite{akrami_multi-field_2021}: The spinning of the fields reduces the sound speed, as the turning rate $\mathfrak{T}$ alters the dispersion relation such that $c_\textnormal{s}=1+4a^2\mathfrak{T}^2/m_\phi^2$, where $m_\phi^2=a^2n^an^b\nabla_a\nabla_bV-a^2\mathfrak{T}^2+R\left(\partial_\tau\phi\right)^2/2$ is an effective mass describing the \gls{de} perturbations depending on the covariant derivative ($\nabla_a$) in the normal direction ($n^a$) of the potential $V$, the scale factor $a$, the Ricci scalar $R$, and the derivative with respect to conformal time of the scalar field $\partial_\tau\phi$.
    With a lower speed of sound, the Jeans scale becomes sub-horizon and the clustering becomes observable \cite{akrami_multi-field_2021,creminelli_spherical_2010,creminelli_effective_2009,takada_can_2006,hu_measuring_2004}.
}
Such a model can simultaneously satisfy the \gls{dsc} and observational constraints on slow-rolling, as the potential can be arbitrarily steep while the effective slow-roll parameter remains small (see also our later discussion in \cref{p:dSC_Inflation}):
\begingroup
\allowdisplaybreaks
\begin{align}
    \frac{\abs{\nabla V}}{V}&=\frac{\sqrt{G^{ab}V_aV_b}}{V}&\geq \mathfrak{s}_1\\
    \sqrt{\epsilon_V}&\sim\frac{\sqrt{G^{ab}V_aV_b}}{V}\\
    \epsilon_\phi&=\epsilon_V\Omega_\phi\left(1+\frac{\mathfrak{T}^2}{9H^2}\right)^{-1}\\
    \mathfrak{T}&=\abs{\nabla_t\vec{T}}\\
    T^a&=\frac{\dot{\phi}^a}{\dot{\phi}}
\end{align}
\endgroup
where we introduced the effective \gls{de} slow-roll parameter $\epsilon_\phi=\frac{3}{2}\left(w_\phi+1\right)=\frac{3}{2}\dot{\phi}^2/\left(\frac{1}{2}\dot{\phi}^2+V\right)$ with $\dot{\phi}^2=G_{ab}\dot{\phi}^a\dot{\phi}^b$ 
and the turning rate $\mathfrak{T}$ that depends on the normalised tangent vector to the field-space trajectory $\vec{T}$,\footnote{
    More details on the turning rate are presented in \cref{p:dSC_Inflation}.
}
and used the relative energy density of \gls{de} $\Omega_\phi=\rho_\phi/\rho_\textnormal{crit}\approx V/\left(3M_\textnormal{P}^2H^2\right)$.\footnote{
    In \cref{p:dSC_Inflation}, we omit $\Omega_\textnormal{Inflaton}=1$, since no other energy component is present during inflation.
    For \gls{de} we see that additional energy components further suppress the rolling-rate of the scalar field, which explains the observed value of $\epsilon_\phi\ll1$ \cite{akrami_multi-field_2021}.
    }
We see that the field can be effectively slow-rolling ($\epsilon_\phi\ll1$), in accordance with observations, while the potential is steep ($\sqrt{\epsilon_V}\sim\order{1}$), in accordance with the \gls{dsc}.

Besides couplings and multi-field settings, also modified gravity or higher-dimensional theories allow for \gls{dsc}-satisfying quintessence models:
\citet{li_effective_2020} state that quintessence can be realised in a \gls{dsc}-compatible way if an \gls{ads} vacua is uplifted by frozen, large-scale Lorentz-violation achieved through a non-trivial Brans\textendash Dicke coupling between the quintessence field and gravity.
Quintessence in a Horndeski setting with a cubic Galileon term $\left(\nabla\phi\right)^2\DAlambert\phi$ in the Lagrangian is found to be compatible with observational constraints and the \gls{dsc} \cite{brahma_dark_2019}.
And quintessence in a supergravity setting is assessed by \citet{chiang_building_2018}:
They propose a model with two separated sectors, a quintessence sector $\phi_q$ and a hidden sector $\phi_h$. The Kähler and superpotential show the separation:
\begin{align}
    K&=\phi_h^*\phi_h+\phi_q^*\phi_q\\
    W&=W_0(\phi_h)+W_1(\phi_q).
\end{align}
The F-term scalar potential is given by
\begin{align}
    V_F&=e^{K/M_\textnormal{P}^2}\left(D_iWK^{i\Bar{j}}D_{\Bar{j}}W^*-\frac{3}{M_\textnormal{P}^2}\abs{W}^2\right)\\
    D_iW&=\frac{\partial W}{\partial\phi_i}+\frac{W}{M_\textnormal{P}^2}\frac{\partial K}{\partial\phi_i},
\end{align}
and yields $V\supset m_\textnormal{3/2}^2\abs{\phi_q}^2$, since the gravitino $m_\textnormal{3/2}$ is related to the superpotential $\expval{\abs{W_0}^2}\sim m_\textnormal{3/2}^2M_\textnormal{P}^4$.
There is a large scale separation between the Hubble scale and the gravitino, which causes the quintessence field to roll down its potential very early, which makes it indistinguishable from a cosmological constant. Since \gls{desi} data suggests that \gls{de} is actually dynamical, this has to be broken, which can either be achieved by a shift symmetry in the quintessence sector or by sequestering the hidden sector. As \citet{chiang_building_2018} show, the former leads to trans-Planckian field ranges (yet satisfies the \gls{dsc}) and the latter violates the \gls{dsc} (yet satisfies the \gls{dc}), as fifth-force constraints require $\left(\abs{\nabla V}/V\right)^2\lesssim\num{e-5}$.

If a model\,\textemdash\,single-field or multi-field\,\textemdash\,satisfies all the constraints, it might do so only for finely chosen values of the free parameters of the model. The fine-tuning problems might be severe, as \citet{hertzberg_quantum_2019} argue: Quintessence could receive quantum corrections from loop effects that are larger than the observed \gls{de} density, i.e. the field potential must be finely tuned to be $V\sim\num{e-120}M_\textnormal{P}^4$ \cite{kolda_quintessential_1999,garny_quantum_2006,choi_string_2000}. Furthermore, the field is supposed to be slow-rolling, which means that also $\abs{\nabla V}\sim\num{e-120}M_\textnormal{P}^3$ has to be achieved. However, the double fine-tuning is only a \textit{double} fine-tuning if $V$ and $\nabla V$ are not correlated in any way.\footnote{
    One example are conformally coupled scalars, where only the quantity $V/\abs{\nabla V}$ matters, such that it is a single fine-tuning problem like in the case of the cosmological constant \cite{hertzberg_quantum_2019}.
    Admittedly, the conformal couplings might not be the best example to make this point, as such couplings invoke fifth forces which are severely constraint by observations and must be either screened, or the coupling must be only active in the dark sector \cite{hertzberg_quantum_2019}. If it is the latter, there are still quantum corrections that could spoil the fine-tuning of the quintessence potential by coupling to the \gls{sm} \cite{hertzberg_quantum_2019}.
}

Another fine-tuning problem is mentioned by \citet{cicoli_sitter_2019}: not only does it require fine-tuning to achieve $V(\phi_0)=\Lambda_\text{cc}^4$ with $\phi_0$ today's quintessence field value, moreover the quintessence field has to be extremely light, $m\simeq\SI{e-32}{\electronvolt}$.

A \gls{dsc}-consistent proposal without fine-tuning for the cosmological constant, but instead for the \gls{ew} scale, has been found in a model of Starobinsky inflation with non-minimal coupling to the Higgs field that thaws later into quintessence \cite{wang_electroweak_2019}.
A quintessence\textendash Higgs coupling has also been investigated by \citet{han_quintessence_2019}. For an exponential quintessence potential $V\sim\exp\left(-\mathfrak{c}\phi\right)$, they derive a lower bound of $\mathfrak{c}\gtrsim\num{0.35(0.05)}$, which supports $\mathfrak{s}_1\sim\order{1}$. They derive their bound by requiring that the \gls{ew} vacuum is stable during inflation, taking into account observational data related to \gls{de} and matter energy density, the Higgs mass and vacuum expectation value, and the mass of the top quark.

As a final note on quintessence, we note that quintessence can play an important role in cyclic cosmology, where the universe undergoes repeated phases of expansion and contraction. If this is the case, the \gls{dsc} puts strong constraints on the number of $e$-foldings per cycle in cyclic models: \citet{coriano_swampland_2020} derive that the number of $e$-folds until the next contraction phase starts is $3\Delta\phi/2M_\textnormal{P}\Omega_{\phi,0}\mathfrak{s}_1\sim\order{1}$, while models of cyclic cosmology predict this number to be of $\order{100}$. The tension vanishes if the \gls{dsc} allowed for $\mathfrak{s}_1\sim\order{\num{e-2}}$ or the \gls{dc} allowed for $\Delta\phi/M_\textnormal{P}\sim\order{100}$. If the swampland conjectures remain strong in their current forms, this puts cyclic cosmology in this form in the swampland.

\subparagraph{Slotheon \gls{de}\footnote{
    \noindent
    From \textit{slow Galilean} \gls{de}, as the slotheon scalar field experiences friction, which makes it moving slower than its canonical counterpart \cite{adak_late_2013}.
    }
}
has an action that slightly differs from quintessence \cite{mukhopadhyay_swampland_2019,mukhopadhyay_evolution_2020,germani_introducing_2012,adak_late_2013}:
\begin{align}
S =& \int\!  \sqrt{-g} \left[ \frac{1}{2} \left.\biggl( M_\textnormal{P}^2 R  - V(\phi) \right.\right.\nonumber\\
&\qquad\qquad\left.\left.
- \left( g^{\mu\nu} - \frac{G^{\mu\nu}}{\Lambda^2} \right) \nabla_\mu \phi \nabla_\nu \phi \right) \right] \, \mathrm{d}^4x,
\end{align}
where $G^{\mu\nu}$ is the Einstein tensor, $\Lambda$ is an energy scale, and the other terms have their usual meaning;
the coupling $\frac{G^{\mu\nu}}{\Lambda^2}  \nabla_\mu \phi \nabla_\nu \phi $ acts as an additional friction that slows the scalar field down.

Evidence supporting the slotheon model could present itself in the following form:
\begin{itemize}
    \item the small-scale structure growth is suppressed around the matter domination\textendash \gls{de} domination transition due to the additional friction term which modifies the \glspl{eom} \cite{mukhopadhyay_evolution_2020,adak_late_2013},
    \item the matter powerspectrum is suppressed at small wavenumbers, but the strength of the suppression declines for increasing wavenumbers \cite{adak_late_2013},
    \item the growth rate $f\sigma_8$ is lowered, and the redshift-space distortion could be observable, as the coupling modifies the Poisson equation \cite{mukhopadhyay_evolution_2020,adak_late_2013}.
\end{itemize}

Considering a thawing field, \citet{mukhopadhyay_swampland_2019} found that the slotheon satisfies the \gls{dsc} and is favoured over quintessence models with a single exponential potential by observational constraints. The slotheon model favours $\mathfrak{s}_1\approx0.8$, while quintessence with a simple single exponential potential favours lower values, which causes tension with the \gls{dsc}. As an aside, it is mentioned that the \gls{dc} is satisfied and that \cref{eq:dScrefined} is violated.

\subparagraph{Thermal \gls{de}} is characterised by the presence of temperature terms in the scalar field potential. The model is presented by \citet{hardy_thermal_2020} and further discussed by \citet{bento_dark_2020}: Conceptually, such a potential is of the form $V\sim V_0+\phi^2T^2$. The temperature dependence can keep the scalar field near a false vacuum and does not require the potential to be flat. When the universe cools down, this mechanism ceases to be effective and the scalar field rolls to its true vacuum. The model makes several potentially observable predictions: The thermal support comes from a hidden sector, which changes $\Delta N_\textnormal{eff}$, which in turn affects \gls{bbn} respectively the \gls{cmb}. The temperature dependence introduces a coupling between the hidden sector and the \gls{sm}, which induces a fifth-force. Thermal phase transitions in the hidden sector might produce a stochastic \gls{gw} background. Since the universe cools down over time, this is rather an \gls{ede} model. See our comments on \gls{ede} elsewhere in this review article regarding \gls{ede} constraints and the feasibility of reducing the Hubble tension.

The compatibility between thermal \gls{de} and the \gls{dsc} depends on the chosen potential: While \cite{bento_dark_2020} claim that this model is compatible with swampland conjectures, and they even mention the \gls{dsc} in a footnote, the potential they use might be in tension with the \gls{dsc}:
For large scalar field values, $\abs{\nabla V}/V\sim1/\phi$, and for small field values $\abs{\nabla V}/V\sim\phi$, i.e. in either case the ratio will be small. There is only a window of compatibility for intermediate field values, such that $\abs{\nabla V}/V\sim\order{1}$. However, $V/\rho_\textnormal{DE}\rightarrow0$ for intermediate field values (see figure 3 in the paper by \citet{bento_dark_2020} respectively figure 1 in the work by \citet{hardy_thermal_2020}).
Yet, an alternative modulus-like zero temperature potential $V(\phi)=m_\phi^2\left(\phi-\phi_1\right)^2/2$ suggested by \citet{hardy_thermal_2020} is actually compatible with the \gls{dsc}.

\subparagraph{Pole \gls{de}}
\citet{linder_pole_2020} presents an Ansatz for \gls{de} with a pole in the kinetic term:
\begin{align}
    L&=\frac{-1}{2}\frac{\varsigma}{\sigma^n}\left(\partial\sigma\right)^2-V(\sigma)\\
    \phi&=\frac{2\sqrt{\varsigma}}{\abs{2-n}}\sigma^{\left(2-n\right)/2}\label{eq:pole_transform}\\
    \sigma&=\left(\frac{\abs{2-n}}{2\sqrt{\varsigma}}\right)^{2/\left(2-n\right)}\phi^{2/\left(2-n\right)}
\end{align}
with $\varsigma$ the residue,
$\sigma\in\left[0,\infty\right]$ the pole position, and
$n$ the order.
He finds that poles can be stilts over the swampland, at least sometimes: The tested models all satisfy the \gls{dc}. For the \gls{dsc}, he expresses \cref{eq:dSc} in terms of the pole $\sigma$:
\begin{equation}
    \frac{\abs{\nabla V}}{V}=\sqrt{\frac{\sigma^n}{\varsigma}}\frac{\abs{\mathrm{d}V/\mathrm{d}\sigma}}{V},
\end{equation}
and uses this to assess the following models where $n>2$.\footnote{
    For $n>2$, monomial potentials get transformed into inverse power-laws, and inverse power-law becomes monomial, which is the more interesting case than $n<2$ where monomials stay monomials and inverse power-law potentials stay inverse power-law potentials.
    }
\begin{description}
    \item[$V\sim\sigma^k$] $\abs{\nabla V}/V=2\abs{k}/\phi\abs{2-n}$, which can be satisfied, since the field excursion is sub-Planckian. 
    \item[$V\sim\exp\left(-\beta\sigma\right),\,n=2$] $\abs{\nabla V}/V=\beta\sqrt{\sigma^n/\varsigma}\rightarrow\beta/\left[\exp\left(-\phi/\sqrt{\varsigma}\right)\sqrt{\varsigma}\right]$, which can satisfy the \gls{dsc} in an observationally viable range where $\phi$ is not too small but $\beta$ is small.
    \item[$V\sim\exp\left(-\beta\sigma\right),\,n=4$] $\abs{\nabla V}/V=\beta\sqrt{\sigma^n/\varsigma}\rightarrow\beta\sqrt{\varsigma}/\phi^2$, which only satisfies the \gls{dsc} for the observationally not viable case with $\beta\sqrt{\varsigma}$ large, as this case drives $\omega$ away from -1.
\end{description}

\paragraph{Fine-Structure Constant}
Studying dS$_{D-p}\cross S^p$ solutions, \citet{montero_ds_2020} find the condition
\begin{equation}
    \left(p-1\right)\abs{\frac{V^\prime}{V}}\geq\abs{\frac{f^\prime}{f}}
\end{equation}
for compatible \glspl{eft}, with $f$ being the inverse gauge coupling of the $\left(p-1\right)$-form field. They apply this constraint on electromagnetism in a quintessence setting, to find a bound for the change in the fine-structure constant $\alpha=1/f$:
\begin{equation}
    \abs{\frac{\dot{\alpha}}{\alpha}}\lesssim \mathfrak{s}_1H\approx\frac{\mathfrak{s}_1}{\num{e10}\text{years}},
\end{equation}
which is consistent with the experimental bound of $\abs{\dot{\alpha}/\alpha}<\num{e-15}\text{year}^{-1}$ respectively $\abs{\dot{\alpha}/\alpha}\lesssim\num{e-5}$ \cite{webb_search_1999,varshalovich_fundamental_1999,webb_further_2001,murphy_further_2001,murphy_does_2003,murphy_further_2003,srianand_limits_2004,chand_probing_2004,murphy_revision_2008,king_spatial_2012,murphy_subaru_2017,welsh_bound_2020,wilczynska_four_2020,milakovic_new_2021,hees_search_2020,webb_indications_2011,berengut_limits_2013,olive_constraints_2002,olive_reexamination_2004,nunes_reconstructing_2004,davoudiasl_variation_2019,damour_oklo_1996}.

\paragraph{Higgs Field}
The Higgs field violates \cref{eq:dSc} as the Higgs potential
\begin{equation}
    V_\textnormal{H}=\lambda_\textnormal{H}\left(\abs{h}^2-v_\textnormal{H}^2\right)^2
\end{equation}
has a local maximum at $h=0$, but since
\begin{equation}
    \frac{\text{min}\left(\nabla_i\nabla_jV\right)}{V}\sim-\frac{10^{35}}{M_\textnormal{P}^2}
\end{equation}
it is compatible with \cref{eq:dScrefined} \cite{fukuda_phenomenological_2019,denef_ds_2018,cheong_higgs_2018,murayama_we_2018,han_quintessence_2019,choi_ds_2018,palti_swampland_2019}.

\citet{cicoli_sitter_2019} investigate if modifications to the Higgs potential or its coupling to quintessence could rescue it from violating \cref{eq:dSc}. 
A Higgs\textendash quintessence coupling is also studied by \citet{denef_sitter_2018,hamaguchi_swampland_2018}.
Such a coupling is problematic, as it generally induces fifth forces that are incompatible with observations (unless fine-tuning takes place) \cite{agrawal_dark_2018,cicoli_sitter_2019,denef_ds_2018,hamaguchi_swampland_2018}. Furthermore, a tension from the time-dependence of the proton-to-electron mass ratio has been shown \cite{hamaguchi_swampland_2018,agrawal_dark_2018}.

\paragraph{Inflation}\label{p:dSC_Inflation}
is often described by a slow-rolling inflaton field. There are various models of inflation, which we discuss in the following. Important parameters to assess inflationary models are the slow-roll parameters 
\begin{align}
    \epsilon_V&\defeq\frac{(\nabla V)^2}{V^2}&&&\geq\mathfrak{s}_1^2\label{eq:epsilonV}\\
    \eta_V&\sim\frac{\nabla^2V}{V}&&&\lesssim -\mathfrak{s}_2.
\end{align}
Slow-roll inflation takes places if $\epsilon_V<1$ and $\eta_V<1$ \cite{cremonini_asymptotic_2023}.
The \gls{dsc} informs us that in \gls{qg} a single scalar field must not be slow-rolling \cite{garg_bounds-dS_2019,dias_primordial_2019} and \citet{agrawal_cosmological_2018} point out that most inflationary models violate \cref{eq:dSc} (cf. \cite{kinney_zoo_2019}).
To rule out inflation based on these preliminary findings would be short-sighted.
Focussing on $\epsilon_V$ and $\eta_V$ might be straightforward, as only the potential has to be evaluated, but the relevant parameters for slow-roll inflation are actually the Hubble slow-roll parameters
\begin{align}
    \epsilon_H&=-\frac{\dot{H}}{H^2}\\
    \eta_H&=\frac{\dot{\epsilon}_H}{H\epsilon_H},
\end{align}
since they depend on the full solutions of the scalar field's \gls{eom} \cite{anchordoqui_s-dual_2021,seo_sitter_2019,brahma_avoiding_2019,liddle_formalising_1994,guleryuz_superuniversal_2023}.
This is for example relevant
in the presence of multiple but non-aligned inflaton fields,
non-\gls{bd} terms \cite{bunch_quantum_1997},\footnote{
    Describing a \gls{bd} vacuum as a finely tuned superposition of spontaneous particle creation that is precisely matched by anti-particles coming from infinity shows the unnaturalness of such a setting, which makes it plausible that non-\gls{bd} terms should be expected \cite{danielsson_quantum_2019}.
}
\gls{dbi} inflation,\footnote{
    In \gls{dbi} inflation, a D-brane moves through a warped throat in an extra-dimension, which leads to non-canonical kinetic terms in the \gls{eft}, and a subluminal speed of sound \cite{guleryuz_non-perpetual_2024,rasouli_warm_2019}.
    This leads to large tensor components in the \gls{cmb}, which are in reach of assessment \cite{alishahiha_dbi_2004}.
} 
or \textit{k}-flation \cite{armendariz-picon_k-inflation_1999}.
In such settings, the tensor-to-scalar ratio can be shifted, such that
\begin{align}
    r_\textnormal{ts}&=16\epsilon_H\lambda_\textnormal{c},\label{eq:bunch-davis_tensor-to-scalar}
\end{align}
where the parameter $\lambda_\textnormal{c}$ could measure the deviation from the \gls{bd} solution, in which case it is constraint by backreaction terms and non-Gaussianity constraints \cite{brahma_avoiding_2019};
represent a subluminal sound speed, as in \gls{dbi} models \cite{kinney_zoo_2019} or \textit{k}-flation \cite{das_note_2019};
or measure the topology deviation through a \gls{gb} term \cite{yi_gaussbonnet_2019,gashti_pleasant_2022}.
Depending on the model, there can be a critical value of $\lambda_\textnormal{c}$ for which $\eta_H$ is minimal \cite{chiang_what_2019}.
The take-home message here is that the \gls{dsc} $\epsilon_V\sim\order{1}$ can be satisfied without violating the Hubble slow-roll condition $\epsilon_H\ll1$ \cite{lin_generalizing_2021}.

In the case of single-field inflation, the Hubble slow-roll parameters are related to the potential slow-roll parameters as follows \cite{chiang_what_2019,guleryuz_superuniversal_2023}:\footnote{
    For analytical studies it can be beneficial to write $\epsilon_H=\frac{3}{2}\left(x^2-y^2+1\right)+\frac{1}{2}z$ with $x=\dot{\phi}/\sqrt{6}M_\textnormal{P}H$, $y=\sqrt{V}/\sqrt{3}M_\textnormal{P}H$, and $z=-\Omega_k=k/a^2H^2$ \cite{alestas_curve_2024}.
}
\begin{align}
    \epsilon_V=\frac{1}{2}\left(\frac{\partial_\phi V}{V}\right)^2&\simeq\frac{\dot{\phi}^2}{2H^2}=\epsilon_H\\
    \eta_V&=2\epsilon_H-\frac{1}{2}\eta_H.
\end{align}

Even if an inflationary model appears to be incompatible with the \gls{dsc}, it might be possible to rescue the model by putting it in a brane-world scenario.\footnote{
    The term \textit{brane-world} loosely refers to any theory where a matter field is localised on a space-filling hypersurface (a brane)\,\textemdash\,the matter field could be exclusively living in the worldvolume of the brane, even though \citet{fichet_braneworld_2020} presents some swampland arguments that favour spacetime-filling matter fields, where only some operators are localised on the brane. Matter fields exclusively localised on a brane lead to global symmetries (violating the no global symmetries conjecture (\cref{sec:nGSym})), and an energy gap of $\abs{p}\sim\left[\Lambda_\textnormal{S}/\sqrt{N_\textnormal{S}},\Lambda_\textnormal{S}\right]$ (see \cref{s:ssc} for the notation), where the \gls{eft} is not valid between the 4d and 5d regime, which highlights an inconsistency \cite{fichet_braneworld_2020}. However, the work of \citet{fichet_braneworld_2020} is criticised by \citet{nortier_comment_2022}: the global symmetries can be avoided by including 5d bilocal operators arising from 5d local operators involving brane form factors. Furthermore, non-local operators that are localised on the brane and couple to the 4d bulk fields avoid global symmetries as well. 
}
A main driver of this observation is that potentials in a brane-world with brane tension $\mathcal{T}$ roll slower than potentials in the standard 4d case, as
\begin{equation}\label{eq:brane_Hubble}
    H^2\simeq\frac{8\pi}{3M_{\textnormal{P;}4}^2}V\left(1+\frac{V}{2\mathcal{T}}\right)
\end{equation}
is proportional to the energy density $\rho$ instead of $\sqrt{\rho}$ \cite{mohammadi_brane_2022,maartens_chaotic_2000,golanbari_brane_2014,mohammadi_constant-roll_2020,elizalde_inflationary_2019,banerjee_inflationary_2017}, which induces an increased Hubble friction compared to the \gls{sm} \cite{osses_reheating_2021}. The Friedmann equation contains terms quadratic in the energy density, which dominate in the high-energy regime, i.e. where the energy density is larger than the brane tension \cite{mohammadi_brane_2022}.
Considering the tension of the brane yields modified slow-roll parameters for the brane:
\begin{align}
    \epsilon_\textnormal{B}&=\epsilon_V\frac{1+V/\mathcal{T}}{\left(1+V/2\mathcal{T}\right)^2}&=&\frac{M_{\textnormal{P};4}^2}{16\pi}\left(\frac{V^\prime}{V}\right)^2\\
    \eta_\textnormal{B}&=\eta_V\frac{1}{1+V/2\mathcal{T}}&=&\frac{M_{\textnormal{P};4}^2}{8\pi}\frac{V^{\prime\prime}}{V},
\end{align}
where $\epsilon_V$ and $\eta_V$ are the standard 4D slow-roll parameters with their usual definition regarding the inflaton potential \cite{osses_reheating_2021,mohammadi_brane_2022}.
Furthermore, the number of $e$-folds is modified as well:
\begin{equation}
    N_e\simeq-\frac{8\pi}{M_{\textnormal{P};4}}\int\!\frac{V}{V^\prime}\left(1+\frac{V}{2\mathcal{T}}\right)\,\mathrm{d}\phi,
\end{equation}
i.e. it takes more $e$-foldings to get from an initial field value $\phi_\textnormal{i}$ to a final field value $\phi_\textnormal{f}$ \cite{osses_reheating_2021}.
The tensor-to-scalar ratio at low energies is not affected, and $r_\textnormal{ts}\simeq16\epsilon_V$ holds;
at high energies, where the massless mode to extend into the 5d bulk, $r_\textnormal{ts}\simeq24\epsilon_\textnormal{B}$ holds \cite{osses_reheating_2021}.
Models of brane inflation to be found compatible with the \gls{dsc} are for example 
power-law inflation of the form $V\sim\phi^n$,
natural inflation of the form $V\sim\left(1-\cos{\phi/f}\right)$, or
T-model inflation (\cref{eq:T-model}) \cite{mohammadi_brane_2022}.

If inflation is not driven by the scalar field but by fluctuations, e.g. induced through non-perturbative effects such as a spinodal instability, the \gls{dsc} looses its constraining power \cite{holman_spinodal_2019}.

In the following, we discuss various (groups of) models of inflation in more detail.

\subparagraph{Slow-roll inflation} is in tension with the \gls{dsc}, if the parameters $\mathfrak{s}_1$ and $\mathfrak{s}_2$ are of $\order{1}$ \cite{matsui_swampland_2020,garg_bounds-dS_2019,agrawal_cosmological_2018,achucarro_string_2019,kinney_zoo_2019,brahma_avoiding_2019,das_note_2019,fukuda_phenomenological_2019,ashoorioon_rescuing_2019,anchordoqui_s-dual_2021,hertzberg_inflationary_2007,caviezel_effective_2009,caviezel_cosmology_2009,flauger_slow-roll_2009,ben-dayan_draining_2019,caviezel_moduli_2010,giddings_dynamics_2006,parameswaran_ds_2024}.\footnote{
    Even before the \gls{dsc} was formalised as a swampland conjecture\,\textemdash\,actually even before the swampland programme itself started\,\textemdash\,\citet{hertzberg_inflationary_2007} showed that inflation in type IIA string theory requires $\epsilon_V>27/13$, i.e. they already presented a bound very similar to the \gls{dsc} that ruled out slow-roll single-field inflation. Also \citet{flauger_slow-roll_2009} present lower bounds on $\epsilon_V\sim\order{1}$ in type IIA settings that rule out slow-roll inflation.
}
The culprits are the scalar spectral index $n_\textnormal{s}$ and the tensor-to-scalar ratio $r_\textnormal{ts}$ that are constrained by observational data to $r_\textnormal{ts}<0.1$ and $n_\textnormal{s}=\num{0.9649(0.0042)}$ \cite{planck_collaboration_Inflation_2020}, while at the same time requested to fulfil $n_\textnormal{s}-1\approx-6\epsilon_V+2\eta_V\gtrsim\left[3\mathfrak{s}_1^2;2\mathfrak{s}_2\right]$ and $r_\textnormal{ts}>8\mathfrak{s}_1^2$ by the \gls{dsc} \cite{kinney_zoo_2019,kehagias_note_2018,matsui_eternal_2019,ben-dayan_draining_2019,agrawal_dark_2018,fukuda_phenomenological_2019,chiang_what_2019}.

This finding seems to hold even in non-minimally coupled matter\textendash non-metricity theory of gravity ($f(\mathcal{Q})$ theory \cite{jimenez_coincident_2018,harko_coupling_2018,heisenberg_review_2023,mandal_constraint_2021}) \cite{mandal_theory_2023}. However, studying a generalised uncertainty principle where $\left[x, p\right] = i\mathcal{F}(p^2)$, $\mathcal{F}$ a function, \citet{garcia-compean_scalar_2024} find that single-field, slow-roll inflation can be realised by obtaining non-standard slow-roll parameters that satisfy the \gls{dsc}. Another possibility to mitigate the constraints by the \gls{dsc} is to lower $r_\textnormal{ts}$ by the introduction of additional factors / mechanisms:
\begin{description}
    \item[Non-\gls{bd} Terms] $r_\textnormal{ts}$ is lowered by a low value of $\lambda_\textnormal{c}$ (see \cref{eq:bunch-davis_tensor-to-scalar} and the works by \citet{brahma_avoiding_2019,ashoorioon_non-bunchdavis_2014,ashoorioon_reconciliation_2014,ashoorioon_rescuing_2019} for a discussion).
    \item[\gls{dbi} Models] The inflaton field is kept near its maximum by an effective sound speed limit, such that $r_\textnormal{ts}\simeq16c_\textnormal{s}\epsilon_V<0.1$, which still requires $\mathfrak{s}_1<\sqrt{2\epsilon_V}<0.4$, i.e. the tension with the \gls{dsc} remains \cite{kinney_zoo_2019}. In a study by \citet{rasouli_warm_2019}, cold \gls{dbi} inflation is shown to be incompatible with Planck data, yet warm \gls{dbi} inflation is compatible with Planck as well as with the \gls{dsc}.
    A multi-field \gls{dbi} model with non-canonical terms satisfies the \gls{dsc} if the speed of sound is sufficiently small \cite{solomon_non-canonical_2020}.
\end{description}

Some authors argue that the \gls{dsc} is actually not violated at all:
\begin{itemize}
    \item \citet{kehagias_note_2018} raise the argument that we do not know how scalar perturbations are created. If curvature perturbations are only created at super-Hubble scales, at the very end of inflation, or after inflation ends, e.g. when a curvaton field decays, the restrictions from the low tensor-to-scalar ratio do not apply and the \gls{dsc} is not violated.
    \item \citet{akrami_landscape_2019} argue that the \gls{dsc} does not even apply to slow-roll inflation, since slow-roll does take place far away from the \gls{ds} regime, as the inflaton perturbations are $\propto1/\abs{\nabla V}$ and observations predict $\abs{\nabla V}\gtrsim\num{e5}V^{3/2}$, i.e. there is no \gls{ds} phase during slow-roll inflation.
\end{itemize}

\subparagraph{Eternal inflation} is not possible if the \gls{dsc} is true \cite{matsui_swampland_2020,brahma_stochastic_2019,kinney_eternal_2019,dimopoulos_steep_2018,matsui_eternal_2019,rudelius_conditions_2019,dvali_quantum_2013,dvali_quantum_2019,wang_eternal_2020,alestas_curve_2024}.
In eternal inflation, quantum fluctuations dominate over classical fluctuations,\footnote{
    This allows the inflaton field to roll upwards in the potential \cite{kinney_eternal_2019}.
    $\delta_q\phi\lesssim\delta_c\phi$ still allows for eternal inflation to happen in some edge cases, as regions where the scalar field moves up the potential expand faster than regions where the scalar field moves down the potential \cite{wang_eternal_2020}.
}
which translates to the following constraints \cite{matsui_swampland_2020,rudelius_conditions_2019,barenboim_eternal_2016,rudelius_dimensional_2021,matsui_eternal_2019,kinney_eternal_2019}:
\begin{align}
    \expval{\delta\phi}_\textnormal{q}&\approx\frac{H}{2\pi}\label{eq:quantum_fluctuation}\\
    \expval{\delta\phi}_\textnormal{c}&\approx\frac{\abs{\dot{\phi}}}{H}\label{eq:classical_fluctuation}\\
    \frac{\expval{\delta\phi}_\textnormal{q}}{\expval{\delta\phi}_\textnormal{c}}=\frac{H^2}{2\pi\abs{\dot{\phi}}}&\gtrsim1\label{eq:eternal_condition}\\
    \Leftrightarrow\frac{H}{M_\textnormal{P}}&\gtrsim2\pi\sqrt{2\epsilon_V}.\label{eq:eternal_slow-roll_condition}%
\end{align}
For eternal inflation to happen, \cref{eq:eternal_condition} has to hold, which is expressed in terms of the slow-roll parameter in \cref{eq:eternal_slow-roll_condition} \cite{matsui_eternal_2019,kinney_eternal_2019}.
Eternal chaotic inflation is incompatible with the \gls{dsc}, since \cref{eq:eternal_slow-roll_condition} yields $H/M_\textnormal{P}\gtrsim2\pi \mathfrak{s}_1$, i.e. a trans-Planckian Hubble scale \cite{matsui_eternal_2019,kinney_eternal_2019,wang_eternal_2020}.

A second issue is that the volume expansion must dominate over the decay rate of the field, which translates into the following condition \cite{matsui_swampland_2020,rudelius_conditions_2019,matsui_swampland_2020,barenboim_eternal_2016,rudelius_dimensional_2021}:
\begin{align}
    \frac{V^{\prime\prime}}{V}&>-\frac{2\left(d-1\right)}{d-2}\frac{1}{M_{\textnormal{P};d}^{d-2}}\label{eq:VolDom}.
\end{align}
This clashes directly with the $\frac{V^{\prime\prime}}{V}<-\mathfrak{s}_2$ constraint from the refined \gls{dsc} \cite{rudelius_dimensional_2021}.\footnote{
    It was argued by \citet{kinney_eternal_2019} that eternal inflation can be made consistent with the refined \gls{dsc}, but this argumentation was debunked by \citet{brahma_stochastic_2019}.
    }

The lifetime of the current phase of accelerated expansion of the Universe can be derived by combing the \gls{dsc} with the \gls{dc} \cite{agrawal_cosmological_2018}:
\begin{equation}
    \tau\lesssim\frac{3\Delta\phi/M_\textnormal{P}}{2\mathfrak{s}_1\Omega_{\phi,0}H_0},
\end{equation}
with $\Delta\phi/M_\textnormal{P}\sim\order{1}$ the field range limit from the \gls{dc},
$\mathfrak{s}_1$ the \gls{dsc} parameter,
$\Omega_{\phi,0}$ today's relative \gls{de} density,
and $H_0$ the Hubble constant. Since this yields a finite number, eternal inflation is ruled out.

A particularly illuminating paper is presented by \citet{rudelius_conditions_2019}: He studies different models of inflation that can lead to eternal inflation and shows that eternal inflation is incompatible with the \gls{dsc}. He even goes one step further and proposes that it is not the case that there cannot be eternal inflation \textit{because} the \gls{dsc} holds, but that there is a \gls{qg} rule that forbids eternal inflation from taking place, and the \gls{dsc} is a mere consequence of this fundamental no-go theorem. We'd like to highlight a few key takeaways from his paper
\begin{itemize}
    \item eternal inflation takes place, unless one of the following conditions holds:
    \begin{align}
        \frac{\abs{\nabla V}}{V}&>\frac{\sqrt{2V}}{2\pi M_\textnormal{P}^3}\label{eq:eternalinflation1}\\
        \frac{\sum_i\nabla_i\nabla_i V}{V}&<-\frac{3}{M_\textnormal{P}}\label{eq:eternalinflation2}\\
        \frac{2\pi^2M_\textnormal{P}^2\abs{\nabla V}^2}{V^2}-\frac{V}{3M_\textnormal{P}^2}\frac{\sum_i\nabla_i\nabla_i V}{V}&>\frac{V}{M_\textnormal{P}^4}\label{eq:eternalinflation3}\\
        \left[-\sign\left(\nabla^pV\right)\right]^{p+1}\frac{\abs{\nabla^pV}}{V^{\left(4-p\right)/2}}&>\mathfrak{c}_pM_\textnormal{P}^{p-4}\\
        \frac{\Gamma}{H^4}&>\frac{9}{4\pi},
    \end{align}
    where $\mathfrak{c}_p\gg1$ is a numerically derived constant and $\Gamma$ the decay rate per unit volume. 
    \Cref{eq:eternalinflation1,eq:eternalinflation2,eq:eternalinflation3} are implied by the \gls{dsc}:
    \begin{itemize}
        \item Since $V<M_\textnormal{P}^4$, \cref{eq:dSc} implies \cref{eq:eternalinflation1} for $\mathfrak{s}_1>\sqrt{2}M_\textnormal{P}/2\pi$.
        \item Since $\abs{\sum_i\nabla_i\nabla_iV}\geq\abs{\min\nabla_i\nabla_jV}$, \cref{eq:dScrefined} implies \cref{eq:eternalinflation2} for $\mathfrak{s}_2>3/M_\textnormal{P}^2$.
        \item The same two observations can also be used to show that \cref{eq:eternalinflation3} is implied by the combined \gls{dsc} (\cref{eq:dSCcombi}) for $2\pi^2\mathfrak{p}_2>1/3$.
    \end{itemize}
    \item Ruling out eternal inflation or \gls{ds} solutions could well be just a lamppost effect. This new insight combined with the above findings would then indicate that eternal inflation takes place in a strong-coupling string regime and is truly a \gls{qg} phenomenon.\footnote{
        \citet{banks_limits_2019} argues that eternal inflation is incompatible with \gls{qg}, as eternal inflation would violate holographic principles and would require an infinite dimensional Hilbert space while the entropy of the Universe appears to be finite.
        }
    \item The motivation to promote a \enquote{no eternal inflation} no-go theorem is that a universe with eternal inflation is qualitatively, fundamentally different from a universe without eternal inflation.\footnote{See e.g. the work by \citet{banks_limits_2019}.} Checking if a model complies with the \gls{dsc} is often parameter-dependent, i.e. there is no fundamental difference between a model that violates the \gls{dsc} and a model that satisfies the \gls{dsc}.
    \item Models of inflation that can but do not necessarily lead to eternal inflation are:
        \begin{itemize}
            \item Power-law inflation with $V\sim\phi^\mathfrak{p}$ can lead to eternal inflation, but parameter-choices that are likely compatible with observations do not lead to eternal inflation.\footnote{
                Even worse, \citet{oikonomou_swampland_2023,oikonomou_rescaled_2022} find models of power-law inflation to be generally incompatible with Planck data, as the observed value of the spectral index of scalar perturbations $n_\textnormal{s}$ requires $1.7\lesssim \mathfrak{p}\lesssim 2.8$ while compatibility with the observed tensor-to-scalar ratio $r_\textnormal{ts}$ requires $0\leq \mathfrak{p}\lesssim0.9$, which is mutually exclusive.
                }
            \item Starobinsky inflation can lead to eternal inflation for super-Planckian field ranges.
            \item Inflection point inflation with a potential of the form $V=V_0+\mathfrak{l}\phi+\frac{1}{6}\mathfrak{c}\phi^3$ is not eternal if compatible with observations.
        \end{itemize}
    \item Models of inflation that necessarily lead to eternal inflation are:
    \begin{itemize}
        \item Quadratic hilltop inflation of the form $V=V_0-\frac{1}{2}m^2\phi^2+\frac{1}{6}\mathfrak{c}\phi^3$ inflate eternally if compatible with observations.
        \item General hilltop inflation with potentials of the form $V=V_0-\frac{1}{p}\mathfrak{c}\phi^\mathfrak{p}$ with $\mathfrak{p}\geq4$ inflate eternally.
    \end{itemize}
\end{itemize}
Similar findings, resulting in a no eternal inflation conjecture are also presented by \citet{russo_dilaton-axion_2022}: having two dilatons and one axion coupled to gravity, they derive bounds on the scalar field space gradient\footnote{
    The scalar field gradient is equal to the dilaton self-coupling in their model.
    }
and curvature that are incompatible with eternal inflation.\footnote{
    The characterising feature of their model is that it reproduces itself under dimensional reduction or truncation (which only happens under the presence of \textit{two} dilatons).
} 
They stress that both criteria are necessary to rule out eternal inflation\,\textemdash\,a bound only on the gradient alone is insufficient. 
Eternal inflation is ruled out by showing that the spacetime metric cannot have a future cosmological event horizon.

Despite these conceptual constraints, there are concrete models of eternal inflation. In the following, we present some, focussing on their interplay with the \gls{dsc}.

\begin{itemize}
    \item Hilltop models of eternal inflation either violate entropy bounds or the \gls{dsc}, and suffer from a graceful exit problem \cite{wang_eternal_2020}.
    Hilltop potentials \cite{lin_topological_2020} as well as models with tachyonic scalar fields \cite{trivedi_rejuvenating_2022} show better consistency in brane-world scenarios.
    \citet{trivedi_rejuvenating_2022} achieved better compatibility between the Gibbons\textendash Hawking entropy bounds and the \gls{dsc} \cite{wang_eternal_2020} by applying a generalised uncertainty principle to derive a lower bound on the potential in such a setting.
    \item Eternal inflation with a steep potential of the form $V(\phi)=V_0\exp\left(\mathfrak{s}_1\phi/M_\textnormal{P}\right)$ is found to be problematic in the light of the \gls{dsc} \cite{dimopoulos_steep_2018,alestas_curve_2024,agrawal_cosmological_2018,akrami_landscape_2019,raveri_swampland_2019,akrami_multi-field_2021}. For $\mathfrak{s}_1\geq1$, there was less than one upwards quantum fluctuation per Hubble time, which essentially means that inflation stops \cite{dimopoulos_steep_2018}.
    \item \citet{matsui_swampland_2020} study a model in the context of the Hartle\textendash Hawking no-boundary proposal:\label{p:EternalHartle}
    An infinitely extended plateau potential violates the \gls{dc} as well as the \gls{dsc} right away, even though it is favoured by observations \cite{planck_collaboration_Inflation_2020}. To get a similar observational behaviour without the theoretical complications, an approximate plateau potential is studied:
    \begin{equation}
        V=V_0\left[\left(\tanh{\frac{\phi}{\sqrt{6\varsigma}}}\right)+\varepsilon\cosh{\frac{\phi}{\sqrt{6\varsigma}}}\right],
    \end{equation}
    with $\varepsilon\ll1$, which breaks the shift symmetry of $\phi$, and $\varsigma$ a dimensionless parameter, which controls whether the model is a small field inflation model or a large field inflation model (since the typical field range is given by $\Delta\phi\sim\sqrt{6\varsigma}$). When $\phi/\sqrt{\varsigma}$ is large, the number of $e$-folds during inflation is high and the slow-roll parameters are small. The value of $V_0$ is bounded by \gls{cmb} observations.\footnote{
        We would expect that also \gls{bbn} observations could provide us with bounds, but this was not taken into consideration by \citet{matsui_swampland_2020}.
        }
    Numerical analysis shows that $\mathfrak{s}_1$ and $\mathfrak{s}_2$ have to be smaller than $\order{10^{-5}}$, a clear contradiction with the \gls{dsc}, which demands $\order{1}$-parameters.
    \item \citet{guleryuz_non-perpetual_2024} presents eternal inflation with a twist that is compatible with observations and the \gls{dsc}: In his model of \textit{non-perpetual} eternal inflation, where a tachyonic instability ends eternal inflation after more than 400 $e$-folds, such that it is locally perceived as eternal inflation, the speed of sound is subluminal and $\mathfrak{s}_2\simeq\left(0.6\sim8\right)$ is found.
\end{itemize}

\subparagraph{Singe-field inflation} in a \gls{gr} based cosmology is ruled out if $N_e\geq60$ and $\abs{\nabla V}>V$ are required \cite{bjorkmo_hyperinflation_2019,dias_primordial_2019,agrawal_cosmological_2018,achucarro_string_2019,kinney_zoo_2019,chiang_what_2019,ben-dayan_draining_2019,kawasaki_primordial_2018,leedom_constraints_2021}.
We discuss various single-field models in more detail in the following.

\citet{herrera_reconstructing_2020} uses Planck data to reconstruct an effective potential with a $k$-essence\footnote{
    In $k$-essence, kinetic terms are present in the Lagrangian of the scalar field, such that $L=-\left(g^{\mu\nu}\partial_\mu\phi\partial_\nu\phi/2\right)\left(1+2g(\phi)\right)-V(\phi)$, with $g(\phi)$ a coupling function.
} Ansatz for the inflaton, and shows that this model has a narrow range of viability, where it satisfies observational constraints as well as the \gls{dsc}:
He expresses the potential in terms of the number of $e$-folds $N_e$, such that
\begin{align}
    V(N_e)&=\frac{N_e}{\lambda+\beta N_e}\\
    \lambda&=\frac{N_e^2}{12\pi^2P_\textnormal{s}},
\end{align}
where the powerspectrum $P_\textnormal{s}\simeq\num{2.2e-9}$ requires $\lambda\simeq\num{e10}$ for $N_e=60$.
For $\beta>0$, the reconstructed effective potential corresponds to natural inflation.
For $\beta<0$, the reconstructed effective potential corresponds to hyperbolic inflation, which yields an exponential respectively power-law function for the effective scalar field potential in different limits.
Taking into account observational constraints as well as the \gls{dsc} yields $\pm\num{1.5e8}<\beta<\pm\num{3.2e8}$, i.e. there is a narrow window for this model to work.

In modified gravity theories, single-field inflation can be accommodated while retaining compatibility with the \gls{dsc} and the \gls{dc} \cite{trivedi_swampland_2022,trivedi_implications_2021,mohammadi_revisiting_2020,sadeghi_swampland_2021}.\footnote{
    It is not a general finding that modified gravity makes inflation compatible with the \gls{dsc} while inflation taking place in a \gls{gr} setting is at odds with the \gls{dsc}.
    In a theory with a non-minimal coupling between matter and curvature, \citet{bertolami_theories_2023,das_distance_2020} find that cold single-field slow-roll inflation is incompatible with the \gls{dsc}. The studied model has the action $S=\int\!\sqrt{-g}\left[f_1(R)M_\textnormal{P}^2/2+f_2(R)\mathcal{L}\right]\,\mathrm{d}^4x$, with two functions that depend on the curvature\,\textemdash\,$f_1(R)$ corresponds to the function known from $f(R)$ gravity that introduces a Yukawa-type \nth{5} force, and $f_2(R)$ introduces an additional force that depends on the spatial gradient of the Ricci scalar \cite{bertolami_theories_2023}.
    Also found to be incompatible with the \gls{dsc} is mimetic $f(G)$ gravity \cite{gashti_further_2023}. 
    Nevertheless, a general statement can be made \cite{trivedi_swampland_2022}: If a theory of modified gravity yields Friedmann equations of the form $F(H)=\rho/3\sim V/3$ during inflation, the slow-roll parameter can be expressed as $\epsilon=F(H)H^\prime \mathfrak{s_1}/H^3$, and the \gls{dsc} is not violated as long as $H^3/F(H)H^\prime\gg\order{1}$ can be obtained during inflation.
        This generalisation relies on two key assumptions \cite{trivedi_swampland_2022}:
        First, the Klein\textendash Gordon equation for the inflaton holds, i.e. $\ddot{\phi}+3H\dot{\phi}+V^\prime=0$.
        Second, the background metric and its perturbations are not modified.
}
Some examples are
\begin{description}
    \item[Inverse Monomial Inflation]
        A potential of the form $V(\phi)=\mathfrak{c}^{4+\mathfrak{p}}/\phi^\mathfrak{p}$ on a brane violates the refined \gls{dsc} and potentially the \gls{dc} but satisfies the combined \gls{dsc} (as well as the (strong) scalar \gls{wgc} (\cref{p:SWGC})) \cite{gashti_exploring_2024}.
    \item[Power-law inflation] 
        $V(\phi)=V_0\phi^{n_1}$ with a coupling to curvature $f(\phi)=f_0\phi^{n_2}$ can simultaneously satisfy observational constraints as well as the \gls{dsc} and the \gls{dc} in scalar\textendash tensor theory \cite{mohammadi_revisiting_2020,gashti_constraints_2022}. The same holds for power-law potentials when a Chern\textendash Simons term\footnote{
            It's not the case that power-law inflation is generally compatible with the \gls{dsc} if there is a Chern\textendash Simons term, e.g. a model with ${n_1}=1$ and ${n_2}=-1$ is found to be compatible, while the case ${n_1}=2$ and ${n_2}=2$ is found to violate the \gls{dsc} as well as the \gls{dc} \cite{fronimos_inflationary_2023}.
            }
        is present \cite{gitsis_swampland_2023,fronimos_inflationary_2023}
        or inverse power-law potentials (in the form of an error function) when a \gls{gb} term is present \cite{gitsis_swampland_2023}.
        A quadratic power-law potential ($V\sim\phi^2$) is found to be compatible with observational constraints, the \gls{dsc}, and the \gls{dc} in a $f(R)=\mathfrak{p}R$ setting for $\mathfrak{p}\leq0.01$ \cite{oikonomou_rescaled_2022}.
        A power-law potential in $f(R,T)$ gravity, with a non-canonical scalar field that couples non-minimally to matter and curvature, is found to satisfy the \gls{dsc} \cite{ossoulian_inflation_2023}.
    \item[Quantum corrected inflation]      
        $V(\phi)=\lambda\phi^4\left(\phi/\Lambda\right)^{4\beta}$, with $\Lambda$ the cutoff scale and $\beta\sim\order{0.1}$ (to satisfy observational constraints) the quantum correction parameter \cite{joergensen_primordial_2014}, satisfies the combined \gls{dsc} in scalar\textendash tensor theory, while violating the refined \gls{dsc}\,\textemdash\,the $\order{1}$ thresholds of both criteria are not satisfied \cite{yuennan_further_2022}. 
    \item[Chaotic Inflation]
        The potential $V(\phi)=m^2\phi^2/2$ with curvature-coupling $f(\phi)=f_0\left(1-\mathfrak{c}\phi^2\right)$, $\mathfrak{c}$ some constant, has a window of compatibility in scalar\textendash tensor theory, where it is compatible with observations, the \gls{dsc}, as well as the \gls{dc} \cite{mohammadi_revisiting_2020}.
    \item[Generalised Chaplygin-like Inflation]
        A form of chaotic inflation (with \gls{gr} as a limit \cite{bertolami_chaplygin_2006}), where the inflaton field is modelled as a fluid with the \gls{eos} $p=-\mathfrak{p}\rho^{-\mathfrak{c}}$ with $0<\mathfrak{c}\leq1$ \cite{bertolami_sitter_2024}. This single-field model satisfies the \gls{dsc} in the $\mathfrak{p}\gg V^{1+\mathfrak{c}}$ regime \cite{bertolami_sitter_2024}, as
        \begin{align}
            \epsilon_V&\ll\left(\frac{\mathfrak{p}}{V^{1+\mathfrak{c}}}\right)^\frac{2+\mathfrak{c}}{1+\mathfrak{c}}\\
            \abs{\eta_V}&\ll\left(\frac{\mathfrak{p}}{V^{1+\mathfrak{c}}}\right)^\frac{1}{1+\mathfrak{c}}.
        \end{align}
    \item[Constant-Roll Inflation]
        The \gls{dsc} with $\eta_V\sim\order{1}$ is satisfied in a brane-world scenario, where $H\propto\rho$ holds, instead of the $H\propto\sqrt{\rho}$ known from 4d cosmology (stemming from terms quadratic in energy density in the Friedmann equation) \cite{mohammadi_brane_2020,stojanovic_constant-roll_2024}.\footnote{See \cref{eq:brane_Hubble}.}
        A constant-roll model with $V\sim\phi^2$ and a coupling $\sim V\phi^2$ to a Chern\textendash Simons term is found compatible with the \gls{dsc} and the \gls{dc} \cite{fronimos_inflationary_2023}.
        In a 4d setting, constant-roll is found to be incompatible with the \gls{dsc} when compatible with observational constraints \cite{shokri_quintessential_2022,shokri_constant-roll_2022}.
    \item[Hilltop Inflation]
        $V^{\prime\prime}/V\sim-\order{\num{e-2}}$ to be compatible with observations, unless put on a brane, where $V^{\prime\prime}/V\sim-\order{1}$ can be achieved \cite{lin_type_2019,osses_reheating_2021}. This is a small-field inflation model, ergo it is compatible with the \gls{dc} \cite{lin_type_2019}.
        In an $f(R)=\mathfrak{p}R$ setting, hilltop inflation was deemed incompatible \cite{oikonomou_rescaled_2022}.
    \item[Logarithmic potentials]
        $V(\phi)=V_0+V_1\log^n\phi$ with coupling $f(\phi)=\sqrt{\log^n\phi}$ are compatible with the \gls{dsc} in scalar\textendash tensor theory \cite{gashti_constraints_2022}.
    \item[Exponential Inflation]
        $V(\phi)=V_0\exp(\mathfrak{c_1}\phi)$ and a coupling to curvature $f(\phi)=f_0\exp(\mathfrak{c_2}\phi)$, $\mathfrak{c_i}$ some constants,\footnote{See \cref{eq:E-model} for the related E-model.} can simultaneously satisfy observational constraints as well as the \gls{dsc} and the \gls{dc} in scalar\textendash tensor theory \cite{mohammadi_revisiting_2020,gashti_constraints_2022}.
        \citet{yuennan_further_2022} find that exponential inflation in scalar\textendash tensor theory violates the refined \gls{dsc} with $\mathfrak{s}_1\sim\order{0.01}$ and $\mathfrak{s}_2\sim\order{0.01}$, yet it satisfies the combined \gls{dsc}.
        Expontential inflation with a \gls{gb} coupling can satisfy the \gls{dsc}, the \gls{dc}, and observational constraint simultaneously \cite{gitsis_swampland_2023}. 
        Exponential inflation in $f(R,T)$ gravity, where a non-canonical scalar field couples non-minimally to matter and curvature, is found to satisfy the \gls{dsc} \cite{ossoulian_inflation_2023}.
    \item[Hyperbolic inflation]
        satisfies the combined \gls{dsc} in scalar\textendash tensor theory, while being in strong tension with the refined \gls{dsc}, as the relative gradient as well as the relative Hessian are both below the required $\order{1}$ values \cite{yuennan_further_2022}.\footnote{See \cref{eq:T-model} for the related T-model.}
    \item[Gauss\textendash Bonnet inflation]
        has the action
        \begin{equation}
            S=\frac{1}{2}\int\!\sqrt{-g}\left[R-\partial_\mu\phi\partial^\mu\phi-2V(\phi)-\mathcal{F}(\phi)\mathfrak{G}\right],
        \end{equation}
        with $\mathfrak{G}=R_{\mu\nu\rho\sigma}R^{\mu\nu\rho\sigma}-4R_{\mu\nu}R^{\mu\nu}+R^2$ the topological \gls{gb} term,
        and $\mathcal{F}(\phi)$ the \gls{gb} coupling function \cite{yi_gaussbonnet_2019,odintsov_viable_2018,chakraborty_inflation_2018,kanti_gauss-bonnet_2015,kanti_singularity-free_1999,rizos_existence_1994,satoh_higher_2008,gashti_pleasant_2022,gitsis_swampland_2023}.
        For the choice $\mathcal{F}(\phi)=3\mathfrak{p}/4V(\phi)$, with $0<\mathfrak{p}<1$, \citet{yi_gaussbonnet_2019} derive
        \begin{align}
            r_\textnormal{ts}&=16\left(1-\mathfrak{p}\right)\epsilon_V\\
            \epsilon_V&=\frac{1-\mathfrak{p}}{2}\left(\frac{V^\prime}{V}\right)^2\\
            \eta_V&=-2\left(1-\mathfrak{p}\right)\left[\frac{V^{\prime\prime}}{V}-\left(\frac{V^\prime}{V}\right)^2\right],
        \end{align}
        which can be compatible with observational constraints regarding $r_\textnormal{ts}$ and simultaneously the \gls{dsc} as well as the \gls{dc}. Similar results are obtained by \citet{gashti_pleasant_2022,gitsis_swampland_2023,odintsov_inflationary_2023,kawai_cmb_2021}.
        A hypergeometric scalar coupling is shown to be incompatible \cite{baddis_swampland_2024,baddis_hypergeometric_2024}.
\end{description}

Lorentz violation can remedy the compatibility between models of single-field inflation and the \gls{dsc}. The relevant equations with $\mathfrak{c_\textnormal{L}}$ parametrising the Lorentz violation are the following \cite{trivedi_lorentz_2022}:
\begin{align}
    \ddot{\phi}+3H\dot{\phi}+\frac{V^\prime}{1+\mathfrak{c_\textnormal{L}}}&=0\\
    H^2&=\frac{1}{3M_\textnormal{P}}\left(\frac{1+\mathfrak{c_\textnormal{L}}}{2}\dot{\phi}^2+V\right)\\
    \epsilon_H&=\frac{M_\textnormal{P}^2}{2\left(1+\mathfrak{c_\textnormal{L}}\right)}\left(\frac{V^\prime}{V}\right)^2\\
    \eta_H&=\frac{M_\textnormal{P}^2}{2\left(1+\mathfrak{c_\textnormal{L}}\right)}\frac{V^{\prime\prime}}{V}\\
    N_e&=\int\!\frac{1+\mathfrak{c_\textnormal{L}}}{M_\textnormal{P}^2}\frac{V}{V^\prime}\,\mathrm{d}\phi.
\end{align}
For $\mathfrak{c_\textnormal{L}}\gg1$ the field can be slow-rolling while satisfying the \gls{dsc} \cite{trivedi_lorentz_2022}. Models that use the breaking of Lorentz invariance include the following:
\begin{description}
    \item[Inflation with Spontaneous Symmetry Breaking] Broken Lorentz invariance
        and a potential of the form $V(\phi)=V_0\left(1+\mathfrak{c}_1\phi^2/M_\textnormal{P}^2+\mathfrak{c}_2\phi^4/M_\textnormal{P}^2\right)$ can be made compatible with observations, the \gls{dsc}, and the \gls{dc} for appropriate choices of $\mathfrak{c}_1$ and $\mathfrak{c}_2$ \cite{trivedi_lorentz_2022}.
    \item[Higgs inflation] 
        is found to be compatible with the \gls{dsc} and the \gls{dc} if Lorentz invariance is broken for $V(\phi)=V_0\left(1-\exp\left(-\sqrt{2/3}\phi/M_\textnormal{P}\right)\right)^2$ \cite{trivedi_lorentz_2022}.
    \item[Radion gauge inflation]
        $V(\phi)=V_0\phi^2/\left[\mathfrak{p}M_\textnormal{P}^2+\phi^2\right]$ with broken Lorentz invariance is compatible with the \gls{dsc} and the \gls{dc} for  $\mathfrak{p}>0.0058$ \cite{trivedi_lorentz_2022}.
\end{description}

\subparagraph{Multi-field inflation} cannot be distinguished from single-field inflation with current observations, as the expected perturbations or non-Gaussian effects are smaller than the current observational thresholds. This does not necessarily change in the near future, as \citet{tokeshi_why_2024} point out. In their study, they found that multi-field models that have a single-field attractor solution typically inflate the most, i.e. volume-selection effects favour scenarios that look like single-field inflation.
However, it is interesting to study multi-field models from a theoretical perspective, as the incompatibility with the \gls{dsc} can be avoided: $\epsilon_V\neq\epsilon_H$ can be obtained \cite{guleryuz_superuniversal_2023}, such that the slow-roll condition potentially can be satisfied without spoiling the \gls{dsc} \cite{achucarro_string_2019,ben-dayan_draining_2019}.\footnote{
    Not all multi-field models satisfy the \gls{dsc}, e.g.
    two-field quintessential Higgs inflation, with a quintessence field coupled to a non-minimally coupled Higgs field \cite{eshaghi_two-field_2022,eshaghi_two-field_2025} doesn't do the trick, as $\mathfrak{s}_1\lesssim\num{8e-3}$ is found to be in agreement with observations.
}
Examples are
axion inflation \cite{damian_two-field_2019,jin_axion_2021,rudelius_conditions_2019};
fat inflation \cite{chakraborty_fat_2020,aragam_rapid-turn_2022};
multi-field \gls{dbi} inflation with non-canonical kinetic terms \cite{solomon_non-canonical_2020},
models of inflation with a flat or negatively curved field space metric \cite{aragam_multi-field_2020};
rapid turn inflation\footnote{
    Rapid turn inflation models are multi-field inflation models where angular momentum prevents the inflaton field from rolling down a steep potential, with the angular momentum either satisfying $\partial_t\left(a^3J\right)=0$, such that the angular momentum redshifts exponentially fast during inflation, or satisfying $\partial_t\left(a^3J\right)=a^3\partial_\theta V(\phi,\theta)$, such that the angular momentum is nearly constant during inflation \cite{kolb_completely_2023}. Models of the former type are also referred to as \textit{hyperbolic inflation}, whereas the latter are models of \textit{monodromy inflation}. Both types satisfy the \gls{dsc}. Moreover, models of monodromy inflation are not only compatible with the \gls{dsc}, they can even produce sufficient \gls{dm} abundance through gravitational particle production \cite{kolb_completely_2023}.
}
such as angular inflation \cite{christodoulidis_angular_2019}, hyperinflation,\footnote{
    \citet{bjorkmo_hyperinflation_2019} found that hyperinflation generalised to an arbitrary number of fields can satisfy either the \gls{dsc} \textit{or} the \gls{dc}, but not both simultaneously while also reheating the universe.
    Small-field hyperinflation compatible with the \gls{dsc} needs to be severely fine-tuned if it is to agree with cosmological observations, such as the powerspectrum, or solve the horizon and flatness problem \cite{bjorkmo_hyperinflation_2019}.
    The potential in hyperinflation agrees with the \gls{dsc} because the steepness of the potential is bound by the curvature of the hyperbolic plane $R$:
    $
        3R<\abs{V^\prime}/V<1/R.
    $
    }
and monodromy inflation;\footnote{
    It is reasoned that axion monodromy inflation is a test of \gls{uv} effects, as non-perturbative effects leave their imprints on \gls{ir} observables such as the powerspectrum and the bispectrum \cite{baumann_inflation_2015,flauger_oscillations_2010,flauger_resonant_2011}.
}
orbital or sidetracked inflation \cite{kolb_completely_2023,garcia-saenz_primordial_2018};
inflation of nilpotent superfields in a supergravity background \cite{guleryuz_superuniversal_2023}; and
warm inflation \cite{das_note_2019,das_warm_2019,motaharfar_warm_2019,berera_warm_1995,bertolami_multi-field_2022}.

To demonstrate how multi-field inflation is constrained by swampland conjectures, we decompose the potential into a tangential and a normal component
\begin{equation}
    V^a=T^aV_\phi+N^aV_N
\end{equation}
\begin{itemize}
    \item the tangential is defined as $T^a=\dot{\phi}^a_0/\dot{\phi}_0$, with $\dot{\phi}_0=\sqrt{G_{ab}\dot{\phi}^a_0\dot{\phi}^b_0}$, using the scalar field space metric $G_{ab}$
    \item $V_\phi$ corresponds to the projection of the gradient along the inflationary trajectory %
    \item the normal component is defined as $N^a=\frac{-D_tT^a}{\abs{D_tT}}$
    \item $D_t$ is the covariant derivative, which is defined as $D_tV^a=\dot{V}^a+\Gamma_{bc}^aV^b\dot{\phi}^c$, using the scalar field space metric $G_{ab}$ to define the Christoffel symbols
\end{itemize}
and focus on the turning rate $\mathfrak{T}$ \cite{achucarro_string_2019}:
\begin{equation}\label{eq:turningrate}
    \mathfrak{T}=\frac{V_{N}}{\dot{\phi}_0}.
\end{equation}
The slow-roll parameters for multi-field inflation can be defined as
\begin{align}
    \epsilon_V&=\frac{1}{2}\frac{V^aV_a}{V^2}\\
    &=\frac{1}{2}\frac{V_\phi^2+V_N^2}{V^2}\\
    &=\epsilon_H\left(1+\frac{\mathfrak{T}^2}{9H^2}\right)\label{eq:e_V-e_H}\\
    \epsilon_H&=-\frac{\dot{H}}{H^2}\\
    &=\frac{1}{2}\frac{V_\phi^2}{V^2}%
\end{align}
where it becomes obvious that we can have situations with $\epsilon_V\sim\order{1}$ while satisfying the current observational \gls{cmb} constraint $\epsilon_H<0.0044$ \cite{achucarro_string_2019}:
\Cref{eq:dSc} boils down to $\mathfrak{s}_1<\sqrt{2\epsilon_V}$ \cite{garg_bounds-dS_2019}, which is satisfied by a high enough turning rate.
If all the fields have masses above the Hubble scale, the slow-roll condition is satisfied if the turning rate is large, i.e. $\mathfrak{T}^2/H^2\gg1$ \cite{chakraborty_fat_2020,aragam_rapid-turn_2022}. This is in contradiction with the common belief that high masses spoil slow-rolling\,\textemdash\,a theme that runs under the name \textit{$\eta$-problem} \cite{chakraborty_fat_2020}. However, settings with a high turning rate seem to be rare from a string theory perspective \cite{aragam_rapid-turn_2022}.\footnote{
    An example of a model with a high turning rate that satisfies the \gls{dsc} is the realisation of \gls{dbi} inflation studied by \citet{solomon_non-canonical_2020}.
    }
For \cref{eq:dScrefined} to hold, the turning rate does not have to be particularly large \cite{bjorkmo_rapid-turn_2019}.
The takeaway is that slow-roll inflationary models are not fully ruled out, but require curved, non-geodesic trajectories \cite{achucarro_string_2019}.

Another Ansatz to satisfy the \gls{dsc}, the \gls{tcc}, as well as observational constraints is presented by \citet{kogai_escape_2020}: In their model, inflation happens in multiple subsequent phases, driven by multiple fields, including spectator fields. By coupling multiple hill-top potentials à la $V=\sum_iV_i(\phi_i)+\sum_{i\neq j}\mathfrak{l}_{ij}V_i(\phi_i)\phi_j^2/2M_\textnormal{P}$, with $\mathfrak{l}_{ij}\sim\order{1}$ coupling constants, a dynamical repetition of inflation can be realised, such that $\epsilon_H\ll1$ and $n_\textnormal{s}-1\simeq-2\epsilon_H+2m_\textnormal{seff}^2/3H^2$, where a slightly tachyonic effective spectator mass $m_\textnormal{seff}^2\sim-0.05H^2$ makes the model compatible with \gls{cmb} observations.

Furthermore, complex fields can be viewed as multi-field models. A complex field can be written as two real fields. \citet{sadeghi_anisotropic_2022} assess anisotropic constant roll inflation with a complex quintessence field. The plots they show indicate that the \gls{dsc} is satisfied for large field ranges (even though a trans-Planckian field excursion violates the \gls{dc}), yet, for field amplitudes between 0 and 1, the \gls{dsc} is violated. They didn't explore if there is an observationally viable phase space window that might be compatible with the swampland conjectures.

\subparagraph{Small-field inflation}\label{sec:SmallFieldInflation} 
in the form of multiphase hilltop inflation with \gls{pbh} production satisfies the second \gls{dsc} criterion (\cref{eq:dScrefined}) for each phase individually \cite{tada_primordial_2019}. Furthermore, there are various models of small-field inflation in brane-world scenarios that are compatible with the \gls{dsc}.
\citet{osses_reheating_2021} show that the \textit{brane} can be slow-rolling as
\begin{align}
    \epsilon_\textnormal{B}&=\epsilon_V\frac{1+V/\mathcal{T}}{\left(1+V/2\mathcal{T}\right)^2}&=&\frac{M_4^2}{16\pi}\left(\frac{V^\prime}{V}\right)^2\\
    \eta_\textnormal{B}&=\eta_V\frac{1}{1+V/2\mathcal{T}}&=&\frac{M_4^2}{8\pi}\frac{V^{\prime\prime}}{V},
\end{align}
where $\mathcal{T}$ is the brane tension, and $\epsilon_V$ and $\eta_V$ are the standard 4D slow-roll parameters, which can be large in this case (to be compatible with the \gls{dsc}), whereas the slow-roll parameters for the brane itself are small and suppressed by a high potential. If a high potential is compatible with other swampland criteria or the inflation model at hand with regard to cosmological observations, needs to be investigated on a case-by-case basis.\footnote{
    For instance, if the potential is very high initially, and remains high, is it still in a high false vacuum today? Can this be probed? Does this have testable/observable effects? Or does the slow-roll approximation just break down towards the end of inflation?
}
Two examples studied by \citet{osses_reheating_2021} are exponential SUSY inflation with
\begin{equation}
    V(\phi)=\Lambda^4\left(1-e^{-\phi/\mu}\right),
\end{equation} 
where the potential is asymptotically flat for $\phi\rightarrow\infty$, the \gls{dc} is satisfied, and the \gls{dsc} is satisfied in a brane-world setting, with
$\Lambda$ a constant mass scale, and
$\mu$ the symmetry breaking scale, and
a hilltop potential of the form 
\begin{equation}
    V(\phi)=\Lambda^4\left[1-\left(\frac{\phi}{\mu}\right)^n\right]
\end{equation}
with the vacuum expectation value $\mu$, about which they conclude that observations rule out $n=2$ as well as $n=4$ for $\mu\lesssim M_\textnormal{P}$, that the potential is compatible with the \gls{dc}, and that in a brane-world scenario the \gls{dsc} is satisfied because the brane can be slow-rolling while the inflaton potential itself can be steep. Furthermore, they find this model to satisfy the \gls{dsc} if a suitable \gls{gb} term is present \cite{osses_reheating_2021}.

Another small-field, single-field inflationary model is power-law plateau inflation with a potential of the form
\begin{equation}
    V=V_0\left(\frac{\phi^{\mathfrak{c}_1}}{\phi^{\mathfrak{c}_1}+m^{\mathfrak{c}_1}}\right)^{\mathfrak{c}_2}
\end{equation}
with $m,\phi$ a mass scale, and ${\mathfrak{c}_i}$ real parameters \cite{dimopoulos_modelling_2016}. The model is compatible with the \gls{dc}, the \gls{dsc}, as well as the \gls{tcc} when put in a brane-world \cite{adhikari_power_2020}. The presence of extra-dimensions in the brane-world scenario modifies the slow-roll parameters, such that all constraints can be satisfied \cite{adhikari_power_2020}.

We make further comments about the feasibility of various other (small-field) inflationary models in brane-world scenarios in other parts of this section.

\subparagraph{Warm inflation}\label{sp:dSC_Warm-Inflation}
converts vacuum energy to radiation through dissipative effects and can be made compatible with the \gls{dsc} \cite{brandenberger_strengthening_2020,berera_warm_1995,bastero-gil_warm_2009,motaharfar_warm_2019,das_warm_2019,bastero-gil_warm_2019,kamali_warm_2020,berera_trans-planckian_2019,kamali_non-minimal_2018,kamali_warm_2019,kamali_reheating_2020,das_swampland_2020,yuennan_warm_2024,das_runaway_2020,rasouli_warm_2019,santos_warm_2022,kamali_recent_2023,berera_thermal_2021,kamali_intermediate_2021,arya_primordial_2024,das_distance_2020}.\footnote{
    Contrary to cold inflation, where Hubble damping dominates, in warm inflation dissipative effects are intrinsic to interactions between the inflaton field and other degrees of freedom \cite{brandenberger_strengthening_2020,berera_warm_2009}. This implies that $T>H$ is required for inflation to be warm \cite{das_swampland_2020}.
    In warm inflation, the origin of perturbations is thermal, contrary to other inflationary models where quantum fluctuations induce the perturbations \cite{motaharfar_warm_2019}.
    }
It does not require a reheating phase at the end of inflation \cite{das_note_2019,kamali_recent_2023}.\footnote{
    \citet{kamali_reheating_2020} found reheating due to an oscillatory phase around the minimum to be incompatible with \cref{eq:dSc}.
}
The inflaton field's \gls{eom} is given by
\begin{equation}
    \ddot{\phi}+3H\left(1+\Upsilon\right)\dot{\phi}+V^\prime=0,
\end{equation}
with $\Upsilon\geq0$ related to the dissipation \cite{das_note_2019}.\footnote{
    $\ddot{\phi}$ is omitted in the original work by \citet{berera_warm_1995}; this is justified as slow-rolling is assured by friction due to the thermal bath \cite{das_swampland_2020}.
}
The radiation energy density evolves as \cite{yuennan_warm_2024}:
\begin{equation}
    \dot{\rho_\textnormal{r}}+4H\rho_\textnormal{r}=3H\Upsilon\dot{\phi}^2.
\end{equation}
Furthermore,
\begin{equation}
    r_\textnormal{ts}=\frac{H}{T}\frac{16\epsilon_V}{\left(1+\Upsilon\right)^{5/2}},
\end{equation}
with $T>H$ the temperature of the thermal bath,\footnote{
    The condition that the inflaton background dynamics as well as the primordial powerspectrum are affected by dissipation\,\textemdash\,$T>H$ \cite{bastero-gil_warm_2009}\,\textemdash\,is not a given \cite{berera_strong_1998,yokoyama_is_1999,rasouli_warm_2019}, but is achieved by various models studied in this section.
}
shows that $r_\textnormal{ts}<16\epsilon$ can be achieved in warm inflation \cite{das_note_2019,bartrum_warming_2013}.\footnote{
    A warm inflation model studied by \citet{yuennan_warm_2024} shows $r_\textnormal{ts}<\num{e-2}$, and for some parameter choices values as small as $r_\textnormal{ts}\sim\num{e-23}$.
    Similarly, a model with strong dissipation by \citet{kamali_warm_2020} yields $r_\textnormal{ts}\sim\num{e-27}$ for their choices of parameters.
}

To assess the viability of warm inflation, the weak and strong dissipation regimes are usually treated separately.
On the one hand, work by \citet{das_warm_2019} indicates that a weak dissipative regime is compatible with observations \textit{and} the \gls{dsc} \cite{das_note_2019}.
On the other hand, strong dissipative effects lead to a scale-dependent scalar powerspectrum \cite{motaharfar_warm_2019} / non-Gaussianities \cite{das_runaway_2020}, and destabilise the inflaton potential \cite{das_swampland_2020}.
Yet, some models of warm inflation seem to circumvent these constraints and are compatible with observational data and swampland conjectures \cite{kamali_warm_2020}.
\citet{bertolami_multi-field_2022} find that strong dissipative effects are required for single- as well as multi-field warm inflation to be compatible with the \gls{dsc}, and work by \citet{bastero-gil_warm_2019} highlights that the weak dissipation regime is in tension with the \gls{dsc} and the \gls{dc}.
Since $\epsilon_V>\mathfrak{s}_1^2/2$, achieving $\Upsilon=0$ requires $\mathfrak{s}_1\sim\order{\num{e-1}}$ \cite{das_warm_2019}, i.e. only a low value of $\mathfrak{s}_1$ allows for weak dissipation.
Strong dissipation allows the inflaton to approach the slow-roll trajectory quicker, which enforces the attractor-like behaviour of the slow-roll solution \cite{kamali_warm_2020}.
Moreover, to solve the $\eta$-problem and to satisfy \cref{eq:dScrefined}, $V^{\prime\prime}\ll H^2$ is required, i.e. Hubble friction has to dominate during inflation \cite{baumann_inflation_2015,kamali_recent_2023}. This problem is naturally avoided in the strong dissipation regime of warm inflation \cite{kamali_recent_2023,berera_warm_2004,das_note_2019}.
\citet{mohammadi_warm_2020} find that compatibility between a model and observational and theoretical constraints depends on the dissipation \textit{function}\,\textemdash\,even in the strong dissipation regime: They study a model where a tachyon field decays into radiation.\footnote{
    See \cref{f:warm-tachyon-inflation} for their model.
    }
The \gls{dsc} is not satisfied if the dissipation coefficient of their model is a power-law function of the tachyon field, neither is the \gls{dc} in that case. However, if a temperature dependence is introduced in the dissipation coefficient, consistency between the model, observational data, and both conjectures is established.
When dissipation is \textit{strong enough} does not have a clear-cut answer: While
warm inflation with a potential of the form $V(\phi)=V_0\left(1-\mathfrak{c}\mathfrak{p}\phi/M_\textnormal{P}\right)^{1/\mathfrak{p}}$ studied by \citet{santos_warm_2022} is found to be generally compatible with the \gls{dsc} for $\Upsilon\geq10$, as well as having a window of compatibility for $\Upsilon\gtrsim1$ depending on the value of $\mathfrak{p}$,
\citet{rasouli_warm_2019} find that their model of warm \gls{dbi} inflation can already be compatible with observations for $\Upsilon\gtrsim0.1$.\footnote{
    Furthermore, they report that cold \gls{dbi} inflation is incompatible with Planck data \cite{rasouli_warm_2019}.
}

\subparagraph{Natural inflation} with a potential of the form
\begin{equation}
    V=V_0\left(1-\cos{\frac{\phi}{f}}\right),
\end{equation}
with $f$ the axion decay constant, faces the constraints
\begin{align}
    f&<\frac{M_\textnormal{P}}{\mathfrak{s}_1}\\
    f&<\frac{M_\textnormal{P}}{\sqrt{\mathfrak{s}_2}},
\end{align}
which are a little weaker than the \gls{wgc} constraints \cite{fukuda_phenomenological_2019}.
Observations favour a trans-Planckian decay constant \cite{kim_completing_2005,freese_natural_1990}, which violates the \gls{dc} \cite{osses_reheating_2021}.\footnote{
    \citet{parameswaran_subleading_2016} suggest that subleading, non-perturbative corrections can change this verdict.
}
Some authors deem this model incompatible with the \gls{dsc} and the \gls{dc} \cite{oikonomou_rescaled_2022}, but others present models with sub-Planckian axion decay constants \cite{albrecht_spinodal_2015,holman_spinodal_2019} or 
non-canonical Lagrangians that are found to be compatible with the \gls{dsc} and the \gls{dc} \cite{heydari_primordial-fast-roll_2024}. Furthermore, a brane-world scenario can offer a viable framework with $\epsilon_\textnormal{B}\ll1$ for $V\gg\mathfrak{T}$ \cite{osses_reheating_2021}.

\subparagraph{Minimal gauge inflation} is embedded into a higher-dimensional theory with a non-Abelian SU(2) gauge symmetry with 5-dimensional gauge coupling constant $g_5$ and compactification radius $r$, such that the inflaton has a potential
\begin{equation}
    V(\phi)=\frac{9}{\left(2\pi\right)^6r^2}\sum_{n=1}^\infty\frac{1}{n^5}\left[1-\cos\left(\sqrt{2\pi r}g_5n\phi\right)\right],
\end{equation}
which can be regarded as a higher-dimensional completion of natural inflation \cite{park_minimal_2019,gong_minimal_2018}.
To be compatible with the \gls{dsc}, $\mathfrak{s}_1\lesssim0.15$ and $\mathfrak{s}_2\lesssim0.01$ are required \cite{park_minimal_2019}, which is in tension with the $\order{1}$ prediction.

\subparagraph{Chaotic inflation} has a potential of the form
\begin{equation}
    V(\phi)=\frac{1}{2}m^{4-n}\phi^n
\end{equation}
and faces the constraints
\begin{align}
    \frac{V^\prime}{V}&=\frac{2}{\phi_\mu}&\geq\frac{\mathfrak{s}_1}{M_\textnormal{P}}\\
    \frac{V^{\prime\prime}}{V}&=\frac{2}{\phi_\mu^2}&\leq\frac{-\mathfrak{s}_2}{M_\textnormal{P}^2}
\end{align}
with
\begin{equation}
    \phi_\mu^2=\frac{3M_\textnormal{P}^2n^2-2n\left(n-1\right)M_\textnormal{P}^2}{1-n_\textnormal{s}^\mu};
\end{equation}
$\mu$ a scale and $n_\textnormal{s}^\mu$ the corresponding spectral index \cite{haque_reheating_2019}.
\citet{haque_reheating_2019} find a tension between their model of chaotic inflation and the \gls{dsc} for parameter choices that are compatible with observational data and the \gls{dc}, since they find $\mathfrak{s}_1^\textnormal{max}=0.1448<\order{1}$.
Smaller than $\order{1}$ values for $\mathfrak{s}_1$ are also found in a study by \citet{matsui_eternal_2019}, in which they explore the possibilities of using chaotic inflation to describe eternal inflation.

A model of chaotic inflation on a brane, where the constraints are expressed as
\begin{align}
    \frac{V^\prime}{V}&=\frac{n}{\phi}&\geq\frac{\mathfrak{s}_1}{M_\textnormal{P}}\\
    \frac{V^{\prime\prime}}{V}&=\frac{n\left(n-1\right)}{\phi^2}&\leq\frac{-\mathfrak{s}_2}{M_\textnormal{P}^2}\\
    r_\textnormal{ts}&=\frac{24n}{N_e\left(n+2\right)}&\lesssim0.1,
\end{align}
is found to be compatible with the refined \gls{dsc}, if the conjecture is satisfied for the initial field value $\phi_\textnormal{i}$ (since the field value is decreasing during inflation) \cite{lin_chaotic_2019}.

\subparagraph{Higgs inflation} \cite{bezrukov_standard_2008} is studied in different flavours:
\begin{itemize}
    \item  Pure Higgs inflation is in tension with the \gls{dsc} (but not the combined \gls{dsc} (\cref{eq:dSCcombi})) \cite{liu_higgs_2021} as well as with asymptotic safety \cite{eichhorn_constraining_2021}.
    \item Palantini\textendash Higgs inflation \cite{liu_palatini_2021} is in tension with the \gls{dsc} (but not the combined \gls{dsc} (\cref{eq:dSCcombi})) \cite{liu_higgs_2021}.
    \item A non-minimal coupling to gravity with \gls{pbh} production \cite{bezrukov_robustness_2018,garcia-bellido_primordial_2017,ezquiaga_primordial_2018} is compatible with the \gls{dsc} and predicts $n_s\approx0.965$ and $r_\textnormal{ts}\approx0.003$ \cite{cheong_higgs_2018}.
    However, it is argued that fine-tuning is required \cite{das_note_2019}.
    \item A Higgs\textendash Dilaton model is found to be compatible with the refined and the combined \gls{dsc} \cite{liu_higgs_2021}.
    \item Higgs-like inflation on a brane is compatible with the \gls{dsc} \cite{osses_reheating_2021}.\footnote{
        $V(\phi)=\Lambda^4\left[1-\left(\phi/\mu\right)^2\right]^2$ is studied by \citet{osses_reheating_2021}, with $\Lambda$ a constant mass scale and $\mu$ the vacuum expectation value in a 5D brane-world cosmology. They note that the Higgs-like potential can be considered a consistent modification of $n=2$ hilltop inflation, where the additional quadratic term avoids negative potential values at high field values, and find the \gls{dsc} as well as \gls{dc} to be satisfied.         
    }
    \item Higgs inflation in a non\textendash Lorentz invariant setting is found to be compatible with the \gls{dsc} and the \gls{dc} \cite{trivedi_lorentz_2022}.
\end{itemize}

\subparagraph{Quintessence inflation} with an inflaton potential of the form $V=V_0\exp\left(-\mathfrak{p}\phi\right)$ is not slow-rolling. 
\citet{cicoli_quintessence-numerically_2022-1} find that quintessence inflation in the form of axions with a hilltop potential requires very finely tuned initial conditions and a very low energy scale of inflation of $H_\textnormal{I}\lesssim\SI{1}{\mega\electronvolt}$ in order to be a viable model of inflation without violated the \gls{dsc}.

\citet{eshaghi_two-field_2022,eshaghi_two-field_2025} study a quintessence inflaton coupled to the Higgs field and find $\mathfrak{s}_1\lesssim\num{8e-3}$ as an upper bound to be consistent with observations, which violates the \gls{dsc}.

\citet{nitta_dynamical_2025} derive a modified \gls{dsc}, assuming the \gls{sec} holds and that the scalar field velocity needs to be taken into account:
\begin{align}
    -\frac{\partial_\phi V}{V}&\geq&&2\sqrt{\frac{D-2}{\left(D-d\right)\left(d-2\right)}}-\lambda\frac{\dot{\phi}}{V}\\
    \lambda&=&&\frac{M_{\textnormal{P};d}^{d-2}}{4}\sqrt{\frac{D-d}{\left(D-2\right)\left(d-2\right)}}\\
    &&&\times\left(-5d^2+d\left(5D+12\right)-11D-2\right).\nonumber
\end{align}
Quintessence inflation violates this bound.

\subparagraph{$\alpha$-attractor models (Starobinsky)} come in two forms \cite{fukuda_phenomenological_2019}:
T-models of the form
\begin{equation}\label{eq:T-model}
    V=V_0\tanh^{2n}\left(\frac{\phi}{\sqrt{6\alpha}M_\textnormal{P}}\right),
\end{equation}
which are in tension with the \gls{dsc} unless a suitable \gls{gb} term is present \cite{gashti_pleasant_2022} or they take place
in a $f(R)=\mathfrak{p}R$ setting with  $\mathfrak{p}\leq0.008$ \cite{oikonomou_rescaled_2022};
and 
E-models of the form
\begin{equation}\label{eq:E-model}
    V=V_0\left(1-e^{-\sqrt{\frac{2}{3\alpha}}\frac{\phi}{M_\textnormal{P}}}\right)^{2n}.
\end{equation}
The Starobinsky model is the special case with $n=1$ and $\alpha=1$, ergo $\mathfrak{s}_1\lesssim1.6$ \cite{fukuda_phenomenological_2019}.
The single-field Starobinsky model in an unmodified \gls{gr} setting is likely part of the swampland \cite{guleryuz_superuniversal_2023}.
But also in a $f(R)=R+\mathfrak{p}R^2$ and a $f(R)=R+\mathfrak{c}_1R^2+\mathfrak{c}_2R^2\log R$ setting, the Starobinsky model is deemed incompatible \cite{artymowski_fr_2019}.
Observational constraints and the \gls{dc} restrict $\mathfrak{s}_1\leq0.025$, which renders the model incompatible with the \gls{dsc} \cite{haque_reheating_2019}.
However, in a $f(R)$ setting, a window of compatibility with observational constraints, the \gls{dsc}, and the \gls{dc} is found by \citet{oikonomou_rescaled_2022,oikonomou_rescaled_2021}, and 
in non-critical string cosmology, where new dynamics are introduced by identifying cosmic time with a dynamic Liouville mode, the \gls{dsc} contains non-conformal contributions, which renders Starobinsky inflation compatible with this modified \gls{dsc} \cite{ellis_supercritical_2020}.
Both types are found to be compatible with the combined \gls{dsc} in a scalar\textendash tensor theory setting (while violating the refined \gls{dsc}, as the $\order{1}$ thresholds cannot be met) \cite{yuennan_further_2022}.

\subparagraph{Ekpyrotic scenario \cite{khoury_big_2002,khoury_ekpyrotic_2001}}\label{sp:ekpyrotic} 
or \textit{slow-contraction scenario} is an alternative to inflation \cite{cook_supersmoothing_2020},
where a canonically normalised scalar field with a negative exponential potential dilutes anisotropies and spacial curvature, solves the flatness and horizon problems, provides a causal mechanism for structure formation, and is consistent with the \gls{tcc} \cite{bernardo_contracting_2021}.
While inflation takes place when $H>0$ and $\epsilon>1$, to have accelerated expansion, the ekpyrotic scenario describes a contracting phase with $H<0$ and $\epsilon>3$, such that the ekpyrotic field energy density ($\propto a^{-2\epsilon}$) grows faster than the small metric anisotropies ($\propto a^{-6}$) \cite{lehners_small-field_2018}.

Since the ekpyrotic scenario involves a negative potential, the \gls{dsc} does not apply \cite{ben-dayan_draining_2019,lehners_small-field_2018}, but replacing $V$ with $\abs{V}$ in \cref{eq:dSc,eq:dScrefined} leads to conjectures that show a compatibility with the ekpyrotic scenario \cite{bernardo_contracting_2021}\,\textemdash\,something that should not tremendously surprise us, since \gls{ads} space is very popular in string theory settings.
However,
\citet{shiu_collapsing_2024} present some conjectures against fast-roll and an anti\textendash trans-Planckian censorship conjecture that indicate that string theory might not provide potentials steep enough to realise the ekpyrotic scenario.\footnote{
    Furthermore, constraints based on a generalised distance measure for the appplication of the \gls{dc} to the ekpyrotic scenario are discussed by \citet{debusschere_distance_2025}.
    }

\subparagraph{Tachyacoustic models}
propose a superluminal sound speed in the early universe to solve the horizon problem and maintain an approximately scale-invariant power spectrum \cite{lin_consistency_2019,magueijo_speedy_2008,bessada_tachyacoustic_2009,babichev_k-essence_2007,lin_trans-planckian_2021}. \citet{lin_consistency_2019} show that tachyacoustic Lagrangians can be compatible with the \gls{dsc}. However, no known string theory offers $c_\textnormal{s}>1$ \cite{lin_consistency_2019,kinney_quantum_2008}. Moreover, such models violate the \gls{tcc} \cite{lin_trans-planckian_2021}.

\paragraph{Particle Physics}
The \gls{sm} quark masses are compatible with the \gls{dsc}, but a large number of additional light quarks is in increasing tension with the conjecture, as metastable states appear \cite{march-russell_qcd_2020}. A coupling to quintessence has been investigated and found not to weaken the constraining power of the \gls{dsc} \cite{march-russell_qcd_2020}.

\subsubsection{General Remarks}

\paragraph{Which observables would indicate an unstable \gls{ds} space?}
Exactly stable \gls{ds} space does not allow for precisely measurable observables \cite{rudelius_conditions_2019,witten_quantum_2001,rudelius_asymptotic_2021}. The ($1\sigma$) signature of a cascade of \gls{ds} decays is claimed to be found in the \gls{nanograv} \cite{brazier_nanograv_2019} data set \cite{arzoumanian_nanograv_2020} for the following model \cite{li_is_2021}:
The lifetime of inflation is limited by the \gls{tcc} to $\int\!H\,\mathrm{d}t<\log\left(M_\textnormal{P}/H\right)$ \cite{bedroya_trans-planckian-inflation_2020}.
The lifetime of \gls{ds} space during inflation is limited by the \gls{tcc} to $\Delta t<H^{-1}\log\left(M_\textnormal{P}/H\right)$ \cite{bedroya_trans-planckian_2020,bedroya_sitter_bubbles_2020}.\footnote{
    It was observed that the logarithmic factor enhances $\Delta\phi$, which potentially relaxes the bound on the second derivative (\cref{eq:dScrefined}) around the top of the potential \cite{seo_entropic_2020,cai_refined_2021}.
    }
A cascade of first-order phase transitions produces \glspl{gw} at different frequencies, which are subsequently redshifted by the \gls{ds} expansions of the following short-lived phases of inflation of the cascade.
This leads to a stochastic \gls{gw} background with a red-tilt, which is compatible with the \gls{nanograv} results at the $1\sigma$ level.

\paragraph{How is the \gls{dsc} motivated?}\label{p:dSC_Motivation}
Constructing stable \gls{ds} solutions in string theory\footnote{
    The \gls{dsc} is deeply rooted in string theory and the observation that it is very challenging to construct \gls{ds} spaces in string theory. Other theories of \gls{qg} are easily compatible with \gls{ds} spaces or predict \gls{ds} spaces right away, e.g. causal set theory \cite{barrau_string_2021,bombelli_space-time_1987,sorkin_light_2009,dowker_evolution_2017,yazdi_everything_2024}.
    Asymptotic safety even prefers a cosmological constant with an \gls{eos} parameter $w=-1$, or at least leads to flattening potentials when travelling through field space \cite{pawlowski_higgs_2019,de_brito_link_2019,eichhorn_quantum_2018,percacci_search_2015,labus_asymptotic_2016}, which is incompatible with the \gls{dsc} \cite{eichhorn_constraining_2021}. According to the string lamppost principle, everything that happens in \gls{qg} can be accommodated in string theory. If this is true, then either asymptotic safety and causal set theory are not valid theories of \gls{qg}, or it is possible to construct \gls{ds} vacua in string theory, or the string lamppost principle itself is wrong.
} 
is a hard problem\,\textemdash\,so hard that scholars started to conjecture that it might be impossible.\footnote{
    Reviewing roughly the first two decades of the \nth{21} century, \citet{danielsson_what_2018} discuss various attempts at constructing \gls{ds} vacua in string theory and their failure, as well as some rough ideas about the fate of \gls{de}.
}
A study on the distribution of stationary points, in particular local minima, of random scalar potentials with a large number of scalar fields by \citet{low_distribution_2021} provides some insight: They find that the number of critical points increases with the number of scalar fields, yet the relative number of local minima decreases exponentially with the number of scalar fields, i.e. it becomes exceedingly rare to find a true minimum with a high number of scalar fields. This hints at why it is so hard to find a stable \gls{ds} vacuum.
Yet, it doesn't proof that it is impossible to construct stable \gls{ds} points in string theory.
In fact, there are non-perturbative constructions of \gls{ds} space \cite{dvali_quantum_2017,dvali_exclusion_2019,dvali_quantum_2019,brahma_sitter_2021,brahma_four-dimensional_2021,bernardo_crisis_2021,hohm_duality_2019,krishnan_sitter_2019,bernardo_od_2020,bernardo_-cosmology_2020,nunez_new_2021,leontaris_seeking_2023,leedom_heterotic_2023,bardzell_type_2022},
arguments in favour of \gls{ds} space in \glspl{eft} \cite{dodelson_new_2014,moritz_towards_2018,crino_sitter_2021,marolf_ir_2010,brahma_sitter_2021,danielsson_sitter_2011,caviezel_cosmology_2009,silverstein_simple_2008,brahma_resurgence_2023,bernardo_contracting_2021,landgren_effective_2022},
and some authors conclude that \gls{ds} vacua are in the landscape of string theory \cite{kallosh_4d_2019}.
Furthermore, a paper proposing that \gls{ds} vacua are in the swampland \cite{moritz_towards_2018} is shown to be erroneous \cite{gautason_tension_2019} and various assumptions are shown not to hold \cite{kallosh_gaugino_2019,cicoli_sitter_2019,kallosh_sitter_2019,kallosh_4d_2019,kallosh_ds_2019}.
However, various attempts to construct stable \gls{ds} solutions are flawed:
Several known \gls{ds} constructions involving non-geometric fluxes are shown to be unstable, as positive tadpole charges are involved that lead to runaway behaviour \cite{plauschinn_moduli_2021} (cf. \cref{s:tadpole}).
\citet{dasgupta_how_2021} present a \gls{ds} solution in the context of type IIB string theory with $\mathfrak{s}_1\gg1$ that is only in the landscape if fluxes are time-independent.
\citet{cordova_new_2019,cordova_classical_2019} present an Ansatz to construct stable \gls{ds} solutions using O8-planes, but their solutions were later shown to be unstable \cite{van_riet_beginners_2023,bena_oh_2021,cribiori_no_2019}.
\citet{landgren_effective_2022} presents a \gls{ds} solution in $\left(2+1\right)$ dimensions, yet the low dimensionality is troublesome from a gravity perspective.\footnote{See \cref{f:low-d_gravity}.}
Charting type IIB using machine learning, \citet{damian_metastable_2022} present 170 \gls{ds} vacua that are obtained by uplifting metastable \gls{ads} vacua, though there are several loopholes mentioned in the paper how the \gls{ds} vacua could be destabilised.\footnote{
    \citet{alwis_radiative_2021} argues that the \gls{dsc} might lose predictability in the bulk entirely when radiative corrections uplift an \gls{ads} solution into the \gls{ds} regime in the \gls{ir}.
    One potential issue with uplifting \gls{ads} to \gls{ds} is addressed by the \gls{tpc}: The tadpole required to uplift \gls{ads} with a small supersymmetry breaking scale using an antibrane in a long warped throat to \gls{ds} might be too large \cite{bena_tadpole_2021}.}
In the context of holography, \citet{grieninger_non-equilibrium_2020} presents \gls{ds} solutions, yet, introducing matter fields makes the family of these solutions disappear.

Moreover, there is evidence that it is not possible to construct stable \gls{ds} solutions using a perturbative approach \cite{danielsson_what_2018,emelin_effective_2020,dasgupta_sitter_2021,dasgupta_quantum_2019,sethi_supersymmetry_2018,kutasov_constraining_2015,hertzberg_inflationary_2008,andriot_new_2020,andriot_automated_2022,maldacena_supergravity_2001,hertzberg_inflationary_2007,witten_quantum_2001,caviezel_moduli_2010,giddings_dynamics_2006,alwis_potentials_2003,parameswaran_ds_2024,green_constraints_2012},
as well as indications of a perturbative \gls{ir} instability in \gls{ds} space \cite{mottola_particle_1985,tsamis_quantum_1996,mukhanov_back_1996,brandenberger_back_2002,brandenberger_backreaction_2018,kitamoto_sitter_2020,polyakov_infrared_2012,polyakov_sitter_2008,andriot_exploring_2022,andriot_erratum_2022,bernardo_contracting_2021}
and entropic arguments against stable \gls{ds} vacua \cite{geng_entropy_2019}.
Moreover, the refined \gls{dsc} (\cref{eq:dScrefined}) is backed by the \gls{dc} and the \gls{ceb}, a semi-classical notion of entropy of \gls{ds} space \cite{brahma_stochastic_2019}.

In a 10d setting with 4d \gls{ds} subspaces, \citet{andriot_open_2019} makes the following observations:
First, no classical \gls{ds} solutions with parallel sources were found.
Second, no stable classical \gls{ds} solutions with intersecting sources were found.
Third, no classical \gls{ds} solutions were found that exhibit the following simultaneously: a large internal volume, a small string coupling, quantized fluxes, and a finite number of orientifolds.
Furthermore, 10-dimensional Minkowski space is supersymmetric with vanishing energy, but \gls{ds} space has positive vacuum energy, which indicates the instability of \gls{ds} space \cite{baumann_inflation_2015}.
The instability of \gls{ds} vacua with broken supersymmetry, under the conditions that first, the \gls{wgc} holds; second, the theory has objects of codimension 1 (branes/membranes); third, a three-form chiral multiplet is coupled to supergravity; and fourth, the three-form multiplet is nilpotent and breaks supersymmetry, is shown by \citet{farakos_sitter_2021}.
Charged membranes are also found to be the culprit by \citet{liu_cosmological_2024,kaloper_implications_2023}:
Charged membranes, if present, change the cosmological constant via their fluxes. Since the membranes have a tension, the \gls{wgc} applies, which limits the tension. However, to stop discharge of the membranes, i.e. to form a stable situation, the tension has to go to infinity, which the \gls{wgc} forbids. Therefore, stable \gls{ds} vacua are in the swampland.

\paragraph{Are there any counterexamples to the \gls{dsc}?}\label{p:dSC_Counterexamples}
There are various counterexamples to \cref{eq:dSc} \cite{denef_ds_2018,conlon_sitter_2018,murayama_we_2018,choi_ds_2018,hamaguchi_swampland_2018,roupec_sitter_2019,garg_bounds_2019,andriot_sitter_2018}, which \cref{eq:dScrefined} evades \cite{ooguri_distance_2019}.\footnote{
    A potential counterexample to the refined \gls{dsc} is presented by \citet{blaback_new_2018}.
}
Other counterexamples presented by \citet{caviezel_cosmology_2009,flauger_slow-roll_2009} are dismissed by \citet{danielsson_sitter_2011}.

\paragraph{How does the \gls{dsc} depend on the dimensionality?}

An insightful link between cosmic expansion and \gls{ds} space is presented by \citet{hebecker_asymptotic_2019}:
They conjecture that accelerated cosmic expansion in the asymptotic region of the field space of a $d$-dimensional \gls{eft} is only possible if there is a metastable \gls{ds} vacuum in a higher dimension. They coin this the \textit{no asymptotic acceleration conjecture}. More concretely: If a $d$-dimensional \gls{eft} realises accelerated expansion through a rolling scalar field at an asymptotically large field traversal, then this \gls{eft} stems from a compactification of a higher-dimensional theory with positive vacuum energy.
This conjecture raises several open questions. On the one hand, we currently observe accelerated expansion and we assume that our Universe already underwent a period of accelerated expansion in the past (inflation). This would then indicate that these phases of inflation are not taking place in asymptotic field limits respectively are not caused by a rolling scalar field. The former explanation would also be supported by the \gls{dc}. Furthermore, it indicates that neither inflation nor \gls{de} cause eternal accelerated expansion. This is in line with the \gls{tcc}.
On the other hand, the conjecture could be satisfied if there were a higher-dimensional \gls{ds} vacuum, i.e. the \gls{dsc} would have to be refined to only apply to lower-dimensional \glspl{eft} and a valid \gls{ds} construction in a higher-dimensional setting should be found. While we don't see à priori a good reason to exclude the possibility of inflation being a decompactification, it seems challenging to explain the current phase of accelerated expansion with a decompactification.

\citet{antoniadis_logarithmic_2020} argue that the instability of \gls{ds} vacua does actually not even appear in 4 spacetime dimensions, only in higher-dimensional theories. In 4d, perturbative effects from localised sources of graviton kinetic terms allow for locally stable \gls{ds} vacua\,\textemdash\,this is only the case in 4d, which would explain why our Universe is 4d.
Moreover, to experience a 4d \gls{ds} cosmology, it is not even compulsory to have a (meta-)stable \gls{ds} vacuum: an alternative can be an unstable \gls{ads} vacuum where a brane nucleates such that on the brane a local \gls{ds} patch exists \cite{banerjee_emergent_2018,banerjee_bubble_2021,danielsson_higher-dimensional_2021,banerjee_curing_2021,banerjee_dark_2020}.

\paragraph{Are \gls{ds} vacua fully excluded?}\label{f:dS_lifetime}
The \gls{dsc} allows for meta-stability. 
The conjecture does not exclude \gls{ds} vacua with a lifetime shorter than the quantum breaking time \cite{dias_primordial_2019,dvali_quantum_2014,vafa_swamplandish_2024,dvali_exclusion_2019,dvali_quantum_2019,dvali_quantum_2017,dvali_quantum_2013,Dvali_Quantum-Exclusion_2014}.
The quantum breaking time is the timescale
\begin{equation}
    t_\textnormal{q}\sim M_\textnormal{P}^2/H^3    
\end{equation}
after which a system can no longer be treated classically, regardless of the order of classical non-linearities that are taken into account \cite{ahmed_supersymmetric_2024,dvali_quantum_2019,dvali_exclusion_2019,damian_effective_2024,dvali_quantum_2017}.\footnote{
    Due to non-linearities, a theory will depart from its classical dynamics after a time $t_\textnormal{cl}\sim1/H$ \cite{damian_effective_2024}.
    The quantum breaking time is inversely proportional to the number of species in the theory, which links the \gls{dsc} to the \gls{ssc} \cite{dvali_quantum_2017}: $t_\textnormal{q}\sim M_\textnormal{P}^2/N_\textnormal{S}\Lambda_\textnormal{cc}^{3/2}$.
    The upper phenomenological bound of $\num{e32}$ species sets an upper bound for the lifetime of our Universe of $\num{e100}$ years.
    If there are more than $N_\textnormal{S}=M_\textnormal{P}^2/\Lambda_\textnormal{cc}$ species, the quantum breaking time becomes shorter than the Hubble time.
}

\gls{ds} vacua could be very short-lived due to spontaneous decompactifications \cite{giddings_fate_2003,giddings_spontaneous_2004} and compactifications \cite{dine_catastrophic_2004,draper_bubble_2021,draper_sitter_2021} \cite{draper_snowmass_2022}.
Furthermore, \gls{ds} vacua can decay into bubbles of nothing \cite{draper_bubble_2021,draper_sitter_2021,dine_catastrophic_2004}, or experience a bounce, where the direction of time reverses without encountering a singularity \cite{kiritsis_sitter_2023,emparan_note_2003,bramberger_non-singular_2019}.
For some proposed \gls{ds} solutions, the predicted lifetime could be at odds with the quantum breaking time \cite{arkani-hamed_measure_2007,dvali_quantum_2019,westphal_lifetime_2008,dvali_quantum_2017} respectively the \gls{tcc} \cite{brahma_consistency_2021}.
For instance, the life-time of a \gls{kklt} \gls{ds} saddle point is
\begin{equation}
    \tau\sim\frac{1}{H\log\left(M_\textnormal{P}/H\right)}<\frac{1}{H}<\frac{\log\left(M_\textnormal{P}/H\right)}{H}    
\end{equation}
i.e. the log-corrections are in the opposite direction as in the \gls{tcc} bound and shorten the lifetime \cite{blumenhagen_kklt_2020}.

A finite \gls{ds} lifetime is obtained if \gls{ds} space is realised as a coherent quantum state of gravitons that decohere after the quantum breaking time \cite{dvali_quantum_2017}:
Relatively recently, \citet{brahma_resurgence_2023,brahma_sitter_2021,bernardo_sitter_2021,brahma_four-dimensional_2021,alexander_sitter_2024} report to have successfully constructed non-perturbatively stable 4d \gls{ds} solutions within a controlled temporal domain by viewing \gls{ds} space as a Glauber\textendash Sudarshan state with a shifted interacting vacuum over a supersymmetric Minkowski background. Instead of a fundamental vacuum, \gls{ds} spacetime would be an emerging, meta-stable coherent state. Therefore, this is not a counterexample to the \gls{dsc} per se.

\paragraph{Does the \gls{dsc} also apply to theories of modified gravity?}
We pointed out in \cref{p:dSC_Inflation} that some forms of inflation are incompatible with the \gls{dsc} in a \gls{gr} context, but compatible with the \gls{dsc} in a theory of modified gravity. A valid question is if the \gls{dsc} is even valid in theories of modified gravity, and what implications the \gls{dsc} might have for such theories.
It has been found that the \gls{dsc} is applicable to theories of modified gravity and can constrain them:
\begin{itemize}
    \item $f(R,T)$ gravity provides a viable setting:
    a D-brane model with $V=V_0\left(1-\left(m/\phi\right)^4\right)$,
    an E-model (see \cref{eq:E-model}),
    a T-model (\cref{eq:T-model}), and
    a model of modular inflation with $V=V_0\left(1-\mathfrak{p}\exp(-\mathfrak{c}\phi)\right)$
    are deemed compatible with the \gls{dsc} \cite{oikonomou_swampland_2023}.
    Furthermore, holographic \gls{de} as inflaton with the parametrisation $f(R,T)=R+8\pi G\mathfrak{c}T$ has a window of compatibility for $\mathfrak{c}\gtrsim\order{\num{e2}}$ \cite{taghavi_holographic_2023}.\footnote{
        The \gls{dc} requires $\mathfrak{c}\gtrsim300$ \cite{taghavi_holographic_2023}.
        }
    \item $f(R)$ theories are restricted but allowed by the \gls{dsc} \cite{mata-pacheco_cosmological_2024,garcia-compean_horavalifshitz_2023}. This is not too surprising, since a $f(R)$ model in the Jordan frame correpsonds to \gls{gr} plus scalar fields in the Einstein frame \cite{benetti_swampland_2019}.
    \begin{itemize}
        \item $\partial_Rf>0$ is required to have a well-defined scalar field instead of ghosts \cite{mata-pacheco_cosmological_2024,de_felice_fr_2010}.
        \item $\partial_R^2f>0$ is required to have a positive mass term for curvature fluctuations \cite{mata-pacheco_cosmological_2024,de_felice_fr_2010}.
        \item \Cref{eq:dSc} translates to 
        \begin{flalign*}
            &&\abs{2f(R)-R\partial_Rf(R)}>\sqrt{3/2}\mathfrak{s}_1\left(R\partial_Rf(R)-f(R)\right)
        \end{flalign*}
        for $f(R)$ theories \cite{artymowski_fr_2019}. While in \gls{gr} a potential $V\propto\exp(-\lambda\phi)$ can always satisfy the \gls{dsc} at the cost of the \gls{dc}, in an $f(R)$ theory with $R\partial_Rf>2f$, i.e. $\partial_\phi V<0$, the model is constraint already by the \gls{dsc} alone \cite{artymowski_fr_2019}.\footnote{
            \citet{bajardi_early_2022} present the same bound yet with a different numerical pre-factor of $\order{1}$ (in a somewhat convoluted form with higher-order derivatives that cancel each other if $\partial_Rf$ is invertible \cite{benetti_swampland_2019}).
        }
        \item Various models do not meet observational expectations:
        \begin{itemize}
            \item While $f(R)$ theories per se are not ruled out by the \gls{dsc}, the constraints that are put on such theories do not automatically solve the Hubble tension either \cite{elizalde_swampland_2022}.
            \item \citet{artymowski_fr_2019} found no patches of the parameter space that are compatible with the \gls{dsc} that allow for suitable models of inflation or \gls{de}. If the \gls{dsc} could be relaxed to $\mathfrak{s}_1\sim\order{0.1}$, suitable models of inflation existed \cite{sadeghi_logarithmic_2019,sadeghi_cosmic_2022}.
            \item An Ansatz $f(R)\sim R^\mathfrak{p}$ yields the \gls{dsc} constraint $1<\mathfrak{p}\lesssim1.45$ \cite{mata-pacheco_cosmological_2024,garcia-compean_horavalifshitz_2023}. \gls{gr} itself is ruled out in the \gls{uv} by the \gls{dsc} since it corresponds to $\mathfrak{p}=1$ \cite{mata-pacheco_cosmological_2024,garcia-compean_horavalifshitz_2023}. \citet{benetti_swampland_2019} report that $\mathfrak{p}<2/3$ and $\mathfrak{p}>2$ satisfy the \gls{dsc}.
            \begin{itemize}
                \item \citet{channuie_refined_2022} found the constraints from the refined \gls{dsc} incompatible with observations, while the combined \gls{dsc} allows for observationally viable models.
                \item For $\mathfrak{p}\lesssim\order{\num{e-1}}$, power-law potentials $\propto\phi^2$, the E-model (\cref{eq:E-model}), the T-model (\cref{eq:T-model}), and a D-brane model were found to be compatible with observations, the \gls{dsc}, and the \gls{dc} \cite{oikonomou_rescaled_2022}.
                \item Natural inflation and quadratic hilltop inflation were deemed incompatible if $\mathfrak{p}\lesssim\order{\num{e-1}}$ \cite{oikonomou_rescaled_2022}.
            \end{itemize}
            \item The Starobinsky model with $f(R)\simeq R+\alpha R^2$ is found to violate the \gls{dsc} in the late Universe \cite{benetti_swampland_2019}.
        \end{itemize} 
        \item The presence of matter fields makes it harder to fulfil \cref{eq:dScrefined}, as matter contributes positively to the second derivative of the potential: $V^{\prime\prime}_\textnormal{eff}=V^{\prime\prime}+\left(1-3w\right)^2\rho_m/6$ \cite{artymowski_fr_2019}.
        \citet{mata-pacheco_cosmological_2024} claims that \cref{eq:dScrefined} is always violated, but others find it to hold if \gls{gb} or Chern\textendash Simons terms are present \cite{gitsis_swampland_2023,odintsov_inflationary_2023}.
    \end{itemize}
    \item A coupling to a parity-violating Chern\textendash Simons term affects the tensor-to-scalar ratio as tensor modes behave differently based on their chirality, but since only tensor modes are affected, the \gls{dsc} remains applicable \cite{fronimos_inflationary_2023}.
    \item Topological massive gravity, more precisely a spin-2 truncation of a higher-spin theory with Chern\textendash Simons terms, fermions, and scalar fields, is found to satisfy the refined \gls{dsc}: with a dimensionless coupling constant $g_6$ for the conformal, sixth-order self-coupling of the scalar field, $\mathfrak{s}_2<3/2$ is found for $V=3H^2\phi^2/8-\abs{g_6}\phi^6/6$ \cite{alvarez-garcia_swampland_2022}.
    \item Lorentz invariance seems not to be a requirement for the applicability of the \gls{dsc} \cite{mata-pacheco_cosmological_2024,trivedi_lorentz_2022,garcia-compean_horavalifshitz_2023}.
    \begin{itemize}
        \item $F(\Bar{R})$ Hořava\textendash Lifshitz theories are constraint but compatible with \cref{eq:dSc}, which shows on the one hand that Hořava\textendash Lifshitz theories are promising \gls{uv} extensions of theories of gravity, but on the other hand it also indicates that the \gls{dsc} might be applicable to theories that break Lorentz invariance \cite{mata-pacheco_cosmological_2024,garcia-compean_horavalifshitz_2023}. However, $F(\Bar{R})$ theories violate \cref{eq:dScrefined} \cite{mata-pacheco_cosmological_2024}.
        \item A constant Hubble parameter is incompatible with \gls{gr} and $f(R)$ theories, but not with more general $F(\Bar{R})$ Hořava\textendash Lifshitz theories that break Lorentz invariance in the \gls{uv} \cite{weinfurtner_projectable_2010,sotiriou_horava-lifshitz_2011,wang_horava_2017,mukohyama_horavalifshitz_2010,calcagni_cosmology_2009,chaichian_modified_2010,elizalde_unifying_2010,horava_quantum_2009} \cite{mata-pacheco_cosmological_2024,garcia-compean_horavalifshitz_2023}.
    \end{itemize}
    \item Applying the \gls{dsc} to exponential, power-law and logarithmic inflaton potentials in scalar\textendash tensor theory leaves a window of compatibility that partially overlaps with observational constraints \cite{gashti_constraints_2022}.
    \item Horndeski theory, a further generalisation of scalar\textendash tensor theories with derivative self-interactions and non-minimal couplings, is constrained by but compatible with the \gls{dsc} \cite{heisenberg_horndeski_2019,brahma_dark_2019}.
\end{itemize}

\paragraph{What are the exact values of $\mathfrak{s}_1$ and $\mathfrak{s}_2$?}\label{p:dSC_c_value}

The literature on speculations about the value of $\mathfrak{s}_1$ is rich, even though any value of $\mathfrak{s}_1$ would do to exclude (meta-)stable \gls{ds} vacua from the landscape \cite{akrami_landscape_2019}.
To just exclude \gls{ds} vacua, $\abs{\nabla V}>\mathfrak{c}$, with $\mathfrak{c}$ some constant, would be sufficient \cite{liu_palatini_2021}. However, there are supersymmetric vacua with flat directions that pose as counterexamples to this constraint \cite{liu_palatini_2021}. To include those in the landscape, $\mathfrak{c}$ has to be promoted to $\mathfrak{c}(\phi)\leq0$, of which $\mathfrak{c}(\phi)=\mathfrak{s}_1V(\phi)$ is one possibility \cite{liu_palatini_2021}. The value of $\mathfrak{s}_1$ is the subject of ongoing debate. We give an overview in the following, but first make some remarks about $\mathfrak{s}_2$, which is less often discussed:
$\mathfrak{s}_2^2=-V^{\prime\prime}/V\gtrsim\mathcal{O}(1)$ is required to have a concave potential with a lower bound on the curvature \cite{raveri_swampland_2019,storm_swampland_2020},
the $\order{1}$ bound pops up in numerical analysis \cite{garg_bounds_2019},
and $\mathfrak{s}_2=\frac{2\left(d-1\right)}{\left(d-2\right)}$ is necessary for potentials that are linear or quadratic in $\phi$ to allow for eternal inflation \cite{rudelius_dimensional_2021}.
That the parameter $\mathfrak{s}_2$ should be of $\order{1}$ can be highlighted by the following approximation using the \gls{dc} \cite{liu_palatini_2021}:
\begin{equation}
    M_\textnormal{P}^2\frac{\nabla_j\nabla_j V}{V}\sim-\frac{M_\textnormal{P}^2}{\left(\Delta\phi\right)^2}\leq-\order{1}.
\end{equation}
The \gls{tcc} yields
\begin{equation}
    \frac{\abs{\Delta V}}{V}>\frac{16}{\left(d-1\right)\left(d-2\right)\log^2\!\left(V\right)}
\end{equation}
as a bound \cite{bedroya_sitter_2020}.
\citet{raveri_swampland_2019} rule out $\mathfrak{s}_2>1.4$ at the 95\% confidence level based on \gls{cmb} and \gls{sn} observations for \gls{de} potentials of the form $V=k\cos(\mathfrak{s}_2\phi)$, with $k$ a numerical factor.\footnote{
    Note that such potentials always fail \cref{eq:dSc}.
}

Regarding the value of $\mathfrak{s}_1$, we start by presenting a concrete yet rather general setting where $\mathfrak{s}_1=0$ is ruled out:
\citet{shiu_late-time_2023} study multi-field multi-exponential potentials for scaling solutions of late-time attractors and find a lower relative bound on the potentials that support the \gls{dsc}, without relying on a specific model.\footnote{
    It is important to note that they do not assume slow-roll.
} 
The potentials with $n$ exponential terms and $k$ scalar fields $\phi^a$ (such that $a=1,\dots,k$) are of the general form
\begin{equation}
    V=\sum_{i=1}^n\Lambda_ie^{-\mathfrak{c}_{ia}\phi^a},
\end{equation}
with $\Lambda_i>0$ and $\mathfrak{c}_{ia}$ constants that depend on the microscopic origin of the scalar potential.\footnote{
    A universal upper bound for multi-field multi-exponential potentials that are positive-definite is presented by \citet{shiu_accelerating_2023}: $\epsilon_H\leq d-1$.
}
The couplings $\mathfrak{c}_{ia}$ span a vector $v_i$ with $\left(v_i\right)_a=\mathfrak{c}_{ia}$. These vectors then span a hypersurface, the convex hull of the exponential couplings $v_\textnormal{CH}$.
As \citet{shiu_late-time_2023} analytically show, 
\begin{equation}
    \epsilon_H\geq\frac{d-2}{4}v_\textnormal{CH}^2
\end{equation}
holds for any late-time solution with cosmic acceleration.
Now, with $\frac{d-2}{4}v_\textnormal{CH}^2\leq\epsilon_H\leq d-1$ it can be shown that $\mathfrak{s}_1>0$, as predicted by the \gls{dsc}, even though without quantifying the exact value of $\mathfrak{s}_1$:
$\abs{\nabla V}/V=0$ is only possible for $\left(d-1\right)-\epsilon_H=\eta_H/2$.\footnote{
    See the work by \citet{shiu_late-time_2023} for the derivation of this result.
}
This equation can be restated in terms of the Hubble parameter as $d-1+\ddot{H}/2\dot{H}H=0$, and then
be recast as 
\begin{equation}
    \frac{\mathrm{d}}{\mathrm{d}t}\left[\left(d-1\right)H^2+\dot{H}\right],
\end{equation}
which is solved by $\left(d-1\right)H_0^2+\dot{H}_0=\mathfrak{c}_0\in\mathbb{R}$.
From the bounds on $\epsilon_H$ and the expectation that $\lim_{t\rightarrow\infty}H(t)=0$, it follows that $\lim_{t\rightarrow\infty}\dot{H}(t)=0$, which yields $\mathfrak{c}_0=0$.
This finding allows us to integrate the solution, to find $H_0=1/\left[\left(d-1\right)t+\mathfrak{c}_1\right]$. This solution corresponds to pure kination and can only be achieved for $\epsilon_H=d-1$, in which case we find $\eta_H=0$.
The conclusion here is that for $\epsilon_H>0,\neq d-1$, the gradient condition $\mathfrak{s}_1>0$ is a strict inequality. This holds for (multi-) field (multi-) exponential potentials of accelerating late-time attractor solutions.

That $\mathfrak{s}_1$ should be of $\order{1}$ comes from several assumptions \cite{calderon-infante_asymptotic_2023}:
First, it is assumed that even in a multi-field case $\epsilon_\textnormal{H}=\epsilon_V$.
Second, the value is obtained in the asymptotic limit of perturbative string theory, but assumed to hold in any asymptotic limit of moduli space, outside the lamppost of known string theory.
Third, if applied to cosmology, this asymptotic limit is applied in the bulk, i.e. in the middle of moduli space.
And fourth, it is assumed the condition holds under dimensional reduction.
We elaborate on all the steps in the following, following closely the work by \citet{calderon-infante_asymptotic_2023}.
\subparagraph{\nth{0}}
We start with an effective action of the following form:
\begin{align}
    S=\int\!\sqrt{-g}\left(\frac{R}{2}+\frac{1}{2}g^{\mu\nu}G_{ab}\partial_\mu\phi^a\partial_\nu\phi^b-V(\phi)\right)\,\mathrm{d}^dx,
\end{align}
with $R$ the Ricci scalar, $g_{\mu\nu}$ the $d$-dimensional spacetime metric, and $G_{ab}$ the moduli space metric.
The Friedmann equation and \gls{eom} are:
\begin{align}
    \frac{\left(d-1\right)\left(d-2\right)}{2}H^2-\frac{1}{2}G_{ab}\dot{\phi}^{a}\dot{\phi}^{b}-V(\phi)&=0\\
    \ddot{\phi}^a+\tilde{\Gamma}^a_{bc}\dot{\phi}^b\dot{\phi}^c+\left(d-1\right)H\dot{\phi}^a+\partial^aV(\phi)&=0,
\end{align}
where the Christoffel symbols $\tilde{\Gamma}^a_{bc}$ are with respect to the moduli space metric $G_{ab}$, $\partial_a$ are derivatives with respect to the fields $\phi^a$, and indices are raised and lowered with the moduli space metric.
The slow-roll conditions are that the potential dominates over the kinetic term and that the friction term dominates over the second-order terms:
\begin{AmSalign}
    V(\phi)&\gg\frac{1}{2}G_{ab}\dot{\phi}^{a}\dot{\phi}^{b}\\
    \left(d-1\right)H\dot{\phi}^a&\gg\ddot{\phi}^a+\tilde{\Gamma}^a_{bc}\dot{\phi}^b\dot{\phi}^c\\
    \Rightarrow H^2&=\frac{2}{\left(d-1\right)\left(d-2\right)}V(\phi)\\
    \Rightarrow\dot{\phi}^a&=\frac{-1}{\left(d-1\right)H}\partial^aV(\phi)\\
    \Rightarrow\frac{1}{2}G_{ab}\dot{\phi}^{a}\dot{\phi}^{b}&=\frac{1}{2}G_{ab}\frac{-1}{\left(d-1\right)H}\partial^aV(\phi)\frac{-1}{\left(d-1\right)H}\partial^bV(\phi)\\
    &=\frac{d-2}{4\left(d-1\right)}\mathfrak{s}_1^2V(\phi).
\end{AmSalign}
Slow-rolling holds for $\mathfrak{s}_1<2\sqrt{\left(d-1\right)/\left(d-2\right)}$.
\subparagraph{\nth{1}}
The slow-roll parameter for the field reads
\begin{align}
    \epsilon_V&=\frac{d-2}{4}\frac{\abs{\nabla V}^2}{V^2}\\
    &=\epsilon_\textnormal{H}\left(1+\frac{\Omega^2}{\left(d-1\right)^2H^2}\right)\\
    \Rightarrow \mathfrak{s}_1&<\frac{2}{\sqrt{d-2}}\left(1+\frac{\Omega^2}{\left(d-1\right)^2H^2}\right)^{1/2}.
\end{align}
For a single field, the slow-roll conditions\footnote{
    See also our discussion around \cref{eq:e_V-e_H}.
    }
becomes $\epsilon_\textnormal{H}=\epsilon_V$ and therefore 
\begin{equation}
    \mathfrak{s}_1<\frac{2}{\sqrt{d-2}}.
\end{equation}
\subparagraph{\nth{2}}
The value of $\abs{\nabla V}/V$ varies with $\phi$, but $\mathfrak{s}_1$ is generally assumed to correspond to the value in the asymptotic limit where $\int_{t_0}^t\!\sqrt{G_{ab}\dot{\phi}^a\dot{\phi}^b}\,\mathrm{d}t\rightarrow\infty$ holds.
Furthermore, it is assumed that this bounds holds in any asymptotic regime, not only in the string theory lamppost of weak string coupling. This is a strong assumption, mainly motivated by the absence of counterexamples.\footnote{
    See e.g. the work by \citet{cicoli_quintessence-parametrically_2022,bento_dark_2020,garg_bounds_2019} for examples.
    }
\subparagraph{\nth{3}}
In the bulk, where cosmology happens, this does not apply. There can even be \gls{ds} vacua and the value of $\mathfrak{s}_1$ is less constrained / unconstrained by theory (so far).
\subparagraph{\nth{4}}
The strong form of the \gls{dsc} assumes that $\mathfrak{s}_1=2/\sqrt{d-2}$ holds as a strict bound. This is motivated by the finding that this bounds remains consistent under dimensional reduction.%

Other constraints on the value of $\mathfrak{s}_1$ are the following:
\begin{itemize}
    \item \gls{cmb} data is compatible with high values of $\mathfrak{s}_1$ (if \gls{de} is modelled with a single scalar field), but when combined with distance\textendash redshift data, only small values remain compatible with observations (as large $\mathfrak{s}_1$ lead to a substantial deviation from $\omega_\text{DE}=-1$) \cite{raveri_swampland_2019}.
    \item \gls{desi} data indicates $\mathfrak{s}_1\gtrsim\order{1}$ at present time and even higher values in the past \cite{arjona_swampland_2024}.
    \item Using Gaussian process reconstruction and observational data on $H(z)$, \citet{elizalde_swampland_2019} derive upper bounds on $\mathfrak{s}_1$ without assuming a form for the potential $V(\phi)$. They use two different kernels, a squared exponential kernel, and a Matern kernel with $\nu=9/2$ that allows for higher derivatives and therefore more smoothness, to reconstruct $\abs{\nabla V}/V$ from $H(z)$. We summarise their obtained bounds in \cref{tab:dSC}.
    \item \citet{andriot_exponential_2024} found $\mathfrak{s}_1<\sqrt{3}$ as a requirement to explain the current accelerated expansion with quintessence.
    \item $\mathfrak{s}_1^\textnormal{sugra}=6/\sqrt{\left(d-2\right)\left(11-d\right)}$ from 11d supergravity \cite{obied_sitter_2018}
    \item $\mathfrak{s}_1^\textnormal{SEC}=2\sqrt{\left(D-2\right)/\left[\left(D-d\right)\left(d-2\right)\right]}$ in 11d supergravity regarding the \gls{sec}\footnote{
        The \gls{sec} can be violated in string theory. \citet{obied_sitter_2018} deemed this bound too strong, as a $d$-dimensional universe needs $\abs{\nabla V}/V<\sqrt{4/\left(d-2\right)}$ to undergo accelerated expansion.
        The violation of the \gls{sec} in our Universe tells us that we are in the bulk of moduli space, and not in an asymptotic boundary region \cite{rudelius_asymptotic_2021}.
        } \cite{obied_sitter_2018,sun_notes_2021}
    \item For \gls{de}, \citet{wang_multi-feature_2020} find $\mathfrak{s}_1\lesssim1.62$ for an extended \gls{lcdm} cosmology with 18 parameters, respectively $\mathfrak{s}_1\lesssim1.60$ for a \gls{cpl} parametrisation.
    \item Every known string theory potential satisfies
    \begin{equation}\label{eq:dS-dim-red}
        \frac{\norm{\nabla V}}{V}\geq\frac{2}{\sqrt{d-2}}
    \end{equation}
    in the asymptotic limit \cite{castellano_iruv_2022,castellano_emergence_2023,van_riet_beginners_2023,rudelius_asymptotic_2021,alestas_curve_2024,mcallister_moduli_2023,grimm_asymptotic_2020,obied_sitter_2018,andriot_open_2019,andriot_web_2020,calderon-infante_asymptotic_2023,shiu_late-time_2023,shiu_accelerating_2023,cremonini_asymptotic_2023,hebecker_no_2023,rudelius_asymptotic_2022,rudelius_dimensional_2021,bedroya_trans-planckian_2020,andriot_accelerated_2023,rudelius_asymptotic_2022,etheredge_sharpening_2022,rudelius_asymptotic_2021,rudelius_dimensional_2021}.\footnote{
        \citet{calderon-infante_asymptotic_2023} present counterexamples to the bound $\mathfrak{s}_1=2/\sqrt{d-2}$ in \gls{cy} flux compactifications with competing terms of the potential in the asymptotic limit by studying the gradient flow trajectories among more general infinite distance limits, besides the usual large volume and weak string coupling limits. However, those counterexamples have not been rigorously tested, as only a subsector of the moduli field space was explored.
    }
    \begin{itemize}
        \item With this bound, the \gls{dsc} is exactly conserved under dimensional reduction \cite{rudelius_asymptotic_2021,rudelius_dimensional_2021}.
        \item It can be obtained by combining the \gls{dsc} with the \gls{ep} \cite{lee_emergent_2022} and the refined \gls{dc} \cite{etheredge_sharpening_2022}.
        \item This bound still allows \gls{ds} space with \gls{de}; however, its lifetime $\tau$ is limited to
            \begin{equation}
                e^{\tau\sqrt{\Lambda}}\lesssim\frac{1}{\sqrt{\Lambda}} \Rightarrow \tau\lesssim\frac{1}{\sqrt{\Lambda}}\log\left(\frac{1}{\sqrt{\Lambda}}\right)
            \end{equation}
            as the horizon scales as $\frac{1}{\sqrt{\Lambda}}$ and regions smaller than the Planck scale cannot exit the horizon and freeze \cite{vafa_swamplandish_2024}.\footnote{
                \citet{vafa_swamplandish_2024} states that this also explains the puzzling observation that $t_0\sim\frac{1}{\sqrt{\Lambda}}$, i.e. that the current age of our Universe corresponds to the \gls{de} horizon, since this is the typical time in such a universe. This comes with the worrying consequence that the \gls{ds} phase is about to end. However, this could be circumvented if \gls{de} is dynamical and not a cosmological constant.
                }
        \item The bound $\frac{\norm{\nabla V}}{V}\geq\frac{2}{\sqrt{d-2}}$ does not directly translate into a bound for the value of $\mathfrak{s}_1$ in the bulk.
        \item Such a strong bound does not allow for accelerated expansion, unless the Universe inhibits negative spatial curvature \cite{andriot_accelerated_2023,alestas_curve_2024}. However, the required curvature needs to be so strong that $\Omega_k$ would be a dominant contribution to the Universe's energy budget, which is incompatible with observations \cite{alestas_curve_2024}.
        \item A simple scalar field model with $V\sim\exp\left(-\mathfrak{p}\phi\right)$ has the \gls{eos} $w=-1+\frac{\mathfrak{p}^2}{2}\frac{d-2}{d-1}$, such that $\mathfrak{p}>\mathfrak{s}_1$ yields the \gls{sec} that rules out an accelerating expansion of the Universe \cite{rudelius_asymptotic_2021}.
        \citet{rudelius_asymptotic_2021} concludes that this condition, and therefore a strong application of the \gls{dsc}, is only viable in the asymptotic limits of the scalar field itself.
    \end{itemize}
    \item Slow-roll inflation suggests $\mathfrak{s}_1<\sqrt{2}$, as otherwise $\epsilon_V\ll1$ is violated \cite{garg_bounds-dS_2019}.
    \item \citet{heisenberg_dark_2018} claim that $\mathfrak{s}_1\leq1.35$ satisfies both, observational and theoretical constraints.\footnote{
        \citet{akrami_landscape_2019} consider the approach of extending existing $1\sigma$ and $2\sigma$ contours to $3\sigma$ used by \citet{heisenberg_dark_2018} incorrect, and suggest a full statistical analysis. \citet{Heisenberg_Dark_II_2018} reply that the purpose of their work was to check if the discrepancy is strong (order of magnitude difference) or more subtle, for which they deemed their approach of extrapolating the $1\sigma$ and $2\sigma$ contours sufficient. No such strong tension is found.
        }
    \item $\mathfrak{s}_1^\textnormal{NEC}=2\sqrt{\left(D-d\right)/\left[\left(D-2\right)\left(d-2\right)\right]}$ respecting the \gls{nec}\footnote{
        \citet{obied_sitter_2018} deem this bound too strong, as observational bounds indicate that $\mathfrak{s}_1\leq0.6$ \cite{agrawal_cosmological_2018}. Furthermore, the \gls{nec} can be violated quantum mechanically. However, see \cref{s:nnNECC} for a conjecture that states that theories violating the \gls{nec} are in the swampland.
    }
    \cite{obied_sitter_2018,sun_notes_2021}.
    \item $\order{1}$ values for $\mathfrak{s}_1$ were also derived studying a 10d string background with 4d \gls{ds} subspaces \cite{andriot_open_2019}.
    \item \citet{akrami_landscape_2019} show that $\mathfrak{s}_1\lesssim1.02$ for all viable quintessence models with $V\sim\exp(\mathfrak{s}_1\phi)$ (note that such models always fail \cref{eq:dScrefined} \cite{raveri_swampland_2019}).
    \item The \gls{tcc} provides a lower bound on the variable $\mathfrak{s}_1\geq2/\sqrt{\left(d-1\right)\left(d-2\right)}$ \cite{basile_sitter_2020,bedroya_trans-planckian_2020,calderon-infante_asymptotic_2023,bedroya_sitter_2020,andriot_web_2020,andriot_quasi-_2023}. It is not violated in any string theory compactification studied by \citet{andriot_web_2020,andriot_quasi-_2023} (\cite{maldacena_supergravity_2001,shiu_stability_2011,andriot_open_2019,brandenberger_new_2019,blaback_smeared_2010,wrase_classical_2010,andriot_refining_2017,quigley_gaugino_2015,kutasov_constraining_2015,gautason_cosmological_2012,green_constraints_2012}).
    The same bound is derived from Hodge theory \cite{bastian_weak_2021,bedroya_trans-planckian_2020}.
    The bound stems from trajectories driven by exponential potentials combined with positive energy contributions from a tower of states\,---\,leaving the tower aside yields the $\mathfrak{s}_1\geq2/\sqrt{d-2}$ discussed above \cite{agmon_lectures_2023}.
    \item \citet{casas_cosmology_2025} link the \gls{ep} and the \gls{ssc} to the \gls{dc} and the \gls{dsc}, to find a bound that links the \gls{dsc} to the \gls{dc}: $\mathfrak{s}_1\geq2/\mathfrak{a}\left(d-2\right)$, with $\mathfrak{a}\geq1/\sqrt{d-2}$ the $\order{1}$ parameter of the \gls{dc}.
    \item \citet{agrawal_cosmological_2018} find a lower bound of $\mathfrak{s}_1<0.6$, (and an upper bound of $1+\omega\ll2/3$) for $z<1$, assuming $\Omega_\phi(z=0)\approx0.7$ and $\Omega_\phi(z>1)\ll1$.
    \item For \gls{dbi} models, \citet{wang_multi-feature_2020} found an upper bound of $\mathfrak{s}_1\lesssim0.58$, whereas for inflationary models with concave potentials $\mathfrak{s}_1\lesssim0.14$ is obtained.
    \item In a string gas setting, \citet{laliberte_string_2020} find $\abs{V^\prime_\textnormal{eff}}/V_\textnormal{eff}\geq1/\left(D-d\right)$.
    \item \citet{agrawal_h_0_2021} use $\mathfrak{s}_1=0.1$, which is one order of magnitude away from $1$.
    \item Even a value of $\mathfrak{s}_1=0.01$ is not explicitly ruled out, according to  \citet{palti_swampland_2019}.
\end{itemize}
To summarise, we can state that the jury on the exact value of $\mathfrak{s}_1$ is still out, as the applicability of an exact bound derived for the asymptotic limit in the bulk, where cosmology happens, is questionable. However, it can act as a guiding principle for model building.

\begin{table*}[htb]
    \centering
$
\begin{tblr}{
 colspec = {XXXXX}
}   \toprule
    & \SetCell[c=2]{l} \text{Squared Exponential} && \SetCell[c=2]{l} \text{Matern ($\nu=9/2)$} &\\\midrule
    H_0 & \expval{\mathfrak{s}_1} & 1\sigma & \expval{\mathfrak{s}_1} & 1\sigma  \\
    \num{67.66(0.42)} & 0.52 & 1.012 & 0.51 & 1.02\\
    \num{71.28(3.74)} & 0.785 & 1.786 & 1.051 & 1.886\\
    \num{73.52(1.62)} & 0.649 & 1.167 & 1.129 & 1.691\\
    \bottomrule
\end{tblr}
$
    \caption[dSC bounds from Gaussian Process reconstruction]{Using observational data on $H(z)$, \citet{elizalde_swampland_2019} derive bounds on $\mathfrak{s}_1$ using a Gaussian process reconstruction. We summarise the derived mean and upper $1\sigma$ bound values.}
    \label{tab:dSC}
\end{table*}

\begin{figure}[htb]
  \begin{center}
    \includegraphics[width=\linewidth]{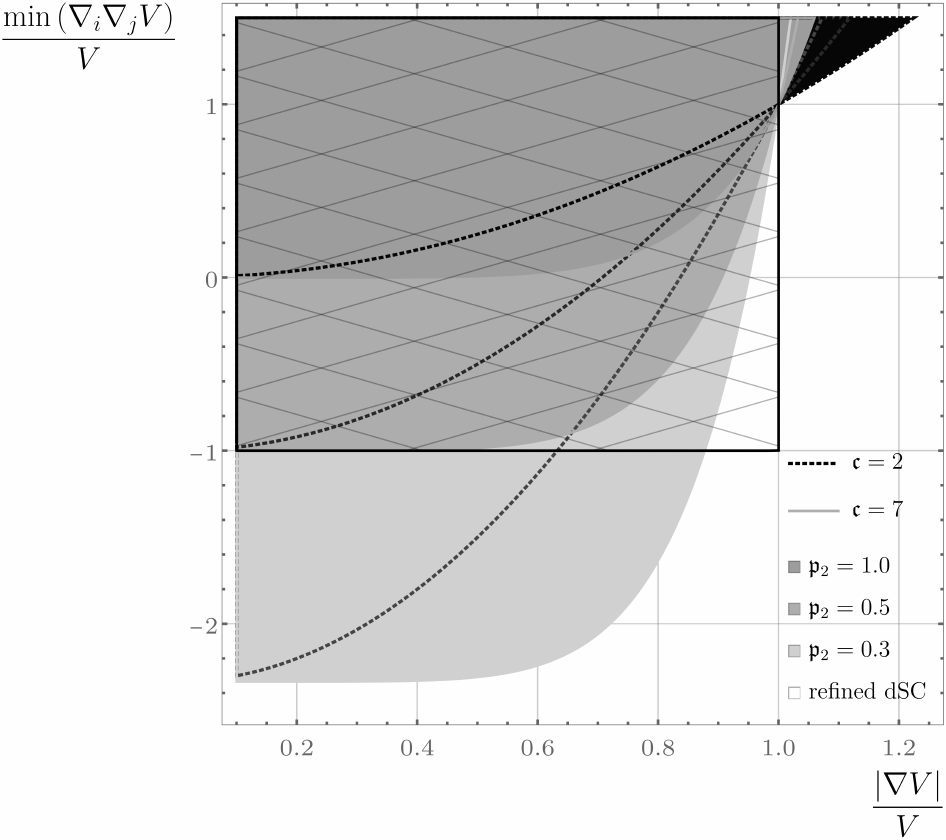}
  \end{center}
  \caption[Combined dSC]{The refined \gls{dsc} (meshed box) and the combined \gls{dsc} (shaded regions) overlap for a large fraction of the parameter space, but do not exactly match. The landscape of \glspl{eft} is to the right and below the curves respectively outside the box. The shaded regions are the swampland. The combined \gls{dsc} has free parameters, which allow for different curves:
  The minimal  $\mathfrak{c}=2$ is depicted as dashed lines. A higher exponent makes the swampland bigger as it rules out more intermediate $\{\mathfrak{s}_1,\,\mathfrak{s}_2\}$-pairs\,\textemdash\,the dashed curve remains flat for larger values of $\mathfrak{s}_1$ (actually for larger values of $\abs{\nabla V}/V$, for which we use $\mathfrak{s}_1$ as a shorthand here).
  A higher value of $\mathfrak{p}_2$ (darker shades) rules out more high-value pairs, where $\mathfrak{s}_1>1$ and $\mathfrak{s}_2>1$\,\textemdash\,a region considered save by the refined \gls{dsc}\,\textemdash, but allows values of $\mathfrak{s}_2$ closer to $0$ when $\mathfrak{s}_1$ is small.
  In the lower-right corner of the refined \gls{dsc} box is a region that is allowed by the combined \gls{dsc}. \citet{roupec_flux_2021} found \gls{ds} saddle points in this region.}\label{f:dSCcombi}
\end{figure}

\paragraph{Combined Conjecture}\label{p:dSC_combined}
The motivation to formulate a combined conjecture (\cref{eq:dSCcombi}) comes from the critique that the refined \gls{dsc} comes as two distinct conditions on two different quantities: $\epsilon_V$ and $\eta_V$.
The constraint on $\epsilon_V$ is derived in a semi-classical, weak-coupling limit, while the second constraint comes directly from requiring $\eta_V\geq-1$ \cite{liu_higgs_2021}.
The combined \gls{dsc} combines the two slow-roll parameters into one condition \cite{andriot_further_2019,sadeghi_sitter_2023,roupec_flux_2021}:
\begin{equation}
    \left(2\epsilon_V\right)^{\mathfrak{c}/2}-\mathfrak{p}_2\eta_V\geq 1-\mathfrak{p}_2.
\end{equation}
Both variations of the \gls{dsc} rule out slow-roll scenarios with small $\epsilon_V$ and positive $\eta_V$, but make different assessments for intermediate values \cite{roupec_flux_2021}.
\begin{itemize}
    \item Inflationary models with $\mathfrak{p}_2\simeq1$ (and $\mathfrak{p}_1\ll1$) are allowed by the combined \gls{dsc}, i.e. even single-field slow-roll models, especially if they satisfy $\epsilon_V\leq-\eta_V\simeq \mathfrak{p}_1\ll1$ \cite{roupec_flux_2021}.
    \item $V\left(\phi\right)=V_0e^{-\mathfrak{s}_1\phi}$ is not in agreement with the combined conjecture but could be in agreement with \cref{eq:dSc} if $\mathfrak{s}_1<0.6$ was allowed \cite{agrawal_cosmological_2018}.
    \item $V\left(\phi\right)=\frac{V_0}{2}\left(1-\tanh\left(\mathfrak{p}_2\phi\right)\right)$ with $V_0,\,\mathfrak{p}_2>0$ would be in agreement with the combined conjecture but not necessarily with the original \gls{dsc} \cite{roupec_flux_2021}.
    \item \citet{sadeghi_sitter_2023} present a model of inflation using non-local Friedmann equations that is compatible with the combined bound \cref{eq:dSCcombi} but violates \cref{eq:dSc,eq:dScrefined}.
    \item \citet{yuennan_composite_2023} examine four different models of inflation with composite scalar fields and find that the models they tested are in tension with the refined \gls{dsc}, but not necessarily with the combined bound.
    \item \citet{liu_higgs_2021} investigates the compatibility between the combined \gls{dsc} and Higgs inflation, Palantini\textendash Higgs inflation, and a Higgs\textendash Dilaton model. While the former two are compatible with the combined \gls{dsc}, only the Higgs\textendash Dilaton model is compatible with the refined \gls{dsc}.\footnote{In later work, \citet{liu_higgs_2022} presents the constraints on the potentials coming from the scalar \gls{wgc} (see \cref{p:SWGC}).}
    \item Inflationary models in mimetic\footnote{
        In mimetic gravity, inflation, \gls{de}, and \gls{dm} are explained without additional scalar fields but as inherent characteristics of gravity. See \citet{sebastiani_mimetic_2017} for a review.
    }
    $f(R,T)=R+\delta T$ gravity, with $T$ the trace of the stress-energy tensor, respectively in $f(\mathfrak{G})$ gravity\footnote{$\mathfrak{G}$ being the \gls{gb} term from \cref{eq:Gauss–Bonnet}.}, are shown to be incompatible with the refined \gls{dsc} (as $\mathfrak{s}_1,\mathfrak{s}_2\sim\order{0.1}$) but have a window of compatibility with the combined \gls{dsc} \cite{gashti_swampland_2022,gashti_further_2023}.\footnote{
    For $f(R)$ theories with $f(R)\propto R^\mathfrak{l}$ the combined \gls{dsc} with $\mathfrak{c}>2$ reads 
\begin{equation*}
    \frac{2\left(\mathfrak{l}-2\right)^2}{3\left(\mathfrak{l}-1\right)^2}\left[\left(\frac{2\left(\mathfrak{l}-2\right)^2}{3\left(\mathfrak{l}-1\right)^2}\right)^{\frac{\mathfrak{c}}{2}-1}-\mathfrak{p}_2\right]>1-\mathfrak{p}_2,
\end{equation*}
which holds for $1<\mathfrak{l}<\sqrt{6}-1$ \cite{artymowski_fr_2019}.
    }
\end{itemize}
We depict the parameter space for different values of $\mathfrak{c}$ and $\mathfrak{p}_2$, and compare it to the swampland region predicted by the refined \gls{dsc} in \cref{f:dSCcombi}.
Some authors consider the combined \gls{dsc} as too loose, given the success of various inflation models that are deemed incompatible with the refined \gls{dsc}.\footnote{
    For example \citet{liu_higgs_2021}, or \citet{roupec_flux_2021}, who critiques that the combined \gls{dsc} puts slow-roll (single-field) inflation in the landscape.
}

\subsubsection{Evidence}
The \gls{dsc} was first proposed by \citet{obied_sitter_2018},
inspired by \citet{brennan_string_2018,danielsson_what_2018},
later refined by \citet{ooguri_distance_2019,garg_bounds-dS_2019,murayama_we_2018},
then combined by \citet{andriot_further_2019,roupec_flux_2021},
and is studied in
4d supergravities \cite{ferrara_sitter_2020,andriot_exploring_2022,andriot_erratum_2022},
4d $\mathcal{N}=2$ \glspl{eft} \cite{cecotti_special_2020},
10d string models with no or broken supersymmetry \cite{basile_sitter_2020},
type I string theory with broken supersymmetry \cite{abel_stability_2019},
supersymmetry-breaking in type II string theory \cite{sethi_supersymmetry_2018,parameswaran_ds_2024},
type II string theory \cite{roupec_flux_2021,roupec_sitter_2019,blaback_new_2018},
10d type II supergravities \cite{andriot_tachyonic_2021},
$\mathcal{N}=1$ type II non-geometric flux compactifications \cite{shukla_rigid_2021},
type IIA orientifold flux compactifications \cite{banlaki_scaling_2019,prieto_moduli_2024,blumenhagen_note_2019,quirant_aspects_2022},
type IIA flux compactifications with (anti-)O6-planes, (anti-)D6-branes
or \gls{kk} monopoles \cite{junghans_weakly_2019},
type IIA supergravities compactified onto 3 dimensions \cite{farakos_classical_2020},
type IIB flux compactifications \cite{ishiguro_sharpening_2021,blanco-pillado_racetrack_2019,bizet_testing_2020,hebecker_f_2019,plauschinn_moduli_2021},
type IIB supergravities \cite{andriot_classical_2024},
type IIB compactifications on \gls{cy}$_3$ \cite{antoniadis_logarithmic_2020},
a non-geometric IIB compactification without Kähler moduli \cite{cremonini_asymptotic_2023},
K-theory in type IIB toroidal compactifications \cite{damian_remarks_2023},
F-theory on \gls{cy} fourfolds \cite{grimm_asymptotic_2020,calderon-infante_asymptotic_2023},
M-theory \cite{dasgupta_quantum_2019,bernardo_sitter_2021,deffayet_moduli_2024,deffayet_stable_2024},
string theory constructions of stable \gls{ds} points \cite{dine_obstacles_2021},
exotic string theories \cite{blumenhagen_ds_2020},
$O(16)\cross O(16)$ heterotic string compactified on a torus \cite{fraiman_non-supersymmetric_2024},
heterotic toroidal orbifolds \cite{leedom_heterotic_2023},
O6 planes in supergravities \cite{maldacena_supergravity_2001},
Glauber\textendash Sudarshan states \cite{brahma_sitter_2021},
the \gls{kklt} mechanism \cite{moritz_uplifts_2019,akrami_landscape_2019},
\gls{kklt} models with anti-brane uplifts \cite{gautason_tension_2019,bena_uplifting_2019},
the \gls{lvs} \cite{sun_large_2024,junghans_topological_2022},
warped throats \cite{dudas_update_2021},
the light of holography \cite{bedroya_holographic_2022},
the context of thermodynamics \cite{seo_thermodynamic_2019},
the presence of all-order effects in $\alpha^\prime$ \cite{basile_string_2021},
relation to asymptotic safety \cite{basile_asymptotic_2021,eichhorn_absolute_2024},
quantum breaking \cite{dvali_quantum_2019,damian_effective_2024},
non-critical string cosmologies \cite{ellis_supercritical_2020}, and
string gases \cite{laliberte_string_2020}.
Furthermore, relations to other swampland conjectures are highlighted in \cref{rel:dSC_DC,rel:TCC_dSC,rel:TPC_dSC,rel:WGC_dSC}.

\subsection{Distance Conjecture}\label{sec:distance}%
The \textit{distance conjecture} is often referred to as the \textit{Swampland Distance Conjecture (SDC)}. To avoid confusion with the \textit{de~Sitter Conjecture (dSC)} we use the abbreviation \textit{DC} for the distance conjecture.

The \gls{dc} says that in a theory that couples to gravity, with a moduli space\footnote{The moduli space $\mathcal{M}$ cannot be compact and has at least one boundary point at infinite distance \cite{grimm_tameness_2022} (see also \cite{hamada_finiteness_2022}).} $\mathcal{M}$ parametrised by the expectation values of
a scalar field $\phi\in\mathcal{M}$ with no potential,
\begin{equation}
    \forall \phi\in \mathcal{M}\,\exists\, \phi_\text{b}\in \partial\mathcal{M}
\end{equation}
such that the geodesic distance $d(\phi,\phi_\text{b})$ between $\phi_\text{b}$ on the boundary  and $\phi$ in the bulk is infinite,
and that there is an infinite tower of states with an associated mass scale $m$, where
\begin{equation}\label{eq:SDC}
    m(\phi)\sim m(\phi_0)\exp(-\mathfrak{a} d(\phi,\phi_0)),
\end{equation} where $\phi_0$ is sufficiently far away from $\phi$ and $\mathfrak{a}\geq0$ is a real \cite{grimm_tameness_2022}
constant \cite{palti_swampland_2019,ooguri_distance_2019}
of order one \cite{grana_swampland_2021,basile_domain_2023,blumenhagen_refined_2018,hebecker_flat_2017,ashmore_moduli-dependent_2021,ooguri_geometry_2007}.
Not only does the \gls{dc} state that such an infinite tower of states becomes massless at infinite distance, it also quantifies this behaviour as exactly exponential \cite{grimm_infinite_2019,ooguri_geometry_2007} and specifies this exponential suppression as the geodesic distance \cite{grimm_tameness_2022}.\footnote{
    It is worth noting that $m(\phi)>M_\textnormal{P}$, $m(\phi_0)>M_\textnormal{P}$ is possible. In that case, the \gls{dc} places no meaningful constraints on the \gls{eft} \cite{rudelius_revisiting_2023}.
    }

\subsubsection{Implications for Cosmology}\label{sss:DC_Cosmology}
If you have a scalar field that travels a long range in its field space, you will get a tower of states, i.e. particles, of which the lowest mass scale is given by
\begin{equation}
    m(\phi)=M_\textnormal{P}e^{-\mathfrak{a}\Delta\phi}
\end{equation}
with $\sqrt{1/2}\leq\mathfrak{a}\leq\sqrt{3/2}$.
This finding is applied to a plethora of settings, as we elucidate in the following.

\paragraph{Boson stars} are found to satisfy the \gls{dc} \cite{choi_probing_2019}.

\paragraph{Dark Sector}
A model independent analysis of observational data found the \gls{dc} to be compatible with observations at low redshift \cite{arjona_machine_2021}.\footnote{
    A parametrisation-dependent Bayesian machine learning algorithm that reconstructs $H(z)=H_0+H_1z^2/\left(1+z\right)$ finds that the \gls{dc} is well respected by \gls{de}, yet, a sign switch in the \gls{de} \gls{eos} takes place within the redshift range $z\in\left[0,\,5\right]$, i.e. \gls{de} is in the phantom regime for some time \cite{elizalde_interplay_2021}. The reconstructed model satisfies the gradient \gls{dsc} criterion \cref{eq:dSc} but violates the Hessian criterion \cref{eq:dScrefined}.
    }
We present more details on concrete models in the following.

\subparagraph{Quintessence} generally satisfies the \gls{dc}, as a model-agnostic reconstruction of quintessence yields $\abs{\Delta\phi}<0.23$ at $2\sigma$ \cite{park_reconstructing_2021}.
Various formulations of single-field quintessence cosmologies agree well with the \gls{dc}, as stated by \citet{schoneberg_news_2023}.
\citet{emelin_axion_2019} find that for axion-like quintessence with a hilltop potential stemming from a Kähler modulus, a modified potential that shows a minimum along the axionic direction is needed to satisfy the \gls{dc} (as well as the \gls{dsc}).

\citet{thompson_beta_2018,thompson_beta_2019} presents \textit{beta function quintessence} that satisfies the \gls{dc} yet violates the \gls{dsc}, as the $\order{1}$ of $\mathfrak{s}_1$ cannot be reached. The refined \gls{dsc} is not tested as no value for $\mathfrak{s}_2$ is derived. The model introduces a function $\beta(\phi)=\mathrm{d}\phi/\mathrm{d}\log a$ and repackages the scalar field potential as
\begin{equation}
    V(\phi)=\frac{-6H_0}{4}e^{-\int\!\beta(\phi)\,\mathrm{d}\phi}\left(1-\frac{\beta^2(\phi)}{6}\right),
\end{equation}
which achieves that the Hubble constant as a cosmological observable is directly part of the potential.
\citet{thompson_beta_2018,thompson_beta_2019} studies four explicit cases with $\beta_i$ constants:
\begin{description}
    \item[Exponential] $V(\phi)=\frac{-6H_0}{4}e^{-\beta_\textnormal{e}}\left(1-\frac{\beta_\textnormal{e}^2}{6}\right)$%
    \item[Logarithm] $V(\phi)=\frac{-6\beta_\textnormal{l}H_0}{4}\log\left(\phi\right)\left(1-\frac{\beta_\textnormal{l}^2}{6\left(\phi\log\phi\right)^2}\right)$%
    \item[Power-Law] $V(\phi)=\frac{-6H_0}{4}\phi^{\beta_\textnormal{p}}\left(1-\frac{\beta_\textnormal{p}^2}{6\phi^2}\right)$%
    \item[Inverse Power-Law] $V(\phi)=\frac{-6H_0}{4}\phi^{-\beta_\textnormal{i}}\left(1-\frac{\beta_\textnormal{i}^2}{6\phi^2}\right)$%
\end{description}
For a \gls{de} \gls{eos} of $\omega_0=0.98$ each model satisfies the \gls{dc} with $\Delta\phi\in\left(0.45,\,0.58\right)$, yet violates the first \gls{dsc}, as the gradient is too flat with $\nabla V/V\sim0.1$.

\subparagraph{Thawing Quintessence} \citet{storm_swampland_2020} find that the \gls{dc} becomes redundant for thawing quintessence models (or hilltop quintessence) that are in agreement with observational data. \citet{montefalcone_dark_2020} partially agree, yet reject hilltop potentials (as well as scenarios with a turning point): in such models, the kinetic energy changes too rapidly, violating observational constraints from \gls{lss} and \gls{bbn} (as such a quintessence field would dominate over matter and radiation far earlier than observed) as well as the refined \gls{dsc}. %
\citet{tada_quintessential_2024} derive a thawing quintessence model and axion-like quintessence by reconstructing $\omega_0\omega_a$CDM from \gls{desi}-data. Both potentials satisfy the \gls{dc} (as well as the \gls{dsc} and the \gls{wgc}).

\subparagraph{Multi-field quintessence} seems to naturally satisfy the \gls{dc} bound: 
For a setting with two fields and action
\begin{AmSalign}
    S&=&\int\!\sqrt{-g}&\left(\frac{RM_\textnormal{P}^2}{2}-V(\phi_1)\right.\\
    &&&\left.\quad-\frac{\left(\partial\phi_1\right)^2}{2}-\frac{\lambda(\phi_1)^2\left(\partial\phi_2\right)^2}{2}\right)\,\mathrm{d}^4x,\nonumber
\end{AmSalign}
\glspl{eom}
\begin{align}
    \ddot{\phi}_1+3H\dot{\phi}_1-\lambda\lambda^\prime\dot{\phi}_2^2+V^\prime&=0\\
    \ddot{\phi}_2+3H\dot{\phi}_2+2\lambda^\prime\dot{\phi}_2\dot{\phi}_1/\lambda&=0,
\end{align}
and geodesic distance 
\begin{equation}
    \Delta\phi=\int\!\sqrt{G_{ab}\dot{\phi}^a\dot{\phi}^b}\,\mathrm{d}t,
\end{equation}
the field traversal is given by 
\begin{equation}
    \frac{\Delta\phi}{M_\textnormal{P}}=\int\!\sqrt{-\left(\frac{3}{2M_\textnormal{P}}\frac{\lambda}{\lambda^\prime}\right)^2+3\frac{V^\prime}{V}\frac{\lambda}{\lambda^\prime}}\,\mathrm{d}N,
\end{equation}
which, for $N_e\leq10$ (during a \gls{de} dominated epoch), yields $\Delta\phi/M_\textnormal{P}\sim N_e/\sqrt{-M_\textnormal{P}\lambda^\prime/\lambda}<1$, which satisfies the \gls{dc} bound \cite{cicoli_out_2020}.\footnote{
    $\lambda$ is a model-defining function, an overdot represents a derivative with respect to time, and a prime a derivative with respect to the scalar field $\phi$.
}
However, the models are constraint by their turning rate $\mathfrak{T}$ \cite{freigang_cosmic_2023}:
\begin{equation}
    \mathfrak{T}<\mathfrak{l}\sqrt{-\dot{H}},\label{eq:turnr}
\end{equation}
with $\mathfrak{l}$ an $\order{1}$ constant that depends on the curvature of the moduli space and the decay rate of the mass scale of the tower of particles, and $H$ the Hubble parameter. This means that a field that drives an accelerated cosmic expansion cannot undergo rapid turns or deviate from geodesics too strongly. The definition of the turning rate $\mathfrak{T}$ needs some context, which is all explained by \citet{freigang_cosmic_2023}: Assume a setting with $a=1,\,\dotsc,\, n$ scalar fields $\phi^a$. When the fields are free, massless, and homogeneous, the Lagrangian $L$ in Minkowski spacetime $\eta$ can be written as 
\begin{equation}
    L=-\frac{1}{2}\eta^{\mu\nu}G_{ab}\partial_\mu\phi^a\partial_\nu\phi^b,
\end{equation}
with $G_{ab}$ the internal field space metric. The \gls{eom} can then be written as follows:
\begin{align}
    \ddot{\phi}^a+\Gamma^a_{bc}\dot{\phi}^b\dot{\phi}^c=&\,0\\
    \Gamma^a_{bc}=&\,\frac{1}{2}G^{ad}\left(\frac{\partial G_{bd}}{\partial\phi^c}+\frac{\partial G_{cd}}{\partial\phi^b}-\frac{\partial G_{bc}}{\partial\phi^d}\right)\\
    \mathrm{D}_tv^a\defeq&\,\dot{v}^a+\Gamma^a_{bc}v^b\dot{\phi}^c\\
    \mathrm{D}_t\dot{\phi}^a=&\,0\label{eq:eombrief},
\end{align}
where $\Gamma^a_{bc}$ are the Christoffel symbols of the moduli space, $v^a$ is any vector, $\mathrm{D}_t$ is the covariant derivative on the field space, and \cref{eq:eombrief} is the now much shorter \gls{eom}. The turning rate is now defined as
\begin{equation}
    \mathfrak{T}\defeq\abs{\mathrm{D}_tT}=\sqrt{G_{ab}\mathrm{D}_tT^a\mathrm{D}_tT^b},
\end{equation}
where we introduced the tangent vector
\begin{equation}
    T^a=\frac{\dot{\phi}^a}{\sqrt{G_{ab}\dot{\phi}^a\dot{\phi}^b}}.    
\end{equation}
Seeing the definition of $\mathfrak{T}$ allows us to make another interpretation of it: it's a measure of how \enquote{well} parallel transport of $T^a$ along $\phi^a$ works, i.e. $\mathfrak{T}$ measures the \textit{non-geodesity}. The importance of this will become clear when studying the general remarks on distance measures and geodesics below.

\citet{eskilt_cosmological_2022} present a spinning two-field model of \gls{de} that satisfies the \gls{dc}. We discuss it in \cref{sp:dSC_Quintessence}.

\citet{payeur_swampland_2024} study rapid-turn \gls{de}, a multi-field \gls{de} model in highly curved field space, and find that the \gls{dc} as well as the \gls{tcc} are violated, except for a small region in the parameter space, where both conjectures, as well as the \gls{dsc} and observational constraints can be met. In their model, expansion is realised as non-geodesic motion in a highly curved field space. An attractor solution with an \gls{eos} of $\omega\simeq-1$ keeps the accelerated expansion eternal (which violates the \gls{tcc}) as the field can keep rolling (which violates the \gls{dc} when it rolls a trans-Planckian distance). Their model is related to hyperinflation and has a two-field action $S$ with field space metric $G_{ab}$ and potential $V(\phi)$:
\begin{align}\label{eq:rapid-turn-DE}
    S&=\int\!\sqrt{-g}\left(\frac{G_{ab}(\phi)\partial_\mu\phi^a\partial_\nu\phi^b}{2}-V(\phi)\right)\,\mathrm{d}^4x\\
    G_{ab}&=\begin{pmatrix}
        1 & 0 \\
        0 & e^{-\phi/M_\textnormal{P}L}
    \end{pmatrix}\\
    L&=\sqrt{-1/M_\textnormal{P}^2R}\\
    V(\phi)&=V_0e^{-\mathfrak{c}\phi/M_\textnormal{P}}
\end{align}
with $R$ the Ricci scalar of the field space. Observational constraints require $\mathfrak{c}L\leq0.0265$. The \gls{eos} is $\omega\approx-1+2\mathfrak{c}L$. The \gls{dc} and the \gls{tcc} are violated at the same time if
\begin{equation}
    690L\left(\mathfrak{c}L+1\right)=-2\mathfrak{c}L+1,
\end{equation}
i.e. the field curvature is bounded by $L\approx1/690$.

\citet{brinkmann_stringy_2022} discuss that multi-field quintessence is tricky to implement in string theory: if the system starts in a matter dominated phase, the system does typically \textit{not} evolve into a state with $\omega_\textnormal{DE}\simeq-1$ and $\Omega_\textnormal{DE}\simeq0.7$. This can be achieved if the system starts in an epoch of kinetic domination. However, even then, fine-tuning is required to match today's Universe.

\subparagraph{\gls{dm} and \gls{ede}}\label{p:DC_EDE} 
\gls{ede} is widely studied as a remedy to the Hubble tension, yet it appears to be incompatible with the observed \glspl{lss} and their formation in the concordance model \cite{schoneberg_h_0_2022,di_valentino_realm_2021}.
\citet{kaloper_dark_2019} takes inspiration from the \gls{dc}\footnote{
    He incorrectly calls it the \gls{wgc} in his paper, yet the \gls{wgc} is concerned with charged particles. However, he refers to the mass and towers of states\,\textemdash\,exactly what the \gls{dc} is about.
    }
to propose a model of \gls{ede} as an axion-like field that rapidly decays into radiation at a redshift of $z\sim3000\sim5000$ to solve the Hubble tension. The tower of states is a testable prediction by this model.

Furthermore, it is suggestive to interpret the predicted light tower of states as \gls{dm}, in particular \gls{dm} coupled to \gls{de}. Various studies come to positive conclusions regarding the compatibility of such models with observational and swampland constraints.
For example, \citet{mcdonough_early_2022} are inspired by the \gls{dc}, and investigate if it is possible to construct a \gls{lss} formation\textendash compatible \gls{ede} model that reduces the Hubble tension by introducing a coupling of \gls{dm} to \gls{de},\footnote{
    \citet{vafa_swamplandish_2024} states that the unification of \gls{de} and \gls{dm} is a natural expectation from the \gls{dc}, and \citet{carroll_quintessence_1998} even shows that there is a coupling unless a symmetry prevents it.
    }
where the \gls{dm} mass is given by $m_\text{DM}=m_0e^{-\mathfrak{a}\Delta\phi/M_\textnormal{P}}$.\footnote{
    In their paper, they write $m_\text{DM}=m_0e^{c\phi/M_\textnormal{P}}$ and mention that $c>0$, and the initial field value, $\phi_i>0$, \textit{decreases} over cosmic time. We have a leading minus sign in our notation, as we use $\mathfrak{a}>0, \,\Delta\phi>0$.
    }
The \gls{de} potential resembles an axion potential $V(\phi)=V_0\left[1-\cos\left(\phi/f\right)\right]^n$. The model successfully reduces the Hubble tension to $1\sigma\sim2\sigma$. \gls{lss} growth is also addressed: $\mathfrak{a}>0$ increases \gls{lss} growth, as matter\textendash radiation equality takes places earlier and the \gls{dm} from their model is self-interacting, i.e. it has an attractive fifth force. To reduce the $S_8$ tension, $\mathfrak{a}<0$ works best. Using different datasets, they find $\mathfrak{a}<0.1$, which is in tension with the $\order{1}$ predictions of the \gls{dc} for the value of this parameter. However, we suspect that the fit could be improved if only a fraction of \gls{dm} were coupled to \gls{de}. Such a model is studied by \citet{wang_does_2023}: Having only a fraction of \gls{dm} interacting with \gls{de} helps alleviate the $S_8$ tension, but a vanishing coupling agrees with observational data as well. The issue is that matter clusters in the matter-dominated era, and a higher Hubble constant $H_0$ increases $S_8$.
Another (\gls{dc}- and \gls{dsc}-inspired) model that addresses the Hubble tension by coupling \gls{de} to \gls{dm} is presented by \citet{agrawal_h_0_2021}:
\begin{AmSalign}
   m_\text{DM}(\phi)&=m_\text{CDM}
   \begin{cases}
       1 & 0<\phi<\phi_0\\
       \exp[-\mathfrak{c}_1\left(\phi-\phi_0\right)] & \phi\geq\phi_0
   \end{cases}\\
   V(\phi)&=\mathfrak{c}_0\left[e^{-\mathfrak{c}_2\phi}e^{-\phi_0\left(\mathfrak{c}_3-\mathfrak{c}_2\right)}+e^{-\mathfrak{c}_3\phi}\right].
\end{AmSalign}
The \gls{dm} is the tower of states that becomes light for large field excursions, with $m_\text{CDM}$ the \gls{dm} mass from the \gls{lcdm} model, and the \gls{de} is in the form of a quintessence field $\phi$. The $\mathfrak{c}_i$ are constants.\footnote{
Even though not intended by \citet{agrawal_h_0_2021}, their model corresponds to an \gls{ede} model presented by \citet{sabla_no_2021}\,\textemdash\,for the appropriate choice of the constants.
}
Since the exponential suppression of the \gls{dm} mass only starts at late time, when $\phi$ rolled enough, the early universe cosmology corresponds to the concordance model. The strength of the coupling between \gls{dm} and \gls{de} is constraint by late-time observations such as tidal tails of satellite galaxies and warping of stellar disks within galaxies, as the coupling introduces a fifth force in the dark sector that affects the \gls{dm}\,\textemdash\,if only a fraction of the \gls{dm} participates in the interaction, an observational effect can be illusive.\footnote{
    However, \citet{bruck_dark_2019} find that introducing a coupling can generally help to comply with the \gls{dc} and the \gls{dsc}.
}
Also \citet{baldes_forays_2019} state observational probes for \gls{dm}\textendash\gls{de} coupling of different models:
\begin{description}
    \item[Quintessence] with a single- or a multi-exponential potential can satisfy the \gls{dc} and the \gls{dsc} if fine-tuning takes place.
    \item[Self-interacting \gls{dm}] with a light scalar mediator whose mass depends on quintessence matches constraints from dwarf galaxies if the mediator mass varies by a factor of $\num{e-9}$  per year.
    \item[Kinetic mixing] between U(1) sectors that is controlled by quintessence is hard to detect as the late-time mixing is suppressed, yet a lower bound on \gls{dm} mass in the early Universe constrains the model.
    \item[Quintessence\textendash Higgs coupling] shifts the Higgs parameter, which in turn modifies the timescale for vacuum decay.
\end{description}

\subparagraph{Further \gls{de} Findings}
The following work mislabels the \gls{dc} as the \gls{wgc}, yet derives conclusions about \gls{de} based on the idea that field ranges have to be sub-Planckian:\footnote{
    Sub-Planckian field ranges are not a general feature of the \gls{wgc}. The electric \gls{wgc} could be satisfied by trans-Planckian \glspl{bh} (in a very mild form of the \gls{wgc} at least), and the magnetic \gls{wgc} can be satisfied by a very weak coupling.
    However, in the case of axions, the \gls{wgc} actually limits the field range to be sub-Planckian, as we discuss in \cref{p:WGC_axion}.
}
\citet{ma_possible_2008} show that holographic \gls{de} \cite{li_model_2004} with a density given by $\rho=3\mathfrak{p}^2M_\textnormal{P}^2/\Lambda^2$ is compatible with the \gls{dc} for $\mathfrak{p}\geq1$ and corresponds to quintessence in this case. This does not indicate that quintessence is generally in the landscape, as \citet{wu_limits_2008} show that a Chaplygin-gas model in the form of quintessence does not satisfy the \gls{dc}. It also does not prove that holographic \gls{de} generally works: \citet{chen_constraints_2008,liu_theoretical_2010} show that holographic \gls{de} is incompatible with observational data if compatible with the \gls{dc}. This is contrasted by work of \citet{sun_weak_2011,SunYue_Weak_2011}, which shows that holographic \gls{de} \textit{is} compatible with the \gls{dc}, if it has interaction terms and the Universe is curved. \citet{sun_weak_2011,chen_constraining_2011} suggest that \gls{de} that decays into \gls{dm} is compatible with the \gls{dc}.

\paragraph{Gravitinos}\label{p:DC_Gravitinos}
The limit when the gravitino mass $m_\textnormal{3/2}$ goes to zero is explored by
\citet{castellano_gravitino_2021,cribiori_gravitino_2021}.
They observe that when the gravitino mass vanishes, the \gls{eft} breaks down and an infinite tower with $m\sim m_\textnormal{3/2}^\mathfrak{g}$ appears. \citet{castellano_gravitino_2021} show a direct link to the \gls{dc} and claim that this goes as $\sim e^{-\mathfrak{a} d(\phi,\phi_0)}$, such that $\mathfrak{g}$ is related to $\mathfrak{a}$ and is in the range of $\frac{1}{3}\leq\mathfrak{g}\leq1$ for 4d type IIA orientifolds.\footnote{
    $m=\mathfrak{c}m_\textnormal{3/2}^\mathfrak{g}=\Tilde{\mathfrak{c}} \exp\left(-\mathfrak{a} d(\phi,\phi_0)\right)$ with $\mathfrak{c}$ and $\Tilde{\mathfrak{c}}$ two constants would be an exact expression. For vanishing gravitino mass and infinite distance, the constant contributions can be neglected, and we find that $\mathfrak{a}=-\mathfrak{g}\log\left(m_\textnormal{3/2}\right)/d(\phi,\phi_0)=\sqrt{3/2}\mathfrak{g}$, where in the last step an explicit relation between the gravitino mass and the distance along this path from \citet{castellano_gravitino_2021} was used. Together with the discovered range for $\mathfrak{g}$, this leads to bounds for $\mathfrak{a}$: $1/\sqrt{6}\leq\mathfrak{a}\leq\sqrt{3/2}$ \cite{castellano_gravitino_2021}, which is consistent with the lower \gls{dc} bound in 4d of $\mathfrak{a}\geq1/\sqrt{2}$. The lower bound of $1/\sqrt{6}$ is also confirmed in a type IIB string theory setting, using Hodge theory \cite{bastian_weak_2021}.
    }
Moreover, they %
make phenomenological predictions: To have supersymmetry at the \SI{1}{\tera\electronvolt}\textendash scale, a \gls{kk} tower $\order{10^8}$ is required.
Furthermore, the Hubble constant (after inflation) is bounded by
\begin{equation}
    H\lesssim m_\textnormal{3/2}^\mathfrak{g} M_\textnormal{P}^{\left(1-\mathfrak{g}\right)}.\label{eq:DCgravitino}
\end{equation}
Their main idea is that in an \gls{eft} with a consistent extension into \gls{qg}, the gravitino mass is not arbitrarily separated from the \gls{uv} scale and the limit $m_\textnormal{3/2}\rightarrow0$ is at infinite distance.\footnote{
    Even before the swampland programme was formally initiated, \citet{kallosh_landscape_2004} showed that $H\lesssim m_\textnormal{3/2}$ has to hold in \gls{kklt}, and that if \gls{ads} is uplifted to \gls{ds}, $m_\textnormal{3/2}\approx V_\textnormal{AdS}/3$.
} %
The bound is compatible with the \gls{ssc} and the \gls{adsdc}. Further constraints come from the magnetic \gls{wgc} ($H<\Lambda_\textnormal{S}<g_\textnormal{3/2}M_\textnormal{P}$) and the \gls{flb} ($m_\textnormal{3/2}>\sqrt{q_\textnormal{3/2}g_\textnormal{3/2}M_\textnormal{P}H}$).

A vanishing gravitino mass leading to inconsistencies is also observed by \citet{coudarchet_geometry_2021,cribiori_supergravity_2023}.
A tower of states becoming light, and therefore missing in the \gls{eft} description, is observed when the gravitino sound speed vanishes and particles with arbitrary momentum (above the cutoff) are produced \cite{kolb_gravitino_2021} (see our section on the \gls{gsc}).%

\paragraph{Horizons}
The \gls{dc} also sheds some light on asymptotic observables and the question of what we can see in the infinite future or on the cosmic horizon\,\textemdash\,if such asymptotic observables exist in cosmology\,\textemdash, as the work by \citet{rudelius_asymptotic_2021} shows: In \gls{ads} space or asymptotically flat spacetimes, we can define asymptotic observables via correlation functions of conformal field theories using the \gls{adscft} correspondence or via the S-matrix elements, respectively. However, in \gls{flrw} cosmology, we are facing a challenge: starting with a singularity in the infinite past prevents an S-matrix description, and in the infinite future, we do not expect a zero energy density on a fixed Cauchy slice of spacetime, which leaves us with geometry fluctuations that might extend infinitely. Furthermore, there is, for any observer, always an unobservable part of the Universe, which can contain an infinite amount of energy and information, i.e. the global state of the Universe is not observable and cannot be described by an S-matrix. If well-defined observables in the asymptotic limit of our Universe exist is unclear; however, having light massive states would make it more likely than having massless modes. The \gls{dc} predicts the existence of such light massive states.

\citet{rudelius_asymptotic_2022} studies late-time cosmologies under the assumption that late-time cosmologies are governed by the infinite-distance limits of scalar fields, which feature exponential decays of their potential 
\begin{equation}
    V(\phi)\sim\exp\left(-\lambda\phi\right),
\end{equation}
such that the Hubble scales goes like 
\begin{equation}
    H\sim\sqrt{\dot{\phi}^2+2V(\phi)}\sim\exp\left(-\mathfrak{a}_H\phi\right),
\end{equation}
with ($d$ the number of large dimensions)
\begin{equation}
    \mathfrak{a}_H=\min\left(\frac{\lambda}{2}\,,\,\sqrt{\frac{d-1}{d-2}}\right).
\end{equation}
If $\lambda\geq2\sqrt{\left(d-1\right)\left(d-2\right)}$, the kinetic energy of the scalar field dominates its potential energy. Now, we have $m\sim\exp\left(-\mathfrak{a}\phi\right)$ from the \gls{dc} and $H\sim\exp\left(\mathfrak{a}_H\phi\right)$.
If $\mathfrak{a}>\mathfrak{a}_H$, the late-time cosmology is either dominated by the \gls{kk} tower or the fundamental string (whichever of the two is the lightest tower).\footnote{
    According to the \gls{ep}, these are the only two options.
}
If it is the \gls{kk} tower, $m_\text{KK}<H$, which implies that the compactified dimension $m_\text{KK}^{-1}$ becomes larger than the Hubble horizon $H^{-1}$, i.e. we are no longer in a $d$-dimensional \gls{flrw} universe. If it is the fundamental string, then the universe is dominated by the stringy effects.
If $\mathfrak{a}<\mathfrak{a}_H$, the cosmology remains a $d$-dimensional \gls{flrw} cosmology with expanding spacetime\,\textemdash\,accelerated expansion for $\mathfrak{a}_H<1/\sqrt{d-2}$, decelerated expansion $\mathfrak{a}_H>1/\sqrt{d-2}$. The light tower of states remains at or above the Hubble scale, and the strong and dominant energy conditions are both satisfied. Unless there is an asymptotic boundary at future spacelike infinity, the Universe cannot accelerate indefinitely in asymptotic scalar field space. Therefore, $\mathfrak{a}_H\geq1/\sqrt{d-2}$ is required. These considerations can be summarised in the following bounds:
\begin{equation}
    \sqrt{\left(d-1\right)/\left(d-2\right)}\geq\mathfrak{a}_H\geq\mathfrak{a}\geq1/\sqrt{d-2}.
\end{equation}

\paragraph{Inflation,}\label{p:DC_Inflation} especially large-field inflation, is probably the most striking field of applicability of the \gls{dc} from a cosmologist's perspective.\footnote{
    \citet{rudelius_revisiting_2023} points out that this is not obvious. The \gls{dc} is concerned with infinite field traversals, which are not necessarily implemented into inflation. The inflaton field might travel only a finite field range or even oscillate, like in the case of axion-like models (see for instance \cite{valenzuela_backreaction_2017}, \cref{p:dSC_Axions,rel:WGC_DC,p:WGC_axion}).
    }
\citet{scalisi_swampland_2019} provide a bound for the inflaton field range that also takes into account the exponential decay of the mass scale:
\begin{equation}
    \Delta\phi<\frac{1}{\mathfrak{a}}\log\frac{M_\textnormal{P}}{\Lambda_\textnormal{S}}.
\end{equation}
A higher cutoff $\Lambda_\textnormal{S}$ implies a smaller allowed field range $\Delta\phi$, such that infinite field ranges become incompatible with the theory. If we describe inflation with an \gls{eft}, then we find that
\begin{equation}
    H\leq\Lambda_\textnormal{S},
\end{equation}
and we can deduce that in a successful model of inflation
\begin{equation}
    \Delta\phi<\frac{1}{\mathfrak{a}}\log\frac{M_\textnormal{P}}{H}
\end{equation}
must hold \cite{scalisi_swampland_2019,matsui_swampland_2020}. For slow-roll inflation with $\mathfrak{a}=1$, this translates into $\Delta\phi<10M_\textnormal{P}$ regarding Planck 2018 \gls{cmb} data \cite{scalisi_swampland_2019}, which is in line with the \textit{refined \gls{dc}}, which claims that $d(\phi,\phi_0)<10 M_\textnormal{P}$ \cite{palti_swampland_2019,rudelius_revisiting_2023}. A smaller value of $\mathfrak{a}$ would allow for a bigger $\Delta\phi$. Some authors claim that large-field inflation is ruled out by the \gls{dc} if $\Delta\phi<M_\textnormal{P}$ is required \cite{bjorkmo_hyperinflation_2019,blumenhagen_large_2018,blumenhagen_swampland_2017}. However, we do not agree with this notion, even though we have not heard of a model that follows this idea: the \gls{dc} says that if a field travels a trans-Planckian range, a tower of light states emerges with exponentially suppressed masses. Assuming the inflaton rolls a trans-Planckian range, a tower of light states emerges, with masses much smaller than the Planck mass. This is what happens at the end of inflation, namely during reheating: quarks and leptons emerge, towers of states, with masses much smaller than the Planck scale ($m_e\sim\num{4e-23}M_\textnormal{P}\sim e^{-51}M_\textnormal{P}$, $m_\text{up}\sim\num{2e-22}M_\textnormal{P}\sim e^{-50}M_\textnormal{P}$). Even applying the very stringent (and not particularly well justified or tested) bounds $1/\sqrt{2}<\mathfrak{a}<\sqrt{6}$ allows $\Delta\phi\sim\order{10M_\textnormal{P}}$. If the emerging tower was equidistant, we could even deduce the number of states: $N=\frac{M_\textnormal{P}}{m}=e^{\Delta\phi}$ \cite{bonnefoy_swampland_2021}. 
The \gls{ep} states that the tower is either a \gls{kk} tower that corresponds to a decompactification, or a tower of string oscillator modes. In the latter case, inflation and the very early Universe have to be described by string cosmology. In the former case, inflation would either describe a decompactification from 3 to 4 spacetime dimensions, or from 4 to 5, if the dark dimension proposal (see \cref{p:AdSDC_darksector}) holds.
A challenge such a model has to overcome is that otherwise suppressed operators of $\order{\Delta\phi/M_\textnormal{P}}$ have to be properly incorporated in the theory. Furthermore, it is a conceptual challenge to explain why the spacing would be in such a way that we only discovered the lightest states of the \gls{sm} towers\,\textemdash\,there is no continuing electron tower in the entire energy range probed so far. Furthermore, such a model would have to be tested for its compatibility with other swampland conjectures and observational constraints.

In the following, we discuss the implications of the \gls{dc} for more established models of inflation.

\subparagraph{Slow-roll inflation} gets constraint by combining the \gls{dc} with the Lyth bound: The \gls{dc} gives an upper bound on the geodesic distance $\Delta\phi_\text{g}$ that separates two points along the inflationary trajectory,
and the Lyth bound\,\textemdash\,determined by the tensor-to-scalar\footnote{
    Scalar fluctuations are induced by matter inhomogeneities, while tensor fluctuations correspond to \glspl{gw} \cite{brandenberger_trans-planckian_2021}.
}
ratio $r_\textnormal{ts}$\,\textemdash\,gives a lower bound on the non-geodesic distance the inflaton traverses between those two points, $\Delta\phi_\text{ng}$;\footnote{
    The background solution trajectory in multi-field inflation can locally bend, which generates interactions between primordial curvature perturbations and field fluctuations orthogonal to the inflaton trajectory, i.e. isocurvature fluctuations \cite{bravo_tip_2020}.
    }
the two distance measures respect the relation $\Delta\phi_\text{g}\leq\Delta\phi_\text{ng}$ \cite{bravo_tip_2020} by definition.
The Lyth bound \cite{lyth_what_1997} sets a lower limit for the field excursion
\begin{equation}
    \frac{\Delta\phi}{M_\textnormal{P}}\geq\Delta N_e\times\sqrt{\frac{r_\textnormal{ts}}{8}},
\end{equation}
which violates the \gls{dc} in the single-field inflation case, where $\Delta\phi$ corresponds to the geodesic distance and has to be sub-Planckian \cite{landete_mass_2018},\footnote{
    Already \citet{ooguri_geometry_2007} pointed out that the \gls{dc} rules out slow-roll inflation.
}
but can be compatible with the \gls{dc} in the multi-field case, where the $\Delta\phi$ relevant for the Lyth bound does not necessarily correspond to a geodesic \cite{bravo_tip_2020}.

\subparagraph{Constant-roll inflation,} where terms quadratic in the energy density are present in the Friedmann equation, takes place in a 5d brane-world scenario, and allows for sub-Planckian field excursion while retaining $\eta_V\sim\order{1}$, is compatible with the \gls{dc} as well as the \gls{dsc} \cite{mohammadi_brane_2020}.

A model of anisotropic constant roll inflation with a complex quintessence field tested by \citet{sadeghi_anisotropic_2022} violates the \gls{dc}.

\subparagraph{Eternal inflation} is in strong tension with the \gls{dc} or swampland conjectures in general \cite{matsui_eternal_2019,rudelius_conditions_2019,wang_eternal_2020,blanco-pillado_eternal_2020,dimopoulos_steep_2018,kinney_eternal_2019,brahma_stochastic_2019,seo_implication_2022,matsui_swampland_2020}.\footnote{
    See the work by \citet{lin_topological_2020} for a one-sentence summary of most of those sources. 
    In \cref{p:EternalHartle}, we discuss eternal inflation in the context of the Hartle\textendash Hawking no-boundary proposal \cite{hartle_genesis_2022,hartle_wave_1983} in the light of the \gls{dc} and the \gls{dsc}.
    } 
\citet{seo_implication_2022} argues that eternal inflation is incompatible with the \gls{dc}:
a rapidly changing inflaton field causes the appearance of a tower of light states, which increases the entropy and the energy inside the cosmological horizon. If this happens quickly enough, the energy density is high enough to produce \glspl{bh}, i.e. \gls{qg} effects are important to describe the particle interactions. The \gls{eft} breaks down, when the entropy inside the cosmological horizon exceeds the \gls{ceb}.
\citet{cohen_effective_1999} quantify when the \gls{eft} breaks down due to such a particle production in flat space, namely when
\begin{equation}
    l^3\Lambda_\text{UV}^4\lesssim lM_\textnormal{P}^2 \Rightarrow \Lambda_\text{UV}\lesssim\sqrt{\frac{M_\textnormal{P}}{l}},
\end{equation}
with $\Lambda_\text{UV}$ the \gls{eft} cutoff and $l$ the size of the region.\footnote{
    This bound is more stringent that the Bekenstein\textendash Hawking bound $l^3\Lambda_\text{UV}^3\lesssim S_\text{BH}=\pi l^2M_\textnormal{P}^2$ such that $\Lambda_\text{UV}\lesssim\left(M_\textnormal{P}^2/l\right)^{1/3}$.
}
\citet{seo_implication_2022} combines this bound
with the \gls{dc}, to derive a bound on the slow-roll parameter:
\begin{equation}
    \epsilon\gtrsim\frac{H^2}{M_\textnormal{P}^2}+\frac{\mathfrak{a}}{6\sqrt{\pi}}\frac{H^3}{M_\textnormal{P}^3}+\order{\left(\frac{H}{M_\textnormal{P}}\right)^4},
\end{equation}
which almost rules out eternal inflation, for which
\begin{equation}
    \epsilon\lesssim\left(2n+1\right)H^2/M_\textnormal{P}^2
\end{equation}
has to hold, $n$ being the quantum number if excitations are considered \cite{seo_eternal_2020}.

However, eternal inflation is not ruled out completely. \citet{lin_topological_2020} presents a model of topological inflation with a hilltop potential \textit{on a 3-brane}:
For inflation to continue eternally, the field has to remain near the top of its potential, which happens if it is close to a topological defect (a monopole, a cosmic string, or a domain wall) \cite{vilenkin_topological_1994,linde_monopoles_1994,linde_topological_1994}. In the inflated region, those defects (such as magnetic monopoles) are diluted away. \citet{lin_topological_2020} examines the case of a domain wall. Since we do not observe domain walls, their thickness has to be larger than the observable Universe. %
The Friedmann equation tells us how the Hubble parameter and the inflaton potential relate:
\begin{equation}
    H^2=\frac{V}{3M_\textnormal{P}^2}\left(1+\frac{V}{2\Lambda_\text{B}}\right),
\end{equation}
with $\Lambda_\text{B}=6\pi\frac{M_{\textnormal{P};5}^6}{M_\textnormal{P}^2}\gtrsim\left(\SI{1}{\mega\electronvolt}\right)^4$ (the bound stemming from \gls{bbn}), $M_{\textnormal{P};5}$ the reduced 5-dimensional Planck mass;
the second term in the brackets is only present in a brane-world \cite{lin_topological_2020}.
In the double-well potential studied by \citet{lin_topological_2020}, the \gls{dc} and the \gls{dsc} are only satisfied with the extra brane term.

\subparagraph{Small-field inflation} appears to be compatible with the \gls{dc}, as \citet{osses_reheating_2021} show, studying different models.\footnote{
    We present the models (and implications from the \gls{dsc}) in \cref{sec:SmallFieldInflation}.
    }
However, they do not consider precise bounds on the parameter $\mathfrak{a}$ for the \gls{dc}.

\citet{mohammadi_brane_2022} study several models of inflation on a brane, and find that power-law inflation of the form $V\sim\phi^n$, natural inflation of the form $V\sim\left(1-\cos{\frac{\phi}{f}}\right)$, and T-model inflation (\cref{eq:T-model}) are compatible with the \gls{dc}.

\citet{cribiori_supergravity_2023} study single-field D-term inflation, chaotic inflation, and Starobinsky inflation in supergravity and find that if a massless gravitino is present in the theory, the models are in tension with the gravitino \gls{dc} and the \gls{flb}, and last only a few $e$-foldings ($N_e=13$ in the case of chaotic inflation, $N_e=7$ in the case of Starobinsky inflation).

\subparagraph{Warm inflation} is characterised by the inflaton field $\phi$ decaying into radiation at the rate $\Upsilon$, which depends on $\phi$ and the temperature of the photon bath, $T$ \cite{das_runaway_2020}:
\begin{align}
    \Ddot{\phi}+3H\dot{\phi}+V^\prime&=-\Upsilon(\phi, T)\dot{\phi}\\
    \rho_r+4H\rho_r&=\Upsilon(\phi,T)\dot{\phi}^2\\
    H^2&=\frac{1}{3M_\textnormal{P}^2}\left(\frac{\dot{\phi}^2}{2}+V(\phi)+\rho_r\right).
\end{align}
To evaluate the compatibility with the \gls{dc} it is beneficial to express the bound in terms of the slow-roll parameter $\epsilon_V$ \cite{motaharfar_warm_2019}:
\begin{equation}
    \frac{\Delta\phi}{M_\textnormal{P}}=\frac{\dot{\phi}}{H}N_e\simeq\frac{\sqrt{2\epsilon_V}}{1+\Upsilon}N_e,
\end{equation}
which can per se be satisfied in the strong as well as the weak dissipation regime (small or large $\Upsilon$) \cite{das_warm_2019,motaharfar_warm_2019}.
The Lyth bound for warm inflation,
\begin{equation}
    \frac{\Delta\phi}{M_\textnormal{P}}\gtrsim\sqrt{\frac{r_\textnormal{ts}}{8}\frac{T}{H}\sqrt{1+\Upsilon}}N_e,
\end{equation}
can as well be satisfied in the strong as well as in the weak dissipation regime: for $N_e\sim60$, $H/T\sim\num{e-1}$, and $r_\textnormal{ts}<\num{e-2}$, the strength of dissipation is not particularly constraint, i.e. even the regime of weak dissipation is completely allowed \cite{das_warm_2019}. 
Overall, models of warm inflation tend to satisfy the \gls{dc}:
\begin{itemize}
    \item Warm inflation scenarios with dissipative effects with a cubic temperature dependence are compatible with the \gls{dc}, as well as with the \gls{dsc} and the \gls{tcc} \cite{das_runaway_2020,santos_warm_2022}.
    \item Models with strong dissipation satisfy the \gls{dc}, the \gls{tcc}, as well as the \gls{dsc} \cite{kamali_warm_2020,das_runaway_2020,berera_trans-planckian_2019,berera_thermal_2021,kamali_intermediate_2021,arya_primordial_2024,das_distance_2020,santos_warm_2022}.
    \item Not all models of warm inflation satisfy the \gls{dc} automatically: a model studied by \citet{bastero-gil_warm_2019} is in tension with the \gls{dc} as well as with the \gls{dsc} in the regime of weak dissipation. Their distributed mass model requires the inflaton to scan over many mass states, ergo, it needs a large field traversal, which violates the \gls{dc}.
    \item \citet{mohammadi_warm_2020} found that it depends on the dissipation function itself: They study a model where a tachyon field decays into radiation. If the dissipation coefficient of their model is a power-law function of the tachyon field, neither the \gls{dc}, nor the \gls{dsc} is satisfied. However, if the dissipation coefficient also depends on the temperature of the photon bath, both conjectures are satisfied for observationally allowed parameter values in the strong dissipation regime.\footnote{\label{f:warm-tachyon-inflation}
        The model tested by \citet{mohammadi_warm_2020} has the following conservation equations:
        \begin{align*}
            \dot{\rho}_\phi+3H\left(\rho_\phi+p_\phi\right)&=-\Xi\left(\rho_\phi+p_\phi\right)\\
            \dot{\rho}_\textnormal{r}+3H\left(\rho_\textnormal{r}+p_\textnormal{r}\right)&=\Xi\left(\rho_\textnormal{r}+p_\textnormal{r}\right)\\
            \Xi\left(\rho_\phi+p_\phi\right)&=3\Xi H^2\dot{\phi}^2\\
            \Upsilon&=\Xi/3H,
        \end{align*}
        where the second-last equation stems from the Friedmann equation.
    }
\end{itemize}

\subparagraph{Inflation including higher spin states} faces additional constraints from the \gls{dc}, if combined with the Higuchi bound \cite{higuchi_forbidden_1987}, as shown by \citet{scalisi_inflation_2020}:
The Higuchi bound ensures unitarity of the \gls{eft} and demands for a state with spin $s$ that
\begin{equation}
    m^2>s\left(s-1\right)H^2.
\end{equation}
Combined with the \gls{dc}, this limits the allowed field range:
\begin{equation}
    \Delta\phi<\frac{1}{\mathfrak{a}}\log\left[\frac{m_0}{H}\frac{1}{\sqrt{s\left(s-1\right)}}\right],
\end{equation}
i.e. a tower where all spins are allowed is incompatible with inflation, as this bound is violated for high-spin states and exponentially suppressed masses. Therefore, a maximum spin value
\begin{equation}
    \frac{m_0}{H}>\sqrt{s_\text{max}\left(s_\text{max}-1\right)}
\end{equation}
must exist. The masses are always super-Hubble for spin states with $s>1$ with a minimum value of $m_0>\sqrt{2}H$.
For the tower related to the string, the Regge trajectory gives a relation for the string scale $M_\textnormal{s}$:
\begin{equation}
    m^2=sM_\textnormal{s}^2.
\end{equation}
Combining this with the Higuchi bound and the \gls{dc} gives a bound for the field range of
\begin{equation}
    \Delta\phi<\frac{1}{\mathfrak{a}}\log\left[\frac{M_\textnormal{s}(0)}{H}\frac{1}{\sqrt{s-1}}\right],
\end{equation}
$M_\textnormal{s}(0)$ being the highest value of the string scale at $\Delta\phi=0$.
In (quasi) \gls{ds} background, the string scale comes with an additional cutoff, this time a cutoff for the applicability of the Higuchi bound, which helps to circumvent the just presented constraint:
the length of the string $l_\textnormal{s}\sim\sqrt{s}/M_\textnormal{s}$ has to be smaller than the Hubble radius $1/H$, which gives an upper limit for the spin state for which the Higuchi bound is applicable: 
\begin{equation}
    s_\text{max}=\left(\frac{M_\textnormal{s}}{H}\right)^2.
\end{equation}
We also have a bound for strings shorter than the Hubble radius; their energy scale has to be super-Planckian:
\begin{equation}
    M_\textnormal{s}>\sqrt{HM_\textnormal{P}},
\end{equation}
which limits the energy scale of inflation, which is expected to be below the string energy \cite{noumi_string_2020,lust_note_2019}. Combining all this gives a refined bound on the field excursion:
\begin{equation}
    \Delta\phi<\frac{1}{\mathfrak{a}}\log\frac{M_\textnormal{s}(0)}{\sqrt{HM_\textnormal{P}}}.
\end{equation}

\subparagraph{Mutli-field inflation} can circumvent the issues presented by the Lyth bound and slow-rolling by only having the combined, effective field slow-rolling, but not the individual fields.\footnote{
    See \cref{p:dSC_Inflation} about slow-rolling and multi-field inflation.
}

Multi-field hyperinflation\footnote{
    In hyperinflation \cite{brown_hyperinflation_2018,mizuno_primordial_2017}, the field space is a hyperbolic plane with constant curvature $\ll1$ and the inflaton never slow-rolls. Instead, it orbits the bottom of the potential, supported by a centrifugal force.
    }
can be brought into agreement with the \gls{dc} \textit{or} the \gls{dsc}, but not simultaneously with the \gls{dsc} \textit{and} the \gls{dc}, while also reheating the universe at the end of inflation \cite{bjorkmo_hyperinflation_2019}.

\subparagraph{Inflation in Modified Gravity}
Various models of inflation satisfy the \gls{dc} in a $f(R)$ theory\footnote{
    \citet{bajardi_early_2022,benetti_swampland_2019} express the \gls{dc} for $f(R)$ gravity in the Jordan frame as $\abs{\Delta\phi}=\abs{\Delta\log\abs{\partial_Rf}}/2=\abs{\Delta R\left(\partial_R^2f\right)/2\partial_Rf}\approx\order{1}$.
}
setting with either a \gls{gb} term incorporating higher-order curvature terms ($R^2-4R_{\mu\nu}R^{\mu\nu}+R_{\mu\nu\rho\sigma}R^{\mu\nu\rho\sigma}$), or a parity-violating Chern\textendash Simons term ($\varepsilon^{\mu\nu\rho\sigma}R_{\mu\nu}^{\quad\alpha\beta}R_{\rho\sigma\alpha\beta}$) coupling to the inflaton \cite{gitsis_swampland_2023,fronimos_inflationary_2023}. 
\citet{benetti_swampland_2019} find that power-law models with $f(R)\propto R^{1+\mathfrak{c}}$ can satisfy the \gls{dc}.

A window of compatibility for holographic \gls{de} as inflaton is found for a $f(R,T)=R+8\pi G\mathfrak{c}T$ theory with $S=\left(16\pi G\right)^{-1}\int\!\sqrt{-g}\left(f(R,T)+L_m\right)\,\mathrm{d}^4x$: the \gls{dc} requires $\mathfrak{c}\gtrsim300$ \cite{taghavi_holographic_2023}.\footnote{
    The \gls{dsc} is satisfied for $\mathfrak{c}\gtrsim\order{\num{e2}}$ \cite{taghavi_holographic_2023}.
}
Two models of slow-roll inflation in $f(R,T)$ gravity with a non-canonical scalar field (a power-law and an exponential potential) studied by \citet{ossoulian_inflation_2023} were found to satisfy the \gls{dc} as well as the \gls{dsc}, but to violate the \gls{tcc}.

\paragraph{Spacetime}
Examining \gls{rn} \glspl{bh} in \gls{ds} space with a dynamic cosmological constant, \citet{luben_black_2021} derive the result that different spacetimes are separated by an infinite distance if and only if they are separated by infinite proper time: All spacetimes with at least one mass parameter are a finite distance away from each other, and the Minkowski limit of infinite mass (or charge) is infinitely far away from every other spherically-symmetric and static spacetime.\footnote{
    For instance, the size of the smallest possible \gls{bh} is fixed by applying the \gls{dc} to the spacetime solutions \cite{angius_intersecting_2023}.
}

\subsubsection{General Remarks}

A rough but directly applicable expression of the \gls{dc} is
\begin{equation}\label{eq:scdc}
    \frac{\Delta \phi}{M_\textnormal{P}}\lesssim\order{1},
\end{equation}
i.e. the field excursion is small in Planck units over the cosmic history \cite{raveri_swampland_2019}. In the following, we'd like to answer questions that might help to get a refined understanding of the \gls{dc}.

\paragraph{What is $\mathfrak{a}$ and which values can it take?}
$\mathfrak{a}$ can be thought of as a \textit{decay rate}, as it regulates how fast the mass decreases in field space \cite{freigang_cosmic_2023} or, in the context of \gls{cy} decompactifications, how much of the space decompactifies \cite{perlmutter_cft_2021}. In the literature, usually a constant $\mathfrak{a}$ is studied. If we allow a varying $\mathfrak{a}$, e.g. allowing lower values, the mass scale decays more slowly, but we are no longer following geodesics and the constraining power of the \gls{dc} is weakened \cite{freigang_cosmic_2023}. 
A lower bound of 
\begin{equation}\label{eq:DC_Bound}
    \mathfrak{a}\geq1/\sqrt{d-2}
\end{equation}
for the lightest\footnote{
    A heavier tower might be present as well and violate this bound, but not the lightest. The lightest tower is the one with the largest $\mathfrak{a}$, as this tower suffers the strongest exponential suppression.
    }
tower of states is found to be valid in various string theory and M-theory compactifications \cite{etheredge_sharpening_2022,etheredge_running_2023,van_de_heisteeg_bounds_on_species_2023,bedroya_holographic_2022,rudelius_asymptotic_2022,castellano_stringy_2023}, derived from the \gls{ep} \cite{agmon_lectures_2023,etheredge_taxonomy_2024,van_de_heisteeg_bounds_on_species_2023,van_de_heisteeg_bounds_on_field_2023}, and saturated in \glspl{cft}\footnote{
    For 2-dimensional \glspl{cft}, \citet{ooguri_universal_2024} derive $1/\sqrt{c}<\mathfrak{a}<1$, with $c$ the central charge.
    }
\cite{perlmutter_cft_2021} and for fundamental strings \cite{bedroya_non-bps_2023}.
\citet{van_de_heisteeg_asymptotic_2022} derives $\mathfrak{a}\geq1/\sqrt{6}$ using asymptotic Hodge theory. %
\citet{calderon-infante_entropy_2023} find, considering the \gls{ceb} in \gls{ds} space,  
\begin{equation}\label{eq:DC_a}
    \sqrt{\frac{d-1}{d-2}}\geq\mathfrak{a}\geq\frac{1}{\sqrt{\left(d-2\right)\left(d-1\right)}}.
\end{equation}
The lower bound is also motivated by the \gls{tcc} \cite{bedroya_sitter_2020,bedroya_trans-planckian_2020,andriot_web_2020}, and
the upper bound by an observation by \citet{casas_cosmology_2025}: if a tower of states decays too fast, no feasible \gls{flrw} cosmology can be obtained, in particular, all the solutions they find violate the Higuchi bound for \gls{ds} solutions.\footnote{
    \citet{casas_cosmology_2025} find a link to the \gls{dsc} and state that the actual upper bound for the tower decay rate is $\mathfrak{a}\leq\min\{\mathfrak{s}_1/2,\sqrt{\left(d-1\right)/\left(d-2\right)}\}$.
}
Due to the dimension-dependence, the bounds weaken in spaces with many large dimensions. \citet{bonnefoy_swampland_2021} propose to introduce a dependence on the dimensionality of the higher-dimensional space: $\mathfrak{a}=\sqrt{\frac{D-2}{\left(D-d\right)\left(d-2\right)}}$, which agrees with the upper bound in \cref{eq:DC_a} for $D=d+1$, as well as with the bound derived from the \gls{ep}:
The \gls{ep} states that the lightest towers that emerge in the infinite distance limit are either \gls{kk} towers that correspond to a decompactification
or towers of string oscillators that corresponds to a fundamental string becoming tensionless. The two cases limit the range of $\mathfrak{a}$:
\begin{equation}
    \sqrt{\frac{D-2}{\left(D-d\right)\left(d-2\right)}}\geq\mathfrak{a}\geq\frac{1}{\sqrt{d-2}},
\end{equation}
where the lower bound corresponds to the emergent string limit and the upper bound to a decompactification, with the highest value for $D=d+1$ \cite{van_de_heisteeg_bounds_on_field_2023}.

Bounds on $\mathfrak{a}$ are generally derived for massless scalar fields. It hasn't been shown yet if such bounds are exactly saturated for massive fields as well. For strings and membranes charged under a single gauge field, \citet{lanza_swampland_2021} motivate the value $\mathfrak{a}=\left.\frac{1}{2}\frac{q}{\mathcal{T}}\right|_\text{extremal}$, with $q$ the charge and $\mathcal{T}$ the tension of the object, whereas the value to be taken is the value from the extremal object that saturates the \gls{wgc} (see \cref{sec:gravity}).\footnote{
    If there are multiple gauge charges, the bound weakens to $\mathfrak{a}\geq1/\sqrt{2n_h}$, with $n_h$ the no-scale/homogeneity factor from the Kähler potential.
    }
\citet{etheredge_distance_2024} propose the \textit{brane distance conjecture}
\begin{align}
    \mathcal{T}&\sim e^{-\mathfrak{a}\Delta\phi}\\
    \mathfrak{a}&\geq\frac{1}{\sqrt{d-p_\text{max}-1}}
\end{align}
with $p_\text{max}\in\{1,\,2,\,\dots,\,d-2\}$; this bound depends solely on the minimum codimension $d-p_\text{max}$ of the brane that satisfies the conjecture.

\paragraph{Where does the exponential suppression come from?}
An example can be given in the context of 5D Einstein gravity, with the metric
\begin{equation}
    G_{MN}\mathrm{d}X^M\mathrm{d}X^N=g_{\mu\nu}(x)\mathrm{d}x^\mu \mathrm{d}x^\nu+r(x)^2\mathrm{d}y^2
\end{equation}
and an effective action for the modulus field $r(x)$
\begin{equation}
    S=\int\!\sqrt{g}\frac{1}{\left(\lambda r\right)^2}\partial_\mu r\partial^\mu r\,\mathrm{d}^4x,
\end{equation}
with $\lambda$ a numerical constant \cite{blumenhagen_large_2018}. For the canonically normalised field $\rho(x)$ we find
\begin{equation}
    \rho=\frac{1}{\lambda}\log r
\end{equation}
and for $\rho>1/\lambda$ there are infinitely\footnote{
    \citet{bedroya_non-bps_2023} argue that the number of string limits is actually countable.
    See \cref{s:fnomf} for other finiteness arguments.
    }
many \gls{kk} modes for which
\begin{equation}
    m_\text{KK}\sim\frac{n}{r}\sim ne^{-\lambda\rho}
\end{equation}
holds \cite{bonnefoy_infinite_2020}, i.e. their masses are exponentially suppressed and the 4-dimensional \gls{eft} with the radial modulus breaks down \cite{blumenhagen_large_2018}.

In the case of Kähler manifolds, the predicted exponential suppression becomes obvious when treating the case of a complex field with 
\begin{equation}\label{eq:distancecomplex}
    d(\phi,\phi_0)=\int_\gamma\!\sqrt{g_{ij}\frac{\partial\phi^i}{\partial s}\frac{\partial\bar{\phi}^j}{\partial s}}\,\mathrm{d}s,
\end{equation}
with definitions analogue to the real case from \cref{eq:moddistance} \cite{blumenhagen_refined_2018}.
The asymptotic leading behaviour of the Kähler potential
\begin{equation}
    K\sim-\sum_{i=1}^3\left(d_i-d_{i-1}\log\left(\Im{\phi^i}\right)\right)
\end{equation}
leads to the Kähler metric
\begin{equation}
    g_{ij}\sim\text{diag}\left(\frac{d_i-d_{i-1}}{\left(\Im{\phi^i}\right)^2}\right),
\end{equation}
which in turn leads, for fixed $\Re{\phi^i}$, to
\begin{AmSalign}
    d(\phi,\phi_0)&=\\
    &\int_\gamma\!\left[\sum_{i=1}\left(d_i-d_{i-1}\right)\left(\frac{\mathrm{d}\log\left(\Im{\phi^i}\right)}{\mathrm{d}s}\right)^2\right]^{1/2}\!\mathrm{d}s,\nonumber
\end{AmSalign}
i.e. for a sufficiently simple path a logarithmic change in the distance is achieved \cite{grimm_infinite_2019}. Taking into account quantum corrections makes it difficult to solve the integral from \cref{eq:distancecomplex}, but the geodesic equation
\begin{equation}
    \frac{\mathrm{d}^2x^\rho}{\mathrm{d}s^2}+\Gamma^\rho_{\mu\nu}\frac{\mathrm{d}x^\mu}{\mathrm{d}s}\frac{\mathrm{d}x^\nu}{\mathrm{d}s}=0
\end{equation}
can still be solved numerically \cite{erkinger_refined_2019}.
For alternative derivations of the logarithmic behaviour of the field distance, see the work by \citet{grimm_infinite_2018,lanza_eft_2021}.

\citet{heidenreich_emergence_2018} show that the same exponential suppression follows from two assumptions: (i) there is a tower of states that becomes light near a particular point in field space, (ii) the scalar field and gravity have a common strong coupling energy scale.\footnote{
    The common strong coupling scale also leads to the finding that trans-Planckian field ranges yield mass changes of the order of the cutoff scale of the \gls{eft}, indicating that trans-Planckian field ranges cannot be accommodated within an \gls{eft} \cite{heidenreich_emergence_2018}.
}

\paragraph{What does it mean for states to become massless?}
A state becoming massless means that the ratio of the mass scale of this state to the Planck mass $M_\textnormal{P}$ becomes zero \cite{font_swampland_2019}. It means that those states have been integrated out\footnote{
    To \textit{integrate out} a state means that a Lagrangian has been re-expressed such that massive states are derivatives that don't contribute to the integral, i.e. to the action. Assume that you have a Lagrangian with two bosonic fields, $L\sim\frac{1}{2}\left(\partial\phi\right)^2+\frac{1}{2}\left(\partial\varphi\right)^2+m^2(\phi)\varphi^2$. Let us write $\phi=\expval{\phi}+\delta\phi$. For the action, we find now that $S=\int\! L\sim\int\! L_0+\delta\phi^2$, i.e. as long as our \gls{eft} is below energy ranges of $m_0$, we can ignore the $\order{\delta\phi^2}$ terms. A bosonic\,\textemdash\,or rather scalar\,\textemdash\,mass state can be expanded as $\frac{1}{2}m^2(\phi)\varphi^2=\frac{1}{2}m^2(\expval{\phi})\varphi^2+\left(m\partial_\phi m\right)|_{\expval{\phi}}\delta\phi\varphi^2+\dots$, where the second term gives rise to the 1-loop wave function renormalisation \cite{seo_stability_2023}. For fermions, we find $m(\phi)\Bar{\Psi}\Psi=m(\expval{\phi})\Bar{\Psi}\Psi+\partial_\phi m|_{\expval{\phi}}\delta\phi\Bar{\Psi}\Psi$ \cite{seo_stability_2023}. This idea is often generalised in the expansion of the Lagrangian into $L=L_\text{EFT}+\sum_n\frac{\mathcal{O}_{n+4}}{\Lambda_\textnormal{S}^{n-4}}$, where $\mathcal{O}$ are some higher-order operators and $\Lambda_\textnormal{S}$ is the high-energy cutoff of the \gls{eft}\,\textemdash\,the questions arise what that cutoff shall be, what can be part of $L_\text{EFT}$, and what can go into $\mathcal{O}$ \cite{de_biasio_geometric_2023,valenzuela_swampland_2023}.
}
\textit{illegally} \cite{baume_instanton_2020}. They should be part of the \gls{eft}, but have been neglected, assuming they were only subleading, higher-order terms that don't significantly contribute. The appearance of massless states can be thought of as inducing a running gauge coupling, which vanishes in the infinite limit, where the states are massless \cite{lee_tensionless_2018,grimm_infinite_2018,klaewer_super-planckian_2017,harlow_wormholes_2016,heidenreich_emergence_2018,heidenreich_weak_2018}.

In the context of \gls{cft} and holography, \citet{conlon_moduli_2020} make the observation that heavy states appear if one goes in the opposite direction with $\abs{\Delta\phi}\gg M_\textnormal{P}$ and $\Delta\phi<0$. Duality would imply that in this case the winding mode should become light\,\textemdash\,but in the case they study, winding modes only become light in the infinite limit, not in the limit of very large field displacements. This is a conundrum. However, they find hints that this might point to a more foundational notion of the \gls{dc}: a negativity condition for the sign of anomalous dimensions for mixed double trace operators, which works in either direction.

\paragraph{How is \textnormal{distance} defined?}
How far apart two theories are can be examined using different measures: based on the information metric  \cite{stout_infinite_2021,stout_infinite_2022}, the number of microstates of a \gls{bh} within one \gls{eft} \cite{li_black_2021}, the low-velocity distance that measures the physical space-time distance between moving \gls{bps} objects, and the de~Witt distance\,\textemdash\,with or without a volume factor\,\textemdash\,that measures the distance between \glspl{eft} in moduli space \cite{li_alliance_2022}.\footnote{
    For the \gls{dc}, typically the de~Witt distance including a volume factor is used\,\textemdash\,neglecting the volume factor leads to light towers of states already at finite distance \cite{li_alliance_2022}\,\textemdash, with $\mathfrak{l}$ an $\order{1}$ constant, $\lambda$ parametrising the one-dimensional path, and $g$ the Riemann metric on the manifold $\mathcal{M}$ \cite{demulder_navigating_2024}:
\begin{equation*}
    d_\textnormal{dW}=\mathfrak{l}\int_{\lambda_\textnormal{i}}^{\lambda_\textnormal{f}}\sqrt{\frac{1}{\textnormal{Vol}\left(\mathcal{M}\right)}\int_\mathcal{M}\sqrt{g}g^{MN}g^{OP}\frac{\partial g_{MO}}{\partial\lambda}\frac{\partial g_{NP}}{\partial\lambda}}\mathrm{d}\lambda.
\end{equation*}
}
It is even possible that two theories are an infinite distance apart according to one measure, but a finite distance according to another one \cite{li_alliance_2022}. Furthermore, different measures do not necessarily measure the distance in spaces of equal dimension: e.g. the low-velocity distance is measured in a higher-dimensional moduli space. The notion of \textit{distance} is typically based on  a measure of \textit{energy};\footnote{
    Defining a (quasi-)distance can be done using the domain wall tension, which is even valid if there is a potential \cite{mohseni_measuring_2024}.
}
however, \textit{entropy} can be used if not all axioms that define a distance measure are required and a \textit{quasi-distance} is sufficient \cite{gusin_measures_2024}.

A general definition of the distance is given by the integral over $\gamma$, the shortest geodesic connecting $\phi$ and $\phi_0$, with $\mathrm{d}s$ the line element along that geodesic:
\begin{equation}\label{eq:moddistance}
    d(\phi,\phi_0)\defeq\int_\gamma\!\sqrt{g_{ij}\frac{\partial\phi^i}{\partial s}\frac{\partial\phi^j}{\partial s}}\,\mathrm{d}s,
\end{equation}
for real scalar fields $\phi^i$, which are coordinates on $\mathcal{M}$ and whose kinetic terms define a metric $g_{ij}$ on $\mathcal{M}$ \cite{palti_swampland_2019,grimm_infinite_2018,de_biasio_geometric_2023}.

\begin{figure}[htb]
  \begin{center}
    \includegraphics[width=\linewidth]{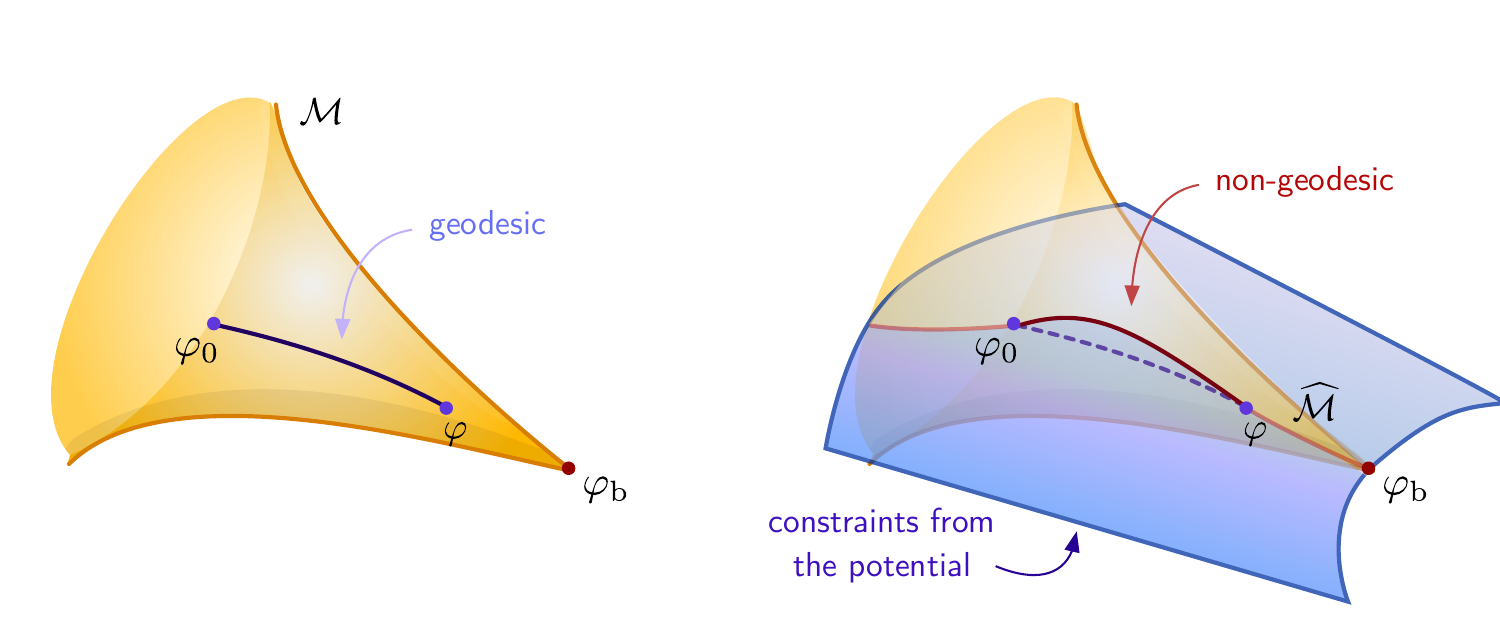}
  \end{center}
  \caption[Moduli Space with Potential]{On the left we see the unrestricted moduli space $\mathcal{M}$ with the geodesic going from $\phi$ to $\phi_0$. On the right, we see the same space, but this time with a potential, which selects a subspace $\widehat{\mathcal{M}}$. The previous geodesic is not contained in this subspace, i.e. the path connecting $\phi$ and $\phi_0$ is longer. Figure taken from \citet{grimm_tameness_2022}.}\label{f:MPot}
\end{figure}

As shown in \cref{f:MPot}, the presence of a potential can change the distance between two points, as the shortest available path between the two points is modified by the potential. This can also be understood as a modification of the distance measure, e.g.
\begin{equation}
    d(\phi,\phi_0)=\int_\gamma\!\sqrt{g_{ij}\frac{\partial\phi^i}{\partial s}\frac{\partial\phi^j}{\partial s}+\left(\frac{1}{\Lambda^3}\frac{\partial\phi^i}{\partial s}\frac{\partial V}{\partial \phi^i}\right)^2}\,\mathrm{d}s,
\end{equation}
with $\Lambda$ some overall scaling,
incorporates the potential into the definition of distance, with a larger gradient of the potential leading to a bigger deviation from the case without potential \cite{schimmrigk_swampland_2019}.
Another proposal to incorporate potentials is
\begin{equation}
    d(\phi,\phi_0)=\int_\gamma\!\sqrt{2E_\text{kin}+2V}\,\mathrm{d}s,
\end{equation}
where $\gamma$ is the on-shell trajectory; this does not necessarily correspond to a geodesic, and $\gamma$ can only connect points that belong to an attractor trajectory \cite{debusschere_distance_2025}.\footnote{
    Attractors are independent of the initial conditions and the total energy \cite{debusschere_distance_2025}.
} If the cosmological attractor is a geodesic, this distance measure closely corresponds to our ordinary understanding of distance \cite{debusschere_distance_2025}.

\citet{basile_domain_2023} treat the case of a discrete set of vacua. To define the distance between discrete vacua, they rely on domain walls to interpolate between the isolated vacua and to get a finite-energy solution:
\begin{align}
    \frac{\Delta(\phi,\phi_0)}{M_\textnormal{P}(\phi_0)^{\frac{d-2}{2}}}&=\label{eq:discretedistance}\\
    &\min\sum_k\frac{d(\phi+\sum_{l<k}\mathcal{E}_l,\phi+\sum_{l\leq k}\mathcal{E}_l)}{M_\textnormal{P}^\frac{d-2}{2}(\phi+\sum_{l\leq k}\mathcal{E}_l)},\nonumber
\end{align}
where they explicitly use the dependence of the Planck mass on the vacuum\,\textemdash\,i.e. the position in moduli space\,\textemdash\,and gradually adjust it along the path, and where $d()$ is meant to be the distance between two points in moduli space as defined in \cref{eq:moddistance}. Since this is a discretised space, we cannot just take a path along the geodesic, which would automatically minimise the path, but we have to explicitly minimise the path under the constraint that $\sum_j\mathcal{E}_j=\phi_0-\phi$, and sum over all steps along a domain wall that leads from $\phi$ to $\phi_0$, by modifying the charges by a basis vector of the lattice $\mathcal{E}$ \cite{basile_domain_2023}.

 \paragraph{What does it mean to investigate the infinite distance limit of moduli space?}
Moduli are massless scalar fields. Each field value corresponds to a vacuum expectation value in string theory, and therefore to the value of a continuous\footnote{
    For example, the 1-sphere $\mathcal{S}^1$, where the maximum distance is $2\pi$ as $\phi\sim\phi+2\pi$, violates the \gls{dc}. Periodic scalars must be part of a larger moduli space, to be allowed in \gls{qg} \cite{palti_swampland_2019}. Moduli spaces are non-compact \cite{brennan_string_2018,hamada_finiteness_2022} if non-trivial.
    }
parameter in an \gls{eft} \cite{castellano_stringy_2023,ooguri_geometry_2007}. The infinite distance limit in field space corresponds to the perturbative regime of the \gls{eft} \cite{castellano_stringy_2023}, where \gls{qg} effects might become observable but are unaccounted for in the theory \cite{calderon-infante_emergence_2024,corvilain_swampland_2019}, i.e. the \gls{eft} breaks down.
This implies that not every thinkable field is also suited to be part of a consistent theory of \gls{qg} \cite{ooguri_geometry_2007}, and that the distance a field can travel is limited, i.e. the energy scale that the \gls{eft} can accommodate is finite \cite{corvilain_swampland_2019,lanza_eft_2021}.

\paragraph{What happens in the infinite field limit?}
The \gls{dc} states that whenever a field develops a large expectation value, decoupling in effective \gls{qft} breaks down at exponentially lower energy scales than expected \cite{palti_swampland_2019}.
To turn the logic around, you can also say that you cannot follow a geodesic in moduli space for a distance bigger than $\sim M_\textnormal{P}$, as after that, your \gls{eft} breaks down \cite{blumenhagen_refined_2018}.
The basic light degrees of freedom are directed by towers of particles with masses $m(\phi_0)$ \cite{ooguri_distance_2019}.
In the infinite field limit, the theory either decompactifies, which causes the appearance of a tower of light \gls{kk} modes or winding states that are unaccounted for in the \gls{eft} \cite{cicoli_string_2023,baldes_forays_2019,lee_emergent_2022,ooguri_geometry_2007,heisenberg_model_2021}, 
or it reduces to a weakly coupled, asymptotically tensionless string theory \cite{lee_emergent_2022,castellano_stringy_2023,xu_tcs_2020,baume_instanton_2020,lee_towers_2021,corvilain_swampland_2019}.\footnote{
    Some infinite distance limits yield a tower of instantons, but the instantons modify the action, such that the limit turns into a finite distance point in moduli space \cite{agmon_lectures_2023,marchesano_instantons_2019}.
    In \cref{sec:emerge} we elaborate on why it is either a \gls{kk} tower or a fundamental string.
} 
The tower becoming light acts as a censorship mechanism:\footnote{
    \citet{remmen_exploration_2021} studies singular isothermal sphere solutions\,\textemdash\,\enquote{a special class of singular solutions for a self-gravitating perfect fluid in general relativity} with an \gls{eos} $w=p/\rho$ and a $1/r^2$ density profile\,\textemdash, and shows that the tower of states predicted by the \gls{dc} shields the naked singularity that would otherwise appear and violate cosmic censorship.
}
it appears at the boundary of field space, where we would encounter an exact global symmetry \cite{castellano_universal_2023,lee_tensionless_2018,hamada_finiteness_2022,rudelius_symmetry-centric_2024}\,\textemdash\,something we cannot have, as is further discussed in \cref{sec:nGSym}.\footnote{
    See also the work by \citet{cordova_generalized_2022} who motivate the \gls{dc} from the no global symmetries conjecture.
} 
The \gls{dc} quantifies \textit{how} the tower of states becomes light, which means it quantifies how to approximate a global symmetry \cite{castellano_universal_2023}.

\paragraph{Is there more than one infinite distance limit?}
Spaces with a richer topology can have different points that are infinitely far away from each other and 
different paths that lead to different towers of light states \cite{grimm_infinite_2019}\,\textemdash\,there are several open questions regarding this aspect: how many towers are there and does each path come with its own individual tower of states, or can paths be deformed or grouped together, such that they lead to the same tower of states, i.e. is there also an infinite number of different towers or are there families of different towers \cite{grimm_tameness_2022}?
In different regions of moduli space, different towers may be the first to emerge new states \cite{lanza_machine_2024}.
Furthermore, in the absence of a clear mass hierarchy, i.e. if there are competing towers, it might no longer be possible to safely model the path of a single field through moduli space as a geodesic; physical paths that solve the \glspl{eom} are required \cite{landete_mass_2018}.
When there are multiple charges, a \textit{convex hull conjecture},\footnote{In analogy to the convex hull conjecture presented in \cref{sec:gravity}.} might be the guiding principle instead \cite{calderon-infante_convex_2021}.
To only consider the lightest tower would be insufficient, as there could be an arbitrary number of towers below the \gls{eft} cutoff above the lightest tower \cite{castellano_iruv_2022}.
However, the Tameness Conjecture (\cref{sec:tame}) implies that it suffices to consider a finite number of different types of states to always be able to define a cutoff for the \gls{eft} \cite{lanza_machine_2024}.

\paragraph{Does the \gls{dc} also apply to scalar fields with a potential?}
A potential acts as a constraint on the available field space in $\mathcal{M}$, as we are now faced with an energy scale dependence, which means that only the subspace $\widehat{\mathcal{M}}\subset\mathcal{M}$ is accessible \cite{grimm_tameness_2022}.
There are three equivalent ways to understand this:
first, the field space gets reduced because we have to integrate out massive scalars at a now lower energy scale;
second, the potential acts as a constraint on the available paths, which means that the shortest available path $d_{\widehat{\mathcal{M}}}(\phi,\phi_0)$ might now be longer than the geodesic through the full moduli space $d_{\mathcal{M}}(\phi,\phi_0)$, i.e. $d_{\mathcal{M}}(\phi,\phi_0)\leq d_{\widehat{\mathcal{M}}}(\phi,\phi_0)$ (e.g. in \cite{hebecker_flat_2017});\footnote{
    This can be understood as the effect of the potential as an external force that drives the motion to non-geodesic paths \cite{demulder_navigating_2024,calderon-infante_convex_2021}.
}
and third, graphically, as in \cref{f:MPot} \cite{grimm_tameness_2022}.
A longer path length means a stronger exponential suppression of the mass scale, i.e. a different tower, which decays faster, might be required in this case, and the \gls{dc} might no longer hold. This can be viewed as a constraint on the scalar potential: only potentials are allowed that provide paths that are compatible with the \gls{dc} \cite{grimm_tameness_2022}. \citet{calderon-infante_convex_2021} show that paths that asymptotically approach a geodesic in the potential-less (parent) moduli space are in agreement with the \gls{dc}. %

It is important to note that the \gls{dc} is formulated regarding the distance in \textit{moduli spaces}, i.e. it is a notion about \textit{theories}. If it's also directly applicable to the field space of a physical field is less clear. 
In the literature, the \gls{dc} is often assumed to hold in the field space as well, even if the fields have a potential. 
Some argue that the mass scale suppression may deviate from its exponential form in this case \cite{freigang_cosmic_2023,grimm_tameness_2022,hebecker_flat_2017}.
Others argue that also massive scalar fields with a potential show an exponential suppression in the asymptotic limit \cite{rudelius_asymptotic_2022,hebecker_asymptotic_2019,ooguri_geometry_2007,dine_is_1985,klaewer_super-planckian_2017,obied_sitter_2018}, however, the value of $\mathfrak{a}$ is less constraint.
A possible counterexample to the applicability of the \gls{dc} in the bulk of the field space is discussed by \citet{hebecker_large_2019}: studying axions, they find towers of states with approximately constant masses\,\textemdash\,and not, as required by the \gls{dc}, exponentially suppressed masses\,\textemdash\,under the assumption that the field distance is parametrically large, but not infinite. Their idea of axion (mis)alignment is also discussed by \citet{palti_swampland_2019}. Contrarily, \citet{reig_stochastic_2021} states that maximally misaligned quintessence always satisfies the \gls{dc}, as $Nf^2\ll M_\textnormal{P}^2$, with $N$ the number of axions and $f$ the decay constant, holds always.
The jury on the applicability of the \gls{dc} to scalar fields with potentials is still out. What can be said is that if we deviate from geodesic paths or allow the scalar field to have a potential, the constraining power of the \gls{dc} is weakened.

\subsubsection{Evidence}

The \gls{dc} is proposed by \citet{ooguri_geometry_2007},
and studied for
nilmanifolds \cite{andriot_laplacian_2018},
the \gls{kklt} scenario \cite{blumenhagen_swampland_2019},
the complex structure moduli space of elliptic K3 surfaces \cite{lee_elliptic_2022},
the complex structure moduli space of \gls{cy} manifolds \cite{grimm_infinite_2019,grimm_infinite_2018,alvarez-garcia_analytic_2022},
the one-parameter family of quintic \gls{cy} manifolds \cite{ashmore_moduli-dependent_2021},
exotic one-parameter \gls{cy} threefolds with pseudo-hybrid points \cite{erkinger_refined_2019}, 
Kähler moduli spaces on \gls{cy} threefolds \cite{corvilain_swampland_2019,lee_emergent_2022},
non-geometric phases of Kähler moduli spaces of \gls{cy} manifolds \cite{blumenhagen_refined_2018},
heterotic Kähler gravity theories \cite{ibarra_heterotic_2025},
type II \gls{cy} compactifications \cite{marchesano_instantons_2019},
type II compactifications on 1-parameter \gls{cy} threefolds \cite{joshi_swampland_2019},
type II flux compactifications \cite{junghans_weakly_2019,valenzuela_backreaction-swampland_2017},
type II $\mathcal{N}=1$ \gls{cy} orientifolds \cite{font_swampland_2019},
type IIA orientifolds with fluxes \cite{shiu_ads_2023},
type IIA supergravity on G$_2$ orientifolds \cite{farakos_scale-separated_2023,farakos_off_2023},
the vector multiplet moduli space of 4D type IIA $\mathcal{N}=2$ compactifications \cite{klaewer_modular_2021},
\enquote{massive IIA theory, M-theory on CY threefolds, and 10d non-supersymmetric strings} \cite{buratti_dynamical_2021_DC},
5d supergravities arising from M-theory compactified on \gls{cy} threefolds \cite{heidenreich_infinite_2021},
M-theory compactifications on \gls{cy} threefolds that allow an infinite sequence of flops\footnote{\enquote{A flop is a birational morphism [that] relates two manifolds by contracting curves on either manifold to arrive at the same (singular) manifold at a common boundary of their Kähler cones.} \cite{brodie_swampland_2021}\label{f:flop}} \cite{brodie_swampland_2021},
M-theory compactifications on G$_2$-manifolds \cite{langlais_analysis_2023,xu_tcs_2020},
heterotic M-theory \cite{deffayet_stable_2024},
type IIB $\mathcal{N}=1$ orientifolds with O3/O7 planes \cite{enriquez-rojo_swampland_2020},
type IIB $\mathcal{N}=2$ \gls{cy} compactifications to 4d \cite{baume_instanton_2020},
type IIB $\mathcal{N}=2$ on $S^5$ and its orbifolds \cite{baume_tackling_2021},
Type IIB isotropic toroidal compactification with non-geometric fluxes \cite{bizet_leaving_2019},
six-dimensional type IIB and F-theories \cite{lee_emergent_2019},
F-theory \cite{lee_towers_2021,lee_physics_2022,lee_tensionless_2018,grimm_asymptotic_2020,chen_massive_2024},
4d $\mathcal{N}=1$ \glspl{eft} \cite{lanza_eft_2021,lanza_large_2022},
4d $\mathcal{N}=2$ \glspl{eft} \cite{cecotti_special_2020},
9d $\mathcal{N}=1$ string theories with running decompactifications \cite{etheredge_running_2023},
high-dimensional spacetimes \cite{bonnefoy_swampland_2021},
\gls{bps} states in six-dimensional supergravities \cite{hayashi_spectra_2023},
string theories with broken supersymmetry \cite{basile_emergent_2022,basile_sitter_2020},
\glspl{cft} \cite{perlmutter_cft_2021,baume_higher-spin_2023,basile_infinite_2023},
hyperbolic spaces \cite{freigang_cosmic_2023},
T-dualities in curved spaces and non-trivial fluxes supporting the background \cite{demulder_topology_2023},
in the light of holography \cite{bedroya_holographic_2022,conlon_moduli_2020,geng_distance_2020},
string dualities \cite{bedroya_dualities_2023},
non-associativity and non-commutativity \cite{bubuianu_nonassociative_2023},
non-invertible symmetries \cite{heckman_fate_2024},
and \glspl{bh} \cite{kehagias_note_2020,angius_small_2023,delgado_black_2023,cribiori_large_2022,hamada_finiteness_2022,luben_black_2021,bonnefoy_infinite_2020}.
The similarities between the \gls{dc} and mathematical flow equations in \gls{gr}, in particular the Ricci flow, were highlighted in \cite{kehagias_swampland_2020,de_biasio_geometric_2023,de_biasio_-shell_2022,demulder_navigating_2024,de_biasio_geometric_2022}.\footnote{
    Working with Ricci flows has the advantage that the problems induced by the presence of potentials are entirely circumvented \cite{demulder_navigating_2024}. A disadvantage is that not all infinite distance points are accessible by geometric flows \cite{demulder_navigating_2024}.
}
A bottom-up perspective by applying the \gls{ceb} on an \gls{eft} that also yields insights into the \gls{ssc} and relations to the \gls{dsc} is verified in M-theory toroidal compactifications \cite{calderon-infante_entropy_2023}.
Further support comes from the finding that the volume of moduli space is finite\footnote{
See \cref{s:fnomf} for various discussion around the topic of finiteness.
}
\cite{douglas_finiteness_2005,cicoli_geometrical_2018}, which limits (but does not exclude) the possibilities of infinite distances within moduli space, and from the finding that trans-Planckian field ranges could cause gravitational backreactions strong enough to make space collapse into a \gls{bh} \cite{nicolis_super-planckian_2008}.\footnote{
    There are claims that trans-Planckian field excursions are not problematic in asymptotic safety \cite{eichhorn_constraining_2021}.
}
Furthermore, the relations to other swampland conjectures are highlighted in \cref{rel:TC_DC,rel:dSC_DC,rel:DC_Cobordism,rel:TCC_DC,rel:WGC_DC,rel:fnomfC_DC,rel:EP_SDC}.

\subsection{Emergence Proposal}\label{sec:emerge}%

The strong form of the \gls{ep} states the following \cite{blumenhagen_demystifying_2024,castellano_emergence_2023,gendler_merging_2021,castellano_towers_2023}:
In \gls{qg}, light particles in the perturbative regime have no kinetic terms in the \gls{uv}. In the \gls{ir}, kinetic terms emerge due to loop corrections involving the sum over a tower of massless states, i.e. kinetic terms emerge from integrating out massive states up to the species scale (\cref{s:ssc}, \cref{eq:speciess}).

A weaker form states that for each singularity at an infinite distance in moduli space of an \gls{eft}, there is an associated infinite tower of states that becomes light and that induces quantum corrections to the metrics, which match the tree-level\footnote{
    Feynman diagrams at lower order look like a tree, whereas higher-order diagrams include loops (and additional factors of $\hbar$).
}
singular behaviour \cite{blumenhagen_demystifying_2024,castellano_emergence_2023}.

A more hands-on definition of the \gls{ep} is the prediction that every infinite-distance limit in moduli space corresponds either to a decompactification, and therefore to a tower of \gls{kk} modes; or to a fundamental string becoming tensionless, and therefore to a tower of string oscillator modes \cite{rudelius_gopakumar-vafa_2024,kaufmann_asymptotics_2024}.

\subsubsection{Implications for Cosmology}
\paragraph{Cosmological Constant}
\citet{castellano_towers_2023} use the \gls{ep} to motivate the idea that the cosmological constant $\Lambda_\text{cc}$ is a fundamental scale, and that all other energy scales can be expressed as
\begin{equation}\label{eq:EP_Scales}
    m\sim \Lambda_\text{cc}^{1/n} M_\textnormal{P}^{1-1/n},
\end{equation}
which corresponds to the \gls{adsdc} (\cref{eq:AdSDC}), is compatible with the \gls{dc}, the no non-supersymmetric theories conjecture (\cref{sec:nononSUSY}), and the (magnetic) \gls{wgc} (\cref{eq:mwgc}), and leads to the relations between the scales presented in \cref{tab:emergence_scales}.

\begin{table*}[htb]
    \centering
$
\begin{tblr}{XXXX
}   \toprule
    \text{Scale} & \text{Symbol} & \text{Relation} & \text{Energy} \\\midrule
    \text{Species Scale} & \Lambda_\textnormal{S} & \sim10\Lambda_\text{cc}^{1/6}M_\textnormal{P}^{5/6} & \sim\SI{e14}{\giga\electronvolt}\\
    \text{Extra Dimension} & m_\textnormal{t} & \sim10^3\Lambda_\text{cc}^{1/2}M_\textnormal{P}^{1/2} & \sim\SI{e6}{\giga\electronvolt}\\
    \text{Gravitino} & m_\textnormal{3/2} & \lesssim10^3\Lambda_\text{cc}^{1/2}M_\textnormal{P}^{1/2} & \lesssim\SI{e6}{\giga\electronvolt}\\
    \text{Electroweak} & m_\text{EW} & \sim10^{-1}\Lambda_\text{cc}^{1/2}M_\textnormal{P}^{1/2} & \sim\SI{e2}{\giga\electronvolt}\\
    \text{Dirac $\nu$} & m_{\nu_1} & \lesssim \Lambda_\text{cc} & \lesssim\SI{e-12}{\giga\electronvolt}\\
    \bottomrule
\end{tblr}
$
    \caption[Scales of Emergence]{The fundamental scales can all be expressed in terms of the cosmological constant $\Lambda_\text{cc}$ and the Planck mass $M_\textnormal{P}$, using the relation $m\sim \Lambda_\text{cc}^{1/n} M_\textnormal{P}^{1-1/n}$, with different values for the integer $n$ \cite{castellano_towers_2023}.}
    \label{tab:emergence_scales}
\end{table*}

\paragraph{Neutrinos}
\Cref{eq:EP_Scales} indicates that the lightest neutrino $\nu_1$ is a Dirac neutrino \cite{castellano_towers_2023}.\footnote{
    While a Majorana neutrino gets its mass through the seesaw mechanism \cite{gell-mann_complex_2013,yanagida_horizontal_1980,dienes_light_1999}, the towers of states predicted by the \gls{ep} lead to suppression of Yukawa couplings that yield light Dirac neutrinos, or put differently: \enquote{the right-handed components receive large wave-function renormalisation effects through their coupling to such an infinite tower of states} \cite{castellano_towers_2023}.
    }
The small mass is explained by an anomalous extra dimension that opens up due to a tower of \gls{sm} singlet states of mass $m_\textnormal{t}\simeq Y_{\nu_3}M_\textnormal{P}\simeq\SI{7e5}{\giga\electronvolt}$ that relates the cosmological constant $\Lambda_\text{cc}$ to the \gls{ew} scale $m_\text{EW}$ as $m_\textnormal{t}m_\text{EW}\lesssim10^2\Lambda_\text{cc}M_\textnormal{P}$ \cite{castellano_towers_2023}.\footnote{
    In early work, \citet{arkani-hamed_predictive_2005} thought that the discovery of such a relation between the three different energy scales, relating the cosmological constant, \gls{ew} scale, and the Planck scale, would be based on an anthropic principle, not on underlying dynamics.
}
Since the \gls{ew} scale and the cosmological constant are constrained by observations, and the Planck scale is a definition, the neutrino mass has to be light to satisfy the (swampland) constraint(s).
Furthermore, to be consistent with observations, the tower mass scale $m_\textnormal{t}\lesssim\SI{e6}{\giga\electronvolt}$, i.e. at \SI{e6}{\giga\electronvolt} a \nth{5} dimension opens up that is only felt by the right-handed neutrinos, since no other particle/field couples to that tower \cite{castellano_towers_2023}.\footnote{
    It is important to note that in this extra dimension scenario, the fundamental scale of gravity is still many orders of magnitude above the \gls{ew} scale \cite{castellano_towers_2023}, unlike e.g. in refs. \cite{arkani-hamed_hierarchy_1998,dvali_phenomenology_2009}.
    A similar scenario is also invoked by \citet{sabir_fermion_2025}: neutrino mixing is explained by a higher-dimensional 4-point operator.
    }

\paragraph{Scalar field} potentials emerge from integrating out towers of states \cite{hamada_finiteness_2022}. A typical scale for 1-loop Casimir potentials that arise from integrating out particle scales is $m^d$ \cite{hamada_finiteness_2022,gonzalo_ads_2021,rudelius_dimensional_2021}, which is compatible with the Higuchi bound\footnote{
    See also our comments on inflation in \cref{p:DC_Inflation}.
    }
that demands that the exponent is larger than 2.
For scalar potentials with a Lagrangian of the form
\begin{equation}
    L=\frac{1}{2}\dot{\phi}-V(\phi),
\end{equation}
and $\lim_{\phi\rightarrow\infty}V(\phi)=0$,\footnote{
    Note that for asymptotically divergent potentials, infinite distance limits of the \gls{eft} are obstructed by the potential and no issue arises \cite{hamada_finiteness_2022}.
    Furthermore, the potential cannot asymptote a finite positive value, as this is restricted by the Higuchi bound \cite{higuchi_forbidden_1987,hamada_finiteness_2022}.
    }
we find that the cosmological constant $\Lambda_\text{cc}$ vanishes in the infinite distance limit where the tower of states becomes light:
\begin{equation}
    \Lambda_\text{cc}=H^2\sim m_\textnormal{t}^\mathfrak{p},
\end{equation}
with $\mathfrak{p}>2$ as demanded by the Higuchi bound \cite{hamada_finiteness_2022}.

\paragraph{\gls{sm} of Particle Physics}
The kinetic terms of all \gls{sm} particles arise from loop corrections and the coupling of \gls{kk} or string towers to the \gls{sm} particles \cite{castellano_towers_2023}. The details of these kinetic terms depend on the interactions between the particles and the specific towers or combinations of towers the particle in question couples to \cite{castellano_towers_2023}.

\citet{kawamura_flavor_2023} shows that the quark flavor structure can be obtained by applying the \gls{ep}: the structure emerges from radiative corrections to towers of massive states.

\subsubsection{General Remarks}
Emergence can be understood as a form of \gls{ir} duality: light fields describe the same physics as towers of states, but in different variables \cite{hattab_particle_2023,palti_swampland_2019,harlow_wormholes_2016,grimm_infinite_2018,hattab_emergence_2024}.\footnote{
    For example, \citet{gendler_merging_2021} write that \enquote{[t]he emergence proposal \textelp{} suggests that the small gauge coupling and the infinite field distance are generated from quantum corrections of integrating out the towers of states becoming light. Hence, the relations between (gauge and scalar) charges and masses that the Swampland conjectures predict would simply be a consequence of the renormalization group flow equations.}
    }
While the towers of states predicted by the \gls{ep} are infinite (in accordance with the \gls{dc}), the states integrated out in an \gls{eft} are below the species scale cutoff (\cref{s:ssc}) \cite{blumenhagen_demystifying_2024}.

\citet{castellano_emergence_2023} present a deeper connection between emergence and other swampland conjectures:
If the corresponding (dual) flux in the vacuum is turned on, kinetic terms that emerge for $\left(d-1\right)$-forms create scalar potentials that are in agreement with the \gls{adsdc}, the \gls{dsc}, the \gls{dc}, as well as with the magnetic \gls{wgc} (\cref{eq:mwgc}).
This once more shows the mutual compatibility between the different swampland conjectures, even though they are independently motivated.

There is evidence that the precise formulation of the \gls{ep} in \gls{qg} is as yet unclear \cite{blumenhagen_demystifying_2024,blumenhagen_emergence_2023}: \gls{qft} computations lead to different results than \gls{bh} physics, ranging from different numerical factors and higher-order terms to additional logarithmic enhancement factors for emergent string limits.
Despite the open questions regarding a quantitative understanding of the \gls{ep}, we'd like to address some questions to gain a better qualitative understanding of the \gls{ep} in the following.

\paragraph{What is actually emerging?}
Dynamics. Kinetic terms emerge in the \gls{ir} by integrating out towers of states from the \gls{uv} species scale $\Lambda_\textnormal{S}$ \cite{blumenhagen_reflections_2024,palti_swampland_2019}.
A gauge coupling $g$ emerges at 1-loop as
\begin{equation}
    \frac{1}{g^2}\sim\sum_{n=1}^{N_\textnormal{S}}q_n^2\log\left(\frac{m_n}{\Lambda_\textnormal{S}}\right)
\end{equation}
from integrating out a tower of charged states with $m_n=n\Delta m$ and $q_n=nq$, such that $\Lambda_\textnormal{S}=N_\textnormal{S}\Delta m$ \cite{blumenhagen_demystifying_2024}.
\citet{blumenhagen_reflections_2024} present another example, referring to the work of \citet{heidenreich_emergence_2018,heidenreich_weak_2018,grimm_infinite_2018,castellano_emergence_2023}:
Given an effective $d$-dimensional action of
\begin{align}
    S=&M_\textnormal{P}^{d-2}\int\!\frac{1}{2}G_{\phi\phi}\partial_\mu\phi\partial^\mu\phi\nonumber\\
    &\qquad+\sum_n\frac{1}{2}\partial_\mu h\partial^\mu h+\frac{1}{2}m_n^2(\phi)h_n^2\,\mathrm{d}^dx
\end{align}
with
$G_{\phi\phi}$ the field space metric,
$\phi$ a light modulus,
and $h_n$ a tower of massive \gls{kk} states with mass $m_n=n\Delta m(\phi)$ that induces a 1-loop correction to the kinetic term for $\phi$ through the 3-point couplings $y=\left[m_n\partial m_n\right]h_n^2\phi$,
the 1-loop corrections 
\begin{equation}
    G_{\phi\phi}^\text{1-loop}\simeq\frac{\Lambda_\textnormal{S}^{d-1}}{M_\textnormal{P}^{d-2}}\frac{\left(\partial_\phi\Delta m\right)^2}{\left(\Delta m\right)^3}+\dots,
\end{equation}
are obtained by
integrating out these modes up the species scale $\Lambda_\textnormal{S}$.
For a \gls{kk} tower, we find $G_{rr}^\text{1-loop}\simeq 1/r^2$, which shows the same functional dependence on the modulus $r$ as the tree-level metric from dimensional reduction of the Einstein\textendash Hilbert action \cite{blumenhagen_reflections_2024}.
In the \gls{uv} regime, fields are not dynamical, and the 1-loop renormalisation group contributions to the moduli field metric, which arise from integrating out a tower of states that are lighter than the natural cutoff of the effective theory, are proportional to the tree-level metric and dominate over the classical contributions to the kinetic terms, i.e. the kinetic terms vanish above $\Lambda_\textnormal{S}$ \cite{blumenhagen_quantum_2020,palti_swampland_2019}.

At the core of the \gls{ep} is the statement that the lightest tower at an infinite distance in moduli space is either a \gls{kk} tower, stemming from a decompactification, or a tower of string oscillator modes, stemming from a fundamental string becoming tensionless \cite{bedroya_dualities_2023,lee_emergent_2019,lee_emergent_2022,lee_towers_2021,lee_physics_2022,lee_tensionless_2018,blumenhagen_demystifying_2024,kaufmann_asymptotics_2024}.
The implications for an \gls{eft} are profoundly different in the two scenarios: in the case of the \gls{kk} tower, the \gls{qg} cutoff is at an energy scale $g^{1/3}M_\textnormal{P}$, whereas in the case of the fundamental string, the cutoff is already at $gM_\textnormal{P}$ \cite{draper_snowmass_2022}.
In the case of decompactification, multiple string-like objects can emerge, which are emergent (wrapped) branes that are becoming asymptotically tensionless \cite{blumenhagen_demystifying_2024}.
In the case of the fundamental string, there is exactly one string-like object: the fundamental string \cite{blumenhagen_demystifying_2024}.
A fundamental string has a graviton as a string state, and the scattering amplitude of the string state is given by string perturbation theory \cite{bedroya_dualities_2023}.
Since there is always a unique critical string that becomes tensionless at the fastest rate, there is at most one unique massless graviton as part of the spectrum of the light states, never multiple massless gravitons \cite{castellano_quantum_2024}.

To give one concrete example of the tensionless fundamental string and the decompactification limit, we refer to \citet{bedroya_tale_2024}: In 10d type IIA supergravity, new \gls{qg} effects appear when the first-order perturbative expansion breaks down\,\textemdash\,the 1-loop four-graviton scattering amplitude is at the tree-level. This happens at the species scale (\cref{s:ssc})
\begin{AmSequation}\label{eq:10dSpeciesScale}
    \Lambda_\textnormal{S}=\frac{1}{\left(2\pi\right)^{1/8}}\left(\frac{3\zeta(3)}{\pi^2}e^{3\phi/\sqrt{2}}+e^{-\phi/\sqrt{2}}\right)^{-1/6}M_{\textnormal{P};10},
\end{AmSequation}
with the two infinite distance limits
\begin{empheq}[left ={\Lambda_\textnormal{S}\sim\empheqlbrace}]{alignat = 3}
    e^{\phi/6\sqrt{2}}M_{\textnormal{P};10}  &\propto M_{\textnormal{P};11}\qquad   &&\phi\rightarrow-\infty\\
    e^{-\phi/2\sqrt{2}}M_{\textnormal{P};10} &\propto M_\textnormal{s}\qquad        &&\phi\rightarrow\infty,
\end{empheq}
i.e. the $\phi\rightarrow-\infty$ limit corresponds to a decompactification\footnote{
    With $-\phi$ being the canonically normalised volume modulus:\\$\Lambda_\textnormal{S}\sim M_{\textnormal{P};D}\sim\exp\left(-\phi\sqrt{\left(D-d\right)/\left[\left(D-2\right)\left(d-2\right)\right]}\right)$ \cite{bedroya_tale_2024}.
    },
and the $\phi\rightarrow\infty$ limit corresponds to a weakly coupled string\footnote{
    With $-\phi$ being the canonically normalised dilaton:\\$\Lambda_\textnormal{S}\sim m_\text{s}\sim\exp\left(-\phi/\sqrt{d-2}\right)$ \cite{bedroya_tale_2024}.
    }.
As an aside, we notice that the species scale is always sub-Planckian.

\paragraph{Where should the kinetic terms come from in the \gls{ir}?}
Assume that there is an infinite-distance limit with a tower of states becoming light.
Furthermore, assume that there is a fermion that couples to this tower.
The wave function $\Psi$ of the fermion is enhanced by factors of $\Psi\sim\left(\Lambda_\text{QG}/m_\text{t}\right)^\mathfrak{l}$ with $\Lambda_\text{QG}$ the \gls{uv} cutoff scale, $m_\text{t}$ the mass scale of the tower, and $\mathfrak{l}$ of $\order{1}$,
and the canonical kinetic terms respectively the Yukawa couplings $Y^{ijk}$\,\textemdash\,which we assume to be of $\order{1}$ in the \gls{uv}\,\textemdash\,get suppressed $Y^{ijk}\rightarrow\Psi_i\Psi_j\Psi_kY^{ijk}$ and generate hierarchies of fermion masses
\cite{castellano_towers_2023,castellano_emergence_2023}.\footnote{The limit $Y\rightarrow0$ corresponds to an infinite distance limit with vanishing coupling and a tower of \textit{gonions} \cite{aldazabal_intersecting_2001}\,\textemdash\,gonions are charged chiral states in massive vector-like representations of the D6-brane gauge group, where every state has the same bifundamental charge \cite{casas_yukawa_2024}.\label{foo:gonions}}

\paragraph{Could there be other towers than \gls{kk} modes and string oscillators?}
In a bottom up approach based on considerations regarding the smallest possible \gls{bh} radius, \citet{bedroya_density_2024} find that in a weak-coupling limit of Einstein gravity in asymptotically flat space, the lightest tower is either a \gls{kk} tower, or a tower with exponential density of states, $\rho(E)\sim e^{E/\Lambda_\textnormal{S}}$, with $E$ the energy\,\textemdash\,a behaviour known from the fundamental string, but not exclusive to the fundamental string. This acts as evidence in favour of the \gls{ep}, but not as conclusive proof, as also other towers of states could show the exponential density.
Other towers discussed in string theory involve \cite{etheredge_distance_2024}:
\gls{kk} modes,
\gls{kk} monopoles,
more general particle towers,\footnote{
    Particle towers can be generated by oscillation modes of unwrapped branes \cite{etheredge_distance_2024}.
    }
winding modes of branes that fully wrap a torus,
oscillation modes of branes that partially wrap a torus,
wrapped branes carrying compact momentum,
or asymptotically tensionless membranes \cite{castellano_quantum_2024}.
There is some evidence that the lightest object must not correspond to a higher-dimensional $p$-brane with $p\geq2$, as this would spoil the consistency of the conjecture under dimensional reduction \cite{castellano_quantum_2024,alvarez-garcia_membrane_2022}, and
\citet{alvarez-garcia_membrane_2022} discuss censorship against emergent membrane limits.

\paragraph{What is the difference between the strong and the weak form of the \gls{ep}?}
The definitions seem to have only a slight difference, namely whether the metric evaluated at the \gls{qg} cutoff vanishes entirely (strong proposal) or just becomes subleading (weak proposal) \cite{castellano_quantum_2024}. Conceptually, however, the implication of this slight difference is tremendous: the statement that graviton dynamics completely emerge from the \gls{uv} fundamental degrees of freedom is accurate regarding the strong proposal but inaccurate regarding the weak proposal \cite{castellano_quantum_2024}.

The strong form of the proposal indicates that there could be a topological fundamental theory, where particles do not propagate, as there are no kinetic terms \cite{castellano_emergence_2023,grimm_infinite_2018,palti_swampland_2019,harlow_wormholes_2016,agrawal_topological_2020}.

\paragraph{A Tale of \sout{Three} \sout{Four} \sout{Five} \sout{Six} Seven Scales}\label{p:tale-of-scales}
The vast corpus of literature brings up different energy scales, which are not always obviously distinct. The main idea is always the same: to have a \gls{qg} cutoff below or at the Planck mass $\Lambda^{d-2}N(\Lambda)\lesssim\order{1}M_\textnormal{P}^{d-2}$ \cite{caron-huot_gravity_2024}. But the details are often obfuscated.
There is a trinity of four scales with five meanings, six flavours, and seven names.
We'd like to present an attempt to disentangle the eight scales, and use them with clearly defined meanings:
\begin{description}
    \item[Planck Scale $M_{\text{P};d}$] $M_{\textnormal{P};d}^{d-2}=\hbar^{d-3}c^{5-d}/8\pi G_\text{N}$, which is equal to $M_{\textnormal{P};D}^{D-2}\left(2\pi r_{D-d}\right)^{D-d}$ in higher dimensional theories with $r_{D-d}$ the (dimensionless) radius (in Planck units) of an extradimensional torus \cite{bedroya_tale_2024,castellano_emergence_2023,castellano_quantum_2024}.
    \item[Non-Locality Scale $\Lambda_\text{\sout{QFT}}$ \cite{caron-huot_gravity_2024}] At length scales shorter than $1/\Lambda_\text{\sout{QFT}}$, fields become strictly non-local. This scale coincides with the scale at which higher-spin states can no longer be neglected in graviton scattering processes.
        In decompactification limits, $\Lambda_\text{\sout{QFT}}\sim M_{\textnormal{P};D}$, whereas in the weakly coupled fundamental string limit, the scale corresponds to the mass of spin $J\geq4$ string states that strongly couple to two gravitons, $M_\textnormal{s}$.
    \item[Emergence Scale $\Lambda_\text{QG}$] In the emergent string limit, string gravity becomes \textit{strong}.
        The \gls{ep} conjectures that this happens either at the string scale or at the \gls{kk} scale, whichever scale is lower in a given setting.
        The majority of texts would set the emergence scale equal to the species scale. However, we would advocate the notion that the species scale is often used as a more general umbrella term, as we will outline below.
        Furthermore, \citet{hattab_particle_2023} note that emergence, as an \gls{ir} phenomenon, cannot come from integrating \textit{out} states that are fundamental in the frame of the emergent field, e.g. integrating out \gls{kk} modes of a fundamental field cannot lead to the emergence of \gls{kk} photons, and integrating out string oscillator modes cannot lead to an emergent string limit. Therefore, string oscillator modes cannot be the lightest tower of states at the string scale by definition.
        At the string scale, string theory is weakly coupled.
        At the emergence scale, gravitational interactions become strong, which could be the ones described by string gravity.
        As a technical definition for the emergence scale, we would propose to use a definition that is often used as one of the\,\textemdash\,non-identical\,\textemdash\,definitions of the species scale: The scale of emergence is the scale at which the 1-loop four-graviton scattering amplitude is at the tree-level \cite{dvali_black-bound_2008,basile_minimal_2024,blumenhagen_demystifying_2024,bedroya_tale_2024,cribiori_species_2023,cribiori_note_2023,caron-huot_gravity_2024}.
        As a motivation for this definition, we see the notion that the 1-loop propagator of the graviton gives a \enquote{worst-case scenario} \cite{aoufia_species_2024} of the scale at which perturbation theory fails.
    \item[Species Scale $\Lambda_\text{S}$] According to a rather loose definition, the species scale $\Lambda_\textnormal{S}$ is the scale after which new gravitational dynamics appear. %
        There is an overlap with the emergence scale, but that the two are not always equal \cite{hattab_particle_2023} can for example be seen in figure 2 by \citet{bedroya_tale_2024}.
        Furthermore, as we will see in \cref{s:ssc}, the cutoff scale of a lower-dimensional theory can be represented by the \gls{bh} horizon radius of the smallest \gls{bh} that can be described within this \gls{eft}. This cutoff does not necessarily correspond to the cutoff given by higher-curvature terms of the higher-dimensional theory \cite{bedroya_density_2024}.
        We propose to consider the species scale as a wider umbrella term that corresponds, in a given setting, to the lowest of the following scales in that setting.
    \item[String Scale $M_\text{s}$] A tower of string oscillator modes emerges at $M_\textnormal{s}$. At this scale, a fundamental string becomes tensionless \cite{etheredge_taxonomy_2024} respectively \enquote{an infinite tower of high-spin states on the graviton Regge trajectory become[s] light} \cite{caron-huot_gravity_2024}.
    \item[\gls{kk} Scale $M_{\text{P};D}$] The mass scale of a tower of \gls{kk} states. At this scale, $\mathfrak{p}$ dimensions decompactify \cite{etheredge_taxonomy_2024}, and we find $N(\Lambda)\sim\Lambda^\mathfrak{p}\mathcal{V}_\mathfrak{p}$ \cite{caron-huot_gravity_2024}, i.e. in the infinite volume limit ($\mathcal{V}\rightarrow\infty$) we get an infinite number of states unless the validity of the theory vanishes ($\Lambda\rightarrow0$). The number of light states corresponds to the number of \gls{kk} modes, and the cutoff scale corresponds parametrically with the higher-dimensional Planck mass $\Lambda^{d+\mathfrak{p}-2}\sim M_\textnormal{P}^{d-2}/\mathcal{V}_\mathfrak{p}\defeq M_{\textnormal{P};d+\mathfrak{p}}^{d+\mathfrak{p}-2}$ \cite{caron-huot_gravity_2024,etheredge_taxonomy_2024}.
    \item[\gls{bh} Scale $\Lambda_\text{BH}$] \citet{bedroya_tale_2024} propose a scale, which they coin the \textit{\gls{bh} scale}, at which \glspl{bh} encounter a Gregory\textendash Laflamme instability \cite{gregory_black_1993,reall_classical_2001,gregory_instability_1994} and undergo a phase transition into a state of lower free energy that is not visible in the \gls{eft}.\footnote{
        A weaker cutoff related to \glspl{bh} is presented by \citet{brustein_shortest_2002}: the quasi-classicality scale for \glspl{bh} of $M_{\textnormal{P};d=4}/N^{1/4}$, which coincides with a 3-level unitarity bound on the production of species in one-graviton exchange amplitudes \cite{dvali_phenomenology_2009}, whereas the scale at which gravitational interaction of elementary particles are strongly coupled is $M_{\textnormal{P};D}/N^{1/\left(D-2\right)}$ \cite{dvali_strong_2009}.
        }
        They find two cases for their \gls{bh} scale:
        In the emergent string limit, the emergence scale corresponds to the Hagedorn temperature, while the \gls{bh} description breaks down at a lower scale, the Horowitz\textendash Polchinski \cite{horowitz_self_1997,horowitz_correspondence_1997} temperature, where the \gls{bh} transitions into self-gravitating strings. Ergo, this is the string scale.
        In the decompactification limit, the emergence scale corresponds to $M_{\textnormal{P};D}$, while the $d$-dimensional \gls{bh} transitions into a $D$-dimensional \gls{bh} already at a lower energy scale, where the \gls{kk} tower emerges. Ergo, this is the \gls{kk} scale.
        \citet{castellano_quantum_2024,bedroya_tale_2024} reasons that the \gls{bh} scale and the species scale have a deeper connection and fulfil the relation
        \begin{equation}
            \frac{\nabla\Lambda_\text{BH}}{\Lambda_\text{BH}}\frac{\nabla\Lambda_\textnormal{S}}{\Lambda_\textnormal{S}}=\frac{1}{d-2},
        \end{equation}
        which can also be understood as a relation between the \gls{bh} and the species entropy
        \begin{equation}
            \nabla\log\mathcal{S}_\text{BH}\cdot\nabla\mathcal{S}_\heartsuit=d-2.
        \end{equation}
    \item[\gls{uv} Cutoff Scale $\Lambda_\text{UV}$] The smallest scale that suppresses all\,\textemdash\,but finitely many\,\textemdash\,higher-derivative operators in the effective action, respectively kinematic forms on the \gls{ir} expansion of scattering amplitudes / \gls{ads} boundary correlators \cite{aoufia_species_2024}. This scale coincides with the parametric curvature scale of the smallest possible \gls{bh} \cite{aoufia_species_2024,bedroya_density_2024}.
\end{description}

That some confusion appears is understandable: In general, we find $\Lambda_\text{UV}\lesssim\Lambda_\text{QG}\lesssim\Lambda_\textnormal{S}\lesssim M_{\textnormal{P};d}$, e.g. for the weakly coupled critical string, we find an inequality \cite{aoufia_species_2024}: $\Lambda_\text{UV}=M_\textnormal{s}$, $\Lambda_\text{QG}=M_\textnormal{s}\sqrt{\log g_s^{-2}}$, $\Lambda_\textnormal{S}=M_\textnormal{s}\log g_s^{-2}$, whereas in decompactification limits, the scales are all equal.

To give an example of the relative sizes of those scales,\footnote{
    An absolute scale for \gls{qg} effects is for example presented by \citet{aoki_is_2021}: taking into account the \gls{sm} of particle physics and \gls{gr}, scattering amplitudes become inconsistent before \SI{e16}{\giga\electronvolt}, i.e. new physics has to emerge below this energy scale.
}
we refer to the dark dimension scenario presented in \cref{sss_AdSDC_Cosmology} \cite{bedroya_tale_2024}:
\begin{align}
    M_\textnormal{P}&\sim\Lambda_\text{cc}^0\\
    \Lambda_\text{S}&\sim\Lambda_\text{cc}^{1/12}\\
    \Lambda_\text{EW}&\sim\Lambda_\text{cc}^{2/12}\\
    \Lambda_\text{BH}&\sim\Lambda_\text{cc}^{3/12}.
\end{align}

\subsubsection{Evidence}
The \gls{ep} or \textit{emergent string conjecture} emerges from the work by \citet{lee_emergent_2019,lee_emergent_2022,grimm_infinite_2018,corvilain_swampland_2019,heidenreich_weak_2018,palti_swampland_2019,heidenreich_emergence_2018,harlow_wormholes_2016}
and was studied for
\gls{kklt} \cite{blumenhagen_quantum_2020},
M-theory \cite{alvarez-garcia_membrane_2022,blumenhagen_emergence_2024},
5d supergravities arising from M-theory compactified on \gls{cy} threefolds \cite{heidenreich_infinite_2021,rudelius_gopakumar-vafa_2024},
5d $\mathcal{N}=1$ supergravities \cite{kaufmann_asymptotics_2024},
F-theory \cite{lee_tensionless_2018,wiesner_light_2022,alvarez-garcia_non-minimal_2023,collazuol_affine_2023},
type IIA, F-, and M-theory \cite{castellano_emergence_2023},
type IIA \gls{cy} compactifications \cite{hattab_emergence_2024},
type IIA 4d $\mathcal{N} = 1$ \gls{cy} orientifolds with chiral matter \cite{casas_yukawa_2024},
type IIA 4d $\mathcal{N} = 2$ \gls{cy} compactifications \cite{marchesano_eft_2023},
type IIA 4d $\mathcal{N} = 2$ string vacua \cite{blumenhagen_demystifying_2024},
type IIA 6d toroidal orbifolds \cite{blumenhagen_emergence_2023},
type IIB $\mathcal{N}=1$ orientifolds with O3/O7 planes \cite{enriquez-rojo_swampland_2020,rojo_swampland_2019},
open strings ending on M5 branes inside \gls{cy} conifolds \cite{hattab_emergent_2024},
compactifications of maximal supersymmetry in flat space \cite{etheredge_sharpening_2022},
10d string models with no or broken supersymmetry \cite{basile_sitter_2020},\footnote{\citet{basile_sitter_2020} did not test the \gls{ep} \textit{on purpose}, but their realisations of the \gls{dc} produces \gls{kk} towers that become massless in the infinite distance limit.}
Schwinger integrals \cite{hattab_particle_2023},
holography \cite{basile_emergent_2022,baume_tackling_2021},
conformal manifolds \cite{perlmutter_cft_2021,baume_higher-spin_2023},
\glspl{bh} \cite{basile_shedding_2024},
the compatibility of different towers with species scale thermodynamics (see \cref{sss:SSC_Remarks}) \cite{herraez_origin_2024}, and
the density of one-particle state and gravitational scattering amplitudes \cite{bedroya_density_2024}.
Furthermore, relations to other swampland conjectures are highlighted in \cref{rel:EP_SDC,rel:WGC_EP}.

\subsection{Festina Lente}\label{sec:Festina}
Even though the \gls{flb}\footnote{
    \textit{Festina Lente} is the Latin translation of the Greek 
    \textit{\textsigma\textpi\textepsilon\textupsilon\textdelta\textepsilon\ \textbeta\textrho\textalpha\textdelta\textepsilon\textomega\textvarsigma}
    and is the imperative \textit{hasten slowly}. The German translation \textit{eile mit Weile} is a widely used proverb. It is the advice to do things quickly but with sufficient diligence. 
    Prominent personae, such as the emperors Augustus and Titus or Cosimo I de' Medici,
    made this proverb to their motto \cite{suetonius_tranquillus_suetonius_2007}.
    Among others, it is used in Goethe's \textit{Hermann und Dorothea} \cite{goethe_hermann_2000} and Bram Stoker's \textit{Dracula} \cite{stoker_dracula_2020}.
    In the context of the swampland programme, it is used to summarise the expected behaviour of charged \glspl{bh} in \gls{ds} space: they should haste to decay, but not too quickly and reveal a naked singularity.
}
is an original and independent conjecture, most of the current studies are the reconciliation of results derived in the context of the \gls{wgc}. To become familiar with many of the concepts we discuss in this section, we would advise reading this section after \cref{sec:gravity} about the \gls{wgc}.

The \gls{flb} states the following:
A particle of mass $m$ under a U(1) gauge field with coupling $g$ and charge $q$ in a \gls{ds} vacuum of energy density $V$ has a mass of at least
\begin{equation}\label{eq:FL}
    m^4\gtrsim2\left(gq\right)^2V=8\pi\alpha V=6\left(gqM_\textnormal{P} H\right)^2,
\end{equation}
where we used the fine-structure constant $\alpha=g^2q^2/4\pi$ and $V=\Lambda/8\pi G=3M_\textnormal{P}^2H^2$ with the Hubble parameter $H=\sqrt{\Lambda/3}$;
today, with $M_\textnormal{P}\sim\SI{e27}{\electronvolt}$ and $H\sim\SI{e-33}{\electronvolt}$, the bound holds for every charged \gls{sm} particle \cite{gonzalo_swampland_2022,montero_fl_2021,montero_festina_2020,chrysostomou_reissner-nordstrom_2023}.\footnote{
    Note that the \gls{flb} is independent of the dimension of the decompactified space because the gauge coupling scales with $E^{2-d/2}$ \cite{montero_fl_2021}.
    While the electron with $m_\textnormal{e}\sim\SI{0.5}{\mega\electronvolt}\gg\sqrt{eM_\textnormal{P}H}\sim\SI{e-3}{\electronvolt}$ satisfies the \gls{flb} with 8 orders of magnitude, it is noteworthy that the bound itself is realised by combining two quantities that are 60 orders of magnitude apart \cite{grana_swampland_2021}.
    }

\subsubsection{Implications for Cosmology}

\paragraph{Axions}
The allowed number of axions in a theory is limited\footnote{
    See the text above \cref{foo:FL_NAxions} for further details.
    }
to
\begin{equation}
    N<\frac{M_\textnormal{P}}{H}.
\end{equation}

In analogy to the axionic \gls{wgc} (\cref{p:WGC_axion}), \citet{guidetti_axionic_2023} propose an axionic \gls{flb}\,\textemdash\,which is consistent under dimensional reduction\,\textemdash\,on purely geometric grounds, without finding direct evidence for such a bound from wormholes or \glspl{bh} with an axion charge.\footnote{
    In analogy to the non-axionic \gls{flb}, \citet{guidetti_axionic_2023} investigate the evaporation of back holes with axionic charges: The evaporation of axionic \glspl{bh} does not produce a singularity, as the discharging process creates string networks with an \gls{eos}-parameter $\omega=-1/3$, i.e. it \textit{enhances} the spacetime acceleration and avoids a big crunch.
}
Their proposed bound is
\begin{equation}
    S_\iota f\gtrsim\sqrt{M_\textnormal{P}H}\sim V^{1/4},
\end{equation}
with $S_\iota$ the instanton action and $f$ the axion decay constant.

\paragraph{Black holes} have a maximum size in \gls{ds} space: the cosmological horizon \cite{montero_fl_2021}.
The \gls{flb} ascertains that \gls{rn} \glspl{bh} stay below this limit \cite{montero_fl_2021,lee_festina-lente_2022}.
A \gls{bh} saturating this bound\,\textemdash\,such that the \gls{bh} horizon coincides with the cosmological horizon\footnote{
    The \gls{bh} horizon can exchange charge and mass with the cosmological horizon \cite{montero_festina_2020,bousso_pair_1996,bousso_charged_1997,bousso_anti-evaporation_1998,bousso_proliferation_1998,bousso_quantum_1999,bousso_adventures_2002,belgiorno_massive_2009,belgiorno_quantum_2010,belgiorno_tunneling_2017,kim_schwinger_2016}.
    Furthermore, the $d$-dimensional \gls{ds} radius $l_d$ is connected to the vacuum energy $\Lambda$ by
    $
        \left[\left(d-1\right)\left(d-2\right)\right]/\left(2l_d^2\right)=M_\textnormal{P}^{2-d}\Lambda,
    $
    where $l_d=1/H$, with $H$ the Hubble parameter, holds \cite{montero_fl_2021}.
}\textemdash\,is called a Nariai \gls{bh} \cite{nariai_static_1999,romans_supersymmetric_1992,montero_festina_2020}.
If the \gls{flb} does not hold, the electric field of charged Nariai \glspl{bh} is screened by Schwinger pairs and the \glspl{bh} crunch on a timescale $1/H$, which produces arbitrarily high curvatures and superextremal \glspl{bh} that have too much mass to fit in a static \gls{ds} patch \cite{montero_fl_2021}.

\begin{figure}[htb]
  \begin{center}
    \includegraphics[width=\linewidth]{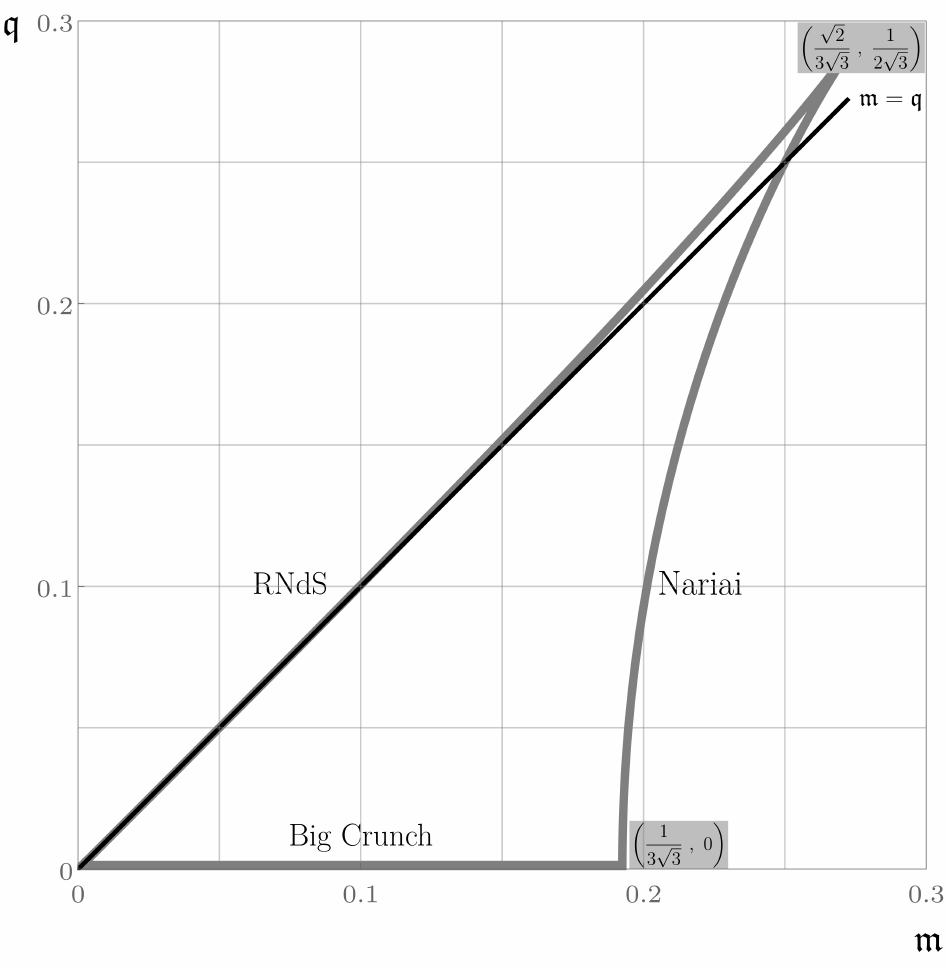}
  \end{center}
  \caption[Black Hole Boundary]{Phase space diagram of extremal \gls{bh} solutions in 4d Einstein\textendash Maxwell\textendash de~Sitter \cite{montero_festina_2020}:
  \glspl{bh} on the upper \gls{rnds} extremality branch (RNdS) have AdS$_2\cross S^{d-2}$ geometry and zero temperature.
  \glspl{bh} on the diagonal line $\mathfrak{m}=\mathfrak{q}$ have a temperature that corresponds to the temperature of the cosmological horizon. These are extremal \gls{rn} \glspl{bh} in an asymptotically flat background with vanishing cosmological constant \cite{hassan_sitter_2025}.
  \glspl{bh} on the lower extremality branch (Nariai) are charged Nariai \glspl{bh} with dS$_2\cross S^{d-2}$ geometry, an area that corresponds to the area of the cosmological horizon, and a temperature that is in equilibrium with the cosmological horizon.
  The two boundaries intersect at the \enquote{ultracold} \gls{bh} \cite{romans_supersymmetric_1992} $\left(\frac{\sqrt{2}}{3\sqrt{3}}\ ,\ \frac{1}{2\sqrt{3}}\right)$ with vanishing temperature and M$_2\cross S^{d-2}$ geometry, where M represents Minkowski space.
  The lower boundary corresponds to the big crunch scenario, which ends at the point of maximum mass for a neutral \gls{bh} $\mathfrak{m}_\text{max}^\text{neutral}=1/3\sqrt{3}$.}\label{f:BHBoundary}
\end{figure}

To avoid naked singularities, superextremal \gls{bh} solutions have to be avoided. In 4d Einstein\textendash Maxwell\textendash de~Sitter, the action is given by
\begin{equation}
    S=\int\mathrm{d}^4x\sqrt{-g}\left[\frac{1}{16\pi G}\left(-R+\frac{6}{l^2}\right)+\frac{1}{4g^2}F_{\mu\nu}F^{\mu\nu}\right],
\end{equation}
with $g$ the determinant of the metric (and not the coupling, as in the remainder of this section), $G$ the gravitational constant, $R$ the Ricci scalar, $l$ the \gls{ds} length scale, and $F_{\mu\nu}$ the field strength tensor of the U(1) gauge field \cite{montero_festina_2020}.
Charged \gls{bh} solutions are described by the \gls{rnds} metric
\begin{align}
    \mathrm{d}s^2&=-U(r)\mathrm{d}t^2+\frac{\mathrm{d}r^2}{U(r)}+r^2\mathrm{d}\Omega\\
    U(r)&=1-\frac{2\mathfrak{m}}{r}+\frac{\mathfrak{q}^2}{r^2}-\frac{r^2}{l^2}
\end{align}
with the parameters $\mathfrak{m}$ and $\mathfrak{q}$ of dimension length \cite{montero_festina_2020,danielsson_charged_2024}.\footnote{
    The parameters $\mathfrak{m}$ and $\mathfrak{q}$ can be expressed in terms of the \gls{ds} length scale $l$,
    the outer \gls{bh} horizon $r_\textnormal{o}$,
    and the cosmological horizon $r_\textnormal{c}$,
    as
    $\mathfrak{m}=\left[r_\textnormal{o}^2\left(r_\textnormal{o}^2-l^2\right)-r_\textnormal{c}^2\left(r_\textnormal{c}^2-l^2\right)\right]/\left[\left(r_\textnormal{c}-r_\textnormal{o}\right)l^2\right]$
    and
    $\mathfrak{q}^2=\left[r_\textnormal{c}\left(r_\textnormal{c}^2-l^2\right)-r_\textnormal{o}\left(r_\textnormal{o}^2-l^2\right)\right]/\left[\left(1/r_\textnormal{c}-1/r_\textnormal{o}\right)l^2\right]$ \cite{aalsma_extremal_2023}.
    In these units, the \gls{wgc} bound can be written in the compact form $\mathfrak{m}\leq2\mathfrak{q}$ \cite{aalsma_extremal_2023}.
    To physical mass $m$ and electromagnetic charge $q$, the parameters relate as
    $\mathfrak{m}=m\hbar c l_{\textnormal{P};4}^2/8\pi$, and
    $\mathfrak{q}^2=q^2 l_{\textnormal{P};4}^2/32\pi^2\epsilon_0\hbar c$, with $\hbar$ the reduced Planck constant, $c$ the speed of light in vacuum, $\epsilon_0$ the electric constant, and $l_{\textnormal{P};4}$ the 4d Planck length \cite{danielsson_charged_2024}.
}
This system is ill-defined for $U(r)=0$. In $\left(\mathfrak{m},\mathfrak{q}\right)$ phase space, the boundary of extremal \gls{bh} solutions is given by
\begin{equation}
    l^4\mathfrak{m}^2-l^4\mathfrak{q}^2-27l^2\mathfrak{m}^4+36l^2\mathfrak{m}^2\mathfrak{q}^2-8l^2\mathfrak{q}^4-16\mathfrak{q}^6=0,
\end{equation}
which we depict in \cref{f:BHBoundary} for Hubble units with $l=1$ \cite{montero_festina_2020}:\footnote{
    \citet{chrysostomou_reissner-nordstrom_2023} provide an \href{https://github.com/anna-chrys/RNdS_QNMs}{interactive \texttt{Mathematica} notebook} to reproduce plots and values of \gls{rnds} \gls{bh} solutions.
}
Everything within the \textit{shark fin} is allowed. Solutions outside the shark fin are superextremal / naked singularities.\footnote{
    The $\mathfrak{q}=0$ bottom boundary is a big crunch singularity \cite{montero_festina_2020}.
}
Solutions on the boundary are extremal, i.e. at least two horizons coincide \cite{hassan_sitter_2025}.\footnote{
    Each solution has three horizons: a cosmological horizon as well as an inner and an outer \gls{bh} horizon \cite{hassan_sitter_2025}.
    At the ultracold point in \cref{f:BHBoundary}, all three horizons coincide.
}
The upper boundary on the left of the fin are the superextremal \gls{rn} \glspl{bh}, which have an important role in \cref{sec:gravity}.\footnote{
    For \gls{rn} \glspl{bh}, the inner and outer \gls{bh} horizon coincide \cite{hassan_sitter_2025}.
}
The lower boundary on the right consists of the charged Nariai \glspl{bh},\footnote{
    For Nariai \glspl{bh}, the outer \gls{bh} horizon coincides with the cosmological horizon \cite{hassan_sitter_2025}.
}
which will take a prominent role in this section.

\Cref{f:BHBoundary} allows for an interesting observation: In \gls{ds} space, there is a maximum mass for neutral \glspl{bh} at
$\frac{1}{3\sqrt{3}}\mathfrak{m}$ in $l=1$ Hubble units.
If the \gls{bh} grows further in mass, it needs to obtain charge as well, to avoid becoming superextremal.
This is a unique \gls{ds} effect, which has a thermodynamical analogy:\footnote{
    A \gls{bh} in \gls{ds} space can be seen as a thermal system of finite size \cite{montero_festina_2020,anninos_sitter_2012,dinsmore_schottky_2020,johnson_sitter_2019}:
    If a static patch is not in equilibrium, the system will re-equilibrate in a finite amount of time, like a system of equilibrium temperature $H/2\pi$. The \gls{bh} will discharge by a particle tunnelling beyond the cosmic horizon. This picture breaks down if there is a light, charged particle \cite{montero_festina_2020}: charged Nariai solutions are more massive than neutral Nariai solutions. A discharge of a Nariai \gls{bh} happens almost instantaneously, leading to a big crunch instead of empty \gls{ds} space, violating the thermodynamic picture.
}
the static patch of \gls{ds} has a finite-dimensional Hilbert space \cite{witten_quantum_2001}, i.e. every observable in the theory has a maximum \cite{montero_festina_2020}.

\paragraph{Dark energy} is bounded by the \gls{flb} to
\begin{equation}
    \Lambda\lesssim\frac{m^4}{4\pi\alpha}\sim\num{3e-89},
\end{equation}
where the numerical value comes from evaluating the bound for the electron\,\textemdash\,and is in agreement with the observed value $\Lambda\sim\num{e-120}$ \cite{montero_fl_2021}.

\paragraph{Dark Matter}
\citet{nam_implications_2023} presents a model of branon \gls{dm}, which we explain around \cref{eq:branon}. The model is incompatible with the \gls{flb}, as it yields a tower of \gls{kk} states with $m_n^2\gtrsim\sqrt{6}\kappa nM_\textnormal{P}H_0$ that violates the \gls{flb} for vanishing gauge coupling $\kappa$.

\paragraph{Inflation}
seems to be a suitable candidate for applications of the \gls{flb}, given it is a \gls{ds} phase and the large body of work that has been created over the last few decades in constraining its parameters and collecting (indirect) observational evidence. However, since inflation only lasts a finite amount of time (roughly 60 $e$-folds), the \gls{flb} could be avoided entirely if Nariai \glspl{bh} simply would not have enough time to discharge and form a naked singularity before inflation ends.
Nariai \glspl{bh} discharge\footnote{
    The near-horizon electric field of a Nariai \gls{bh} is $E\sim qgM_\textnormal{P}H$, with $E=\sqrt{6}qgM_\textnormal{P}H$ at the ultracold point in \cref{f:BHBoundary} \cite{chrysostomou_reissner-nordstrom_2023}.
}
over a time of
\begin{align}
    t_\text{BH}&\sim1/\sqrt{qE}=1/\sqrt{qgM_\textnormal{P}H}\\
    &=\frac{1}{H}\left(\frac{H}{gM_\textnormal{P}}\right)^{1/2}=\left(\frac{1}{g^2V}\right)^{1/4},
\end{align}
where the second line holds of the non-Abelian case with $q=1$ \cite{mishra_confinement_2023,montero_fl_2021,montero_festina_2020}.
The \gls{tcc} sets an upper bound on the lifetime of \gls{ds} space \cite{mishra_confinement_2023}:
\begin{equation}
    t_\text{dS}<t_\text{TCC}=\frac{\log\left(M_\textnormal{P}/H\right)}{H}.
\end{equation}
If we demand that the timescale for a \gls{bh} to decay
is larger than the duration of inflation $N_e/H$,
we find the bound\footnote{
    The bound is incompatible with the magnetic \gls{flb} \cref{eq:fl_magnet} \cite{montero_swampland_2022}, unless inflation happens in less than 1 $e$-fold.
} \cite{montero_swampland_2022}
\begin{align}
    \frac{1}{\sqrt{qgM_\textnormal{P}H}}&>\frac{N_e}{H}\\
    \Rightarrow g&\lesssim\frac{H}{qN_e^2M_\textnormal{P}}.\label{eq:FL_e-fold}
\end{align}
\citet{mishra_confinement_2023} reasons that the Fokker\textendash Planck equation can be used to estimate how long it takes a scalar field to fluctuate away from the top of its potential by displacing the field by $\order{H}$, and calculates how long it takes to violate the slow-roll conditions. Based on these considerations, he finds a bound of applicability for the \gls{flb} of
\begin{equation}
    \frac{M_\textnormal{P}^2\abs{V^{\prime\prime}}}{V}\lesssim\frac{\sqrt{g}M_\textnormal{P}}{\left(3V\right)^{1/4}},
\end{equation}
which would be in tension with the \gls{dsc} for small coupling constants.\footnote{
    The tension is avoided by applying the magnetic \gls{flb} (\cref{eq:fl_magnet}), which demands that
    \begin{equation}
        g^2\gtrsim\frac{3}{2}\left(\frac{H}{M_\textnormal{P}}\right)^2\defeq g_\text{min}^2,
    \end{equation}
    and using $V=3M_\textnormal{P}^2H^2$, such that the bound of applicability becomes
    \begin{equation}
         \frac{M_\textnormal{P}^2\abs{V^{\prime\prime}}}{V}\lesssim\left(\frac{g}{g_\text{min}}\right)^{1/2}\left(\frac{3}{2}\right)^{1/4},
    \end{equation}
    which is consistent with the \gls{dsc} \cite{mishra_confinement_2023}.
}
This makes it even less likely to find a coupling small enough to satisfy \cref{eq:FL_e-fold}.
We conclude that it is not trivial to circumvent the \gls{flb} during inflation. The \gls{flb} is likely a meaningful constraint for inflation.

Thermal effects could limit the energy scale of inflation to $\Lambda_\text{infl}\ll\SI{e5}{\giga\electronvolt}$, as we show around \cref{eq:FL_Inflation}.

For singe-field slow-roll inflation, the bounds
\begin{align}
    h&\geq h\left(\frac{16\pi^3g_\text{re}g_\text{em}}{60m_e^4}\right)^{1/4}T_\text{rh}&>\SI{e4}{\giga\electronvolt}\\
    H&\leq\frac{2m_e^2}{\sqrt{96g_\text{em}} M_\textnormal{P}}&<\SI{e7}{\giga\electronvolt}\\
    r_\textnormal{ts}&\lesssim\num{3e-15},
\end{align}
with $h$ the value of the Higgs field,
$g_\text{rh}$ the effective degrees of freedom during reheating,
$g_\text{em}\simeq\num{9.5e-3}$ the electromagnetic coupling,
$m_e$ the electron mass,
$T_\text{rh}\gtrsim T_\text{BBN}\sim\SI{10}{\mega\electronvolt}$ the temperature during reheating,
and $r_\textnormal{ts}$ the tensor-to-scalar ratio,
can be obtained by noting that the Higgs vacuum expectation value does not correspond to the \gls{ew} scale during inflation \cite{lee_festina-lente_2022}.

If the model of inflation solely consists of the \gls{sm} particles plus an inflaton field, the charged \gls{sm} particles violate the \gls{flb} for high enough inflaton field energies \cite{montero_swampland_2022}.
The \gls{flb} can be satisfied in Higgs inflation, if the vacuum expectation value of the Higgs field is high enough to make gluons, quarks, and leptons sufficiently massive \cite{montero_swampland_2022,montero_festina_2020}.
Another option to satisfy the \gls{flb} is by coupling the inflaton to the kinetic terms of the gauge fields, as this reduces the gauge coupling\,\textemdash\,one has to make sure though that compatibility with the \gls{wgc} is retained: this means that $m^2\geq\sqrt{6}gqM_\textnormal{P}H\geq3H^2$ \cite{montero_swampland_2022}. Since the expected mass of particles during inflation is of $\order{H}$ this is likely satisfied \cite{montero_swampland_2022}.

\citet{cribiori_supergravity_2023} show that single-field D-term inflation is in tension with the \gls{flb}, but also with the gravitino \gls{dc}.

\paragraph{Particle Physics}
Since the \gls{flb} for the U(1) gauge field in our Universe is satisfied, the photon is massless.

Charged states obtain mass bounds:
There can be no massive, charged state with mass below $\Lambda^2 M_\textnormal{P}^2$, which is satisfied today but can be an important constraint during inflation \cite{montero_fl_2021}.
Massless charged particles are only compatible with the \gls{flb} if
the \gls{de} potential vanishes entirely \cite{lee_festina-lente_2022},
or the particles arise from confined or Higgsed non-Abelian gauge fields, i.e.
\begin{align}
    m_\text{gauge field}\gtrsim H\\
    \Lambda_\text{c}\gtrsim H,
\end{align}
with $\Lambda_\text{c}$ the confinement scale \cite{mishra_confinement_2023,montero_fl_2021,montero_swampland_2022}.
This bound relates a pure \gls{qft} quantity ($\Lambda_\text{c}$), with a quantity ($H$) closely related to the \gls{ds} spacetime that acts as a background \cite{mishra_confinement_2023}.\footnote{
    \citet{venken_cosmological_2023} argues that the confinement scale should be higher than a scale proportional to the vacuum energy and that the Higgs mass should be higher than a scale proportional to the Higgs vacuum scale when thermal effects are taken into account. We discuss this in more detail in our paragraph on thermal effects.
}

Using the \gls{flb}, a non-zero Higgs vacuum expectation value is considered by \citet{gonzalo_swampland_2022,montero_fl_2021}.
\citet{lee_festina-lente_2022} study the applicability of the \gls{flb} for various forms of the Higgs potential: The \gls{uv} behaviour is yet undetermined, but allows for three different scenarios.
First, the potential could be monotonically increasing, in which case the \gls{flb} is not applicable.
Second, the potential could have degenerate Higgs vacua\footnote{
    For degenerate Higgs vacua, the \gls{flb} is trivially satisfied, as the effective potential vanishes.
}
or inflection points.\footnote{
    In the latter case, the vacuum energy at the deflection points becomes too large to satisfy the \gls{flb}, unless quantum effects allow a tunnelling faster than \gls{bh} decay, which renders the \gls{flb} inapplicable.
}
The \gls{flb} is applicable and sets an upper limit on the quartic coupling.
And third, the potential could have a local \gls{uv} minimum that is negative, i.e. the space would be \gls{ads} and the \gls{flb} not applicable (unless another source of vacuum energy lifts the combined effective potential into \gls{ds}).
An argument against a cowboy hat potential with a local minimum at the origin is that in this setting, the electron, and other hypercharged states, remain classically massless \cite{montero_fl_2021}.
To be compatible with the \gls{flb}, any long-lived minimum of the Higgs potential must break \gls{ew} symmetry \cite{montero_fl_2021}.

Millicharged particles\footnote{
    \textit{Millicharged}, \textit{minicharged}, or \textit{nanocharged} particles are terms used for hypothetical particles with a charge smaller than the electron charge.
}
are constrained by observations\footnote{
    Probes and constraints for particles with charge less than the electron charge come from
    \SI{21}{\centi\meter}\textendash line dispersion \cite{caputo_constraints_2019,munoz_insights_2018},
    \gls{bbn} \cite{vogel_dark_2014,brust_new_2013,berezhiani_cosmological_2009,davidson_updated_2000,davidson_limits_1991},
    \gls{cmb} \cite{vogel_dark_2014,berezhiani_cosmological_2009,melchiorri_new_2007,dubovsky_narrowing_2004,adshead_dark_2022},
    galaxy interactions \cite{stebbins_new_2019,kadota_new_2016,cruz_astrophysical_2023},
    lab experiments \cite{gluck_lamb_2007,davidson_updated_2000,della_valle_first_2014,ahlers_particle_2007}
    particle accelerators \cite{de_montigny_minicharged_2023,davidson_limits_1991,prinz_search_1998}, and
    stellar evolution \cite{davidson_updated_2000,davidson_limits_1991}
    including \glspl{sn} \cite{davidson_astrophysical_1994,mohapatra_astrophysical_1990,chang_supernova_2018}.
    Millicharged particles could help explain the \SI{511}{\kilo\electronvolt} anomaly observed in the galactic bulge \cite{huh_galactic_2008}.
    See \citet{de_montigny_minicharged_2023} for a recent review, including theoretical considerations and phenomenological implications.
}
but the constraints can be improved by taking into account the \gls{flb}: \citet{montero_swampland_2022} suggest $q\geq\left(m/\SI{1.6}{\milli\electronvolt}\right)^2$, or even $q\geq\left(m/\SI{10}{\milli\electronvolt}\right)^4$ if kinetic mixing takes place.
\citet{ban_phenomenological_2023} find $m\gtrsim\SI{5}{\milli\electronvolt}$ as a limit for a single U(1) in the dark sector.

As \citet{kobayashi_schwinger_2014} show, particle production through Schwinger pair production during magnetogenesis is constrained by a strong suppression of the magnetic field by \gls{ir} hyperconductivity: as shown by \citet{frob_schwinger_2014}, light charged particles with $m^2\ll H^2$ can induce strong currents, $J\sim H^3/E$ for $m/H\ll qE/H^2\ll 1$, which end magnetogenesis. However, the regime of \gls{ir} hyperconductivity cannot be reached if the \gls{flb} holds \cite{grewal_characters_2022}. This weakens the constraints on magnetogenesis coming from hyperconductivity, yet not those coming from other considerations, e.g. that the magnetic field must not dominate the energy budget of the Universe.

\subsubsection{General Remarks}
The \gls{flb} does not apply if the symmetry is broken, but if there is a phase where the symmetry is intact, the bound applies, e.g. a phase of unbroken \gls{ew} symmetry with vanishing masses is incompatible with a positive cosmological constant \cite{agmon_lectures_2023}.

\paragraph{What can we learn from taking the limits of the \gls{flb}?}
\begin{itemize}
    \item In the limit $H\rightarrow0$ the \gls{flb} becomes trivial\,\textemdash\,as expected, for a conjecture about \gls{ds} space \cite{montero_festina_2020}.
    \item The limit $M_\textnormal{P}\rightarrow\infty$ forbids all charged particles. This indicates that gravity cannot be decoupled in a \gls{ds} theory \cite{montero_festina_2020}.
    \item In the limit $m^2\ll gqM_\textnormal{P}H$, the Schwinger pair production is exponentially suppressed, and the \gls{bh} looses its charge slowly, eventually vaporising completely, without ever leaving the shark fin region from \cref{f:BHBoundary} \cite{montero_festina_2020}. This corresponds to reaching equilibrium of the thermal bath and leaving empty \gls{ds} space \cite{montero_festina_2020}.
    \item In the limit $m^2\gg gqM_\textnormal{P}H$, the Schwinger pairs screen the electric field of the \gls{bh}, i.e. the electric field is replaced by photons, which do not support the solution, and the \gls{bh} collapses into a big crunch \cite{montero_festina_2020}.
\end{itemize}

\paragraph{What features are distinct for dS?}
Other than in flat or \gls{ads} space, where a particle that tunnels out of a \gls{bh} will eventually fall back into it (unless gravity is the weakest force, as demanded by the \gls{wgc}), in \gls{ds} space, a particle can tunnel into a region where the cosmological expansion carries it away from the \gls{bh}, eventually\footnote{
    The lifetime of a \gls{bh} in \gls{ds} space is exponentially long in the \gls{ds} length scale $l$ \cite{montero_festina_2020}.
    }
evaporating the \gls{bh} completely\,\textemdash\,in \gls{ds} the decay of \glspl{bh} is not a \gls{qg} prediction, but already part of the theory in a semi-classical picture \cite{montero_festina_2020}.

An important observation in \cref{sec:AdSDC} was that \gls{ads} spaces are likely not scale separated. For \gls{ds}, the \gls{flb} implies the opposite \cite{montero_fl_2021}: \gls{ds} spaces are scale separated with
\begin{equation}
    M_\text{KK}\gtrsim\sqrt{\Lambda}.
\end{equation}
This positions Minkowski space as a \textit{great divider}, where the behaviour on each side is qualitatively distinct \cite{montero_fl_2021}.

\paragraph{What are the implications of \textnormal{thermal effects}?}
An upper bound on the energy scale of inflation of $\Lambda_\text{I}\ll\SI{e5}{\giga\electronvolt}$ can be obtained by studying thermal effects in \gls{ds} space, as  \citet{venken_cosmological_2023} shows:
First of all, when studying thermal effects in \gls{ds} space, one has to consider cooling and dilution due to cosmic expansion. 
However, unless the coupling is miniscule ($g\lesssim H/M_\textnormal{P}$), charged Nariai \glspl{bh} decay within 1 $e$-fold,\footnote{
    This can be obtained by combining the magnetic \gls{flb} (\cref{eq:fl_magnet}) with \cref{eq:FL_e-fold}.
}
such that the temperature of the thermal radiation can be considered constant.
Then, one has to express the energy density of thermal radiation $\rho_r$ as
\begin{equation}
    \rho_r=\sigma T^d,
\end{equation}
with $d$ the number of spacetime dimensions, and $\sigma$ a dimensionless coefficient that depends on $d$ and on the thermally excited degrees of freedom.
As a next step, one has to consider that the \gls{flb} is only applicable when the vacuum energy is dominating, which allows Nariai \glspl{bh} to exist. The exact cutoff of applicability is yet unknown. The best we can do is require
\begin{align}
    \frac{\rho_r}{V}&\leq \mathfrak{l}\\
    \Rightarrow T&\leq\left(\frac{\mathfrak{l}V}{\sigma}\right)^{1/d} \label{eq:FL_temperature}
\end{align}
where the factor $\mathfrak{l}\sim\order{1}$ depends on the specific model in question.
It is expected that $\sigma\sim N_T$, with $N_T$ the number of species that are part of the thermal radiation, i.e. species that are lighter than $T$. $N_T$ is certainly smaller than $N_\textnormal{S}$, the total number of species part of the \gls{eft} as set by the species scale (\cref{s:ssc}), but since the species scale is certainly higher than the Hubble scale (since cosmic expansion is definitely part of the \gls{eft}), the bound that takes thermal effects into account is always stronger than the bound that ignores thermal effects.
This is of particular importance during inflation, where the bound could be recast as
\begin{equation}
    \Lambda_\text{QCD}<\left(\mathfrak{l}V_\text{I}/\sigma\right)^{1/d},
\end{equation}
which would either mean that $V_\text{I}<\Lambda_\text{QCD}$ or that there is a large number of extra species present during inflation (a recurring theme in the swampland programme):
\begin{equation}
    N_T\gtrsim \mathfrak{l}\left(\frac{\Lambda_\text{I}}{\Lambda_\text{QCD}}\right)^{d},
\end{equation}
with $\Lambda_\text{I}=V_\text{I}^{1/d}$.
We can combine the bounds to make a statement about the energy scale of inflation:
\begin{AmSalign}
    N_T&\gtrsim \mathfrak{l}\left(\frac{\Lambda_\text{I}}{\Lambda_\text{QCD}}\right)^{d}\\
    \Lambda_\textnormal{S}&=\frac{M_{\textnormal{P};d}}{N_\textnormal{S}^{\frac{1}{d-2}}}\\
    \Rightarrow N_\textnormal{S}&=\left(\frac{M_{\textnormal{P};d}}{\Lambda_\textnormal{S}}\right)^{d-2}\\
    N_T&<N_\textnormal{S}\\
    \Rightarrow \mathfrak{l}\left(\frac{\Lambda_\text{I}}{\Lambda_\text{QCD}}\right)^d&<\left(\frac{M_\textnormal{P}}{\Lambda_\textnormal{S}}\right)^{d-2}\\
    \Lambda_\text{I}&\ll\Lambda_\text{S}\\
    \Rightarrow \Lambda_\text{I}&\ll M_\textnormal{P}^{\left(d-2\right)\left(2d-2\right)}\Lambda_\text{QCD}^{d/\left(2d-2\right)},\label{eq:FL_Inflation}
\end{AmSalign}
which yields $\Lambda_\text{I}\ll\SI{e5}{\giga\electronvolt}$. This indicates that inflation itself is not a high-energy process.
This is not unambiguously a welcomed finding, for example, the overshoot problem in string cosmology could possibly be solved by a large energy scale for inflation \cite{conlon_catch-me-if-you-can_2022}.
The constraint could be circumvented if inflation is not described by the same \gls{eft} as the late-time Universe.

\paragraph{Is the \gls{flb} exact?}
\citet{abe_black_2023} question the leading numerical factor in \cref{eq:FL}: Backreactions might modify the exact value of the $\order{1}$ numerical factor.
\citet{aalsma_extremal_2023} go a step further and investigate if backreactions could avoid the creation of a naked singularity entirely, and find this indeed to be the case.\footnote{
    It is argued that it is impossible to overspin or overcharge an extremal 4d Kerr\textendash Newman \gls{bh} \cite{sorce_gedanken_2017}.
    } 
They find that there are unsuppressed decay channels %
that do not end in a big crunch, but in a \gls{rnds} geometry, without creating a singularity, i.e. the observer remains outside the horizon during the entire discharge process.\footnote{
    \citet{hassan_sitter_2025} raise the concern that multi-particle states\,\textemdash\,in particular particle\textendash anti-particle annihilation\,\textemdash\,have not been considered in the study by \citet{aalsma_extremal_2023}, which might open up additional decay channels.}

\paragraph{How can the \gls{flb} be expressed for scalar fields?}
\citet{montero_fl_2021} argue that in a 4d theory with a U(1) gauge field, and a scalar field $\phi$ with potential $V(\phi)>0$, such that the scalar field couples to the gauge field as in
\begin{align}
    S=&\int\!\sqrt{\abs{g}}\left(\frac{1}{2}M_\textnormal{P}^2R-\frac{1}{2}\left(\partial\phi\right)^2\right.\nonumber\\
    &\left.-\frac{1}{4}\lambda(\phi)F_{\mu\nu}F^{\mu\nu}-V(\phi)\right)+\text{matter}
\end{align}
the inequality
\begin{equation}\label{eq:NariaiCond}
    \abs{\frac{V^\prime}{V}}<\abs{\frac{\lambda^\prime}{\lambda}}
\end{equation}
needs to be fulfilled to assure that a Nariai branch exists.
They note that this can be in tension with the \gls{dsc}, unless $\mathfrak{s}_1<\abs{\lambda^\prime/\lambda}$ ($\mathfrak{s}_1$ the \gls{dsc} $\order{1}$ constant).
We would object to the existence of the tension: The \gls{flb} makes sure extremal Nariai \glspl{bh} can decay and do not grow unbounded to scales bigger than the cosmological horizon. If there is no Nariai solution, one does not have to worry about the unbounded growth of a Nariai \gls{bh} at all, as there simply is no such \gls{bh}.

\paragraph{Can the \gls{flb} be applied to multiple fields?}
A big difference between the \gls{wgc} and the \gls{flb} is that the \gls{flb} applies to \textit{every} particle; to satisfy the \gls{wgc}, it suffices if \textit{a} particle exists that satisfies the \gls{wgc} bound \cite{montero_fl_2021,montero_swampland_2022}.
This leads to the multi-field generalisation of the \gls{flb}\,\textemdash\,the analogue of the convex hull conjecture (see text along \cref{f:ConvexHull}):
For a system with multiple gauge fields %
with Lagrangian
\begin{equation}
    L=\frac{M_\textnormal{P}^2}{2}R-\Lambda-\frac{1}{4}f_{AB}F_{\mu\nu}^AF^{B\mu\nu}%
\end{equation}
with $R$ the Ricci scalar, $\Lambda$ the \gls{ds} \gls{de} density, $f_{AB}$ the mixing tensor, and $F_{\mu\nu}$ the gauge field strength tensor, 
the \gls{flb} can be written as
\begin{equation}\label{eq:FL_mulicharge}
    m^4\gtrsim2q_Ag_A\left(f^{-1}\right)^{AB}q_Bg_B\Lambda M_\textnormal{P}
\end{equation}
which simplifies to the single-field case for $f^{-1}=1$ \cite{ban_phenomenological_2023,montero_fl_2021,montero_swampland_2022}.
A simple example can be a model with two mixing Abelian gauge fields, a \gls{sm} U(1) gauge field and a hidden field $h$ with U$_h$(1) gauge symmetry \cite{ban_phenomenological_2023}: Let's say the kinetic term in the Lagrangian is of the form
\begin{equation}
    L\sim-\frac{1}{4}F_{\mu\nu}F^{\mu\nu}-\frac{1}{4}\mathcal{F}_{\mu\nu}\mathcal{F}^{\mu\nu}-\frac{\chi}{2}\mathcal{F}_{\mu\nu}F^{\mu\nu},
\end{equation}
with $F_{\mu\nu}$ referring to the \gls{sm} U(1),
$\mathcal{F}_{\mu\nu}$ referring to the hidden gauge field,
and $\chi$ the mixing parameter between the electromagnetic U(1) and the hidden U$_h$(1),
such that
\begin{equation}
    f=\begin{pmatrix}
        1&\chi\\
    \chi&1
    \end{pmatrix}.
\end{equation}
Then, the \gls{flb} reads as
\begin{equation}
    m^4\gtrsim\frac{6\left(M_\textnormal{P}H\right)^2}{1-\chi^2}\left(g_hq_h\right)^2
\end{equation}
or, if there are more than two gauge symmetries in the hidden sector, as
\begin{equation}
    m^4\gtrsim6\left(M_\textnormal{P}H\right)^2\sum_{j=1}^N\left(g_{hj}q_{hj}\right)^2.
\end{equation}

Using the formalism of
the charge-to-mass ratio vector $\Vec{z}=\frac{1}{m}\left(q_1,\dots,q_N\right)$ and an approximation of \cref{eq:FL_mulicharge}, \citet{guidetti_axionic_2023} express the range of allowed values as
\begin{equation}
    1<\norm{\Vec{z}}<\sqrt{\frac{M_\textnormal{P}}{H}}.
\end{equation}
They use this bound to find a maximum number of axions that are allowed in a theory by the following reasoning:
Assume you have a theory with $N$ axions.
The largest individual axionic charge that is allowed by the \gls{flb} in the theory (if the axionic \gls{flb} holds in this form), is $z_\text{max}=M_\textnormal{P}/fS_\iota$.\footnote{
    Here, $f$ is the axion decay constant, as usual, which is unrelated to the mixing tensor $f^{AB}$.
}
This in turn gives us an upper bound for the charge-to-mass ratio vector of $\norm{\Vec{z}}=\sqrt{N}M_\textnormal{P}/fS_\iota$.
It follows that the number of axions in the theory
\begin{equation}
    N<\frac{f^2S_\iota^2}{M_\textnormal{P}H}.
\end{equation}
Furthermore, if the axionic \gls{wgc}\footnote{$fS_\iota<M_\textnormal{P}$\label{foo:FL_NAxions}} holds, then
\begin{equation}
    N<\frac{M_\textnormal{P}}{H}.
\end{equation}

\paragraph{How can the \gls{flb} constrain general $p$-forms?}
In \cref{p:WGC_p-form} we explain how the \gls{wgc} can be generalised for $p$-forms. \citet{hassan_sitter_2025} generalise the \gls{flb} for $p$-forms:
\begin{align}
    \mathcal{T}^{\frac{1}{p+1}+\frac{\beta-p}{4}}&\gtrsim gHM_\textnormal{P}^\frac{d-2}{2}\\
    \mathcal{T}^{\frac{\beta-p}{4}}&\gtrsim gM_\textnormal{P}^\frac{d-2}{2},
\end{align}
with $\mathcal{T}$ the tension and $\beta=4$ for even $p\neq4$, and $\beta=3$ for odd $p\neq3$.
For $p>4$, these are upper bounds on the tension, and lower bounds for $p<4$. This can be understood as follows: The charge on the brane scales as $\mathcal{T}^{p/4}$. For $p>4$, the charge on the brane increases with increasing tension, which increases the nucleation rate (in analogy to the Schwinger pair production rate). Since the \gls{flb} requires that decays happen not too fast, it is natural to understand this as an upper bound on the tension.
Furthermore, for 2-branes without world-volume gauge fields and in the absence of axions, the bound
\begin{equation}
    \mathcal{T}\gtrsim\left(gHM_\textnormal{P}\right)^{3/2}
\end{equation}
is derived.

\paragraph{Is there a \textnormal{magnetic} counterpart?}
In analogy to the magnetic \gls{wgc} (\cref{eq:mwgc}) that demands that the magnetic monopole is not a \gls{bh} and sets an upper bound on the cutoff scale of an \gls{eft}, one can demand that the magnetic monopole is not a Nariai \gls{bh}, which yields a lower bound on the gauge coupling \cite{montero_fl_2021}:
\begin{equation}\label{eq:fl_magnet}
    g^2\geq\frac{3}{2}\left(\frac{H}{M_\textnormal{P}}\right)^2.
\end{equation}
This bound assures that a Nariai solution exists, and that the magnetic monopole (which is a fundamental particle and not a \gls{bh}) fits in the static \gls{ds} patch. Furthermore, this bound is equivalent to demanding that the \gls{eft} cutoff $gM_\textnormal{P}$ from the flat-space magnetic \gls{wgc} is not below the Hubble scale.

\subsubsection{Evidence}
The \gls{flb}
was first studied by \citet{montero_festina_2020}, and later discussed in the context of
type II string theory \cite{guidetti_axionic_2023} and
holography \cite{mishra_confinement_2023}.
A meticulous overview article is presented by \citet{montero_fl_2021}. \citet{hassan_sitter_2025} provide a compact discussion and extend the \gls{flb} to more general $p$-forms.
Furthermore, relations to other swampland conjectures are highlighted in \cref{rel:FLB_SSC,rel:WGC_FLB,rel:AdSDC_FLB}.

\subsection{Finite Number of Massless Fields Conjecture}\label{s:fnomf}

The number of massless fields that appear in a $d$-dimensional \gls{eft}, which is coupled to (Einstein) gravity, is finite \cite{brennan_string_2018}.

\subsubsection{Implications for Cosmology}
\paragraph{Inflation}
An aspect where the number of massless fields is of particular relevance is inflation. Some models explain the exponential inflation in the early universe by a collection of multiple fields, each described by an \gls{eos} of the form of
\begin{equation}\label{eq:FNomF_KG}
    \ddot{\phi}+3H\dot{\phi}+V^\prime=0,
\end{equation}
where each field experiences a restoring force from its potential, but also the Hubble friction \cite{dimopoulos_n-flation_2008}.
The Hubble term $3H^2=\sum_\phi V\left(\phi\right)$ depends on the collection of scalar field potentials \cite{dimopoulos_n-flation_2008}. Models that require an infinite number of fields belong to the swampland.

As we will elaborate in \cref{sss:FNomFC_Remarks}, a finite number of massless scalar fields is related to the finiteness of moduli space. \citet{cecotti_moduli_2020} argues that a finite moduli volume should be understood as a condition for having a finite action. Requiring a finite action for the Universe has interesting consequences: If finiteness from the past to the future is demanded, the lifetime of the Universe has to be finite, the Universe needs to be spatially compact, and a cosmological constant, massless scalar fields, eternal inflation, the ekpyrotic scenario, as well as bounce and cyclic models are ruled out \cite{barrow_finite_2020}. If only finiteness up to now, without making predictions about the future, is required, then inflation is allowed as a transient phase, eternal inflation is still ruled out, yet the ekpyrotic scenario (though requiring an infinite field displacement, which violates the \gls{dc}) as well as cyclic models (though highly constrained and requiring a singular beginning) can be accommodated \cite{jonas_cosmological_2021}.

\paragraph{Dark Energy and Dark Matter}\label{p:FNomF_DE_DM}
A finite number of vacua can rule out the anthropic principle \cite{marsh_exacerbating_2017}: The number of vacua simply might be too small to explain the necessary fine-tuning. This is in particular suggested for models where \gls{dm} interacts with \gls{de}. The variation in the \gls{de} density $\rho_\Lambda$ is related to the variation in the \gls{dm} mass as $\delta\rho_\Lambda\sim\delta m_\text{DM}m_\text{DM}^3$, which indicates that a small shift in \gls{dm} mass yields a large shift in the \gls{de} density. For the anthropic principle to hold, the number of available vacua must be larger than the required fine-tuning. \citet{marsh_exacerbating_2017} finds fine-tuning of the order of $10^{10^{10}}$ necessary for simple examples of \gls{dm}\textendash\gls{de} interactions, which vastly exceeds even the most generous estimates of the number of string theory vacua of $\num{e500}$ for type IIB string theory or $\num{e272000}$ in F-theory \cite{taylor_f-theory_2015}.\footnote{
    Going into the opposite direction, \citet{bousso_geometric_2011} reason that the number of vacua that allow for observers to exist, $\mathfrak{N}$, is related to the time \gls{de} begins to dominate, $t_\Lambda$, the time it takes for observers to exist, $t_\textnormal{obs}$, as well as to the timescale related to spatial curvature, $t_\textnormal{c}$: $\log t_\Lambda\approx\log t_\textnormal{obs}\approx\log t_\textnormal{c}\approx\log\sqrt{\mathfrak{N}}$\,\textemdash\,this yields  $\mathfrak{N}\sim\num{e123}$, which is smaller than the number of flux vacua in string theory. Here, no coupling between \gls{de} and \gls{dm} was assumed. It is more to be understood as a naïve upper bound on the size of the landscape.
}
Observational evidence against the anthropic principle would include the observation of energy transfer between \gls{de} and \gls{dm}, which constrains masses\footnote{
    If \gls{dm} has a decay channel into two gamma photons, the mass can be constrained \cite{marsh_exacerbating_2017}. If such a decay is present during a redshift range of $20<z<1100$, the \gls{cmb} spectrum is damped at small scales due to the extra energy injection (which could ionise hydrogen and helium and increase the optical depth to recombination)\,\textemdash\,that such a damping is not observed puts upper bounds on the energy injection from this decay channel \cite{anchordoqui_cosmological_2024,slatyer_general_2017}.
}, couplings, and the relative energy density of \gls{dm}, $\Omega_\text{DM}$.
Making the anthropic principle implausible is not a problem per se, as a proper theory that predicts the observations might be preferred over the anthropic principle. Besides, \citet{dimopoulos_n-flation_2008} explain that having multiple fields could actually make fine-tuning obsolete, as the fields can dampen each other.

\citet{tomboulis_multigravity_2021} investigates if there are flat metric solutions in multigravity theory if there is a cosmological constant the size of the Planck mass\footnote{
    Multigravity needs Planck-sized cosmological constants (which arise from integrating out matter fields) in order to be consistent up to the Planck scale \cite{tomboulis_multigravity_2021}.
    }
and concludes that without fine-tuning there are no such solutions for a finite number of fields, but that there are flat solutions with an infinite-dimensional parameter space if there is an infinite number of fields. The fields considered are $n$ massive spin-2 fields, $n+1$ rank-2 tensor fields, but only one massless field. Having infinitely many fields indicates that the underlying fundamental theory is non-local \cite{tomboulis_multigravity_2021}.

\paragraph{Particle physics} produced only a limited number of fundamental particles, which agrees well with the notion that there should only be a finite number of massless scalar fields, as every parameter entering the Lagrangian has to be viewed as the vacuum expectation value of a field \cite{brennan_string_2018,vafa_cosmic_2019}.

\subsubsection{General Remarks}\label{sss:FNomFC_Remarks}

In earlier work, \citet{vafa_string_2005} states that this conjecture was not based on a mathematical fact, rather on experience from working in string theory. A concern could be that there might be solutions that successfully describe \gls{qg} but are not realised in string theory. For 10 dimensions, it can be shown that this is not the case: every solution free of inconsistencies is realised in string theory \cite{adams_string_2014}. Furthermore, there is more rigorous mathematical support for such a conjecture nowadays:
\begin{itemize}
    \item In all known examples, the dimension of the cohomology of the manifold is finite \cite{vafa_string_2005}. The cohomology class provides an upper bound for the number of massless modes on the manifold \cite{vafa_string_2005}.\footnote{
        Since there are no known string realisations without any extra scalar fields (there is always at least the string coupling or the volume modulus), a lower bound might also exist \cite{vafa_string_2005}.
    }
    \item For a $d$-dimensional theory, the rank of the gauge group is $r_G\leq26-d$ \cite{kim_four_2020}, ergo, the number of matter contents is finite \cite{montero_cobordism_2021}.
    \item Compactness of moduli space.
\end{itemize}
The last point needs to be debated.
The idea that there is only a finite number of massless modes is based on the conjecture that for a specific number of dimensions (and supersymmetry), only a finite number of \gls{cy} compactifications exists \cite{tarazi_finiteness_2021,heckman_fine_2019}.\footnote{
    The predominance of \gls{cy} manifolds in string theory can be explained by an observation by \citet{candelas_vacuum_1985}: To have chiral fermions and a number of supercharges that can reasonably be dealt with, a Ricci-flat manifold with SU(3) holonomy is sufficient \cite{tadros_low_2022}\,\textemdash\,\gls{cy} manifolds are Ricci-flat Kähler manifolds, and as such are smooth manifolds with a complex structure and a Riemann metric, i.e. a scalar product exist, and the space is differentiable.
}
\citet{blumenhagen_basic_2013} note that no exact \gls{cy} metric on a compact manifold has been constructed (only algorithmic\footnote{
    More recently also through machine learning \cite{anderson_moduli-dependent_2021}.
    }
approximations), whereas for non-compact manifolds, exact \gls{cy} metrics exist.
For a compact region of moduli space, the number of flux vacua is finite \cite{ashok_counting_2004}, respectively, a moduli space of finite volume has a finite number of different \gls{qg} vacua \cite{vafa_string_2005,hamada_finiteness_2022,acharya_finite_2006,douglas_finiteness_2005,kim_finite_2024}.\footnote{
    The number of massless scalars emerging from compactification is proportional to Hodge numbers of the internal manifold\,---\,it is assumed but not proven that there is an upper bound on Hodge numbers for compact \gls{cy} manifolds \cite{brennan_string_2018}.
}
While \citet{douglas_geometry_2006} say that \gls{cy} spaces are not compact,
\citet{kumar_comments_2021} mention that it is often assumed that the number of \gls{cy} threefolds is finite, even though it is not proven.\footnote{
    A finite number of \gls{cy} threefolds indicates a finite number of tuneable parameters in \glspl{eft} that are part of the landscape \cite{heckman_fine_2019}.
} 
\citet{gross_finiteness_1993} shows that there is only a finite number of \gls{cy} manifolds that admit an elliptic fibration \cite{kumar_bound_2009}, and the tameness conjecture (\cref{sec:tame}) implies that there is only a finite number of topologically distinct compact \gls{cy} manifolds.
Further support comes from the \gls{dc}: the moduli space for an \gls{eft} with a finite cutoff scale must be compact, as otherwise infinite distance limits exist, where massless towers of states emerge that cause the breakdown of the \gls{eft} \cite{hamada_finiteness_2022}.\footnote{
    For example, if the moduli space of a probe brane is not compact, there is an infinite number of internal states, which violates the Bekenstein bound \cite{hamada_finiteness_2022,bedroya_compactness_2022,hamada_8d_2021}.
}
The discussion could be resolved by having a finite landscape of theories and an infinite swampland \cite{kim_finite_2024,taylor_infinite_2018}.\footnote{
    Having an infinite swampland and a finite landscape is to be expected: The landscape is spanned by the different moduli respectively the different compactifications that can be performed, leading to a discrete set, while the swampland contains all the possible choices of couplings or operator coefficients to build \glspl{eft}, which are continua \cite{hebecker_lectures_2023}.
}

\citet{long_desert_2021} state that the following reasoning in favour of the conjecture that the number of massless fields is finite is incorrect (yet we will reason that their counterargument is toothless):
Given a cutoff scale for an \gls{eft}, it is reasonable that there is a finite number of string vacua that can act as a background, which also means that the number of massless or light species modes allowed in the theory is finite. Each such mode can end up in a \gls{bh}, increasing its entropy. Without the cutoff, the number of such modes is infinite, which renders the \gls{bh} entropy infinite as well.
According to \citet{long_desert_2021}, this reasoning is incorrect, as the \gls{bh} entropy could be finite even in the case of an infinite number of modes:
The radius of a \gls{bh} that is fully described by the \gls{eft} is $r>1/\Lambda$.
The entropy of such a \gls{bh} is given by $\mathcal{S}>\left(M_\textnormal{P}/\Lambda\right)^{d-2}$ and should be bigger than the number of species in the theory, as each mode can be a constituent of the \gls{bh}: $\mathcal{S}>N_\textnormal{S}$.
It follows that there is no contradiction between an infinite number of species and the finiteness of \gls{bh} entropy if $N_\textnormal{S}<\left(M_\textnormal{P}/\Lambda\right)^{d-2}$ respectively $\Lambda\leq\Lambda_\textnormal{S}=M_\textnormal{P}/N_\textnormal{S}^{1/\left(d-2\right)}$ holds \cite{dvali_black-bound_2008,dvali_black_2010,dvali_species_2010}.
However, we'd like to point out that in this case the cutoff scale of the \gls{eft} is going to zero, which makes the \gls{eft} invalid everywhere, effectively ruling out \glspl{eft} with an infinite number of massless fields.

For \gls{ads} space, there is an argument against an infinite number of massless fields invoking the \gls{adsdc}, presented by \citet{lust_ads_2019}:
String theory backgrounds of the form \gls{ads}$_d\cross S^{D-d}/Z_k$ allow for an infinite number of massless fields by sending $k\rightarrow\infty$. In the absence of separation of scales, $S^{D-d}$ cannot be made arbitrarily small compared to the \gls{ads}$_d$ radius. The \gls{adsdc} states that there is no separation of scales. Therefore, $S^{D-d}$ is always part of the full spacetime, and the infinite number of massless modes is localised on the codimension-${\left(D-d\right)}$ defect, which is allowed in \gls{qg}. This is not possible in pure \gls{ads} space. This leaves us with two possible conclusions: either, \gls{ads} space cannot be purely flat, i.e. there is always an internal space $Y_p$, such that \gls{ads}$_d\cross Y_p\rightarrow\text{Mink}_d$, or, there is only a finite number of massless fields.

\paragraph{How could an infinite number of vacua even emerge?}
\citet{acharya_finite_2006} state three possibilities:
\begin{description}
    \item[Topology] The extra-dimensions could be described by an infinite number of different topologies.%
    \item[Fields] The solutions to the \glspl{eom} could be infinitely numerous, ranging from different metrics to brane configurations or gauge fields.%
    \item[Fluxes] The number of fluxes could be infinite.%
\end{description}
Regarding topologies, there is a theorem by \citet{cheeger_finiteness_1970}, according to which a sequence of Riemannian manifolds with sectional curvatures\footnote{
    Without the bound, there might not be an \gls{eft} for the specific compactification of the manifold \cite{hamada_finiteness_2022}.
    }
and diameters bounded from above and volumes bounded from below,\footnote{
    Without bounds on the diameters and the volumes, infinite towers of states emerge in the infinite distance limits, causing the \glspl{eft} to break down \cite{hamada_finiteness_2022}.
    }
has only a finite number of diffeomorphisms \cite{hamada_finiteness_2022}.
This implies that there can only be a finite number of topologies in a sequence, if the manifolds have a non-zero volume and are not too elongated, and if all components of the Riemann tensor are bounded.
\citet{acharya_finite_2006} present explicit examples of infinite sequences of Einstein 7-manifolds as infinite topologies that fail because of Cheeger's theorem:
On the one hand, all components of the Riemann curvature tensor are bounded because every closed geodesic is longer than the string scale. As long as strings have a non-zero length, the volume of the extra dimensions will always be positive, i.e. there is a lower bound on the volume. If a manifold is too elongated, it had an infinite diameter, which contradicts the \gls{dc}.
On the other hand, an infinite sequence of manifolds with non-negative volume leads to infinite volumes in the curled-up dimensions, i.e. a contradiction in itself, since the extra-dimensions are defined to be small, which means that they cannot be infinitely large. Only a finite number leads to finite volumes of the extra dimensions.

Regarding the number of fields and fluxes, \citet{acharya_finite_2006} point out that distinct solutions are distinct points in moduli space, but if they lead to the same \gls{eos}, they should be close together, so close in fact that they do not physically present different vacua. We think that this is not necessarily true and that a more elaborate proof is needed. On the one hand, largely separated points in moduli space could represent theories that have aspects that cancel each other in the \gls{eos}. On the other hand, even a small shift in moduli space could mean a large difference\,\textemdash\,think of phase transitions \cite{acharya_finite_2006}.
Moreover, \citet{bakker_finiteness_2023} show that in order to exclude the pathological examples of \citet{acharya_finite_2006}, it suffices to show the definability of the period mapping in the complex structure moduli space.
Furthermore, support from a broader string-theoretical context for a finite number of fields and especially fluxes is listed in \cref{sss:FNomFC_Evidence}.

\paragraph{What goes wrong at infinity?}
An infinite number of fields within an \gls{eft} does not have the same implications as an infinite number of different vacua in moduli space.
On the one hand, the latter might not be too problematic, as the real world would only occupy a finite ball in the infinite-dimensional parameter space, with a radius shrinking with more precise experimental data \cite{kumar_review_2006}, i.e. the landscape would be finite, while the swampland would be infinite.
On the other hand, an infinite number of vacua would imply a non-compact configuration space, and a diverging partition function, as scalar fields probe all the vacua, even if the vacua are disconnected (which would contradict the cobordism conjecture (\cref{sec:cobordism})) \cite{hamada_finiteness_2022}. This is also supported by Cheeger's theorem \cite{cheeger_finiteness_1970}, as well as by the work of \citet{cecotti_moduli_2020}, who argues that a finite moduli volume should be understood as a condition for having a finite action.

An infinite number of fields in an \gls{eft} is more problematic: it leads to non-renormalisability, infinite entropy, and diverging partition functions.
(i) If there is an infinite number of couplings, Wilsonian renormalisability is not given \cite{chaudhuri_chl_2005}.
(ii) The entropy of a system is given by
\begin{align}
    \mathcal{S}&=\mathfrak{N}T^3L^3\\
    &=\mathfrak{N}^{1/4}E^{3/4}L^{3/4}\\
    &=\mathfrak{N}^{1/4}L^{3/2},
\end{align}
with $\mathfrak{N}$ the number of fields, $T$ the temperature, $L$ the size of the system, $E=\mathfrak{N}T^4L^3$ the energy in the system \cite{hamada_finiteness_2022}, which implies that an infinite number of fields yields an infinite entropy (which is also true for a system of infinite extent).
(iii)
A partition function
\begin{align}
    Z&=\sum_{\mathfrak{N}=-\phi_\text{max}\sqrt{\Lambda/2\pi^2}}^{\phi_\text{max}\sqrt{\Lambda/2\pi^2}}e^{-2\mathfrak{N}^2\pi^2/\phi_\text{max}T}\\
    &\simeq\frac{\phi_\text{max}}{\sqrt{2\pi^2}}\int_{-\sqrt{\Lambda}}^{\sqrt{\Lambda}}\!e^{-p^2/T}\,\mathrm{d}p\\
    &\simeq\frac{\phi_\text{max}}{\sqrt{2\pi/T}}\erf\left(\sqrt{\Lambda/T}\right),
\end{align}
with $T$ the temperature,
$\phi_\text{max}$ a maximum field value (the field could be periodic with $\phi\simeq\phi+\phi_\text{max}$),
and $\Lambda$ the theory cutoff, might diverge if the number of fields, $\mathfrak{N}$, is infinite, unless the \gls{dc}, the \gls{ssc}, and the \gls{wgc} are satisfied, such that $\phi_\text{max}$ is finite and that $\Lambda(\phi_\text{max})\leq1/\phi_\text{max}^2$ \cite{hamada_finiteness_2022}.

Studying charged \glspl{bh}, \citet{hong_causal_2010} find strong cosmic censorship and \gls{bh} complementarity\footnote{
    They show that an observer could see a quantum clone of information: once as in-falling and once in the form of Hawking radiation, if there is an exponentially large number of fields present in the theory.
}
violated, if the number of massless fields is exponentially large. In this case, they were able to extend semi-classical results behind the inner horizon, which indicates that the singularity becomes observable and the strong \gls{ccc} is violated. A finite number of massless fields conserves cosmic censorship, as trans-Planckian curvatures occur around the horizon, keeping the singularity a physical singularity, while the breakdown of the \gls{eft} is guaranteed, making it compatible with our expectations about \gls{qg}.

\subsubsection{Evidence}\label{sss:FNomFC_Evidence}
That the number of massless fields is finite was conjectured long ago, e.g. by \citet{vafa_string_2005,acharya_finite_2006}, and is sometimes also referred to as the \textit{Finite Flux Vacua (FFV) Conjecture} \cite{kumar_comments_2021}.\footnote{
    The number of vacua scales with the distance in flux space \cite{conlon_explicit_2004}.
}
It is supported by a broader string theory context \cite{grimm_exact_2024,douglas_chirality_2004,gross_finiteness_1993,martinez-pedrera_finding_2013,eguchi_distribution_2006,douglas_geometry_2006,grimm_moduli_2021,antoniadis_flux_2024,torroba_finiteness_2007,kumar_review_2006,plauschinn_flux_2023,ashok_counting_2004,denef_distributions_2004,douglas_critical_2006,douglas_spaces_2010,gannon_monstrous_1999,cappelli_-d-e_2009,douglas_spaces_2010,kumar_bound_2009,hamada_enumerating_2024,douglas_landscape_2007,blumenhagen_statistics_2005,lee_string_2017,grimm_taming_2021,acharya_finite_2006,douglas_chirality_2004,douglas_landscape_2007,eguchi_distribution_2006,douglas_geometry_2006,grimm_taming_2021,grimm_moduli_2021,antoniadis_flux_2024,braun_restrictions_2011,milanesi_type_2007,grimm_exact_2024,torroba_finiteness_2007,kumar_review_2006,plauschinn_flux_2023,ashok_counting_2004,denef_distributions_2004,douglas_critical_2006,kumar_comments_2021,kim_finite_2024,kumar_bound_2009,kumar_global_2010,grimm_finiteness_2024,raghuram_orders_2022,loges_134_2022,he_complete_2022,carifio_machine_2017,sammani_finiteness_2024,bedroya_non-bps_2023},
the tameness conjecture (\cref{sec:tame,rel:fnomfc_TC}),
the \gls{tpc} (\cref{rel:TPC_tame}),
and Hodge theory \cite{bakker_finiteness_2023},
as well as shown for vacua that resemble our Universe \cite{douglas_statistics_2003}.\footnote{
    Having infinitely many string vacua that resemble our Universe would cast serious doubt on the predictive power of string theory, respectively M-theory \cite{ashok_counting_2004,douglas_statistics_2003}.
    }
Relations to other swampland conjectures are highlighted in \cref{rel:fnomfC_DC,rel:TPC_finite,rel:SSC_fnomfC,rel:WGC_fnomfC}.

\subsection{Gravitino Swampland Conjecture}\label{sec:gravitino}%

\begin{displayquote}
    The sound speed\footnote{\enquote{In a general theory of many interacting fields, the scalar sound speed $c_\textnormal{s}$ may be understood as the determinant of the matrix of sound speeds of all fields kinetically coupled to the gravitino and with mass below the \gls{uv} cutoff} \cite{kolb_gravitino_2021,kolb_catastrophic_2021}.} of gravitino(s) must be positive-definite, $c^2_\textnormal{s} > 0$, at all points in moduli space and for all initial conditions, in all 4d effective field theories that are low-energy limits of quantum gravity \cite{kolb_catastrophic_2021}.
\end{displayquote}

\subsubsection{Implications for Cosmology}
\citet{kolb_gravitino_2021} present a null hypothesis and three possible outcomes, based on the \gls{gsc}:
The null hypothesis is that there is \enquote{a constant mass gravitino, no light superpartners at the end of inflation that kinetically couple with the gravitino, and a conventional inflationary thermal history.}
The three possible outcomes are:
\begin{itemize}
    \item The gravitino is observed. This will make it impossible to observe the B-mode polarization of the \gls{cmb}.
    \item B-mode polarization of the \gls{cmb} is observed. Then, it will be impossible to observe a gravitino in collider experiments.\footnote{Contrary \citet{cribiori_gravitino_2021}, which state that the detection of B-modes just sets a lower bound of \SI{100}{\giga\electronvolt} for the gravitino mass.}
    \item Gravitinos and B-mode polarization are observed. The null hypothesis has to be rejected.
\end{itemize}

\subsubsection{General Remarks}
For simple models of cosmic acceleration in \gls{ds}, the gravitino is well described by Rarita\textendash Schwinger \cite{rarita_theory_1941} models, in which the sound speed of the gravitino is given by 
\begin{equation}
    c_\textnormal{s}^2=\left(\frac{m_\textnormal{3/2}^2-\omega H^2}{m_\textnormal{3/2}^2+H^2}\right)^2,
\end{equation}
such that, in a radiation dominated universe, the sound speed vanishes when $H=\sqrt{3}m_\textnormal{3/2}$ \cite{kolb_gravitino_2021}.
The given definition of the \gls{gsc} reduces to the statement that the gravitino mass\footnote{
    The gravitino mass is sometimes interpreted as a measure of the supersymmetry breaking scale \cite{anchordoqui_scale_2023} $\Lambda_\text{SUSY}^2\simeq m_\textnormal{3/2}M_\textnormal{P}$\,\textemdash\,the absence of supersymmetry in particle accelerators and a high gravitino mass go hand in hand \cite{cribiori_sitter_2022}.
    }
$m_\textnormal{3/2}$ has to be above the value of the Hubble constant at the end of inflation $H_\textnormal{e}$:\footnote{
    Generally, the following holds: if $m<H$, i.e. when the \gls{kk} tower drops below the Hubble scale, the \gls{eft} would experience decompactification, which means that it is no longer a \gls{flrw} cosmology. If there is no \gls{kk} tower, but strings, then the universe would undergo a stringy phase when $m<H$, so, again, the \gls{eft} is no longer applicable, but a string theory is needed \cite{rudelius_revisiting_2023}. %
}
\begin{equation}
    m_\textnormal{3/2}>H_\textnormal{e}\label{eq:gravitinomass}
\end{equation}
for certain, but not necessarily all, supergravity models \cite{kolb_gravitino_2021,kolb_catastrophic_2021}.

\paragraph{What happens if the gravitino is \textnormal{too light}?}
The massive spin-3/2 state (a gravitino) carries a helicity-3/2 state and a helicity-1/2 state with decoupled \glspl{eom}.
For $m_\textnormal{3/2}<H$ the sound speed of the helicity-1/2 gravitino vanishes, which causes a divergent production of gravitationally produced particles with arbitrarily high wavenumber $k$ \cite{kolb_catastrophic_2021} (since with vanishing sound speed, the energy per field excitation is independent of the momentum \cite{kolb_gravitino_2021}).

The particle number of non-relativistic particles at late times is a physical observable, and a divergent production of gravitinos is therefore pathological \cite{kolb_catastrophic_2021}.
However, there are supergravities free of this pathology \cite{kolb_catastrophic_2021}, e.g. a cubic nilpotent superfield studied by \citet{terada_minimal_2021}.
\citet{dudas_slow_2021} find that neither $c_\textnormal{s}=0$ is a general feature of supergravities, nor that $c_\textnormal{s}=0$ always causes catastrophic gravitino production:
A nilpotent superfield eliminates a scalar from the theory and can be seen as an \gls{ir} limit of a heavy sgoldstino (the scalar field associated with supersymmetry breaking) \cite{ferrara_cosmology_2016}.
The mixing of the inflatino and the goldstino prevents the catastrophic gravitino production \cite{antoniadis_salvage_2021,hasegawa_gravitino_2017}.
The culprit is an additional constraint on the higher-derivative operators, which eliminates the pseudoscalar, fermionic, and auxiliary partners of the inflaton \cite{ferrara_cosmology_2016,dallagata_origin_2016}\,\textemdash\,these constraints introduce additional pathologies such as $c_\textnormal{s}>1$.

This catastrophic gravitino problem differs from the conventional gravitino problem \cite{khlopov_is_1984,ellis_cosmological_1984}, as the catastrophic particle production is driven by \gls{uv} modes and the conventional gravitino problem takes place during inflation/the period of reheating after inflation \cite{eberl_gravitino_2021,nilles_coupled_2001,nilles_nonthermal_2001,dalianis_constrained_2017}, i.e. within the \gls{ir} \gls{eft} \cite{kolb_catastrophic_2021}.

\paragraph{What supports the \gls{gsc}?}
The \gls{gsc} demands that the gravitino mass is above the Hubble scale. 
Other swampland conjectures rule out a vanishing gravitino mass:
The \gls{flb} says that charged gravitinos cannot be massless.
The no non-supersymmetric theories conjecture (\cref{sec:nononSUSY}) says that the superpotential $W$ in $\mathcal{N}=1$ supergravities cannot vanish, but this also means that the gravitino mass cannot vanish, as $m_\textnormal{3/2}^2=e^KW\Bar{W}$ ($K$ the Kähler potential) \cite{cribiori_sitter_2022}.
\citet{cribiori_gravitino_2021} show that the limit of vanishing gravitino mass corresponds to an infinite tower of states, a behaviour we know from the \gls{dc} that indicates the breakdown of the \gls{eft} (\cref{sec:distance}).
\citet{cribiori_sitter_2022} shows that for some models, the gravitino mass vanishes in the points of infinite volume\,\textemdash\,the same points where a global symmetry (\cref{sec:nGSym}) is restored.
\citet{cribiori_weak_2021} show that at \gls{ds} critical points the determinant of the gravitino mass matrix vanishes, and that the \gls{wgc} puts these points in the swampland.
Also \citet{coudarchet_geometry_2021} find inconsistencies if the gravitino mass vanishes.

\paragraph{Is the relation between the gravitino mass and the Hubble scale an exact inequality?}
No, the inequality as presented in \cref{eq:gravitinomass} certainly has missing $\order{1}$ parameters. However, we can gain more insights using the \gls{dc} (see text around \cref{eq:DCgravitino}).

Furthermore,
for supersymmetric \gls{ads} vacua, there is a direct relation between the gravitino mass and the cosmological constant
$\Lambda_\textnormal{cc}\geq-3m_\textnormal{3/2}^2$,\footnote{
    \citet{castellano_gravitino_2021} argue that the bound is saturated and the expression is an equality, whereas \citet{cribiori_gravitino_2021} state it as a bound, which would allow for finite gravitino mass in the $\abs{\Lambda}\rightarrow0$ limit.
    }
and it follows that $m\sim m_\textnormal{3/2}^\mathfrak{g}\sim\abs{\Lambda}^{\mathfrak{g}/2}$ \cite{castellano_gravitino_2021,cribiori_gravitino_2021}.\footnote{
    The parameter $\mathfrak{g}$ is defined in \cref{sec:distance} around \cref{eq:DCgravitino}.
}
The last expression strongly resembles the \gls{adsdc} from \cref{sec:AdSDC} \cite{cribiori_gravitino_2021,lust_ads_2019} (even though it is on an entirely different background and yields different predictions \cite{cribiori_sitter_2022}).\footnote{
    The \gls{gsc} applies to \gls{ads}, Minkowski, and \gls{ds}.
    On the one hand, the \gls{adsdc} does not require the gravitino mass to go to zero in the limit of vanishing potential. It could be another tower, and the gravitino mass could remain finite in this limit \cite{cribiori_gravitino_2021}.
    On the other hand, the \gls{gsc} is independent of the potential (when supersymmetry is broken) \cite{cribiori_gravitino_2021}.
    However, the limit $m_\textnormal{3/2}\rightarrow0$ always implies that the potential vanishes as well \cite{cribiori_gravitino_2021}.
}

\subsubsection{Evidence}
The \gls{gsc} was introduced by \citet{kolb_catastrophic_2021,kolb_gravitino_2021} and studied in the context of
supergravities \cite{terada_minimal_2021}, and
4d $\mathcal{N}=1,2$ supersymmetries \cite{cribiori_gravitino_2021}.

\subsection{No Global Symmetries Conjecture}\label{sec:nGSym}

In the presence of gravity, there are no continuous global symmetries \cite{banks_constraints_1988}.

In this article, the no global symmetries conjecture is an argument we often refer to, e.g. when mentioning that an infinite distance limit or the limit of vanishing coupling is problematic because a global symmetry arises in such a limit.\footnote{
    For example, \citet{cordova_generalized_2022}, who motivate the \gls{dc} as well as the \gls{wgc} from the no global symmetries conjecture, and
    \citet{montero_fl_2021} show that if the no global symmetries conjecture (\cref{sec:nGSym}) is violated in supergravities, the \gls{flb} is violated as well.
}
Yet, the conjecture that there are no global symmetries in \textit{\gls{qg}} is not necessarily itself a \textit{swampland} conjecture, as it does not really constrain \glspl{eft}: An \gls{eft} can have global symmetries. It is only expected that those symmetries are explicitly broken or gauged at higher energies \cite{grana_swampland_2021,lee_tensionless_2018,montero_chern-simons_2017}.\footnote{
    This is not a string theory\textendash exclusive result, e.g. unbroken chiral symmetries are discussed in loop \gls{qg} \cite{barnett_fermion_2015,gambini_no_2015}, and truncated renormalisation group flows in asymptotic safety scenarios in Euclidean space show no symmetry-violating terms \cite{eichhorn_light_2011,eichhorn_quantum_2017,eichhorn_viability_2017,eichhorn_nonminimal_2018,eichhorn_constraining_2021}, yet at higher order or in Lorentzian spacetime, these symmetries might well be broken.
}
Still, given the importance of the conjecture for the swampland programme, we would like to make a few humble comments, without attempting to deliver a complete picture or comprehensive review.

\subsubsection{Implications for Cosmology}
Some ideas rely on the existence of global symmetries, e.g. global or semilocal strings, global monopoles, textures \cite{burgess_continuous_2008,kamionkowski_planck-scale_1992}, and domain walls, which can either be ruled out based on the no global symmetries conjecture alone, or in combination with observational data \cite{afzal_nanograv_2023}.

\paragraph{Axions} might be required in \gls{qg} to break global symmetries \cite{draper_snowmass_2022,mcnamara_cobordism_2019,heidenreich_chern-weil_2021}.

\paragraph{Black holes}\label{p:nGSym_BH} are expected to decay via Hawking radiation and their entropy is expected to be finite. Moreover, they are expected to have no hair. Furthermore, it is expected that there are no naked singularities.
None of those expectations would be met if there were global symmetries.

If we had a continuous global symmetry, there would be \glspl{bh} carrying global charge.
If \glspl{bh} evaporate via Hawking radiation, the evaporation process is unitary and consistent with the Bekenstein\textendash Hawking entropy formula ($S=A/4G$) \cite{harlow_global_2021}.
Since Hawking radiation is insensitive to charge\footnote{
    The degree of global symmetry violation by Hawking radiation can be quantified using relative entropies \cite{chen_signatures_2021}.
},
either charge conservation is violated \cite{brennan_string_2018}, or
we would end up with a remnant\footnote{
    We make more remarks on remnants in \cref{p:WGC_Remnants}.
    }
carrying the global charge \cite{harlow_weak_2023,rudelius_topological_2020,banks_symmetries_2011,zeldovich_new_1976,hawking_particle_1975}.\footnote{
    This would happen in a finite time and might already be observable after the Page time (see \cite{almheiri_entropy_2019,almheiri_page_2020,penington_entanglement_2020,penington_replica_2020}), i.e. it is not necessary to wait until the \gls{bh} fully evaporates \cite{chen_signatures_2021}.
    }
Such a remnant represents a naked singularity, which is incompatible with the \gls{ccc}\footnote{
    The \gls{ccc} is a classical conjecture about \gls{gr} and not a string-theoretical swampland conjecture, and is therefore not treated in this review. However, a brief discussion can be found in \cref{p:WGC_CCC} in the context of the \gls{wgc}.
    }.
Each remnant corresponds to a microstate in the theory.
The \gls{bh} information paradox tells us that a \gls{bh} of finite size can only hold a finite amount of information, and therefore the number of microstates is finite\footnote{
    The Coleman\textendash Mermin\textendash Wagner theorem \cite{coleman_there_1973,mermin_absence_1966} states that there is no spontaneous symmetry breaking of global symmetries in vacuum states or thermal states \cite{montero_chern-simons_2017}. The theorem applies to the IR of the Euclidean solution\,\textemdash\,where a \gls{bh} can be regarded as a thermal state (despite the Hawking radiation)\,\textemdash, and predicts that the expectation value of any charged operator vanishes, which in turn renders the global charge completely unobservable outside the \gls{bh} and the \gls{bh} microstate infinitely degenerate \cite{montero_chern-simons_2017}.
}
as well\,\textemdash\,with a continuous global charge we would end up with an infinite number of different remnants and therefore infinite entropy \cite{harlow_global_2021,banks_symmetries_2011,bousso_covariant_1999,reece_tasi_2023,harlow_symmetries_2019,hsin_violation_2021}.\footnote{
    This argument about \gls{bh} thermodynamics does not only rule out continuous global symmetries, but applies, in a similar fashion, to non-compact continuous gauge symmetries as well, which indicates that Abelian gauge groups in \gls{qg} are U(1) and not $\mathbb{R}$, ergo, electric charge is quantised \cite{reece_tasi_2023}.
    As a further aside, we note that gravitons are neutral under the global symmetry, which means that gravitational bremsstrahlung cannot radiate away the global charge either \cite{van_beest_lectures_2022}.
    }

A too-naive way out of the dilemma is to assume that global symmetries exist but that it is Hawking radiation, which is the culprit and the wrong description of the decay process: 
the global charge of a \gls{bh} is a one-point function of a simple operator\,\textemdash\,being wrong about this indicates that the \gls{eft} cannot be trusted at all \cite{reece_tasi_2023}.
\citet{reece_tasi_2023} advances the notion that it is more likely to discard continuous global symmetries than to abandon several decades of successful and self-consistent research about \gls{bh} thermodynamics. We would like to raise the point that paradigmatic shifts keep happening in physics. Fortunately, there are stronger arguments against continuous global symmetries and this is probably not the paradigmatic shift we require.

Besides continuous symmetries, we could also consider discrete global symmetries, as in this case, the number of stable remnants might be large but finite. However, at least in the context of \gls{adscft}\footnote{
    In \gls{adscft}, \gls{bh} evaporation is unitary \cite{harlow_global_2021}.
}
\cite{harlow_constraints_2019,harlow_symmetries_2019}, the compatibility with discrete global symmetries with \gls{qg} is generally questioned \cite{harlow_global_2021}.\footnote{
    \glspl{qg} in less than $\left(3+1\right)$ dimensions with global symmetries do not have \glspl{bh} described by the Bekenstein\textendash Hawking entropy, ergo, the argument does not apply \cite{harlow_global_2021}.
    }

Another issue that arises in the context of global charges and \glspl{bh} is the no-hair theorem. To be compatible with the no-hair theorem, the global charge is only allowed to show as quantum hair \cite{dvali_black_2013}.\footnote{\label{foo:BHhair}Quantum hair arises when a continuous gauge group is Higgsed to a discrete subgroup \cite{dvali_black-bound_2008,krauss_discrete_1989}.
    Classically, one can build \glspl{bh} with arbitrary charge $Q=\int_{S^2}\!B$ \cite{montero_chern-simons_2017}:
    If we have a field $B=Q\mathrm{d}\Omega$ with $Q$ the charge and $\mathrm{d}\Omega$ the unit sphere $S^2$ volume form, where $B$ is not a local observable, no issue arises, as any contractible region can be gauged away.
    Furthermore, quantum hair from massive higher-spin gauge fields can be realised \cite{dvali_black_flavors_2006,dvali_black_spin_2006,dvali_black-bound_2008}.
    The charges related to Aharonov\textendash Bohm phases are only detectable quantum-mechanically \cite{dvali_nature_2008,dvali_black-flavors_2006}.
}
This could mean that the no-hair theorem is only semi-classically true, but not in \gls{qg}, respectively it only shows at higher energy scales.
Without being able to measure the charge, the uncertainty regarding that charge is infinite, which means that the entropy is also infinite, which is in contradiction with our expectation that the entropy of a \gls{bh} is finite and related to its mass squared \cite{palti_swampland_2019}.

Contrary to the presented points, \citet{dvali_skyrmion_2016} argue that semi-classical arguments based on \gls{bh} remnants do not hold: global charges can be conserved by \glspl{bh} and show as classical skyrmion hair.\footnote{
    A skyrmion \gls{bh} is a baryonic \gls{bh} where the \gls{bh} horizon is inside the baryon \cite{dvali_skyrmion_2016}.
}
This does not rule out the no global symmetries conjecture, but it voids arguments in favour of it based on semi-classical \gls{bh} physics.
Also \citet{dvali_black_2013} argue that the \gls{bh} argument cannot be used to rule out global symmetries, as higher-order quantum effects always allow deducing the global charge of a \gls{bh}.

To summarise, we can say that throwing global charge in a \gls{bh} can lead to three different outcomes:
\begin{itemize}
    \item The \gls{bh} swallows the global charge and keeps it, eventually leading to stable remnants, at odds with \gls{bh} thermodynamics.
    \item The \gls{bh} destroys the global charge, being at odds with charge conservation.
    \item  The \gls{bh} radiates the global charge away, being at odds with the no-hair theorem.
\end{itemize}

\paragraph{Inflation}
Some models of inflation might show (approximate) global symmetries regarding some aspect of the model, e.g. Starobinsky inflation has a global scale invariance related to the $R^2$-term, which is broken by the Einstein\textendash Hilbert term in the action \cite{ketov_starobinsky_2024}.

\paragraph{Wormholes}
Early work by \citet{kallosh_gravity_1995} indicates that pure Einstein-gravity in 4d is incompatible with global symmetries, but that the addition of axions or stringy effects might remove those incompatibilities, allowing for global symmetries in the context of wormholes.

If one includes Euclidean wormholes\,\textemdash\,closed connected spatial components of the universe (baby universes\footnote{
    A baby universe carries only a global charge, no gauge charge \cite{mcnamara_baby_2020,coleman_black_1988}. 
    })\,\textemdash\,in
the gravitational path integral, global charge can end up in those regions, inaccessible to the observer in the parent universe, which violates charge conservation \cite{giddings_loss_1988,abbott_wormholes_1989,kallosh_gravity_1995,harlow_weak_2023,hsin_violation_2021}. This is not an issue for gauge charge, as gauge charge must be zero in the closed universe \cite{harlow_weak_2023}.\footnote{
    If the no global symmetries conjecture respectively the cobordism conjecture (\cref{sec:cobordism}) hold, the Hilbert space of a baby universe is one-dimensional \cite{agmon_lectures_2023}.
}

\subsubsection{General Remarks}
\paragraph{Between which symmetries should we distinguish?}
Global symmetries describe unitary transformations of quantum states and act non-trivially on the Hilbert space \cite{grana_swampland_2021,van_beest_lectures_2022} and commute with the Hamiltonian \cite{prieto_moduli_2024}, while gauge symmetries describe mathematical redundancies in the description of the Hilbert space of the theory \cite{fichet_approximate_2020}.
A continuous symmetry leads to a conserved worldsheet current, which in turn serves as the vertex operator for a gauge boson \cite{banks_constraints_1988,fichet_approximate_2020}.
A symmetry can be discrete or continuous, and it can be gauged or spontaneously\footnote{
    Spontaneously broken symmetries keep their topological operators, which we assume to be forbidden in \gls{qg} \cite{heidenreich_chern-weil_2021}.
}
or explicitly broken.\footnote{
    \citet{montero_chern-simons_2017} reason that Chern\textendash Simons terms\,\textemdash\,describing the most generic breaking mechanism\,\textemdash\,must be present in a theory that can be consistently extended into the UV.
    \citet{garcia-valdecasas_non-invertible_2023} opposes this idea and states that Chern\textendash Simons terms are only a signal of non-invertibility of symmetries, not of their breaking.
}
In particle physics, ((continuous) global) symmetries play an important role, as the couplings and interactions are determined by (approximate (continuous)) global symmetries \cite{burgess_continuous_2008}, and the very weak couplings of local/gauge symmetries can be treated as global symmetries \cite{cicoli_string_2023,harlow_weak_2023,daus_towards_2020,fichet_approximate_2020}.

\paragraph{Where do global symmetries come from?}
Models with approximate global symmetries are not continuous deformations of a theory with exact global symmetries, instead, such models emerge as \glspl{eft}, where the small parameters that describe the approximate global symmetries are of dynamical origin and not free parameters \cite{fichet_approximate_2020}.
Global symmetries often emerge in the infinite distance limit of an \gls{eft}, but are censored by the appearance of an infinite tower of light states \cite{lee_tensionless_2018}.\footnote{
    A behaviour known from the \gls{dc}.
}

A global symmetry can arise from compactification. Compactifying a $D$-dimensional theory with a $\left(p+1\right)$-form to $d=p+3$ dimensions can lead to global symmetries \cite{montero_chern-simons_2017}. This makes the actual application of the no global symmetries conjecture as a swampland conjecture troublesome, as it doesn't allow us to rule out a theory of \gls{qg} from the \gls{eft} perspective.
Yet, it highlights why global symmetries on an \gls{eft}-level are gauge symmetries on the full \gls{qg}-level \cite{brennan_string_2018}:
Global symmetries usually originate from symmetries of compact extra dimensions, which are gauged since diffeomorphisms of the internal space are part of the gravitational gauge symmetry (cf. \cite{tadros_low_2022,polchinski_string-theory-volume-2_1998}).

\paragraph{To what degree are global symmetries allowed in \glspl{eft}?}
The \gls{wgc} provides us with a breaking scale, respectively a limit for the \enquote{globalness} of a symmetry.
The size of the symmetry-violating effects in the IR \gls{eft} are estimated to be $\exp\left(-8\pi^2M_\textnormal{P}^2/\Lambda\right)$ with $\Lambda$ the cutoff scale of the \gls{eft} \cite{draper_snowmass_2022,kallosh_gravity_1995,fichet_approximate_2020,daus_towards_2020}. This gives rise to the critique that this conjecture does not yield meaningful phenomenological constraints, as the gauging / breaking scale is not specified \cite{grana_swampland_2021}.
Furthermore, there are examples of lower-dimensional theories with global symmetries \cite{borissova_non-local_2024,harlow_global_2021}.\footnote{
    But those theories are no \gls{qg}-candidates to actually describe our Universe \cite{borissova_non-local_2024}.
}

\paragraph{What are the holographic arguments against global symmetries?}
If we have a \gls{cft} with a global symmetry, it will have a massless particle that corresponds to the gauge field of that symmetry, but this means that the symmetry is local in spacetime and not global \cite{banks_constraints_1988,cicoli_string_2023}, a contradiction with the initial assumption that the symmetry was global, ergo, global symmetries on \glspl{cft} are self-contradictory.

\citet{harlow_symmetries_2019,harlow_constraints_2019} use the \gls{adscft} correspondence to show that there are no global symmetries in (holographic) \gls{qg}. This agrees with the notion that a global symmetry on the worldsheet is always gauged away in the target space \cite{palti_swampland_2019,banks_constraints_1988}.\footnote{
    A global symmetry on the string world-sheet would lead to global spacetime symmetries, which do not exist \cite{banks_symmetries_2011}. 
}
This does not mean that \gls{qg} cannot have gauge symmetries. However, \citet{palti_swampland_2019} notes that a \enquote{theory with a gauge symmetry, coupled to gravity, must have states of all possible charges (consistent with Dirac quantization) under the gauge symmetry.}\footnote{This is the \gls{cc}.}

\subsubsection{Evidence}
The no global symmetries conjecture is highly intertwined with the cobordism conjecture (\cref{sec:cobordism}) and the \gls{cc}, as well as other swampland conjectures \cite{rudelius_symmetry-centric_2024}, as we point out in \cref{rel:CC_nGSymC,rel:Cobord_nGSym,rel:WGC_nGSC,rel:nGSC_AdSDC}.
However, as we stated in the beginning, we do not consider the no global symmetries conjecture a \textit{swampland conjecture}.
The literature on global symmetries is too extensive and beyond the scope of this article to be exhaustively reviewed here.
An introduction to the topic and related questions is presented by \citet{reece_tasi_2023,agmon_lectures_2023,van_beest_lectures_2022}.
An overview of arguments supporting the no global symmetries conjecture is given by \citet{harlow_weak_2023}.
Support for this conjecture comes in the form of
\gls{bh} physics \cite{banks_symmetries_2011,baek_global_2024}, and
topological \cite{harlow_global_2021,chen_signatures_2021,hsin_violation_2021,abbott_wormholes_1989,yonekura_topological_2021},
heuristic \cite{susskind_trouble_1995},
worldsheet \cite{banks_constraints_1988}, and
holographic \cite{harlow_constraints_2019,harlow_symmetries_2019,montero_are_2017,ashwinkumar_duality_2023,mcnamara_baby_2020,belin_charged_2021}
arguments.
Furthermore, the conjecture is studied in the context of asymptotic safety by \citet{eichhorn_absolute_2024}.
Moreover, there is a growing body of work on
non-invertible \cite{heidenreich_non-invertible_2021,choi_non-invertible_time_2023,garcia-valdecasas_non-invertible_2023,cordova_generalized_2022,koide_non-invertible_2021,choi_non-invertible_global_2022,choi_non-invertible_duality_2022,choi_non-invertible_condensation_2023,choi_non-invertible_Gauss_2023,kaidi_kramers-wannier-like_2022,benini_factorization_2023,roumpedakis_higher_2023,bhardwaj_non-invertible_2023,arias-tamargo_non-invertible_2023,hayashi_non-invertible_2022,kaidi_non-invertible_2022,cordova_non-invertible_2022,antinucci_continuous_2022,bashmakov_6d_2022,bashmakov_non-invertible_2023,damia_non-invertible_2023,damia_continuous_2023,bhardwaj_universal_2022,bartsch_non-invertible_2024,lin_decomposition_2022,etxebarria_branes_2022,apruzzi_non-invertible_2023,heckman_branes_2023,freed_topological_2024,niro_exploring_2023,kaidi_symmetry_2023,mekareeya_mixed_2023,antinucci_holography_2022,chen_solitonic_2023,karasik_anomalies_2023,cordova_neutrino_2022,decoppet_gauging_2023,etxebarria_goldstone_2022,nguyen_semi-abelian_2021,kaidi_higher_2022,wang_gauge_2022}
and generalised global symmetries \cite{montero_chern-simons_2017,gaiotto_generalized_2015,cordova_snowmass_2022,mcgreevy_generalized_2023,cvetic_generalized_2023}
and topological operators \cite{rudelius_topological_2020,wang_fermionic_2023}.\footnote{
    A technical argument against topological operators in \gls{qg} is presented by \citet{reece_tasi_2023}: \enquote{the gravitational path integral sums over all spacetimes, including those of nontrivial topology, and admits topology-changing transitions. Topology is not expected to be an invariant property in quantum gravity, and so it should not be possible to construct well-defined operators that correspond to topological invariants.}
}

\subsection{Non-Negative Null Energy Condition Conjecture}\label{s:nnNECC}%
Every 4d theory in flat or closed space with a completion to M-theory satisfies the \gls{nec}, i.e.
\begin{equation}
    T_{\mu\nu}l^\mu l^\nu\geq0
\end{equation}
for a light-like vector $l^\mu$ and the matter stress\textendash energy tensor $T_{\mu\nu}$ \cite{bernardo_four-dimensional_2021}.

\subsubsection{Implications for Cosmology}
Violating the \gls{nec} allows for wormholes\footnote{
    A wormhole solution that satisfies the \gls{nec} but requires closed timelike curves is presented by \citet{engelhardt_holographic_2016}.
} \cite{visser_traversable_2003}, warp-drives, and time machines \cite{visser_energy_2000}.
Forbidding \gls{nec}-violation rules these out.

\citet{folkestad_penrose_2023} studies the Penrose inequality\footnote{
    The Penrose inequality $m_\text{ADM}\geq\sqrt{A/16\pi}$ guarantees that the \gls{adm} mass\,\textemdash\,measuring the total mass\textendash energy content of an asymptotically flat spacetime\,\textemdash\,is always bigger than the mass of a \gls{bh} that would cover the same surface area as the local spacetime patch. A loose interpretation of this might be that a universe cannot be smaller than a \gls{bh} of the same mass as the universe, i.e. a universe will always be bigger than a \gls{bh} of the same mass. It also relates to the \gls{ccc} and rules out naked singularities.
}
and finds that satisfying the \gls{nec} is not a sufficient condition to guarantee that the Penrose inequality is satisfied: moreover, the theory has to be supersymmetric or satisfy the stronger dominant energy condition.\footnote{
    A theory that satisfies the dominant energy condition $T_{\mu\nu}v^\mu u^\nu\geq0$ for any time-like vectors $u^\mu$, $v^{\nu}$ automatically satisfies the \gls{nec}.
}

\citet{montefalcone_dark_2020} claim that for cosmic acceleration to happen, one of the following three conditions must be violated:
\begin{enumerate}
    \item The \gls{nec}
    \item The compact dimensions are of fixed size
    \item The metric is (conformally) Ricci-flat
\end{enumerate}
They argue that for models that satisfy all three conditions, Newton's gravitational constant would have to vary more strongly than compatible with observational bounds. Furthermore, they note that if the \gls{nec} is violated, the source of the \gls{nec}-violation has to reside in the compact dimension and vary in such a way that it precisely tracks the observed \gls{eos} of \gls{de}, $w_\textnormal{DE}(z)$. On the one hand, this sounds feasible, given the popularity of tracking solutions for scalar fields. On the other hand, it sounds unnecessary: The size of compact dimensions is parametrised by dynamical moduli. Furthermore, decompactifications occur on finite timescales, which allows transient phases of cosmological acceleration.

\paragraph{Dark energy} models with $w<-1$ \cite{sawicki_hidden_2013,arefeva_null_2008,qiu_null_2008,creminelli_starting_2006}, e.g. phantom fields or some forms of $k$-essence, violate the \gls{nec} \cite{bernardo_four-dimensional_2021}, and are therefore in the swampland.

\paragraph{Inflation} with a cosmological bounce in flat spacetime violates the \gls{nec}\footnote{
    \citet{molina-paris_minimal_1999} present a model that satisfies the \gls{nec} if the \gls{sec} is violated instead.
    }
\cite{qiu_bouncing_2011,lin_matter_2011,cai_bouncing_2007,brandenberger_bouncing_2017,de_rham_unitary_2017,easson_g-bounce_2011,creminelli_starting_2006} and is therefore in the swampland.
The same holds for \textit{Superinflation}, where the Hubble parameter grows during inflation \cite{baldi_inflation_2005,bernardo_four-dimensional_2021}.

\subsubsection{General Remarks}
Since any inequality of the form $f(u_1,u_2)\geq0$ with timelike vector fields $u$ and continuous function $f$ implies $f(l_1,l_2)\geq0$ for null vector fields as well, stronger energy conditions, like for example the \gls{sec}, imply the \gls{nec} \cite{bernardo_inheritance_2023,maeda_energy_2020}.

The \gls{nec} itself is neither derived from fundamental principles \cite{curiel_primer_2017,parikh_derivation_2015,parikh_thermodynamic_2017}, nor does it uphold in higher-dimensional theories (including M-theory), and there are various \glspl{eft} violating it \cite{bernardo_four-dimensional_2021,qiu_bouncing_2011,lin_matter_2011,cai_bouncing_2007,brandenberger_bouncing_2017,de_rham_unitary_2017,easson_g-bounce_2011,kobayashi_inflation_2010,creminelli_galilean_2010,alberte_relaxing_2016,visser_energy_2000,nicolis_erratum_2011,nicolis_energys_2010,rubakov_null_2014,sawicki_hidden_2013,arefeva_null_2008,qiu_null_2008,creminelli_starting_2006,buniy_null_2006,dubovsky_null_2006}, e.g. theories with superluminal propagation, pathological instabilities, or unitarity respectively causality violations \cite{wesley_oxidised_2009,koster_no-go_2011,carroll_can_2003,hsu_gradient_2004,cline_phantom_2004,dubovsky_null_2006,buniy_null_2006,tipler_causality_1976,eling_lorentz_2007,lee_dynamics_2011}.
\citet{das_constraints_2019} show that to have \gls{ds} expansion in 4d with static compactified dimensions, either the \gls{nec} is violated everywhere in the compactified dimensions, the extra-dimensions have positive curvature, or the 4d expansion sources warping. This follows from the geometry, i.e. the metric, irrespective of matter content or the Einstein field equations.
Imposing the \gls{nec} to higher-dimensional spaces yields geometric restrictions that cannot be produced from the lower-dimensional spacetimes, as this would require null vectors in the higher-dimensional space with vanishing components in external directions \cite{bernardo_four-dimensional_2021}. 
In M-theory, the \gls{nec} does not have to be satisfied in 11 dimensions, since higher-order terms allow the M-theory action to be free of pathologies respectively to cancel anomalies \cite{wesley_oxidised_2009}.\footnote{
    Yet if an energy condition is satisfied in a higher-dimensional manifold\,\textemdash\,given some reasonable assumptions, e.g. constant internal volume and positive well-defined internal metric (meaning Newton's constant is constant, and the internal metric is non-singular)\,\textemdash, the lower-dimensional energy condition is respected as well \cite{bernardo_inheritance_2023}.
    \citet{steinhardt_dark_2009,wesley_new_2008,wesley_oxidised_2009} argue that if the higher-dimensional theory satisfies the \gls{nec}, then accelerated expansion in 4d is only a transient phenomenon of a few $e$-folds\,\textemdash\,ruling out most models of inflation and forbidding various forms of \gls{de}. However, \citet{koster_no-go_2011} show that this only holds if one makes overly strong assumptions, namely it holds only in the absence of branes and localised matter sources for \gls{kk} compactifications that have a warp factor in the metric that depends on the volume modulus of the compact space\,\textemdash\,the last assumption is incompatible with a time-dependent setup, ergo cannot lead to accelerated expansion in 4d.
    \citet{deffayet_stable_2024} note that if a universe becomes trapped in a meta-stable phase without extra-dimensions decompactifying, then the \gls{nec} must be violated with a inhomogeneously distributed, time-varying component that is synchronised with the \gls{eos} of the decompactified \gls{eft}. However, none of their tested potentials in heterotic M-theory shows such a behaviour that would require \gls{nec}-violation.
}
\citet{bernardo_four-dimensional_2021} show that this cancellation takes place if there is a hierarchy among the higher-order terms, such that the M-theory action is well-defined, which imposes the following constraint on the string coupling:
\begin{align}
    \frac{\mathrm{d}g_\textnormal{s}}{\mathrm{d}\tau}&\propto g_\textnormal{s}^{1+1/n}\\
    {1+1/n}&\geq0\\
    \Rightarrow\frac{1}{n}&\geq-1.\label{eq:nnNECC}
\end{align}
This can be compared to the \gls{nec}:
\begin{align}
    -\frac{\mathrm{d}^2\log t^\mathfrak{c}}{\mathrm{d}t^2}+kt^{-2\mathfrak{c}}&\geq0\\
    \Rightarrow \mathfrak{c}+kt^{2\left(1-\mathfrak{c}\right)}&\geq0,
\end{align}
with $k\in\{0,\pm1\}$ the curvature.
For a 4d \gls{flrw} metric with $a(t)\propto t^\mathfrak{c}$ respectively $a(\tau)\propto \tau^n$, we find that $\mathfrak{c}=n/\left(1+n\right)$.
For flat and closed spacetimes ($k\in\{0,+1\}$), the \gls{nec} corresponds to \cref{eq:nnNECC}, yet for negatively curved space, \cref{eq:nnNECC} does not imply the \gls{nec}. This observation leads to the swampland conjecture.
The work by \citet{bernardo_four-dimensional_2021} rests on four assumptions:
\begin{enumerate}
    \item The external \gls{flrw} spacetime is parametrised by $a(\tau)\propto\tau^n$.\footnote{This holds for example for perfect fluids in 4d with $p=w\rho$ \cite{bernardo_four-dimensional_2021}.}
    \item The Wilsonian \gls{eft} description is well-defined, i.e. the quantum corrections have a hierarchy in the underlying M-theory.\footnote{This means that the quantum corrections are of different order and the action allows for truncation \cite{bernardo_four-dimensional_2021}.}
    \item $G_\textnormal{N}=\text{cst.}$\footnote{This does not rule out time-dependent fluxes or internal dimensions \cite{bernardo_four-dimensional_2021}.}
    \item The type IIB dilaton is time-independent.\footnote{This should not be too difficult to relax \cite{bernardo_four-dimensional_2021}.}
\end{enumerate}

Introducing a non-minimal coupling, e.g. through a contribution to the Lagrangian of the form $\delta\mathcal{L}=\Xi R\phi^2$, where $\Xi$ is the coupling strength, $R$ the Ricci scalar and $\phi$ the field strength, leads to a violation of the \gls{nec} \cite{fliss_non-minimal_2023}. This violation disappears if one goes (using a conformal transformation) from the Jordan frame, where the field strength $\phi$ appears as a product with the Ricci scalar $R$, to the Einstein frame, where no such multiplicity appears \cite{fliss_non-minimal_2023}. \citet{fliss_non-minimal_2023} raise the question if not both frames should lead to similar spacetime geometries. In the \gls{eft} view they adopt, the metrics are close. They show that in their example, the \gls{nec} is only violated for large field values. And since this only happens in the Jordan frame, they conclude that the \gls{eft} breaks down at large field values, which is predicted by the \gls{dc} as well. Furthermore, they derive a bound on the maximum number of particles in the \gls{eft}:
\begin{equation}
    N_\text{max}\leq\frac{5}{8}\frac{\pi^{d-2}}{\Gamma(d-2)V_{d-3}}\frac{\phi_\text{max}^2}{\lambda^{d-2}_\bot},
\end{equation}
with $\lambda_\bot$ an arbitrary parameter and $V_{n-3}$ the volume of unit sphere $S^{n-3}=\frac{2\pi^{\frac{n-2}{2}}}{\Gamma\left(\frac{n-2}{2}\right)}$. This bound is interesting regarding the \gls{ssc}, as it places a lower bound on the cutoff scale than the \gls{ssc} itself:
\begin{equation}
    \Lambda<\frac{M_\textnormal{P}^d}{\left(\frac{5}{8}\frac{\pi^{d-2}}{\Gamma(d-2)V_{d-3}}\frac{\phi_\text{max}^2}{\lambda^{d-2}_\bot}\right)^{\frac{1}{d-2}}}<\frac{M_\textnormal{P}^d}{N^{\frac{1}{d-2}}}.
\end{equation}

\subsubsection{Evidence}
\citet{bernardo_four-dimensional_2021} promote the \gls{nec} to a swampland conjecture. Their paper is to our knowledge the first and so far only paper that does this. The \gls{nec} itself is too broad to be reviewed here in depth; therefore, we mostly focus on the paper by \citet{bernardo_four-dimensional_2021}, the sources they cite, as well as the papers that cite \citet{bernardo_four-dimensional_2021}. A relation to the \gls{tcc} is mentioned in \cref{rel:TCC_nnNECC}.

\subsection{No Non-Supersymmetric Theories Conjecture}\label{sec:nononSUSY}%

The no non-SUSY conjecture states that there are no non-supersymmetric stable \gls{ads} vacuum solutions in 4 dimensions \cite{ooguri_non-supersymmetric_2017,freivogel_vacua_2016,hebecker_winding_2021}. It's been suggested that even metastable vacua are ruled out, but there are discussions about and counterexamples to this more stringent version of the conjecture \cite{hebecker_winding_2021,fischbacher_new_2010,guarino_brane-jet_2020,guarino_stable_2021,almuhairi_magnetic_2014,narayan_stability_2010,horowitz_nonperturbative_2008,ooguri_non-supersymmetric_2017,seo_kaluza-klein_2024}.

\subsubsection{Implications for Cosmology}
\paragraph{Big Bang}
Avoiding a big bang and a past-incomplete beginning of our Universe, \citet{nam_non-singular_2024} proposes that our Universe could have started as the transition of an unstable, non-supersymmetric \gls{ads} vacuum into a supersymmetric \gls{ads} vacuum. This predicts that our Universe is closed. A thin bubble-wall is mediating the decay, and there is an \gls{adm} mass term that decays into radiation and therefore takes care of reheating. Their model goes into inflation right after the tunnelling.

\paragraph{Black Holes}
\gls{ads} describes the near-horizon geometry of extremal \glspl{bh}. The instability of \gls{ads} goes hand in hand with the conjectured instability of \glspl{bh} as described in \cref{sec:gravity}.

\paragraph{Dark Energy}
If a stronger form of the no non-SUSY conjecture were true, namely that no stable nor metastable \gls{ads} vacua were allowed, then late-time cosmologies with a small negative cosmological constant were forbidden \cite{hebecker_winding_2021}.

\paragraph{Neutrinos}\label{p:nnSUSYC_neutrinos}
Minkowski and \gls{ds} vacua are generally allowed if there is a surplus of light fermions present; surplus meaning that there are more (light) fermions than (massless and light) bosons, light meaning $m\lesssim\Lambda^{1/d}$ \cite{gonzalo_ads_2021}. This leads to interesting predictions for neutrinos:
On the one hand, the mass of the lightest neutrino is constrained to $m_\nu\lesssim\Lambda^{1/4}$ \cite{gonzalo_swampland_2022}; on the other hand, the neutrino is either a Dirac particle \cite{ooguri_non-supersymmetric_2017,hamada_weak_2017,gonzalo_ads-phobia_2018}; or a Majorana particle accompanied by a sufficiently large number of axions, a very light Weyl fermion, or a light gravitino in Minkowski or \gls{ds} space \cite{anchordoqui_darkdimension_2023,anchordoqui_dark_2024,gonzalo_ads_2021,ibanez_constraining_2017,grana_swampland_2021}.\footnote{
    See also \cref{sss:WGC_Cosmology} on neutrinos with similar conclusions.
}
If the lightest neutrino had a mass $m_\nu>\Lambda^{1/4}$, a stable \gls{ads} potential would form,\footnote{
    Compactification of the \gls{sm} yields a potential for the radion field depending on two terms: the cosmological constant and a neutrino-mass-dependent one-loop Casimir potential \cite{gonzalo_swampland_2022}.
}
which would push the \gls{sm} into the swampland\,\textemdash which does not happen if the lightest neutrino has a mass $m_\nu\lesssim\Lambda^{1/4}$ \cite{gonzalo_swampland_2022}.
If we write the neutrino mass in terms of it's Yukawa coupling $g_\textnormal{Y}$ and the Higgs vacuum expectation value $h$, we find that the naturalness problems of the cosmological and \gls{ew} energy scales are related \cite{van_beest_lectures_2022}: $\expval{h}\lesssim\Lambda^{1/4}/g_\textnormal{Y}$. The swampland programme might remedy the hierarchy problem by showing that \glspl{eft} with higher \gls{ew} energy scales simply do not admit \gls{uv} completions into \gls{qg}.

In a generalisation of the neutrino case, \citet{nakayama_revisiting_2019} explore a \gls{dm} model with two additional light chiral fermions: a quasi-stable fermion and an ultralight stable fermion. The quasi-stable fermion in their model acts as \gls{dm} with two dominant decay channels: into three neutrinos and into a neutrino and a photon. The latter decay channel could explain the observed x-ray line at \SI{3.55}{\kilo\electronvolt} \cite{boyarsky_checking_2015,boyarsky_unidentified_2014,bulbul_detection_2014}. The ultralight stable fermion destabilises \gls{ads} vacua.

\paragraph{Particle Physics}
Very reassuring findings regarding the \gls{sm} are presented by \citet{gonzalo_fundamental_2018}: Assuming that \gls{ads} vacua are unstable, they show the need for a Higgs boson, that its experimentally determined mass lies in the narrow theoretically allowed range, and that there have to be three or more generations of particles. Without the Higgs, there would be more bosonic degrees of freedom (36) than fermionic ones (24, assuming Dirac neutrinos), rendering the potential negative, leading to a stable, non-supersymmetric \gls{ads} vacuum \cite{grana_swampland_2021}.

\subsubsection{General Remarks}
To stabilise a vacuum, \citet{etxebarria_nothing_2020} see two angles:
topology and dynamics\,\textendash\,the former does not stabilise vacua in \gls{qg} and the latter requires supersymmetry.
An example of a topological obstruction is the absence of global symmetries.\footnote{
    See \cref{sec:nGSym} for why there are no global symmetries in \gls{qg}.
}
An example of a dynamical obstruction is the \gls{wgc}, which quantifies how close we can get to restoring a global symmetry, which puts a constraint on the dynamics of the theory, e.g. in the form of certain energy conditions.
The no non-SUSY conjecture is an offspring of the \gls{wgc}: The \gls{wgc} says that gravity is weak, such that \glspl{bh} can decay. This happens when the electromagnetic forces overpower the gravitational forces.
An extension of the \gls{wgc} states that the equality between gravitational attraction and electromagnetic repulsion only occurs if the theory is supersymmetric and the states in question are \gls{bps} states \cite{cicoli_string_2023}. For non-supersymmetric states, this leads to an instability \cite{ooguri_non-supersymmetric_2017,cicoli_string_2023,ibanez_constraining_2017,banerjee_sitter_2019}: Assume we have a brane charged with respect to a flux, which supports a non-supersymmetric vacuum, and a tension less than this charge. Such a brane will nucleate in \gls{ads}, the flux-induced Coulomb repulsion wins over the tension, and the brane expands, which reduces the flux when the brane reaches the \gls{ads} boundary in finite time. Only an infinite number of fluxes could stabilise a non-supersymmetric \gls{ads} solution, which would contradict the \textit{finite number of massless fields conjecture} (see \cref{s:fnomf}) \cite{gonzalo_swampland_2022,cicoli_string_2023}. 
To put it more concise: The \gls{wgc} predicts a codimension one brane with a tension smaller than its charge, which corresponds to an instability in \gls{ads} \cite{van_beest_lectures_2022}.
Another approach to show why this is problematic comes from holography: In the gravitational dual description, the decay can be slow and the brane be quasi-stable. At the horizon, however, the decay is instantaneous and there is no \gls{ads} dual description. If there are no stable non-supersymmetric \gls{ads} solutions, then there are no non-supersymmetric \glspl{cft} that admit a gravity dual \cite{narayan_stability_2010,freivogel_vacua_2016}.
The no non-SUSY conjecture generalises these observations and rules out non-supersymmetric states entirely \cite{harada_further_2022}.
However, a word of caution is in order:
The no non-SUSY conjecture is motived by \gls{bh} physics and is therefore based on \gls{gr}\,\textemdash\,non-supersymmetric theories based on topological field theories are therefore \textit{not} ruled-out \cite{ooguri_non-supersymmetric_2017}.

\citet{freivogel_vacua_2016} generally question the stability of vacua. They claim that every non-supersymmetric \gls{ads} and Minkowski vacuum can decay, and that every supersymmetric vacuum is marginally unstable and connected to at least one vacuum through a stability bound\textendash saturating domain wall. Every \gls{ds} vacuum is meta-stable and decays on scales smaller than the horizon \cite{freivogel_vacua_2016}. Their train of thought runs as follows: If the Universe lives long enough to produce Boltzmann brains, Boltzmann brains would outnumber \enquote{ordinary observers}. Since this is not the case, the vacuum must decay on timescales smaller than the timescale that leads to Boltzmann brains, i.e. the decay process occurs on a scale smaller than the horizon scale. If the vacuum is stable, the time frame where Boltzmann brains live is so much bigger than the time span before that, that it is very unlikely to live in the pre\textendash Boltzmann brain period. This reasoning is very similar to the anthropic principle, and we leave it to the reader to judge to what degree they want to follow this reasoning, which is challenged by \citet{banks_note_2016} (in the appendix of the paper):
The stability of \gls{ds} is an issue if our Universe began as a low entropy fluctuation of a finite ensemble, whose time development is determined by a time-independent Hamiltonian. Since the observed cosmology has no time-like Killing vector, a time-dependent Hamiltonian is more accurate \cite{banks_note_2016,banks_holographic_2015}.
Furthermore, in complete contradiction with \citeauthor{freivogel_vacua_2016}'s work is the work by \citet{danielsson_universal_2016}, which shows that a perturbatively stable \gls{ads} vacuum is automatically non-perturbatively stable\,\textemdash\,this forbids tunnelling within the landscape from one stable vacuum to another one.\footnote{
    It is argued that the observed instabilities in non-supersymmetric \gls{ads} vacua are of perturbative nature and originate from the interplay between open and closed strings \cite{danielsson_fate_2017}.
}
A vacuum can only decay into a lower-energy vacuum, if the energy difference is bigger than the tension in the domain wall between the two vacua. The tension is proportional to the surface area of the domain wall, which is, for \gls{ads}, proportional to the volume. Volume and surface growing at the same rate prohibits tunnelling, as the energy to tunnel is always too high. \cite{narayan_stability_2010}
An example of a stabilised 4d $\mathcal{N}=1$ Minkowski vacuum is presented by \citet{bardzell_type_2022}, arising from a non-geometric type IIB setup.
\citet{banks_limits_2019} argues that no \gls{ds} nor Minkowski vacuum can decay into an \gls{ads} vacuum, as this would lead to a big crunch singularity, where no tunnelling process exists for. This agrees with the \gls{adsdc}, which states that \gls{ads} and \gls{ds} space are an infinite distance apart.

\subsubsection{Evidence}
This conjecture was first proposed by \citet{ooguri_non-supersymmetric_2017,freivogel_vacua_2016}.
There is supportive evidence for this conjecture from the study of charged \glspl{bh} \cite{aalsma_extremal_2018} and bubbles of nothing \cite{etxebarria_nothing_2020}.
Relations to other swampland conjectures are highlighted in \cref{rel:WGC_nnSUSYC,rel:AdSDC_nnSUSYC,rel:Cobordism_nnSUSYC}.

Instability of non-supersymmetric solutions is for example shown for
non-simply connected Ricci-flat\footnote{These are Minkowski spacetimes, not \gls{ads}.} manifolds \cite{acharya_supersymmetry_2020},
systems of fractional D3-branes at singularities \cite{buratti_supersymmetry_2019},
various AdS$_4$ vacua \cite{quirant_aspects_2022},
AdS$_2\cross S^2$ \cite{maldacena_anti-sitter_1999},
the SO(3) × SO(3)\textendash invariant AdS$_4$ vacuum of 11-dimensional supergravity \cite{malek_tachyonic_2020,bena_brane-jet_2020},
classical flux compactifications to 6d and 4d in type II supergravities \cite{andriot_automated_2022},
AdS$_5$ solutions in M-theory compactified on positive Kähler\textendash Einstein spaces \cite{ooguri_new_2017},
AdS$_6$ solutions in Type IIB string theory \cite{apruzzi_non-supersymmetric_2021},
AdS$_6$ and AdS$_7$ \cite{suh_non-susy_2020},
AdS$_7$ in type IIA string theory \cite{apruzzi_ads7_2020,junghans_susy-breaking_2019},
AdS$_7$ and AdS$_4$ solutions arising from compactifications of massive type IIA supergravity \cite{danielsson_swamp_2017},
massive type IIA supergravity in the presence of two O8 sources \cite{bena_oh_2021}, and
AdS$_5\cross S^5/\mathbb{Z}_k$ \cite{horowitz_nonperturbative_2008}.
\citet{marchesano_new_2022,casas_membranes_2022,prieto_moduli_2024} investigate massive type IIA orientifold AdS$4\cross X6$ compactifications where the X6 admits a \gls{cy} metric and show that most branches are unstable, but some leave some room for a possible stability, without coming to a clear conclusion about the actual stability.

Claims\footnote{
    The listed examples test for some forms of instabilities, but not for all of them. A truly stable non-supersymmetric vacuum would have no brane-jet instabilities, the full \gls{kk} spectrum would have no unstable modes (no tachyons), and tunnelling to a (possibly supersymmetric) vacuum would be forbidden. No such non-supersymmetric vacuum is known.
}
of non-supersymmetric but stable vacua are for example presented for
gauged 4d supergravities \cite{andriot_exploring_2022,andriot_erratum_2022},
massive IIA supergravity with regard to the \gls{kk} spectrum \cite{guarino_stable_2021},
massive IIA supergravity with flux with regard to tunnelling \cite{narayan_stability_2010},
massive type IIA flux compactifications of the form AdS$_4\times$X$_6$, where X$_6$ admits a \gls{cy} metric and O6-planes wrapping three-cycles \cite{carrasco_new_2024},\footnote{
    This model shows scale separation, and is therefore also a potential counterexample to the \gls{adsdc}, yet it has a non-universal mass spectrum of light fields and the non-perturbative stability is untested.
}
AdS$_8$ solutions of type IIA supergravity with an internal $S^2$ with a U(1) isometry against bubble instabilities \cite{cordova_ads_8_2019},
type IIB AdS$_3$ flux vacua \cite{arboleya_type_2024},
dynamical supersymmetry breaking in 10d type IIB\footnote{This solution is asymptotically no longer \gls{ads}.} \cite{argurio_octagon_2022},
ISO(7) supergravity with regard to brane-jet stability \cite{guarino_brane-jet_2020},
non-supersymmetric embeddings of a magnetic U(1) \cite{almuhairi_magnetic_2014},
non-supersymmetric \glspl{cft} obtained by supersymmetry-breaking boundary conditions in \gls{ads} geometries built from \gls{bps} branes \cite{giombi_double-trace_2018},
\gls{ads} in heterotic string theory \cite{baykara_non-supersymmetric_2023},
and using machine learning in type IIB string theory \cite{damian_metastable_2022}.
\citet{karlsson_compactifications_2021} assesses the instability of 11-dimensional supergravity compactified on the squashed 7-sphere, with incomplete results.
Neither list is exhaustive.

\subsection{Positive Gauss\textendash Bonnet Term Conjecture}\label{s:pGBTC}%

The \gls{gb} term has to be positive:
\begin{align}\label{eq:Gauss–Bonnet}
    \mathfrak{G}=R_{\mu\nu\rho\sigma}R^{\mu\nu\rho\sigma}-4R_{\mu\nu}R^{\mu\nu}+R^2>0.
\end{align}

\subsubsection{Implications for Cosmology}
\paragraph{Black Holes}
A negative contribution from the \gls{gb} term during Hawking evaporation might lead to a naked singularity \cite{boulware_string-generated_1985}, violating the \gls{ccc} \cite{ong_holographic_2022} and various swampland conjectures that disfavour the existence of stable remnants (\cref{sec:cobordism,sec:nGSym,sec:gravity}).

\paragraph{Dark Energy}
The \gls{gb} term affects the Hubble constant, as can be seen in the spatial components of the stress-energy tensor \cite{hu_renormalizable_2024}:\footnote{\citet{hu_renormalizable_2024} allow for torsion in their work.}
\begin{align}
    S&=\int\!\sqrt{-g}\left(\frac{R}{16\pi G}+f\left(\mathfrak{G}\right)\right)\,\mathrm{d}^4x\\
    T_{ij}&=\left[-a^2\left(2\dot{H}+3H^2\right)+8\left(\ddot{f}\ddot{\mathfrak{G}}+\dddot{f}\dot{\mathfrak{G}}^2\right)\right.\\
    &\qquad\left.-a^2H^2-10a^2\ddot{f}\dot{\mathfrak{G}}H^3\right]\delta_{ij}.
\end{align}
The first term shows the standard cosmic expansion dynamics and is linked to the Friedmann equations.
The second term represents higher-order gravitational corrections.
The last term shows the impact of the \gls{gb} term on the Hubble parameter and the spacetime geometry.

\paragraph{Inflation}\label{p:pGBTC_Inflation}
While the introduction of a \gls{gb} term does not introduce vector modes, it modifies the dispersion relation of scalar and tensor modes \cite{glavan_einstein-gauss-bonnet_2020}. By imposing that \glspl{gw} travel at the speed of light,\footnote{
    In Einstein\textendash Gauss\textendash Bonnet theory, the \gls{gw} speed is given by $c_\textnormal{T}^2=1-8\mathfrak{c}\left(\ddot{g}_\text{GB}-H\dot{g}_\text{GB}\right)/2\left(1/\left(8\pi G\right)^2-8\mathfrak{c}\dot{g}_\text{GB}H\right)$, which is observed to be equal to the speed of light \cite{odintsov_swampland_2020}.
    }
which yields $\ddot{g}_\text{GB}=H\dot{g}_\text{GB}$,
\citet{odintsov_swampland_2020} link the \gls{gb} coupling function to the tensor-to-scalar ratio and the scalar field dispersion:
\begin{align}
    \dot{g}_\text{GB}&=g^\prime_\text{GB}\dot{\phi}\\
    \Delta\phi&>\Delta N_e\frac{g^{\prime}_\text{GB}}{g^{\prime\prime}_\text{GB}}\\
    r_\textnormal{ts}&=8\left(\frac{g^\prime_\text{GB}}{g^{\prime\prime}_\text{GB}}\right)^2,
\end{align}
and therewith to the \gls{dc} as well as to the \gls{dsc} and the \gls{tcc}.

\subsubsection{General Remarks}
It is often assumed that the \gls{gb} term is a total derivative in 4d that does not contribute to gravitational dynamics (acting as a topological term), as the contribution to the \glspl{eom} is proportional to $d-4$ \cite{glavan_einstein-gauss-bonnet_2020}. However, if the coupling constant is taken to be $g_\text{GB}\rightarrow g_\text{GB}/\left(d-4\right)$, this factor cancels and the \gls{gb} term does contribute \cite{glavan_einstein-gauss-bonnet_2020}. Even if it is acting as a topological term without introducing additional degrees of freedom, the \gls{gb} invariant can facilitate renormalisation  \cite{hu_renormalizable_2024}. The quadratic Ricci term helps with stability in the \gls{uv} regime of a theory, where it improves the divergence behaviour \cite{hu_renormalizable_2024}.

An important argument why the \gls{gb} term should be positive comes from string theory: In Einstein\textendash Gauss\textendash Bonnet theory, the \gls{gb} term corresponds to the inverse string tension, which is positive \cite{ong_holographic_2022}.

If the \gls{gb} term is considered a perturbation to the on-shell action, then the contribution is found to be positive in the $\mathcal{N}=1$ axiverse settings studied by \citet{martucci_wormholes_2024}.

Other works do not constrain the \gls{gb} term itself but its coupling function,
while \citet{charmousis_instability_2008} argue that \gls{gb} gravity is generally unstable,
\citet{camanho_causality_2016} argue that $g_\text{GB}\gg l_\textnormal{P}^2$ violates causality;
\citet{sircar_extending_2016} find that a negative coupling leads to anomalies;
\citet{yaida_landscape_2009,buchel_holographic_2010,brigante_viscosity_2008} find $-7/36\leq g_\text{GB}\leq9/100$;
and \citet{odintsov_swampland_2020} find $\ddot{g}_\text{GB}=H\dot{g}_\text{GB}$ as we discussed in \cref{p:pGBTC_Inflation}.

\subsubsection{Evidence}
This conjecture is only recently discussed in a swampland context, e.g. by \citet{odintsov_swampland_2020,ong_holographic_2022}.
Motivation comes from
holography \cite{ong_holographic_2022},
considerations of unitarity in pure 4D gravity \cite{cheung_positivity_2017},
stability of simple dilatonic models \cite{etxebarria_nothing_2020},
as well as the suppression of problematic wormhole effects \cite{martucci_quantum_2023,kallosh_gravity_1995}.

\subsection{Species Scale Conjecture}\label{s:ssc}%

In a $d$-dimensional theory coupled to gravity with $d$-dimensional Planck mass $M_{\textnormal{P};d}$,
there are $N_\textnormal{S}$ particle states below a cutoff scale
\begin{equation}\label{eq:speciess}
    \Lambda_\textnormal{S}=\frac{M_{\textnormal{P};d}}{N_\textnormal{S}^{\frac{1}{d-2}}};
\end{equation}
after this scale, new gravitational dynamics appear, e.g. in the form of higher-derivative operators \cite{dvali_black_2010,palti_swampland_2019,castellano_emergence_2023,basile_minimal_2024,dvali_evaporation_2009,dvali_black-entropy_2008}.

An earlier definition that is recently regaining popularity is the following:\footnote{
    It has recently been noted that this definition allows for interpreting the number of species as a species entropy, which leads to the \nth{0}\textendash\nth{3} law of species thermodynamics.
}
If the smallest \gls{bh} that can be semi-classically\footnote{
    Quantum-corrections \textit{increase} the \gls{bh} horizon, ergo, the bound is absolute and the species scale can only increase due to quantum-corrections \cite{dvali_species_2010,brustein_bound_2009}.
}
described in an \gls{eft} has a Schwarzschild radius of $r_\heartsuit$, then there are $N_\textnormal{S}$ weakly coupled elementary particles with a decay width much less than their mass in the theory, and the following relation holds:
\begin{equation}
    r_\heartsuit>l_\textnormal{S}\defeq N_\textnormal{S}^{1/\left(d-2\right)}l_{\textnormal{P};d},
\end{equation}
with $l_{\textnormal{P};d}=1/M_{\textnormal{P};d}$ the $d$-dimensional Planck length and $l_\textnormal{S}$ the \textit{species scale} \cite{dvali_species_2010,brustein_bound_2009,dvali_strong_2009,dvali_quantum_2009,dvali_nature_2010,dvali_black-bound_2008,dvali_black_2010,antoniadis_strings_2014,anchordoqui_dark_2022_GKZ}. The species scale is the physical resolution limit, below which the $N_\textnormal{S}$ bits of information that correspond to the $N_\textnormal{S}$ particle species can no longer be stored \cite{gomez_string_2013,dvali_quantum_2009,dvali_strong_2009}.
This earlier definition, and the more recent one above that focuses on an energy scale instead of a length scale, are related by noting that $l_\textnormal{S}^{-1}=\Lambda_\textnormal{S}$ and $N_\textnormal{S}\simeq\left(l_\textnormal{S}M_\textnormal{P}\right)^{d-2}\simeq\mathcal{S}_\textnormal{S}$, where we already introduce the species entropy that will be of interest later.

The species scale gets weaker and vanishes in the limit of infinite (large) dimension $d$ \cite{bonnefoy_swampland_2021}.

\subsubsection{Implications for Cosmology}

\paragraph{Axions}\label{p:SSC_Axions} have non-renormalisable interactions that lead to a cutoff scale for the validity of the description \cite{seo_axion_2024}:
\begin{equation}
    f\frac{8\pi^2}{g^2}\lesssim M_\textnormal{P},
\end{equation}
with $f$ the axion decay constant and $g$ the coupling to the gauge field.\footnote{
    See \cref{p:WGC_axion} for further explanations about axions.
    }
For non-Abelian gauge interactions, we can introduce the instanton action $S_\iota=\frac{8\pi^2}{g^2}$ \cite{reece_extra-dimensional_2024} to recover the axionic \gls{wgc} \cite{seo_axion_2024}.
Furthermore, \citet{seo_axion_2024} derives
\begin{equation}
    f\gtrsim\frac{\Lambda_\textnormal{S}}{\sqrt{N_\textnormal{S}}}=\frac{M_\textnormal{P}}{N_\textnormal{S}},
\end{equation}
and \citet{reece_extra-dimensional_2024} motivates
\begin{equation}
    \Lambda_\text{UV}\sim\frac{f}{g}.
\end{equation}
The latter cutoff is not an \gls{eft} cutoff, but a \gls{uv} cutoff where the Chern\textendash Simons description of the axion\textendash gluon interaction breaks down \cite{reece_extra-dimensional_2024}.

\paragraph{Black hole} quantities can often be expressed in terms of the \gls{bh} radius, as long as we can approximate the 
\gls{bh} as a $d$-dimensional Schwarzschild \gls{bh} \cite{cribiori_species_2023,basile_shedding_2024}:
\begin{align}
    m_\text{BH}&=\left(r_\text{BH}\right)^{d-3}\label{eq:m_BH-r_BH}\\
    T_\text{BH}&=\left(r_\text{BH}\right)^{-1}\label{eq:T_BH-r_BH}\\
    \mathcal{S}_\text{BH}&=\left(r_\text{BH}\right)^{d-2}\label{eq:S_BH-r_BH}\\
    \Rightarrow \mathcal{S}_\text{BH}T_\text{BH}^{d-2}&=M_{\textnormal{P};d}^{d-2}.
\end{align}
The lowest value for each quantity is obtained for the minimal \gls{bh} of the theory, which leads to definitions for the species' quantities \cite{basile_minimal_2024,brustein_bound_2009,dvali_nature_2008,blumenhagen_emergence_2023,antoniadis_strings_2014}:
\begin{align}
    r_\heartsuit&\simeq N_\textnormal{S}^\frac{1}{d-2}M_{\textnormal{P};d}^{-1}&&\simeq\Lambda_\textnormal{S}^{-1}\\
    m_\heartsuit&\simeq N_\textnormal{S}^\frac{d-3}{d-2}M_{\textnormal{P};d}&&\simeq\Lambda_\textnormal{S}^{3-d}\label{eq:SSC_m_minBH}\\
    T_\heartsuit&\simeq N_\textnormal{S}^\frac{-1}{d-2}M_{\textnormal{P};d}&&\simeq\Lambda_\textnormal{S}\\
    \mathcal{S}_\heartsuit&\simeq N_\textnormal{S}&&\simeq\Lambda_\textnormal{S}^{2-d}.\label{eq:SSC_entropy-number-equality}
\end{align}

Below $r_\heartsuit$, gravity is no longer weakly coupled \cite{brustein_bound_2009}, and the \gls{bh} description can no longer be semi-classical.
For example, quantum hair\footnote{See \cref{foo:BHhair}.} should show during the evaporation of smaller \glspl{bh}:
large \glspl{bh} radiate their charge away through neutral Hawking-radiation.
Small \glspl{bh} have to get rid of their quantum hair eventually, which means that there is a length scale at which \glspl{bh} get a haircut\,\textemdash\,and this length scale is given by the species scale \cite{dvali_nature_2008}.
Furthermore, the minimal \gls{bh} quantities lead to implications for all species. For example, let us start with a 4d theory with $N_\textnormal{S}$ species, all of mass $m$, and each with a $Z_2$ symmetry. Let us then take a \gls{bh} and add one particle of each species to the \gls{bh}. The $N_\textnormal{S}$ units of the different $Z_2$ charges will be radiated away once the Hawking temperature reaches the mass scale $m\sim T_\text{H}$ \cite{dvali_black-entropy_2008}. The \gls{bh} is then of mass $m_\text{BH}\sim M_\textnormal{P}^2/m\sim E_\text{BH}\sim N_\textnormal{S}m$
and a bound on the maximum number of species follows \cite{dvali_black-entropy_2008,dvali_power_2008}:
\begin{equation}\label{eq:SSC_N_max}
    N_\text{max}\lesssim M_\textnormal{P}^2/m^2.
\end{equation}
Moreover, the minimal size of a \gls{bh} is relevant for theory building, as the minimal \gls{ds} horizon size we can describe with an \gls{eft} corresponds to the size of the smallest \gls{bh}, $r_\heartsuit\sim1/\Lambda_\textnormal{S}$ \cite{vafa_ray-singer_2024}.

As we will show later, there is an upper bound on the change of the species scale.\footnote{
    See around \cref{eq:upperSpeciesChange}.
    }
Together with \cref{eq:SSC_entropy-number-equality}, we find the \gls{bhedc} from \cref{p:AdSDC_BH} again \cite{cribiori_species_2023}:
\begin{align}
    \abs{\frac{\mathcal{S}_\textnormal{S}^\prime(\phi)}{\mathcal{S}_\textnormal{S}(\phi)}}^2&\lesssim\frac{\mathfrak{l}^2}{M_\textnormal{P}^{d-2}}\label{eq:SSCEntropyChange}\\
    \Rightarrow \mathcal{S}_\textnormal{S}(\phi)&\lesssim\mathcal{S}_\textnormal{S}(\phi_0)e^{\mathfrak{l}\Delta\phi}\label{eq:SSCEntropy}\\
    m(\phi)&\sim e^{-\mathfrak{a}\Delta\phi}\label{eq:SSCDC}\\
    \Rightarrow \mathcal{S}_\textnormal{S}(\phi)&\lesssim m(\phi)^{-\mathfrak{b}}\label{eq:SSCBHEDC}
\end{align}
where we introduced the unspecified $\order{1}$ constant $\mathfrak{l}$ in \cref{eq:SSCEntropyChange},
integrated over the positive region of the derivative to get \cref{eq:SSCEntropy},
introduced the \gls{dc} in \cref{eq:SSCDC},
and found the \gls{bhedc} in \cref{eq:SSCBHEDC} by identifying $\mathfrak{b}=\mathfrak{l}/\mathfrak{a}>0$.

\paragraph{Early Universe}
Having a tower of states that emerges means that the number of (relativistic) degrees of freedom (usually denoted as $g_*$) increases. As we will show later, an increase in the temperature results in an increase in the number of species. This has consequences:
\begin{itemize}
    \item The density of relativistic species $\rho_\text{r}=\frac{\pi^2}{30}g_*(T)T^4$ decreases over time; $g_*=106.75$ in the early Universe, where all \gls{sm} particles are relativistic \cite{baumann_cosmology_2015}. However, the \gls{ssc} might shed new light on this number and increase it for early times.
    \item In the radiation dominated epoch, the Hubble parameter scales like $H\propto \sqrt{g_*}$, which affects the freeze-out temperature for \gls{bbn}: increasing $g_*$ increases the freeze-out temperature, which increases the neutron\textendash proton ratio at freeze-out, which in turn increases the final helium abundance \cite{baumann_cosmology_2015}.
\end{itemize}

\paragraph{Gravitinos}
\citet{cribiori_note_2023} identify the entropy of the species with their number, $\mathcal{S}_\textnormal{S}=N_\textnormal{S}$, show for the gravitino mass $m_\textnormal{3/2}$ that
\begin{equation}
    N_\textnormal{S}\simeq-\log m_\textnormal{3/2}^2,
\end{equation}
and generalise the relation for $V\leq0$:
\begin{equation}
    N_\textnormal{S}\leq-\log\left(-\frac{V}{3}\right)
\end{equation}
respectively
\begin{equation}
    e^{-N_\textnormal{S}}\gtrsim -V\geq0.
\end{equation}

\begin{figure}[htb]
  \begin{center}
    \includegraphics[width=\linewidth]{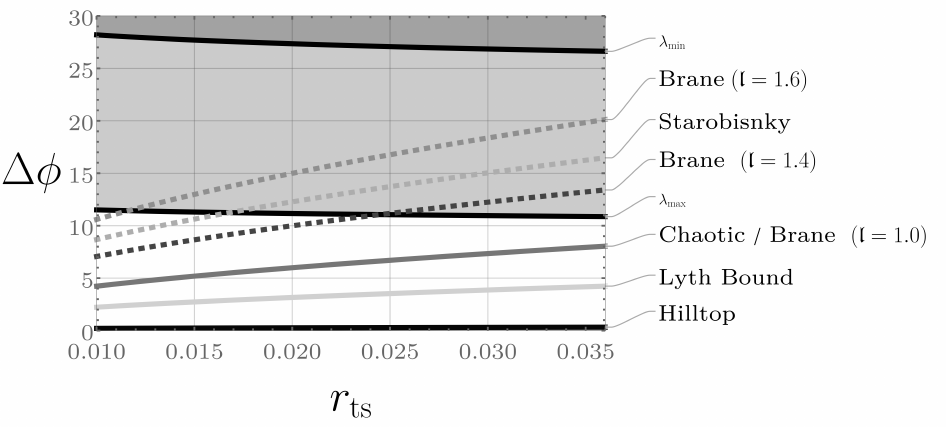}
  \end{center}
  \caption[Species Scale Inflation Bounds]{Depicted are the upper (\cref{eq:SSClambdamin}) and lower (\cref{eq:SSClambdamax}) bounds derived from the species scale for the maximal field range for various inflationary models. An observational bound on the tensor-to-scalar ratio $r_\textnormal{ts}<0.036$ \cite{bicepkeck_collaboration_improved_2021} has been applied. Once more, we see that large field displacement is generally in tension with swampland considerations.
  }\label{f:SSCInflation}
\end{figure}

\paragraph{Inflation}
is still an open field for model building. The species scale can be used to assess specific inflationary models.
For flat regions of the potential,\footnote{
    The flat regions of the potential correspond approximately to \gls{ds} space with a maximal horizon size of $r_{\text{h}_\text{max}}\sim1/\sqrt{V}$ \cite{vafa_ray-singer_2024}.
}
there are proposed upper bounds of $V(\phi)<\Lambda_\textnormal{S}^2(\phi)$ \cite{vafa_ray-singer_2024,van_de_heisteeg_bounds_on_field_2023}
respectively $V<\Lambda_\textnormal{S}^d$ \cite{casas_modular_2024} to ensure that the potential is described within the \gls{eft}.
Since the number of states $N_\textnormal{S}\sim m^{-\mathfrak{l}}M_\textnormal{P}^\mathfrak{l}$, with $\mathfrak{l}$ some $\order{1}$ parameter, we find $\Lambda_\textnormal{S}\sim m^{\mathfrak{l}/2}M_\textnormal{P}^{-\mathfrak{l}/2}$; this, and the moduli dependence of the species scale together with the exponential mass suppression predicted by the \gls{dc}, motivate
\begin{equation}\label{eq:SSC_exponential}
\Lambda_\textnormal{S}=\Lambda_0e^{-\lambda_\textnormal{S}\Delta\phi},
\end{equation}
with $\Lambda_0<M_\textnormal{P}$ corresponding to the starting point with zero field displacement \cite{scalisi_species_2024}.
This yields an upper bound on the scalar field range
\begin{equation}
    \Delta\phi\leq\frac{1}{\lambda_\textnormal{S}}\log\frac{M_\textnormal{P}}{\Lambda_\textnormal{S}}-\mathfrak{l},
\end{equation}
with
\begin{equation}
    \lambda_\textnormal{S}\defeq\abs{\frac{\Lambda_\textnormal{S}^\prime}{\Lambda_\textnormal{S}}},
\end{equation}
which measures how fast the species scale $\Lambda_\textnormal{S}$ becomes light \cite{etheredge_distance_2024,scalisi_species_2024}
and $\mathfrak{l}\sim\order{1}$ a correction factor (which we will omit in the following) that can be introduced to correct for an overestimation of the diameter of the moduli space in the asymptotic limit \cite{van_de_heisteeg_species_2024}.
In the limit $\Lambda_\textnormal{S}\rightarrow0$, the \gls{eft} is nowhere valid, and in the limit $\Lambda_\textnormal{S}\rightarrow M_\textnormal{P}$, no field displacement is allowed.
We know that $\Lambda_\textnormal{S}\geq H=\sqrt{\pi^2A_sr_\text{ts}/2}M_\textnormal{P}$, and with the amplitude of the scalar mode $A_s\approx\num{2.1e-9}$ \cite{planck_collaboration_Cosmological-Parameters_2020} it follows that
\begin{equation}
    \Delta\phi\lesssim\frac{1}{2\lambda_\textnormal{S}}\log\frac{10^8}{r_\text{ts}},
\end{equation}
where we have used that $\log1/\sqrt{r}=\frac{1}{2}\log1/r$ \cite{scalisi_species_2024}.
Since the tensor-to-scalar ratio from primordial \glspl{gw} has a known bound of $r_\textnormal{ts}<0.036$ \cite{bicepkeck_collaboration_improved_2021},
and $\lambda_\textnormal{S}$ has a
lower bound \cite{calderon-infante_entropy_2023,castellano_quantum_2024,van_de_heisteeg_species_2024} of
\begin{equation}\label{eq:SSClambdamax}
    \lambda_\textnormal{S}\geq\frac{1}{\sqrt{\left(d-1\right)\left(d-2\right)}}
\end{equation}
respectively
\begin{equation}
    \lambda_\textnormal{S}\geq\frac{1}{\sqrt{d-2}}
\end{equation}
in the emergent string limit\footnote{
    \citet{casas_cosmology_2025} propose $\lambda_\textnormal{S}\geq1/\mathfrak{a}\left(d-2\right)$ to reflect that the \gls{ssc}, the \gls{ep}, and the \gls{dc} go hand in hand to describe the predicted emerging tower of light states.
}
\cite{casas_modular_2024,van_de_heisteeg_bounds_on_species_2023,calderon-infante_entropy_2023,castellano_quantum_2024,ishiguro_stabilization_2024}
and an upper bound \cite{cribiori_note_2023,van_de_heisteeg_bounds_on_species_2023} of
\begin{equation}\label{eq:SSClambdamin}
    \lambda_\textnormal{S}\leq\frac{\order{1}}{M_\textnormal{P}^{d-2}},
\end{equation}
inflationary models are once more constrained from a swampland perspective.
In \cref{f:SSCInflation}, we present the following inflationary models and their constraints on the field displacement for different tensor-to-scalar ratios $r_\textnormal{ts}$ \cite{scalisi_species_2024}:
\begin{description}
    \item[Lyth Bound] $\Delta\phi=\sqrt{2\epsilon_V} N_e\gtrsim\sqrt{\frac{r_\textnormal{ts}}{0.002}}$. The Lyth bound is considered to be a generic lower bound on the required field displacement to successfully inflate the Universe enough within $N_e=60$ $e$-folds \cite{lyth_what_1997,scalisi_species_2024,boubekeur_hilltop_2005,garcia-bellido_lyth_2014,van_de_heisteeg_bounds_on_field_2023}.
    \item[Starobinsky Inflation] $\Delta\phi\simeq15\log\left(60\right)\sqrt{2r_\textnormal{ts}}$. For most values of $r_\textnormal{ts}$, Starobinsky inflation is in tension with the species scale. We elaborate on Starobinsky inflation and a similar model below.
    \item[Hilltop Models] $\Delta\phi\simeq\left(\frac{\mathfrak{p}-2}{\sqrt{2}\left(\mathfrak{p}-1\right)}\right)^{1-2/\mathfrak{p}}\mu^{2-2/\mathfrak{p}}\sim r_\textnormal{ts}^{1/\mathfrak{p}}$ stemming from potentials of the form 
    \begin{equation*}
        V(\phi)=V_0\left(1-\left(\frac{\phi}{\mu}\right)^\frac{2\left(1-\mathfrak{p}\right)}{2-\mathfrak{p}}+\dots\right),
    \end{equation*}
    where the ellipsis indicates higher-order terms in $\phi/\mu$. Hilltop models are unproblematic from a species scale perspective, as well as compatible with the \gls{dc} and the \gls{dsc}, but incompatible with the \gls{wgc} and the Lyth bound.
    \item[Chaotic Inflation]  $\Delta\phi\simeq30\sqrt{2r_\textnormal{ts}}$. We discuss chaotic inflation in more detail in \cref{p:WGC_chaotic-inflation} and below.
    \item[Brane Inflation / Inverse Hilltop Models] $\Delta\phi\simeq30\sqrt{2r_\textnormal{ts}}/\left(2-\mathfrak{l}\right)$; $1<\mathfrak{l}<2$ controls the model  \cite{dvali_brane_1998,dvali_infrared_1999,dvali_d-brane_2001,burgess_inflationary_2001,kachru_towards_2003,hertzberg_inflationary_2008}. For $\mathfrak{l}=1$ we find chaotic inflation and for $\mathfrak{l}\simeq1.51$ Starobinsky inflation. For $\mathfrak{l}<1.26$ no tension with the species scale arises, whereas for $\mathfrak{l}>1.7$ tension with the species scale bound ($\lambda_\text{min}$) cannot be avoided.
\end{description}

\subparagraph{More on Starobinsky Inflation}
\citet{lust_starobinsky_2023} show that Starobinsky inflation belongs to the swampland: In the Starobinsky model, the standard Einstein\textendash Hilbert action has an extra term proportional to the square of the scalar curvature:
\begin{equation}
    S=\int\!\sqrt{-g}\frac{M_\textnormal{P}^2}{2}\left(R+\frac{R^2}{M^2}\right)\,\mathrm{d}^4x
\end{equation}
where the scale $M\sim\SI{e14}{\giga\electronvolt}$ is fixed by \gls{cmb} measurements. The extra term can be explained by / be interpreted as
\begin{itemize}
    \item scalar fields
    \item additional matter fields
    \item towers of states.
\end{itemize}
In their paper, they show that the $R^2$-term can be regarded as a renormalisation effect of a tower of states, which implies that the scale $M$ is related to the species scale $\Lambda_\textnormal{S}$. Furthermore, they reason that $H\sim M\sim\Lambda_\textnormal{S}\sim\SI{e14}{\giga\electronvolt}$ and deduce that \num{e10} species are present during inflation, which has severe phenomenological consequences for inflation, for example, \gls{kk} modes should be present, even in the \gls{eft}, as $m_\text{KK}\sim\SI{e4}{\giga\electronvolt}<H$. Therefore, the Starobinsky model is an inconsistent \gls{eft} and lies in the Swampland.

\subparagraph{Modular Inflation}
A potential with similar features as Starobinsky inflation, without sharing its incompatibility with the \gls{ssc}, is the following \cite{casas_modular_2024}:
\begin{equation}
    V(\phi)=\mu^{d-2}M_\textnormal{P}^2G^{ij}\frac{\left(\partial_i\Lambda_\textnormal{S}\right)\left(\partial_j\Lambda_\textnormal{S}\right)}{\Lambda_\textnormal{S}^2}.
\end{equation}
\begin{itemize}
    \item There are no singularities at finite distance in moduli space, i.e. it is described within the \gls{eft},
    \item the potential depends solely on the species scale and its derivatives,
    \item a sufficiently large period of inflation can be realised due to the flat plateau that can be reached asymptotically, and
    \item the potential is invariant and automorphic to the duality symmetries of the theory.
\end{itemize}
The potential can be expressed in a modular form,\footnote{
    The potential can be expressed as
    $V\simeq\left(\text{Im}{\tau}\right)^2\abs{\Tilde{G}_2}^2/N_\textnormal{S}^2,
    $
    with $\Tilde{G}_2$ the Eisenstein modular form of weight 2,
    and $N_\textnormal{S}\simeq-\log\left(\text{Im}\tau\abs{\eta(\tau)}^4\right)$, $\eta$ being the Dedekind function, and
    $\tau$ being a complex scalar modulus \cite{casas_modular_2024}.
} 
that offers some noteworthy advantages \cite{casas_modular_2024},\footnote{
    First, for large moduli, the model resembles the Starobinsky model,
    second, the number of $e$-folds corresponds to the number of species, $N_e\simeq N_\textnormal{S}$,
    and third, the tower of states seems to play an important role during inflation, as also $\epsilon_V\simeq\Lambda_\textnormal{S}^4$ holds \cite{casas_modular_2024}.
} 
but, moreover, the model can be generalised for $SL(2,\mathbb{Z})$ modular symmetries in such a way that the species scale is directly connected to observables \cite{aoki_inflationary_2024}:
The generalised Lagrangian is of the form
\begin{align}
    \frac{L\left(\tau,\Bar{\tau}\right)}{\sqrt{-g}}&=\frac{R}{2}M_\textnormal{P}^2-\frac{M_\textnormal{P}^2\partial\tau\partial\Bar{\tau}}{2\beta^2\tau_2^2}-V\left(\tau,\Bar{\tau}\right)\\
    \tau&=\tau_1+i\tau_2
\end{align}
with $\beta$ a model-parameter.
Inflation takes place if $\tau_1=0$ and $\tau_2\gg1$, such that the potential will be of the form
\begin{equation}
    V\left(\tau,\Bar{\tau}\right)\simeq V_0\left(1-\frac{\varsigma}{\tau_2}+\dots\right),
\end{equation}
with the model parameter $\varsigma$ related to the normalisation of the potential,\footnote{
    The case $\varsigma=1$, $\beta=\sqrt{2/3\alpha}$ is related to the $\alpha$-attractor model \cite{aoki_inflationary_2024}.
}
which, in case of canonically normalised inflation with $\tau_2=\exp\left(\beta\phi/M_\textnormal{P}\right)$, can be written as
\begin{equation}
    V_\textnormal{I}\left(\phi\right)\simeq V_0\left(1-\varsigma e^{-\beta\phi/M_\textnormal{P}}\right).
\end{equation}
As a next step, the slow-roll parameters and the number of $e$-folds can be computed:
\begin{align}
    \epsilon_V&=\frac{\varsigma^2\beta^2e^{-2\beta\phi/M_\textnormal{P}}}{2}\\
    \eta_V&=-\beta^2\varsigma e^{-\beta\phi/M_\textnormal{P}}\\
    N_e&=\frac{1}{M_\textnormal{P}}\int_{\phi_\textnormal{rh}}^{\phi_*}\!\frac{1}{\sqrt{2\epsilon_V}}\,\mathrm{d}\phi\\
    &=\frac{e^{\beta\phi_*/M_\textnormal{P}}-e^{\beta\phi_\textnormal{rh}/M_\textnormal{P}}}{\beta^2\varsigma}.
\end{align}
The last equation gives the number of $e$-foldings between the end of reheating and the time the \gls{cmb} exits the horizon.
Now, we use $\phi_*\gg\phi_\textnormal{rh}$ and $\tau_2\gg1\Rightarrow N_\textnormal{S}\simeq2\pi\tau_2$ to find
\begin{align}
    \phi_*&\simeq\frac{\log\left(\beta^2\varsigma N_\textnormal{S}\right)}{\beta}\\
    \tau_2^*&\simeq\beta^2\varsigma N_\textnormal{S}\\
    N_e&\simeq\frac{\tau_2^*}{\beta^2\varsigma}\\
    &=\frac{N_\textnormal{S}}{2\pi\beta^2\varsigma}\\
    &\simeq\frac{1}{2\pi\beta^2\varsigma}\left(\frac{M_\textnormal{P}}{\Lambda_\text{S}}\right)^2\\
    n_s&=1-\frac{2}{N_e}\\
    &\simeq1-4\pi\beta^2\varsigma\left(\frac{\Lambda_\textnormal{S}}{M_\textnormal{P}}\right)^2\\
    r_\textnormal{ts}&\simeq\frac{8}{\beta^2N_e^2}\\
    &\simeq32\pi^2\beta^2\varsigma^2\left(\frac{\Lambda_\textnormal{S}}{M_\textnormal{P}}\right)^4\\
    A_s&=\frac{V}{24\pi^2\epsilon_V}\\
    &\simeq\frac{V_0\beta^2N_e^2}{12\pi^2},
\end{align}
where the expression for the scalar powerspectrum can be used with the observed value of $A_s\simeq\num{2.1e-9}$ to fix $V_0$.
Observational constraints on $n_s=0.9649$ then fix the species scale to
\begin{equation}
    \Lambda_\textnormal{S}\simeq\frac{1}{\beta}\sqrt{\frac{2}{\varsigma}}\SI{8.97e16}{\giga\electronvolt}.
\end{equation}
Observational constraints on $r_\textnormal{ts}<0.035$ fix the species scale to
\begin{equation}
    \Lambda_\textnormal{S}\lesssim\SI{1.74e17}{\giga\electronvolt},
\end{equation}
for $\beta=1$ and $\varsigma=2$.
\citet{aoki_inflationary_2024} further generalise these considerations and derivations for \gls{cy} threefolds, to fix the species scale with inflationary observables to $\SI{e15}{\giga\electronvolt}<\Lambda_\textnormal{S}<\SI{e17}{\giga\electronvolt}$.

\subparagraph{More on Chaotic Inflation}
Requiring that a \gls{bh} consisting of $N_\textnormal{S}$ heavy species can decay within the lifetime of \gls{ds} space, \citet{dvali_power_2008} present a bound on the species mass
\begin{equation}\label{eq:SSC_mass-limit-dS}
    m\lesssim\frac{M_\textnormal{P}}{\left(M_\textnormal{P}/H\right)^{1/3}}.
\end{equation}
This bound applies during inflation, where the species are massive \cite{dvali_power_2008}:
In chaotic inflation, we have a scalar field with $V(\phi)=\frac{1}{2}m^2\phi^2+g\phi\Bar{\psi}_i\psi_i$, where the last term with coupling constant $g$ describes the interaction with the $N_\textnormal{S}$ species, and the field can reach arbitrarily large values, as long as $m^2\phi^2\ll M_\textnormal{P}^4$ holds.
A slow-rolling field follows the equation
\begin{equation}
    \Ddot{\phi}+3H\dot{\phi}+V^\prime(\phi)=0
\end{equation}
with
\begin{equation}
    H^2\simeq V(\phi)/3M_\textnormal{P}^2
\end{equation}
and in the case of chaotic inflation
$V^\prime(\phi)=m^2\phi$ and $H^2=\frac{m^2\phi^2+\dot{\phi}^2}{6M_\textnormal{P}^2}$.
During inflation, the species mass is $m=g\phi$, and \cref{eq:SSC_mass-limit-dS} implies
\begin{equation}
    g\phi\lesssim\frac{M_\textnormal{P}}{\left(M_\textnormal{P}^2/\left(m\phi\right)\right)^{1/3}},
\end{equation}
which limits the coupling to
\begin{equation}
    g\lesssim\left(\frac{M_\textnormal{P}}{\phi}\right)^{2/3}\left(\frac{m}{M_\textnormal{P}}\right)^{1/3}.
\end{equation}
For chaotic inflation with 60 $e$-foldings, $\phi\lesssim10M_\textnormal{P}$, and $m\sim\SI{e12}{\giga\electronvolt}$ (derived from density perturbations), we find $g<\num{e-3}$.

\paragraph{Particle Physics}
An obvious question is: What kind of additional species should we expect? Observations do not suggest that there are additional neutrino species, yet the observational constraints allow for up to $N_\textnormal{S}<\order{30}$ for normal ordering and for up to $N_\textnormal{S}<\order{100}$ for inverted ordering \cite{ettengruber_testing_2024}. Furthermore, additional Higgs bosons or a zoo of \gls{dm} particles can be in agreement with the current observational constraints. There are suggestions that entire copies of the \gls{sm} could exist \cite{ettengruber_testing_2024,dvali_phenomenology_2009}.\footnote{
    The idea that a huge number of species ($N_\textnormal{S}=\num{e32}$ copies of the \gls{sm}) could explain the hierarchy problem, i.e. why the \gls{ew} sector / (squared) Higgs mass are 32 orders of magnitude below the (squared) Planck scale, was put forward early on by \citet{dvali_phenomenology_2009}. However, since the \gls{lhc} did not show any signs of microscopic \glspl{bh} at the TeV scale, this solution is ruled out.
    }

\subsubsection{General Remarks}\label{sss:SSC_Remarks}
The species scale plays an important role for the \gls{ep} and in tying together the \gls{ep} and the \gls{wgc} (see \cref{rel:WGC_EP}) respectively the \gls{ep} and the \gls{dc} (see \cref{rel:EP_SDC}). Furthermore, there is an interplay between the \gls{ssc} and the \gls{flb} (\cref{rel:FLB_SSC}). In the remainder of this section, we will focus on the motivation behind the \gls{ssc}.

\paragraph{Why would the number of species play a role regarding the validity of a theory?} 
The effects of having multiple species become clear by studying a concrete example by \citet{grimm_infinite_2018}. Let us start with the Lagrangian for heavy fermions:
\begin{equation}\label{eq:L:fermion}
    L=\Bar{\psi}\partial_\mu\gamma^\mu\psi+\frac{1}{2}\left(\partial\phi\right)^2-m(\phi)\Bar{\psi}\psi
\end{equation}
and integrate out $\psi$ to obtain a low-energy \gls{eft}\textendash metric that only depends on $\phi$:
\begin{equation}
    g_{\phi\phi}=g_{\phi\phi}^\text{UV}+\frac{\left(\partial_\phi m\right)^2}{8\pi^2}\log\frac{\Lambda_\text{UV}^2}{m^2},
\end{equation}
where the logarithmic term comes from 1-loop corrections when integrating out the heavy state. In general, higher-order corrections $\propto\left(\partial_\phi m\right)^4$ are expected to be present but subleading and therefore not of relevance to the argument here. Our original theory (\cref{eq:L:fermion}) was no theory of \gls{qg}. Therefore, it ceases to be valid at $\Lambda_\text{UV}$. Now, when do quantum corrections become dominant? Since we are using a perturbative approach, we continue to assume that $\left(\partial_\phi m\right)\ll1$. The limit $m\rightarrow0$ is a divergence, but it is a single-point locus in moduli space up to which the field space distance remains finite. To obtain infinite distances, we actually have to consider a model with multiple particles! We can continue with the fermionic case and study the multi-field metric
\begin{equation}
    g_{\phi\phi}=g_{\phi\phi}^\text{UV}+\sum_i^{N_\textnormal{S}}\frac{\left(\partial_\phi m_i\right)^2}{8\pi^2}\log\frac{\Lambda_\text{UV}^2}{m_i^2}.
\end{equation}
Now, the quantum part can actually dominate the \gls{eft} metric, when $\sum_i^{N_\textnormal{S}}\left(\partial_\phi m_i\right)^2\gg g_{\phi\phi}^\text{UV}$. The number of states in the theory, $N_\textnormal{S}$, is fixed by the upper cutoff scale $\Lambda_\text{UV}$. Above this scale, gravity becomes strongly coupled, i.e. $\Lambda_\text{UV}\sim\Lambda_\text{S}$. 
This shows that the number of states is of relevance when assessing the validity of a theory.
To quantify this further, we can use that there is also a lower cutoff, $\Lambda_0\sim m_0$, where the lightest massive state is being integrated out. Our tower of states will have a mass separation scale $\Delta m$, such that $m_k(\phi)=m_0(\phi)+k\Delta m(\phi)$.
\textit{If} $N_\textnormal{S}\Delta m\gtrsim m_0$, then
\begin{align}
    \Lambda_\text{S}&\simeq\left(M_\textnormal{P}^2\Delta m\right)^\frac{1}{3}\\
    N_\textnormal{S}&\simeq\left(\frac{M_\textnormal{P}}{\Delta m}\right)^\frac{2}{3}\\
    \Rightarrow \Lambda_\textnormal{S}&=\frac{M_\textnormal{P}}{N_\textnormal{S}^{\frac{1}{2}}}.
\end{align}
\Cref{eq:speciess} is the $d$-dimensional generalisation \cite{grimm_infinite_2018}.

\paragraph{What happens if the species scale is violated?}
Generally, an \gls{eft} breaks down before reaching the Planck scale because the Planck mass of the higher-dimensional theory, $M_{\textnormal{P};D}\sim R^\frac{1}{2-D}$, is smaller than $M_{\textnormal{P};d}$: gravity \textit{feels}\footnote{
    Gravity cannot be localised on a submanifold of codimension zero, therefore, on a 4D spacetime embedded in a higher dimensional space with additional, curled-up, dimensions, gravity seems to \textit{leak out} of the 4D \gls{eft} \cite{de_biasio_geometric_2023}. Given a $\left(D-d\right)$-dimensional internal subspace of volume $V$, we find that $M_{\textnormal{P};d}^2=M_{\textnormal{P};D}^{2+D-d}V$ \cite{vafa_swamplandish_2024}.
    }
the extra dimensions, and is diluted, i.e. quantum gravitational effects become important earlier than expected \cite{palti_swampland_2019}.
This corresponds to the initial interpretation or definition by \citet{dvali_black-bound_2008}: They conjectured that $M_\textnormal{P}^2\gtrsim N_\textnormal{S}\Lambda_\textnormal{S}^2$, such that a large number of species weakens gravity.\footnote{
    The motivation for this comes directly from the derivation \cite{dvali_black-bound_2008}: each of the $N_\textnormal{S}$ species contributes $\Lambda_\textnormal{S}^2$ to the normalisation of the graviton wave function, which results in an effective contribution to $M_\textnormal{P}$ of $N_\textnormal{S}\Lambda_\textnormal{S}^2$.
    }

A concrete example of a violation of the \gls{ssc} is presented by \citet{kaplan_species_2019}:
If the species scale in a 4d large $N_\textnormal{S}$ confining gauge theory with glueballs and mesons of spin $J>2$ is violated, i.e. if $N_\textnormal{S}>\frac{M_\textnormal{P}}{\Lambda_\text{QCD}}$, causality is violated and gravity might be stronger than the gauge forces between the hadrons, which would also violate the \gls{wgc}.

Other implications of a \gls{ssc} violation are that
\glspl{bh} could become larger than the \gls{ds} horizon, and that
\glspl{bh} could have hair.\footnote{If the entropy of a \gls{bh} was smaller than then species entropy, we knew that not all species were part of this particular \gls{bh}, ergo we had information about its constituents we wouldn't have otherwise, i.e. the no-hair theorem was violated.} 

\paragraph{How can the species bound be derived?}
\Cref{eq:speciess} arises from perturbative and non-perturbative considerations \cite{cribiori_note_2023,cribiori_species_2023,dvali_black-bound_2008}:
Perturbatively, the species scale arises from loop corrections to the graviton operator.
When the first-order corrections are comparable to the tree-level term, respectively the inverse propagator becomes divergent,
the perturbation breaks down, which happens at the species scale.
Alternatively, one can study the \gls{eft} action: higher curvature corrections become relevant around the species scale, as they control the $R^2$ term \cite{cribiori_species_2023}, cf. \cite{calmet_matching_2024}.
Non-perturbatively, one can study the entropy of \glspl{bh}:
The smallest \gls{bh} needs to be able to emit or absorb all $N_\textnormal{S}$ species of the theory, i.e. the entropy of the \gls{bh} should correspond to the number of species in the theory: 
\begin{equation}
    \mathcal{S}_\textnormal{S}\simeq\mathcal{S}_\heartsuit\simeq N_\textnormal{S}.
\end{equation}
The smallest \gls{bh} has a radius of $r_\heartsuit\simeq1/\Lambda_\textnormal{S}$, and the Bekenstein\textendash Hawking area law yields $\mathcal{S}_\textnormal{S}\simeq r_\heartsuit^{d-2}$ \cite{cribiori_species_2023}.
After this executive summary, we now present the three approaches in more detail.

\subparagraph{Perturbative Arguments}
The species scale can be interpreted as the scale that suppresses higher-derivative corrections to the \gls{eft} \cite{calderon-infante_emergence_2024}.
That the perturbation breaks down for higher energies can be seen at the example of the graviton propagator \cite{castellano_quantum_2024}:
The resummed 1-loop propagator of the graviton in Lorentzian signature in a flat, 4d background is given by
\begin{equation}
    i\Pi^{\mu\nu\rho\sigma}=i\left(P^{\mu\rho}P^{\nu\sigma}+P^{\mu\sigma}P^{\nu\rho}-P^{\mu\nu}P^{\rho\sigma}\right)\pi(p^2)
\end{equation}
with $P^{\mu\nu}=\eta^{\mu\nu}-\frac{p^\mu p^\nu}{p^2}$ satisfying $P^\rho_\sigma P^\sigma_\kappa=P^\rho_\kappa$ a projection operator onto polarisation states transverse to $p^\sigma$, 
$p$ the momentum of the external graviton state,
$\mu$ an energy scale related to the renormalisation of $R^2$ and $R_{\mu\nu}R^{\mu\nu}$ curvature terms \cite{aydemir_self-healing_2012,castellano_emergence_2023},
and
\begin{equation}
    \pi^{-1}(p^2)=2p^2\left(1-\frac{N_0p^2}{120\pi M_{\textnormal{P};4}^2}\log(-p^2/\mu^2)\right),
\end{equation}
with 
\begin{equation}
    N_0=N_\textnormal{s}/3+N_\textnormal{W}+4N_\textnormal{v}
\end{equation}
the weighted sum of light degrees of freedom (scalars, Weyl spinors, and vectors in 4 dimensions),
such that the perturbation expansion breaks down when the second term in the bracket is comparable to $1$.
This happens at the energy scale
$\Lambda_\textnormal{S}$ \cite{cribiori_species_2023,castellano_emergence_2023,aydemir_self-healing_2012}:%
\begin{equation}
    \frac{M_\textnormal{P}^{d-2}}{N_0}\simeq\Lambda_\textnormal{S}^{d-2}\log\left(\Lambda_\textnormal{S}/\mu\right).
\end{equation}
A solution to this equation can be found by using the (-1)-branch of the Lambert $W$ function \cite{cribiori_species_2023,castellano_emergence_2023,corless_lambertw_1996}:
\begin{align}
    \Lambda_\textnormal{S}^{d-2}&\simeq\frac{-\left(d-2\right)M_\textnormal{P}^{d-2}}{N_0W_{-1}\left(-\frac{d-2}{N_0}\left(\frac{M_\textnormal{P}}{\mu}\right)^{d-2}\right)}\\
    &\simeq\frac{M_\textnormal{P}^{d-2}}{N_0\log N_0}\\
    &\simeq\frac{M_\textnormal{P}^{d-2}}{\log N_0!},
\end{align}
where the second last expansion is a large $N_0$ approximation, which yields, using Stirling's formula, the last equation.
Note that it is a weighted sum over the number of massless states $N_0$ that appears in the equations, not the number of species $N_\textnormal{S}$ itself. This allows us to define the species entropy as
\begin{equation}
    \mathcal{S}_\textnormal{S}\simeq\log N_0!,
\end{equation}
paving the way for a bridge to entropic arguments \cite{cribiori_species_2023}.

The appearance and relevance of logarithmic corrections is subject to ongoing debate:
On the one hand,
the logarithmic corrections appear whenever starting from 1-loop corrections to graviton propagators \cite{dvali_black-bound_2008,basile_minimal_2024,blumenhagen_demystifying_2024,cribiori_species_2023,castellano_emergence_2023,aydemir_self-healing_2012,castellano_quantum_2024}, and
\citet{cribiori_note_2023} show that the logarithmic corrections arise from requiring that the species scale is modular invariant.
On the other hand,
there are arguments that the logarithmic corrections are to be ignored:
\begin{itemize}
    \item logarithmic corrections do not appear in the \gls{bh} picture \cite{blumenhagen_demystifying_2024,cribiori_species_2023,blumenhagen_emergence_2023,agmon_lectures_2023,van_de_heisteeg_bounds_on_species_2023},
    \item logarithmic corrections only arise, when the \gls{qft} is extended to energies above the string scale\,\textemdash\,a regime where the approach is no longer fully reliable \cite{agmon_lectures_2023,van_de_heisteeg_bounds_on_species_2023,blumenhagen_emergence_2023},
    \item logarithmic corrections only arise in the purely massless case, and do not appear with massive species \cite{basile_minimal_2024,van_de_heisteeg_species_2024}.    
\end{itemize}
At this point, we would like to refer to our \textit{tale of scales}\footnote{See \cref{p:tale-of-scales}} and venture the guess that the logarithmic terms are part of the emergence scale, but not necessarily of the species scale. The confusion arises from not distinguishing between the two scales.

\subparagraph{Entropic Arguments}
To show that pathologies are only avoided if the \gls{ssc} holds, we can study a large $N_\textnormal{S}$ \gls{qcd} theory in a flat 4d spacetime region of radius $l$ and temperature $T$, with
\begin{equation}
    l=\frac{1}{4\pi T},
\end{equation}
which is the Hawking temperature of a \gls{bh} of radius $l$ \cite{kaplan_species_2019}.
Dimensional analysis tells us that the entropy of such a region is given by
\begin{equation}
    \mathcal{S}\sim l^3T^3N_\textnormal{S}^2,
\end{equation}
which is supported by finite temperature lattice computations above the critical deconfinement temperature \cite{osborn_implications_1994}.
For
\begin{equation}
    l\sim\frac{N_\textnormal{S}}{M_\textnormal{P}}
\end{equation}
the spherical region approaches the entropy of a \gls{bh}
\begin{equation}
    \mathcal{S}=8\pi^2l^2M_\textnormal{P}^2,
\end{equation}
but for large $N_\textnormal{S}$ and $l<N_\textnormal{S}/M_\textnormal{P}$, the entropy of the spherical region would be larger than the entropy of a \gls{bh} of the same size, which can only be avoided if
\begin{equation}
    N_\textnormal{S}\lesssim\frac{M_\textnormal{P}}{\Lambda_\textnormal{S}},
\end{equation}
i.e. the species bound holds \cite{kaplan_species_2019}.

Given a thermodynamic system of size $l$, \citet{herraez_origin_2024} show that there are three different regimes with regard to the temperature $T$ of the system and the species scale:
\begin{description}
    \item[$l^{-1}<T<\Lambda_\text{S}$] If gravitational collapse is avoided and the \gls{ceb} is not violated, the entropy scales with the volume, $\mathcal{S}\sim l^3$.
    \item[$l^{-1}\simeq T<\Lambda_\text{S}$] If gravitational collapse is avoided and the \gls{ceb} is not violated, the system transitions from one regime to the other, while the entropy is given by counting the species contributing to the thermodynamic ensemble.
    \item[$l^{-1}\simeq T\rightarrow\Lambda_\text{S}$] The rules of species thermodynamics apply, and the entropy scales like the area of the box. In the limit $l^{-1}\simeq T\simeq\Lambda_\textnormal{S}$, the \gls{ceb} is saturated, the system should collapse gravitationally, and $\mathcal{S}\rightarrow N_\textnormal{S}$ \cite{castellano_iruv_2022,herraez_origin_2024}.
\end{description}

Interesting from a swampland perspective is, of course, the last case, where we are approaching limiting energy (densities).
In analogy to the laws of \gls{bh} thermodynamics, the laws of species thermodynamics can be formulated by noting that time derivatives can be expressed in terms of changes in the scalar field\footnote{
    In the swampland we mostly consider adiabatic motion of a scalar field through moduli space, and we can, for time-dependent string backgrounds, identify the scalar field with time, i.e. $\phi=t$.
}
\cite{cribiori_species_2023}:
\begin{description}
    \item[\nth{0} law of species thermodynamics] $\forall\phi_0,\phi_1,\phi_2\in\mathcal{M}$ s.t. $T_\textnormal{S}(\phi_1)=T_\textnormal{S}(\phi_0)$ and $T_\textnormal{S}(\phi_2)=T_\textnormal{S}(\phi_0)$, it follows that $T_\textnormal{S}(\phi_1)=T_\textnormal{S}(\phi_2)$.
    \item[\nth{1} law of species thermodynamics] Two neighbouring stationary towers are related to each other by $\delta E_\textnormal{S}=T_\textnormal{S}\delta\mathcal{S}_\textnormal{S}+V_e\delta q+\dots$, with $V_e$ the electrostatic potential. While the entropy takes the place of the horizon area, the temperature, which is linearly related to the species scale $\Lambda_\textnormal{S}$, takes the place of the surface gravity.
    \item[\nth{2} law of species thermodynamics] The species entropy is a non-decreasing function of scalar field displacement towards the boundary of moduli space. $\forall\phi_1,\phi_2\in\mathcal{M}$ s.t. $\min_{\phi\in\partial\mathcal{M}}\Delta(\phi_1,\phi)>\min_{\phi\in\partial\mathcal{M}}\Delta(\phi_2,\phi)$, we have $\Lambda_\textnormal{S}(\phi_2)<\Lambda_\textnormal{S}(\phi_1)$, $\mathcal{S}_\textnormal{S}(\phi_2)>\mathcal{S}_\textnormal{S}(\phi_1)$, and $\delta N_\textnormal{S}\geq0$, i.e. the number of species must not decrease,\footnote{
        This finding can be expressed in simple and obvious terms: The number of species below temperature $T$ must not decrease with an increase in temperature \cite{herraez_origin_2024}.
    }
    which is aligned with other swampland conjectures such as the \gls{dc} \cite{herraez_origin_2024,cribiori_species_2023}. It is also aligned with the \gls{ep}: for \gls{kk} towers, we go in the direction of compactification, such that the volume increases ($\delta\mathcal{V}\geq0$); and for string towers, we go into the direction of the tensionless string, such that the coupling strength decreases ($\delta g_s\leq0$). Using the \gls{adsdc}, we find that the cosmological constant can only decrease ($\delta\Lambda_\text{cc}\leq0$).
    \item[\nth{3} law of species thermodynamics] The boundary of moduli space $\partial\mathcal{M}$ cannot be reached by a finite number of steps, i.e. it is at infinite distance. No physical process can reduce the species temperature $T_\textnormal{S}$ to zero by a finite number of operations. The limit of a vanishing cosmological constant cannot be reached by a finite sequence of operations.
\end{description}

\subparagraph{\gls{bh} Arguments}

In a semi-classical approximation, the decay rate of a \gls{bh} is given by
\begin{equation}
    \frac{\mathrm{d}m_\text{BH}}{\mathrm{d}t}\approx-N_\textnormal{S}T^2_\text{H};
\end{equation}
using that the Hawking temperature scales like $T_\text{H}\sim1/m$
together with \cref{eq:SSC_entropy-number-equality,eq:m_BH-r_BH,eq:S_BH-r_BH,eq:T_BH-r_BH,eq:SSC_m_minBH}
yields a lifetime of
\begin{align}
    \tau&\sim\frac{1}{N_\textnormal{S}}\int_0^{1/\Lambda_\textnormal{S}}\!\frac{1}{T_\text{H}^2}\mathrm{d}m\\
    &\sim\frac{1}{m_\heartsuit^2}\int_0^{1/\Lambda_\textnormal{S}}\!m^2\mathrm{d}m\\
    &\sim m_\heartsuit\\
    &\sim\frac{1}{T_\text{H}}
\end{align}
which shows that the semi-classical description breaks down if we allow for smaller \glspl{bh} than allowed by the \gls{ssc}, as then the lifetime of such a \gls{bh} would be longer than the timescale over which the Hawking temperature is approximately constant, $1/T_\text{H}$
\cite{dvali_black-bound_2008,castellano_quantum_2024,dvali_quantum_2009,castellano_emergence_2023,dvali_black-entropy_2008}.\footnote{We worked in 4 dimensions and Planck units to shorten the notation. The general expression is $\tau\sim\frac{1}{N_\textnormal{S}}\int_0^{M_{\textnormal{P};d}^{d-2}/\Lambda^{d-3}}\!\left(\frac{m}{M_{\textnormal{P};d}^{d-2}}\right)^\frac{2}{d-3}\,\mathrm{d}m\sim1/\Lambda$ as $T_\text{H}=1/r\sim\left(M_{\textnormal{P};d}^{d-2}/m\right)^{1/\left(d-3\right)}\sim\Lambda$.}

The occurrence of the species scale can also be understood by the following considerations presented by \citet{van_de_heisteeg_bounds_on_species_2023}:
In the Einstein\textendash Hilbert action, higher-order curvature terms are suppressed by additional powers of $M_\textnormal{P}$. If $R\sim M_\textnormal{P}^2$, the higher-order terms are of $\order{1}$ and the \gls{eft} breaks down.
The size of the smallest \gls{bh} can be derived from entropy considerations: it would be too naive to assume that the smallest \gls{bh} has $r_\heartsuit\sim1/M_\textnormal{P}$, as even the smallest \gls{bh} can carry all states of the theory, i.e. $\mathcal{S}_\heartsuit\geq\mathcal{S}_\textnormal{S}=\Lambda_\textnormal{S}^{2-d}M_\textnormal{P}^{d-2}$ and $r_\heartsuit\geq l_\textnormal{S}=1/\Lambda_\textnormal{S}$ \cite{cribiori_species_2023},
with $\Lambda_\textnormal{S}=M_\textnormal{P}/N_\textnormal{S}^{\frac{1}{d-2}}$ the suppression factor for curvature terms in the \gls{eft}.

Furthermore, at least the smallest \gls{bh} must fit into the \gls{ds} horizon:
Given a scalar field $\phi$ with potential $V(\phi)$ in an approximately flat (quasi-)\gls{ds} space with Hubble radius
\begin{equation}
    r_H\sim\frac{1}{H}=\frac{M_\textnormal{P}^{\left(d-2\right)/2}}{\sqrt{V}},
\end{equation}
the Gibbons\textendash Hawking entropy
\begin{equation}
    \mathcal{S}_\text{dS}\sim\left(r_HM_\textnormal{P}\right)^{d-2}
\end{equation}
is certainly larger than the number of light states $N_\textnormal{S}$ \cite{van_de_heisteeg_bounds_on_field_2023}.
Using the species scale (\cref{eq:speciess}), we find that
\begin{equation}
    \frac{\sqrt{V}}{M_\textnormal{P}^{\left(d-2\right)/2}}\leq\Lambda_\textnormal{S}=\frac{1}{r_\text{min}},
\end{equation}
i.e. the smallest \gls{bh} fits into the \gls{ds} horizon \cite{van_de_heisteeg_bounds_on_field_2023}.\footnote{
    Curvature terms in an \gls{eft} are $V\sim R\lesssim\Lambda_\textnormal{S}^2$; using \cref{eq:SSC_exponential} then yields a bound on the field range for flat potentials: $\Delta\phi\lesssim\sqrt{\left(d-1\right)\left(d-2\right)}\log\left(1/V_0\right)$ \cite{vafa_origin_2025}.
}

\paragraph{Is the species scale a constant?}
When moving through moduli space, the number of species, and therewith the species scale, changes.\footnote{
    For example, the number of light species is higher in the early Universe than it is today.
    }
The change of the species scale can be quantified and happens within bounds.
Upper bounds of the form
\begin{equation}\label{eq:upperSpeciesChange}
    \abs{\frac{\nabla\Lambda_\textnormal{S}}{\Lambda_\textnormal{S}}}^2<\frac{\mathfrak{l}}{M_\textnormal{P}^{d-2}},
\end{equation}
with $\mathfrak{l}>\frac{1}{d-2}$ an $\order{1}$ parameter\footnote{
    \Citet{van_de_heisteeg_species_2024} find $\mathfrak{l}=1$ except in a 4d $\mathcal{N}=2$ theory arising from a type II compactification on \gls{cy}$_3$ and a $d=8$ maximal supergravity setting, where corrections to the $R^2$ respectively $R^4$ setting force the slope to approach the asymptotic value from above.
}
that depends on $d$,
and $\abs{\nabla\Lambda_\textnormal{S}}^2=G^{ij}\partial_{\phi_i}\Lambda_\textnormal{S}\partial_{\phi_j}\Lambda_\textnormal{S}$,
$G^{ij}$ the inverse moduli field space metric,
hold everywhere in moduli space \cite{scalisi_species_2024,cribiori_note_2023,van_de_heisteeg_bounds_on_species_2023,cribiori_species_2023}
and mean
that integrating out the short-distance modes of moduli does not change the obtained \gls{eft} any further when the massive states are already integrated out, since the massive states always dominate the short-distance modes of other fields, in particular the modes of massless scalar fields, as a single massless field cannot account for the entropy of the smallest \gls{bh} \cite{van_de_heisteeg_bounds_on_species_2023}.
An upper bound indicates that finite scalar field variations must be allowed \cite{scalisi_species_2024},
but also that the cosmological constant experiences an exponential suppression \cite{cribiori_note_2023}.
\citet{andriot_bumping_2023} shows that $\nabla\Lambda_\textnormal{S}/\Lambda_\textnormal{S}$ does not grow monotonically, but has a maximum before it approaches its asymptotic value (which lies below the maximum).
The origin of the bump is not entirely clear and presumably model dependent; \citet{andriot_bumping_2023} finds that axions do not generate such a bump,
nor do the exponential potentials of dilatonic and volume factors,
but saxionic exponential potentials do.
The maximum point of $\Lambda_\textnormal{S}$\,\textemdash\,the desert point \cite{long_desert_2021} or centre of moduli space \cite{van_de_heisteeg_moduli-dependent_2024}\,\textemdash\,gives an estimate of the smallest number of light states that a \gls{qg} theory can have \cite{van_de_heisteeg_species_2024}.

A lower bound of the form
\begin{equation}
    \abs{\frac{\Lambda_\textnormal{S}^\prime(\phi)}{\Lambda_\textnormal{S}(\phi)}}\geq\frac{1}{\sqrt{\left(d-1\right)\left(d-2\right)}},
\end{equation}
is derived by
\citet{calderon-infante_entropy_2023},
which represents a connection to the \gls{dc}, as a vanishing value corresponds to an infinite field distance, which is forbidden according to the \gls{dc} \cite{scalisi_species_2024}.

Furthermore, an equality of the form
\begin{equation}\label{eq:mass-SSC-dynamics}
    \frac{\grad m}{m}\frac{\grad \Lambda_\textnormal{S}}{\Lambda_\textnormal{S}}=\frac{M_{\textnormal{P};d}^{d-2}}{d-2},
\end{equation}
is derived by
\citet{castellano_stringy_2023,castellano_universal_2023,castellano_quantum_2024},
where the nabla operator indicates that they have multiple species in mind.
By using $\Lambda_\textnormal{S}=M_{\textnormal{P};d}/N_\textnormal{S}^{1/\left(d-2\right)}$, \cref{eq:mass-SSC-dynamics} can be written in a form that does not explicitly depend on the number of large spacetime dimensions \cite{castellano_quantum_2024}:
\begin{equation}
    \frac{\grad m}{m}\frac{\grad N_\textnormal{S}}{N_\textnormal{S}}=-M_{\textnormal{P};d}^{2-d}.
\end{equation}
This form constrains the density variation of states below the species scale and the rate at which they are becoming light: the denser the spectrum becomes, the faster the species scale vanishes \cite{castellano_quantum_2024}.

\paragraph{How can the species scale be understood if there are multiple towers?}\label{p:SSC_multiple-towers}
An individual tower has a spectrum of
\begin{equation}
    m_n=n^{1/p}m_\textnormal{t},
\end{equation}
with $m_\textnormal{t}$ the mass of the lightest state,
$n$  representing the step of the state within the tower,
and $p$ representing the number of towers with identical mass gap \cite{seo_axion_2024,castellano_iruv_2022}.
There can be multiple \gls{kk} towers, but only one fundamental string \cite{castellano_emergence_2023,castellano_iruv_2022}.
For a string tower, we find 
\begin{align}
    m_\textnormal{t}&=m_\textnormal{s}=1/l_\textnormal{s}\\
    m_n&\sim\sqrt{n}m_\textnormal{s}\\
    p&\rightarrow\infty;
\end{align}
$m_n$ are \textit{Regge modes} with oscillator number $n$ \cite{dvali_evaporation_2009,seo_axion_2024}.
There is an upper bound on the mass scale of towers,
\begin{equation}
    m\lesssim\Lambda_\text{IR}^{2\lambda_d}M_{\textnormal{P};d}^{1-2\lambda_d},
\end{equation}
coming from the size of the universe, $1/\Lambda_\text{IR}$, as the lowest energy that can occur must correspond to a wavelength that fits into the universe \cite{castellano_emergence_2023,castellano_iruv_2022}.
In general, $\lambda_d=\left(d-2+p\right)/\left[2p\left(d-1\right)\right]$, but for a single \gls{kk} tower, \citet{castellano_iruv_2022} find, using the \gls{ceb}, that in \gls{ds} and \gls{ads} space $\lambda\geq1/2$, which corresponds to the \gls{adsdc} (\cref{eq:AdSDC}).
For multiple towers or stringy towers, they find $\lambda\geq1/d$; a lower value would lead to gravitational collapse.

If there are multiple towers, we have
\begin{equation}
    \Lambda_{\textnormal{S},i}\simeq N_i^{1/p_i}m_i\simeq\frac{M_{\textnormal{P};d}}{N_i^\frac{1}{d-2}},
\end{equation}
with $N_i$ the number of states of the $i$-th tower \cite{castellano_emergence_2023,castellano_quantum_2024}.
If there is no mixing between the states, i.e. the fields are independent and there are no states with mixed charges,
we find
$\Lambda_\textnormal{S}=\min\{\Lambda_{\textnormal{S},i}\}$ \cite{castellano_emergence_2023,castellano_quantum_2024}. This means that the species scale is dominated by the tower with the lightest mass scale \cite{castellano_emergence_2023}. The total number of states is $N_\text{tot}=\sum_iN_i.$

If the towers mix, the leading tower alone is insufficient to give the species scale, as heavier towers can still contribute with a diverging number of states, and we find \cite{castellano_quantum_2024,castellano_emergence_2023}:
\begin{align}
    p_\text{eff}&=\sum_ip_i\\
    m_\text{eff}&=\left(\prod_i m_i^{p_i}\right)^{1/p_\text{eff}}\\
    N_\textnormal{S}&\simeq\prod_iN_i\\
    &\simeq\left(\frac{M_{\textnormal{P};d}}{m_\text{eff}}\right)^\frac{\left(d-2\right)p_\text{eff}}{d-2+p_\text{eff}}\\
    \Lambda_\textnormal{S}&\simeq M_{\textnormal{P};d}\left(\frac{M_{\textnormal{P};d}}{m_\text{eff}}\right)^\frac{-p_\text{eff}}{d-2+p_\text{eff}}.
\end{align}
Which towers have to be included can be determined by an iterative process \cite{castellano_quantum_2024}:
You start with the lightest tower and determine its species scale.
If the second lightest towers starts below this scale, the second-lightest tower has to be included in the calculation of the species scale.
Then, you compare the mass scale of the third-lightest tower to this newly obtained species scale, and if this mass scale is below the species scale based on the two lightest tower, the third tower has to be included as well. And so forth.
In short: if $m_i\geq m_{i+1}^{\left(d-2+p_{\text{eff},i}\right)/p_{\text{eff},i}}$, 
the $i+1$-th tower has to be included.

In \cref{p:WGC_multiparticle}, we will encounter the convex hull conjecture of the \gls{wgc}. A conceptually similar conjecture is formulated for the \gls{ssc}: \enquote{The convex hull of species vectors $\{\Vec{\mathcal{Z}}\}$ defined at infinity should contain the ball of radius $\lambda_{\text{S,min}}=1/\sqrt{\left(d-1\right)\left(d-2\right)}$} with the species vector given by $\mathcal{Z}^a=-\delta^{ab}e_b^i\partial_i\log\Lambda_\textnormal{S}$ with $e_i^a(\phi)$ a vielbein in field space that satisfies $\delta_{ab}e_i^a(\phi)e_j^b(\phi)=G_{ij}(\phi)$ \cite{castellano_quantum_2024}.
The species vector for \gls{kk} modes is given by $\mathcal{Z}_{\text{KK},n}=\sqrt{n/\left[\left(d+n-2\right)\left(d-2\right)\right]}$,
whereas the species vector for string oscillator modes is given by $\mathcal{Z}_\textnormal{s}=1/\sqrt{d-2}$ \cite{castellano_quantum_2024}.
A relation to the mass can be obtained by defining the scalar charge-to-mass vector $\zeta^i=-\partial^i\log m$ and $\zeta^a=e_i^a\zeta^i$, which is $\zeta_{\text{KK},n}=\sqrt{\left(d+n-2\right)/\left[n\left(d-2\right)\right]}$ for the \gls{kk} modes,
respectively $\zeta_\textnormal{s}=1/\sqrt{d-2}$ for the string oscillator modes \cite{castellano_quantum_2024}.

\paragraph{How are charged towers to be treated?}
When studying the stability and abundance of species, we have to differentiate between two types of species (charged and neutral) and we have to consider two different phases (before and after the first half-life).
Charged species have a lower temperature than their neutral counterparts, as $T_\text{S,c}=\Lambda_\textnormal{S}^2$ but $T_\text{S,n}=\Lambda_\textnormal{S}$ \cite{basile_minimal_2024}.
During the first half-life, species undergo thermal decay with
\begin{equation}
    \Gamma_{\textnormal{S},\text{th}}\simeq T_\textnormal{u}>T_\textnormal{u}^2\simeq H,
\end{equation}
where we introduced the temperature of the universe $T_\textnormal{u}$ and used the finding that in the early, radiation dominated universe, we have, according to Stefan's law, an energy density $\rho\sim T_\textnormal{u}^4$ and a scale factor $a\sim T_\textnormal{u}^{-1}$, such that $H\simeq T_\textnormal{u}^2$ \cite{basile_minimal_2024}.
After the first half-life, the species is decaying quantum-mechanically, with
\begin{equation}
    \Gamma_{\textnormal{S},\text{qm}}\simeq T_\textnormal{u}^3<T_\textnormal{u}^2\simeq H,
\end{equation}
where the species decouple and the expansion of the universe happens in a background with constant \gls{uv} cutoff scale $\Lambda_\textnormal{S}$ \cite{basile_minimal_2024}.
This indicates that charged species are more stable than neutral ones.
Furthermore, we can use this to check for the compatibility of other proposals, such as the microscopic dark dimension proposal presented in \cref{sss_AdSDC_Cosmology}:
The proposal predicts a tower of $m_\text{KK}\simeq\Lambda_\text{cc}^{1/4}\simeq\SI{1}{\electronvolt}$.
Our discussion suggests that the corresponding species scale is
\begin{equation}
    \Lambda_\textnormal{S}\simeq m_\text{KK}\simeq\Lambda_\text{cc}^{1/4},
\end{equation}
ergo a decoupling temperature of
\begin{align}
    T_{\textnormal{S}}&=\left(\left(\Lambda_\text{cc}^{1/4}\right)^\beta\right)^{1/3}&=\Lambda_\text{cc}^{\beta/12}%
\end{align}
where $\beta=1$ ($\beta=2$) correspond to the neutral (charged) case, and we took the third root because we are interested in the decoupled behaviour.
The charged case with $T_\text{S,c}=\Lambda_\text{cc}^{1/6}\simeq\SI{1}{\giga\electronvolt}$ corresponds to the prediction from \citet{gonzalo_dark_2022} that the initial temperature for the cosmology with the correct \gls{dm} abundance and 4 stable large dimensions in a dark dimension scenario is $T=\Lambda_\text{cc}^{1/6}\simeq\SI{1}{\giga\electronvolt}$ \cite{basile_minimal_2024}.

\subsubsection{Evidence}
The species bound was first\footnote{
    Similar consideration were also discussed by \citet{veneziano_large-n_2002}.
    }
introduced in the context of \glspl{bh} by \citet{dvali_black-bound_2008,dvali_black_2010,dvali_nature_2008,dvali_black-entropy_2008},
and is derived in the context of
\gls{bh} physics \cite{brustein_bound_2009},
higher curvature corrections to extremal \gls{bh} entropy \cite{cribiori_black_2023},
small \glspl{bh} in 4d $\mathcal{N}=2$ supergravity and 10d examples \cite{calderon-infante_emergence_2024},
4d $\mathcal{N}=2$ type II \gls{cy} threefolds \cite{van_de_heisteeg_moduli-dependent_2024},
the $\mathcal{N}=1$ axiverse \cite{martucci_wormholes_2024},
type IIB string theory \cite{ishiguro_stabilization_2024},
type II toroidal orbifolds \cite{cribiori_note_2023},
various dimensions and amounts of supersymmetry \cite{van_de_heisteeg_species_2024},
the dominance of local over non-local perturbations \cite{arfaei_locality_2023},
causality on gravitational interactions between higher-spin massive composite particles \cite{kaplan_species_2019}, and
quantum information theory \cite{dvali_quantum_2009}.
Relations to other swampland conjectures are highlighted in \cref{rel:FLB_SSC,rel:SSC_fnomfC,rel:WGC_EP,rel:EP_SDC}.
\citet{castellano_quantum_2024} gives a thorough and up-to-date overview that gives more insights into the string theoretical intricacies than this review aims to provide.

\subsection{Tadpole Conjecture}\label{s:tadpole} %
The flux contributions to the D3-tadpole\footnote{
    A tadpole is a 1-loop Feynman diagram with only one external leg \cite{evans_diagramology_2018}. If the tadpoles do not cancel, the vacuum respectively the background is not stable and does not solve the \glspl{eom} \cite{kitazawa_tadpole_2008}.
    According to \citet{coleman_notes_2013}, the alternative name \textit{spermion} was rejected by the editors of physical review.
}
that stabilise $n_\text{mod}=h^{3,1}$ complex-structure moduli of a \gls{cy} fourfold in a smooth F-theory compactification
respectively $n_\text{mod}=2\left(h^{2,1}+1\right)$ complex-structure moduli of a \gls{cy} threefold in a smooth type IIB string theory compactification
grow at least linearly with $n_\text{mod}$, such that
\begin{equation}
    q_{D3}^\text{stabilisation}>\mathfrak{t}n_\text{mod},
\end{equation}
for large $n_\text{mod}$ with $\mathfrak{t}$ the flux-tadpole constant \cite{bena_tadpole_2021,plauschinn_tadpole_2022}.
The strong form of the conjecture assumes that $\mathfrak{t}>1/3$ \cite{bena_tadpole_2021}.\footnote{
    The lower bound $\mathfrak{t}=1/3$ is questioned in studies by \citet{coudarchet_symmetric_2023,lust_tadpole_2022} (without providing a rigorous proof that it actually is violated).
    Nevertheless, \citet{becker_fluxes_2022} as well as \citet{rajaguru_fully_2024} present counterexamples to this stronger form, yet not to the \gls{tpc} per se, as $\mathfrak{t}>1/4$ would suffice.
    }

\subsubsection{Implications for Cosmology}
While no direct implications for cosmology are discussed in the literature, the \gls{tpc} acts as supporting evidence for other swampland conjectures, as we note in \cref{rel:TPC_dSC,rel:TPC_tame,rel:TPC_AdSDC,rel:TPC_finite}.

\subsubsection{General Remarks}
Compactification gives rise to (unphysical) massless scalar fields, e.g. Kähler moduli or complex-structure moduli \cite{bena_algorithmically_2021}.
If there was a mechanism to give those moduli mass, it would explain why we do not observe them\,\textemdash\,such a mechanism is called moduli stabilisation \cite{prieto_moduli_2024}.
Flux can stabilise moduli \cite{van_riet_beginners_2023,bena_algorithmically_2021}:
the moduli become massive, if topologically non-trivial magnetic fluxes along the cycles of the internal geometry are turned on.
The flux contributes to the tadpole as $\int H_3\wedge F_3=q_\text{flux}=q_\text{O3}-q_\text{D3}$ \cite{van_riet_beginners_2023} respectively $q_\text{flux}+q_\text{D3}=6\left(8+h^{1,1}+h^{3,1}-h^{2,1}\right)$ in F-theory, where for a large number of moduli, the $h^{3,1}$-term dominates \cite{prieto_moduli_2024}.
Tadpole charges have to be negative \cite{plauschinn_moduli_2021}: empirically, settings with positive tadpole charges show runaway behaviour.
The electric brane charges have to sum up to zero on a compact manifold, i.e. the brane\textendash tadpole cancellation condition places an upper bound on the number of (integer quantised) fluxes \cite{bena_algorithmically_2021,lust_tadpole_2022,van_riet_beginners_2023,lust_large_2021}.
According to the \gls{tpc}, the fluxes needed to stabilise the moduli carry a charge proportional to the number of stabilised moduli, if the number of moduli is large.
This presents a tension to settings with a large number of moduli, i.e. there will always remain flat directions / moduli that are not stabilised \cite{prieto_moduli_2024}.\footnote{
    This agrees with the \textit{Massless Minkowski Conjecture} presented by \citet{andriot_exploring_2022,andriot_erratum_2022}, which states that \enquote{10d supergravity solutions compactified to 4d Minkowski always admit a 4d massless scalar}. A massless scalar field often\,\textemdash\,though not only\,\textemdash\,appears along a flat direction \cite{andriot_erratum_2022}.
}

Counterexamples to the \gls{tpc}, where the flux contribution to the tadpole does not scale with the number of moduli, i.e. $q_{D3}^\text{stabilisation}$ is independent of $h^{3,1}$, are presented by \citet{prieto_moduli_2024} and \citet{marchesano_f-theory_2021}, and debunked by \citet{lust_large_2021} and \citet{grimm_moduli_2022}.
However, it has been noted that spaces containing singularities might violate the \gls{tpc} \cite{bena_tadpole_2021,gao_lvs_2022}.

\subsubsection{Evidence}
The \gls{tpc} is presented by \citet{bena_tadpole_2021}
and studied in the context of
M-theory \cite{bena_algorithmically_2021,lust_tadpole_2022},
type IIB and F-theory \cite{tsagkaris_moduli_2023,bena_d7_2022,lust_large_2021,becker_fluxes_2022,gao_lvs_2022,marchesano_f-theory_2021,plauschinn_tadpole_2022,coudarchet_symmetric_2023,grimm_moduli_2022,grana_tadpole_2022,braun_tadpoles_2023,junghans_topological_2022},
the $1^9$ Landau\textendash Ginzburg model \cite{becker_stabilizing_2024},
the $2^6$ Landau\textendash Ginzburg model \cite{rajaguru_fully_2024},
the $2^6$ Gepner model \cite{becker_tadpole_2024}, and
higher-order effects \cite{smith_all-orders_2024}.
Furthermore, relations to other swampland conjectures are highlighted in \cref{rel:TPC_dSC,rel:TPC_tame,rel:TPC_AdSDC,rel:TPC_finite}.

\subsection{Tameness Conjecture}\label{sec:tame}%

A valid \gls{eft} below a given cutoff scale is labelled by a definable parameter space, and has a scalar field space and coupling functions that are definable under an o-minimal structure, which is $\mathbb{R}_\text{an,exp}$ \cite{grimm_tameness_2022,grimm_taming_2021,douglas_tameness_2023}.\footnote{
    $\mathbb{R}_\text{an,exp}$ is an o-minimal structure generated by $\mathbb{R}_\text{alg}$\,\textemdash\,the semi-algebraic set of all polynomial (in)equalities\,\textemdash, i.e. $\mathbb{R}_\text{an,exp}$ is the structure of all restricted real analytic functions $\mathbb{R}_\text{an}$, and the real exponential function $\exp:\mathbb{R}\rightarrow\mathbb{R}$ \cite{grimm_taming_2021}.
    The set of all restricted real analytic functions and exponential polynomial equations and its projections generate this structure \cite{grimm_taming_2021}, which can be understood as the subset of $\mathbb{R}^n$ described by the polynomials $P\left(x_1,\dots,x_m,f_1,\dots,f_m,e^{x_1},\dots,e^{x_m}\right)=0$, with $f_i$ a restricted real analytic function \cite{grimm_tameness_2022}.
    An analytic function is a function that at any point of its domain can be expressed as a power series that converges at that point \cite{douglas_tameness_2024}.
    A structure is o-minimal if every definable subset of $\mathbb{R}$ is a finite union of points and open intervals \cite{grimm_taming_2021,douglas_tameness_2023,douglas_tameness_2024}.
}

\subsubsection{Implications for Cosmology}
Scalar field potentials have only a finite number of minima, as tame functions do not allow for an infinite number of oscillations by definition \cite{grimm_taming_2021}.\footnote{
    Tame functions do not show pathological behaviour like infinite oscillations or accumulation of singularities, for example, $V(\phi)=\sin\left(\phi^{-1}\right)$ cannot appear in a string compactification \cite{grimm_taming_2021,douglas_tameness_2024}.
}

\subsubsection{General Remarks}
The tameness conjecture imposes finiteness constraints on the geometry of the \gls{eft} \cite{grimm_tameness_2022},\footnote{
    The parameter space of an \gls{eft} is not only assumed to exist in general, but also to be restricted, as certain components, e.g. infinite discrete components like a lattice, are excluded \cite{grimm_tameness_2022}.
} 
i.e. it constrains all the couplings $g$ \cite{grimm_taming_2021}, the field space metrics $G_{ab}$, the scalar potentials $V$, kinetic matrices \cite{grimm_tameness_2022}, as well as all the correlators, partitions \cite{lanza_machine_2024}, amplitudes, and so forth \cite{douglas_tameness_2023}. A reason to expect physical theories to be tame stems from the observation that observables are usually real analytic functions or (a finite number of) phase transitions \cite{douglas_tameness_2023}.

Some authors state that assessing the tameness of a theory requires knowledge about the \gls{uv} properties of said theory \cite{douglas_tameness_2023,grimm_taming_2021,douglas_tameness_2024}. The top-down prediction is that an \gls{eft} with a finite cutoff that is part of the landscape is tame \cite{douglas_tameness_2023}, since, as long as the number of loop corrections is finite, tameness is preserved when integrating out heavy fields \cite{douglas_tameness_2024}.\footnote{
    Feynman amplitudes are tame \cite{douglas_tameness_2024}.
}
An \gls{eft} that is not tame is therefore certainly in the swampland, whereas an \gls{eft} that is tame might or might not be in the landscape.

\subsubsection{Evidence}
The tameness conjecture is proposed by \citet{grimm_taming_2021},
and discussed in the context of
type IIB string theory \cite{grimm_tameness_2022,grimm_moduli_2022},
\glspl{qft} \cite{douglas_tameness_2024},
\glspl{qft} and \glspl{cft} \cite{douglas_tameness_2023}, and using
machine learning techniques \cite{lanza_machine_2024}.
Furthermore, relations to other swampland conjectures are highlighted in \cref{rel:TPC_tame,rel:TC_DC,rel:fnomfc_TC}.

\subsection{Trans\textendash Planckian Censorship Conjecture}\label{sec:tpcc}%

The \gls{tcc} is presented by \citet{bedroya_trans-planckian_2020}:
Trans-Planckian quantum fluctuations remain quantum.
Trans-Planckian quantum fluctuations should always remain smaller than the Hubble horizon, and therefore do not freeze in an expanding universe. Otherwise, there would be classical observations of trans-Planckian quantum modes.
For a phase of accelerated expansion, this means that
a mode with physical wavelength $\lambda<l_\textnormal{P}=1/M_\textnormal{P}$ at $t_\textnormal{i}$ will always remain smaller than the Hubble scale and the following bound on length scales leaving the horizon holds:
\begin{align}
    \frac{a_\textnormal{f}}{a_\textnormal{i}}&<\frac{M_\textnormal{P}}{H_\textnormal{f}}\label{eq:TCC_sub-Planckian-Hubble}\\
    \int_{t_\textnormal{i}}^{t_\textnormal{f}}\!H(t)\,\mathrm{d}t&<\log\frac{M_\textnormal{P}}{H_\textnormal{f}},
\end{align}
with $a_\textnormal{f}$ and $H_\textnormal{f}$ the scale factor respectively the Hubble factor at the end of the phase, and $a_\textnormal{i}$ the scale factor at the beginning of the phase \cite{shlivko_trans-planckian_2023,heisenberg_model_2021,bedroya_holographic_2022,das_runaway_2020,brandenberger_trans-planckian_2021,bedroya_trans-planckian_2020,blumenhagen_quantum_2020}.
This is a global, non-local statement, i.e. it must be true everywhere \cite{blumenhagen_quantum_2020}.
As it should be with any swampland conjecture, this statement becomes trivial when gravity decouples, i.e. for $l_\textnormal{P}\rightarrow0$ respectively $M_\textnormal{P}\rightarrow\infty$.

\subsubsection{Implications for Cosmology}
An important quantity in the \gls{tcc} is the Hubble parameter. Since $H$ corresponds to the energy density in the Universe, the bound $H<M_\textnormal{P}$ should naturally hold at all times \cite{bedroya_trans-planckian_2020}.
In models without accelerating expansion, the Hubble radius grows faster than the wavelength of any mode, i.e. the \gls{tcc} is always satisfied \cite{brandenberger_trans-planckian_2021}.\footnote{
    For example, in string gas cosmology \cite{brandenberger_superstrings_1989}, the early phase of the universe is quasistatic, and in the ekpyrotic scenario \cite{khoury_ekpyrotic_2001} (see \cref{sp:ekpyrotic}), it is contracting, ergo, the \gls{tcc} constraints are easily satisfied \cite{bedroya_trans-planckian-inflation_2020}.
    }
The same holds for emergent cosmologies or scenarios with a bounce \cite{brandenberger_bouncing_2017,finelli_generation_2002,gasperini_pre-big-bang_1993,khoury_big_2002,battarra_cosmological_2014,bramberger_quantum_2017,farnsworth_spinor_2017,ijjas_fully_2017,gielen_perfect_2016,easson_g-bounce_2011,creminelli_smooth_2007,buchbinder_new_2007}, as long as the energy scale of the bounce respectively the energy scale of the emergent phase is below the Planck scale \cite{brandenberger_trans-planckian_2021,bedroya_trans-planckian-inflation_2020}.\footnote{
    Modes leaving the Hubble radius have a length of $\lambda\sim M_\textnormal{P}/\Lambda^2$, with $\Lambda$ the energy scale of the bounce respectively the emergence scale \cite{brandenberger_trans-planckian_2021}.
}
In the following, we study cases where the \gls{tcc} is more constraining.

\paragraph{Lifetime}
The \gls{tcc} limits the lifetime of a metastable \gls{ds} solutions to be \cite{bedroya_trans-planckian_2020,rasulian_swampland_2021}
\begin{equation}
    \tau<\frac{1}{H_\textnormal{f}}\log\frac{M_\textnormal{P}}{H_\textnormal{f}};
\end{equation}
since 
\begin{equation}
    \dot{H}=-\frac{\left(1+w\right)\rho}{\left(d-2\right)},
\end{equation}
$H$ is monotonically decreasing with time for an \gls{eos} parameter $w>-1$, such that the \gls{tcc} implies
\begin{align}
    H_\textnormal{f}\left(t_\textnormal{f}-t_\textnormal{i}\right)\leq\int^{t_\textnormal{f}}_{t_\textnormal{i}}\!H\,\mathrm{d}t<\log\frac{M_\textnormal{P}}{H_\textnormal{f}}.
\end{align}
For the lifetime of our Universe, this yields an upper\footnote{
    For example, \citet{tsukahara_life-time_2023} studies a setting where D3-branes cause a decay of \gls{ds} space before the \gls{tcc} bound is reached.
    } 
bound of $\order{\num{e12}}$ years \cite{bedroya_trans-planckian_2020}.
\citet{agmon_lectures_2023} notes that our Universe passes only thanks to the log term, and
\citet{vafa_origin_2025} notes that the accelerated expansion we see today could indicate that we are the start of \enquote{the end of our universe}.
If the universe lives longer than the lifetime predicted by the \gls{tcc}, the \gls{eft} breaks down, the cosmological constant dramatically changes in value, or we tunnel to a different vacuum solution \cite{bedroya_sitter_2020}.

It is insightful to compare this lifetime bound with other bounds:
\citet{aalsma_chaos_2020} find that the scrambling time\footnote{
    \enquote{The scrambling time of a system is the time that it takes for the information of a generic pure state to disperse among all microscopic degrees of freedom} \cite{bedroya_sitter_2020}. For example, if a state falls into a \gls{bh} and perturbs the equilibrium of the horizon, the dispersion time is the time it takes for that external perturbation to thermalise and to be radiated away \cite{bedroya_sitter_2020}. For \gls{ds} space, it is the time it takes something trans-Planckian to stretch beyond the Hubble horizon \cite{bedroya_sitter_2020}.
    }
allows for $N_e<\log\mathcal{S}_\textnormal{dS}$ $e$-foldings, with $\mathcal{S}_\textnormal{dS}$ the entropy of \gls{ds} space. Since $\mathcal{S}_\textnormal{dS}\sim M_\textnormal{P}^2/H^2$, this allows for twice as many $e$-foldings as the \gls{tcc}. \citet{aalsma_chaos_2020} stress that inflation doesn't have to stop after that, but backreaction effects become important. This limit assures that the entanglement entropy between sub- and super-Hubble modes remains below the thermal entropy after inflation, respectively the Gibbons\textendash Hawking entropy of \gls{ds} space \cite{brandenberger_string_2021,brahma_entanglement_2020,aalsma_shocks_2021,brahma_constraints_2022}.
Up to logarithmic corrections \cite{blumenhagen_sitter_2021}, the scrambling time of \gls{ds} space is therfore equal to (twice \cite{andriot_web_2020,aalsma_chaos_2020}) the \gls{tcc} bound \cite{bedroya_sitter_2020}.\footnote{
    A scrambling time that corresponds to the lifetime predicted by the \gls{tcc} is for example found in a holographic model where the vacuum is dominated by \textit{quantum hair} \cite{addazi_holographic_2020}: trans-Planckian modes get entangled with the quantum hair (and decohere) within $\tau\sim\sqrt{N}l_\textnormal{P}\log N$, with $N\sim \mathcal{S}$ the number of quantum hairs that correspond to the entropy of the vacuum $\mathcal{S}$, which saturates the \gls{tcc} bound. %
    }
Alternatively, treating particles as wave packets leads to a \gls{uv} cutoff scale that imposes a lifetime for the \gls{eft} description, after which new physics is to be expected \cite{blamart_tcc_2023}: $\tau_\textnormal{th}\sim 2H^{-1}\log\left(H^{-1}\right)$ for an eternal \gls{ds} phase, respectively $\tau_\textnormal{th}\sim H^{-1}\log\left(H^{-1}\right)$ for a \gls{ds} phase preceded by a matter- or radiation-dominated phase.
Similarly, \citet{danielsson_thermalization_2004} derive an upper bound for the lifetime of \gls{ds} space, based on the idea that if non-gravitational interactions are frozen, a local observer will see that everything thermalises within $\tau\sim R^3/l_\textnormal{P}^2$. The bound relies on weaker assumptions than the \gls{tcc} and allows for a longer lifetime of \gls{ds} space.
Stronger assumptions yield a thermalisation time of \gls{ds} space of $\tau_\textnormal{th}\sim H^{-1}\log\left(H^{-1}\right)$, which is equal to the \gls{tcc} prediction, ergo, by the time a \gls{ds} space thermalises, it ceases to exist: it is too short-lived to be considered a thermal background \cite{bedroya_sitter_2020}.

\citet{andriot_negative_2023} look into the future of a possible crunch of the universe, and investigate the implications of the \gls{tcc} in this context:
Their guiding idea is that even during a crunch, no relativistic mode that corresponds to the typical energy scale of the universe at a time $t_\textnormal{i}$ can be blue-shifted to become trans-Planckian.
To investigate the implications of this idea, they study the quantity $\sqrt{V}/M_\textnormal{P}$ instead of $H$. In \gls{ds} space, $H$ and $\sqrt{V}/M_\textnormal{P}$ represent the same quantity, but in other spacetimes, $H$ does not provide a meaningful length scale. This change to energy scales allows studying such systems more generally. For \gls{ads} spaces, conditions for the scalar field potential very similar to the \gls{dsc} are obtained:
\begin{align}
    \mathfrak{c}&=\frac{2}{\sqrt{\left(d-1\right)\left(d-2\right)}}\\
    0>V(\phi)&\geq-e^{-\mathfrak{c}\abs{\phi-\phi_i}}\\
    \expval{-\frac{V^{\prime}}{V}}_{\phi\rightarrow\infty}&\geq \mathfrak{c}\\
    \expval{\frac{V^{\prime\prime}}{V}}_{\phi\rightarrow\infty}&\geq \mathfrak{c}^2.
\end{align}
While the \gls{tcc} yields a maximum lifetime of the expansion phase, this more general treatment yields a lifetime for a contracting phase:
\begin{equation}
    t_\textnormal{f}-t_\textnormal{i}<\frac{1}{\abs{H_\textnormal{i}}}\log\frac{M_\textnormal{P}^2}{\sqrt{\abs{V_\textnormal{i}}}}.
\end{equation}
The closer to the crunch we start, the larger is $\abs{H_\textnormal{i}}$, which indicates the consistency of the statement.

\paragraph{Scalar field potentials} are constrained in the asymptotic limit of moduli space \cite{bedroya_trans-planckian_2020,heisenberg_model_2021,bedroya_trans-planckian_2020,van_beest_lectures_2022,basile_sitter_2020}:
\begin{equation}\label{eq:tcc}
    \left.\frac{\abs{\nabla V}}{V}\right|_\text{asympt}\geq\frac{2}{\sqrt{\left(d-1\right)\left(d-2\right)}}.
\end{equation}
The bound becomes trivial for many large dimensions \cite{bonnefoy_swampland_2021}.
In 4d, the bound is also derived using Hodge theory \cite{bastian_weak_2021}.
Since the bound holds only in the asymptotic limit, it is weaker than for example the \gls{dsc} \cite{basile_sitter_2020}, but it is analytically more powerful, as it does not contain any unknown $\order{1}$-constants \cite{bedroya_trans-planckian_2020}.
We can derive this bound as follows \cite{bedroya_trans-planckian_2020}:
The \gls{tcc} informs us that
\begin{equation}
    \int_{\phi_\textnormal{i}}^{\phi_\textnormal{f}}\!\frac{H}{\dot{\phi}}\,\mathrm{d}\phi=
    \int_{t_\textnormal{i}}^{t_\textnormal{f}}\!H(t)\,\mathrm{d}t<\log\frac{M_\textnormal{P}}{H_\textnormal{f}}.
\end{equation}
The Friedmann equation
\begin{equation}
    \frac{\left(d-1\right)\left(d-2\right)}{2}H^2=\frac{\dot{\phi}^2}{2}+V(\phi)
\end{equation}
yields, for positive $V(\phi)$,
\begin{AmSalign}
    \frac{H}{\abs{\dot{\phi}}}&>\frac{1}{\sqrt{\left(d-1\right)\left(d-2\right)}}\\
    \Rightarrow \frac{\abs{\phi_\textnormal{f}-\phi_\textnormal{i}}}{\sqrt{\left(d-1\right)\left(d-2\right)}}&<-\log H_\textnormal{f}\label{eq:TCC-scalar-field}\\
    \Rightarrow H_\textnormal{f}&<e^{-\frac{\abs{\phi_\textnormal{f}-\phi_\textnormal{i}}}{\sqrt{\left(d-1\right)\left(d-2\right)}}}\\
    \Rightarrow V(\phi)&<\frac{\left(d-1\right)\left(d-2\right)}{2}e^{-\frac{2\abs{\phi-\phi_\textnormal{i}}}{\sqrt{\left(d-1\right)\left(d-2\right)}}},
\end{AmSalign}
where we used the \gls{tcc} to get the second line, then rearranged the second line to get the third line, and used that $V(\phi)$ is bounded from above from the Friedmann equation in the last line. Please note that we omitted factors of $M_\textnormal{P}=1$.
To see why it only holds in the asymptotic field limit, we can come from the other direction, assuming $V^\prime<0$ for definiteness:
\begin{align}
    \left.\expval{\frac{-V^\prime}{V}}\right\rvert_{\phi_\textnormal{i}}^{\phi_\textnormal{f}}&=\frac{1}{\Delta\phi}\int_{\phi_\textnormal{i}}^{\phi_\textnormal{f}}\!\frac{-V^\prime}{V}\,\mathrm{d}\phi=\frac{\log\left(V_\textnormal{i}\right)\log\left(V_\textnormal{f}\right)}{\Delta\phi}\\
    &>\frac{\log\left(V_\textnormal{i}\right)-\log\left(\left(d-1\right)\left(d-2\right)/2\right)}{\Delta\phi}\nonumber\\
    &\qquad+\frac{2}{\sqrt{\left(d-1\right)\left(d-2\right)}},
\end{align}
which yields \cref{eq:tcc} in the infinite field limit.

A bound on the scalar field excursion using \cref{eq:tcc} is derived by \citet{van_de_heisteeg_bounds_on_field_2023}:
\begin{equation}
    V(\phi)\leq \mathfrak{c}\exp\left(\frac{-2\Delta\phi}{\sqrt{\left(d-1\right)\left(d-2\right)}}\right),
\end{equation}
which implies that a scalar field potential decays exponentially for sufficiently large scalar field displacements, i.e. a scalar field potential can only be flat for a finite scalar field range.

\paragraph{Axions}\label{p:tPCC_Axion}
with decay constant $f$ and potential $V\propto\cos(\phi/f)$ satisfy the \gls{tcc} if $f\sqrt{\mathfrak{N}}\sim0.6$, $\mathfrak{N}$ an arbitrary number of identical axion fields $\phi$, as numerical simulations by \citet{shlivko_trans-planckian_2023} show. This presents a dilemma for inflation, as observational constraints on the spectral tilt of the \gls{cmb} powerspectrum, $n_s=0.9649\pm0.0042\approx1-\frac{2}{f\sqrt{\mathfrak{N}}}$ \cite{planck_collaboration_Inflation_2020} require $f\sqrt{\mathfrak{N}}>1$.

Axions as quintessential \gls{de} with an instanton-generated potential $V(\phi)=m_\text{SSB}^2\cdot e^{-S_\iota \cos(\phi/f)}$ that is of the order of today's critical energy density, $m_\text{SSB}\gtrsim\order{\textnormal{TeV}}$ the mass scale of supersymmetry breaking and $S_\iota$ the instanton action, 
faces its own challenges, as \citet{shlivko_trans-planckian_2023} numerically shows:
\begin{align}
    m_\text{SSB}^2\cdot e^{-S_\iota }&\lesssim H_0^2\\
    \Rightarrow S_\iota &\gtrsim2\ln\left(\frac{m_\text{SSB}}{H_0}\right)\\
    f&\sim1 / S_\iota \sim{10^{-2}}\\
    \Rightarrow \mathfrak{N}&\sim10^4.
\end{align}
While $\mathfrak{N}\sim10^4$ is technically possible in string theory, \citet{shlivko_trans-planckian_2023} finds that it requires fine-tuning of the initial conditions for the fields to satisfy the \gls{tcc}.

\paragraph{Primordial Black Holes}
\citet{cai_mass_2020} work out mass constraints for \glspl{pbh} under the assumption that \glspl{pbh} are formed from large curvature perturbations at small scales after entering a matter- or radiation-dominated phase. If a \gls{pbh} forms during the radiation dominated era, the mass can be estimated from the horizon mass as
\begin{equation}
    m_\textnormal{PBH}=\varsigma\frac{4}{3}\pi H_\textnormal{c}^{-3}\left(3M_\textnormal{P}^2H_\textnormal{c}^2\right),
\end{equation}
with $\varsigma$ the collapse efficiency.
Then, assuming that the Hubble parameter during the formation of the \gls{pbh} $H_\textnormal{c}$ is smaller than the Hubble parameter at the end of inflation $H_\textnormal{f}$, a lower bound on the mass can be obtained:
\begin{align}
    m_\textnormal{PBH}&=\frac{4\pi\varsigma M_\textnormal{P}^2}{H_\textnormal{c}}\\
    &\geq\frac{4\pi\varsigma M_\textnormal{P}^2}{H_\textnormal{f}}\\
    &\geq4\pi\varsigma M_\textnormal{P}e^{N_e}\\
    m_\textnormal{PBH,r}&>\varsigma\left(\frac{H_\textnormal{f}}{\SI{e9}{\giga\electronvolt}}\right)^2M_\odot\\
    m_\textnormal{PBH,m}&>\varsigma\left(\frac{\SI{e9}{\giga\electronvolt}}{H_\textnormal{f}}\right)^4M_\odot.
\end{align}
The bound $m_\textnormal{PBH,r}$ assumes a radiation-dominated phase after inflation. The bound $m_\textnormal{PBH,m}$ assumes a short matter-dominated phase between inflation and reheating.
If the Hubble scale during inflation is $H>\SI{10}{\tera\electronvolt}$, then \glspl{pbh} cannot account for the entirety of \gls{dm} \cite{cai_mass_2020}.

\paragraph{Dark Energy}
The \gls{tcc} forbids any trans-Planckian mode to ever cross the horizon, which constrains the number of $e$-folds during the \gls{de} dominated epoch \cite{mizuno_universal_2020}.\footnote{
    Given the fact the the \gls{de}-dominated phase is not even ongoing for 1 $e$-fold so far, this is admittedly a rather theoretical constraint.
}
\citet{li_trans-planckian_2020} argue that the \gls{de} dominated phase should be shorter than naively assumed:
\begin{align}
    N_\textnormal{DE}&<\log\left(\frac{M_\textnormal{P}}{H_\textnormal{I,ini}}\right)\\
    \tau&<\frac{1}{H_0}\log\left(\frac{M_\textnormal{P}}{H_\textnormal{I,end}}\right),
\end{align}
i.e. they claim that today's Hubble parameter $H_0$ as well as the Hubble parameter at the beginning of inflation $H_\textnormal{I,ini}$ are of relevance, instead of just the Hubble parameter at the end of inflation $H_\textnormal{I,end}$.
They make three strong assumptions to derive this bound:
\begin{itemize}
    \item $H_\textnormal{DE}=H_0$ is constant during the \gls{de} era.
    \item The present horizon is $k_0=a_0H_0=a_\textnormal{I,ini}H_\textnormal{I,ini}$.
    \item The \gls{dsc} has to be applied to their chosen \gls{ads} potential.
\end{itemize}
We leave it to the reader to assess how well motivated those assumptions are, but we see especially the last assumption critically, as the \gls{dsc} should not be necessary for \gls{ads} potentials (see \cref{sec:deSitter} for further details).

\subparagraph{Quintessence}
models show a large difference between the Planck scale and today's Hubble scale, which means that the \gls{tcc} is easily satisfied \cite{heisenberg_model_2021}.\footnote{
    For scalar fields, we derive the bound $\frac{\abs{\phi_\textnormal{f}-\phi_\textnormal{i}}}{\sqrt{\left(d-1\right)\left(d-2\right)}}<-\log \frac{H_\textnormal{f}}{M_\textnormal{P}}$ in \cref{eq:TCC-scalar-field}.
    }
The lifetime bound of the \gls{tcc} implies that quintessence has to discharge until the time $\tau_\textnormal{TCC}$ \cite{bedroya_sitter_2020}, which requires that the quintessence potential has a minimal gradient of $\abs{V^\prime}/V\gtrsim\order{\log^{-2}V}$ in the bulk respectively $\abs{V^\prime}/V\gtrsim2/\left[\left(d-1\right)\left(d-2\right)\right]$ asymptotically \cite{bedroya_sitter_2020}.
If we want to extend a \gls{de} dominated phase into the infinite future, the constraint from \cref{eq:TCC-scalar-field} strengthens, and a tension with the \gls{dc} arises, as a large field displacement develops \cite{heisenberg_model_2021}. Since the \gls{tcc} already predicts a finite lifetime of the universe, we wouldn't give too much weight to this concern.

\citet{koga_catalytic_2021} study a model of bubble nucleation in 5 dimensions.
A bubble has an inside and an outside, which can have \gls{ds}, Minkowski, or \gls{ads} geometry.
In their study, \citet{koga_catalytic_2021} found that if both sides are \gls{ads}, positive cosmological constants yield eternal 4d \gls{ds} spaces on the bubble that violate the \gls{tcc}. They reason that the bubble tension, which they treated as a free parameter, is actually constrained, and taking into account this constraint on the bubble tension rules out solutions where a decay process leads to a bubble with a positive cosmological constant. %
Moreover, 4d \gls{ads} spaces cannot be created on the bubble, as there exists no solution that leads to such a decay, i.e. the cosmological constant cannot be negative. %
They conclude that the cosmological constant must vanish, 
and propose a thawing and a freezing quintessence field instead, one acting as inflation, the other as \gls{de}.

A multi-field \gls{de} model in curved field space is studied by \citet{payeur_swampland_2024}. They find that either the \gls{tcc} is violated, as the epoch of accelerated expansion is eternal, or the \gls{dc}, as the field eventually passes a Planckian field range. We present their model around \cref{eq:rapid-turn-DE}. They claim that a small patch in the parameter space is compatible with the \gls{tcc}, the \gls{dc}, the \gls{dsc}, as well as observational constraints.

\subparagraph{Generalised Proca vector fields} are appealing because they can successfully reduce the Hubble tension \cite{heisenberg_model_2021}. \citet{heisenberg_model_2021} study a model with three fixed points: a radiation dominated phase, a matter dominated phase, and a \gls{de} dominated phase. The \gls{de} dominated phase shows phantom-like behaviour, with an \gls{eos} approaching $\omega=-1$ from below. They show that this fixed point is inconsistent with the \gls{tcc}. An issue with Proca fields could be that they are ruled out by the \gls{tcc} if they approach a \gls{ds} solution asymptotically: since the \gls{tcc} restricts the lifetime of \gls{ds} space, a cosmological model should not asymptote a \gls{ds} attractor solution \cite{brahma_consistency_2021}.

\paragraph{Dark matter} produced by the amplification of quantum fluctuations during inflation could suffer from the low energy scale of inflation predicted by the \gls{tcc}, as a small Hubble scale during inflation keeps quantum fluctuations small \cite{tenkanen_trans-planckian_2020}.

\paragraph{Pre-Inflationary Phase}
\citet{mizuno_universal_2020} derive a bound for the epoch before inflation:
\begin{equation}
    \frac{H_\textnormal{I}}{M_\textnormal{P}}\lesssim\left(\frac{T_0}{T_\textnormal{rh}}\right)^{1+\frac{2}{1+3w}},
\end{equation}
with $w$ the \gls{eos} before inflation. For $w<1/3$, this bound becomes even more stringent than the \gls{tcc} bound.
A radiation-dominated phase before inflation leads to an energy scale of inflation of \SI{e4}{\giga\electronvolt} and a bound on the tensor-to-scalar ratio of $r_\textnormal{ts}<\num{e-47}$ \cite{brandenberger_strengthening-TTC_2020}.
A contracting phase before inflation relaxes the constraints \cite{brandenberger_strengthening-TTC_2020}.
A radiation-dominated epoch before brane inflation leads to inconsistencies: inflation would start at $t_\textnormal{i}=\SI{e23}{\second}$, and $r_\textnormal{ts}<\num{e-54}$ \cite{mohammadi_brane_2021}. If the Hubble parameter is allowed to vary as $H(N_e)=H_\textnormal{i}e^{\mathfrak{c}N_e}$, the pre-inflation phase lasts for \SI{e-8}{\second} and $r_\textnormal{ts}<\num{e-2}$ is allowed \cite{mohammadi_brane_2021}.

\paragraph{Inflation}\label{p:TCC_Inflation}
might be in the swampland, according to
\citet{bedroya_tcc_2024}, who motivate the \gls{tcc} from a bottom-up perspective.
However, while they show that a certain class of models is likely part of the swampland, inflation itself is not conceptually ruled out: if the cosmology at hand either will never transition into a decelerating phase or will have phases that are dominated by quantum effects, such as bubble nucleation, tunnelling events, or large quantum fluctuations, the model can evade their crushing conclusion.
The \gls{tcc} makes sure that no trans-Planckian modes that describe microscopic \gls{uv} modes are redshifted into macroscopic \gls{ir} observables described by an \gls{eft}.
However,
the relevance of such constraints on time-dependent \glspl{eft} has been questioned \cite{kaloper_quantum_2019,dvali_inflation_2020,burgess_cosmological_2021,komissarov_cosmology_2023,lacombe_multi-scalar_2023} \cite{cicoli_string_2023}. In the following, we assume their relevance to 
derive some quantitative bounds, assess constraints, and study different classes of models in more detail.

The inflaton potential has an upper bound, which we derive by following the evolution of a mode \cite{bedroya_trans-planckian-inflation_2020}:
\begin{itemize}
    \item Initially, we start with the Hubble radius $1/H_\textnormal{i}$.
    \item This scale grows by the factor of $\exp N_e$ during inflation.
    \item Between the end of inflation and the beginning of the radiation dominated epoch, i.e. during reheating, the scale grows by a factor of $a_\textnormal{r}/a_\textnormal{f}$.
    \item Since the end of reheating, the scale has grown by a factor of $\frac{T_\textnormal{r}g_*(T_\textnormal{r})^{1/3}}{T_\textnormal{0}g_*(T_\textnormal{0})^{1/3}}$, where we account for the change in the effective number of relativistic degrees of freedom.\footnote{
        Conservation of entropy informs us that the temperature $T$ is related to the effective number of relativistic degrees of freedom  $g_*(T)$ as $T\propto g_*^{-1/3}a^{-1}$ \cite{baumann_cosmology_2015}.
    }
    \item This approximates the evolution of this length scale until today, ergo, it should roughly correspond to today's Hubble scale.
\end{itemize}
This motivates the following relation:
\begin{equation}
    \frac{1}{H_\textnormal{i}}e^{N_e}\frac{a_\textnormal{r}}{a_\textnormal{f}}\frac{T_\textnormal{r}g_*(T_\textnormal{r})^{1/3}}{T_\textnormal{0}g_*(T_\textnormal{0})^{1/3}}\simeq\frac{1}{H_0}.
\end{equation}
\begin{itemize}
    \item For the inflaton potential, we know that $H=\sqrt{V}/\sqrt{3}M_\textnormal{P}$.
    \item If we assume that reheating takes place rapidly enough, $a_\textnormal{r}/a_\textnormal{f}\sim1$ holds.
    \item Assuming that reheating lasts less than a Hubble time, $T_\textnormal{r}\approx V^{1/4}$ holds \cite{bedroya_trans-planckian-inflation_2020}.\footnote{A more accurate description would not significantly change the result, as the temperature during reheating can be expressed in terms of $H$ as $T_\textnormal{r}=\left(45/4\pi^3g_*(T_\textnormal{r})\right)^{1/4}\sqrt{H_\textnormal{r}M_\textnormal{P}}$ \cite{brandenberger_stringy_2024}.
    }
    \item To get an order of magnitude estimate, we can also assume that $g_*(T_\textnormal{r})^{1/3}\sim g_*(T_\textnormal{0})^{1/3}$.\footnote{
        Since for the \gls{sm} $g_*(T_\textnormal{r})$ varies between 10.56 (with a reheating temperature of \SI{1}{\mega\electronvolt}) and 106.75 (with a reheating temperature of \SI{170}{\giga\electronvolt}) \cite{mizuno_universal_2020,husdal_effective_2016}, this assumption is justified.
        $T_\textnormal{r}\approx\SI{1}{\mega\electronvolt}$ is the lower bound for the reheating temperature from \gls{bbn} \cite{mizuno_universal_2020,hasegawa_mev-scale_2019,hannestad_what_2004,kawasaki_mev-scale_2000,kawasaki_cosmological_1999}.
        }
    \item The Friedmann equation informs us that the Hubble scale is given by the current energy density: $1/H_0=\sqrt{3}M_\textnormal{P}/\sqrt{\rho_0}$.
    \item The energy density today corresponds roughly to the following expression: $\rho_0\approx T_0^4\frac{T_\textnormal{eq}}{T_0}\frac{1}{\Omega_\textnormal{m}}$, with $T_\textnormal{eq}$ the temperature during radiation\textendash matter equality, and $\Omega_\textnormal{m}$ today's relative matter density.
\end{itemize}
This leaves us with:
\begin{align}
    \frac{\sqrt{3}M_\textnormal{P}}{\sqrt{V}}e^{N_e}\frac{V^{1/4}}{T_\textnormal{0}}&\simeq\frac{\sqrt{3}M_\textnormal{P}}{\sqrt{T_0^4\frac{T_\textnormal{eq}}{T_0}\frac{1}{\Omega_\textnormal{m}}}}\\
    \Rightarrow e^{N_e}&\simeq\frac{V^{1/4}}{\sqrt{T_0T_\textnormal{eq}}}\sqrt{\Omega_\textnormal{m}}.
\end{align}
Then, we use $H\simeq\sqrt{V/3M_\textnormal{P}^2}$ again, apply the \gls{tcc} bound $e^{N_e}<M_\textnormal{P}/H$, and solve for $V$ to find
\begin{align}
    V^{3/4}&<\sqrt{3}M_\textnormal{P}^2\sqrt{T_0T_\textnormal{eq}}\\
    V^{1/4}&<\SI{e9}{\giga\electronvolt}\sim\num{e-10}M_\textnormal{P}.\label{eq:TCC_energy}
\end{align}
If we express this bound again in terms of $H$ to find $H\lesssim\order{\num{e-20}}M_\textnormal{P}$, and assume a scale invariant scalar powerspectrum $P_\zeta=H^2/8\pi^2M_\textnormal{P}^2\epsilon_Vc_\textnormal{s}\sim\num{e-9}$ we find the bound on the tensor-to-scalar ratio $r_\textnormal{ts}=16\epsilon_Vc_\textnormal{s}\lesssim\num{e-31}$ \cite{shi_large_2021,brahma_trans-planckian_2020-Inflation,bedroya_trans-planckian-inflation_2020,kamali_relaxing_2020,dhuria_trans-planckian_2019,brandenberger_trans-planckian_2021,brandenberger_stringy_2024,van_de_heisteeg_bounds_on_field_2023,santos_warm_2022,kamali_recent_2023}.

Allowing the reheating process to last for $N_\textnormal{rh}$ $e$-foldings leads to a weaker bound \cite{shi_large_2021,mizuno_universal_2020}:
\begin{align}
    e^{N_\textnormal{rh}}&=\frac{a_\textnormal{rh}}{a_\textnormal{f}}\\
    &=\left(\frac{\rho_\textnormal{f}}{\rho_\textnormal{rh}}\right)^\frac{1}{3\left(1+w\right)}\\
    &=\left(\frac{\sqrt{3}M_\textnormal{P}H_\textnormal{f}}{T_\textnormal{rh}^2}\right)^\frac{2}{3\left(1+w\right)}\\
    \Rightarrow \frac{H}{M_\textnormal{P}}&<\frac{M_\textnormal{P}H_0}{T_\textnormal{rh}T_0}\left(\frac{M_\textnormal{P}H}{T_\textnormal{rh}}\right)^{\frac{2}{3\left(1+w\right)}-1}\\
    r_\textnormal{ts}&<\num{e-8}\left(\frac{\SI{1}{\mega\electronvolt}}{T_\textnormal{rh}}\right)^2,
\end{align}
where \gls{bbn} puts a constraint of $T_\textnormal{rh}>\SI{1}{\mega\electronvolt}$ \cite{hasegawa_mev-scale_2019,hannestad_what_2004,kawasaki_mev-scale_2000,kawasaki_cosmological_1999}. %

These constraints can be mitigated, e.g. by 
\begin{itemize}
    \item assuming $H$ to be dynamic during inflation, which leads to the relaxed constraints $V^{1/4}<\num{e-5}M_\textnormal{P}$ and $r_\textnormal{ts}<\num{e-10}$ \cite{kamali_relaxing_2020};
    \item a non-standard cosmology after the end of inflation \cite{kamali_relaxing_2020,dhuria_trans-planckian_2019,torabian_non-standard_2020}, for example, the phase between the end of inflation and \gls{bbn} could be a phase with an \gls{eos} of $w=-1/3$, which yields \cite{dhuria_trans-planckian_2019}:
    \begin{itemize}
        \item $r_\textnormal{ts}\lesssim\num{e-8}$
        \item $H_\textnormal{I}\approx\SI{e10}{\giga\electronvolt}$
        \item $V^{1/4}\approx\SI{e14}{\giga\electronvolt}$;
    \end{itemize}
    \item non-standard initial conditions for fluctuations \cite{kamali_relaxing_2020}: Tensor modes could also be produced outside of inflation by non-\gls{bd} terms, e.g. by modified gravity, thermal fluctuations, or excited initial states \cite{brahma_trans-planckian_2020-Inflation,agullo_loop_2015}.
    Non-\gls{bd} states for tensor modes can increase the upper bound for $r_\textnormal{ts}$ by about a factor of \num{e6}\,\textemdash\,further remedy comes from allowing scalar modes to also be sourced from non-\gls{bd} states \cite{brahma_trans-planckian_2020-Inflation}.
    Even though the enhancement from non-\gls{bd} states on scalar modes allows $\epsilon_H<\num{e-9}$, it cancels in a calculation of $r_\textnormal{ts}$\,\textemdash\,but it allows the tensor modes to be enhanced even further, which then eventually allows for $r_\textnormal{ts}<\num{e-3}$.
\end{itemize}

\subparagraph{Single-field inflation} in a \gls{gr} setting seems to be only consistent with the \gls{tcc} if fine-tuning takes place \cite{trivedi_rejuvenating_2022,bedroya_trans-planckian-inflation_2020,jin_axion_2021}.
Also, non-\gls{bd} scalar or tensor modes lead to viable solutions that satisfy the \gls{tcc} as well as observational constraints \cite{naskar_generic_2022}:
The powerspectra for scalar and tensor modes are given by
\begin{align}
    P_\textnormal{s}(k)&=\frac{H^2}{8\epsilon M_\textnormal{P}^2}\abs{u_k^\textnormal{s}+w_k^\textnormal{s}}^2\\
    P_\textnormal{t}(k)&=\frac{2H^2}{M_\textnormal{P}^2}\abs{u_k^\textnormal{t}+w_k^\textnormal{t}}^2\\
    r_\textnormal{ts}&=\frac{P_\textnormal{t}(k)}{P_\textnormal{s}(k)}=16\epsilon\frac{\abs{u_k^\textnormal{t}+w_k^\textnormal{t}}^2}{\abs{u_k^\textnormal{s}+w_k^\textnormal{s}}^2}
\end{align}
with $\epsilon=\epsilon_V=\epsilon_H$ (since it is the single-field case) the slow-roll parameter,
and $u$ and $w$ the Bogolyubov coefficients that characterise the non-\gls{bd} states, such that $u_k=1$ and $w_k=0$ corresponds to the \gls{bd} states.\footnote{
    The Bogolyubov coefficients appear in the definition of the bispectra / 3-point correlation functions.
    Studying three-point correlation functions has several benefits \cite{naskar_generic_2022}:
    (i) they are sensitive to the initial state;
    (ii) autocorrelation of tensor modes respectively non-Gausssianities show in the spectrum of primordial \glspl{gw};
    (iii) tensor\textendash scalar\textendash scalar correlators probe the diffeomorphism breaking during inflation \cite{bartolo_distinctive_2016};
    (iv) mixed correlators show as the quadrupole moment in the \gls{cmb} spectrum, and might be observable in future \gls{cmb} missions \cite{abazajian_cmb-s4_2016}.
}
The Bogolyubov coefficients satisfy the Wronskian condition $u_k^2-w_k^2=1$.
The \gls{tcc} puts the following constraint on the backreactions \cite{greene_decoupling_2005,holman_enhanced_2008,naskar_generic_2022}:
\begin{equation}
    w_0\leq\sqrt{\epsilon\eta}\frac{HM_\textnormal{P}}{\Lambda},
\end{equation}
with $\Lambda>H$ a cutoff scale.
It is now no longer $\epsilon<\num{e-31}$ that has to hold, but $\epsilon/\abs{u_k^\textnormal{s}+w_k^\textnormal{s}}^2<\num{e-31}$. This relaxes the bound on $r_\textnormal{ts}$.

Among others, the \gls{tcc} was explicitly used to by \citet{okada_inflection-point_2021} to constrain an inflection point scenario by demanding $H_\textnormal{I}<\SI{1}{\giga\electronvolt}$,
as well as by \citet{mohammadi_brane_2022}, who derived $\rho<3M_\textnormal{P;4}M_\textnormal{P;5}^3/4\pi e^{N_e}\approx\SI{e35}{\giga\electronvolt^4}$ as an upper bound for the energy density at the end of brane inflation,
where they explicitly set the 5-dimensional Planck mass to $M_\textnormal{P;5}\sim\SI{2e14}{\giga\electronvolt}$, which rules out the typical \gls{gut} scale ($\left(\SI{e16}{\giga\electronvolt}\right)^4\sim\SI{e64}{\giga\electronvolt^4}$) models.

\subparagraph{Slow-roll large-field inflation} is in tension with the \gls{tcc}.
While \citet{sanna_trans-planckian_2021} found constant roll inflation and some models in $f(R)$ gravity that suppress the amplitude of primordial \glspl{gw} to be compatible with the \gls{tcc},
\citet{ossoulian_inflation_2023} found that even models that are compatible with the \gls{dc} and the \gls{dsc} can violate the \gls{tcc} constraints, such as a power-law and an exponential model in $f(R,T)$ gravity with a non-canonical scalar field.
Let us highlight why large field excursions and slow-rolling are problematic.
In terms of field excursion, we can make the following observation \cite{bedroya_trans-planckian-inflation_2020}:
Let us first write the field range as
\begin{equation}
    \abs{\Delta\phi}=\abs{\dot{\phi}\Delta t}.
\end{equation}
Then, we use the slow-roll condition $3H\dot{\phi}=-V^\prime$ and the \gls{tcc} condition $\Delta t\leq H^{-1}\log H^{-1}$ to find
\begin{equation}
    \abs{\Delta\phi}\leq\abs{\frac{V^\prime}{3H^2}\log H^{-1}}.
\end{equation}
With the slow-roll parameter $\epsilon_V=\left(V^\prime/V\right)^2/2$ and the \gls{tcc} energy bound on the potential $V$ we find then $\abs{\Delta\phi}\leq\num{e-13} M_\textnormal{P}$, which conflicts with large-field inflation.

That slow-rolling is problematic can also be seen by focussing on the amplitude of the primordial powerspectrum respectively the tensor-to-scalar ratio \cite{sanna_trans-planckian_2021}:
We start with a vanilla slow-roll model of inflation:
\begin{align}
    \frac{3H^2M_\textnormal{P}^2}{8\pi}&=\frac{\dot{\phi}^2}{2}+V(\phi)\\
    \ddot{\phi}+3H\dot{\phi}&=-\frac{\mathrm{d}V}{\mathrm{d}\phi}\\
    w&=\frac{\dot{\phi}^2-2V}{\dot{\phi}^2+2V}\\
    \epsilon_H&=-\frac{\dot{H}}{H}\\
    \eta_H&=\frac{\dot{\epsilon}_H}{H\epsilon_H}\\
    P&=\frac{1}{8\pi^2\epsilon_H}\frac{H^2}{M_\textnormal{P}^2}&\sim&\num{e-9}\label{eq:TCC_powerspectrum}\\
    n_\textnormal{s}&=1-2\epsilon_H-\eta_H&=&\num{0.9649(0.0042)}\\
    r_\textnormal{ts}&=16\epsilon_H&<&0.06.
\end{align}
The slow-roll approximation $\epsilon_H,\abs{\eta_H}\ll1$ yield then
\begin{align}
    \epsilon_H&\approx\frac{3\left(1+w(N_e)\right)}{2}\\
    \eta_H&\approx-\frac{\mathrm{d}\log\left(1+w(N_e)\right)}{\mathrm{d}N_e}\\
    1-n_\textnormal{s}&=3\left(1+w(N_e)\right)-\frac{\mathrm{d}\log\left(1+w(N_e)\right)}{\mathrm{d}N_e}\\
    r_\textnormal{ts}&=24\left(1+w(N_e)\right).
\end{align}
Solving \cref{eq:TCC_powerspectrum} for $\epsilon_H$, and using the \gls{tcc} bound on $H$ yields
\begin{equation}
    \epsilon_H<\frac{e^{-2N_e}}{8\pi^2P}.
\end{equation}
Applying this bound to the slow-roll approximation for $\epsilon_H$ yields $1+w<\num{e-45}$ which in turn yields $r_\textnormal{ts}<\num{e-44}$, basically eradicating any gravitational B-modes.

\subparagraph{Small-Field Models} 
\citet{kadota_trans-planckian_2020} study small-field inflation and find $N_e^\textnormal{min}\sim2N_e^\text{CMB}$: the minimally required number of $e$-folds (to be compatible with the \gls{tcc}) is typically twice or more than what is required to explain \gls{cmb} fluctuations.

The following models are found to violate the \gls{tcc}:
\begin{itemize}
    \item The Starobinsky model\footnote{See around \cref{eq:E-model}.} \cite{sanna_trans-planckian_2021}
    \item Small-field inflation in a brane-world\footnote{
    See \cref{sec:SmallFieldInflation} for further details. $r_\textnormal{ts}\sim\left(\num{e-2}\sim\num{e-1}\right)$ is too large to be compatible with the \gls{tcc} \cite{osses_reheating_2021}.
    } \cite{osses_reheating_2021}
    \item Supersymmetric hybrid inflation\footnote{
    Models that are a hybrid between chaotic inflation with dominant quantum fluctuations, and spontaneous phase transitions, such that inflation ends due to a fast-roll of a second field that is initiated by the first field \cite{ahmed_supersymmetric_2024,linde_hybrid_1994,copeland_false_1994}.
    Unless a non-minimal coupling to the Kähler potential is introduced, the predicted $r_\textnormal{ts}$ is too large to be compatible with the \gls{tcc} \cite{ahmed_supersymmetric_2024}.
    } \cite{ahmed_supersymmetric_2024}
\end{itemize}

\subparagraph{Eternal inflation} is on the brink of incompatibility with the \gls{tcc} \cite{bedroya_sitter_2020,bedroya_sitter_bubbles_2020,blumenhagen_sitter_2021}:
on the one hand, any local \gls{ds} minimum has a limited lifetime \cite{cai_refined_2021,seo_entropic_2020,dvali_quantum_2019,rudelius_conditions_2019,dvali_exclusion_2019},
and on the other hand, if inflation lasted forever, a Hubble patch might evolve where trans-Planckian modes classicalise \cite{bedroya_sitter_bubbles_2020}. Furthermore, applied to individual tunnelling events, the \gls{tcc} implies that $\abs{V^\prime}>V^{3/2}$, an inequality that must be violated in eternal inflation \cite{agmon_lectures_2023}.

\subparagraph{Hilltop inflation} generally has a small slow-roll parameter $\epsilon_V$, as required by the \gls{tcc} \cite{brahma_swampland_2020}.

\subparagraph{Warm inflation} is challenged by the \gls{tcc} \cite{brandenberger_strengthening_2020,kamali_warm_2020,berera_trans-planckian_2019,aalsma_chaos_2020,berera_role_2020}.
Warm inflation can satisfy the \gls{tcc} if dissipation is strong \cite{kamali_warm_2020,das_runaway_2020,santos_warm_2022,kamali_recent_2023,kamali_intermediate_2021,arya_primordial_2024,kamali_minimal_2021,das_distance_2020}
or if it takes place in a multiphase model of inflation, where inflation is interrupted by a radiation dominated phase \cite{berera_trans-planckian_2019,berera_thermal_2021}.

\subparagraph{Higgs-inflation} with a plateau shaped potential, extended by an $R^2$ term in the context of Palatini gravity, is found to be incompatible with the \gls{tcc} and observational \gls{cmb} data \cite{tenkanen_initial_2020}. Another (asymptotically) plateau model in Palatini gravity, with a more general $F(R)$-term, is found to be compatible with the \gls{tcc} and \gls{cmb} constraints \cite{tenkanen_trans-planckian_2020}.

\subparagraph{Chromonatural inflation\footnote{See \cref{p:chromonatural_inflation} for further details.},} where an axion field is coupled to a non-Abelian gauge field via a Chern\textendash Simons coupling, such that the coupling effectively acts as friction, obtains an upper bound for the energy scale of inflation by the \gls{tcc}, but is not ruled out \cite{berera_thermal_2021}.

\subparagraph{D-term hybrid inflation} \cite{halyo_hybrid_1996,binetruy_d-term_1996,domcke_inflation_2018,domcke_unified_2017,schmitz_axion_2018} is compatible with the \gls{tcc} \cite{schmitz_trans-planckian_2020}:
it has a scalar power spectrum with an amplitude of the order of the \gls{gut} scale $\sim\SI{e16}{\giga\electronvolt}$, and sets an upper bound on the gravitino mass scale of $m_\textnormal{3/2}\lesssim\SI{10}{\mega\electronvolt}$ that would allow \gls{dm} to consist of thermally produced gravitinos.
The model is particularly interesting, as it can reduce the Hubble tension: unstable gravitinos (\gls{dm}) decay into an axion quintessence field (radiation) and an axino (warm \gls{dm}). The Hubble parameter shows a slower decrease in this model than in \gls{lcdm} and shows a higher value of the Hubble constant today, which reduces the Hubble tension.\footnote{
    Hybrid inflation ends with a rapid second-order phase transition, which directly points to a potential tension with the refined \gls{dsc} (\cref{eq:dScrefined}), which was indeed found \cite{schmitz_trans-planckian_2020}: $\epsilon\ll\eta\simeq-0.02$. The \gls{dc}, however, is satisfied, as well as the \gls{wgc} \cite{schmitz_trans-planckian_2020}.
}

\subparagraph{$k$-flation} with a more general Lagrangian and a time-dependent \gls{eos} differs from inflation described by a canonical scalar field: modes do not freeze out at the Hubble radius, but at the freeze-out radius
\begin{equation}
    r_\textnormal{f}=\sqrt{\frac{\mathcal{F}}{\mathcal{F}^{\prime\prime}}},
\end{equation}
with
\begin{equation}
    \mathcal{F}=\frac{a\sqrt{2\epsilon_H}}{\sqrt{c_\textnormal{s}}},
\end{equation}
with $a$ the scale factor, $\epsilon_H=-\dot{H}/H^2$ the Hubble slow-roll parameter (see \cref{p:dSC_Inflation}), and $c_\textnormal{s}$ the speed of sound \cite{lin_trans-planckian_2020,khoury_rapidly-varying_2009,geshnizjani_general_2011}.
For canonical single-field inflation ($c_\textnormal{s}=1$, $\epsilon_V\ll1$), the freeze-out radius corresponds to the comoving Hubble length $\left(aH\right)^{-1}$,\footnote{$\mathcal{F}^{\prime\prime}/\mathcal{F}\sim a^{\prime\prime}/a\sim2a^2H^2$} while for $k$-flation, it corresponds to the acoustic length $r_\textnormal{f}\sim c_\textnormal{s}/aH$ \cite{lin_trans-planckian_2020}. This finding motivated \citet{lin_trans-planckian_2020} to formulate a \gls{tcc} that depends on the speed of sound:
\begin{equation}
    \log\frac{c_\textnormal{s}(a)M_\textnormal{P}}{H(a)}>\log\frac{a}{a_\textnormal{i}},
\end{equation}
with $a_\textnormal{i}$ the initial scale factor at the beginning of inflation. This bound has to hold to avoid the classicalisation of trans-Planckian modes. This brings more free parameters, which means that additional assumptions have to be made to constrain models, respectively, more observational constraints have to be taken into account to assess the validity of a model.

\subparagraph{Multi-phase inflation}
happening in several stages, each with a relatively low energy scale, can help to accommodate the \gls{tcc} restrictions \cite{li_trans-planckian_2020,torabian_breathing_2020}. Inflation at high energy scales cannot last long enough (without spoiling the \gls{tcc}) to explain the horizon problem, but a succession of several phases of inflations can \cite{torabian_breathing_2020}. To solve the horizon problem,
\begin{align}
    \frac{a_\textnormal{I}}{a_0}\frac{H_\textnormal{I}}{H_0}&=\frac{a_\textnormal{I}}{a_\textnormal{f,1}}\frac{a_\textnormal{f,1}}{a_\textnormal{i,2}}\cdots\frac{a_{\textnormal{i,}m-1}}{a_{\textnormal{f,}m}}\frac{a_{\textnormal{f,}m}}{a_\textnormal{r}}\frac{a_\textnormal{r}}{a_0}\frac{H_\textnormal{I}}{H_0}\\
    &=\exp\left({-\sum_{i=1}^mN_i}\right)\frac{T_0}{T_\textnormal{r}}\frac{H_I}{H_0}\frac{a_{\textnormal{f,}m}}{a_\textnormal{r}}\\
    &=\exp\left({-\sum_{i=1}^mN_i}\right)\frac{T_0}{T_\textnormal{r}}\frac{H_I}{H_0}\left(\frac{H_\textnormal{r}}{H_m}\right)^{\frac{2}{3\left(1+w_m\right)}}\nonumber\\
    &\qquad\cdot\prod_{i=1}^{m-1}\left(\frac{H_{i+1}}{H_i}\right)^{\frac{2}{3\left(1+w_i\right)}}\\
    &\leq1
\end{align}
has to hold, with $H_\textnormal{I}$ the initial Hubble parameter of the first phase of inflation \cite{torabian_breathing_2020}.
Using
\begin{equation}
    \frac{a_i}{a_{i+1}}=\left(\frac{H_{i+1}}{H_i}\right)^{2/3\left(1+w\right)},
\end{equation}
the \gls{tcc} can be expressed as
\begin{equation}
    H_\textnormal{I}^{2-\frac{2}{3\left(1+w\right)}}<M_\textnormal{P}H_\Lambda T_0^{-1}T_\textnormal{r}H_1^{\frac{-2}{3\left(1+w\right)}}\frac{a_2}{a_1}\cdots\frac{a_n}{a_{n-1}}\frac{a_\textnormal{r}}{a_n},
\end{equation}
which allows us to examine different scenarios \cite{torabian_non-standard_2020}:
\begin{description}
    \item[$H_\textnormal{I}<\SI{e14}{\giga\electronvolt}$] If the phase after inflation and before \gls{bbn} is dominated by kinetic terms with $w=1$ like $k$-essence, then $H_\textnormal{I}^{5/2}<M_\textnormal{P}H_\Lambda T_0^{-1}T_\textnormal{r}H_\textnormal{r}^{-1/3}$ holds, which yields $H_\textnormal{I}\lesssim\left(\SI{10}{\mega\electronvolt}/T_\textnormal{r}\right)^{1/5}\,\textnormal{MeV}$.
    \item[$H_\textnormal{I}<\SI{0.1}{\giga\electronvolt}$] If the phase after inflation and before matter\textendash radiation equality is radiation dominated, then $H_\textnormal{I}^{3/2}<M_\textnormal{P}H_\Lambda T_0^{-1}T_\textnormal{r}H_\textnormal{r}^{-1/2}$ holds.
    \item[$H_\textnormal{I}<\SI{100}{\giga\electronvolt}$] If the phase after inflation and before \gls{bbn} is dominated by matter with $w=0$, then $H_\textnormal{I}^{4/3}<M_\textnormal{P}H_\Lambda T_0^{-1}T_\textnormal{r}H_\textnormal{r}^{-2/3}$ holds, which yields $H_\textnormal{I}\lesssim100\left(\SI{10}{\mega\electronvolt}/T_\textnormal{r}\right)^{1/4}\,\textnormal{GeV}$.
    \item[$H_\textnormal{I}<\SI{e15}{\giga\electronvolt}$] If the phase after inflation and before \gls{bbn} is dominated by exotic matter with $w=-1/3$ like cosmic strings, then $H_\textnormal{I}<M_\textnormal{P}^2H_\Lambda T_0^{-1}T_\textnormal{r}^{-1}$ holds, which yields $H_\textnormal{I}\lesssim\SI{e14}{\giga\electronvolt}\left(\SI{10}{\mega\electronvolt}/T_\textnormal{r}\right)$.    
    \item[$H_\textnormal{I}<\SI{100}{\giga\electronvolt}$] If the phase after inflation and before \gls{bbn} is dominated by moduli\footnote{See \cref{p:TCC_moduli} why such a phase is incompatible with the \gls{tcc}.} with the modulus decay rate $H\sim\Gamma_\phi\sim m_\phi^3/M_\textnormal{P}^2$ and a lower bound on the modulus mass of \SI{10}{\tera\electronvolt}, then $H_\textnormal{I}^{3/2}<M_\textnormal{P}H_\Lambda T_0^{-1}T_\textnormal{r}m_\phi^{-1/2}\left(m_\phi/\Gamma_\phi\right)^{2/3}$ holds, which yields $H_\textnormal{I}^{3/2}<M_\textnormal{P}^{11/6}H_\Lambda T_0^{-1}m_\phi^{-1/3}$.
\end{description}
\gls{tcc}-compatible multi-stage models include
warm multiphase inflation, where inflation is interrupted by a radiation-dominated phase \cite{berera_trans-planckian_2019,berera_thermal_2021},
and chain inflation, where inflation undergoes a rapid succession of tunnelling events from the minimum of one vacuum to the next \cite{winkler_power_2021}. In chain inflation, bubbles of the new vacuum percolate in each step, releasing radiation \cite{winkler_power_2021}.\footnote{
    If an axion potential is used, the \gls{cmb} powerspectrum puts an upper limit on the axion decay constant of chain inflation, which agrees with the notion of several swampland conjectures that do the same, i.e. the \gls{dsc}, the \gls{ssc}, the \gls{tcc}, the \gls{wgc}, and possibly the \gls{flb}.%
}

\subparagraph{String gas cosmology} \cite{battefeld_string_2006} produces a scale-invariant powerspectrum \cite{nayeri_producing_2006} and a nearly scale-invariant spectrum of \glspl{gw}, with a slight blue tilt \cite{brandenberger_tensor_2007}. While the horizon problem is absent in string gas cosmology, the flatness problem needs to be addressed \cite{kamali_creating_2020}. \citet{kamali_creating_2020} combine string gas cosmology with power-law inflation to do so, and find that it is compatible with the \gls{tcc}. %

\subparagraph{Tachyacoustic cosmology} solves the horizon problem by allowing a superluminal speed of sound \cite{lin_consistency_2019,magueijo_speedy_2008,bessada_tachyacoustic_2009,babichev_k-essence_2007,lin_trans-planckian_2021}.\footnote{
    See the work by \citet{ellis_causality_2007} for arguments against superluminal speeds of sound.
    }
Modes freeze out when they cross the acoustic horizon $c_\textnormal{s}k=aH$, and solve the horizon problem if
\begin{equation}
    \frac{c_\textnormal{s}(a_\textnormal{i})}{a_\textnormal{i}H_\textnormal{i}}\geq\frac{1}{a_0H_0},
\end{equation}
which yields a period of trans-Planckian energy densities and violates the \gls{tcc} \cite{lin_trans-planckian_2021}.

\paragraph{Moduli-Dominated Phase}\label{p:TCC_moduli}
Observational data allows for a phase between the end of inflation and \gls{bbn} that is not radiation-dominated, i.e. the \gls{eos} for that phase could differ from $w=1/3$, such as in the case of a moduli-dominated phase \cite{dhuria_trans-planckian_2019}:
The moduli are of mass $m_\textnormal{mod}\ll H_\textnormal{I}$, which start to oscillate when $H\approx m_\textnormal{mod}$, and decay into radiation when the Hubble parameter corresponds to their decay width $\Gamma_\textnormal{mod}\approx H$. Such a moduli-dominated epoch is incompatible with the \gls{tcc}: The moduli need to be lighter than $H_\textnormal{I}$ to get displaced from their minima and to start oscillating, and the reheating temperature must be above the \gls{bbn} temperature. The upper bound on the energy scale of inflation from the \gls{tcc} does not allow for such a window.

\subsubsection{General Remarks}
The \gls{tcc} should be viewed as a necessary but not a sufficient condition to forbid classicalisation of quantum modes:\footnote{
    See \cref{p:TCC_Quantum-Effects} for an explanation that we do not mean that all quantum effects are forbidden by the \gls{tcc}.
    }
in principle, non-linear modes could develop below the Hubble scale and become classical \cite{brandenberger_trans-planckian_2021,berera_role_2020}.
Large quantum fluctuations are always a possibility that would influence the background trajectory stochastically \cite{saito_is_2020}.
\citet{burgess_cosmological_2021} argue that it is not even a necessary condition, as trans-Planckian modes would not necessarily spoil an \gls{eft} description: if the vacuum evolves continuously and adiabatically into the ground state of the \gls{eft}, the description remains valid.
\citet{berera_role_2020} add that crossing the horizon does not necessarily lead to classicalisation (or decoherence) either: the wave needs to interact with the environment to decohere. %
\citet{cai_pre-inflation_2020} argue that sub-Planckian modes could have interacted with trans-Planckian modes before inflation, which would alter their state (to a non-\gls{bd} state), which would result in the observability of trans-Planckian quantum effects on classical scales. They argue that this can only be avoided if the pre-inflationary phase is \enquote{past-complete} and Minkowski, i.e. there are no singularities, not even in the infinite past. This is for example the case in certain bounce cosmologies, as they argue.
However, we think that this would not lead to the observability of trans-Planckian physics: if the modes start in a non-\gls{bd} state early on, but decohere during inflation, we will not be able to fully trace back the evolution of the mode to a time before the interaction with the trans-Planckian mode. The \gls{tcc} is then sufficient to shield us from these trans-Planckian effects, and our \gls{eft}-description remains unaffected by the earlier trans-Planckian physics. Decoherence is likely to complete already during inflation \cite{burgess_minimal_2023,lombardo_decoherence_2005,barvinsky_decoherence_1999,martineau_decoherence_2007}. However, \citet{burgess_decoherence_2008} argue that modes much shorter than the Hubble scale do not decohere longer modes (that are still shorter than the Hubble scale) as easily, as those, being pure, can be approximated as \gls{bd} states (as \gls{bd} states are attractor solutions of the squeezing equations \cite{albrecht_inflation_1994}).
\citet{brahma_trans-planckian_2020-Inflation} argues that assuming a \gls{bd} state at the onset of inflation is not justified: \gls{bd} states might be the initial states of each mode, but they might lay in the infinite past. An \gls{eft} description has a cutoff scale and the \gls{eft} description might break down before a classical mode is blue-shifted to a sub-Planckian length (when going back in time), as the mode might reach an energy scale that lies above the \gls{eft} cutoff, before it reaches the Planck scale \cite{brahma_sitter_2021}.
This cutoff can even be described by another swampland conjecture. This reminds us that the swampland programme is a web of interconnected conjectures. 
The \gls{tcc} gives conceptual guidance and helps to strengthen other conjectures such as the \gls{dsc} by specifying the $\order{1}$ constant.\footnote{
    If non-perturbative correction are present in the potential, the \gls{tcc} could obtain log-corrections, as for example the \gls{adsdc} does. This could lift the bound to $\abs{\nabla V}\geq V\left(\mathfrak{c}_1\log(\abs{V})+\mathfrak{c}_2\right)$, with $\mathfrak{c}_{1,2}$ constants, which would allow for stable \gls{ds} vacua, e.g. for $V=e^{-\mathfrak{c}_2/\mathfrak{c}_1}$ \cite{blumenhagen_quantum_2020}.
}
However, the \gls{tcc} is also more forgiving than other conjectures \cite{deffayet_stable_2024}: For example, it allows the \gls{dsc} to be violated for short periods or individual points. To violate the \gls{tcc}, the strong \gls{dsc} must be violated for some time, namely
\begin{equation}\label{eq:tPCC-dSC-violation}
    \frac{\abs{\nabla V}}{V}\leq\frac{2}{\sqrt{\left(d-1\right)\left(d-2\right)}}
\end{equation}
has to hold over a field range
\begin{equation}
    \Delta\phi\geq\frac{\sqrt{\left(d-1\right)\left(d-2\right)}}{2}\log\left(\frac{\left(d-1\right)\left(d-2\right)}{2V_\textnormal{max}}\right),
\end{equation}
with $V_\textnormal{max}$ the maximum value of the potential within the field range.

\paragraph{What motivates the \gls{tcc}?}

On purely mathematical grounds, Grönwall's inequality\footnote{
    Or \textit{Gronwall's} inequality, if you follow his spelling, as he dropped the umlauts after immigrating to the USA.
    }
implies that $a_\textnormal{f}/a_\textnormal{i}<M_\textnormal{P}/H_\textnormal{f}$ holds if $\int_{t_\textnormal{i}}^{t_\textnormal{f}}\!H(s)\,\mathrm{d}s<\log\left(M_\textnormal{P}/H_\textnormal{f}\right)$ holds, assuming $\dot{a}(t)/a(t)\leq H(t)$ \cite{cotsakis_trans-planckian_2023}.

An \gls{eft} describing our Universe should have a valid vacuum solution. This is at least indicated by holography and perturbative string theory. \citet{bedroya_tcc_2024} show that \glspl{eft} that violate the \gls{tcc} do not have asymptotic vacuum solutions, respectively they show that a certain class of cosmologies do not have an \gls{eft} description of the vacuum in the asymptotic future boundary spacetime, namely cosmologies that
(i) violate the \gls{tcc},
(ii) have a transition to a deceleration phase,
(iii) are flat or negatively curved,
(iv) have a classical background evolution, i.e. are not dominated by quantum effects such as bubble nucleation or large quantum fluctuations. 
The strong implication of their work is that the \gls{tcc} holds not only on the infinite boundary of moduli space, but even deep inside the bulk, i.e. cosmology has to satisfy the \gls{tcc}.

The \gls{tcc} assures independence of our \gls{eft} description from \gls{qg} effects \cite{schneider_trans-planckian_2021}.
We do not have to speculate about initial conditions on trans-Planckian scales to understand \gls{lss} formation \cite{wolf_explanatory_2023}.
If trans-Planckian physics were to affect \gls{lss} growth, low-energy physics would provide us with a possibility to examine high-energy physics in a way that is different from \gls{ir}/\gls{uv} mixing as we usually understand it. The \gls{tcc} not only secludes trans-Planckian physics from our examination, it even renders it impossible for us to figure out if there is trans-Planckian physics in the early Universe at all \cite{schneider_trans-planckian_2021}.

If the \gls{tcc} were violated, 
information got lost in the asymptotic limit of field space, e.g. in the infinite future \cite{bedroya_holographic_2022},
and all dynamical quantum fluctuations $H<k<l_\textnormal{P}^{-1}$ became classical \cite{bedroya_sitter_2020}.
Even though the observation of \gls{qg} phenomena in the \gls{cmb} or in \glspl{lss} would be a \enquote{unique opportunity} to learn more about \gls{qg} \cite{komissarov_cosmology_2023}, the \enquote{classical observation of a sub-Planckian quantum mode, [\dots] is a bit strange!} \cite{bedroya_trans-planckian_2020}.

Besides these mathematical, quantum-gravitational, and string-theoretical considerations, there is also observational support that the \gls{tcc} holds:
The \gls{tcc} bound (\cref{eq:TCC_sub-Planckian-Hubble}) constrains the number of $e$-foldings that are allowed to happen during inflation:
\begin{equation}
    N_e<\log\left(\frac{M_\textnormal{P}}{H_\textnormal{f}}\right).
\end{equation}
Another bound on the number of $e$-foldings comes from structure formation and the horizon problem: the current comoving Hubble radius must have been inside the Hubble horizon at the beginning of inflation, i.e.
\begin{align}
    a_\textnormal{i}H_\textnormal{i}&<a_\textnormal{0}H_\textnormal{0}\\
    N_e&=\log\left(a_\textnormal{f}/a_\textnormal{i}\right)\\
    \Rightarrow e^{N_e}&>\frac{H_\textnormal{i}}{H_0}\frac{a_\textnormal{i}}{a_0}.
\end{align}
Furthermore, a similar bound can be derived based on entropy considerations \cite{brahma_entanglement_2020}:
the entropy from quantum fluctuations in the early universe can be understood as momentum space entanglement entropy between sub- and super-Hubble modes. Since (globally) entropy is increasing, one can assume that the entropy at the beginning of inflation is smaller than the entropy at the beginning of a radiation-dominated phase. This yields the bound $N_e<\log\left(M_\textnormal{P}/H\right)^{1/4}+\log\epsilon_H^{-1/2}$, which is, at least for the observed values from the powerspectrum, numerically close to the \gls{tcc} bound. %

\paragraph{What is special about the Hubble radius?}
The Hubble radius acts as a length scale threshold, where the dynamical behaviour of modes differs fundamentally:
\begin{description}
    \item[Sub-Hubble] Modes act like independent harmonic oscillators and stretch with the expansion of space \cite{schneider_trans-planckian_2021,kiefer_quantum--classical_1998,kamali_relaxing_2020,brandenberger_trans-planckian_2013}.
    \item[Super-Hubble] Modes act like overdamped oscillators \cite{schneider_trans-planckian_2021},
    fluctuations with a canonical kinetic term become squeezed and classicalise \cite{kiefer_quantum--classical_1998,kamali_relaxing_2020,brandenberger_trans-planckian_2013} respectively
    modes freeze and become non-dynamical \cite{bedroya_sitter_2020,brandenberger_trans-planckian_2013,brandenberger_trans-planckian_2021}.\footnote{
    Scalar and tensor fluctuations might stop oscillating, when exiting the Hubble radius, but they can still grow in amplitude \cite{brandenberger_trans-planckian_2021}.
    }
    Super-Hubble modes experience decoherence, which makes them equivalent to classical, stochastic perturbations that remain classical when re-entering the Hubble horizon \cite{polarski_semiclassicality_1996,bedroya_sitter_2020}.
\end{description}

\paragraph{Why is there not just a simple energy cutoff to exclude trans-Planckian modes from an \gls{eft}?}
The \gls{tcc} can be understood as an expression of the non-renormalisability of \gls{gr} \cite{bedroya_sitter_2020}: In flat space, a theory can be renormalised if classical high-momentum perturbations with $k>\Lambda$ ($\Lambda$ the momentum cutoff) do not influence the $k<\Lambda$ modes. In \gls{gr}, this naive cutoff ceases to exist, as expanding space can stretch $k>\Lambda$ modes into $k<\Lambda$ modes.
To relax this sharp cutoff, two different energy scales need to be defined, a \gls{uv} cutoff $\Lambda_\textnormal{UV}$ and an \gls{ir} cutoff $\Lambda_\textnormal{IR}$, such that \gls{uv} modes cannot stretch to become \gls{ir} modes. The \gls{tcc} summarises this as the condition that $k>l_\textnormal{P}^{-1}$ modes cannot become $k<H$ modes.
This preserves unitarity in the \gls{eft} description: no new Fourier modes have to be added to the Hilbert space, since all trans-Planckian modes remain super-Hubble \cite{brandenberger_trans-planckian_2021,brandenberger_stringy_2024,brandenberger_string_2021,brandenberger_unitarity_2022,brahma_sitter_2021}. A simple \gls{uv} cutoff would have to be constantly adjusted with expanding space, respectively the associated Hilbert space would be time-dependent, with the degrees of freedom increasing over time, which leads to a non-unitary evolution \cite{berera_role_2020}.

The finding that in an expanding background trans-Planckian modes should be hidden, as there cannot be a \gls{uv} cutoff in an expanding background, leads us to a \gls{ccc} analogy \cite{brahma_trans-planckian_2020}:
While in the \gls{ccc} the naked singularity behind the \gls{bh} horizon is shielded, the \gls{tcc} assures that trans-Planckian modes are and remain behind the Hubble horizon \cite{brandenberger_trans-planckian_2021,brandenberger_fundamental_2019,brandenberger_string_2021}.

\paragraph{How are quantum effects to be interpreted in the light of the \gls{tcc}?}\label{p:TCC_Quantum-Effects}
Quantum effects are essential for various models of inflation and large structure formation, as well as particle physics and the description of our Earthly environment\,\textemdash\,they can even be observed in a lab\,\textemdash; to exclude them would miss the point in trying to find an \gls{eft} that describes our Universe \cite{saito_is_2020}.
The \gls{tcc} explicitly rules out \textit{trans-Planckian} physics. But how is this rather vague statement to be interpreted? There are different venues to explore \cite{saito_is_2020}:
\begin{itemize}
    \item The \gls{tcc} excludes trans-Planckian effects for a certain time, such that trans-Planckian quantum fluctuations would only classicalise after the \gls{tcc}-safe phase, where the \gls{eft} is not assumed to hold any longer.\footnote{
        The \gls{tcc} assumes that a problematic mode actually existed in the far past. But if the corresponding fluctuation simply did not exist in the past, it cannot exit the Hubble horizon and the trans-Planckian problem does not arise in practise \cite{dvali_inflation_2020}. 
    }
    \item The \gls{tcc} excludes the observability of trans-Planckian effects, i.e. \gls{tcc}-violating processes are allowed, but are not (yet) observable.
    \item The \gls{tcc} gives trans-Planckian effects a negligible probability, i.e. \gls{tcc}-violations are sufficiently suppressed.
\end{itemize}
If not stated otherwise, our interpretation of the \gls{tcc} throughout this paper is that trans-Planckian effects are allowed, but do not classicalise within what we call the lifetime of the universe, i.e. trans-Planckian effects might become observable in the future, at a time the \gls{eft} description breaks down.

\citet{saito_is_2020} argue that a probabilistic treatment, where \gls{tcc}-violations just have to be sufficiently suppressed, is not promising: even though it works as a swampland condition, as it constrains \glspl{eft}, it does not work as a reliable censorship of trans-Planckian effects from a quantum mechanical standpoint. In particular, such a probabilistic view does not noticeably change predictions for expectation values of observables, as the suppressed outliers do not contribute enough to averages.
\citet{dvali_inflation_2020} state that \gls{ds} quantum fluctuations are typically produced with wavelength $\sim1/H$, much shorter modes are only produced at a rate $\sim\exp\left(-1/H\lambda\right)$. This makes trans-Planckian modes extremely rare: for $H\ll M_\textnormal{P}$ there is one such mode per Hubble patch per time interval $\tau\sim 1/\left(H\exp\left(-1/H\lambda\right)\right)$ \cite{dvali_inflation_2020}.
Nevertheless, \citet{mondal_wave_2022} uses such considerations to make an informed modification of the wave function of the universe:
The Einstein\textendash Hilbert action is complemented by a complex boundary term. This extra term yields a lower Hubble parameter after inflation, which changes the primordial powerspectrum in a non\textendash scale invariant fashion. Since current observations indicate a scale-invariant powerspectrum, this modification has to be weak, ergo, the energy scale of inflation has to be low, which is in line with \gls{tcc} predictions. The real part of the extra term yields a phase factor to the wave function, while the complex part yields an exponentially suppressed addition. This can be interpreted as an exponentially suppressed probability to end up in a \gls{tcc}-violating state.

\paragraph{How is the \gls{tcc} to be understood in other frames?}

The \gls{tcc} can be formulated in the Jordan frame, instead of in the Einstein frame \cite{guleryuz_trans-planckian_2021}:
In the Jordan frame, the Planck mass can be regarded as dynamical, respectively, the cutoff scale is not necessarily the Planck mass but, for example, the string scale, such that the \gls{tcc} reads as $a_\textnormal{f}H_\textnormal{f}<a_\textnormal{i}M_\textnormal{P}(\phi)$ respectively as $a_\textnormal{f}H_\textnormal{f}<a_\textnormal{i}M_\textnormal{S}$, which leads to the modified constraints 
\begin{AmSalign}
    L_\textnormal{J}&=\sqrt{-g}\left(\frac{M_\textnormal{P}^2+\xi f(\phi)}{2}R-\frac{\left(\partial\phi\right)^2}{2}-V_\textnormal{J}(\phi)\right)\\
    M_\textnormal{P}(\phi)^2&=M_\textnormal{P}^2+\xi f(\phi_\textnormal{i})\\
    V_\textnormal{J}(\phi)&\sim f(\phi)^2\\
    &\lesssim\num{e-40}\left(M_\textnormal{P}^2+\xi f(\phi_\textnormal{i})\right)^2\\
    r_\textnormal{ts}&=16\epsilon_V\frac{f(\phi_\textnormal{i})}{f(\phi)\left(1-3\epsilon_V\right)}\\
    &<\frac{f(\phi_\textnormal{i})}{f(\phi)}\num{e-30}
\end{AmSalign}
with 
$M_\textnormal{P}$ the Planck constant in the Einstein frame,
$\xi$ the non-minimal coupling constant,
$f(\phi)$ a generic function of the scalar field $\phi$, and
$\epsilon_V$ the slow-roll parameter in the Einstein frame.
The \gls{tcc} acts then as a selection rule in the Jordan frame on possible theories \cite{guleryuz_trans-planckian_2021}.
The same holds in Scalar\textendash Tensor theory, where the Planck mass is variable in the same fashion as in the Jordan frame, and the \gls{tcc} constraints on the Hubble parameter respectively on the number of $e$-foldings can be written as \cite{shi_large_2021}:
\begin{align}
    e^{N_e}&<\frac{M_\textnormal{P}\sqrt{f(\phi_\textnormal{i})}}{H_\textnormal{f}}\\
    \frac{H}{M_\textnormal{P}\sqrt{f(\phi_\textnormal{i})}}&<\frac{H_0T_0}{H_\textnormal{f}T_\textnormal{rh}}\frac{a_\textnormal{rh}}{a_\textnormal{f}}.
\end{align}

The \gls{tcc} can also be applied to holographic cosmology, since the cosmic time evolution is dual to the inverse renormalisation group flow \cite{strominger_inflation_2001}, which yields the bounds
\begin{align}
    \left(\frac{\phi_\textnormal{i}}{\phi_\textnormal{f}}\right)^{1/\mathfrak{c}}&<\frac{M_\textnormal{P}}{H_\textnormal{IR}}\\
    \phi_\textnormal{f}&>\frac{\mathfrak{c}}{2\pi c}\left(\frac{H_\textnormal{IR}}{M_\textnormal{P}}\right)^\mathfrak{c},
\end{align}
with $\phi=\mathfrak{c}/2\pi c+\order{\mathfrak{c}^2}$, where $\mathfrak{c}\simeq-\eta_V/2$ is related to the second slow-roll parameter, $c$ is the central charge in the operator product expansion and is related to the scale of inflation, and $H_\textnormal{IR}$ is the \gls{ir} fixed point in the holographic model \cite{bernardo_trans-planckian_2020}.

\subsubsection{Evidence}
The \gls{tcc} from \citet{bedroya_trans-planckian_2020,bedroya_trans-planckian-inflation_2020} is studied in
10 different type II and heterotic string theories \cite{andriot_web_2020}, 
K-theory in type IIB toroidal compactifications \cite{damian_remarks_2023},
M-theory \cite{bernardo_sitter_2021,deffayet_stable_2024},
in the light of holography \cite{bedroya_holographic_2022},
asymptotic safety \cite{basile_asymptotic_2021},
and
considering entropy bounds \cite{sun_notes_2021}.
Furthermore, relations to other swampland conjectures are highlighted in \cref{rel:TCC_DC,rel:TCC_dSC,rel:WGC_TCC,rel:TCC_nnNECC}.

\subsection{Unique Geodesic Conjecture}\label{s:UGC}%

A \textit{marked}\footnote{
    A marked moduli space is a moduli space of a physical theory with a given basis choice for the observables\,\textemdash\,such a space is also known as Teichmüller space \cite{raman_swampland_2024}.}
moduli space $\hat{\mathcal{M}}$ has at most one geodesic between any two points in $\hat{\mathcal{M}}$ and exactly one shortest path, which is piece-wise geodesic, between any pair of points in $\hat{\mathcal{M}}$ that are separated by a finite distance \cite{raman_swampland_2024}.
This implies that $\hat{\mathcal{M}}$ is contractible,\footnote{
    This is also an implication of the cobordism conjecture (\cref{sec:cobordism}) \cite{raman_swampland_2024}.
}
and that there is at most one geodesic between each pair of points in $\hat{\mathcal{M}}$ at asymptotically infinite distance, i.e. if there is a geodesic between two points, this geodesic is unique \cite{raman_swampland_2024}.

\subsubsection{Implications for Cosmology}
No direct cosmological implications of this conjecture have been worked out as of today. We can only hint at some very speculative ideas of our own. 

\paragraph{Dark Energy}
If moduli spaces are contractible, this limits the number of possible vacua. This could put strong constraints on the possible values for a cosmological constant. On the one hand, this could explain the specific value we measure today for the cosmological constant. On the other hand, it could render the number of available vacua so small that the anthropic principle cannot be invoked to explain its smallness (see also our discussion in \cref{p:FNomF_DE_DM}).

\paragraph{Inflation}
If moduli space is contractible, eternal inflation cannot take place, as this would require non-trivial cycles through moduli space, which might be incompatible with contractability. Contractability and unique geodesics make it more likely that cosmology evolves towards unique and clearly defined attractor solutions. This agrees with the cobordism conjecture (\cref{eq:cobordism}).

\subsubsection{General Remarks}
\citet{ruehle_attractors_2024} found that \textit{un}marked spaces allow for an infinite number of geodesics between two points via flop\footnote{See \cref{f:flop}.} transitions. Furthermore, they show a correspondence between geodesics\,\textemdash\,which are second-order \glspl{pde}\,\textemdash\,and attractor flows\,\textemdash\,which are first-order \glspl{pde} given by the gradient flow along a potential. The flow lines are geodesics \cite{ruehle_attractors_2024}. Showing uniqueness of a first-order \gls{pde} is simpler \cite{ruehle_attractors_2024}: attractor points are dense in moduli space \cite{lam_attractors_2024} and the flows are unique; therefore, the corresponding geodesics are unique on a dense set of points. Smoothness of moduli space metrics implies then that the geodesics are unique on the entire (marked) moduli space, except for the boundaries of the moduli space. However, flow lines could intersect multiple times between two points in a marked moduli space. This would spoil the uniqueness.

\citet{ooguri_geometry_2007} conjecture that moduli spaces have negative scalar curvature in the asymptotic limits. This was found not to be true \cite{trenner_asymptotic_2010,marchesano_moduli_2024,wilson_geometry_2006,marchesano_asymptotic_2024}: the curvature in the asymptotic limit can be positive, and even diverge to $+\infty$. \citet{marchesano_moduli_2024} conjecture that in the case of a divergence, the \gls{eft} decouples from gravity, indicating that the moduli space is compatible with \gls{qg}. \citet{marchesano_asymptotic_2024} found that in the case of curvature divergence, a \textit{rigid} 4d \gls{eft}\footnote{
    Consider an \gls{eft} with cutoff scale $\Lambda_\text{EFT}$. When approaching this energy scale, this \gls{eft} looses validity. However, it is possible that a sub-sector of this \gls{eft} decouples from gravity, keeping some non-gravitational gauge interactions and kinetic terms finite. This sub-sector does not include the infinite distance limit along which the \textit{full} \gls{eft} brakes down. \citet{marchesano_asymptotic_2024,marchesano_moduli_2024} refer to a theory that corresponds to such a sub-sector as a \textit{rigid field theory}.
}
decouples from gravity and the moduli field curvature is $R\sim\left(\Lambda_\text{WGC}/\Lambda_\text{S}\right)^{2n}$, connecting the \gls{ssc} cutoff $\Lambda_\text{S}$ with the \gls{wgc} cutoff $\Lambda_\text{WGC}=g_\text{rigid}M_{\textnormal{P};d}^{\left(d-2\right)/2}$.\footnote{
    The integer $n\in\{1,\,2,\,3\}$ indicates what kind of \gls{uv} completion the rigid theory might have ($n=1$ a 6d little string theory, $n=2$ a 6d super\gls{cft}, $n=3$ a 5d super\gls{cft}). Having a simple quantity like scalar curvature encoding the \gls{uv} behaviour of an \gls{eft} is a remarkable finding that raises hopes about what might be possible by considering more complicated quantities like the curvature tensor \cite{marchesano_asymptotic_2024}.
}
The scalar moduli field curvature can therefore be regarded as a measure of the relative strength of the gauge interactions in terms of gravitational interactions, i.e. a gauge interaction that is parametrically stronger than gravity leads to a positive divergence in moduli field space \cite{marchesano_asymptotic_2024}.
\citet{cecotti_swampland_2021} note that the divergence disappears when choosing a different metric (the Hodge metric instead of the Weil\textendash Petersson metric) \cite{marchesano_moduli_2024}.

\subsubsection{Evidence}
This conjecture was recently proposed by \citet{raman_swampland_2024}.

\subsection{Weak Gravity Conjecture}\label{sec:gravity}
Gravity is the weakest force. No observation has been made so far to contradict this statement. The \gls{wgc} states that this is true for all string theoretical compactifications, and that there is always a\footnote{
    Gravity is not necessarily the weakest force for \textit{all} particles in the theory, e.g. the strong force acts more weakly on the electron than gravity \cite{palti_brief_2020}.
    }
stable particle, which repulsive gauge force exceeds the attractive gravitational force \cite{cota_asymptotic_2022}.
The general requirement of gravity being the weakest force can be concretised in various ways, with different $\order{1}$ constants as a threshold for the charge-to-mass ratio
\begin{subequations}
    \begin{empheq}[left ={\frac{q}{m}\geq \empheqlbrace}]{align}
        &1  \label{eq:wgcarkani}\\
        &\sqrt{G}  \label{eq:wgcG}\\
        &\left.\frac{\abs{q}}{m}\right|_\text{ext BH}.  \label{eq:wgcextBH}
    \end{empheq}
\end{subequations}
\Cref{eq:wgcarkani} is presented in the original paper \cite{arkani-hamed_string_2007}.
\Cref{eq:wgcG} attempts to take more care of the $\order{1}$ constants involved.
\Cref{eq:wgcextBH} states that there is a (superextremal) particle with a charge-to-mass ratio greater than (or equal to) the charge-to-mass ratio of an extremal \gls{bh}.\footnote{
    For an extremal \gls{bh} the charge (or spin) equals the mass in appropriate units ($Q=M$), and the inner and outer horizons overlap ($r_\heartsuit=M\pm\sqrt{M^2-Q^2}$).
    }

A more formal definition of the \gls{wgc} reads as follows:
In a theory with a $U(1)$ gauge symmetry; gauge coupling $g$; and action
\begin{equation}
    S=\int\!\mathrm{d}^dx\sqrt{-g}\left[\left(M_{\textnormal{P};d}\right)^{d-2}\frac{R_d}{2}-\frac{F^2}{4g^2}+\dots\right],
\end{equation}
with $d$ the number of dimensions of the \gls{eft}, $R_d$ the $d$-dimensional Ricci scalar, $F^2=\frac{1}{q!}F_{\mu_1...\mu_q}F^{\mu_1...\mu_q}$ the strength of the gauge field $A_{\nu...}$ \cite{heidenreich_sharpening_2016}, e.g. with the relation $F_{\mu\nu}=\frac{1}{2}\partial_{[\mu}A_{\nu]}$, and $M_{\textnormal{P};d}$ the $d$-dimensional Planck mass; a particle with mass
\begin{equation}\label{eq:ewgc}
    m\leq\sqrt{\frac{d-2}{d-3}}gq\left(M_{\textnormal{P};d}\right)^{\frac{d-2}{2}},
\end{equation}
$q$ the charge of the particle; exists \cite{palti_swampland_2019,heidenreich_sharpening_2016,lee_tensionless_2018}.\footnote{
    It is conjectured that the equality can only be reached by \gls{bps} states under supersymmetry \cite{ooguri_new_2017,ooguri_non-supersymmetric_2017,cota_asymptotic_2022,arkani-hamed_string_2007,banks_note_2006}.
    \citet{freivogel_vacua_2016} state that
    \enquote{[t]he weak gravity conjecture is a kind of anti-BPS bound: it says the mass of a certain charged particle must be smaller or equal to the minimum value it could have in a supersymmetric theory.}
    The reason it can only be saturated under supersymmetry comes from the link between the mass and the gauge coupling that the \gls{wgc} establishes: it connects Poincaré and internal symmetries, which is, according to the Coleman\textendash Mandula theorem \cite{coleman_all_1967}, only allowed in supersymmetry \cite{prieto_moduli_2024}.
} 
\Cref{eq:ewgc} is called the \textit{electric weak gravity conjecture}. Applying it to the magnetic dual field\footnote{
    We show this in \cref{p:WGC_electric-magnetc}.
}
yields the \textit{magnetic weak gravity conjecture}
\begin{equation}\label{eq:mwgc}
    \Lambda\lesssim g\left(M_{\textnormal{P};d}\right)^\frac{d-2}{2},
\end{equation}
which states the magnetic Coulomb force acts more strongly on the monopole of the smallest magnetic charge than the Newtonian gravitational force \cite{furuuchi_weak_2018}, and that the \gls{eft} is bounded from above by the gauge coupling \cite{palti_swampland_2019,van_beest_lectures_2022,arkani-hamed_string_2007,grana_swampland_2021,palti_brief_2020,saraswat_weak_2017}.\footnote{
    This has the interesting consequence that an \gls{eft} in the limit of vanishing coupling is valid nowhere.
}

There are no known counterexamples to \cref{eq:ewgc,eq:mwgc}, which makes the critique plausible that the conjecture might be too weak to constrain anything. %
There are various attempts to strengthen the \gls{wgc}:
\begin{itemize}
    \item Requiring that the \gls{wgc}-satisfying particle is part of the \gls{eft}, i.e. $m<\Lambda$ \cite{heidenreich_weak_2015}.\footnote{This is coined the \textit{effective} \gls{wgc} \cite{heidenreich_weak_2015}.}
    \item Requiring that this particle is the lightest charged particle \cite{arkani-hamed_string_2007}.\footnote{For multiple charges, there is some support from the \gls{ceb} for this conjecture \cite{brown_fencing_2015}, but there are also counterexamples \cite{aharony_convexity_2021,heidenreich_evidence_2017,heidenreich_weak_2015}.}
    \item Requiring that this particle is of fundamental charge \cite{arkani-hamed_string_2007}.\footnote{\citet{arkani-hamed_string_2007} claimed to have found a counterexample in the heterotic string, but this counterexample was later rejected \cite{harlow_wormholes_2016,heidenreich_sharpening_2016}. Another counterexample comes from the inclusion of higher-derivative terms in the Lagrangian, which allows large \glspl{bh} to decay into smaller \glspl{bh}, instead of into fundamental particles \cite{saraswat_weak_2017,kats_higher-order_2007}.}
    \item Requiring that the lightest charged particle is superextremal \cite{arkani-hamed_string_2007}.\footnote{There are counterexamples to this strong form \cite{heidenreich_evidence_2017,heidenreich_weak_2018,harlow_weak_2023}. In particular, the statement that the lightest charged particle is superextremal is not conserved under Higgsing \cite{harlow_weak_2023}.\label{foo:counterex}}
    \item Requiring that the particle with the smallest charge is superextremal \cite{harlow_weak_2023,arkani-hamed_string_2007}.\footnote{See \cref{foo:counterex}.}
    \item Requiring that a superextremal particle exists, which itself is not a \gls{bh} \cite{harlow_weak_2023}.
    \item Requiring that a superextremal particle exists for every charge in charge lattice \cite{heidenreich_sharpening_2016}.\footnote{This is the (ruled-out) \gls{lwgc}. See \cref{foo:counterex}, the arguments presented by \citet{heidenreich_weak_2015}, and the counterexample presented by \citet{harlow_weak_2023}.}
    \item Requiring that there is always an infinite tower of superextremal states with $g^2q_i^2\gtrsim M_i/M_\textnormal{P}$ \cite{cota_asymptotic_2022,montero_weak_2016,heidenreich_evidence_2017,heidenreich_sharpening_2016,heidenreich_repulsive_2019,harlow_weak_2023}.\footnote{This is the \gls{twgc}.}
    \item Requiring that for every \gls{qg} theory with massless gauge fields, a sub-lattice of the charge lattice exists, with superextremal particles at every site \cite{heidenreich_weak_2018,heidenreich_evidence_2017,hayashi_spectra_2023,lee_tensionless_2018,heidenreich_repulsive_2019,harlow_weak_2023}.\footnote{This is the \gls{slwgc}.}
    \item Requiring that gravity is the weakest force for all possible states \cite{palti_weak_2017}.\footnote{This is the super \gls{wgc} \cite{palti_weak_2017}).}
    \item Extending the conjecture to multi-particle states\footnote{This is done in the form of the convex hull conjecture \cite{cheung_naturalness_2014,harlow_weak_2023,palti_swampland_2019,saraswat_weak_2017,palti_brief_2020,van_beest_lectures_2022}. See \cref{p:WGC_multiparticle}.} or general $p$-forms.\footnote{See \cref{p:WGC_p-form}.}
    \item Specialising the conjecture for scalar fields\footnote{See \cref{p:SWGC}.} or higher-spin states.\footnote{See \cref{p:WGC_spin-stes}.}
\end{itemize}

We see that there is a whole family of conjectures, each with the goal of establishing the general rule of \enquote{gravity is the weakest force} with a twist, some are debunked with counterexamples, others have more evidence in their favour, and all make slightly different predictions for possible observables. %
In the following, we discuss the implications for cosmology.

\subsubsection{Implications for Cosmology}\label{sss:WGC_Cosmology}
The \gls{wgc} is of particular interest, as it might be a conjecture with testable implications for low-energy physics \cite{harlow_weak_2023}. For the \gls{sm}, the implications are feeble \cite{harlow_weak_2023}: The electron satisfies the \gls{wgc}-bound by more than 20 orders of magnitude, and since the electron charge is almost of order one, the tower of charged particles predicted by the \gls{twgc}/\gls{slwgc} could start around the Planck mass.
Some implications are reassuring: the neutrino is electrically neutral, photons are massless\footnote{
    See \cref{p:WGC_Photons} for further details.
}
\cite{reece_photon_2019}, and electric charge is quantised \cite{abu-ajamieh_implications_2024}.
More exciting predictions are the upper bound on the proton lifetime of \num{e42} years \cite{abu-ajamieh_implications_2024}, and the existence of magnetic monopoles. Because the mass of the magnetic monopole could be around the Planck scale, its existence is currently not falsifiable.
Furthermore, an additional aspect that might be of interest is the finding that the Bell inequalities \cite{bell_einstein_1964} cannot be violated if the \gls{wgc} does not hold, indicating that there were hidden variables in quantum theory if gravity was not the weakest force \cite{sinha_bell_2023}.\footnote{
    Violating the Bell inequalities might give rise to cosmological observations \cite{choudhury_bell_2017,maldacena_model_2016}.
}

\paragraph{Axion models,}\label{p:WGC_axion}
be it for inflation, hyperinflation, \gls{ede}, or quintessence, are constrained by the axionic \gls{wgc}
\begin{equation}
    S_\iota \leq M_\textnormal{P}/f
\end{equation}
with
$S_\iota$ the instanton\footnote{
    Instantons are objects charged under axions, lack time-evolution, and their action is the equivalent to the tension \cite{heidenreich_sharpening_2016}. Instantons are not physical objects like particles or branes, but events that are introduced as a tool for computation \cite{stout_instanton_2020}.
    }
action\footnote{
    The action term $S_\iota $ plays the role of the mass \cite{junghans_large-field_2016}.
}, and
$f$ the axion\footnote{
    By an axion we mean any shift-symmetric periodic scalar. The axion period is called the axion decay constant $f$. The inverse axion decay constant plays the role of the gauge coupling \cite{junghans_large-field_2016}.
}
decay constant \cite{blumenhagen_large_2018,gonzalo_strong_2019,harlow_weak_2023,hebecker_euclidean_2018,grimm_infinite_2020,montero_transplanckian_2015,heidenreich_instanton_2020,hebecker_can_2017,kooner_warping_2016,choi_recent_2021}.
This bound is motivated from multiple angles:
focussing on $S_\iota$, focussing on $f$, and focussing on the combined bound $fS_\iota$.

\subparagraph{$S_\iota$} needs to be large enough to guarantee perturbative control over the \gls{eft}, as in a string compactification, $S_\iota $ is set by the volume in string units of the internal cycle wrapped by the instanton \cite{cicoli_early_2023}.
The instanton expansion breaks down when $S_\iota <1$, and new, light states appear \cite{harlow_weak_2023,stout_instanton_2020}.\footnote{
    \citet{stout_instanton_2020} notes that it poses an issue that the interesting limit of small instanton action is ill-defined if one wants to study how \gls{qg} prevents global symmetries. He finds that it might be better to opt for a different formalism based on the Yang\textendash Lee theory of phase transitions, which allows studying the different energy levels of the instantons: There will be an axion field value $\phi_*$ for which the energy gap $\Delta E$ between the energy of a state and the vacuum is minimal. The \gls{wgc} can then be expressed as $\sqrt{\Delta E(\phi_*)/\Delta E^{\prime\prime}(\phi_*)}\lesssim M_\textnormal{P}/f$, which remains well-defined in the limit of vanishing instanton action.
    }

$S_\iota$ plays a key role in what could be considered a magnetic form of the axionic \gls{wgc}:
Studying global U(1) symmetries Higgsed under axions, \citet{daus_towards_2020} find the condition
\begin{equation}
    e^{-S_\iota }\gtrsim e^{-M_\textnormal{P}^2/\Lambda^2},
\end{equation}
which they call the \textit{Swampland Global Symmetries Conjecture}. The bound relates the cutoff scale $\Lambda$ to the action of the instanton $S_\iota$.%

\subparagraph{$f\leq M_\text{P}$} is another condition to be in a regime of perturbative control \cite{brown_fencing_2015,dolan_transplanckian_2017,hebecker_euclidean_2018,goswami_enhancement_2019}.
This rules out trans-Planckian field ranges \cite{draper_virtual_2018}:
Axions fragment into inhomogeneous modes on a timescale $\log(t)\sim m/fqg$. A slowly fragmenting scalar moves a distance $\Delta\phi\lesssim M_\textnormal{P}m/fqg$. Applying the \gls{wgc} yields $\Delta\phi\lesssim M_\textnormal{P}^2/f$. Requiring $f<M_\textnormal{P}$ rules out trans-Planckian field ranges.

At first, it seems that models with multiple fields can circumvent the constraint by having the individual decay constants on a sub-Planckian scale, but the effective decay constant trans-Planckian.
The alignment of multiple fields  \cite{kaplan_clockwork_2016,choi_realizing_2016,choi_natural_2014,higaki_natural_2014,hebecker_large_2019,kim_completing_2005,reig_stochastic_2021}, where the parent \gls{qg} theory satisfies the \gls{wgc} \cite{saraswat_weak_2017}, would be such a mechanism.
However, while \citet{montero_transplanckian_2015} show that kinetic alignment yields field ranges $\sim\sqrt{\mathfrak{N}}f_\text{max}$ (with $f_\text{max}$ the maximal axion periodicity and $\mathfrak{N}$ the number of axions), which corresponds to travelling diagonally through the charge lattice, and
\citet{bachlechner_planckian_2016} find $S_\text{min}\gtrsim\sqrt{\mathfrak{N}}M_\textnormal{P}/f_\text{max}$, such that gravitational instanton contributions with $f_\text{max}=M_\textnormal{P}$ are still under control,
\citet{rudelius_possibility_2015,rudelius_constraints_2015} shows that in a theory with $\mathfrak{N}$ axions and $\sum_{i=1}^\mathfrak{N}\left(f_iS_i\right)^2\leq1$, such that the convex hull conjecture is satisfied,
the effective axion decay constant $f_\text{eff}\sim f_iS_i$ scales like $1/\sqrt{\mathfrak{N}}$,
because one also aligns the charge vectors $\Vec{z}_i=\sum_j\frac{M_\textnormal{P}}{f_{ij}S_i}\Vec{e}_j$ (with $\left\{\Vec{e}_j\right\}$ an orthonormal basis of the vector space), such that the charge vectors have to increase in length to still span the unit ball at smaller angles separating them.
This cancels the effect of alignment and a trans-Planckian field range is required (but forbidden), unless there are additional potential contributions or multiple windings of the instanton, such that the moduli space can be traversed multiple times. The observation can be summarised in the statement that the moduli space for axions has a radius $r\lesssim\pi M_\textnormal{P}$ \cite{rudelius_constraints_2015,bachlechner_planckian_2015}.
Even when travelling diagonally through axion space, the axion moduli space is bounded by the \gls{wgc}, and large decay constants are simply in the swampland.
Moreover, \citet{conlon_quantum_2012} finds that the \gls{ceb} is violated if $\sum_if_i^2>\order{1}M_\textnormal{P}$.
Furthermore, having a large number of axions $\mathfrak{N}$ seems problematic in itself: either, the number of instantons grows larger than $\mathfrak{N}^2$, probably even exponentially, or there has to be a loophole that allows the axions to traverse a trans-Planckian field range \cite{junghans_large-field_2016}.
For a U(1) gauge theory with a 2-form gauge potential, the bound $\Lambda\lesssim\sqrt{fM_\textnormal{P}}$ is motivated in analogy to the magnetic \gls{wgc} \cite{reece_photon_2019,craig_rescuing_2018}.

\subparagraph{$fS_\iota\sim1$} is a bound that should hold in string theory, since the axion decay constant and the instanton action are related to the saxion $\varphi$, such that
\begin{align}
    f&\sim\frac{1}{\varphi}\\
    S_\iota &\sim\varphi\\
    \Rightarrow fS_\iota &\sim1,
\end{align}
which shows that the large-field regime $f>1$ is inaccessible in the instanton expansion with $S_\iota >1$ \cite{blumenhagen_large_2018}.
The same relation can also be understood as an analogy to the (non-axionic) \gls{wgc}:
\begin{equation}
    \left.\frac{q_\iota}{S_\iota}\right|_\text{WGC}\geq\left.\frac{q_\iota}{S_\iota}\right|_\text{ext}\sim\frac{f}{M_\textnormal{P}},
\end{equation}
with $q_\iota$ the instanton charge; to strengthen the analogy, one can require that this relation is satisfied by the instanton with the smallest action\footnote{
    \citet{hebecker_large_2019} suggest that axion potentials with large monotonic regions\,\textemdash\,which they construct in a type IIB setting\,\textemdash\,might parametrically violate this strong form of the axionic \gls{wgc}.
}
\cite{kooner_warping_2016} or the one of lowest charge\footnote{
    This immediately opens up the loophole that the \gls{wgc}-instanton is not the one of lowest charge \cite{brown_axionic_2016,hebecker_winding_2015,brown_fencing_2015,rudelius_constraints_2015}.
} \cite{hebecker_euclidean_2018,grimm_infinite_2020,montero_transplanckian_2015,heidenreich_instanton_2020,harlow_weak_2023}.

\citet{heidenreich_weak_2015} mention that in a 4d theory with a U(1) plus a \gls{kk} U(1), the axion decay constant $f$ grows like $1/\sqrt{S_\iota}$, which leaves a small loophole for trans-Planckian decay constants compatible with the \gls{wgc} (and in particular with the convex hull conjecture).

\subparagraph{Counterexamples} to the axionic \gls{wgc} are presented in the form of Einstein\textendash Maxwell\textendash dilaton theory \cite{loges_thermodynamics_2020} and Einstein\textendash axion\textendash dilaton theory \cite{andriolo_duality_2020}, which violate the axionic \gls{wgc}, unless positivity bounds on higher-derivative operators \cite{adams_causality_2006}, and in the latter case additionally symmetries on the higher-order derivatives (which are inherited from the \gls{uv} theory), are taken into account \cite{loges_duality_2020} (see \citet{andriolo_duality_2020} for more comments on the necessity but insufficiency of positivity bounds).

Furthermore, warped geometries with open strings might provide a counterexample to the axionic \gls{wgc} \cite{kooner_warping_2016}: in the limits of supergravity, where instanton effects are exponentially suppressed, large axionic decay constants are generated with $M_\textnormal{P}/f_\iota<S_\iota$. Either, the instantons couple to the open string axions with a suppressed effective axion coupling $f^\prime<f$ or new stringy effects appear in that limit, such that the effective coupling is smaller\,\textemdash\,or the axionic \gls{wgc} is incorrect \cite{kooner_warping_2016}.

\paragraph{Black Holes}\label{p:WGC_BH}
The magnetic \gls{wgc} demands that the minimally charged magnetic object is not a black hole/brane \cite{hebecker_what_2017}, i.e. the magnetic monopole is not a \gls{bh} \cite{harlow_weak_2023,heidenreich_evidence_2017}.
The cutoff scale of the theory is related to the classical radius of the magnetic monopole \cite{heidenreich_evidence_2017}.\footnote{The technicalities are laid out by \citet{saraswat_weak_2017}:
    Since a \gls{bh} is a classical concept, all charges carried by \glspl{bh} are macroscopic and continuous \cite{rojo_swampland_2019}.
    Continuous gauge symmetries are compact in \gls{qg} \cite{banks_symmetries_2011}.
    Therefore, the magnetic charge in \gls{bh} solutions obeys the Dirac quantisation $q_\textnormal{m}=2\pi n/g$ with $n\in\mathbb{Z}$ (to avoid exact global symmetries \cite{de_la_fuente_natural_2015}), and the extremal magnetically charged \gls{bh} at zero temperature has a mass of $m_\text{BH}=q_\textnormal{m}$ and a large entropy $S\sim g^{-2}$, i.e. it has a large number of degenerate states. There is no apparent reason to believe that magnetic charge can only appear in such a high-entropy setting. Therefore, it is expected that there is a magnetic monopole with mass $m_\textnormal{m}\sim1/g^2l$ that is not a \gls{bh}, i.e. it is an extended object\,\textemdash\,meaning that it is bigger than the Schwarzschild radius, ergo $l\gtrsim1/g^2lM_\textnormal{P}^2$. It follows that $1/l\lesssim gM_\textnormal{P}$, which shows the connection to the magnetic \gls{wgc}: $\Lambda_\text{EFT}\sim1/l\lesssim gM_\textnormal{P}$.
    }

The electric \gls{wgc} requires \glspl{bh} to decay \cite{fisher_semiclassical_2017,cota_asymptotic_2022,arkani-hamed_string_2007,banks_note_2006}:\footnote{
    Studying the full non-linear effects of charged extremal \glspl{bh} in Einstein\textendash Born\textendash Infeld Gravity, \citet{wu_superradiant_2024} find that \enquote{large extremal particle-like black holes may break up and decay even when gravity is not the weaker force}.
}
Assume you start with a large \gls{bh}. The \gls{bh} emits particles through Hawking radiation. Since Hawking radiation is a charge-neutral process, the \gls{bh} sheds its mass, yet not its charge, i.e. at one point it will be so strongly charged that the electromagnetic repulsion overcomes the gravitational attraction\,\textemdash\,if gravity remains the weakest force.\footnote{
    Forming a superextremal \gls{bh} by overcharging an extremal \gls{bh}, i.e. by throwing in an additional charged particle, does not work \cite{anand_analyzing_2024}: the Coulomb repulsion is too strong, and the particle will never reach the horizon \cite{izumi_gedanken_2024}.
}
If the \gls{wgc} is violated, \glspl{bh} do not decay and form stable remnants:
The number of particles of mass $m$ and charge $q$ a \gls{bh} can emit is $n=Q/q$, with $Q$ the \gls{bh} charge, resulting in a total mass of $m\cdot n=mQ/q<M$, where the \gls{bh} mass $M$ cannot be reached due to conservation of (binding) energy. When the extremality $Q/M=1$ is reached, the \gls{bh} temperature is zero\footnote{
    In theories with a dilaton, this is not necessarily the case, but the Bekenstein\textendash Hawking entropy approaches zero in the extremal limit \cite{heidenreich_black_2020}.
    },
Hawking radiation stops, and the \gls{bh} is semiclassically stable \cite{montero_weak_2016}.
Assuming a very weak gauge coupling of $g\sim\num{e-100}$ and a \gls{bh} of several Planck masses, e.g. $M\sim10M_\textnormal{P}$, the \gls{bh} can have any charge in between 0 and $Q\sim\num{e100}$ and still satisfy $M>gQM_\textnormal{P}$ \cite{arkani-hamed_string_2007}. The quantum-mechanical totalitarian principle\footnote{
    The quantum-mechanical totalitarian principle \enquote{everything that is not forbidden is compulsory} (which is actually not the totalitarian principle, but the principle of plenitude) is often attributed to Gell-Mann, even though he wrote \enquote{[a]nything that is not compulsory is forbidden} \cite{gell-mann_interpretation_1956} (which is the actual totalitarian principle) \cite{kragh_physics_2019}.
}
would assure that each possible state gets realised, and we end up with $\sim\num{e100}$ remnants \cite{arkani-hamed_string_2007}.
Remnants represent naked singularities\,\textemdash\,which violate the \gls{ccc}\,\textemdash\,and as such global symmetries\,\textemdash\,which violate the no global symmetries conjecture (\cref{sec:nGSym}).\footnote{
    The compatibility of the existence of remnants with \gls{qg} is still under debate \cite{giddings_black_1992,palti_swampland_2019,van_beest_lectures_2022,heidenreich_evidence_2017}, e.g. disfavoured by \citet{susskind_trouble_1995,banks_note_2006} but found to be unproblematic by \citet{casini_relative_2008}.
    We further discuss the question of stable remnants in \cref{p:WGC_Remnants,p:Cobordism_BH,p:nGSym_BH}.
    Moreover, new research on the possible decay paths of \glspl{bh} might indicate that a \gls{bh} is never at risk of becoming superextremal and remnants might be avoided entirely \cite{brown_evaporation_2024,davies_nonsingular_2024}.
}
The \gls{wgc} assures that this situation is avoided: The \gls{wgc} predicts the existence of a particle with $q/m>1$, which the \gls{bh} can emit through Schwinger pair production.\footnote{
    Hawking radiation is the dominant process for \gls{bh} discharge if the temperature is high compared to the \gls{bh} mass; if the temperature is low compared to the \gls{bh} mass, Schwinger pair production is dominating \cite{furuuchi_weak_2018}.
    Both processes can be interpreted as a form of quantum tunnelling \cite{parikh_hawking_2000,furuuchi_weak_2018}. This means that even an exponentially suppressed decay rate represents an instability of the \gls{bh}, which satisfies the \gls{wgc}.
    The decay does not need to be efficient \cite{furuuchi_weak_2018}: if arbitrarily large \gls{rn} \glspl{bh} were requested to decay efficiently, charged particles with arbitrarily small mass were required to be present in the theory, which is not even true for the \gls{sm}, as there is no massless charged particle.
    \citet{lin_schwinger_2024} find that the Schwinger rate declines and vanishes when a particle's charge-to-mass ratio approaches the \gls{bh} extremality bound, satisfying the \gls{wgc}.
}
The \gls{bh} can shed its charge and ceases to be extremal.

For a \gls{bh} in an Einstein\textendash Maxwell universe, with an action \cite{kats_higher-order_2007} given by
\begin{AmSalign}
    S=&\int \mathrm{d}^4x\sqrt{g}\left[\frac{M_\textnormal{P}^2}{4}R-\frac{1}{4}F_{\mu\nu}F^{\mu\nu}\right.\nonumber\\
    &+\left.\mathfrak{l}\left(F_{\mu\nu}F^{\mu\nu}\right)^2+\mathfrak{p}\left(F_{\mu\nu}\tilde{F}^{\mu\nu}\right)^2+\mathfrak{c} F_{\mu\nu}F_{\rho\sigma}W^{\mu\nu\rho\sigma}\right],
\end{AmSalign}
with $W^{\mu\nu\rho\sigma}$ the Weyl tensor,
the \gls{wgc} bound reads as
\begin{equation}
    \frac{M_\textnormal{P}^2q^2}{m^2}\leq1+\frac{4}{5q^2}\left(2\mathfrak{l}-\mathfrak{c}\right)+\order{\frac{1}{q^4}},
\end{equation}
where the leading-order correction\textendash term is likely positive, as various considerations show \cite{mcpeak_holography_2021}:
\gls{bh} thermodynamics\footnote{
    Considerations of \gls{bh} entropy lead to the conclusion that the charge-to-mass ratio increases with decreasing size, i.e. larger extremal \glspl{bh} can decay into smaller extremal \glspl{bh}, as the latter have a higher charge-to-mass ratio \cite{cheung_proof_2018}.
    The extreme case is the extremal \gls{bh} of minimal charge, which has vanishing area and entropy\,\textemdash\,if a \gls{bh} is extremal and its charge is the elementary electric charge, it contains only one microstate and the entropy vanishes \cite{cano_exact_2021}.
}
\cite{cheung_proof_2018,vafa_cosmic_2019,harlow_weak_2023,fisher_semiclassical_2017,andriolo_tower_2018,cottrell_weak_2017},\footnote{
    There might be a loophole in the entropy argument used by \citet{cheung_proof_2018}, arising from allowed field redefinitions, as is pointed out by \citet{hamada_weak_2019}.
}
positivity of the S-matrix\footnote{
    Positivity bounds on the scattering amplitude show that extremal \glspl{bh} are self-repulsive \cite{bellazzini_amplitudes_2019}.
}
\cite{bellazzini_amplitudes_2019},
unitarity and causality \cite{hamada_weak_2019,arkani-hamed_causality_2022},
and running of the renormalisation group\footnote{
    Renormalisation of one-loop divergences in Einstein\textendash Maxwell theories with additional massless matter fields leads to positivity bounds on higher-derivative terms that enable \glspl{bh} to decay \cite{charles_weak_2019}.
}
\cite{charles_weak_2019}.
The higher-order corrections to the bound can be interpreted as a reduction in the mass\footnote{
    Being open to the possibility of above\textendash \gls{ir} cutoff \glspl{bh}, \citet{aalsma_new_2021} derive a bound on the higher-order correction\textendash induced mass change, which excludes pathological matter contributions:
    \begin{equation}\label{eq:wgcspin}
        m^\prime=\int_\Sigma\mathrm{d}^{d-1}x\sqrt{h}\delta T^\text{eff}_{ab}\xi^an^b\leq0,
    \end{equation}
    with $\Sigma$ a Cauchy slice of constant Killing time $t$ with induced metric $h_{ab}$,
    $\delta T^\text{eff}_{ab}=\delta T_{ab}+F_{ac}\delta F_b^c$ the effective stress\textendash energy tensor,
    $\xi^a$ a timelike Killing vector,
    and $n^b$ a unit normal vector.
}
of an extremal \gls{bh} that saturates the bound \cite{heidenreich_sharpening_2016,kats_higher-order_2007,aalsma_new_2021,cremonini_higher_2010,goon_universal_2020,sadeghi_weak_2022-1,cano_non-supersymmetric_2022} respectively as an increase
of its charge-to-mass ratio\footnote{
    This increase in the charge-to-mass ratio is equivalent to an increase in entropy \cite{aalsma_weak_2019,hamada_weak_2019}.
    The positivity bound can be studied on the level of the Lagrangian.
    That contributions of the form $\left(F_{\mu\nu}\right)^2$ or $\left(F_{\mu\nu}\tilde{F}^{\mu\nu}\right)^2$ yield positive contributions is obvious, however terms like $R_{\mu\nu\rho\sigma}F^{\mu\nu}F^{\rho\sigma}$ or how to eliminate the $t$-channel graviton pole in dispersion relations \cite{alberte_positivity_2020,henriksson_bounding_2022} is less obvious \cite{de_rham_snowmass_2022}.
    In principle, finite-size effects of massive particles, loops of massless particles or bare higher-derivative couplings in the Lagrangian could increase or decrease the charge-to-mass ratio of finite-sized \glspl{bh}\,\textemdash\,the latter case would violate the \gls{wgc}, if there are no light, superextremal particles \cite{heidenreich_repulsive_2019}.
    \citet{arkani-hamed_causality_2022} reason that the corrections to the Wilson coefficients are enforced by unitarity.
}
\cite{mezerji_correlation_2022,arkani-hamed_string_2007,kats_higher-order_2007,cheung_proof_2018,bellazzini_amplitudes_2019,cheung_entropy_2019}.\footnote{
    \citet{cano_extremality_2020} note that according to the existing literature, this statement seems to hold, but is mainly studied for Einstein\textendash Maxwell theory. \citet{cano_alpha_2019,cano_extremality_2020} find that in Einstein\textendash Maxwell\textendash Dilaton theory or more generally heterotic string theories, the corrections are either positive or null (meaning that the next higher-order term is absent).
    As an interesting aside, in their setup, the Wald entropy ceases to be correlated to the charge-to-mass ratio. This implies that in a setting where the perturbation of the entropy is positive, the \gls{wgc} is not automatically implied.
    The same observation is made by \citet{aalsma_corrections_2022}, when studying corrections to extremal \glspl{bh} using the Iyer-Wald formalism.
    Furthermore, for black branes, it can be shown that corrections to entropy are not necessarily correlated with corrections to extremality \cite{sadeghi_weak_2022}, but if four-derivative couplings obey scattering positivity bounds, higher-order derivative corrections decrease the tension-to-charge ratio of extremal black branes \cite{noumi_higher_2022}.
    Taking into account higher-order corrections to the static, spherically symmetric, extremal solution of non-vanishing horizon area, \citet{etheredge_derivative_2022} show that mass, entropy, and self-force corrections are of independent and arbitrary sign. Furthermore, they note\,\textemdash\,referring to \citet{mcpeak_entropy_2022}\,\textemdash\, that the entropy corrections at fixed mass and charge near extremality can diverge and are an independent quantity from extremal entropy corrections, which are finite.
}
\citet{henriksson_bounding_2022} claim that a small violation of the bound might be allowed, referring to work that shows a reduced \gls{eft} cutoff scale \cite{bellazzini_amplitudes_2019,alberte_positivity_2020}, but also stating that a small violation of the bound itself would indicate a weakening of causality, i.e. superluminality\footnote{
    \enquote{Roughly, the resolution seems to be that gravitational interactions universally cause a time delay, so EFT operators that cause a time advance are allowed in principle as long as the advance is smaller than the gravitational time delay} \cite{henriksson_bounding_2022}.
}
\cite{de_rham_speed_2020,hollowood_causality_2016,de_rham_causality_2020,drummond_qed_1980,goon_superluminality_2017} would be part of the theory, yet without spoiling the S-matrix consistency \cite{alberte_qed_2021}.
Furthermore, \citet{aalsma_corrections_2022} worries that small \glspl{bh} might experience different corrections than large \glspl{bh}: On the one hand, large \glspl{bh} could be above the \gls{eft} cutoff scale. On the other hand, the ratio between the local and the non-local corrections to the gravitational \glspl{eom} is $\sim\left(r_\heartsuit/l_\textnormal{P}\right)^2$, with $r_\heartsuit$ the radius of the \gls{bh} horizon and $l_\textnormal{P}$ the Planck length scale, i.e. for large \glspl{bh} the classical / local corrections dominate \cite{arfaei_locality_2023}.
If higher-derivative terms make small \glspl{bh} superextremal, then they also have to make them self-repulsive \cite{gendler_merging_2021}: If not, two such superextremal \glspl{bh} could form a gravitationally bound state with twice the charge and less than twice the mass, i.e. with a higher charge-to-mass ratio. This is an important observation for the \gls{rfc}.
At least for small \glspl{bh} in \gls{ads} space, the extremal mass decreases in a canonical ensemble if unitarity and the \gls{nec} are given \cite{aalsma_corrections_2022}.

Large \glspl{bh} containing multiple charges are able to decay when the convex hull of the charge-to-mass ratio of small \glspl{bh} contains the unit ball\,\textemdash\,regarding four-derivative corrections, the space of allowed charge-to-mass ratios is convex, as the corrections are smaller than the two-derivative terms; this simplifies the convex hull requirement to requiring a positive shift to extremality \cite{jones_black_2020}.\footnote{
    See \cref{p:WGC_multiparticle} for comments on the convex hull conjecture and multi-particle states.
    }

\glspl{bh} do not exist in isolation. The implications of the presence of additional fields is studied by various authors:
\begin{itemize}
    \item Studying \gls{rn} \gls{bh} solutions in \gls{ds} spacetime with a varying cosmological constant, \citet{luben_black_2021} find that $m^2>qgM_\textnormal{P}H$ is necessary to avoid super-extremality of the \gls{bh}: such a \gls{bh} discharges with the Schwinger rate $\Gamma\sim \exp\left(-m^2/qgE\right)$ \cite{frob_schwinger_2014,ai_schwinger_2020,brown_schwinger_2019,gould_observing_2019,qiu_schwinger_2020}, i.e. it discharges rapidly for $m^2\ll qE$ and evolves towards a superextremal neutral Nariai solution, which lies outside of physical phase space (as they show); therefore it violates the \gls{ccc}. This observation was motivated by a \gls{ds} version of the \gls{wgc}\,\textemdash\,the \gls{flb}.
    \item \citet{chen_universal_2020} study \gls{rn} and Bardeen \glspl{bh} in \gls{ads} spacetimes with quintessence, and find that larger \gls{ads} radii make it easier to satisfy the \gls{wgc}. Furthermore, they note that negative corrections to the quintessence density decreases the charge-to-mass ratio of the \gls{bh}.
    \item For extremal Kerr\textendash Newman \gls{ads} \glspl{bh}, \citet{sadeghi_weak_2022-1} find that the presence of quintessence decreases the charge-to-mass ratio, as the added correction is inversely proportional to the entropy.
\end{itemize}

There are a few additional aspects where insights about \glspl{bh} from the \gls{wgc} can be gained:
\begin{itemize}
    \item In 4-dimensional asymptotically flat \gls{rn} or \gls{gb} spacetimes, an innermost stable circular orbit exists, unless the \gls{wgc} is violated and the charge-to-mass ratio is greater than unity \cite{paul_iscos_2024}.
    \item \citet{alipour_weak_2024} study \gls{rn} \glspl{bh} in Kiselev spacetime and find that smaller photon spheres are associated with larger charge-to-mass ratios $Q/M$ of the \glspl{bh}, which yields a potential testing ground for the \gls{wgc}, given enough observational precision of \gls{bh} photon spheres, respectively the absence of a photon sphere violates the \gls{wgc} in their setting.
    \item The deformability of \glspl{bh} can be studied using tidal Love numbers, as work in an early stage by \citet{de_luca_implications_2023} indicates that there is a relation between positivity of tidal Love numbers and higher-order corrections through the \gls{wgc}.
    \item Another open question is how rotation affects the \gls{wgc} bounds:
    On the one hand, rotation seems to have an impact on the corrections to positivity bounds \cite{aalsma_corrections_2022,reall_higher_2019,cano_leading_2019}.
    On the other hand, \glspl{bh} can lose angular momentum through superradiance \cite{aalsma_rotating_2022}. Furthermore, in pure gravity in six or more dimensions, the angular momentum of \glspl{bh} is unconstrained \cite{myers_black_1986}.
    \citet{aalsma_rotating_2022} examine a class of \glspl{bh} that allow a mapping of rotating Kerr\textendash Newman solutions to charged but non-rotating \gls{kk} \glspl{bh}, without finding consistent corrections to \gls{wgc} bounds.
    \citet{ma_negative_2023} discuss rotating black strings, where the higher-order corrections induce a \textit{negative} entropy shift, spoiling the mutually re-enforcing bond between the \gls{wgc} and the \gls{ccc}.
\end{itemize}

\paragraph{Cosmic Censorship}\label{p:WGC_CCC}
The \gls{ccc}
requests that naked singularities must be behind a horizon \cite{harlow_weak_2023,wald_gravitational_1997,cui_weak_2024,penrose_golden_2002}.
No naked singularity has been observed, and numerical simulations suggest that even in higher-dimensional spaces, quantum corrections would restore the censorship for proposed counterexamples \cite{harlow_weak_2023,gregory_black_1993,lehner_final_2011,emparan_predictivity_2020,crisford_violating_2017,horowitz_evidence_2016,crisford_testing_2018,horowitz_further_2019,figueras_end_2016,figueras_end_2017,lehner_black_2010}.
To satisfy the \gls{ccc} respecetively the \gls{wgc}, a particle of a certain minimum charge has to be present in the theory \cite{cui_weak_2024,horowitz_further_2019}. However, the charge bounds are not necessarily equal, as examples exist where lower bounds to the \gls{ccc} exist but do not agree to the \gls{wgc} bound \cite{horowitz_further_2019,song_weak_2021,song_weak_2024,hu_weak_2020}.
Some weaker forms of the \gls{wgc} allow the superextremal particles to be \glspl{bh} themselves, yet superextremal \glspl{bh} would represent naked singularities, and are therefore incompatible with the \gls{ccc}, which rules out those weaker forms of the \gls{wgc} \cite{sadeghi_weak_2024,sadeghi_reissner-nordstrom_2024}.
\citet{sadeghi_strong_2023} come to the conclusion that for charged \glspl{bh} in \gls{ds} space, the \gls{ccc} is (not) satisfied whenever the \gls{wgc} is (not) obeyed.
For \gls{ads} spaces, \citet{anand_analyzing_2024} find that the \gls{wgc} prevents a violation of the \gls{ccc}: The \gls{ccc} could be violated if one was able to overcharge a \gls{rn} \gls{bh} by throwing in an additional particle. As they show, the \gls{wgc} assures that all available particles that could be thrown into the \gls{bh} increase the entropy and enlarge the horizon, such that the \gls{ccc} is always satisfied. Exploring the other direction, they also show that if the central charge of a radiating extremal \gls{bh} surpasses the electric charge, superradiant \gls{wgc}-particles are emitted that move the \gls{bh} away from the extremal state, protecting the \gls{ccc}.
Also \citet{crisford_violating_2017} adumbrate that the mechanism they propose to violate cosmic censorship fails if the \gls{wgc} holds.
For quintessence, the compatibility of the \gls{ccc} and the \gls{wgc} is shown \cite{alipour_wgc_strings_2023,alipour_wgc_RPS_2023}.
We conclude that gravity's weakness seems to prevent naked singularities \cite{crisford_testing_2018}, in \gls{ds} as well as in \gls{ads} spaces.\footnote{
    Implicitly, we were referring to the weak \gls{ccc} in this section. There is a strong form that states that all processes described by \gls{gr} remain predictable throughout the entire spacetime. \citet{tu_short-hair_2024} find that processes that violate the strong \gls{ccc} require \gls{wgc}-violating particles and \glspl{bh} with short hair (deviations from the \gls{rn} metric near a \gls{bh} horizon, i.e. there are localised deviations for interactions near the horizon).
}
\begin{displayquote}
    It is perhaps ironic that for many years people hoped that cosmic censorship would fail so that we had the possibility of observing effects of quantum gravity. Now we find that a conjecture about quantum gravity is preserving cosmic censorship. It appears that quantum gravity wants to remain hidden. \cite{horowitz_further_2019}
\end{displayquote}

\paragraph{Dark energy}\label{p:WGC_DE} faces model-dependent constraints. Especially, models that couple \gls{de} to \gls{dm} have to make sure that the coupling is in such a way that gravity remains the weakest force.
In the following, we start with general statements and transition into more concrete findings about the implications of the \gls{wgc} for \gls{de}.

From \cref{eq:magtension} we can see that the mass scale of a magnetic monopole\footnote{
    A 't~Hooft\textendash Polyakov monopole \cite{agmon_lectures_2023}.
}
is $m\sim\Lambda/g^2$; its size is $\sim1/\Lambda$ \cite{huang_weak_2006,huang_weakgravity_2008}. If one assumes that the magnetic monopole is smaller than the universe, $L\geq1/\Lambda$, and that the minimally charged monopole is lighter than a Nariai \gls{bh}, $m\lesssim L/G$, we can set bounds on the \gls{de} density $\rho_V$, by using the size of pure \gls{ds} space with a positive cosmological constant ($L=\sqrt{3/8\pi G\rho_V}$) \cite{huang_weak_2006}:
\begin{align}
    L&=\sqrt{3/8\pi G\rho_V}\\
    &\geq1/\Lambda\sim1/mg^2\\
    &\gtrsim G/Lg^2\\
    \Rightarrow 3/8\pi G\rho_V&\gtrsim G/g^2\\
    \Rightarrow \rho_V&\lesssim g^2/G^2\sim g^2M_\textnormal{P}^4.
\end{align}
A small coupling would therefore explain a small cosmological constant. This makes the smallness of the cosmological constant more natural, but one has to explain the smallness of the coupling.

One possibility to realise a near-Minkowski universe is by starting with an unstable \gls{ds} universe, which decays through brane nucleation into a stable Minkowski solution. This idea is presented by \citet{bousso_quantization_2000}, and makes a flat Universe probable, which circumvents anthropic arguments. However, their realisation is in tension with the \gls{wgc} \cite{liu_cosmological_2023}. Another realisation is presented by \citet{kaloper_quantum-mechanical_2022}. In this realisation, the tension with the \gls{wgc} can be avoided \cite{liu_cosmological_2023}.

\gls{de} in the form of a dynamical scalar field is constraint by the scalar \gls{wgc}. \Cref{tab:wgc} gives an overview of various potentials and the constraints they face from the scalar \gls{wgc}.

Axions as \gls{ede} \cite{poulin_cosmological_2018} were promising candidates for lowering the Hubble tension \cite{poulin_early_2019} and have been studied in a string-theoretical context \cite{alexander_axion-dilaton_2019} before, yet it appears that the \gls{wgc} limits the axion decay constant \textit{in this context} to $f<0.008 M_\textnormal{P}$, which makes this form of \gls{ede} less likely from a theoretical standpoint \cite{rudelius_constraints_2023}. In a general axion context, we mentioned the possible loophole of having a very small instanton action, which would allow the axion decay constant to be large. This loophole is not available in the \gls{ede} context, as the axion field needs to be light \cite{rudelius_constraints_2023}.
\citet{cicoli_early_2023} do not share these conclusions: they find that axionic \gls{ede} with $f\simeq0.2$ can present a viable model which can be embedded in type IIB string theory. However, the \gls{wgc} makes fine-tuning of the model necessary, which, from our standpoint, speaks rather against such a model, even though it does not rule it out.

Axionic hilltop quintessence that is in agreement with the \gls{dsc}, violates the \gls{wgc} and is therefore in the swampland, unless there are heavy particles and supersymmetry that enable the axion to be slow-rolling by reducing the instanton action (\textit{reducing} meaning to be of $\order{1}$ instead of $S_\iota \sim\order{100}$, which is the value in the absence of heavy particles) \cite{ibe_quintessence_2019}. However, \citet{cicoli_quintessence-numerically_2022-1} find that a sub-Planckian axion decay constant is easier to realise near the maximum of the axion potential, where the supersymmetry breaking scale is extremely low, i.e. axionic quintessence is rather viable in a non-supersymmetric Minkowski vacuum. \citet{tada_quintessential_2024} derive axion-like quintessence by a reconstructing \gls{desi}-data for $\omega_0\omega_a$CDM. The potential satisfies the \gls{wgc} (as well as the \gls{dsc} and the \gls{dc}).

\citet{sadeghi_can_2023} put forward the idea that \glspl{bh} surrounded by quintessence or a cloud of strings might experience a repulsive force, indicated by the \gls{wgc}, which would contribute to the expansion of the Universe. \glspl{bh} as drivers of cosmic expansion have been studied previously \cite{farrah_observational_2023,parnovsky_can_2023,avelino_can_2023,mistele_comment_2023,lei_black_2024}. We'd like to remark that to explain the expansion of space, it is not enough to have repulsive objects that travel through space.

\paragraph{Dark Matter} 
If \gls{dm} is charged under a massless Abelian gauge field, i.e. there is a dark photon\footnote{
    See \cref{p:WGC_Photons} for further details.
}
\cite{long_dark_2019,bastero-gil_vector_2019,dror_parametric_2019,co_dark_2019,reece_searching_2009,fabbrichesi_dark_2021,fan_double-disk_2013,agrawal_relic_2020,graham_vector_2016}, important consequences for the \gls{uv} cutoff scale arise \cite{harlow_weak_2023,craig_weak_2019}:
Observations indicate that \gls{dm} is approximately collisionless, but collective dark plasma effects, which can lead to density fluctuations, are possible nevertheless \cite{heikinheimo_evidence_2015,ackerman_dark_2009}.
The frequency of the fluctuations in the plasma is 
\begin{equation}
    \omega_\text{plasma}=\sqrt{\frac{g^2\rho_\text{DM}}{m_\text{DM}^2}}\gtrsim\frac{\sqrt{\rho_\text{DM}}}{M_\textnormal{P}},
\end{equation}
with $\rho$ the energy density, and the inequality arises from applying the \gls{wgc} bound.
The timescale for those density fluctuations ($\sim1/\omega_\text{plasma}$) corresponds roughly to the timescale of galaxy cluster mergers. This means that galaxy mergers might provide evidence for the effects of the weak gauge forces between \gls{dm} particles that approximately saturate the \gls{wgc} bounds. The \gls{dm} distribution in galaxy cluster collisions acts as an indicator of the gauge coupling.

Interacting dark radiation in the form of dark gluons as part of a non-Abelian gauge group in the dark sector \cite{buen-abad_non-abelian_2015} bears a cutoff from the \gls{twgc}/\gls{slwgc} that potentially induces a tension between inflationary models and cosmological observations associated to interacting dark radiation \cite{harlow_weak_2023}.

\textit{Branon \gls{dm}} violates the \gls{slwgc} \cite{nam_implications_2023}:
Branons are fluctuations of our Universe's brane, which is embedded into a higher-dimensional bulk. The model is specified by the following set of equations:
\begin{align}\label{eq:branon}
    S&=S_\text{bulk}+S_\text{brane}\\
    S_\text{bulk}&=\frac{M_{\textnormal{P};5}^3}{2}\int\!\sqrt{-G}\left(R_5-2\Lambda_{\textnormal{cc};5}\right)\,\mathrm{d}^5X\\
    S_\text{brane}&=\int\!\sqrt{-\tilde{g}}\left(\mathcal{T}^4+\mathcal{L}_\textnormal{SM}+\dots\right)\,\mathrm{d}^4x\\
    \mathrm{d}s_\textnormal{bulk}^2&=G_{MN}\mathrm{d}X^M\mathrm{d}X^N\\
    &=g_{\mu\nu}\mathrm{d}x^\mu\mathrm{d}x^\nu-r^2\left(\mathrm{d}\theta+X_\mu\mathrm{d}x^\mu\right)^2\\
    \mathrm{d}s_\text{brane}^2&=\tilde{g}_{\mu\nu}\mathrm{d}x^\mu\mathrm{d}x^\nu\\
    \tilde{g}_{\mu\nu}&=G_{MN}\partial_\mu Y^M\partial_\nu Y^N\\
    &=g_{\mu\nu}-r_0^2\left(X_\mu+\partial_\mu Y^\theta\right)\left(X_\nu+\partial_\nu Y^\theta\right)\\
    Y^M&=\left(x^\mu\,,\,Y^\theta\left(x\right)\right),
\end{align}
with
$G$ ($\tilde{g}$) the determinant of the bulk (brane) metric,
$R_5$ the 5d Ricci scalar,
$\Lambda_{\textnormal{cc};5}$ the 5d cosmological constant,
$X_M$ ($x_\mu$) bulk (3-brane) coordinates,
$\mathcal{T}$ the brane tension,
$\mathcal{L}_\textnormal{SM}$ the \gls{sm} Lagrangian,
$g_{\mu\nu}$ the 4d tensor of the bulk metric,
$r$ the scalar component of the bulk metric,
$\theta$ an angle parametrising $S^1$,
$X_\mu$ the 4d vector component of the bulk metric,
$Y^M$ the static gauge,
$Y^\theta$ the branon that describes the fluctuations on the 3-brane along $S^1$, and
$r_0=\expval{r}$ the stabilised radion field.
Dimensional reduction of the bulk action yields an effective action in 4 dimensions:
\begin{equation}
    S_4=\int\sqrt{g}\left(\frac{M_{\textnormal{P};4}^2}{2}R_4-\frac{X_{\mu\nu}X^{\mu\nu}}{4\kappa^2}\right),
\end{equation}
with all 4d quantities, and
\begin{equation}
    \kappa=\frac{\sqrt{2}}{M_{\textnormal{P};4}r_0}
\end{equation}
the \gls{kk} gauge coupling.
The \gls{slwgc} implies that $\Lambda\lesssim\kappa^{1/3}M_{\textnormal{P};4}$, i.e. the \gls{eft} breaks down when the \gls{kk} field decouples, which puts this model of \gls{dm} in the swampland.

\paragraph{Hierarchy Problem}
Another interesting feature of the \gls{wgc} is the predicted upper bound on particle masses, which can explain why the \gls{ew} scale is so much lower than the Planck scale \cite{harlow_weak_2023,cheung_naturalness_2014,lust_scalar_2018,craig_weak_2019}:
The difference between the baryon number and the lepton number might be a gauge symmetry $B-L$.\footnote{
    Fermi explained the neutron decay with a four-fermion interaction Lagrangian of the form $L_\beta\sim G_F\Bar{n}pe^-\Bar{\nu}+\text{h.c.}$ \cite{reece_tasi_2023}.
    This beta decay Lagrangian is compatible with two global symmetries, the baryon number, and the lepton number \cite{reece_tasi_2023}.
    Discovering that the neutron is unstable, opened the question if the proton is unstable as well, which lead to the decay Lagrangian $L_\text{dec}=y_ppe^-\pi^0+\text{h.c.}$, which explicitly breaks the two global symmetries, but allows for the $B-L$ symmetry \cite{reece_tasi_2023}.
}
If neutrinos are the \gls{wgc} particles and acquire their mass through \gls{ew} symmetry breaking, the Dirac neutrino mass is then predicted to be $m_\nu=y_\nu h/\sqrt{2}<\sqrt{2}g_{B-L}M_\textnormal{P}$, with $y_\nu\sim\num{e-12}$ the \gls{sm} value and $h$ the Higgs scale.\footnote{
Invoking the more stringent \gls{twgc} and \gls{slwgc}, additional predictions arise \cite{harlow_weak_2023}: The expected towers of $B-L$-charged particles should start at $m\sim g_{B-L}M_\textnormal{P}\lesssim\order{\text{keV}}$, which would indicate the existence of billions of yet undetected particles that interact with ordinary matter below the TeV-scale (if this gauge symmetry is unbroken). This is not ruled out by data, yet surprising. In the case of an unbroken $B-L$-symmetry, the cutoff scale for the \gls{eft} is around $\SI{e10}{\giga\electronvolt}$.
}
Experimental bounds on the coupling $g_{B-L}\sim\num{e-28}$ predict then $h\lesssim\num{e-16}M_\textnormal{P}$, which is in agreement with the observed \gls{ew} hierarchy.\footnote{
\citet{abu-ajamieh_implications_2024} disagree: they use experimental constraints from the \gls{lep} and the \gls{wgc} to derive a lower bound of $g\gtrsim\num{8.7e-17}$; this means that an extra U(1) cannot solve the hierarchy problem.}

This is a very nice, experimentally confirmed prediction based on a swampland conjecture, unfortunately one with a big shortcoming: We just replaced the unexplained smallness of the \gls{ew} hierarchy with another unexplained small number, namely the smallness of the $B-L$ coupling $g_{B-L}\sim\num{e-28}$. This shift might be even more problematic, as it might hint at inconsistencies within the theory itself: A small coupling restores a global symmetry (see \cref{sec:nGSym})!
The issue might arise because of the attempt to relate the neutrino mass to the Higgs mass\,\textemdash\,which seems to be the obvious thing to do, but not a necessity: to relate the \gls{wgc} bound to the \gls{ew} scale, \textit{a} new Abelian gauge group needs to be introduced with charged matter that obtains some of its mass from the Higgs field \cite{craig_weak_2019}, but this has not necessarily to be done in the form of a neutrino.

\paragraph{Inflation}\label{p:WGC_Inflation}
The \gls{wgc} predicts a suppression of the scalar powerspectrum at scales $k\gtrsim\SI{100}{\per\mega\parsec}$ \cite{winkler_probing_2020}.

Large-field inflation\footnote{
    An interesting aspect of large-field inflation, i.e. inflationary models where the inflaton field traverses a trans-Planckian range, is that the potential is subject to all Planck suppressed corrections (because of the trans-Planckian field range), which means that large-field inflation is always sensitive to \gls{uv} physics \cite{brown_axionic_2016}.
    See the work by \citet{banks_possibility_2003} on the viability of trans-Planckian field ranges for inflation.
}
seems to be in tension with the \gls{wgc} \cite{heidenreich_weak_2015}, as the following examples will highlight. A reoccurring theme will be that the necessary trans-Planckian field ranges are in tension with the magnetic \gls{wgc}. The constraints can sometimes be circumvented by adding complexity to the model, e.g. in the form of additional fields\footnote{
    Models invoking additional fields are for example pole $\mathfrak{N}$-flation (where the inflaton field are $\mathfrak{N}$ open string moduli such as D3-branes) \cite{dias_pole_2019} or multi axion natural inflation (see \cref{p:WGC_axion}).
    If a model is compliant with the \gls{wgc}, it still might make predictions, e.g. an oscillating scalar powerspectrum \cite{wang_natural_2005,pahud_oscillations_2009,flauger_oscillations_2010,kobayashi_running_2011,easther_planck_2014}, that are in tension with observations \cite{de_la_fuente_natural_2015}.
}
\cite{saraswat_weak_2017}.
We discuss various models of inflation in the following.

\subparagraph{Axionic inflation,} such as natural inflation, where the inflaton is an axion with a super-Planckian decay constant \cite{freese_natural_1990,kim_completing_2005}, is in strong tension with the \gls{wgc} \cite{blumenhagen_large_2018,kooner_warping_2016,brown_axionic_2016,harlow_weak_2023,arkani-hamed_string_2007,hebecker_large_2019,benakli_revisiting_2020,brown_fencing_2015,heidenreich_axion_2016,gonzalo_strong_2019,palti_natural_2015,rompineve_sorbello_imprints_2017},
so is hyperinflation in its simplest form with two-fields, as the axion decay constant gets exponentially large during hyperinflation \cite{bjorkmo_hyperinflation_2019}.
This can be remedied by generalising the models \cite{bjorkmo_hyperinflation_2019},
e.g. by
\begin{itemize}
    \item coupling the kinetic term of the inflaton to the Einstein tensor \cite{germani_uv-protected_2011},
    \item additional symmetries in the Higgs sector \cite{ibanez_inflaton_2014},
    \item localising 3-forms that give rise to axions near strong coupling limits in F-theory \cite{grimm_axion_2014},
    \item letting the D-branes move around warped throats \cite{kooner_warping_2016,kenton_d-brane_2015},
    \item allowing the metric to be dynamic, and the string solution to be non-static (which leads to topological inflation\footnote{
        Models of \textit{topological inflation} with closed axion string loops and $nf>M_\textnormal{P}$, where the axion winding number $n$ is large, might violate the \textit{topological censorship theorem} by collapsing into a \gls{bh}, such that no observer could measure the field excursion \cite{dolan_transplanckian_2017,friedman_topological_1995,harlow_weak_2023}. %
        \citet{hebecker_what_2017} warn that a horizon could also arise in solutions with $f<M_\textnormal{P}$, which means that either no string solution would exist, or that horizons are not an argument against topological inflation. %
    } with trans-Planckian axion decay constants) \cite{hebecker_what_2017},
    \item replacing the axionic cosine potentials in natural inflation with modular functions\footnote{
        Modulated natural inflation introduces spectral distortion of the \gls{cmb} (wiggles in the tensor mode) through the higher harmonics, which in turn causes the scalar power spectrum to deviate from a power-law form and helps the model to be in agreement with the \gls{wgc} if multiple axion fields are present \cite{kappl_modulated_2016}.
    },
    \item considering subleading, non-perturbative corrections \cite{parameswaran_subleading_2016},
    \item introducing multiple axion fields \cite{baumann_inflation_2015,shiu_large_2015,shiu_widening_2015,choi_natural_2014,long_aligned_2014,dimopoulos_n-flation_2008,kim_completing_2005,bachlechner_planckian_2016,de_la_fuente_natural_2015,rudelius_constraints_2015,hebecker_winding_2015,higaki_axion_2015,bachlechner_new_2015,berg_dantes_2010,bachlechner_planckian_2015,cicoli_n-flation_2014,kappl_aligned_2014,ben-dayan_hierarchical_2014,ben-dayan_towards_2015,ruehle_natural_2015,grimm_axion_2008,gao_combining_2014,kappl_natural_2015,ali_natural_2015,blumenhagen_string_2016,kappl_modulated_2016,choi_aligned_2016,angus_aligned_2021,nath_enhancement_2019,reig_stochastic_2021,rompineve_sorbello_imprints_2017}\footnote{
        \citet{heidenreich_instanton_2020} study some of those models regarding their compatibility with the \gls{twgc}/\gls{slwgc} and find
        that isotropic $\mathfrak{N}$-flation \cite{dimopoulos_n-flation_2008} cannot produce an effective trans-Planckian axion decay constant, unless a modification, coined \textit{stretched} $\mathfrak{N}$-flation is invoked;
        that random matrix $\mathfrak{N}$-flation seems to be in tension with the \gls{slwgc};
        and that \gls{knp} alignment seems to be missing a \gls{uv}-completion and needs to invoke two (or more) instanton species to satisfy the \gls{slwgc} bounds \textit{and} create large-field inflation.
    },
    \item disentangling the axion field range from the decay constant,\footnote{
        Studying domain walls ($p=d-1$-forms), \citet{hebecker_axion_2016} discover that for inflationary axion models with the Hubble constant much smaller than the cutoff ($H\ll\Lambda$), large field displacements are actually allowed up to $\phi\lesssim m^{-2/3}f^{1/3}M_\textnormal{P}^{4/3}$, but the decay constant $f$ cannot be small. This bound is derived following the idea that to realise slow-rolling, the tension of a membrane is decreasing, which causes the height of the axionic oscillations to decrease, i.e. slow-roll is a continuous cosmic bubble nucleation, which is constrained by the magnetic \gls{wgc} by the bound of $\Lambda^3\sim mfM_\textnormal{P}$, but unconstrained by the electric \gls{wgc}.
    }
    as in axion monodromy\footnote{
        A broken discrete shift symmetry, e.g. by introducing non-perturbative effects such as branes that add a non-periodic part (which dominates at large field displacements) on top of the periodic part (which gets suppressed at large volume) \cite{guidetti_axionic_2023}.
        }
    inflation \cite{blumenhagen_towards_2014,blumenhagen_large_2018,palti_natural_2015,silverstein_monodromy_2008,kaloper_ignoble_2011,mcallister_gravity_2010,mcallister_powers_2014,hebecker_d7-brane_2014,marchesano_f-term_2014,franco_axion_2015,ibanez_relaxion_2016,hebecker_axion_2016,blumenhagen_challenge_2015,hebecker_tuning_2015,escobar_d6-branes_2016,escobar_large_2016,palti_towards_2014,hebecker_towards_2016,blumenhagen_flux-scaling_2016,retolaza_bifid_2015,blumenhagen_towards_2015,westphal_string_2015,dolan_transplanckian_2017,blumenhagen_swampland_2017,rompineve_sorbello_imprints_2017}, or by
    \item clockwork \cite{kaplan_clockwork_2016, choi_realizing_2016,choi_natural_2014} mechanisms.\footnote{
        Clockwork mechanisms can satisfy the conjectures, but suffer from statistical improbability and need to invoke two (or more) instanton species to satisfy the \gls{slwgc} bounds \textit{and} create large-field inflation \cite{heidenreich_instanton_2020}.
    }
\end{itemize}
See the work by \citet{heidenreich_weak_2015,junghans_large-field_2016,brown_fencing_2015,rudelius_constraints_2015} for more cautious notes. For instance, subleading instanton effects should give rise to resonant non-Gaussianity, which would be detectable in the \gls{cmb} or in \glspl{lss} \cite{rudelius_constraints_2015}.

The following example of axion monodromy is a large-field inflation model compatible with the \gls{wgc} \cite{rompineve_sorbello_imprints_2017}:
The Lagrangian and potential of the model are given by
\begin{align}
    L&=\left(\partial\phi\right)^2-V\left(\phi/f\right)\\
    V\left(\phi\right)&=\frac{m^2\phi^2}{2}+\beta(\phi)\cos\left(\frac{\phi}{f}\right),
\end{align}
with $\phi$ an axion that couples to instantons,
$f$ the axion decay constant,
$\beta$ a parameter.
Slow-roll inflation starts in a regime of large $\phi$, where the quadratic term dominates. 
The field rolls into the regime where the wiggles from the cosine-terms dominate: there, domain walls separate local minima, which the field tunnels through with a non-vanishing probability.\footnote{
    The wiggling of the field in the late phase of inflation causes \glspl{gw} that are a fingerprint of this model that might become observable with the upcoming space-borne \gls{gw} detectors \cite{rompineve_sorbello_imprints_2017}.
} 
This gives a discrete set of $n\in\mathbb{Z}$ vacua of energy
\begin{equation}
    E\simeq\frac{m^2\phi^2_\textnormal{min}}{2}\simeq\frac{m^2n^2\left(2\pi f\right)^2}{2}\simeq\frac{n^2q^2g^2}{2}.
\end{equation}
For $p=3$ domain walls in 4 dimensions, the \gls{wgc} is $\mathcal{T}\leq gqM_\textnormal{P}$.
The tension can also be expressed as $\mathcal{T}=\sqrt{\beta}f$.
To satisfy the electric \gls{wgc}, the tension has to be small. A small tension indicates that the parameter $\beta$ is small, i.e. the wiggles from the cosine term are small, which makes inflation last longer. This means a long-lasting slow-roll phase of inflation is preferred by the electric \gls{wgc}.
Since there are no $\left(-2\right)$-branes, we can to obtain the magnetic \gls{wgc} for domain walls
\begin{equation}
    \Lambda\lesssim\left(qg\right)^{1/3}M_\textnormal{P}^{1/3}\simeq\left(2\pi m f\right)^{1/3}M_\textnormal{P}^{1/3}
\end{equation}
by analytic continuation of \cref{eq:mwgc_p-form}.
During inflation, the Hubble scale is below the cutoff scale. This helps us to constrain the allowed field range during inflation:
\begin{align}
    H&=\sqrt{\frac{V}{3M_\textnormal{P}^2}}\\
    &=\frac{m\phi}{\sqrt{6}M_\textnormal{P}}\\
    &\lesssim\Lambda\\
    \Rightarrow \frac{\phi}{M_\textnormal{P}}&\lesssim\left(\frac{M_\textnormal{P}}{m}\right)^{2/3}\left(\frac{2\pi f}{M_\textnormal{P}}\right)^{1/3},
\end{align}
which rules out parametrically large field ranges, but leaves a comfortable cushion of $\phi/M_\textnormal{P}\lesssim\num{e3}$, i.e. this form of axion monodromy inflation is an example of large-field inflation that is compatible with the \gls{wgc}.\footnote{
    An unresolved issue with this model is that instantons will couple to the axions, which does not leave a flat direction in the potential to inflate \cite{rompineve_sorbello_imprints_2017}. See the work by \citet{hebecker_winding_2015} for a possible solution to this.}

\subparagraph{Extranatural Inflation\footnote{
    \unboldmath{\noindent
    Natural inflation needs trans-Planckian field excursion, in order to explain the 50 to 60 $e$-folds together with the observed bounds on the spectral tilt and the tensor-to-scalar ratio \cite{rompineve_sorbello_imprints_2017,van_beest_lectures_2022}. Extranatural inflation sacrifices the gauge coupling, which can be very tiny ($\sim\order{\num{e-3}}$), or the locality of Wilson lines \cite{furuuchi_weak_2018,furuuchi_u1_b-l_2014}.}
}}\!%
\cite{arkani-hamed_extranatural_2003} is ruled out by the \gls{wgc} \cite{saraswat_weak_2017}: The Wilson line\footnote{
    Since a Wilson line is a non-local gauge operator, this non-local interaction is a manifestation of \gls{uv}/\gls{ir} mixing in the theory \cite{lust_scalar_2018}.
}
from a U(1) gauge field around a circular extra-dimension $S^1$ of radius $r$ gives rise to the inflaton field $\phi$. Since the U(1) is gauge invariant and compact, the potential of the inflaton is of the form $\cos\left(n\phi/f\right)$ with $n\in\mathbb{Z}$ and $f=1/2\pi r g$. In such a setting, inflation lasts for $N_e\sim f/M_\textnormal{P}$ $e$-foldings, which requires the gauge coupling $g$ to be small enough to explain enough $e$-foldings. However, the magnetic \gls{wgc} presents a cutoff $\Lambda<gM_\textnormal{P}$, and the size of the compactification manifold has to be larger than $1/\Lambda$. This leads to the following constraint:
\begin{equation}
    1\lesssim2\pi r\Lambda\lesssim2\pi r gM_\textnormal{P}=M_\textnormal{P}/f,
\end{equation}
which rules out single field extranatural inflation, as it requires a super-Planckian decay constant \cite{de_la_fuente_natural_2015,heidenreich_weak_2015,huang_constraints_2007}. We see that\,\textemdash\,while natural inflation is ruled out by the electric \gls{wgc}\,\textemdash\,extranatural inflation is constrained by the magnetic \gls{wgc} \cite{rompineve_sorbello_imprints_2017}.

\subparagraph{Chromonatural inflation}\label{p:chromonatural_inflation}
with
\begin{equation}
    L=-\mu^4\left[1+\cos(\chi/f)\right]-\frac{\lambda}{8f}\chi F^a_{\mu\nu}\Tilde{F}^{a\mu\nu},
\end{equation}
$\mu\lesssim\Lambda_\text{QG}$ an energy scale,
$\chi$ an axion,
$f$ the axion decay constant,
$\lambda$ the coupling, which could be quantised ($\lambda=n\frac{g^2}{4\pi^2}$, $n\in\mathbb{Z}$),
and $F^a_{\mu\nu}$ the field strength tensor,
is in tension with the \gls{slwgc} combined with observational constraints: the number of $e$-folds needs to be high enough to explain observations and the bounds on the scalar power spectrum amplitude and spectral tilt need to be respected, which requires the coupling to be small ($g\sim\num{e-6}$), but $\mu$ to be large, which spoils the \gls{slwgc}-requirement $\mu\lesssim\Lambda_\text{QG}$ \cite{heidenreich_weak_2018,dimastrogiovanni_low-energy_2013,adshead_chromo-natural_2012,adshead_gauge_2013,adshead_perturbations_2013,agrawal_clockwork_2018}.
Claimed to be compatible with the \gls{wgc} is \textit{spectator chromonatural inflation}, where the axion field takes a spectator role besides an inflaton field \cite{holland_chromonatural_2020,bagherian_inflated_2023}: Both fields have a sub-Planckian decay constant, where it is important to note that for the axionic gauge field, this is not a decay constant in the usual understanding, as the field is not normalised and therefore not coupled in the usual sense. The coupling is instantaneous, as the coupling is instantly activated when the inflaton field reaches a certain threshold of stabilisation.

\subparagraph{Chaotic inflation}\label{p:WGC_chaotic-inflation}
\cite{linde_chaotic_1983}
requires trans-Planckian field excursion to inflate enough \cite{kaloper_natural_2014,madsen_chaotic_1988}, which can be realised e.g. by identifying a complex inflaton field with a Wilson line \cite{ibanez_higgs-otic_2015} or membrane emission \cite{kaloper_natural_2009}, but to be compatible with the scalar \gls{wgc}, the potentials have to be asymptotically linear \cite{gonzalo_strong_2019}.
(Single-field and assisted \cite{liddle_assisted_1998}) eternal chaotic inflation appears to be incompatible with the \gls{wgc} \cite{huang_eternal_2007,huang_weak_2007,huang_weak_2008}.\footnote{
    \citet{huang_eternal_2007} present a concrete example with the potential $V(\phi_i)=\frac{1}{2}m^2\phi_i^2+\frac{1}{4}\lambda\phi_i^4$:
    If the field is dominated by the quartic contribution, $H\sim\sqrt{\mathfrak{N}\lambda}\phi^2/M_\textnormal{P}$, and the \gls{wgc}\,\textemdash\,acting as an \gls{ir} cutoff of the theory\,\textemdash\,puts a bound of $\mathfrak{N}\lambda\phi^4/M_\textnormal{P}^2\leq H^2\leq \Lambda^2\sim\lambda M_\textnormal{P}^2$, which yields $\phi<M_\textnormal{P}/\mathfrak{N}^{1/4}$. As the number of fields $\mathfrak{N}\geq1$, this is in direct contradiction with our initial finding that for chaotic inflation, trans-Planckian field excursion is necessary to yield enough $e$-foldings while maintaining slow-rolling.
    If the field is dominated by the mass term, $H=\sqrt{\mathfrak{N}}m\phi/M_\textnormal{P}$; analogue reasoning yields $\phi<M_\textnormal{P}^2/\mathfrak{N}m=\frac{M_\textnormal{P}}{\mathfrak{N}}\frac{M_\textnormal{P}}{m}$. For eternal inflation to happen, $\mathfrak{N}<M_\textnormal{P}/m$. \citet{huang_eternal_2007} claim that this leads to a contradiction with the initial assumption that the field traverses a trans-Planckian range, as $m<M_\textnormal{P}$. We don't see this contradiction, as this just yields $\phi<\frac{M_\textnormal{P}}{\mathfrak{N}}\frac{M_\textnormal{P}}{m}>\frac{M_\textnormal{P}}{M_\textnormal{P}/m}\frac{M_\textnormal{P}}{m}=M_\textnormal{P}$.
    The same potential is studied by \citet{huang_weak_2008}, but their example might be incompatible with the \gls{dsc} and would not hold for large $\mathfrak{N}$.
}
Early work by \citet{huang_weak_2007} proposes a lower cutoff bound for the inflaton field of $\Lambda\leq H$, with $H$ the Hubble scale, which would not only rule out trans-Planckian field excursion, but even trans-Hubbleian. The bound is derived for polynomial potentials, based on the idea that gravitational contributions to the running of scalar coupling should be less than the contribution of scalar fields themselves, to comply with the \gls{wgc} (\citet{huang_gravitational_2007} elaborates further on this idea, and \citet{huang_constraints_2007} makes some notes). 

\subparagraph{Higgs inflation and Starobinsky inflation} \cite{starobinsky_new_1980} of the form $V(\phi)=\left(1-\exp(-\sqrt{2/3}\phi/M_\textnormal{P})\right)^2$ are inconsistent with the scalar \gls{wgc} (see \cref{p:SWGC}) in their most simple form\,\textemdash\,perturbations need to be included \cite{gonzalo_strong_2019}. The inconsistency of Starobinsky inflation without the inclusion of perturbations comes from the observation that the gravitational scattering amplitudes are larger than the self-scattering amplitudes of the inflaton at large background field values \cite{dudas_testing_2023,benakli_revisiting_2020,gonzalo_strong_2019}. To realise Starobinsky inflation, two chiral fields are necessary \cite{ellis_starobinsky-like_2013}, of which one needs to be stabilised \cite{dudas_testing_2023}. This, however, does not guarantee that the \gls{wgc} (or other swampland conjectures) are satisfied. Depending on the concrete realisation of the potential, there might be a tension between the model and the \gls{wgc} \cite{dudas_testing_2023}.

\citet{liu_higgs_2022} presents the constraints coming from the (strong) scalar \gls{wgc} (\cref{p:SWGC}) for Higgs inflation, Higgs\textendash Dilaton inflation, and Palatini Higgs inflation.\footnote{
    The equations are lengthy, and are not used by \citet{liu_higgs_2022} to compare with observational data. Therefore, we won't reproduce the results here.
    }

\subparagraph{Inverse monomial inflation} is a single field model on a brane with a potential of the form $V(\phi)=\mathfrak{c}^{4+\mathfrak{p}}/\phi^\mathfrak{p}$ that satisfies the (strong) scalar \gls{wgc} (as well as the combined \gls{dsc}) \cite{gashti_exploring_2024}. However, it violates the refined \gls{dsc} and potentially the \gls{dc} \cite{gashti_exploring_2024}.

\subparagraph{Running Vacuum Model (Vacuumon)}
\citet{mavromatos_string-inspired_2020,basilakos_quantum_2020,basilakos_gravitational_2020,basilakos_scalar_2019,basilakos_starobinsky-like_2016} present this model, which depicts the vacuum as a fluid with a scalar field description:\footnote{
    The vacuumon depicts a running vacuum solution and uses a scalar field to represent this. However, the scalar field does not undergo quantum fluctuations and does not generate cosmological perturbations in the standard way of an inflaton field.
}
The scalar field itself is not slow-rolling during inflation, yet the model allows for an axion field that is slow-rolling. There is no singularity at the beginning of the Universe. The field exits inflation gracefully and enters a radiation-dominated epoch without reheating. The model satisfies the strong scalar \gls{wgc} (\cref{eq:strongswgc}) as well as the \gls{dc} and the \gls{dsc}.
The following equations define the model:
\begin{AmSalign}
    \rho&=\frac{\dot{\phi}^2}{2}+V(\phi)&\sim\mathfrak{c}_0+\mathfrak{c}_1H^2+\mathfrak{c}_2H^4\\
    p&=\frac{\dot{\phi}^2}{2}-V(\phi)\\
    \dot{\phi}^2&=-2\dot{H}M_\textnormal{P}^2\\
    V&=3H^2M_\textnormal{P}^2\left(1+\frac{a}{6H^2}\frac{\mathrm{d}H^2}{\mathrm{d}a}\right)
\end{AmSalign}
where we indicated in the first line that the density contains terms proportional to $H^2$ as well as proportional to $H^4$ with undefined constants $\mathfrak{c}_i$. Due to the  $H^4$-term, the model shows Starobinsky inflation\textendash like behaviour at early times. At late times, the vacuumon takes the role of dynamic \gls{de}.
To test the model for compatibility with the swampland conjectures, the potential can be expressed in terms of the vacuumon $\phi$. \citet{mavromatos_string-inspired_2020} compare two cases: (a) a generic running vacuum model with relativistic matter, and (b) a string-inspired running vacuum model with stiff gravitational axions.
\begin{align}
    (a)&
    \begin{cases}
        V(\phi)&=V_0\frac{2+\cosh^2\phi}{\cosh^4\phi}\\
        V_0&=\frac{H_\textnormal{I}^2}{\lambda}\\
        \phi(a)&=\sinh^{-1}\left(\sqrt{\beta}a^2\right)
    \end{cases}\\
    (b)&\begin{cases}
        V(\phi)&=V_0\frac{\frac{2}{3}+\cosh^2\phi}{\cosh^4\phi}\\
        V_0&=\frac{9H_\textnormal{I}^2}{\lambda}\\
        \phi(a)&=\sqrt{\frac{2}{3}}\sinh^{-1}\left(\sqrt{\beta}a^3\right)
    \end{cases}
\end{align}
with $\lambda$ a parameter that defines the scale of inflation and $\beta$ a parameter that defines how quickly the vacuumon transitions between the different phases. These can then be tested for consistency with the swampland conjectures, which \citet{mavromatos_string-inspired_2020} confirm.

\paragraph{Neutrinos:}
The scalar \gls{wgc} requires normal mass hierarchy for neutrinos \cite{gonzalo_strong_2019}.
Furthermore, the lightest neutrino is Dirac with $m_{\nu_1}\lesssim\Lambda_4^{1/4}$ \cite{gonzalo_strong_2019,ibanez_constraining_2017}.\footnote{
    See \cref{p:nnSUSYC_neutrinos,p:AdSDC_particles} for other conjectures that have a preference for Dirac neutrinos.
    In the context of the \gls{wgc}, the mass bound is explained by \citet{lust_scalar_2018} using their findings about scalar fields: The tower of bound states of the lightest particle is protected from decaying. In the \gls{sm} these are neutrinos. Over long distances, neutrinos feel only gravity and can form a bound gravitational state. However, if there is a cosmological constant, which acts as a repulsive force, the force acting on two neutrinos is modified:
$
    F=m_\nu\left(-m_\nu/r^2+\Lambda r/3\right).
$
To avoid bound states, $\Lambda>m_\nu^4$ is required.
}
It's either a lower bound on the cosmological constant $\Lambda_\textnormal{cc}$ or an upper bound on the neutrino mass (of the lightest neutrino) $m_{\nu_1}$. The latter indicates the scale of \gls{ew} symmetry breaking and gives therefore the mass of the Higgs boson \cite{lust_scalar_2018}.

Adopting a ruled-out form of the \gls{wgc}, namely that the \gls{wgc} particle has to be the lightest particle in the theory, \citet{das_neutrino_2020} derive a lower bound on the electric charge of neutrinos of $q_\nu\gtrsim\num{e-28}\frac{m_\nu q_e}{\SI{0.1}{\electronvolt}}$.
Since the lightest particle does not need to be the \gls{wgc}-particle, nor does the \gls{wgc} only apply to the electric charge, this bound does not hold. Neutrinos are likely electrically neutral \cite{abu-ajamieh_implications_2024}.

\paragraph{Photons}\label{p:WGC_Photons}
Observations put stringent bounds on the mass of (\gls{sm}) photons \cite{goldhaber_photon_2010,wu_constraints_2016,bonetti_photon_2016,bonetti_frb_2017}.
The \gls{wgc} indicates that \gls{sm} photons are massless \cite{reece_photon_2019}.
If the photon gained a mass through a mechanism (other than Higgsing), the string scale would be $M_\textnormal{s}\lesssim2\pi\sqrt{m_\gamma M_\textnormal{P}/e}$, which would mean that stringy effects should be visible at the \gls{lhc}, unless the elementary charge is actually not the electron charge $e$, but a small quantity, which would change the denominator to $e/N$ \cite{draper_snowmass_2022,craig_rescuing_2018}.
\citet{craig_rescuing_2018} propose that the arguments used by \citet{reece_photon_2019} in favour of massless photons could be parametrically violated in the \gls{ir}, allowing massive photons to be part of \glspl{eft}.

Beside a massless \gls{sm} photon, there could be an additional massive dark photon.
A dark photon is a (ultra-light) \gls{dm} candidate  that gains mass through Higgsing or by eating an axion (Stückelberg photon) \cite{goodsell_naturally_2009,cline_status_2024}, and is studied for a range of phenomenological implications:
    dark massive photons could alter the \gls{cmb} \cite{pospelov_new_2018}, 
    heat the early Universe \cite{kovetz_bounds_2019},
    and act as an extra radio background \cite{montero_swampland_2022,garcia_effective_2020,caputo_dark_2020}.
Dark photons could be produced by quantum fluctuations during inflation in the form of massive vector bosons with a mass of $m_\gamma\approx\SI{e-5}{\electronvolt}\left(\SI{e14}{\giga\electronvolt}/H_\textnormal{I}\right)^4$, with $H_\textnormal{I}$ the Hubble parameter during inflation \cite{graham_vector_2016,craig_rescuing_2018}. The proposed mechanism forbids the presence of scalar fields with $m_\phi<H_\textnormal{I}$, as otherwise isocurvature perturbations would occur that are incompatible with observations \cite{craig_rescuing_2018,graham_vector_2016,fox_probing_2004}.
This constraint together with the \gls{twgc} translates into a mass bound for dark photons,
\begin{align}
    H_\textnormal{I}\lesssim\Lambda\lesssim g^{1/3}M_\textnormal{P}&\sim\left(\frac{m_\gamma}{H_\textnormal{I}}\right)^{1/3}M_\textnormal{P}\\
    m_\gamma&\gtrsim\SI{10}{\electronvolt}
\end{align}
that can be lowered in a broken clockwork construction,
\begin{align}
    H_\textnormal{I}&\lesssim\left(\frac{g^nm_\gamma}{H_\textnormal{I}}\right)^{1/3}M_\textnormal{P}\\
    m_\gamma&\gtrsim\frac{\SI{10}{\electronvolt}}{q^{n/2}},
\end{align}
which would allow for smaller dark photons masses, depending on the parameters $q$ and $n$ ($g\sim q^nm_\gamma/H_\textnormal{I}$) \cite{craig_rescuing_2018}.
Another lower mass bound on dark photons is derived from positivity bounds on the scattering amplitude: a dark photon must have a mass larger than its U(1) gauge coupling, as otherwise the \gls{eft} breaks down in already observable energy regimes \cite{noumi_phenomenological_2023}. Furthermore, there is a lower bound on the cross-section for non-gravitational interactions between dark sector particles $X$ and \gls{sm} photons $\gamma$ of $\sigma_{\gamma-X}^\textnormal{ng}\gtrsim\order{\left(qg\Lambda/M_\textnormal{P}m_e\right)^2}$ \cite{noumi_phenomenological_2023}.

\paragraph{Quarks}
Under the caveat that holography is applied in a regime of strong gravity, \citet{mcinnes_density_2022} presents a peculiar finding, which might indicate that quark matter \textit{cannot} exist, not even in the core of neutron stars, as he derives a lower bound on density:
\begin{equation}
    \rho_\text{enth}(T=0)>\frac{3\sqrt{3}}{4\left(32\pi N_\text{c}^4\right)^{1/6}k_5l_5^5}\left(\frac{m}{q}\right)_\text{min},
\end{equation}
with $\rho_\text{enth}$ the enthalpy density,
$N_\text{c}$ the number of colors characterising the boundary field,
$k_5$ controlling the relation between the enthalpy density and the quark chemical potential of cold quark matter, and
$l_5$ the AdS$_5$ length scale.
We won't investigate this idea here any further, but leave it here as a source of inspiration for future work.

\subsubsection{General Remarks}\label{sss:WGC_remarks}
The \gls{wgc} does not only demand that gravity is the \textit{weakest} force \cite{brennan_string_2018,palti_weak_2017}, it is also a \textit{weak} conjecture itself, as there might be a cutoff scale below $\Lambda$, after which the theory changes in character such that the \gls{wgc} does not apply \cite{palti_swampland_2019}. Nevertheless, the \gls{wgc} is a \enquote{prime example of a swampland conjecture \textelp{} [with] a broad consensus that it is actually a correct statement} \cite{cicoli_string_2023}. In the following, we would like to gain some intuition by addressing various questions.

\paragraph{Why should we not believe the \gls{wgc}?}
There are no proofs.\footnote{
    This statement has been repeated in the literature over and over again for the past few years. However, more recently \citet{heidenreich_proving_2024} proofed the \gls{slwgc} (and therefore also the \gls{twgc} or any milder form, such as the general \gls{wgc}) for bosonic strings in $d\geq6$ dimensions. This is still no fully general proof, but one step towards it.
    }
The initial motivation for the \gls{wgc} was the assumption that ((near-)extremal) \glspl{bh} should be able to decay. There is no observation yet that proofs that this is actually the case. Furthermore, in \gls{adscft}\footnote{
    Even though the \gls{wgc} is studied in \gls{cft} context, it is unclear if a counterexample from \gls{cft} ruled out the \gls{wgc} \cite{montero_weak_2016}: On the one hand, weakly coupled gauge fields and \glspl{bh} are present in three spacetime dimensions, on the other hand, gravity works differently in three spacetime dimensions than in higher-dimensional spacetimes, such that the \gls{wgc} might not be applicable to \glspl{cft}, or maybe only to a certain subclass of \glspl{cft}.%
}
this is a statement about \glspl{bh} with a radius smaller than the \gls{ads} radius, which are not well understood \cite{harlow_weak_2023}. There are entropic arguments that also large near-extremal \glspl{bh} need to be unstable \cite{montero_holographic_2019}, but this could be satisfied by un-charged \gls{bh} decays and is therefore no proof of the \gls{wgc} either. However, for large \glspl{bh}, a charged decay channel is possible, which would support the \gls{wgc}.

Another argument against the \gls{wgc} is the absence of magnetic monopoles: The minimally charged magnetic monopole in the theory should have a mass $m_\text{m}\sim\Lambda/g^2$ and a size of $r_\text{m}\sim1/\Lambda$ \cite{arkani-hamed_string_2007}. \glspl{bh} are classical concepts, and therefore the charges should be macroscopic and the minimally charged magnetic monopole should not be a \gls{bh}, i.e. $1\gtrsim m_\text{m}/M_\textnormal{P}^2r_\text{m}$, which yields the magnetic \gls{wgc} \cite{arkani-hamed_string_2007}. But, we haven't observed magnetic monopoles yet, maybe because they actually are \glspl{bh} and the \gls{wgc} does not hold?

Despite the strong evidence from string theory, the \gls{wgc} might simply suffer from a lamppost effect: most known examples from string theory involve either weak coupling or \gls{bps} states\,\textemdash\,no non-supersymmetric example of strong coupling is known \cite{harlow_weak_2023}.

\paragraph{Which observations would rule out the \gls{wgc}?}
The detection of a super-light $B-L$ gauge boson in neutron\textendash antineutron scattering experiments \cite{addazi_super-light_2018}.

Furthermore, if we observed features that are predicted by inflationary models that were deemed incompatible with the \gls{wgc}, the \gls{wgc} would be challenged.

The discovery of a very light magnetic monopole would mean that either the cutoff of the \gls{eft} is very low or that the coupling is strong. Both would be surprising and might lead to incompatibilities with the \gls{sm} (but precise bounds would have to be checked).

If the strong correspondence between the \gls{wgc} and the \gls{ccc} holds, the observation of naked singularities would pose an issue for both conjectures.

\paragraph{What speaks against stable remnants?}\label{p:WGC_Remnants}
Stable \gls{bh} remnants correspond to global charges, which contradicts the no global symmetries conjecture (\cref{sec:nGSym}) as well as the cobordism conjecture (\cref{sec:cobordism}).
One of the issues with a large number of stable remnants is that they contribute to the mean energy density with $\rho(m,g)\exp(-m/T)$, with $T$ the temperature of the thermal bath, such that the integrated contribution diverges, which renders the \gls{eft} pathological \cite{montero_weak_2016}.\footnote{
    In other words, the energy density a uniformly accelerated observer observes would diverge due to the Unruh effect caused by the infinitely many \gls{bh} states we can build at arbitrarily low energy \cite{montero_chern-simons_2017}.
    }

The early Universe had a radiation dominated epoch between inflation and \gls{bbn}, with an \gls{eos} of $\omega\approx1/3$.
If \glspl{pbh} were formed during this time, their \gls{eos} was $\omega=0$.
In order not to spoil \gls{bbn}-predictions, most \glspl{pbh} must evaporate into massless degrees of freedom \cite{solomon_generalizing_2020}.
If they form stable remnants, i.e. evaporation stops at the extremal limit, the \gls{eos} of the universe does not reach $\omega=1/3$ \cite{solomon_generalizing_2020} or \glspl{pbh} overclose the Universe \cite{dai_constraints_2009}.

Another argument disfavouring stable remnants comes from the \gls{ceb}:
\textit{If} there is no charge quantisation nor mass quantisation, there is an infinite spectrum of degrees of freedom, which violates the \gls{ceb} \cite{banks_symmetries_2011}, i.e. the entropy is infinite in a finite volume \cite{dai_constraints_2009}.
We italicised the \enquote{if} because charge quantisation limits the number of degrees of freedom to a large but finite value. Furthermore, the number of stable remnants within a given finite volume would still be finite.

A large but finite number of remnants per se seems not to be a problem, so it is not undisputed that the \gls{wgc} even is a meaningful constraint, as only the vanishing coupling limit $g\rightarrow0$ with an infinite number of remnants poses an actual problem \cite{montero_holographic_2019}. Since in explicit examples the number of \gls{bh} remnants might be large but finite, the tower of states that emerges is finite and the states are labelled by distinct charges and are therefore in principle distinguishable \cite{cottrell_weak_2017,cheung_proof_2018}.
In particular, when charge quantisation is considered, the density of states remains finite \cite{saraswat_weak_2017}.
Once more, we'd like to stress that the swampland programme makes statements about the boundary of theory space, cosmology happens in the bulk and might circumvent certain conjectures. Nevertheless, the remnants argument can provide us with some bottom-up intuition about what can go wrong when an \gls{eft} violates the \gls{wgc} \cite{van_beest_lectures_2022}.

\paragraph{Is there no proof of the \gls{wgc}?}
There is an actual string-theoretical proof presented by \citet{heidenreich_proving_2024}. However, the proof only holds for bosonic string theories in six or more dimensions. In less than six dimensions, other objects such as D-branes or solitons might be part of the charge spectrum. Furthermore, string loop corrections might spoil the bound.
Besides this proof, considerations from \gls{bh} thermodynamics are used to derive the \gls{wgc} bound for specific systems:

For the case of a Kerr\textendash Newman \gls{bh}, described by a \gls{rn} metric, and a massive scalar field,
\citet{hod_proof_2017} starts from Bekenstein's generalised second law of thermodynamics ($S_\text{BH}=k_Bc^3A/4G\hbar$, $A$ the \gls{bh} surface area) \cite{hawking_particle_1975,bekenstein_black_1973,bekenstein_generalized_1974}
to derive a lower bound on the relaxation time of perturbed physical systems, the universal relaxation bound $Tt_\text{relax}\geq\hbar/\pi$, $T$ the temperature \cite{gruzinov_kerr_2007,hod_note_2007,hod_slow_2008,hod_universal_2007,mezerji_correlation_2022}, which, as he shows, implies the \gls{wgc}.

A similar approach is taken by \citet{fisher_semiclassical_2017}, who also start from the generalised second law of thermodynamics: They show that in theories where the \gls{wgc} is violated, \glspl{bh} can grow unbounded. Furthermore, they show that larger \glspl{bh} contain fewer microscopic states, as decreasing quantum corrections outcompete an increasing Bekenstein\textendash Hawking term. As there is no bound in size, very large \glspl{bh} will eventually reach negative entropy states, i.e. the \glspl{bh} itself contains less than 1 microscopic state. It is expected that unitarity forbids this from happening. However, unitarity of the S-matrix alone is insufficient to assure that the \gls{wgc} holds \cite{loges_thermodynamics_2020}.\footnote{
    Large \gls{rn} \glspl{bh} are also violating Lloyd's \cite{lloyd_ultimate_2000} quantum information theory bound on computational speed \cite{cottrell_complexity_2018}.
    }

A result in agreement with the \gls{wgc}, but stronger in nature, is derived by \citet{yu_cosmic_2018} when studying the second law of \gls{bh} thermodynamics: to protect cosmic censorship, a lower bound on the mass-to-charge ratio exists for \textit{all} massive charged particles.

Assuming that \gls{bh} do not leave behind stable remnants, and applying the Pauli exclusion principle, \citet{baddis_swampland-statistics_2024} derive a statistical bound on the \gls{bh} charge-to-mass ratio that strongly resembles the \gls{wgc}:
\begin{equation}
    \frac{Q^2}{M^2}>\frac{gmT}{T_\textnormal{P}M_\textnormal{P}^3},
\end{equation}
with $T_\textnormal{P}$ the Planck-temperature. This bound assures that the formation of stable remnants is suppressed.

Not a proof, but observational evidence from astrophysics is presented by \citet{gasperini_new_2006}:
Ranging from planets to superclusters, spanning 25 orders of magnitude, a relation between the mass $m$ of an object and its angular momentum $J$ appears to hold quite well:
\begin{equation}
    J=\lambda m^\mathfrak{p},
\end{equation}
where $\mathfrak{p}=2$ can be set as a scale free choice if
\begin{equation}
    1/\lambda=\alpha c/G_\textnormal{N},
\end{equation}
with $\alpha$ the fine-structure constant, $c$ the speed of light, and $G_\textnormal{N}$ Newton's gravitational constant \cite{brosche_uber_1963,wesson_self-similarity_1979,wesson_clue_1981,carrasco_density_1982}.
A similar relation is found regarding the charge \cite{sirag_gravitational_1979,surdin_magnetic_1977,arge_magnetic_1995,baliunas_magnetic_1996,vasiliev_gyro-magnetic_2000}:
\begin{align}
    J&=qm\\
    1/q^2&=\alpha G_\textnormal{N}/c^2.
\end{align}
By using $\alpha=e^2/\hbar c$, $\hbar=c=1$ in natural units, where $1/\sqrt{G_\textnormal{N}}=M_\textnormal{P}$, we find that $\lambda$ respectively $q$ give rise to energy scales that are familiar from the \gls{wgc}:
\begin{align}
    \frac{1}{\sqrt{\lambda}}&\sim\Lambda&=&\sqrt{\alpha}M_\textnormal{P}&&&=&eM_\textnormal{P}\\
    q&\sim\Lambda&=&\frac{M_\textnormal{P}}{\sqrt{\alpha}}&=&\frac{M_\textnormal{P}}{e}&=&g_\textnormal{m}M_\textnormal{P}.
\end{align}
The \gls{wgc} follows as an empirical fit to astrophysical data.

\paragraph{What are the implications of the \gls{wgc} for \glspl{eft}?}\label{p:WGC_EFT}
The magnetic \gls{wgc} (\cref{eq:mwgc}) limits the range of validity of an \gls{eft}: If the coupling is weak, the cutoff scale of the \gls{wgc} is low. %
At higher energy scales, new physics appears \cite{heidenreich_weak_2018,harlow_weak_2023}. \citet{craig_weak_2019} give an example:
Assuming the U(1) described by the \gls{eft} is an \gls{ir} remnant from a Higgsed SU(2); the scale $\Lambda$ will then correspond to the mass scale of a massive W boson, which appears for energies above $\Lambda$ (and below the Planck scale).
\citet{abu-ajamieh_generalized_2024} apply the scalar \gls{wgc} (\cref{eq:swgc}) to the only known scalar field in our Universe, the Higgs field, and derive bounds on the scale where new physics is expected to appear:
\begin{itemize}
    \item If the lightest neutrino is Majorana, the lightest fermion is the electron, and $\Lambda\lesssim\SI{1.2e13}{\giga\electronvolt}$.
    \item If the lightest neutrino is Dirac with $m_{\nu_e}=\SI{1.1}{\electronvolt}$, $\Lambda\lesssim\SI{2.6e7}{\giga\electronvolt}$.
    \item If the \gls{wgc} particle is the Higgs itself, the scalar \gls{wgc} predicts that new physics arise before the Higgs field suffers the instability from the quartic coupling \cite{degrassi_higgs_2012,hook_probable_2015,elias-miro_higgs_2012} around $\sim\SI{e10}{\giga\electronvolt}$.
\end{itemize}

It is currently debated whether the \gls{wgc} is even applicable to \glspl{eft} or only to \gls{qg}. On the one hand,
\citet{furuuchi_weak_2018} shows that if one starts with an \gls{eft} that satisfies the \gls{wgc}, spontaneous gauge symmetry breaking cannot lead to a violation of the \gls{wgc}.
On the other hand,
\citet{saraswat_weak_2017} shows that Higgsing a \gls{wgc}-compatible theory of \gls{qg} can lead to a \gls{wgc}-violating \gls{eft}.\footnote{
    For example, magnetic monopoles are evaded when the magnetic confinement in the Higgs phase is taken into account \cite{saraswat_weak_2017}. In the Higgs phase, states of different charges can mix, which destroys the commutation of charge and mass, i.e. the \gls{wgc} is ill-defined \cite{cheung_infrared_2014}. The reason for the low-energy violation of the \gls{wgc} might lie in non-minimal field content in the high-energy regime \cite{saraswat_weak_2017}.
}
However, this \gls{eft} satisfies the much weaker bound $\Lambda_\text{EFT}\lesssim M_\textnormal{P}/\sqrt{-\log g}$.
This bound is motivated by entropy bounds \cite{saraswat_weak_2017} and ambiguous in Abelian theories, as the gauge coupling can be normalised\,\textemdash\,proposing that the coupling of the particle with the weakest charge shall be taken still leaves some wiggle room: the particle of smallest charge could be above the \gls{eft} cutoff \cite{furuuchi_weak_2018,saraswat_weak_2017}.\footnote{
    This loophole is avoided if one requires that the charge remains sub-Planckian \cite{lust_scalar_2018,harlow_wormholes_2016} or by demanding that the slightly stronger \textit{effective \gls{wgc}} holds: the particle that satisfies the \gls{wgc} has to be part of the \gls{eft} \cite{heidenreich_weak_2015}.
}
In non-Abelian theories, the gauge field itself is charged and can be used to normalise the gauge coupling constant \cite{furuuchi_weak_2018}.
This weaker bound, however, is also questioned, as a complete breakdown of the \gls{eft} description is expected when an infinite tower of particles appears, which happens at $\Lambda\sim g^{1/3}M_\textnormal{P}$ \cite{craig_weak_2019,heidenreich_evidence_2017,heidenreich_weak_2018}.

\paragraph{Can the \gls{wgc} be understood in terms of classical physics?}
\Cref{eq:ewgc} can be easily understood in a usual Einstein\textendash Maxwell theory with charged particles as the condition $F_\textnormal{EM}\geq F_\textnormal{G}$ for the particle species with the largest charge-to-mass ratio, with
\begin{align}
    F_\textnormal{G}&=\frac{m^2}{8\pi M_\textnormal{P}^2r^2}\\
    F_\textnormal{EM}&=\frac{(gq)^2}{4\pi r^2},
\end{align}
$r$ the distance between two such particles. If $F_\textnormal{EM}<F_\textnormal{G}$, two such particles would attract each other and form a bound state with $Q=2q$ but $M<2m$, such that $Q/M>q/m$, a violation of the assumption that the initial particle was the one with the highest charge-to-mass ratio \cite{palti_swampland_2019}. As charge and energy are both conserved, this is a stable state. Adding up more of those particles, since they are mutually attractive ($F_\textnormal{EM}<F_\textnormal{G}$), leads to stable bound states with arbitrarily large charge.

\paragraph{What are the implications for our understanding of spacetime?}
In strongly curved (higher dimensional) backgrounds, the \gls{wgc} can be regarded as a constraint on the local scalar curvature, expressed through the Ricci scalar $R$: 
\begin{equation}
    \sqrt{R}<gM_\textnormal{P},
\end{equation}
as the local curvature has to be smaller than the coupling in order to have an effectively four-dimensional theory \cite{klaewer_super-planckian_2017}.

In \cref{sec:nononSUSY}, we discuss the stability of \gls{ads}, Minkowski, and \gls{ds} vacua. \citet{liu_cosmological_2024} show that \gls{ds} vacua are unstable if a charged membrane is part of the \gls{eft}, and that Minkowski vacua are unstable if a \gls{wgc}-satisfying membrane exists. This is in analogy to the stability of \glspl{bh}, which decay when a \gls{wgc}-satisfying particle exists.

The generalisation of the \gls{wgc} to \gls{ads} is not straightforward \cite{nakayama_weak_2015}: First, in \gls{ads}, there is no no-hair theorem for \glspl{bh}, which would allow them to shed charges in processes that do not exist in Minkowski spacetime. Second, the Gedankenexperiment of the infinitely growing \gls{bh} by adding up more and more particles (when the \gls{wgc} is violated), doesn't apply to \gls{ads} because the \gls{bh} will at one point be larger than the \gls{ads} scale. And third, the idea of Hawking radiation is based on the finite temperature of a \gls{bh} from the Unruh effect as seen by an observer at infinity. \gls{ads} is not infinite, ergo, it's not a given that the critical \gls{wgc} bound can be reached. Since refinements to the \gls{wgc} in \gls{ads} tend to approach the conditions from Minkowski space for large \gls{ads} scales, and our Universe is observed to be relatively flat, i.e. the \gls{ads} scale is very large, we refer the interested reader to the literature on holography and \glspl{cft} listed in \cref{sss:WGC_Evidence} for further studies.

\paragraph{How are the \textnormal{magnetic} and the \textnormal{electric} \gls{wgc} related to each other?}\label{p:WGC_electric-magnetc}
\citet{palti_swampland_2019}, referring to \citet{arkani-hamed_string_2007}, shows how \cref{eq:ewgc} and \cref{eq:mwgc} are related: \Cref{eq:ewgc}, applied to the magnetic dual field, gives a constraint on the mass of a magnetic monopole:
\begin{equation}
    m_\text{mag}\leq g_\text{mag}M_\textnormal{P}\sim\frac{M_\textnormal{P}}{g_\text{el}},
\end{equation}
which can serve as a probe of the gauge theory \cite{arkani-hamed_string_2007}.

The monopole field has a cutoff at the scale $\Lambda$, as its field is linearly divergent. Therefore, the mass of the monopole is of the order of the field energy, i.e.
\begin{equation}
    m_\text{mag}\sim\frac{\Lambda}{g_\text{el}^2}.
\end{equation}
\Cref{eq:mwgc} follows right away.

\paragraph{Does the \gls{wgc} apply to scalar fields?}\label{p:SWGC}
The inclusion of massless scalar fields modifies the \gls{wgc} regarding two independent aspects \cite{lee_stringy_2019}: It changes the extremality condition for charged \glspl{bh}\footnote{
    The charge-to-mass ratio of an extremal \gls{bh} looks different, if the gauge coupling has a moduli dependence \cite{lee_stringy_2019}.
}, and Yukawa interactions between the \gls{wgc}-satisfying particles have to be included.\footnote{
    The \gls{bh} must be able to decay into particles that do not form gravitationally bound states, ergo, the gauge force must be strong enough to not only overcome the gravitational attraction, but also a possibly attractive Yukawa potential \cite{lee_stringy_2019}.
}
Interestingly, for a tower of asymptotically massless charged particles, as predicted by the \gls{dc}, the two modifications are actually equivalent \cite{lee_stringy_2019}.\footnote{
    For \gls{bps} states, this is assured by supersymmetry; for non-supersymmetric theories, it represents the finding that highly excited string states can approach extremal \glspl{bh} asymptotically \cite{lee_stringy_2019,sen_extremal_1995}.
}

There are various proposals for the exact quantitative changes of the \gls{wgc} bound. The most widely discussed ones are the \textit{repulsive force conjecture} (which we discuss in \cref{s:rfc}) and the \textit{strong scalar \gls{wgc}} (which is presented by \citet{gonzalo_strong_2019} and generalised by \citet{kusenko_fundamental_2020,gonzalo_pair_2020}).
To cover some preliminaries, we start with a weaker form, the \textit{scalar \gls{wgc}}.

\subparagraph{Scalar \gls{wgc}}
If the scalar field has a potential\footnote{
    The potential cannot have quartic terms \cite{gonzalo_pair_2020,gonzalo_strong_2019}.
    }
but is massless, there is a particle that satisfies the strict inequality
\begin{equation}\label{eq:swgc}
    \abs{m^\prime}^2=g^{ij}\left(\partial_{i}m\right)\left(\partial_{j}m\right)>m^2/M_\textnormal{P}^2,
\end{equation}
which is related to the \gls{dc}\footnote{
    If the field is canonically normalised, i.e. $g^{ij}=\delta^{ij}$, then $m_\phi\sim e^\phi$ \cite{grimm_infinite_2018}.
    See \cref{sec:distance} for further details. 
} \cite{palti_swampland_2019,palti_weak_2017}, but does not consider repulsive interactions \cite{benakli_revisiting_2020}.
\Cref{eq:swgc} says that, regarding massless scalar fields, $2\rightarrow2$\textendash scattering interactions are stronger than gravity \cite{shirai_is_2019}.\footnote{
    The scalar force $F_\text{scalar}=\abs{m^\prime}^2/4\pi r^2$ is always stronger than the gravitational force $F_\text{gravity}=m^2/8\pi M_\textnormal{P}^2r^2$ \cite{shirai_is_2019}.
    }
This form of the scalar \gls{wgc} seems to predict a fifth force, which is already excluded on observational grounds to $\abs{m^\prime}^2<\order{\num{e-5}}\left(m^2/M_\textnormal{P}^2\right)$ \cite{shirai_is_2019,esposito-farese_tests_2004,will_confrontation_2006,bertotti_test_2003}.\footnote{
    Experimental data suggests that a fifth force is no more than \numrange{e-6}{e-3} times as strong as gravity over a range of \qtyrange{1}{e4}{\metre} \cite{will_confrontation_2006}.
    However, there could be screening mechanisms at work like the chameleon mechanism \cite{khoury_chameleon_2004,hinterbichler_towards_2011} or a disformal coupling via the Lagrangian, which leads to a geodesic equation that contains a four-force that depends on the four-velocity, i.e. the force could be screened in the solar system, where the velocity of matter is non-relativistic \cite{boehmer_interacting_2015}.
    Furthermore, a screening of conformal effects by disformal coupling contributions is discussed by \citet{van_de_bruck_disformal_2015}.
}
\citet{etheredge_sharpening_2022} voice some doubts about this form of the scalar \gls{wgc}, as it fails to deliver a compelling argument \textit{why} the gravitational force should be weaker than the attractive force mediated by massless scalar fields. They claim that only the following bound holds for massless scalar fields:
\begin{equation}
    \frac{\left(\partial_\phi m\right)^2}{m^2}\geq\frac{g_{\phi\phi}}{d-2},
\end{equation}
where the $\phi\phi$ component of the moduli space metric $g_{\phi\phi}$ appears, for instance, in the kinetic term $\frac{1}{2}g_{\phi\phi}\mathrm{d}\phi\wedge\star\mathrm{d}\phi$ for $\phi$ in the action.

If a small positive cosmological constant is part of the theory, \citet{antoniadis_weak_2020} propose some corrections to the \gls{wgc}:
\begin{align}
    qg&>\sqrt{4\pi G}\left(1+\left(\frac{Gm}{2l}\right)^2\right)&l\rightarrow\infty\\
    qg&>\left(\frac{32\pi^2}{3}\right)^{1/4}\sqrt{lm}&l\rightarrow0
\end{align}
with $l\rightarrow\infty$ being the low curvature limit and $l\rightarrow0$ the strong curvature limit.
Note that $M_\textnormal{P}=1/\sqrt{8\pi G}$.

A generalisation to massive scalar fields (that is valid around the minimum of the potential of a scalar field) is presented by
\citet{benakli_revisiting_2020} for several popular potentials:
\begin{equation}\label{eq:swgc_pot}
    4m_0^2\abs{\frac{\partial^4V}{\partial^2\phi\partial^2\phi^*}}_{\phi=0}\geq\frac{\Tilde{c}}{M_\textnormal{P}}\abs{\frac{\partial^2V}{\partial\phi\partial\phi^*}}^2_{\phi=0},
\end{equation}
with $\Tilde{c}\approx1$.
The constraints are summarised in \cref{tab:wgc}.
\begin{table*}[ht]
    \centering
$
\begin{tblr}{XXX
} \toprule
 \text{Potential} & V(\phi) & \text{Constraint} \\\midrule
 \text{Mexican Hat} & -m^2\Bar{\phi}\phi+\lambda(\Bar{\phi}\phi)^2 & \lambda\geq\frac{1}{12}\frac{m^2}{M_\textnormal{P}}\\
 \text{Axion-like} & \mu^4\left(1-\cos\left(\frac{\phi}{f}\right)\right) & f^2\leq M_\textnormal{P}^2\\
 \text{Inverse power-law} & \Lambda^{4+p}\phi^{-p} & \phi_0^2\leq\frac{\left(p+2\right)\left(2p+1\right)}{3}M_\textnormal{P}^2\\
 \text{Exponential} & \Lambda e^{-\lambda\phi/f} & f^2\leq\frac{2}{3}\lambda^2M_\textnormal{P}^2\\
 \text{Double Exponential} & \Lambda_1e^{-\lambda_1\phi/f}+\Lambda_2e^{-\lambda_2\phi/f} & \ddag\\
 \text{Starobinsky} & \Lambda^4\left(1-e^{-\sqrt{2/3}\phi/M_\textnormal{P}}\right)^2 & \phi_0\leq\sqrt{\frac{3}{2}}\log\left(\frac{14}{\sqrt{51}-4}\right)M_\textnormal{P}\sim2M_\textnormal{P}\\
 \bottomrule
\end{tblr}
$
    \caption[WGC Potential Constraints]{The scalar \gls{wgc} puts constraints on scalar potentials. \citet{benakli_revisiting_2020} evaluated \cref{eq:swgc_pot} for several popular potentials to derive the \gls{wgc} constraints. The expression for the double exponential potential is too long to fit into the table:
    $\ddag=\lambda_1^4\Lambda_1^2\left(\frac{2}{3}\frac{\lambda_1^2}{f^2}-\frac{1}{M_\textnormal{P}^2}\right)e^{-2\lambda_1\phi_0/f}+\lambda_2^4\Lambda_2^2\left(\frac{2}{3}\frac{\lambda_2^2}{f^2}-\frac{1}{M_\textnormal{P}^2}\right)e^{-2\lambda_2\phi_0/f}+\Lambda_1\Lambda_2\lambda_1^2\lambda_2^2\left(\frac{10/3\lambda_1\lambda_2-\lambda_1^2-\lambda_2^2}{f^2}-\frac{2}{M_\textnormal{P}}\right)e^{-\left(\lambda_1+\lambda_2\right)\phi_0/f}\geq0$.
    }
    \label{tab:wgc}
\end{table*}

A generalisation of the scalar \gls{wgc} to axions (that guarantees that the gauge repulsion exceeds the combined gravitational and scalar attraction) is presented by
\citet{vittmann_axion-instanton_2023}:
\begin{equation}
    S_\iota^2+M_\textnormal{P}^2\partial_\phi S_\iota\Bar{\partial}_\phi S_\iota\leq M_\textnormal{P}^2/f^2,
\end{equation}
which is derived from the more general statement
\begin{equation}
    Q^2\geq S_\iota^2+G^{ab}\nabla_aS_\iota\Bar{\nabla}_bS_\iota,
\end{equation}
with $G^{ab}$ the moduli space metric.\footnote{
    Since the term with the derivatives is positive definite, a lower bound for the axion decay constant can be derived: 
        $f<M_\textnormal{P}/\sqrt{1+\abs{\partial_\phi S_\iota}^2M_\textnormal{P}^2}\leq M_\textnormal{P}$.
}

\subparagraph{Strong Scalar \gls{wgc}}
\citet{gonzalo_strong_2019} present the \textit{strong scalar \gls{wgc}}, which applies to all scalars and implies that every scalar interaction has to be stronger than the gravitational interaction: 
\begin{equation}\label{eq:strongswgc}
    \frac{1}{M_\textnormal{P}^2}\left(\frac{\mathrm{d}^2V}{\mathrm{d}\phi^2}\right)^2\leq2\left(\frac{\mathrm{d}^3V}{\mathrm{d}\phi^3}\right)^2-\frac{\mathrm{d}^2V}{\mathrm{d}\phi^2}\frac{\mathrm{d}^4V}{\mathrm{d}\phi^4},
\end{equation}
and that there is a tower of extremal states that satisfies the \gls{dc}.
\Cref{eq:strongswgc} is equivalent to
\begin{equation}
    M_\textnormal{P}^2m^2\partial_\phi^2\frac{1}{m^2}\geq1, 
\end{equation}
for $m^2=V^{\prime\prime}$ \cite{dudas_testing_2023}.
Based on this conjecture, \citet{gonzalo_strong_2019} derive a set of implications:
\begin{itemize}
    \item The axion decay constant is sub-Planckian: $f<M_\textnormal{P}$.
    \item Scalar field potentials have to be linear and allow for trans-Planckian excursion.\footnote{Applying \cref{eq:strongswgc} to a linear potential $V(\phi)=a\phi+b$ always satisfies the scalar \gls{wgc}, even for trans-Planckian field values $\phi>M_\textnormal{P}$.}
    \item Massive scalars always have interaction channels beside gravity.\footnote{\Cref{eq:strongswgc} is always violated for purely quadratic potentials $V(\phi)=m^2\phi^2$. This rules out tachyonic instabilities in the potential \cite{brahma_relating_2019}.}
    \item If inflation occurs in the form of large single field inflation, the tensor-to-scalar ratio is $r_\textnormal{ts}\approx0.07$.
    \item Neutrinos have normal mass ordering.
    \item The lightest neutrino is Dirac with $m_{\nu_1}\lesssim\Lambda_4^{1/4}$.    
\end{itemize}
The strong scalar \gls{wgc} is disputed by \citet{freivogel_conjecture_2020,benakli_revisiting_2020}.
The derivation of \cref{eq:strongswgc} from first principle is challenging, as it mixes long-range and short-range forces \cite{dallagata_covariant_2020,kusenko_fundamental_2020}. If the fourth derivative was absent, the condition itself were incompatible with the \gls{dsc} \cite{kusenko_fundamental_2020,shirai_is_2019}. Furthermore, a dilute gas of non-relativistic atoms would violate \cref{eq:strongswgc} \cite{freivogel_conjecture_2020,dudas_testing_2023}.

A multi-field extension of the strong scalar \gls{wgc} is presented by \citet{gonzalo_pair_2020}.
Their guiding principle is that any force mediator, e.g. photons or scalars, must be produced at a higher rate than gravitons. This notion is independent of \gls{bh} physics.\footnote{However, pair production is an important aspect of \gls{bh} evaporation.}
The proposal of pair production is stronger than the convex hull conjecture, as in the pair production proposal, actual particles are produced.
The inequality they derive is invariant under holomorphic transformations and valid for $n$ complex moduli:
\begin{equation}
    \frac{g{i\Bar{j}}}{n}\abs{\left(\partial_im^2\right)\left(\partial_{\Bar{j}}m^2\right)-m^2\left(\partial_i\partial_{\Bar{j}}m^2\right)}\geq\frac{m^4}{M_\textnormal{P}^2}.
\end{equation}

\paragraph{Does the \gls{wgc} apply to all spin-states?}\label{p:WGC_spin-stes}
We discuss the different possible spin-states individually.

\subparagraph{Spin-0}
\citet{palti_fermions_2020} presents the magnetic \gls{wgc} (\cref{eq:mwgc}) for spin-0 states as
\begin{equation}
    m_\textnormal{t}\sim\abs{\partial_\phi m_\textnormal{t}}M_\textnormal{P}
\end{equation}
with $m_\textnormal{t}$ the mass scale of an infinite tower of states, which is motivated by the \gls{ep} and the \gls{dc}.

\subparagraph{Spin-1/2}
Explicitly considering the spin, \citet{urbano_towards_2018} shows that the \gls{wgc} also holds for spin-1/2 Dirac fermions.
\citet{palti_fermions_2020} presents the magnetic \gls{wgc} (\cref{eq:mwgc}) for spin-1/2 states as
\begin{equation}
    m_\textnormal{t}\sim Y M_\textnormal{P}
\end{equation}
with $Y$ is a renormalisable Yukawa coupling.\footnote{Gravitinos and Goldstinos do not have a renormalisable Yukawa coupling \cite{palti_fermions_2020}.}

\subparagraph{Spin-1}
\citet{palti_fermions_2020} presents the magnetic \gls{wgc} (\cref{eq:mwgc}) for spin-1 states as
\begin{equation}
    m_\textnormal{t}\sim gM_\textnormal{P}.
\end{equation}

\subparagraph{Spin-2}
\citet{klaewer_spin-2_2019} present the magnetic \gls{wgc} (\cref{eq:mwgc}) for spin-2 states as
\begin{equation}
    m_\textnormal{t}\sim m_2M_\textnormal{P}/M_w,
\end{equation}
with $m_\textnormal{t}$ the cutoff of the mass term of the Stückelberg gauge field, and
$M_w\leq M_\textnormal{P}$ the interaction mass scale.\footnote{
    If there are several spin-2 fields, the cutoff is ambiguous unless the field with the weakest interaction scale is taken (that leads to the largest $M_w$) \cite{kundu_regge_2023}. Assuming $M_w\sim M_\textnormal{P}$ implies $M_\textnormal{t}\sim m_2$, such that there is no parametric gap between the spin-2 state and an infinite tower of states \cite{kundu_regge_2023}. \citet{kundu_regge_2023} generally put theories with a massive spin-2 particle separated from an infinite tower into the swampland.
}
This bound is even applicable to massive gravitons, where the cutoff is of the order of the graviton mass $m_\textnormal{t}\sim m_\textnormal{g}$.
The conjecture rules out a continuous limit, in which \gls{qg} can be taken into a theory with two massless spin-2 fields \cite{klaewer_spin-2_2019}. For that case, i.e. for the graviton mass approaching 0, an infinite tower of (spin-2 or higher-spin) states is expected \cite{bachas_massive_2019}.
\citet{de_luca_cheeger_2021} find the conjecture to hold in \gls{ads} vacua if the cosmological constant is much smaller than $m_2$, but not otherwise.
The conjecture is examined, and its assumptions tested by \citet{de_rham_spin-2_2019}:
\begin{itemize}
    \item To derive the bound, it is assumed that the scaling between the helicity modes and the sources has a smooth massless limit. This is questioned by the observation that the interactions scale with negative powers of the spin-2\textendash mass-term \cite{van_dam_massive_1970,vainshtein_problem_1972}.
    \item The identification $g\sim m_2/M_w$ becomes meaningless in the $m_2\rightarrow0$ limit, as the interaction term blows up for fixed $M_w$.
    \item \citet{klaewer_spin-2_2019} did not consider the possibility of a mass gap between the light spin-2 states and the \gls{uv} tower of states. Having no mass gap between the mass of the spin-2 particle of the \gls{eft} and the infinite tower that completes the \gls{uv} theory, means that the \gls{eft} is not under perturbative control.
    \item The conjecture might be in tension with the \gls{ep}.
    \item The spin-2 \gls{wgc} is invoked to avoid global symmetries, but the helicity-1 mode of the graviton is not a genuine vector field with a global U(1) charge. The U(1) symmetry proposed by \citet{klaewer_spin-2_2019} vanishes in unitary gauge, ergo, there is no global symmetry.
    \item A counterexample to the spin-2 \gls{wgc} is found in diffeomorphism-invariant couplings.
    \item Counterexamples to the proposed cutoff of the spin-2 \gls{wgc} are found in interacting spin-2 theories.
    \item There are counterexamples to the spin-2 \gls{wgc} in massive gravity \cite{bachas_massive_2018,bachas_massive_2019,bachas_quantum_2018,bachas_spin-2_2011}.
\end{itemize}
\Citet{de_rham_spin-2_2019} conclude that the conjecture as presented by \citet{klaewer_spin-2_2019} is invalid. However, they stress that this does not rule out a \gls{wgc} for spin-2 states entirely:
For local and Lorentz-invariant massive spin-2 fields, they propose
\begin{equation}
    m_\textnormal{t}\lesssim \left(m_2^2M_w\right)^{1/3}.
\end{equation}
Furthermore, positivity bounds on the S-matrix yield a bound for the coupling: $g\lesssim\left(m/M_w\right)^{1/4}\ll1$ \cite{de_rham_improved_2018}.

\subparagraph{Spin-J}
\citet{sammani_higher_2024} present a form of the \gls{wgc} for higher-spin states that matches the form of \cref{eq:ewgc}.
A generalisation for higher-spin fields is also presented by \citet{porrati_model_2009}.
\citet{kaplan_closed_2021} present a necessary condition for particles with spin $J\geq3$ in 4d spacetime:
\begin{equation}
    m_\text{gr}\lesssim
    \begin{cases}
        m_J\left(\frac{\abs{g_J}M_\textnormal{P}}{m_J}\right)^{\frac{1}{2\left(J-2\right)}} & \abs{g_J}\gtrsim\frac{m_J}{M_\textnormal{P}}\\
        m_J & \abs{g_J}\lesssim\frac{m_J}{M_\textnormal{P}},
    \end{cases}
\end{equation}
where $m_J$ is the mass of an elementary particle $X_J$ of spin $J\geq3$, $m_\text{gr}$ is the mass of a higher-spin particle in the gravity sector (which implies that the particle $X_J$ of mass $m_J$ can be one of a non-gravitational \gls{qft}). The conjecture is that the particle $X_J$ can only couple to gravity, if there is a higher-spin state in the gravitational sector that satisfies the above bound. If we take $m_\text{gr}$ as a bound $\Lambda_\text{gr}$, which we integrate out, then we obtain a low-energy \gls{qft}, which obviously breaks down for $\abs{g_J}\lesssim\frac{m_J}{M_\textnormal{P}}$, as at this point non-gravitational interactions are weaker than gravitational interactions, and it was not valid to integrate out / decouple gravity in this regime. Above $\Lambda_\text{gr}$, stringy effects appear, ergo, this scale can also be interpreted as the string scale.

\begin{figure}[htb]
  \begin{center}
    \includegraphics[width=0.48\linewidth]{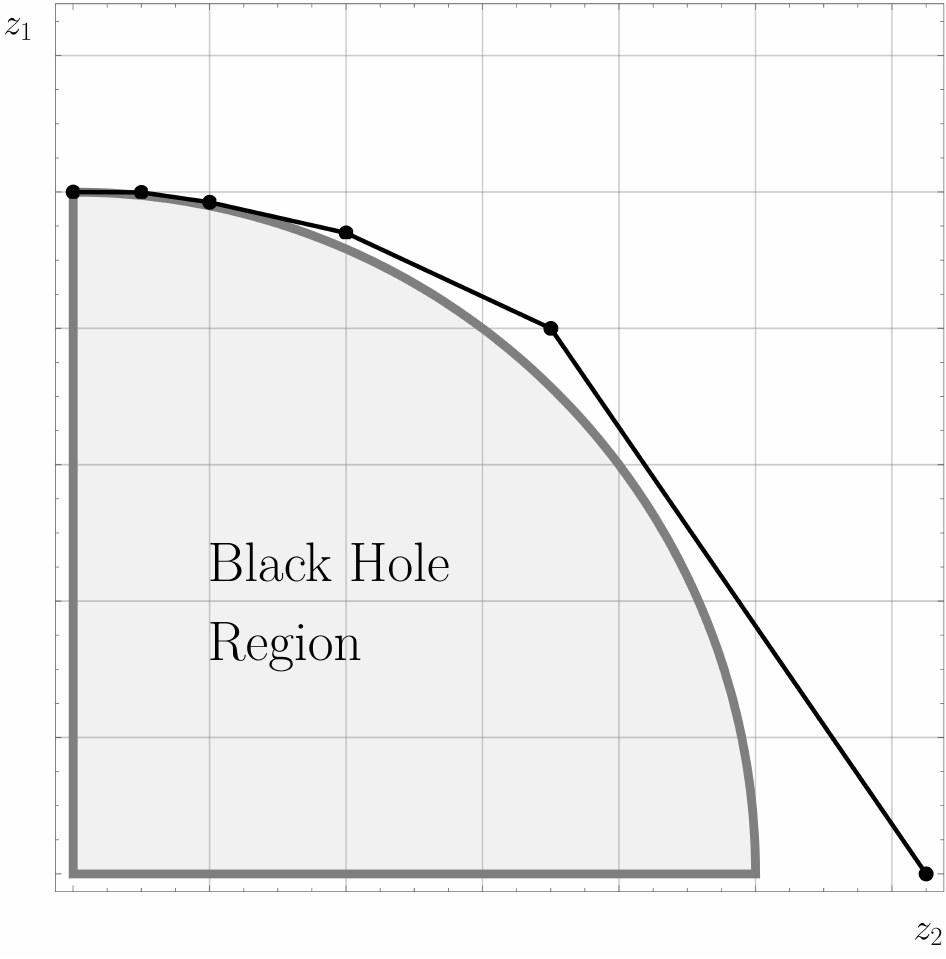}
    \includegraphics[width=0.48\linewidth]{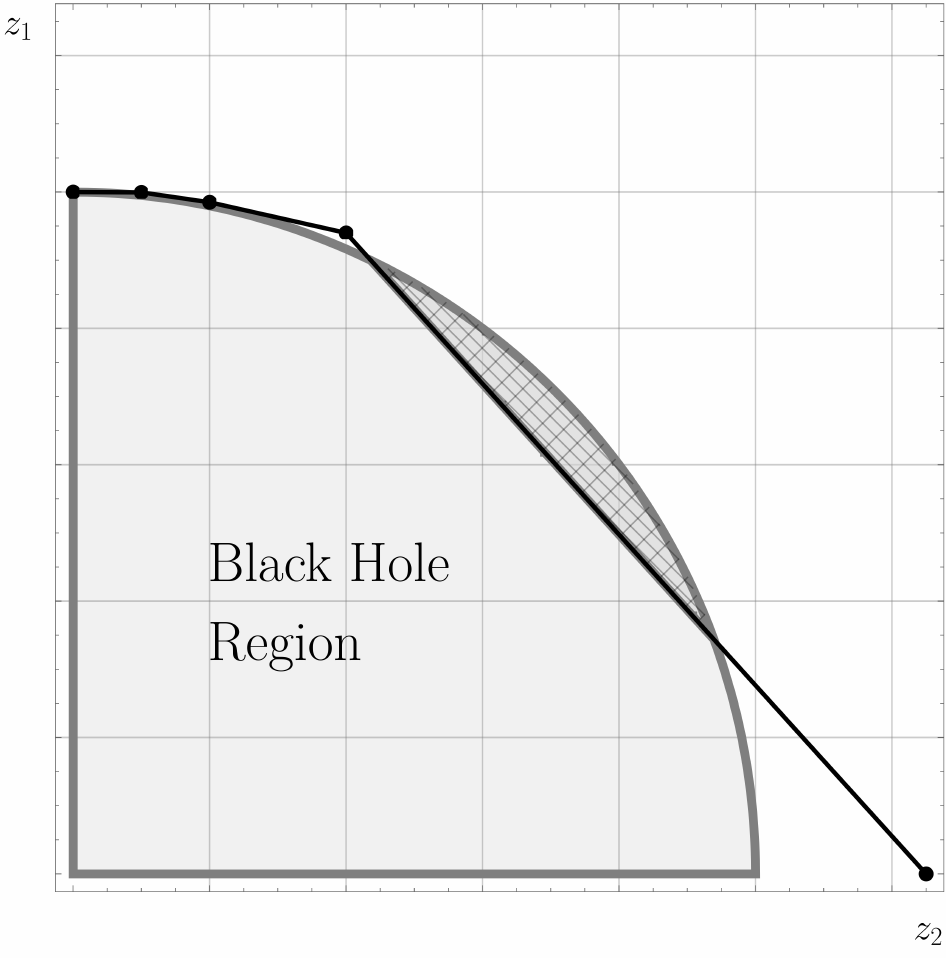}
  \end{center}
  \caption[Convex Hull Conjecture]{Reducing a higher-dimensional parent theory onto a circle $S^1$ leads to superextremal \gls{kk}-modes. If the circle is too small (right figure), intermediate states can violate the convex hull conjecture. Figure adapted from \citet{harlow_weak_2023}.}\label{f:ConvexHull}
\end{figure}

\paragraph{Does the \gls{wgc} also apply to multi-particle states?}\label{p:WGC_multiparticle}
In a theory with $N$ charges, each particle state is associated with a vector $\Vec{z}=\frac{1}{m}\left(q_1,\dots,q_N\right)$.
It is not sufficient that one species $i$ with $\abs{\Vec{z_i}}>1$ exists, as orthogonal states could still form stable \glspl{bh} \cite{cheung_naturalness_2014}.
It is also not sufficient to have a species $i$ for each U(1) individually with $\abs{\Vec{z_i}}>1$, as combined states could exist that would form stable \glspl{bh} \cite{cheung_naturalness_2014,heidenreich_weak_2015}.
Moreover, it is also not sufficient that every \gls{kk} mode on its own is superextremal \cite{harlow_weak_2023}, as you might be able to create multi-particle states that form stable \glspl{bh}, as in \cref{f:ConvexHull}.

In a theory with $N$ charges and different couplings $g_i$, there is a charge vector $\Vec{q}=\left(q_1g_1,...,q_Ng_N\right)$ which yields an effective charge of $q_\text{eff}=\sqrt{q_1^2/g_1^2+\dots+q_N^2/g_N^2}\leq q_1/g_1+\dots+q_N/g_N$ \cite{heidenreich_weak_2015,van_beest_lectures_2022,palti_brief_2020}.
To ascertain that extremal \glspl{bh} can decay, the convex hull of the charge-to-mass ratio vector
\begin{equation}
    \Vec{z}\defeq\Vec{q}\cdot M_\textnormal{P}/m
\end{equation}
has to include the unit ball, i.e. 
\begin{equation}
    \abs{\Vec{z}}>1
\end{equation}
is required \cite{palti_swampland_2019,cheung_naturalness_2014,saraswat_weak_2017,palti_brief_2020,van_beest_lectures_2022}.\footnote{
    With $N$ identical charges, this gives the bound $m\leq q/\sqrt{N}$ \cite{saraswat_weak_2017}. From the species scale (\cref{eq:speciess,s:ssc}) we know that $\delta M_\textnormal{P}^2\sim N\Lambda^2$. It follows that for large $N$, the \gls{wgc} bound is independent of $N$, since the contributions just cancel, as was already noted by \citet{cheung_naturalness_2014}.
    }
This is known as the \textit{convex hull conjecture}.

The magnetic version of the convex hull conjecture demands that a monopole charged under all symmetry groups is not a \gls{bh}, i.e. $\Lambda\lesssim M_\textnormal{P}/\sqrt{\frac{1}{g_1^2}+\dots+\frac{1}{g_N^2}}$ \cite{saraswat_weak_2017}.\footnote{
    For identical gauge couplings, this yields $\Lambda\lesssim M_\textnormal{P}\frac{g}{\sqrt{N}}$ \cite{saraswat_weak_2017}. The latter cutoff was already conjectured early on by \citet{huang_weak_2008}.
    }

\paragraph{What is the Lattice \gls{wgc}?}\label{p:LWGC}
The motivation for the stronger form of the \gls{wgc} comes from the observation that the \gls{wgc} can fail after compactification if there are different charges. The proposed remedy is that the lightest state in any charge-direction is superextremal \cite{heidenreich_sharpening_2016}.
This condition implies the convex hull conjecture, and an infinite number of charged particles with potentially super-Planckian masses\,\textemdash\,which can be interpreted as extremal \glspl{bh}, if the higher-order corrections to the extremality bound reduce the mass of extremal \glspl{bh} \cite{heidenreich_sharpening_2016} (which is likely the case, as we have reasoned in \cref{p:WGC_BH}).

A more formal definition of the \gls{lwgc} is the following:
\begin{displayquote}
    For every point $\Vec{q}$ on the charge lattice, there is a particle of charge $q$ with charge-to-mass ratio at least as large as that of a large, semi-classical, non-rotating extremal black hole with charge $\Vec{q}_\text{BH}\propto\Vec{q}$ \cite{heidenreich_sharpening_2016}.
\end{displayquote}
This means that there is a particle lighter than $qM_\textnormal{P}$, a particle lighter than $2qM_\textnormal{P}$, a particle lighter than $3qM_\textnormal{P}$, a particle lighter \dots \cite{heidenreich_axion_2016}.

This conjecture does not hold, though, as there are counterexamples to this conjecture \cite{heidenreich_weak_2015,harlow_weak_2023,heidenreich_evidence_2017,heidenreich_weak_2018}.

\paragraph{What is the Tower \gls{wgc}?}\label{p:TWGC}
The convex hull conjecture might be violated after circle compactification (onto a small enough circle)\footnote{
    \citet{cota_minimal_2024} question the necessity of a tower of states below a certain compactification size. They argue that a theory remains consistent if it is compactified onto a circle below a certain threshold, even without the appearance of a tower of states. An exception to this exception is the tower of states that appears\,\textemdash\,as requested by the \gls{ep}\,\textemdash\,in the weak-coupling limit of the heterotic string \cite{cota_minimal_2024}.
}
\cite{harlow_weak_2023,heidenreich_sharpening_2016,cota_asymptotic_2022}.\footnote{
    This form of dimensional reduction is not per se a counterexample to the \gls{wgc} because it first needs to be checked if the higher-dimensional parent theory is a valid theory of \gls{qg} at all \cite{harlow_weak_2023}, as an infinite number of charged particles indicates that the theory is only valid up to a cutoff scale, which itself indicates the appearance of new particles, e.g. massive \gls{kk} modes or strings, which might satisfy the \gls{wgc}, i.e. they allow \glspl{bh} to decay \cite{heidenreich_sharpening_2016}.
}
To avoid this violation, an infinite tower of \gls{wgc} satisfying states is necessary.\footnote{
    The tower is predicted to start from a mass scale $\Lambda$ \cite{cota_asymptotic_2023}: if the tower starts below the cutoff scale, the species scale (\cref{s:ssc}) is lower, due to the additional species present. A mass scale below the cutoff scale was compatible with the mild form of the \gls{wgc} itself, i.e. no internal inconsistency would arise; however, there might be a tension with stronger versions, as for higher masses in the tower, superextremal \glspl{bh} were to be expected.
    }
The existence of such a tower is predicted by the \gls{twgc}.\footnote{   
    An infinite tower of states beyond the cutoff scale is also predicted by the \gls{dc} and corresponds well to an expectation from duality: Far away from a theory at $\phi_1$ in moduli space, an infinite tower of heavy particles arises, with a mass $m\sim e^{\abs{\mathfrak{a}}d(\phi_1,\phi_2)}$\,\textemdash\,where we use the same symbols as in \cref{sec:distance} with $d(\phi_1,\phi_2)$ the geodesic distance in moduli space $\mathcal{M}$ between $\phi_1$ and $\phi_2$\,\textemdash\,and $\abs{\mathfrak{a}}\geq1/\sqrt{d-2}$ \cite{etheredge_dense_2023}.
    That at least one of the towers that emerge is one of superextremal states can be motivated from entropy arguments \cite{cota_asymptotic_2022,arkani-hamed_string_2007}. Furthermore, the \gls{ep} predicts the appearance of a charged tower in the asymptotic limit of vanishing gauge coupling \cite{cota_asymptotic_2022,palti_swampland_2019,harlow_wormholes_2016,heidenreich_emergence_2018,grimm_infinite_2018}.
}
A formal definition of the \gls{twgc} reads as follows:
\begin{displayquote}
    For every site in the charge lattice, $\vec{q}\in\Gamma$, there exists a positive integer $n$ such that there is a superextremal particle of charge $n\vec{q}$ \cite{heidenreich_repulsive_2019,harlow_weak_2023}.
\end{displayquote}

The \gls{twgc} is stronger than the convex hull conjecture, as additional constraints are required, but weaker than the \gls{slwgc}, as the tower can have gaps, i.e. it is not required that the (sub-)lattice is filled \cite{andriolo_tower_2018}.

The tower can be narrow and sharply defined at weak coupling, but its meaning becomes blurry for strong coupling \cite{heidenreich_repulsive_2019}. It can be satisfied by unstable resonances \cite{enriquez-rojo_swampland_2020},
but not by multi-particle states \cite{heidenreich_repulsive_2019,harlow_weak_2023}. Unstable resonances do not correspond to states in Hilbert space, but to localised peaks in the S-matrix of a scattering process: for weak coupling this means that the lifetime of the particle is long and the peak itself is localised at a narrow energy scale, which corresponds to the mass of the unstable particle; for strong coupling the peak is spread out, which makes it difficult to define a mass and therefore also challenging to define the meaning of the \gls{twgc} \cite{harlow_weak_2023}.\footnote{
    The mass of a resonance, i.e. the location of a pole in the complex energy plane of the S-matrix, is a complex quantity, which introduces an ambiguity in the interpretation of the conjecture \cite{heidenreich_evidence_2017}.
    }
If it was an infinite tower of stable states, one of the arguments in favour of the \gls{wgc} presented by \citet{arkani-hamed_string_2007} would break down.\footnote{
    Heavy towers can be unstable, and the validity of the tower scalar \gls{wgc} is still under debate \cite{etheredge_dense_2023}.
}
From the claim $m/q<1$, they deduce that the number of exactly stable particles in \gls{qg} in asymptotically flat space is finite (except for \gls{bps} states, but those are bound), which implies that even for neutral particles the number of massless degrees of freedom is finite. We wonder if an infinite tower of stable states would imply infinite degrees of freedom for some particles.

A potential\footnote{
    The full spectrum was not computed, which might remedy the observed disagreement with the conjecture.
    }
counterexample to the \gls{twgc} comes from 4d F-theory, where an infinite tower of states appears that does not satisfy the bounds presented by the \gls{twgc} \cite{heidenreich_repulsive_2019,lee_modular_2019}. The culprit is the dimensionality: The \gls{twgc} in $D$ dimension is closely related to the \gls{wgc} in $D-1$ dimensions, and it is not clear if the \gls{wgc} should hold at all in $d<4$ \cite{heidenreich_repulsive_2019}, as there are no asymptotically flat \glspl{bh} \cite{harlow_weak_2023}.\footnote{
    See also \cref{f:RFC:dimension} about dimensionality.
    }

\paragraph{What is the sub-Lattice \gls{wgc}?}\label{p:sLWGC}
The \gls{slwgc} is a stronger conjecture than the \gls{twgc}, as it requires a superextremal particle at each site of a full-dimensional sub-lattice of the charge lattice \cite{hayashi_spectra_2023,lee_tensionless_2018,heidenreich_evidence_2017}, which means that the integer $n$ from the \gls{twgc} is independent of $\vec{q}$ \cite{heidenreich_repulsive_2019,harlow_weak_2023}.
This reflects the principle of completeness \cite{harlow_wormholes_2016,lee_tensionless_2018,klaewer_super-planckian_2017,enriquez-rojo_swampland_2020} (see \cref{sec:complet}) and implies the \gls{twgc} \cite{alim_weak_2021}.
A formal definition of the \gls{slwgc} is the following:
\begin{displayquote}
    There exists a positive integer $n$ such that for any site in the charge lattice, $\vec{q}\in\Gamma$, there is a superextremal particle of charge $n\vec{q}$ \cite{heidenreich_repulsive_2019,harlow_weak_2023}.
\end{displayquote}

If a theory saturates a \gls{slwgc} bound, gravity and the gauge theory become strongly coupled at the same parametric energy scale, i.e. when the charged particles become increasingly broad, the density of states of different charges must behave nicely, such that gravity and electromagnetism become strong at the same scale\,\textemdash\,this implies that a particle exists which satisfies the \gls{wgc} and it suggests a unification of forces \cite{heidenreich_weak_2018}. This explains how a gauge theory can emerge: the low-energy coupling is small, because of the \gls{uv} dynamics of heavy particles \cite{heidenreich_weak_2018}.
The \gls{slwgc} predicts the existence of an energy scale $\Lambda_\text{gauge}<\Lambda_\text{QG}$ at which the gauge coupling becomes strong\,\textemdash\,the mass of every charged and weakly coupled particle is bounded by 
\begin{equation}
    m<\Lambda_\text{gauge}\lesssim e^2\expval{q^2}_{\Lambda_\text{gauge}}M_\textnormal{P}^{D-2}
\end{equation}
where the mass of every particle is bounded by the \textit{average} charge of the particles, i.e. particles lighter than $\Lambda_\text{gauge}$ are on average superextremal \cite{heidenreich_weak_2018}.

There is a loophole in this variation of the \gls{wgc}: since the lattice can be made arbitrarily sparse, the constraining power of this form of the \gls{wgc} can be arbitrarily diluted, as the charges for arbitrarily sparse lattices become arbitrarily massive \cite{heckman_fate_2024,arkani-hamed_string_2007,harlow_weak_2023,brown_axionic_2016,montero_transplanckian_2015,rudelius_constraints_2015}.\footnote{
    \citet{harlow_weak_2023} state that it's plausible that the coarseness is always of $\order{1}$; they did not encounter any cases with $k>3$, but also mention that it's an open question how exactly the coarseness of the sub-lattice manifests itself in the \gls{ir} physics.
    \citet{hebecker_large_2019} find that (mis-)aligned winding modes can produce settings with arbitrarily large coarseness, which would parametrically violate the \gls{slwgc}.
    \citet{hebecker_thraxions_2019} present a setup with axions in a warped throat, where the sub-lattice of states can be made parametrically coarse.
    A very sparse sub-lattice might also be realised through Higgsing, with superextremal particles with masses even above the magnetic \gls{wgc} \cite{harlow_weak_2023}. However, \citet{heidenreich_evidence_2017} state that the \gls{slwgc} is not robust under Higgsing, which leaves two options:
    First, the \gls{wgc}\,\textemdash\,in general, and not only the \gls{slwgc}\,\textemdash\,is not suitable to constrain \gls{ir} physics because if the conjecture is not robust under Higgsing, it means that the full, un-Higgsed charge lattice needs to be known, i.e. the \gls{uv} physics need to be known.
    Second, the \gls{slwgc} constrains any \gls{ir} phases of a theory because there is no sharp distinction between a Higgsed gauge group and a massive U(1), i.e. the gauge group is well-defined in the \gls{ir}, whereas it is not entirely clear what \textit{full charge lattice} actually means.
}
\citet{heidenreich_evidence_2017,saraswat_weak_2017} suggest that the coarseness $k$ changes the cutoff scale to $\Lambda\lesssim gkM_\textnormal{P}$.

\citet{casas_yukawa_2024} study \textit{gonions}\footnote{See \cref{foo:gonions}.} \cite{aldazabal_intersecting_2001} in a type IIA setting, and find that the towers of gonions do not satisfy the \gls{twgc}/\gls{slwgc}\,\textemdash\,but do also not violate the convex hull conjecture under compactification\,\textemdash, as all the elements of the tower have the same charge. In a setting of one dominant gonion tower with two decompactified dimensions, they find the gonion mass $m_\textnormal{g}$ to scale like $m_\textnormal{g}\sim g^2M_\textnormal{P}$, such that the \gls{wgc} is not saturated, as the mass is suppressed by $g^2$ and not just $g$, i.e. the gonion masses are not extremal.

A potential direction to attack the \gls{slwgc} might be through (quasi-)modular elliptic genera, as they contain gaps in the charge spectrum \cite{lee_modular_2019}.\footnote{
    The work by \citet{lee_modular_2019} is not a robust counterexample to the \gls{slwgc}, as the results are only valid in the weak coupling limit.
    }

The briefly mentioned 4d F-theory counterexample to the \gls{twgc} would also be a counterexample to the \gls{slwgc}. It appears that the possibility of large logarithmic terms in loop diagrams require a modification of the \gls{slwgc} in 4 dimensions, compared to higher-dimensional theories \cite{heidenreich_proving_2024,heidenreich_evidence_2017,heidenreich_weak_2018,lee_modular_2019,klaewer_quantum_2021}.

\paragraph{Does the \gls{wgc} apply to strings and branes as well?}\label{p:WGC_p-form}
\Cref{eq:ewgc,eq:mwgc} can be generalised for arbitrary $p$-form fields in any dimension $d$.\footnote{
    $p$-form gauge fields bring charges to $p-1$-dimensional objects ($p-1$-branes), with 0-branes being point like particles and 1-branes being 1-dimensional strings \cite{rudelius_introduction_2023}.
}
The electric \gls{wgc}, \cref{eq:ewgc}, reads as: 
\begin{equation}\label{eq:pwgc}
    \frac{p\left(d-p-2\right)}{d-2}\mathcal{T}^2\leq g^2q^2M_\textnormal{P}^{d-2}
\end{equation}
with $\mathcal{T}$ the tension\footnote{
    The tension has units of mass per length \cite{rudelius_introduction_2023}.
}
of the state \cite{bonnefoy_swampland_2021}, which reduces to the mass for a $1$-form field.\footnote{
The \gls{wgc} does \textit{not} weaken in higher-dimensional settings.
Furthermore, for $p=0$ and $p=d-2$, no charged black object exists \cite{heidenreich_sharpening_2016}.
}

The magnetic \gls{wgc}, \cref{eq:mwgc}, can be expressed for a $(d-p-3)$-brane charged under a $p$-form gauge field:
\begin{equation}\label{eq:mwgc_p-form}
    \Lambda\lesssim\left(g^2M_\textnormal{P}^{d-2}\right)^{1/2p};
\end{equation}
the tension is approximately
\begin{equation}\label{eq:magtension}
    \mathcal{T}\sim\Lambda^p/g^2\gtrsim\mathcal{T}_\text{BB}\sim M_\textnormal{P}^{d-2}r_\heartsuit^p,
\end{equation}
where $\mathcal{T}_\text{BB}$ is the tension of a black brane with Schwarzschild radius $r_\heartsuit$ \cite{harlow_weak_2023,hebecker_what_2017}.

\paragraph{Further Comments}
We mostly focussed on mass limits, but the mass-to-\textit{charge} ratio has two components.
Imposing an upper bound on the charge has not been extensively explored in string theory \cite{harlow_weak_2023}, but some considerations and implications are presented by \citet{ibanez_note_2018}.

The \gls{wgc} should hold everywhere in physical space \cite{klaewer_super-planckian_2017}: To say that gravity is the weakest force is a local statement. Demanding that \glspl{bh} can decay means that a particle can escape to infinity, i.e. to the horizon. It is natural to expect that the conjecture holds everywhere in between as well.
Studying the \gls{wgc} in a non-local context (in Lee\textendash Wick \gls{qed} \cite{lee_negative_1969,lee_finite_1970,grinstein_lee-wick_2008}), \citet{abu-ajamieh_phenomenological_2024} find that in the regime of non-locality, the \gls{wgc} bounds become even stronger, as the electric force gets modified at high energies.
A similar finding is presented by \citet{abu-ajamieh_aspects_2024}:
The non-local electric force is given by 
\begin{equation}
    F(r)=-\frac{q}{4\pi r^2}\erf\left(\frac{r\Lambda_\text{nl}}{2}\right)+\frac{q\Lambda_\text{nl}}{4\pi^{3/2}r}\exp\left(-\frac{r^2\Lambda_\text{nl}^2}{4}\right),
\end{equation}
with $\Lambda_\text{nl}$ the energy scale of non-locality.
A modified, non-local \gls{wgc} follows from requiring that gravity is weaker than the non-local electric force:
\begin{align}
    m&\leq\sqrt{2}qM_\textnormal{P}\sqrt{\erf\left(\frac{r\Lambda_\text{nl}}{2}\right)-\frac{r\Lambda_\text{nl}}{\sqrt{\pi}}\exp\left(-\frac{r^2\Lambda_\text{nl}^2}{4}\right)}%
\end{align}
In the low-energy limit $r\Lambda_\text{nl}\gg1$, the non-local \gls{wgc} agrees with the \gls{wgc}.
In the high-energy limit, the non-local version is more stringent, as the non-local factor becomes smaller.

The \gls{wgc} holds under dimensional reduction \cite{rudelius_dimensional_2021}.

If we want to be able to measure the change in charge of a \gls{bh}, $\delta Q=qg$, when throwing in an additional particle, $\delta Q$ needs to be larger than the change in quantum fluctuations due to this additional particle, $\delta\sqrt{\expval{Q^2}}\sim Ggm/\delta t$ \cite{schadow_aspects_2018}. This leads to the following inequality, which motivates the \gls{wgc}:
\begin{align}
    \delta\sqrt{\expval{Q^2}}&\sim Ggm/\delta t\\
    &<Ggm/t_\textnormal{P}&=gm/M_\textnormal{P}\\
    &<qg\\
    \Rightarrow m/M_\textnormal{P}&<q.
\end{align}

Exploring a new angle of implications of the \gls{wgc}, \citet{freivogel_conjecture_2020} formulate the \textit{Bound State Conjecture}:
\begin{displayquote}
    In a theory where the heaviest stable particle has mass $m$, it is impossible to construct adiabatically a black hole that is parametrically smaller than the black hole that can be built from free scalars of the same mass $m$.
\end{displayquote}
Gauge forces should not allow for stable bound states that are smaller than a gravitationally bound state would permit, i.e. attractive gauge forces cannot form stars or other objects that decay to collapse into \glspl{bh} that are arbitrarily small. This also means that we cannot use an efficient, adiabatic \gls{ir} process that gathers enough particles to create a \gls{uv} excitation. The conjecture does not rule out bound states per se, it rather stresses that gauge forces cannot parametrically increase the binding force. \glspl{bh} with $r\lesssim1/m$ can only form via Hawking radiation (and we would assume processes like Schwinger pair production, even though the latter is not explicitly stated by the authors). %
The conjecture is weaker than the \gls{rfc} or the scalar \gls{wgc}, as the bound state conjecture does not rule out bound states entirely, but limits them. Attractive forces are allowed, but cannot be parametrically stronger than gravity. However, it is a stronger conjecture regarding other aspects, namely, that it also holds in the absence of gravity. The bound is explicitly independent of $M_\textnormal{P}$. It is also independent of the spacetime dimension $d$.

\subsubsection{Evidence}\label{sss:WGC_Evidence}
The \gls{wgc} was first proposed by \citet{arkani-hamed_string_2007}. Noteworthy overviews are, for example, given by \citet{heidenreich_repulsive_2019,harlow_weak_2023,reece_tasi_2023,palti_brief_2020,rudelius_introduction_2023}.
The \gls{wgc} is studied in the context of
Hodge theory \cite{bastian_weak_2021},
F-theory \cite{lee_tensionless_2018,lee_stringy_2019,lee_modular_2019,lee_emergent_2019},
M-theory compactifications on \gls{cy} threefolds \cite{marrani_non-bps_2022},
toroidal compactifications \cite{heidenreich_sharpening_2016},
circle compactifications \cite{benakli_newton_2023,cota_minimal_2024,collazuol_e_9_2022},
\gls{cy} compactifications \cite{font_swampland_2019,grimm_infinite_2018},
type I models with broken supersymmetry \cite{bonnefoy_weak_2019},
type II orbifolds \cite{heidenreich_evidence_2017},
massive type IIA orientifold AdS$4\cross X6$ compactifications ($X6$ admits a \gls{cy} metric) \cite{marchesano_new_2022,casas_membranes_2022,prieto_moduli_2024},
type IIA 4d $\mathcal{N} = 1$ \gls{cy} orientifolds with chiral matter \cite{casas_yukawa_2024},
type IIB $\mathcal{N}=1$ orientifolds with O3/O7 planes \cite{enriquez-rojo_swampland_2020},
type IIB compactifications on \gls{cy} threefolds \cite{gendler_merging_2021,van_de_heisteeg_asymptotic_2022},
USp(32) and U(32) orientifold projections of type IIB and 0B strings and SO(16)×SO(16) projection of the heterotic string \cite{basile_supersymmetry_2021},
4d $\mathcal{N} = 1$ SU(N) gauge theory \cite{agarwal_large_2020},
$\mathcal{N}=2$ supergravity \cite{eguchi_rigid_2007,cribiori_weak_2021},
perturbative bosonic string theory in six or more dimensions \cite{heidenreich_proving_2024},
non-supersymmetric string setups \cite{bonnefoy_weak_2020},
dimensional reduction \cite{cremonini_nut_2021},
low-dimensional spaces \cite{li_low_2006},
symmetries \cite{loges_duality_2020},
T-duality \cite{brown_fencing_2015},
\glspl{cft} \cite{andriolo_self-binding_2023,palti_convexity_2022,aharony_convexity_2021,antipin_more_2021,urbano_towards_2018,nakayama_weak_2015,bae_modular_2019,sadeghi_weak_2023,montero_are_2017},
holography \cite{montero_holographic_2019,montero_weak_2016,harlow_wormholes_2016},
quantum information theory \cite{cottrell_complexity_2018},
quantum corrections \cite{klaewer_quantum_2021},
asymptotic safe gravity \cite{de_alwis_asymptotic_2019,basile_asymptotic_2021,eichhorn_absolute_2024},
\glspl{bh} \cite{hamada_finiteness_2022,hod_proof_2017,fisher_semiclassical_2017,cheung_proof_2018,cottrell_weak_2017,shiu_weak_2017,paul_iscos_2024,hebecker_weak_2017,arkani-hamed_causality_2022,sadeghi_rps_2022,aspman_decay_2024,gashti_thermodynamic_2024,alipour_weak_2024,chen_deformations_2024,afshar_mutual_2025,anand_stability_2025,baddis_swampland-statistics_2024},
entropy bounds \cite{fisher_semiclassical_2017,hod_proof_2017,schadow_aspects_2018}, %
cosmic censorship \cite{crisford_violating_2017,crisford_testing_2018,horowitz_further_2019},  %
consistency of \glspl{eft} \cite{cheung_infrared_2014,andriolo_tower_2018,hamada_weak_2019,bittar_gravity-induced_2024}, and
string gases \cite{laliberte_string_2020}.

The \textit{axion \gls{wgc}} is studied in the context of
\gls{cy}-compactifications of type IIA string theory \cite{grimm_infinite_2020},
D7-branes \cite{mininno_dynamical_2021},
and type IIA flux vacua \cite{goswami_enhancement_2019}.

The \textit{scalar \gls{wgc}}, specifically targeting scalar fields, is studied by \citet{gonzalo_strong_2019,palti_weak_2017,lust_scalar_2018,gonzalo_pair_2020,benakli_revisiting_2020}.
\citet{vittmann_axion-instanton_2023} presents a summary that outlines many of the mathematical steps.

Generalisation of the \gls{wgc} for multiple particles in the form of the convex hull conjecture \cite{cheung_naturalness_2014}, the \gls{twgc} \cite{andriolo_tower_2018}, or the \gls{slwgc} \cite{harlow_wormholes_2016} are studied in the context of
F-theory \cite{lee_stringy_2019,lee_tensionless_2018,cota_asymptotic_2022},
F-/M-theory compactified on fibered \gls{cy} threefolds \cite{cota_state_2021},
5d supergravities arising from M-theory compactified on \gls{cy} threefolds \cite{heidenreich_infinite_2021,alim_weak_2021,cota_asymptotic_2023},
M-theory compactified on \gls{cy} fourfolds \cite{charkaoui_asymptotic_2024},
the heterotic string \cite{aalsma_weak_2019},
type IIB $\mathcal{N}=1$ orientifolds with O3/O7 planes \cite{enriquez-rojo_swampland_2020},
superconformal field theories \cite{alday_growing_2019},
large $\mathcal{N}$ superconformal gauge theories \cite{agarwal_classification_2021},
effective \gls{kk} field theories \cite{heidenreich_evidence_2017},
non-Abelian gauge groups \cite{heidenreich_weak_2018},
modular invariance in \glspl{cft} \cite{heidenreich_repulsive_2019,heidenreich_evidence_2017,aalsma_weak_2019,montero_weak_2016},
AdS$_3$ \cite{montero_weak_2016},
Gopakumar\textendash Vafa invariants \cite{gendler_moduli_2023}, and
entropy \cite{hamada_finiteness_2022}.

Furthermore, relations to other swampland conjectures are highlighted in \cref{rel:WGC_CC,rel:WGC_DC,rel:WGC_EP,rel:WGC_dSC,rel:WGC_FLB,rel:WGC_TCC,rel:WGC_nGSC,rel:AdSDC_WGC,rel:WGC_fnomfC,rel:WGC_nnSUSYC}.

\subsubsection{Repulsive Force Conjecture}\label{s:rfc}
The \gls{rfc} is a conjecture closely related to, yet distinct from the \gls{wgc}, based on work by \citet{arkani-hamed_string_2007,palti_weak_2017,lee_stringy_2019,lust_scalar_2018} and named by \citet{heidenreich_repulsive_2019}. It is conjectured that there is always a particle that is self-repulsive over long ranges, i.e. the long-range repulsive gauge force is at least as strong as the sum of all attractive forces between two identical copies of the particle \cite{heidenreich_repulsive_2019}.
In the presence of a scalar field, the coupling of the particle with the largest charge-to-mass ratio to the field leads to an additional attractive force \cite{palti_swampland_2019}.\footnote{\label{f:RFC:dimension}
    There appears to be an interesting link between the presence of scalar fields and dimensionality: The bound obtained from the \gls{wgc} for a $d$-dimensional theory \textit{with} a scalar field is equal to the bound of a $d+1$-dimensional theory \textit{without} a scalar field \cite{palti_swampland_2019,lust_scalar_2018}.
}
This generalises \cref{eq:ewgc} by adding a coupling to scalar fields $\mu$:
\begin{equation}
    m^2+\mu^2\left(M_\textnormal{P}^d\right)^{d-2} \leq 2g^2q^2\left(M_\textnormal{P}^d\right)^{d-2};
\end{equation}
which can also be written as 
\begin{equation}
    kq^2-G_\textnormal{N}m^2-G_\phi\left(\frac{\mathrm{d}m}{\mathrm{d}\phi}\right)^2\geq0,
\end{equation}
with $k$ the electromagnetic constant, $G_\textnormal{N}$ the gravitational constant, $G_\phi$ the scalar force constant, and $\frac{\mathrm{d}m}{\mathrm{d}\phi}$ the scalar charge \cite{heidenreich_proving_2024};\footnote{
    In 4 dimensions, the absence of such a particle indicates the presence of an infinite tower of stable Newtonian bound states \cite{arkani-hamed_string_2007}, an observation that does not hold in higher dimensions \cite{heidenreich_proving_2024,heidenreich_repulsive_2019}.
}
and generalised to $d$-dimensions and multiple scalar fields as
\begin{equation}\label{eq:fswgc}
    \left(\frac{q}{m}\right)^2\geq\frac{d-3}{d-2}+\frac{g^{ij}\partial_im\partial_jm}{m^2},
\end{equation}
where $i$ and $j$ run over the canonically normalised massless scalar fields \cite{lee_stringy_2019}.\footnote{
    The $d$-dimensional Planck mass has been omitted in the notation. It has to be included such that scalars are of mass dimension zero, which adds a factor of $M_\textnormal{P}^{d-2}$ to the \gls{rhs} of \cref{eq:fswgc} \cite{lee_stringy_2019}.
    }
This as well can be generalised to $p$-form fields, to constrain the brane tension \cite{van_beest_lectures_2022}:
\begin{equation}
    f^{ab}q_aq_b\geq g^{ij}\left(\partial_i\mathcal{T}\right)\left(\partial_j\mathcal{T}\right)+\frac{p\left(d-p-2\right)}{d-2}\mathcal{T}^2
\end{equation}
using the inverse gauge kinetic function $f^{ab}$.

In the weak coupling limit, it is expected that the scalar \gls{wgc} and \gls{rfc} are equivalent \cite{gendler_merging_2021}: In the infinite field distance limit, the gauge coupling becomes zero, to avoid global symmetries. If there is a tower of states which becomes exponentially massless and saturates the \gls{wgc}, then the scalar dependence of the gauge kinetic function shows an exponential rate as well, such that the \gls{wgc} and the \gls{rfc} approach zero equally fast. 

In 4d, it is even possible to merge the \gls{wgc} and the \gls{rfc} into one conjecture: the convex hull conjecture.
If a theory violates the \gls{wgc}, it violates the convex hull conjecture, as there is an infinite tower of stable \gls{bh} states with increasing charge-to-mass ratio, i.e. there is no upper bound for $\abs{\Vec{z}}$.
If a theory violates the \gls{rfc}, it violates the convex hull conjecture, as there is a direction in charge space for which any multi-particle state is self-attractive, which means that new states of increasing charge-to-mass ratio can be formed, i.e. there is no upper bound for $\abs{\Vec{z}}$. In more than 4 dimensions, this is no longer the case, as mutually attractive particles do not always form bound states.

For higher-dimensional theories, \citet{heidenreich_proving_2024} present a proof for the \gls{wgc} together with the \gls{rfc}: if there is a particle that is self-repulsive throughout the entire moduli space, this particle is superextremal, and both conjectures are satisfied.

For single-particle states in the absence of massless scalar fields or in theories without supersymmetry, the \gls{rfc} and the \gls{wgc} are identical, as self-repulsiveness and superextremality are then the same \cite{heidenreich_repulsive_2019}.\footnote{
    $F_{12}=\left(k^{ab}q_{1a}q_{2b}-G_\textnormal{N}m_1m_2-g^{ij}\mu_{1i}\mu_{2j}\right)/r^{D-2}$, with $k$ the coupling of the gauge charges $q_a$, $r$ the distance between the two particles in $D$ dimensions, $G_\textnormal{N}$ Newton's gravitational constant, $m$ the mass, and $g$ the scalar coupling between the scalar charges $\mu_i$ defines the force between two particles \cite{heidenreich_repulsive_2019}, which, in the absence of massless scalar fields, reduces to the \gls{wgc} for single-particle states.
    }

If there are multiple photons, the \gls{rfc} implies the \gls{wgc}, as a self-repulsive multi-particle state is extremal, however, a superextremal multi-particle state could be self-attractive, which makes the \gls{rfc} the stronger conjecture in the case of multiple photons but no massless scalar fields. In the presence of massless scalar fields, the two conjectures make different predictions. 
The \gls{rfc} is true even in the absence of gravity, which is of particular importance in \glspl{qg} with massless scalar fields which could mediate long-range attractive forces \cite{heidenreich_repulsive_2019}.

For the \gls{wgc}, \glspl{bh} are an important topic.
While mild forms of the \gls{wgc} can be satisfied by the presence of superextremal \glspl{bh}, the same cannot be said about the \gls{rfc}: in particular, \citet{cremonini_repulsive_2022} show some explicit examples with no self-repulsive \glspl{bh} in some direction of the charge space, contradicting the \gls{rfc} while satisfying the \gls{wgc}.
It is argued that maximally charged \glspl{bh} are indeed self-repulsive \cite{kats_higher-order_2007,charles_weak_2019,horne_black_1993}, or at least have vanishing self-force \cite{heidenreich_black_2020}.
However, this seems not to be always true:
\citet{heidenreich_black_2020} shows that static and spherically (and worldvolume translation invariant) symmetric \glspl{bh} (black branes) of vanishing Hawking temperature or vanishing Bekenstein\textendash Hawking entropy have, at the two-derivative level, vanishing long-range force, whereas static and spherically symmetric \gls{bh}s/branes of finite temperature/entropy are self-attractive.
Furthermore,
\citet{cremonini_mass_2023} study the long-range force between non-identical extremal \glspl{bh} in an Einstein\textendash Maxwell\textendash Dilaton setting with dyonic \glspl{bh}, carrying both, magnetic and electric charge, and a Lagrangian of the form
\begin{equation}
    \mathcal{L}=\sqrt{-g}\left(R-\frac{1}{2}\left(\partial\phi\right)^2-\frac{1}{4}e^{a\phi}F^{\mu\nu}F_{\mu\nu}\right):
\end{equation}
for $a>1$, the force is repulsive, and the mass function is convex,
for $a<1$, the force is attractive, and the mass function is concave,
and for $a=1$, the force vanishes and the \gls{bh} is a \gls{bps} state.

A new spin is given to the \gls{rfc} when angular momentum is considered: a co-rotating object of small mass in the vicinity of the event horizon of an extremal \gls{bh} feels an attractive gravitational force that is exactly cancelled by the repulsive centrifugal force. For a particle satisfying \cref{eq:wgcspin}, gravity becomes weaker than the centrifugal force.

As a concluding note, we'd like to mention that the \gls{rfc} is preserved under dimensional reduction \cite{rudelius_dimensional_2021}.

\subsection{Conjecture Relations}\label{sec:Conj-Relation}
The conjectures are not isolated and fully separated from each other. There are relations and connections between different conjectures. Some might even become redundant. We discuss relations between the different conjectures in the following.

\subsubsection{The FLB and the SSC}\label{rel:FLB_SSC}
Combining the \gls{flb} with the \gls{ssc} for axions leads to a general lower limit for a particle's mass of $m\gtrsim H$ that could be saturated by neutrinos \cite{gonzalo_swampland_2022,hamada_weak_2017,ibanez_constraining_2017,casas_small_2024}:
The species entropy has to be smaller than the entropy of the universe, i.e.
\begin{equation}
    N_\textnormal{S}\lesssim\frac{M_\textnormal{P}^2}{H^2},
\end{equation}
and for axions we have
\begin{equation}
    \frac{g^2}{8\pi^2}\gtrsim\frac{1}{N_\textnormal{S}},
\end{equation}
such that
\begin{equation}
    \frac{g^2}{8\pi^2}\gtrsim\frac{H^2}{M_\textnormal{P}^2}
\end{equation}
has to hold \cite{seo_axion_2024}.
From the \gls{flb}, we know that
\begin{equation}
    \frac{g^2}{8\pi^2}\lesssim\frac{m^4}{H^2M_\textnormal{P}^2},
\end{equation}
with $m$ the mass of a particle with U(1) charge;
together, the two bounds yield
\begin{equation}
    m\gtrsim H,
\end{equation}
i.e. the Hubble scale is a lower limit for the mass \cite{seo_axion_2024}, which might be reached by neutrinos, see e.g. the work by \citet{gonzalo_swampland_2022,hamada_weak_2017,ibanez_constraining_2017,casas_small_2024}.

\subsubsection{The SSC and the EP and the WGC}\label{rel:WGC_EP}
The \gls{wgc} can be obtained by using the \gls{ep} together with the species scale (\cref{s:ssc}) \cite{castellano_towers_2023,castellano_emergence_2023} (cf. \cite{kaufmann_asymptotics_2024,cota_minimal_2024,lee_tensionless_2018,lee_emergent_2022}):
For a 4d U(1) gauge theory with a massless field with no kinetic terms, the tower of states (a single \gls{kk} tower) goes as
\begin{equation}
    m_n^2=n^2m_\textnormal{t}^2
\end{equation}
with the charges $q_n=n\in\mathbb{Z}$.
The highest state of the tower that we can consider corresponds to the species scale ($\Lambda_\textnormal{S}^2\lesssim M_\textnormal{P}^2/N_\textnormal{S}$):
\begin{equation}
    \Lambda_\textnormal{S}\simeq N_\textnormal{S}m_\textnormal{t}.
\end{equation}
The 1-loop contribution to the wave function renormalisation by the tower is given by \cite{lee_tensionless_2018,palti_swampland_2019,castellano_towers_2023}
\begin{equation}\label{eq:WGC_EP-field_metric}
    \frac{1}{g^2}\sim\sum_n^{N_\textnormal{S}}n^2\log\left(\frac{\Lambda_\textnormal{S}^2}{m_\textnormal{t}^2n^2}\right)\sim N_\textnormal{S}^3\lesssim\frac{M_{\textnormal{P};d}^{d-2}}{m_\textnormal{t}^2}.
\end{equation}
We find that the bound on the mass scale of the tower corresponds to the magnetic \gls{wgc} $m_\textnormal{t}^2\lesssim g^2M_\textnormal{P}^2$.
The massless field metric is given by $g_{\phi\phi}$, and is in principle divergent at the 1-loop level, unless the cutoff is applied.\footnote{
    It is a legitimate question how we can end up with a seemingly classical result, if we start at the quantum level. The answer is that the \enquote{quantumness} (the factors of $\hbar$) of the \gls{ep} is cancelled by the \enquote{quantumness} of the species scale \cite{castellano_emergence_2023}. See also appendix A in the work by \citet{blumenhagen_emergence_2023}.
    \citet{castellano_quantum_2024} even speculates that this finding might indicate that it does not make sense to distinguish between \textit{classical} and \textit{quantum} gravity.
}
This relation between the coupling and the number of states shows that the number of states is finite\,\textemdash\,this is in agreement with the finite number of massless states conjecture (\cref{s:fnomf}) \cite{castellano_emergence_2023,tarazi_finiteness_2021,hamada_finiteness_2022,vafa_string_2005}.

\subsubsection{The SSC and the EP and the DC}\label{rel:EP_SDC}
\Cref{eq:WGC_EP-field_metric} can be recast as
\begin{equation}
    g_{\phi\phi}\lesssim M_{\textnormal{P};d}^{d-2}\left(\frac{\partial_\phi m_\textnormal{t}}{m_\textnormal{t}}\right)^2,
\end{equation}
from which we can directly compute the distance:
\begin{align}
    \sqrt{M_{\textnormal{P};d}^{2-d}}\Delta\phi&\sim\sqrt{M_{\textnormal{P};d}^{2-d}}\int_{\tau_a}^{\tau_b}\!\sqrt{g_{\phi\phi}\left(\partial_\tau\phi\right)^2}\,\mathrm{d}\tau\\
    &\sim\int_{\phi_a}^{\phi_b}\!\frac{\partial_\phi m_\textnormal{t}}{m_\textnormal{t}}\,\mathrm{d}\phi\\
    &\sim\log\left(\frac{m_\textnormal{t}(\phi_b)}{m_\textnormal{t}(\phi_a)}\right)\\
    \Rightarrow\frac{m_\textnormal{t}(\phi_a)}{m_\textnormal{t}(\phi_b)}&\sim e^{-\mathfrak{a}\sqrt{M_{\textnormal{P};d}^{2-d}}\Delta\phi},
\end{align}
where we qualitatively find the \gls{dc} in the last step \cite{castellano_emergence_2023,palti_swampland_2019}.
Also the inverse direction works: If we have emergence together with the \gls{wgc} and the \gls{dc}, we find the species scale \cite{castellano_emergence_2023}.
If we have the species scale, the \gls{wgc}, and the \gls{dc}, we find emergence \cite{castellano_emergence_2023}.
In more qualitative terms, we can say that the \gls{dc} and the \gls{ep} imply each other:\footnote{
    The tangent vector of the infinite distance limit in moduli space has to lie in the subspace of the tangent space generated by the light tower vectors \cite{etheredge_taxonomy_2024}. If this is not the case, the \gls{ep} does not imply the lower bound, but both conjectures, the \gls{dc} and the \gls{ep}, could still hold.
}
the \gls{dc} says in an infinite distance limit, a massless tower of states will emerge, while the \gls{ep} says that when an infinite tower of particles becomes massless exponentially fast, we see an infinite distance singularity in moduli space \cite{marchesano_eft_2023}.
However, even if one leads to the other, we don't arrive there using the same street: while it is sufficient to focus on a single leading tower to test the \gls{dc}, to test the \gls{ep}, control over the full spectrum of particles below the species scale is needed \cite{marchesano_eft_2023}.
In practice, there will be a cutoff in tower masses that have to be taken into account when calculating the species scale, as taking into account even heavier towers will not lift the species scale above the tower mass of this heavier towers, i.e. they can be neglected.\footnote{See our comments in \cref{p:SSC_multiple-towers}.}
In so far, the \gls{ep} could be regarded as a refinement of the \gls{dc}, as the \gls{ep} specifies what kind of towers emerge in the infinite distance limit, namely \gls{kk}-towers or string oscillator modes \cite{etheredge_taxonomy_2024,kaufmann_asymptotics_2024}.

The \textit{Distant Axionic String Conjecture} states that in a 4d \gls{eft}, infinite distance limits are characterised by axionic strings that become tensionless respectively the infinite towers of states that emerge are accompanied by a string that is magnetically coupled to an axion \cite{grimm_tameness_2022,lanza_eft_2021,lanza_swampland_2021}: Axion strings have an infinite tower of oscillatory modes and generate backreactions on moduli fields. When moving to an infinite distance limit, such a string becomes tensionless, and the oscillatory modes generate the infinite tower of states predicted by the \gls{dc}.\footnote{
    \citet{hamada_axion_2019} note that axion backreactions are always negligible in the \gls{uv}, but have important effects in the \gls{ir}.
    }

\citet{castellano_stringy_2023} provide evidence for a statement that differs from though is closely related to the \gls{dc}:
\begin{align}
    g^{ij}\left(\partial_i\ln m\right)\left(\partial_j\ln\Lambda_\textnormal{S}\right)&=\frac{1}{d-2}\\
    g^{ij}\left(\partial_i\ln m\right)\left(\partial_j\ln N_\textnormal{S}\right)&=-1,
\end{align}
where $g^{ij}$ is the moduli space metric, and $\Lambda_\textnormal{S}=M^d_\textnormal{P}/N_\textnormal{S}^{\frac{1}{d-2}}$
is used to derive the dimension-independent bound that only depends on the number of states $N_\textnormal{S}$\,\textemdash\,a peculiar finding that might hint at some underlying structure that governs the particle towers in the infinite field distance limit \cite{castellano_stringy_2023}. The more fields we get, the slower the lightest tower's mass scale decays \cite{castellano_stringy_2023,castellano_universal_2023}.

How intricately related the \gls{ssc} and the \gls{dc} are, becomes obvious in the relation presented in \cref{eq:mass-SSC-dynamics}: the exponential mass suppression predicted by the \gls{dc} must be matched by an exponential change in the species scale, to fulfil the equality.
In the infinite field distance limit, a tower of states becomes light, such that an infinite number of species has to be taken into account, such that the \gls{eft} description breaks down, as the cutoff scale $\Lambda_\textnormal{S}$ runs to zero \cite{calderon-infante_emergence_2024}.
The underlying reason could be that, according to the equivalence principle, gravity couples to everything in the same fashion and cannot be tuned, as it is given by the energy\textendash momentum distribution of matter \cite{basile_shedding_2024}.
At the species scale, the coupling of the graviton is of order one and an infinite tower of states becomes light, as described by the \gls{dc}.

Expressing everything in terms of entropy might shed some light on the matter: the \gls{dc} may be considered a statement about the species respectively entropy distance \cite{cribiori_species_2023}:
In \cref{eq:SSCDC,eq:SSCBHEDC,eq:SSCEntropy,eq:SSCEntropyChange} we show that the \gls{ssc} yields the \gls{bhedc} when combined with the \gls{dc}.
The tower of light states predicted by the \gls{bhedc} can be regarded as a description of the degrees of freedom for the transitions between \gls{bh} microstates, which are completely degenerated in the large entropy limit; the same holds for the tower of states in the large species scale (entropy) limit.

\subsubsection{The SSC and the Finite Number of Massless Fields Conjecture}\label{rel:SSC_fnomfC}
The \gls{ssc} and the finite number of massless fields conjecture (\cref{s:fnomf}) are perfectly compatible:
\begin{equation}
    \Lambda_\textnormal{S}=\frac{M_{\textnormal{P};d}}{N_\textnormal{S}^{\frac{1}{d-2}}};
\end{equation}
an infinite number of fields predicts a vanishing cutoff scale and the \gls{eft} is not valid anywhere.

\subsubsection{The Finite Number of Massless Fields Conjecture and the DC}\label{rel:fnomfC_DC}
A connection between the finite number of massless fields conjecture (\cref{s:fnomf}) and the \gls{dc} can be found in the work of \citet{corvilain_swampland_2019,hamada_finiteness_2022}: If the number of species increases when a field travels in moduli space, then an infinite tower of light fields is integrated out when the field travels an infinite distance. Both indicate the breakdown of the \gls{eft}. Note that an infinite number of states is not required to obtain an infinite distance \cite{hamada_finiteness_2022}.

\subsubsection{The Finite Number of Massless Fields Conjecture and the Tameness Conjecture}\label{rel:fnomfc_TC}
The tameness conjecture (\cref{sec:tame}) implies that the number of compact \gls{cy} manifolds is finite, in line with the finite number of massless fields conjecture (\cref{s:fnomf}).
Tameness implements a generalised principle of finiteness at a fundamental level \cite{douglas_tameness_2023}. The concept of tameness can be used to prove that the number of flux vacua in certain string compactifications is finite \cite{douglas_tameness_2023,bakker_finiteness_2023}.

\subsubsection{The Tameness Conjecture and the DC}\label{rel:TC_DC}
The tameness conjecture (\cref{sec:tame}) forbids \gls{eft} geometries with an infinite number of singularities, i.e. it guarantees that there is a finite number of partitions respectively a finite number of infinite towers of states \cite{grimm_tameness_2022,lanza_machine_2024}. However, tameness alone does not prove the \gls{dc} \cite{grimm_tameness_2022}. It does, however, indicate that a finite number of towers suffices to always have a defined cutoff scale of the \gls{eft} \cite{lanza_machine_2024}.

\subsubsection{The DC and the TCC}\label{rel:TCC_DC}

\citet{andriot_web_2020} reason that the parameter $\mathfrak{a}$ of the \gls{dc} is equal to half the \gls{dsc} parameter and therefore also equal to half the \gls{tcc} bound: $\mathfrak{a}=\mathfrak{s}_1/2=1/\sqrt{\left(d-1\right)\left(d-2\right)}$.\footnote{
    See \cref{rel:dSC_DC} for further comments on this relation.
}
However, a fundamental reason for this observation cannot be presented, it is, however, shown that this holds in various type IIA and type IIB string theories.

\subsubsection{The TCC and the Non-Negative NEC Conjecture}\label{rel:TCC_nnNECC}
\citet{alexander_sitter_2024} remark that combining the \gls{tcc} with the \gls{nec} strengthens the support for describing \gls{ds} space as a Glauber\textendash Sudarshan state in SO(32) heterotic string theory.

\subsubsection{The TCC and the dSC}\label{rel:TCC_dSC}
While the \gls{tcc} sets a bound on the derivative of the potential of an individual scalar field (taking the derivative with respect to the geodesic in field space), the \gls{dsc} considers all scalar fields in a theory by taking the gradient of the potential \cite{rudelius_asymptotic_2021}. In other words \cite{agmon_lectures_2023}: the \gls{tcc} constrains the shape of the potential over a long field range, while the \gls{dsc} has point-wise implications.
This also indicates that the \gls{dsc} rules out meta-stable \gls{ds} vacua in the regime of parametric control, while the \gls{tcc} constrains the lifetime of such a vacuum away from that regime \cite{van_riet_beginners_2023}.\footnote{
    It also means that the \gls{tcc} allows that the \gls{dsc} is violated point-wise, respectively, over limited field ranges, as we point out around \cref{eq:tPCC-dSC-violation}.
    }

\citet{bedroya_trans-planckian_2020} derive a global version of the \gls{dsc} from the \gls{tcc}:
\begin{equation}
    \left. \expval{\frac{-V^\prime}{V}}\right|_{\phi_\textnormal{i}}^{\phi_\textnormal{f}}>\frac{1}{\Delta\phi}\log(V_\textnormal{i}/m)+\frac{2}{\sqrt{\left(d-1\right)\left(d-2\right)}}
\end{equation}
with $\left. \expval{-V^\prime/V}\right|_{\phi_\textnormal{i}}^{\phi_\textnormal{f}}$ being the average of the relative change in the potential over the interval $\left[\phi_\textnormal{i},\phi_\textnormal{f}\right]$ and $m$ a mass scale \cite{blumenhagen_quantum_2020}.
Furthermore, they show that the \gls{tcc} also supports \cref{eq:dScrefined}: $V^{\prime\prime}$ has to be large around a minimum because if the field stays too long around the minimum, the \gls{tcc} might get violated.
The other direction is shown by \citet{brahma_trans-planckian_2020}: under the assumption that $\Delta\phi\rightarrow\infty$ they derive the \gls{tcc} from the \gls{dsc}, without making any assumptions about quantum fluctuations crossing the horizon.

The combination of the \gls{dsc} and the \gls{tcc} has strong implications for single-field slow-roll inflation, since the slow-roll parameters
\begin{align}
    \epsilon_V&=\frac{M_\textnormal{P}^2}{2}\left(\frac{\partial_\phi V}{V}\right)^2\\
    \eta_V&=M_\textnormal{P}^2\frac{\partial^2_\phi V}{V},
\end{align}
resemble the conditions of the \gls{dsc}, which tells us that one of the two must be large. From the \gls{tcc} we can deduce that $\epsilon$ must be small. Therefore, $\eta$ must be large, which is generally true for plateau models of inflation \cite{lehners_small_2024}.
If a single scalar field with $V\sim e^{-\mathfrak{s}_1\phi}$ is the main driver of expansion, then the \gls{tcc} suggests that
\begin{equation}
    \mathfrak{s}_1\geq2/\sqrt{d-2},
\end{equation} unless a tower of light states that contributes to the expansion is present as well, then
\begin{equation}
    \mathfrak{s}_1\geq2/\sqrt{\left(d-2\right)\left(d-1\right)}
\end{equation} should hold, except the \gls{ep} holds as well, according to which a tower of light states is always heavier than the Hubble scale, such that the tower cannot contribute to the expansion significantly, and
\begin{equation}
    \mathfrak{s}_1\geq2/\sqrt{d-2}
\end{equation} was actually the correct bound \cite{bedroya_holographic_2022}.

\subsubsection{The dSC and the DC}\label{rel:dSC_DC}
\citet{barrau_string_2021} mention a loose chain of arguments that show that the \gls{dsc} follows from the \gls{dc}: The distance a field can travel is limited by the \gls{dc} to a Planckian-distance. A field that travels a trans\textendash Planckian range is interpreted as a decompactification of a dimension. The number of dimensions of the Hilbert space of the theory is, potentially, linked to the entropy of the space. The entropy is used to define temperature in thermodynamics. Requiring a positive temperature is equivalent to putting a bound on the relative change of the potential. The \gls{dsc} describes such a bound. \citet{brahma_trans-planckian_2020} stresses that a monotonically increasing entropy is crucial for this argument.

A more quantitative link between the \gls{dsc} and the \gls{dc} is presented by \citet{etheredge_sharpening_2022}: According to their refined \gls{dc}, a tower of light states emerges when a scalar field $\phi$ travels a distance $\Delta\phi$ in field space. The mass scale of this tower behaves like
\begin{equation}
    m\lesssim\exp\left(-\frac{\abs{\Delta\phi}}{\sqrt{d-2}}\right)M_\textnormal{P}^d.
\end{equation}
For trans-Planckian field ranges, the \gls{eft} has a cutoff of
\begin{equation}
    \Lambda_\textnormal{S}\lesssim\exp\left(-\frac{\abs{\Delta\phi}}{\left(d-1\right)\sqrt{d-2}}\right)M_\textnormal{P}^d,
\end{equation}
which is expected to be above the \gls{ir} Hubble scale, i.e.
\begin{align}
    H\sim\sqrt{V}&\lesssim\Lambda_\textnormal{S}\\
    V&\lesssim\exp\left(-\frac{2\abs{\Delta\phi}}{\left(d-1\right)\sqrt{d-2}}\right)M_\textnormal{P}^d\\
    \Rightarrow\frac{\abs{\Delta V}}{V}&\geq\frac{2}{\left(d-1\right)\sqrt{d-2}},
\end{align}
where we found a slightly weaker lower bound for the \gls{dsc} than in \cref{eq:dS-dim-red}. \citet{etheredge_sharpening_2022} strengthen their result and derive \cref{eq:dS-dim-red} as well, by invoking the \gls{ep} that states that an infinite-distance limit has to be an emergent string limit or a decompactification limit, and that
\begin{equation}
    H\lesssim\exp\left(-\frac{\abs{\Delta\phi}}{\sqrt{d-2}}\right)M_\textnormal{P}^d
\end{equation}
must hold for a consistent \gls{eft}. Ergo, their sharpened \gls{dc}, combined with the \gls{ep}, indicates the strong asymptotic \gls{dsc}.
Based on work by \citet{bedroya_trans-planckian_2020} and \citet{rudelius_asymptotic_2022}, \citet{cremonini_asymptotic_2023} show the reverse, namely that a strong form of the \gls{dsc} would imply the sharpened \gls{dc} (yet only in one-dimensional moduli spaces). They quantify the relation between the \gls{dsc} and the \gls{dc}, and examine the relation $\mathfrak{s}_1=2\mathfrak{a}$.
Such a relation is also studied by \citet{bedroya_sitter_2020,andriot_web_2020,hebecker_asymptotic_2019}.
A fundamental reason for why $\mathfrak{a}=\mathfrak{s}_1/2$ should hold remains elusive, but the observation that it is also equal to half the \gls{tcc} bound (see \cref{rel:TCC_DC,sec:tpcc}) hints at a deeper connection between those swampland conjectures.
\citet{andriot_web_2020} are motivated by this finding to propose a bound on the derivative of the mass of the form
\begin{equation}
    \expval{\frac{\abs{m^\prime}}{m}}_{\Delta\phi\rightarrow\infty}\geq\mathfrak{a},
\end{equation}
in analogy to the \gls{dsc}. Furthermore, this very bound can be seen as a generalisation of the scalar \gls{wgc}, when written as $\left(\partial_\phi m\right)^2\geq\mathfrak{a}^2m^2$ \cite{andriot_web_2020}.
Moreover, they propose a map between the mass and the potential that is supposed to hold in the infinite field distance limit: $m/m_0\simeq\sqrt{\abs{V/V_0}}$, with $m_0$ and $V_0$ constant.
Such a relation is certainly surprising, since the \gls{dc} is mostly treated in \gls{cy} manifolds and their compactification to flat space, while the \gls{dsc} is a statement about curved \gls{ds} space \cite{andriot_web_2020}. 

\citet{seo_stability_2023} investigates the stability of a \gls{ds} solution in the presence of either one extremely light state or several towers of similar mass under the assumption that the \gls{dc} and the \gls{ep} hold, and finds that the \gls{dsc} is satisfied if the lightest tower is fermionic, but not if the lightest tower is bosonic.

Using the Lyth bound, the \gls{dsc}, as well as the \gls{dc}, the following relation can be derived \cite{matsui_eternal_2019}:
From the \gls{dc} we know that $\Delta\phi\lesssim\mathfrak{l}M_\textnormal{P}$, with $\mathfrak{l}\sim\order{1}$.
From the Lyth bound we know that $\Delta\phi=\sqrt{2\epsilon_V} N_e$
From the \gls{dsc} we know that the slow-roll parameter can be expressed as $\epsilon_V>\mathfrak{s}_1^2/2$.
Taking everything together, we find that $N_e\lesssim\mathfrak{l}/\mathfrak{s}_1$, which highlights the incompatibility of eternal inflation with the swampland conjectures. This assumes that $\mathfrak{s}_1\sim\order{1}$ holds as well, but see our discussion in \cref{p:dSC_c_value} about the value of $\mathfrak{s}_1$.

The \gls{dsc} bounds are supported by Bousso's covariant entropy bound \cite{bousso_covariant_1999} as long as entropy is dominated by an increasing number of states, as predicted by the \gls{dc} \cite{seo_sitter_2019,ooguri_distance_2019,ooguri_geometry_2007,brahma_trans-planckian_2020}.

\citet{castellano_gravitino_2021} argue for a link between their gravitino distance conjecture and the \gls{dsc}, when the Higuchi bound is considered: unitarity of massive representations of higher spin particles imposes a bound on the lightest \gls{kk} tower\,\textemdash\,which could be the gravitino tower\,\textemdash\,and the bound reads
\begin{equation}
    m_\textnormal{3/2}^{2\mathfrak{g}}\simeq m_\text{KK}^2\geq2H^2=2\frac{V_0}{M_\textnormal{P}^2}
\end{equation}
and establishes a direct connection between the gravitino mass going to zero and the potential going to zero, which \enquote{comes along with a tower of massless states}. We wouldn't agree with the quoted statement, as the \gls{dsc} makes no statement about a vanishing potential\,\textemdash\,only the derivative of the potential is of relevance to the \gls{dsc}, but the potential itself is allowed to vanish. This breaks the chain of arguments, and this link between the gravitino \gls{dc} and the \gls{dsc} cannot be established.

For warm inflation, a bound on the dissipation rate $\Upsilon$ is given by
\begin{equation}
    \Upsilon\gtrsim\frac{\mathfrak{s}_1N_e}{\Delta\phi/M_\textnormal{P}}-1
\end{equation}
to assure that the \gls{dc} as well as the \gls{dsc} are satisfied \cite{motaharfar_warm_2019}.
A range for the tensor-to-scalar ratio $r_\textnormal{ts}$ is bound by the two conjectures:
\begin{equation}
    \frac{8\mathfrak{s}_1^2}{\left(1+\Upsilon\right)^2}\frac{1}{\mathcal{F}}<r<\frac{8\left(\Delta\phi/M_\textnormal{P}\right)^2}{N_e^2}\frac{1}{\mathcal{F}},
\end{equation}
with
\begin{AmSalign}
    \mathcal{F}(k/k_*)&\defeq\\
    &\left(1+\frac{2}{\exp(H_*/T_*)-1}+\frac{2\pi\sqrt{3}\Upsilon_*}{\sqrt{3+4\pi\Upsilon_*}}\frac{T_*}{H_*}\right)g(\Upsilon_*),\nonumber
\end{AmSalign}
where $g(\Upsilon_*)$ is the radiation\textendash inflaton coupling, and the asterisk indicates the value at horizon crossing \cite{motaharfar_warm_2019}. Cold inflation with $\Upsilon=0$ is obviously in tension with the combination of the two conjectures \cite{motaharfar_warm_2019}.

\citet{storm_swampland_2020} show that whenever the refined \gls{dsc} is satisfied for a slowly rolling thawing quintessence field, the \gls{dc} is satisfied as well (the converse is not true). Similar is shown by \citet{raveri_swampland_2019}, but they also point out that thawing models generally increase the Hubble tension.

A derivation of the \gls{dsc} that connects the \gls{dc} with the \gls{ssc} is presented by \citet{hebecker_asymptotic_2019}:
They start with a single tower of light states of equal spacing with $m=M_\textnormal{P}\exp(-\mathfrak{a}\phi)$, as predicted by the \gls{dc}, below the species scale
\begin{equation}
    \Lambda_\textnormal{S}=M_\textnormal{P}/\sqrt{N}\approx\left(M_\textnormal{P}^2m\right)^{1/3}\approx M_\textnormal{P}e^{-\mathfrak{a}\phi/3}.
\end{equation}
A positive scalar potential $V(\phi)$ introduces a curvature scale $\sim H$ such that $3H^2M_\textnormal{P}^2=V$ holds.
Perturbative control is achieved as long as $\Lambda_\textnormal{S}\gtrsim H$ holds, i.e.
\begin{equation}
    M_\textnormal{P}^2e^{-2\mathfrak{a}\phi/3}\gtrsim V/M_\textnormal{P}^2.
\end{equation}
The \gls{dsc} follows right away, however, as we only worked approximately, with unspecified constants, we find only an upper bound. Furthermore, the derivation relies on strong assumptions about the potential, e.g. that it does not have an oscillating term and that the number of species undergoes an exponential growth.
However, the potential itself faces no constraints about being small or large.\footnote{
    \citet{bena_uplifting_2019} find that the \gls{dsc} is satisfied in their \gls{kklt} scenario with warped throats and anti-brane uplift, not only for large values of $V$, but also for parametrically small values of the potential.
    In such a scenario, the cosmological constant we measure is the difference between negative energy contributions from an \gls{ads} minimum used for moduli stabilisation and positive contributions from an anti-brane, such that $\Lambda_\textnormal{cc}=V_{\overline{D3}}-V_\textnormal{AdS}$ \cite{akrami_landscape_2019}.
}

\subsubsection{The DC and the Cobordism Conjecture}\label{rel:DC_Cobordism}
\citet{angius_at_2022} find that infinite field distance limits of consistent \glspl{eft} can be realised as solutions approaching an \gls{etw} brane, which shows the general compatibility between the \gls{dc} and the cobordism conjecture (\cref{sec:cobordism}).

If the \gls{dc} is recast in terms of Ricci flows, it can be conciliated with the cobordism conjecture, which ultimately links distances in moduli field space to global symmetries \cite{velazquez_cobordism_2022}.\footnote{
    This is a surprising connection: Ricci flows deal with \textit{locally} defined geometric flow equations and properties \cite{velazquez_cobordism_2022}.
    }

\subsubsection{The Cobordism Conjecture and the No non-Supersymmetric Theories Conjecture}\label{rel:Cobordism_nnSUSYC}
An ongoing discussion is the compatibility of metastable \gls{ds} vacua with the no non-supersymmetric theories conjecture (\cref{sec:nononSUSY}). Studying the transition of vacua into bubbles of nothing, \citet{seo_kaluza-klein_2024} find a possible counterexample: Compactifying 6-dimensional Einstein\textendash Maxwell theories onto $S^2$, they find that the \gls{ds} vacuum has allowed transition paths into a bubble of nothing (in accordance with the cobordism conjecture (\cref{sec:cobordism})), which is equivalent to an infinitely large \gls{ads} space, i.e. a transition from \gls{ds} to \gls{ads} without the emergence of a tower of \gls{kk}-states. This would be a contradiction to the \gls{dc}, which requires the emergence of a \gls{kk} tower, and to the \gls{adsdc}, which states that \gls{ds} and \gls{ads} spaces are an infinite distance apart. However, if the conjectures are applicable or not depends on the transition path, respectively on the flux $f$ and uplift $\lambda$ and if the two are related as $f^2\lambda=\text{cst.}$, a question that remains unanswered in the work of \citet{seo_kaluza-klein_2024}.

\subsubsection{The Cobordism Conjecture and the CC}\label{rel:CC_Cobord}
It is claimed that completeness (\cref{sec:complet}) follows from the cobordism conjecture (\cref{sec:cobordism}) \cite{mcnamara_cobordism_2019,mcnamara_gravitational_2021}. 
A bordism preserves fluxes, but to be cobordant to the empty set, the fluxes have to be cancelled by defects (charges), which implies that the charge spectrum must be complete to cancel all possible fluxes, ergo, the cobordism conjecture implies the \gls{cc} \cite{andriot_looking_2022}.
However, this claim is countered by the observation that symmetries arising from cobordism groups are always Abelian and group-like, whereas the symmetries in \glspl{eft} can be non-Abelian and non-invertible \cite{rudelius_topological_2020,heidenreich_non-invertible_2021,mcnamara_gravitational_2021,dierigl_swampland_2021}. \citet{mcnamara_gravitational_2021} shows that the cobordism group measures the global symmetry breaking regardless of topology fluctuations\,\textemdash\,a statement in tension with the claim that all higher-order symmetries are broken by topology fluctuations \cite{yonekura_topological_2021}.

\subsubsection{The Cobordism Conjecture and the no Global Symmetries Conjecture}\label{rel:Cobord_nGSym}
If the \gls{cc} or the no global symmetries conjecture (\cref{sec:nGSym}) is violated, the cobordism class is non-trivial and the cobordism conjecture (\cref{sec:cobordism}) is violated as well \cite{hamada_8d_2021}. A non-trivial bordism group represents a conserved global charge that \textit{labels} different backgrounds in the form of different topological operators, i.e. a vanishing bordism group means no global symmetries \cite{debray_chronicles_2023,blumenhagen_dimensional_2023}, and no global symmetries mean a trivial bordism group \cite{agmon_lectures_2023}.

\subsubsection{The No Global Symmetries Conjecture and the CC}\label{rel:CC_nGSymC}
That there is a relation between the completeness of the spectrum and the absence of global symmetries can be seen when studying the gauge spectrum of a gauge theory, 
e.g. in a U(1) Maxwell theory without dynamical electric charges, there is a U(1) 1-form global symmetry \cite{harlow_symmetries_2019,choi_non-invertible_Gauss_2023}.
Similarly, there is a global 1-form symmetry if there are no fundamental quarks in a SU(N) gauge theory \cite{harlow_symmetries_2019}.

It is commonly acknowledged that the no global symmetries conjecture (\cref{sec:nGSym}) implies the \gls{cc} \cite{draper_snowmass_2022,harlow_symmetries_2019,rudelius_topological_2020,heidenreich_non-invertible_2021}.
The presence of charged matter in every representation of a compact and connected or finite and Abelian gauge group $G$ is equivalent to the absence of a 1-form symmetry under which Wilson lines are charged \cite{heidenreich_non-invertible_2021,harlow_weak_2023}.\footnote{
    In terms of topological defects, this means that a spectrum is complete, if all Wilson lines end on the fields of electrically charged particles, which in turn implies that the electric 1-form symmetry is broken, as otherwise some Wilson lines would link the electric defect \cite{choi_non-invertible_Gauss_2023}.
    }
However, a gauge group can be discrete and non-Abelian or continuous but disconnected, and the spectrum can then be incomplete without a global symmetry arising \cite{rudelius_topological_2020,harlow_weak_2023}.
Furthermore, the opposite direction contains loopholes as well:
\citet{heidenreich_non-invertible_2021} show that the absence of invertible global symmetries does not necessarily imply that the spectrum of the theory is complete. 
However, the absence of non-invertible topological operators implies the \gls{cc} \cite{rudelius_topological_2020,heidenreich_non-invertible_2021,mcnamara_gravitational_2021}.

\subsubsection{The No Global Symmetries Conjecture and the AdSDC/DC}\label{rel:nGSC_AdSDC}
The appearance of a light tower of states is censoring \gls{qg} from having a global symmetry, which is substantiated by the \gls{adsdc}, the \gls{dc}, the \gls{wgc}, and directly addressed by the no global symmetries conjecture (\cref{sec:nGSym}) \cite{montero_pure_2023,harada_further_2022}.
\citet{grimm_infinite_2018} show that the emergence of global symmetries is blocked in string theory because global symmetries would emerge at infinite distance loci in field space, but \glspl{eft} break down at infinite distance singularities, preventing the emergence of global symmetries. Following the reasoning of \citet{mcnamara_cobordism_2019}, the absence of global symmetries enforces a trivial cobordism class for landscape theories. A generic feature of trivial cobordisms are domain walls. This requires us to treat the distances between points in the moduli space as discrete, i.e. we must use the distance definition of \citet{basile_domain_2023}, as outlined in \cref{sec:distance}. A different aspect of the same observation is highlighted by \citet{corvilain_swampland_2019}: The infinite tower of states that is predicted to emerge when a field travels an infinite distance in field space is identified with an initially discrete charge orbit, which becomes a continuous shift symmetry at infinite distance, i.e. a global symmetry. Since the \gls{eft} breaks down as stated by the \gls{dc}, no such global symmetry arises.

\subsubsection{The AdSDC and the FLB}\label{rel:AdSDC_FLB}
The \gls{adsdc} as well as the \gls{flb} make a statement about the cosmological constant and leptons. Applied to electrons and neutrinos, we find that
\begin{equation}
    m_e^2\geq g\Lambda_4^{1/2}\gtrsim gm_\nu^2,
\end{equation}
which puts an upper bound on the gauge coupling $g\lesssim\frac{m_e^2}{m_\nu^2}$ \cite{gonzalo_swampland_2022}.

\subsubsection{The AdSDC, the dSC, and the No Non-SUSY Conjecture}\label{rel:AdSDC_nnSUSYC}
Bringing together the two \gls{ads} conjectures, the \gls{adsdc} and the no non-SUSY conjecture (\cref{sec:nononSUSY}), with the \gls{dsc}, only Dirac neutrinos with normal hierarchy are compatible without fine-tuning\,\textemdash\,Majorana neutrinos or Dirac neutrinos in inverted hierarchy need a fine-tuning of the \gls{dsc} parameter $\mathfrak{s}_1\lesssim0.01$ \cite{gonzalo_swampland_2022}.

\subsubsection{The AdSDC and the WGC}\label{rel:AdSDC_WGC}
\citet{cribiori_weak_2022} motivate the \gls{adsdc} by the magnetic \gls{wgc} (\cref{sec:gravity,eq:mwgc}) and discuss that $\mathcal{N} = 2$ and $\mathcal{N} = 8$ \gls{ads} vacua with a residual gauge group containing an Abelian factor\,\textemdash\,which is needed to apply the \gls{wgc}\,\textemdash\,cannot be scale separated, and that the same reasoning also applies to theories with broken but extended residual supersymmetry\,\textemdash\,such that there are still gravitini charged under the U(1) gauge group\,\textemdash\,but not to certain $\mathcal{N} = 1$ models. They derive the following relation:
\begin{equation}
    \abs{\Lambda_\text{AdS}}\gtrsim q^2g^2M_\textnormal{P}^2\gtrsim q^2\Lambda_\text{UV}^2,
\end{equation}
with $\Lambda_\text{AdS}$ the vacuum energy and $\Lambda_\text{UV}$ the cutoff scale. This makes scale separation unachievable in models with more than 4 spacetime dimensions and more than $\mathcal{N}=1$ supersymmetry \cite{coudarchet_hiding_2024}.

\subsubsection{The WGC and the no non-SUSY Conjecture}\label{rel:WGC_nnSUSYC}
A strong form of the \gls{wgc}\footnote{
    The \gls{wgc} under consideration of massless scalar fields, as in \cref{eq:fswgc} respectively the \gls{rfc}.
    } (\cref{sec:gravity})
says that only \gls{bps} states in supersymmetric theories can saturate the \gls{wgc}-bound \cite{ooguri_non-supersymmetric_2017}, which means that only \gls{bps} \glspl{bh} can be marginally stable \cite{harlow_weak_2023}. The same reasoning can be applied more generally to $p$-form fields, with the conclusion that a non-supersymmetric \gls{ads} vacuum is unstable, when supported by fluxes \cite{harlow_weak_2023}. This is in line with the no non-supersymmetric theories conjecture (\cref{sec:nononSUSY}).
For example, in a 4d \gls{eft}, a D4-brane wrapping a two-cycle or a D8-brane wrapping the internal space could nucleate and trigger a non-perturbative instability if the charge of the brane is larger than its tension \cite{lanza_swampland_2021,lanza_how_2019}\,\textemdash\,which is precisely what the \gls{wgc} demands \cite{coudarchet_hiding_2024}.
Some authors claim that no vacua in string theory are stable \cite{freivogel_vacua_2016}, whereas others point out that the interpretation of the scalar \gls{wgc} / \gls{rfc} as an anti-\gls{bps} bound only holds if there are no \textit{fake supersymmetries} \cite{gendler_merging_2021,palti_weak_2017}.

\subsubsection{The WGC and the dSC}\label{rel:WGC_dSC}
While the strong scalar \gls{wgc} (\cref{eq:strongswgc}) is compatible with but independent of the \gls{dsc} (\cref{eq:dSc}) \cite{gonzalo_strong_2019}, the \gls{dsc} can be written as a convex hull conjecture for the scalar \gls{wgc} for membranes \cite{calderon-infante_asymptotic_2023}:
The \gls{dsc} (\cref{eq:dSc}) for an asymptotic scalar potential $V(\phi)=\sum_kV_k(\phi)$ is satisfied if the convex hull spanned by the \gls{ds} ratios $\upsilon_k^a=-\delta^{ab}e_b^i\partial_i V_k/V_k$, where $e_b^i$ are inverse vielbeins spanning an orthonormal basis, is outside the ball of radius $\mathfrak{s}_1$. While for the \gls{wgc} charged states have to span a convex hull that contains the unit ball, the \gls{ds} ratio points must span a convex hull completely outside of the unit ball, i.e. the unit ball is not contained in respectively excluded / separated / disconnected from the convex hull, as any point inside the unit ball violates the \gls{dsc}. The connection of the convex hull \gls{dsc} and the convex hull scalar \gls{wgc} comes from the observation that the \gls{ds} ratios correspond to the scalar charge-to-mass ratios of membranes,
\begin{equation}
    \vec{\upsilon}_k=-2\vec{e}^i\frac{\partial_i \mathcal{T}_k}{\mathcal{T}_k},
\end{equation}
which measure the ratio between the scalar force and the gravitational force.
The \gls{dsc} corresponds to a scalar \gls{wgc} where the condition that gravity is the weakest force applies to all membranes, which is a much stronger condition than the requirement to have at least one such membrane. This indicates that the \gls{dsc} with the requirement of $\mathfrak{s}_1\sim\order{1}$ might be too strong. Knowing this, it is no surprise that membranes which satisfy the \gls{wgc} generate scalar potentials which satisfy the \gls{dsc}, as shown by \citet{lanza_swampland_2021}.

\subsubsection{The WGC and the FLB}\label{rel:WGC_FLB}
Whereas the \gls{wgc} puts an upper bound $m<gM_\textnormal{P}$ on charged particles, the \gls{flb} is a lower mass bound on charged particles. Using the fine structure constant $\alpha=g^2/4\pi$, we can write the bounds as
\begin{equation}
    \SI{e-3}{\electronvolt}\sim\left(8\pi\alpha\rho_\Lambda\right)^{1/4}<m<\left(8\pi\alpha\right)^{1/2}M_\textnormal{P}\sim\SI{e26}{\electronvolt},
\end{equation}
where the charge was normalised to the \gls{wgc} charge \cite{montero_fl_2021} and the numerical values are obtained for a U(1) charge in our Universe \cite{chrysostomou_reissner-nordstrom_2023}.
This can be re-expressed as a bound on the fine structure constant \cite{montero_fl_2021}:
\begin{equation}
    \alpha\gtrsim\frac{\rho_\Lambda}{8\pi M_\textnormal{P}^4}.
\end{equation}

For axions, the proposed \gls{flb} is
\begin{equation}
    S_\iota f\gtrsim\sqrt{M_\textnormal{P}H},
\end{equation}
while the \gls{wgc} demands
\begin{equation}
    S_\iota f\leq M_\textnormal{P},
\end{equation}
which limits the range of the product of the instanton action and the axion decay constant to
\begin{equation}
    \sqrt{M_\textnormal{P}H}\lesssim S_\iota f\leq M_\textnormal{P}.
\end{equation}

\subsubsection{The WGC and the No Global Symmetries Conjecture}\label{rel:WGC_nGSC}
The \gls{wgc} is motivated by the no global symmetries conjecture (\cref{sec:nGSym}): In the limit of vanishing coupling ($g=0$), a gauge symmetry becomes indistinguishable from a global symmetry. The \gls{wgc} prevents this by indicating that the \gls{eft} ceases to be valid in this limit, as the cutoff scale $\Lambda=0$ indicates that this theory is not valid anywhere.\footnote{
    See also \citet{cordova_generalized_2022} for a motivation of the \gls{wgc} from the no global symmetries conjecture.
}

\citet{fichet_approximate_2020} propose that an \gls{eft} with a global symmetry will be UV-completed into a higher-energy \gls{eft} without global symmetries, where the rate of global symmetry violations by \glspl{bh} is lower than the rate of global symmetry violations by local processes.
The main idea here is that local symmetry-violating effects are at least as important as Boltzmann-suppressed \gls{bh} effects.
They coin it the \textit{Swampland Symmetry Conjecture} and show that it holds whenever the \gls{lwgc} holds. Since there are counterexamples to the \gls{lwgc}, it is important to note that the \gls{lwgc} is a sufficient condition for their conjecture, yet not a necessary condition. The swampland symmetry conjecture can still hold, even in cases the \gls{lwgc} is violated.
In its current form, the swampland symmetry conjecture is not a particularly constraining conjecture. If the rate bound is not satisfied by an \gls{eft}, the swampland symmetry conjecture only demands that it has to be satisfied by a higher-energy extension of this \gls{eft}, which is still a sub-Planckian \gls{eft}.

\subsubsection{The WGC and the CC}\label{rel:WGC_CC}
The \gls{cc} can be viewed as a refinement of the \gls{slwgc}, as each point in the postulated lattice has to be populated \cite{lee_tensionless_2018}. This has further consequences, namely that formulas for the masses of those state should be valid far away from the infinite distance limits \cite{lee_tensionless_2018} and would make the corresponding conjecture applicable in the bulk as well.

\subsubsection{The WGC and the DC}\label{rel:WGC_DC}
As it is often the case in the Swampland, several conjectures make the same predictions for a model from different angles. For example, a charged \gls{kk} bubble, stabilised by flux, can create arbitrarily large scalar field variations \cite{horowitz_tachyon_2005,harlow_weak_2023}. The radion field traverses an infinite distance in field space to the bubble wall \cite{harlow_weak_2023}. The \gls{dc} predicts an infinite tower of states under this behaviour. The \gls{wgc} predicts the instability of the solution due to Schwinger pair production of charged matter \cite{harlow_weak_2023,draper_gravitational_2019}.

\citet{palti_weak_2017,palti_swampland_2019} presents a link between the \gls{wgc} and the \gls{dc}. Applying the \gls{wgc} to a canonically normalised scalar field $\phi$ leads to a relation for the mass (see \cref{eq:swgc}),
\begin{equation}
    \abs{\partial_\phi m}>m,
\end{equation}
which is violated for any power-law $m\sim\phi^k$ and sufficiently large field values. Therefore, $m\sim\exp(-\mathfrak{a}\phi)$ with $\mathfrak{a}>1$ is required: the mass behaviour of the \gls{dc}. He concludes that \enquote{a particle must have exponentially decreasing mass if gravity is to remain the weakest force acting on it even for large scalar expectation values.} This weak gravity conjecture resembles the \gls{dc}.
In all known string compactifications, gauge couplings show an exponential decay in the asymptotic limits, such that
\begin{equation}
    \Lambda\leq g\sim e^{-\mathfrak{a}\phi};
\end{equation}
the cutoff signals an infinite tower of light states, ergo the exponential suppression reproduces the \gls{dc} \cite{hamada_finiteness_2022}.
Similar conclusions are presented by \citet{grimm_infinite_2018,heidenreich_evidence_2017,grana_swampland_2021,heidenreich_emergence_2018}: Both, the \gls{dc} and the \gls{wgc} make a statement about the cutoff scale of an \gls{eft}. Whereas the \gls{dc} makes a statement about a tower of states, the \gls{wgc} makes a statement about individual particles \cite{etheredge_sharpening_2022}. Whereas the \gls{dc} only applies to infinite distances in field limit, the \gls{wgc} applies to all particles, including e.g. axions that can only undergo a finite displacement in field space. This shows that there is an overlap between the two conjectures, but also clear and distinct differences, such that one cannot imply the other \cite{etheredge_sharpening_2022}.
\citet{castellano_gravitino_2021} give another explicit example: a gravitino becoming massless can be interpreted as the membrane in a string compactification becoming tensionless. Applying the \gls{wgc} to membranes in 4 dimensions yields
\begin{equation}
    \frac{\mathcal{T}^2}{M_\textnormal{P}^2}\left(\frac{q}{m}\right)^2_\text{extremal}\leq g^2q^2.
\end{equation}
Therefore, a membrane becoming tensionless indicates that the charge vanishes. Once again, the \gls{dc} and the \gls{wgc} prevent a global symmetry (\cref{sec:nGSym}) from appearing, and do so in the same fashion.
Both conjectures act as censorship mechanisms for global symmetries at infinite distance points in the moduli field space: instead of recovering a global symmetry, the \gls{dc} predicts an infinite tower of light states; this also holds for infinite distance points where the charge of a state goes to 0, i.e. where the \gls{wgc} would saturate the bound \cite{font_swampland_2019}.

Considering multiple species, the \gls{wgc} as well as the \gls{dc} contain a statement about the minimal size of the convex hull encircling the states:
The \gls{wgc} demands that the convex hull of the charge-to-mass ratio vector $\Vec{z}=\Vec{q}\cdot M_\textnormal{P}/m$ contains the unit ball, with $\Vec{q}=\left(q_1g_1,\dots q_Ng_N\right)$.
The \gls{dc} demands that the convex hull of species scale vectors $\mathcal{Z}=-\delta^{ab}e_b^i\frac{\partial_i\Lambda_{\textnormal{S},\beta}}{\Lambda_{\textnormal{S},\beta}}$ with $\Lambda_{\textnormal{S},\beta}\sim m_\beta^{\frac{p_\beta}{d-2+p_\beta}}$ contains the ball with radius $r=1/\sqrt{\left(d-1\right)\left(d-2\right)}$, where $e_b^i(\phi)$ is the vielbein of the field space metric, $\Lambda_{\textnormal{S},\beta}$ is the species scale of the tower $\beta$, and $p$ is the density of the tower ($p=1$ for \gls{kk}, $p=\infty$ for strings) \cite{calderon-infante_entropy_2023}. The \gls{wgc} makes a local statement about one point in moduli space, whereas the \gls{dc} makes a statement about the boundary of moduli space: the \gls{dc} describes the asymptotic validity of the \gls{wgc}, and the \gls{dc} describes the exponential behaviour of a tower along its gradient flow trajectory \cite{etheredge_running_2023}.

A bottom-up perspective on the \gls{dc} and a derivation of the \gls{wgc} using \gls{bh} physics is presented by \citet{hamada_finiteness_2022}: starting with an Einstein\textendash Maxwell action with a scalar in four dimensions (the dilaton $\phi$),
\begin{equation}
    S=\int\!\sqrt{-g}\left(R+2\abs{d\phi}^2+\frac{1}{2g(\phi)^2}\abs{F}^2\right)\,\mathrm{d}^4x,
\end{equation}
with an arbitrary gauge coupling $g(\phi)$ for which $g(\phi)\rightarrow0$ for $\phi\rightarrow\infty$ holds, they show that small \glspl{bh} exist, independent of the particular dependence of $g(\phi)$, as long as it vanishes at infinity. At the core of such a \gls{bh}, the gauge coupling approaches zero. Since such a \gls{bh} becomes extremely small, the dilaton undergoes parametrically large field displacements near the horizon. Unless the \gls{dc} holds, such small \glspl{bh} violate the Bekenstein entropy bound. This can be seen as follows: The \glspl{bh} become almost point-like objects. As such, they represent states, which allows us to define a box of length $L$ and count the number of states that fit inside this box. Near the box boundary, we require gravity to be only weakly coupled. The area of the box grows like $L^2$, which means that the number of particles that fit inside the box cannot grow faster than $L^2$, otherwise entropy bounds were violated (the box cannot contain more entropy than the corresponding Schwarzschild \gls{bh}). Since the \glspl{bh} are considered to be point-like, gravitational effects that depend on their geometrical cross-section are negligible, therefore, each \gls{bh} of charge $Q$ counts as an additional species, and the total entropy is the sum over the contributions of those species. This sum is dominated by the contributions of light species with $m\ll T$, where the temperature is defined by $\frac{1}{T}=\partial_E S$, $E\sim L$ being the energy of the microcanonical ensemble. By going to sufficiently weak coupling, an arbitrary number of species is light enough, since $m=g(\phi_0)Q$.\footnote{
    $\phi_0$ being the asymptotic value of the dilaton \cite{hamada_finiteness_2022}.
    }
In 4 dimensions we have
\begin{align}
    S=NT^3L^3\\
    E=NT^4L^3\\
    \Rightarrow S=N^{1/4}E^{3/4}L^{3/4}=N^{1/4}L^{3/2}\leq L^2\\
    \Rightarrow N=Q_\text{max}\lesssim L^2,\label{eq:qmax}
\end{align}
with $Q_\text{max}$ the charge of the largest light \gls{bh}, which can be taken arbitrarily large for point-like \glspl{bh}, which then violates the bound from \cref{eq:qmax}.\footnote{This bound is not violated in an Einstein\textendash Maxwell theory \textit{without} the dilaton field, as there the extremal \glspl{bh} are \gls{rn} \glspl{bh} of finite size, i.e. $Q_\text{max}\propto L$ \cite{hamada_finiteness_2022}.} The violation is avoided, since the cutoff of the \gls{eft} $\Lambda$ gives the \glspl{bh} a non-zero effective size, i.e. there are no truly point-like \glspl{bh} and the entropy inside the box is bound. For the \gls{bh} solution to make sense, the gradient of the dilaton field has to be below the \gls{eft} cutoff. This means that we are forced us to stop a finite distance from the horizon and that the \gls{bh} has a finite effective area that can depend on the charge. The \gls{eom} for the dilaton is
\begin{equation}
    \Ddot{\phi}=\frac{\mathrm{d}g^2}{\mathrm{d}\phi}Q^2e^{2U},
\end{equation}
with $U$ an independent function, such that $e^{2U}$ is monotonic. Integrating this equation yields
\begin{equation}
    \dot{\phi}^2\geq\Delta g^2Q^2e^{2U},
\end{equation}
with $\Delta g^2$ the change of $g^2$ from its asymptotic value.
Using the Ansatz for the static spherically symmetric metric of the extremal electric solution of charge $Q$
\begin{equation}
    \mathrm{d}s^2=-e^{2U}\mathrm{d}t^2+e^{-2U}\left(\frac{\mathrm{d}\tau^2}{\tau^2}+\frac{1}{\tau^2}\mathrm{d}\Omega_2^2\right),
\end{equation}
where $\tau=0$ at infinity and $\tau=-\infty$ at the \gls{bh} horizon, gives
\begin{equation}
    \abs{\mathrm{d}\phi}^2=\tau^4e^{2U}\dot{\phi}^2\geq\frac{\Delta g^2Q^2}{A^2}
\end{equation}
for the gradient of the dilaton field, which diverges when the \gls{bh} area goes to zero. The \gls{eft} breaks down when
\begin{equation}
    \tau^4e^{2U}\dot{\phi}^2\sim\Lambda^2.
\end{equation}
From that, \citet{hamada_finiteness_2022} derive
\begin{equation}
    \Lambda\leq g\frac{Q}{A},
\end{equation}
and since $Q_\text{max}\lesssim L^2$ and $A\sim L^2$, such that $Q/A\leq1$, they find
\begin{equation}
    \Lambda\leq g,
\end{equation}
which is the magnetic \gls{wgc} \cref{eq:mwgc}.

The tower scalar \gls{wgc} is related to the \gls{dc}: the examined quantity from the tower scalar \gls{wgc}, $-\grad\log m(\phi)$, implies the \gls{dc} \textit{if} this vector is aligned with the infinite distance limit, which is not necessarily true, as the vector (both characteristics, length and direction) depend on the position in moduli space. \citet{etheredge_dense_2023} proposes two new conjectures, which, if both true, imply the \gls{dc} and the tower scalar \gls{wgc}.

\citet{gendler_merging_2021} show how the (scalar) \gls{wgc} can complete the \gls{dc}, and how the \gls{dc} can promote the \gls{wgc} to the \gls{twgc}: On the one hand, the parameter $\mathfrak{a}$ in the \gls{dc} is an unspecified $\order{1}$-parameter, which can be bound by the \gls{wgc} to $\mathfrak{a}>1/\sqrt{6}$. On the other hand, the \gls{wgc} does not specify how many and which states become light in the small coupling limit, but the \gls{dc} suggests that it is an infinite tower of states that becomes light, i.e. the \gls{dc} promotes the \gls{wgc} to the \gls{twgc}. The contribution from massless scalar fields to the \gls{wgc} bound determines the mass dependence in the \gls{dc} \cite{gendler_merging_2021}.

The \gls{slwgc} can be seen as a combination of the \gls{dc} and the \gls{wgc}, as the first particle of the light tower of states that is predicted by the \gls{dc} is the particle predicted by the \gls{wgc}, i.e. the \enquote{initial} mass is given by the \gls{wgc} and this mass is then exponentially suppressed according to the \gls{dc} \cite{heidenreich_sharpening_2016,heidenreich_evidence_2017,klaewer_super-planckian_2017}.

The gauge coupling in the scalar \gls{rfc} in \cref{eq:fswgc} decreases exponentially with the geodesic field distance, which is the behaviour predicted by the \gls{dc} and establishes the same censorship of global symmetries, which would appear when the coupling vanished \cite{gendler_merging_2021}. It is important to remark that the \gls{dc} does not specify whether the particles are charged or not. \citet{gendler_merging_2021} present some arguments why the fusion of the \gls{wgc} and the \gls{dc} is useful anyway:
They believe that it is always possible to find a $p$-form gauge coupling that vanishes in every infinite field distance limit, as global symmetries have to be obstructed.
Since infinite field limits correspond to decompactifications or strings becoming tensionless, there is always a \gls{kk} photon or a 2-form gauge field that becomes weakly coupled \cite{gendler_merging_2021}.
The charged tower does not have to be the lightest tower. A heavier but charged tower can be used to yield an upper bound on the scalar field range \cite{gendler_merging_2021}.

Also \citet{heidenreich_repulsive_2019} present a link between the \gls{wgc} and the \gls{rfc}. Their argument goes as follows: If the \gls{rfc} is violated in four dimensions, a (multi-)particle (state) exists that is not self-repulsive, i.e. there are bound states with twice the charge but less than twice the mass, when both particle(s/states) are combined, such that the charge-to-mass ratio of the bound state is larger than the charge-to-mass ratio of its components. This process can now be iterated ad infinitum, which leads to an infinite tower of states with ever-increasing charge-to-mass ratio. If the \gls{dc} is true, then it supports the \gls{rfc}.

\subsubsection{The WGC and the TCC}\label{rel:WGC_TCC}
Starting from the strong scalar \gls{wgc} (\cref{eq:strongswgc}) \citet{cai_refined_2021} derive the \gls{tcc} up to some $\order{1}$ parameter.

\subsubsection{The WGC and the Finite Number of Massless Fields Conjecture}\label{rel:WGC_fnomfC}
\textit{Finiteness} is the governing principle in the work of \citet{hamada_finiteness_2022}: They observe that the \gls{wgc} limits the mass of charged states, the \gls{dc} limits the distance in field space for the validity of an \gls{eft} in the presence of gravity, and that the number of degrees of freedom (\cref{s:fnomf}) is limited in a valid theory of \gls{qg}. In this work, we also explored additional instances of finiteness: the \gls{tcc} limits the field excursion, and the tameness conjecture (\cref{sec:tame}) imposes finiteness constraints on the geometry of the \gls{eft}. In the presence of gravity, infinities get replaced by finite numbers\,\textemdash\,often by sending $M_\textnormal{P}$ to a finite value, instead of having it suppressing the gravitational coupling\,\textemdash, which gives rise to many Swampland conjectures\,\textemdash\,this raises the notion that the conjectures might be connected and that at a fundamental level, finiteness plays a crucial role in \gls{qg} \cite{hamada_finiteness_2022}.

\subsubsection{The  Finite Number of Massless Fields Conjecture and the TPC}\label{rel:TPC_finite}
The \gls{tpc} (\cref{s:tadpole}) implies that the vast majority of the $\num{e272000}$ possible compactifications in F-theory are in the swampland, as the D3-charge induced by the fluxes that are necessary to stabilise the moduli would be larger than the tadpole bound allows \cite{bena_tadpole_2021}. This agrees nicely with the finite number of massless fields conjecture (\cref{s:fnomf}), which predicts a large swampland and a small landscape. A finite number of fluxes with a tadpole of $\order{1}$ cannot stabilise a growing number of moduli \cite{bena_tadpole_2021}.

\subsubsection{The TPC and the AdSDC}\label{rel:TPC_AdSDC}
The \gls{adsdc} is supported by the \gls{tpc} \cite{bena_tadpole_2021}: A scale separated \gls{ads} vacuum requires a small vacuum expectation value of the flux superpotential, which in turn requires the D3-charge to be large, where the number of moduli is large as well \cite{denef_distributions_2004}.

\subsubsection{The TPC and the dSC}\label{rel:TPC_dSC}
The \gls{tpc} supports the \gls{dsc}: \gls{ds} vacua that are obtained by uplifting \gls{ads} using an antibrane in a long warped throat and a small supersymmetry breaking scale require a large tadpole \cite{bena_tadpole_2021}, which is incompatible with the bounds from the \gls{tpc}. \citet{junghans_topological_2022} finds that \gls{ds} vacua in a \gls{lvs} require a D3 tadpole in the range $\order{500}\sim\order{\num{e16}}$. The \gls{tpc} supports therefore the \gls{dsc} in so far that it puts hard to meet constraints on \gls{ds} vacua.

\subsubsection{The TPC and the Tamness Conjecture}\label{rel:TPC_tame}
\citet{grimm_moduli_2022} note that the \gls{tpc} is compatible with the tameness conjecture (\cref{sec:tame}).

\section{Conclusion}\label{s:summary}%

\begin{displayquote}
    All you really need to know for the moment is that the universe is a lot more complicated than you might think, even if you start from a position of thinking it’s pretty damn complicated in the first place \cite{adams_ultimate_1996}.
\end{displayquote}

\begin{acknowledgments}
We'd like to thank Professor Peter Coles for all the support we received during the creation of this work, in particular for the independence and freedom we were granted to explore the depths of theoretical physics and the vast body of work it has produced so far. We hope to contribute to its accessibility and digestibility.

Kay Lehnert is a recipient of the John and Pat Hume Scholarship and acknowledges support from the Friedrich Naumann Foundation for Freedom, the Bundesministerium für Forschung, Technologie und Raumfahrt, and from the Swiss Study Foundation.
Furthermore, Kay is a member of the COST actions
BridgeQG\footnote{Bridging high and low energies in search of quantum gravity, CA23130},
CosmoVerse\footnote{Addressing observational tensions in cosmology with systematics and fundamental physics, CA21136}, and 
THEORY-CHALLENGES\footnote{Fundamental challenges in theoretical physics, CA22113}.

The energy consumption of the author's devices used for the creation of this work is annually offset at \href{https://climeworks.com/checkout/referral/AqO8j10d}{climeworks}. Therefore, we consider this work as \textit{probably} carbon-neutral.

\end{acknowledgments}

\bibliography{references,manual}

\begin{appendices}\crefalias{section}{appendix}
\section{Symbols}
\begingroup
\allowdisplaybreaks
\begin{AmSalign*}
    A_\mu&&&\text{a gauge field}\\
    \mathcal{A}&&&\text{a manifold}\\
    a&&&\text{an integer}\\
    a(t)&&&\text{scale factor}\\
    \mathfrak{a}&\sim\order{1}&&\text{\gls{dc} constant}\\
    \alpha&&&\text{fine structure constant}\\
    B&&&\text{a field}\\
    \mathcal{B}&&&\text{a manifold}\\
    b&&&\text{a constant}\\
    \mathfrak{b}&>0&&\text{\gls{bhedc} constant}\\
    \beta&&&\text{a parameter}\\
    C&&&\text{a constant}\\
    c&&&\text{a constant}\\
    \mathfrak{c}&&&\text{a constant}\\
    \chi&&&\text{mixing parameter}\\
    D&&&\text{number of dimensions}\\
    D_t&&&\text{covariant derivative}\\
    d&&&\text{distance or}\\
    &&&\text{number of large dimensions}\\
    \mathfrak{d}&\sim\order{1}&&\text{\gls{adsdc} constant}\\
    E&&&\text{energy}\\
    e_a&&&\text{basis vector}\\
    \epsilon_V&\defeq\frac{(\nabla V)^2}{V^2}&&\text{slow-roll parameter}\\
    \varepsilon_0&=\frac{q_e^2}{2\alpha hc}&&\text{vacuum permittivity}\\
    \varepsilon&\ll1&&\text{a small number}\\
    \eta_V&\sim\frac{\nabla^2V}{V}&&\text{slow-roll parameter}\\
    \eta&&&\text{Minkowski metric}\\
    F_{\mu\nu}&=\frac{1}{2}\partial_{[\mu}A_{\nu]}&&\text{field strength tensor}\\
    \mathcal{F}&&&\text{a function}\\
    f&&&\text{axion decay constant}\\
    f_{AB}&&&\text{mixing tensor}\\
    \mathfrak{f}&&&\text{inverse axion decay constant}\\
    \phi&&&\text{scalar field}\\
    \varphi&&&\text{saxion}\\
    G&&&\text{gravitational constant}\\
    G_{\phi_i\phi_j}&&&\text{internal field space metric}\\
    \mathcal{G}&&&\text{space of all possible metrics}\\
    \mathfrak{G}&&&\text{Gauss\textendash Bonnet term}\\
    g&&&\text{gauge coupling or}\\
    &&&\text{metric norm / determinant}\\
    \mathfrak{g}&&&\text{gravitino \gls{dc} constant}\\
    \Gamma^i_{jk}&&&\text{Christoffel symbol}\\
    \Gamma&&&\text{lattice or decay rate}\\
    \gamma&&&\text{geodesic}\\
    H&=\frac{\mathrm{d}\log{a}}{\mathrm{d}t}&&\text{Hubble parameter}\\
    h_{\mu\nu}&&&\text{induced metric}\\
    h&&&\text{value of Higgs field or}\\
    &&&\text{complex structure moduli number}\\
    \hbar&&&\text{reduced Planck constant}\\
    \iota&&&\text{instanton}\\
    J&&&\text{spin}\\
    K&&&\text{degeneracy parameter}\\
    k&&&\text{a constant}\\
    \kappa&=\frac{\sqrt{2}}{M_{\textnormal{P};4}r_0}&&\text{\gls{kk} gauge coupling}\\
    L&&&\text{Lagrangian}\\
    \mathcal{L}&&&\text{Lie derivative}\\
    l&&&\text{length}\\
    \mathfrak{l}&\sim\order{1}&&\text{an $\order{1}$ constant}\\
    \Lambda_\textnormal{cc}&&&\text{cosmological constant}\\
    \Lambda&&&\text{cutoff}\\
    \lambda_\text{S}&&&\text{relative change species scale}\\
    \lambda&&&\text{a parameter}\\
    M_{\textnormal{P};d}^{d-2}&=\hbar^{d-3}c^{5-d}/8\pi G_\text{N}&&\text{$d$-dimensional Planck mass}\\
    M&&&\text{mass}\\
    \mathcal{M}&&&\text{moduli space}\\
    m&&&\text{mass}\\
    \mathfrak{m}&&&\text{mass parameter}\\
    \mu&&&\text{energy scale or}\\
    &&&\text{coupling / charge}\\
    N_e&&&\text{number of $e$-folds}\\
    N&&&\text{number of states}\\
    \mathcal{N}&&&\text{number of supersymmetries}\\
    \mathfrak{N}&&&\text{number of fields}\\
    n^a&&&\text{normal vector}\\
    n_s&&&\text{spectral tilt}\\
    n&&&\text{an integer}\\
    \Omega&&&\text{relative energy density}\\
    \omega&&&\text{\gls{eos}}\\
    P(k)&&&\text{powerspectrum}\\
    p&&&\text{pressure}\\
    \mathfrak{p}&\geq0&&\text{a positive constant}\\
    \Psi&&&\text{wave function}\\
    Q&&&\text{charge}\\
    q&&&\text{charge / deceleration parameter}\\
    \mathfrak{q}&&&\text{charge parameter}\\
    R&&&\text{Riemann/Ricci tensor/scalar}\\
    r_\textnormal{ts}&&&\text{tensor-to-scalar ratio}\\
    r&&&\text{radius}\\
    \rho&&&\text{density}\\
    S&&&\text{action}\\
    \mathcal{S}&&&\text{entropy}\\
    s&&&\text{flow parameter or spin}\\
    \mathfrak{s}_1&\sim\order{1}&&\text{\gls{dsc} constant}\\
    \mathfrak{s}_2&\sim\order{1}&&\text{refined \gls{dsc} constant}\\
    \Sigma&&&\text{Cauchy slice}\\
    \sigma&&&\text{a numerical factor}\\
    \varsigma&&&\text{a parameter}\\
    T^a&=\frac{\dot{\phi}^2}{\sqrt{G_{ab}\dot{\phi}^a\dot{\phi}^b}}&&\text{tangent Vector}\\
    T_{ab}&&&\text{stress\textendash energy tensor}\\
    T&&&\text{temperature}\\
    \mathcal{T}&=M_\textnormal{P}^2e^il^i&&\text{string tension}\\
    \mathfrak{T}&\defeq\abs{D_tT}&&\text{turning Rate}\\
    t&&&\text{time}\\
    \mathfrak{t}&&&\text{flux-tadpole constant}\\
    \tau&&&\text{conformal time}\\
    u&&&\text{Bogolyubov coefficient}\\
    \Upsilon&&&\text{dissipation rate}\\
    \upsilon&&&\text{\gls{ds} ratio}\\
    V&&&\text{potential (of a field)}\\
    \mathcal{V}&&&\text{volume}\\
    v^a&&&\text{a vector}\\
    W^{\mu\nu\rho\sigma}&&&\text{Weyl tensor}\\
    \mathcal{W}&&&\text{a manifold}\\
    w&&&\text{Bogolyubov coefficient}\\
    X^N&&&\text{bulk coordinates}\\
    X&&&\text{a particle}\\
    \Xi&&&\text{coupling strength}\\
    \xi&&&\text{vector field}\\
    Z&&&\text{partition function}\\
    \mathcal{Z}^a&=-\delta^{ab}e_b^i\partial_i\log\Lambda_\textnormal{S}&&\text{species vector}\\
    \Vec{z}&\defeq\Vec{q}\cdot\frac{M_\textnormal{P}}{m}&&\text{charge-to-mass ratio vector}\\
    \zeta^i&=-\partial^i\log m&&\text{scalar charge-to-mass vector}
\end{AmSalign*}
\endgroup
\label{a:sym}
\printnoidxglossaries
\end{appendices}

\end{document}